\documentclass[twoside]{ecsthesis}      % Use the Thesis Style
\usepackage{natbib}            % Use Natbib style for the refs.
\usepackage[pdfpagemode={UseOutlines}, bookmarks=true, bookmarksopen=true, bookmarksopenlevel=0, bookmarksnumbered=true, hypertexnames=false, colorlinks, linkcolor={blue}, citecolor={blue}, urlcolor={red}, pdfstartview={FitV}, unicode, breaklinks=true]{hyperref}
\usepackage{lmodern} %police vectorielles
\usepackage[T1]{fontenc}
 \rmfamily
 \DeclareFontShape{T1}{lmr}{b}{sc}{<->sub*cmr/bx/sc}{}
 \DeclareFontShape{T1}{lmr}{bx}{sc}{<->sub*cmr/bx/sc}{} %allow to write in bold small capital
\usepackage{microtype} %improves the spacing between words and letters
\usepackage[autolanguage]{numprint}
\usepackage{amsmath} %some mathematical symbols 
\allowdisplaybreaks %allow to cut equation in 2 pages
\usepackage{amssymb} %some mathematical symbols
\usepackage{mathrsfs} %some mathematical symbols
\usepackage{wasysym}
\usepackage{multicol}
\usepackage{verbatim}
\usepackage{booktabs} %allows to create tables without vertical separators

\usepackage{feynmp}
\usepackage{cancel} 
\usepackage{bold-extra}
\usepackage{comment} 
\usepackage{tabularx} % gestion avancée des tableaux
\usepackage{psfrag} % remplacement du texte d'une figure ps par du texte latex
\usepackage{psfrag} %pour images EPS
\usepackage{url}
 \usepackage{amsthm} % environement théorème
	\newtheorem{mydef}{Definition}

\usepackage{dcolumn}   % needed for some tables
\usepackage{bm}        % for math

\usepackage{epsfig}
\usepackage{pstricks}
\usepackage{multirow}
\usepackage{graphicx}
\usepackage{axodraw4j}
\usepackage{lipsum}
\usepackage{float}

\usepackage{notoccite}

\newcommand{\BibTeX}{{\rm B\kern-.05em{\sc i\kern-.025em b}\kern-.08em T\kern-.1667em\lower.7ex\hbox{E}\kern-.125emX}}

\newcounter{address}
\newcounter{alg}

\renewcommand{\d}{\mathrm{d}}

\renewcommand{\L}{\mathcal{L}}
\newcommand{\gmu}{\gamma^\mu}

\newcommand{\abs}[1]{\left\vert #1\right\vert}

\newcommand{\hc}{\mathrm{h.c}}

\newcommand{\Lag}{\mathcal L}

\def\slash#1{\rlap{\hbox{$\mskip 1 mu /$}}#1}

\def\ti              {\tilde}
\def\nt              {\ti\chi^0}
\def\st              {\ti t}

\def\PL              {P_L^{}}
\def\PR              {P_R^{}}

\def\ETmiss {E_T^{\rm miss}}
\newcommand{\MET}{E_T^{\rm miss}}

\newcommand{\sfrac}[2] {{\textstyle \frac{#1}{#2}}}

\renewcommand\contentsname{Table of Contents}

\begin{document}
\frontmatter
%\title      {Phenomenology of new fermions in cosmology and collider physics}
\title      {Phenomenology of extra quarks at the LHC}
%\authors    {\texorpdfstring
%             {\href{mailto:hp1y12@soton.ac.uk}{Hugo Prager}}
%             {Hugo Prager}
%            }
\authors    {Hugo Prager}
\department  {Theory group}
\group       {}
\addresses  {\groupname\\\deptname\\\univname}
\date       {\today}
\subject    {}
%\keywords   {Extra quarks, Extra lepton, Vector-like quarks, Dark Matter, Supersymmetry, model independent searches, recasting tools, relic density, interference, offshellness, large width}
\maketitle

\begin{abstract}

In this thesis, we study in a model independent way models of new Physics featuring extra quarks (XQs). These quarks are predicted by several extensions of the Standard Model (SM) but have never been observed yet even though many searches have been designed to find them at the Large Hadron Collider (LHC). 

After an introduction about the SM and the LHC, we present the main properties of these XQs and a model independent parametrisation that can be used to describe their phenomenology with generic hypotheses about their mixing with SM quarks, both in the case of XQ coupling with SM bosons and with Dark Matter (DM) candidates. %We also provide a theoretical description of the Narrow-Width Approximation (NWA) that the experimentalists use in their analysis and which allows to factorise the production and decay of XQs, as well as a with a short description of the most recent phenomenological analyses present in literature. 

In these two cases we study the offshellness effects in pair-production and decay and show that if the Narrow-Width Approximation (NWA), that we describe in detail, is a good approximation of the full result in the small width over mass ratio limit, sizeable differences occur when the XQ width becomes larger. The conclusion of our analysis is that even though the small width assumption 
%made for the experimental searches 
is always conservative it is not possible to trivially rescale the mass bounds obtained considering processes of pair production and decay in the NWA to determine constraints for XQs with large widths. 

We also study the role of interference in the process of pair production of new heavy XQs decaying to SM particles and show that in the NWA the interference contribution can be described by considering a parameter which contains only the relevant couplings and the scalar part of the propagators of the new quarks, both at the cross section level and at the distribution level. 

Finally, we study how various %analyses targeting either $t\bar t +\MET$ or multi-jet + $\MET$ signatures in the context of 
Supersymmetry (SUSY) searches perform for our simplified model with XQs decaying to DM. We show that cross section upper limit maps and efficiency maps obtained for stop simplified models in stop searches can also be applied to analogous XQ models, provided the NWA applies: the bound for XQs can therefore be obtained from the SUSY ones just by rescaling the exclusion with the XQ cross section. 

\end{abstract}

\tableofcontents
\listoffigures
\listoftables

%% -----------------------
%% lstpatch.sty
%% -----------------------
%% lstpatch cannot be distributed with these files. I believe it is only needed if the
%% \lstlistoflistings is used. So this has been turned off by default. Re-add if required:
%% \usepackage{lstpatch}
%% \lstlistoflistings
%% You will need to download lstpatch, possibly from:
%% http://web.mit.edu/texsrc/source/latex/listings/lstpatch.sty
%% -----------------------

%% -----------------------
%% Authorship declaration
%% -----------------------
%% Either include citations like below (as many as required spaced with commas or 'and').
\authorshipdeclaration{\citep{Barducci:2013zaa}, \citep{Kraml:2016eti}, \citep{Moretti:2016gkr}, and \citep{Moretti:2017qby}}
%% Or state no citations like below
%% \authorshipdeclaration{}
%% -----------------------

\acknowledgements{The three years I spent in Southampton were an amazingly enriching adventure, and a lot of people contributed to make this time so great.

My first thank goes to Stefano Moretti and Luca Panizzi 
%and Claire Shepherd-Themistocleous 
for the great supervision they gave me. They were always present when I needed them for any reason and provided me with very good advices and a lot of support during the three years I spent working with them in Southampton which made my work environment very pleasurable!

I also want to thank my collaborators Aldo Deandrea in Lyon, Daniele Barducci in Annecy and Trieste, Sabine Kraml and Ursula Laa in Grenoble. All of them were very good research partners with whom I enjoyed working a lot.

My colleagues also helped me a lot with my research and made my offices nice places to work: thank you Juri, Miguel, Maria, Jason and Marc in Southampton as well as Nicolas, Mickael, Jean-Baptiste and Solene in Lyon. 

I could not forget to mention here my housemate Anita, Juanito, Horacio, Katie, Mike, Marine with whom I spent so many good times during my PhD as well as all the members of the Skunks Ultimate team of Southampton Uni with whom I visited too many sport centres and scout huts in the UK without seeing the rest of the city and got injured too many times!

Finally, I think that a lot of my friends in France and in England also deserve an acknoledgement: thank you Scaro, Dr Taz, Speaker Louis, Pouloud and all your housemates for hosting me so many times in London; thank you Ann, M\'elie and Jeremy for an amazing trip in the New Forest; thank you Dad, Mum, Elsa, Gramouloud, Pipi and Lola for all the good time we had together in England; and thanks to all the friends who hosted me during my many stays in Lyon and thanks to all the others I am forgetting in France, in the UK and in the rest of the world!

And last but not least I want to give a massive thank to Dermot O'Brien who accompanied me almost everywhere during these three years. He has been simultaneously my colleague, my housemate, my festival companion, sometimes my teacher, sometimes my student and most importantly a great friend who will remain one of my best encounter in this country!}

\listofsymbols{ll}{
$M^T$ & The transpose of the matrix $M$ \\
$[A,B]$ & The commutator of the matrix $A$ and $B$ \\
$\{A,B\}$ & The anticommutator of the matrix $A$ and $B$ \\
h.c. & The hermitian conjugated \\
$\delta$ & The Dirac delta distribution \\
$\vec{\sigma}$ & The spin vector \\
$\theta_W$ & The Weinberg angle, $c_W = \cos \theta_W$, $s_W = \sin \theta_W$\\
$v$ & The Higgs Vacuum Expectation Value (VEV) \\
$\sigma_i$ & The Pauli Matrices \\%, 
%$\sigma_1 = \left(\begin{array}{cc} 
%0 & 1 \\
%1 & 0 
%\end{array} \right)$, 
%$\sigma_2 = \left(\begin{array}{cc} 
%0 & -i \\
%i & 0 
%\end{array} \right)$, 
%$\sigma_3 = \left(\begin{array}{cc} 
%1 & 0 \\
%0 & -1 
%\end{array} \right)$ \\$\gamma^\mu$ & The gamma matrices $\gamma^\mu = (\gamma^0, \gamma^1, \gamma^2, \gamma^3)$ \\
%$\gamma^0$ & $= \left(\begin{array}{cccc} 
%1 & 0 & 0 & 0 \\ 
%0 & 1 & 0 & 0 \\ 
%0 & 0 & -1 & 0 \\ 
%0 & 0 & 0 & -1 
%\end{array} \right)$ \\
%$\gamma^1 = \left(\begin{array}{cccc} 
%0 & 0 & 0 & 1 \\ 
%0 & 0 & 1 & 0 \\ 
%0 & -1 & 0 & 0 \\ 
%-1 & 0 & 0 & 0 
%\end{array} \right)$ \\
%$\gamma^2 = \left(\begin{array}{cccc} 
%0 & 0 & 0 & -i \\ 
%0 & 0 & i & 0 \\ 
%0 & i & 0 & 0 \\ 
%-i & 0 & 0 & 0 
%\end{array} \right)$ \\
%$\gamma^3 = \left(\begin{array}{cccc} 
%0 & 0 & -1 & 0 \\ 
%0 & 0 & 0 & 1 \\ 
%-1 & 0 & 0 & 0 \\ 
%0 & 1 & 0 & 0 
%\end{array} \right)$ \\
$\gamma^5$ & The fifth gamma matrix $\gamma^5 = i \gamma^0 \gamma^1 \gamma^2 \gamma^3$ \\%$= \left(\begin{array}{cccc} 
%0 & 0 & 1 & 0 \\ 
%0 & 0 & 0 & 1 \\ 
%1 & 0 & 0 & 0 \\ 
%0 & 1 & 0 & 0 
%\end{array} \right)$ \\
$P_{L/R}$ & The left/right-handed projection operator $P_{L/R} = \frac{1 \mp \gamma^5}{2}$ \\
$U(1)_Y$ and $e$ & The electromagnetic gauge group and its associated coupling \\
$SU(2)_L$ and $g$ & The weak gauge group and its associated coupling \\
$SU(3)_c$ and $g_S$ & The QCD gauge group and its associated strong coupling \\

%$\sigma^- = \frac{\sigma_1 - i \sigma_2}{2}$, 
%$\sigma^+ = \frac{\sigma_1 + i \sigma_2}{2}$
\ \\
LHC & Large Hadron Collider \\
CERN & Organisation Europ\'eenne pour la Recherche Nucl\'eaire \\
SM & Standard Model \\
BSM & Beyond the Standard Model \\
SUSY & Supersymmetry \\
MSSM & Minimal Supersymmetric Standard Model \\
DM & Dark Matter \\
XQ & eXtra Quark \\
ChQ & Chiral Quarks \\ 
VL & Vector-Like \\
VLQ & Vector-Like Quark \\
VEV & Vacuum Expectation Value \\
BR & Branching Ratio \\
NWA & Narrow-Width Approximation \\
FW & Finite Width \\
MC & Monte-Carlo \\
SR & Signal Region

}

\mainmatter

%% ------------- --------------------- --------------------- ---------------------- --------------------- ---------------------- ----------------------
%% Chapter I: Introduction
%% ------------- --------------------- --------------------- ---------------------- --------------------- ---------------------- ---------------------- 

\chapter{Introduction} \label{Chapter:Introduction}

%% ------------- --------------------- --------------------- ---------------------- --------------------- ---------------------- ----------------------

\section{The Standard Model of Particle Physics}

The Standard Model (SM) of Particle Physics is the theory describing three of the four known fundamental forces in the Universe (the electromagnetic, weak, and strong interactions) in term of gauge theories, as well as classifying all known elementary particles. It was developed in stages throughout the latter half of the 20th century with the current formulation being finalized in 1967 upon experimental confirmation of the existence of quarks. Since then, this theory has been more and more validated by experimental evidences such as the discovery of the bottom quark in 1977 \cite{Herb:1977ek}, the weak current mediated by the $W^\pm$ and $Z$ boson in 1983 \cite{Arnison:1983rp,Arnison:1983mk}, the top quark in 1995 \cite{Abe:1995hr}, the $\tau$ neutrino in 2000 \cite{Kodama:2000mp} until the latest discovery of the Higgs boson by the ATLAS \cite{Aad:2012tfa} and CMS \cite{Chatrchyan:2012xdj} collaborations announced at the \textit{Organisation Europ\'eenne pour la Recherche Nucl\'eaire} (CERN) on the 4th of July 2012. This discovery has established the existence of the last missing piece of the SM and has ended a nearly forty years search for this particle that was theorized in 1964 by Peter Higgs, Fran\c cois Englert and Robert Brout \cite{Higgs:1964ia,Englert:1964et,Higgs:1964pj}. The three scientists proposed a mechanism, now commonly called the \emph{Higgs mechanism}, through which the gauge bosons of the SM acquire mass. This idea was rewarded with the Nobel Prize for Physics on the 8th of October 2014, attributed to Peter Higgs and Fran\c cois Englert, Robert Brout having passed away in May 2011.

\begin{figure}[!ht]
\centering
\epsfig{file=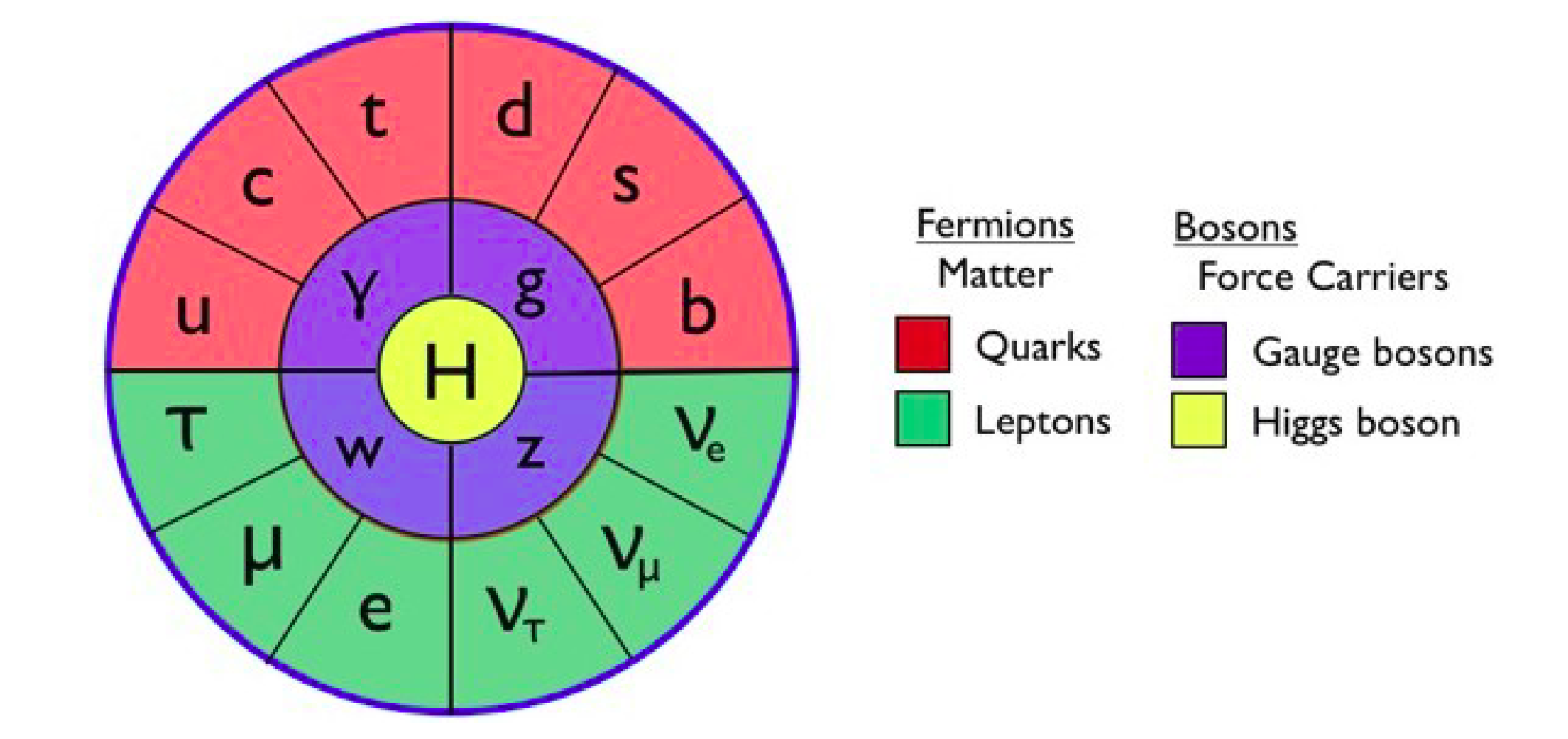, angle=-90,width=0.7\textwidth}
\caption[The particles of the Standard Model.]{The particles of the Standard Model. This picture based on a representation from the \textit{Particle Fever} movie \cite{ParticleFever} was borrowed from \url{http://theoryandpractice.org/2013/08/a-fresh-look-for-the-standard-model}.}
\label{fig:SM-particles}
\end{figure}

The SM is a gauge theory based on the group $SU(3) \times SU(2) \times U(1)$ which are responsible of the strong, weak and electromagnetic interaction. 
These interactions are mediated by four different types of vector gauge bosons: 
\begin{itemize}
\item the gluons which are the mediators of the strong interaction which binds the quarks together inside the hadrons,
\item the $W^\pm$ and $Z$ bosons which carry the weak interaction,
\item the photon which mediate the electromagnetic interaction.
\end{itemize}
On top of these gauge bosons, the SM also predict the existence of 12 fermions (and their 12 associated anti-fermions) that are the constituents of matters. Six of them are the quarks that compose the hadrons while the six others are the leptons which are split in tree charged leptons (electron, muon and tau) and three neutral neutrinos.
Finally, the SM predict the existence of a scalar Higgs boson which is needed to explain why the other elementary particles, except the photon and gluon, are massive. 
All these SM particles are represented on Fig. \ref{fig:SM-particles}. More details about the SM can be found in \cite{Halzen:1984mc,Cottingham:2007zz}.

\begin{figure}[!ht]
\centering
\epsfig{file=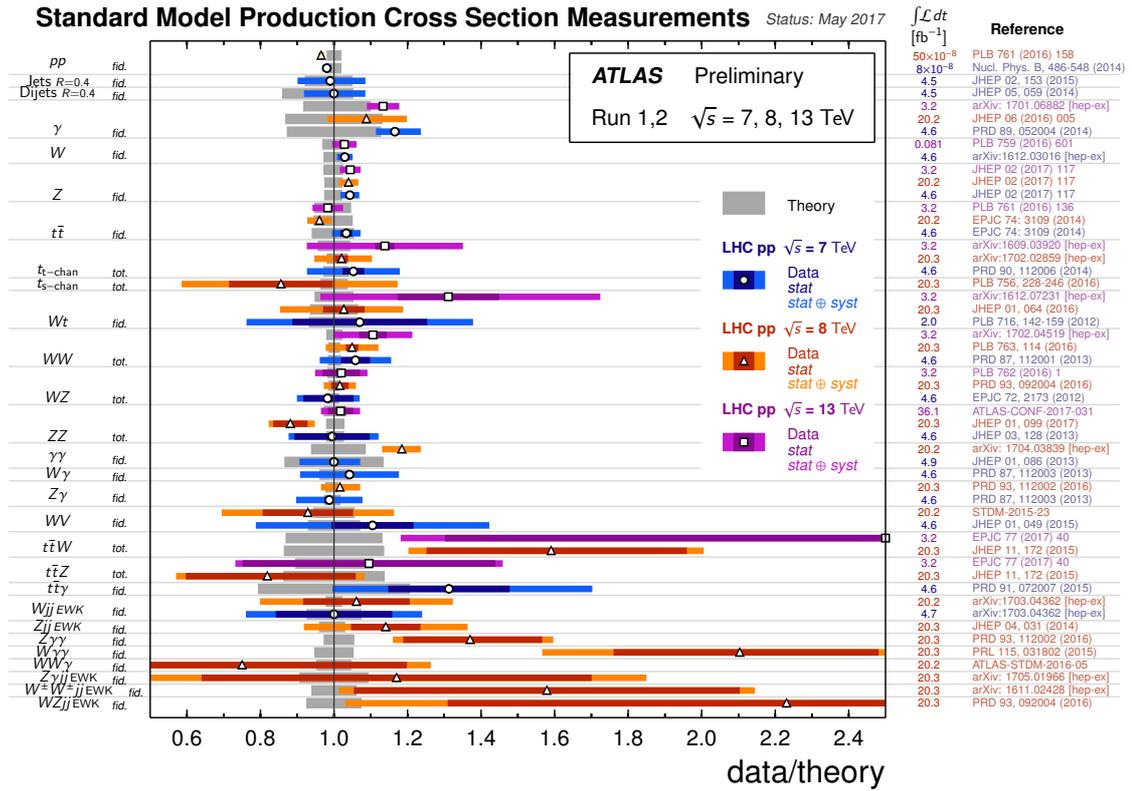, width=1.\textwidth}
\caption[The data/theory ratio for several Standard Model total and fiducial production cross section measurements, corrected for leptonic branching fractions.]{The data/theory ratio for several Standard Model total and fiducial production cross section measurements, corrected for leptonic branching fractions. The dark-colour error bar represents the statistical uncertainly. The lighter-colour error bar represents the full uncertainty, including systematics and luminosity uncertainties. Not all measurements are statistically significant yet.}
\label{fig:SM-test}
\end{figure}

The SM extraordinarily agrees with a large number of data collected so far by various collider experiments (such as LEP, LEP2, Tevatron and LHC) as shown in Fig. \ref{fig:SM-test} which was taken from the ATLAS twiki \cite{AtlasTwiki}.

However, despite all its experimental validations, there are theoretical and experimental indications that the SM cannot be the ultimate theory of Nature.

One of the main experimental observations that the SM fails to explain and that we will address in this thesis is the evidence of \emph{Dark Matter} (DM)\footnote{For more details and recent reviews about particle Dark Matter, see \cite{Bertone:2004pz}.}.

The existence of Dark (i.e., non-luminous and non-absorbing) Matter is by now well established. The earliest evidence for DM came from the observation that various luminous objects (stars, gas clouds, globular clusters, or entire galaxies) move faster than one would expect if they only felt the gravitational attraction of other visible objects. An important example is the measurement of galactic rotation curves. 
The rotational velocity of an object on a stable orbit with radius $r$ around a galaxy should be inversely proportional to $r$ when we lie outside the visible part of the galaxy and mass tracks light but in most galaxies one finds that this velocity becomes approximately constant out to the largest values of $r$ where the rotation curve can be measured. This implies the existence of a dark halo whose mass density increases with the radius. At some point this density will have to fall off faster (in order to keep the total mass of the galaxy finite), but we do not know at what radius this will happen. 
%The rotational velocity $v$ of an object on a stable Keplerian orbit with radius $r$ around a galaxy scales like $v(r) \propto \sqrt{M(r)/r}$, where $M(r)$ is the mass inside the orbit. If $r$ lies outside the visible part of the galaxy and mass tracks light, one would expect $v(r) \propto 1/ r$. Instead, in most galaxies one finds that $v$ becomes approximately constant out to the largest values of $r$ where the rotation curve can be measured. This implies the existence of a dark halo, with mass density $\rho (r) \propto 1/r^2$ , i.e., $M(r) \propto r$; at some point $\rho$ will have to fall off faster (in order to keep the total mass of the galaxy finite), but we do not know at what radius this will happen. 
These observations include measurements of the peculiar velocities of galaxies in the cluster, %which are a measure of their potential energy if the cluster is virialized;
measurements of the X-ray temperature of hot gas in the cluster, %which again correlates with the gravitational potential felt by the gas; 
and most directly studies of (weak) gravitational lensing of background galaxies on the cluster.
A particularly compelling example involves the bullet cluster (1E0657-558) which recently (on cosmological time scales) passed through another cluster. As a result, the hot gas forming most of the cluster's baryonic mass was shocked and decelerated, whereas the galaxies in the clusters proceeded on ballistic trajectories. Gravitational lensing shows that most of the total mass also moved ballistically, indicating that DM self-interactions are indeed weak.

Moreover, the existence of DM is only one of the experimental observations that the SM fails to explain, we can also mention the evidence of \emph{Dark Energy} \cite{Peebles:2002gy}, an unknown form of energy which is hypothesized to permeate all of space, tending to accelerate the expansion of the universe. Dark energy is the most accepted hypothesis to explain the observations since the 1990s indicating that the universe is expanding at an accelerating rate. Assuming that the standard model of cosmology is correct, the best current measurements indicate that Dark Energy contributes 68.3\% of the total energy in the present-day observable universe. The mass–energy of dark matter and ordinary (baryonic) matter contribute 26.8\% and 4.9\%, respectively, and other components such as neutrinos and photons contribute a very small amount. This means that the SM describe less than 5\% of the universe content.

The observation of \emph{neutrino oscillation} \cite{Fukuda:1998mi} by a multitude of experiments in several different contexts is also a phenomemon that the SM does not explain. Indeed the neutrinos are supposed to be massless in the SM but in order to explain these change of flavours at least two of them are required to be massive \cite{GonzalezGarcia:2007ib}. This is of such great theoretical and experimental interest that the experimental discovery of neutrino oscillation, and thus neutrino mass, by the Super-Kamiokande Observatory and the Sudbury Neutrino Observatories was recognized with the 2015 Nobel Prize for Physics.

The \emph{baryon asymmetry problem}, i.e. the imbalance in baryonic matter and antibaryonic matter observed in the Universe \cite{Sakharov:1967dj} also remain unexplained by the SM. 
%A natural assumption that the Universe be neutral with all conserved charges means that the Big Bang should have produced equal amounts of matter and antimatter. Since this does not seem to have been the case, it is likely some physical laws must have acted differently or did not exist for matter and antimatter.
Several competing hypotheses exist to explain the imbalance of matter and antimatter that resulted in baryogenesis but none of them have been confirmed.
CP violation is one of the needed ingredient for generating baryon asymmetry and its only source in the SM is a complex phase in the quark mixing matrix of the weak interaction which, given the limits on baryon number violation, is insufficient to account for the observed baryon asymmetry of the Universe\footnote{There may also be a non-zero CP-violating phase in the neutrino mixing matrix, but this is currently unmeasured.}.

Also, from the theoretical point of view, the non inclusion of a \emph{quantistic description of gravitation} seems the biggest limitation of the SM. Indeed difficulties arise when one attempts to quanticize gravity via graviton bosons: the theory one gets in this way is not renormalizable and therefore cannot be used to make meaningful physical predictions.

The \emph{hierarchy problem} is also unanswered since the Higgs mass cannot even be calculated in the strict context of the SM \cite{Seiberg:1993vc}. Even assuming new Physics at a larger scale, one would expect that the large quantum contributions to the square of the Higgs boson mass would inevitably make the mass huge, comparable to the scale at which new physics appears, unless there is an incredible fine-tuning cancellation between the quadratic radiative corrections and the bare mass.

Finally we can also mention the fact that according to QCD there could be a violation of CP symmetry in the strong interactions. However, no violation of the CP-symmetry is known to have occurred in experiments. As there is no known reason for it to be conserved in QCD specifically, this is another fine tuning problem known as the \emph{strong CP problem} \cite{MANNEL2007170}. \\

A lot of different Beyond the SM (BSM) models have been proposed to solve these problems, such as Supersymmetry (SUSY) and its various extensions \cite{Martin:1997ns,Giudice:1998bp,Cooper:1994eh,Sohnius:1985qm,Fayet:1976cr}, extra dimensions models \cite{ArkaniHamed:1998rs,ArkaniHamed:1998nn,Antoniadis:1998ig,ArkaniHamed:1998vp}, Composite Higgs Models \cite{Contino:2010rs,Agashe:2004rs,Mrazek:2011iu} and many others \cite{Lykken:2010mc}. These models usually features several new particles which could eventually be detected in particles collider such as the LHC.

%% ------------- --------------------- --------------------- ---------------------- --------------------- ---------------------- ----------------------

\section{The Large Hadron Collider}

The Large Hadron Collider (LHC) is the world's largest and most powerful particle collider, as well as the most complex experimental facility ever built, and the largest single machine in the world. It was built at CERN between 1998 and 2008 in the 27 km circumference LEP tunnel and started operations on the 10th of September 2008. Seven different experiments are currently present: ATLAS (A Toroidal LHC Apparatus), CMS (Compact Muon Solenoid), LHCb (LHC-beauty), ALICE (A Large Ion Collider Experiment), TOTEM (TOTal Elastic and diffractive cross section Measurement), LHCf (LHC-forward) and MoEDAL (Monopole and Exotics Detector At the LHC).
The centre of mass energy of the beam has been then gradually increased first to the energy of 7 TeV on the 30 of May 2010, then to the energy of 8 TeV on the 5th of April 2012 and more recently to the energy of 13 TeV on the 20th of May 2015, while the collected integrated luminosity has reached the value of $\simeq$ 40 fb$^{-1}$ in 2016 and is predicted to reach $\simeq$ 300 fb$^{-1}$ in 10 years of operations.
A planned upgrade of the CERN machine, the high luminosity LHC (HL-LHC), plans to bring the integrated luminosity up to the level of $\simeq$ 3 ab$^{-1}$.

\begin{figure}[!ht]
\centering
\epsfig{file=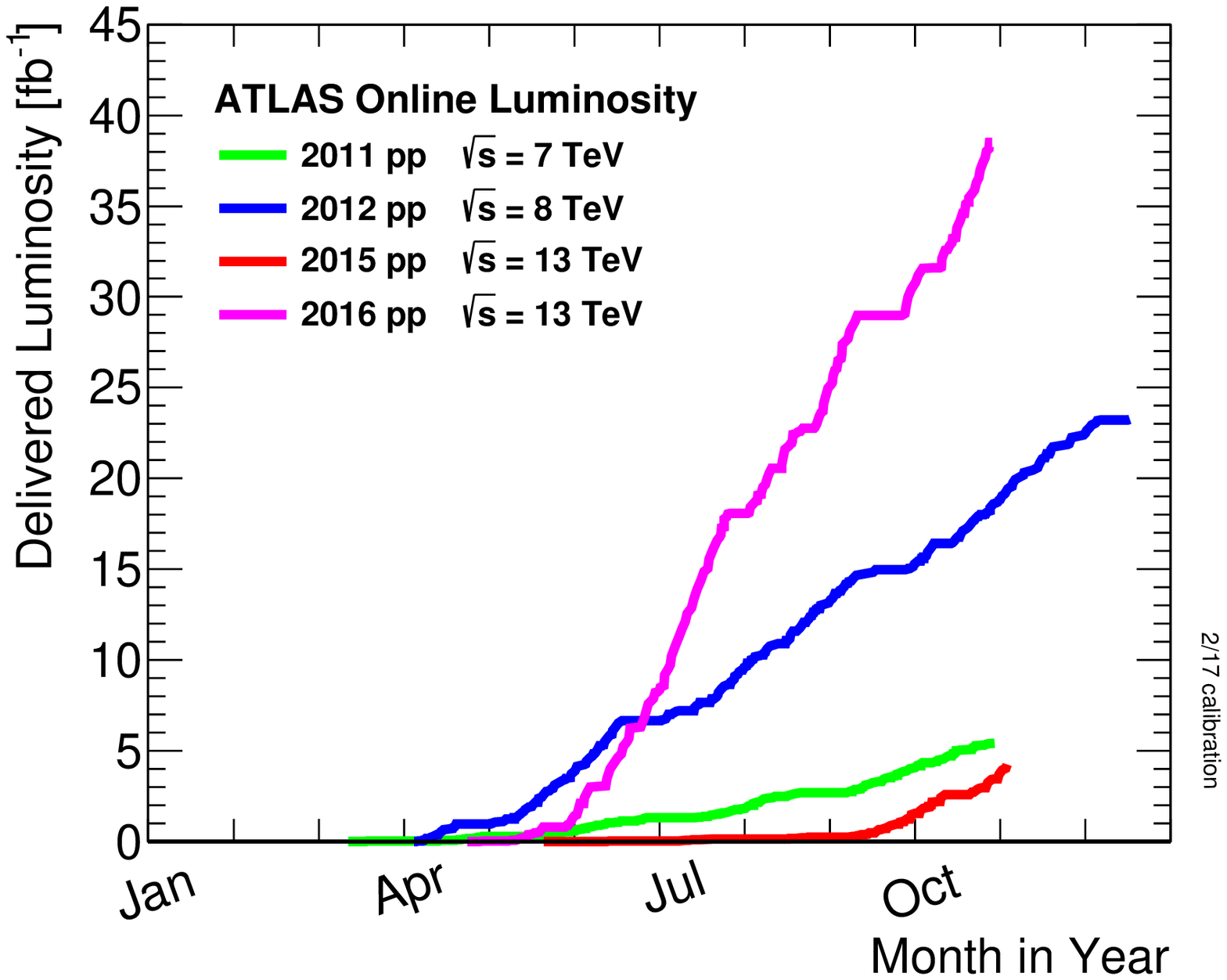, width=.48\textwidth}
\epsfig{file=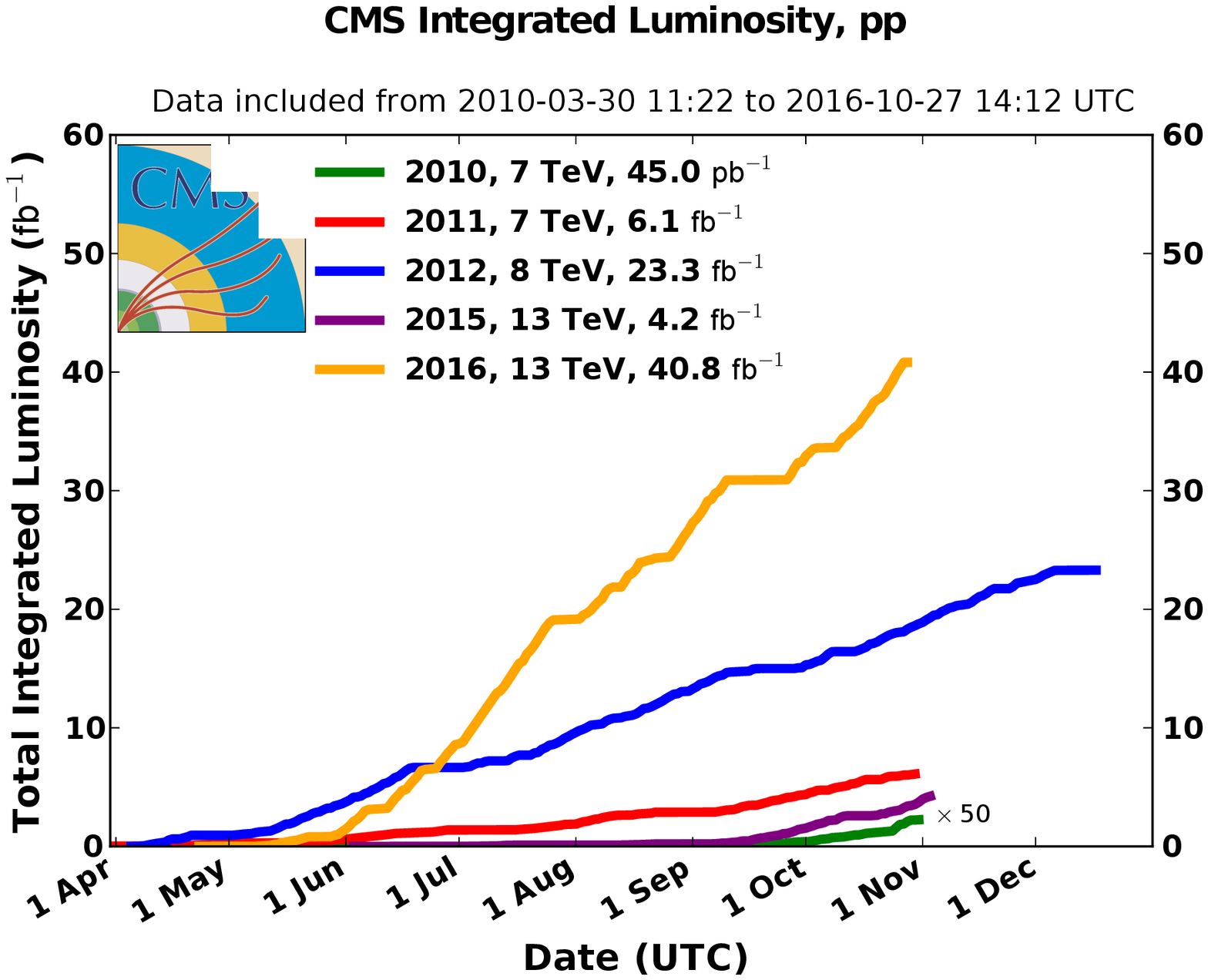, width=.48\textwidth}
\caption[ATLAS and CMS collected luminosity with the 7 TeV and 8 TeV runs of the LHC during the years 2010, 2011 and 2012.]{ATLAS and CMS collected luminosity in fb$^{-1}$ with the 7 TeV and 8 TeV runs of the LHC during the years 2010, 2011 and 2012. These plots were borrowed from the ATLAS and CMS twiki \cite{AtlasTwiki,CMSTwiki}.}
\label{fig:LHC-lumi}
\end{figure}

Besides the already mentioned discovery of the Higgs boson, the LHC has also discovered the bottomonium meson $\chi_b$(3P), multiple exotic hadrons, including pentaquarks or tetraquarks and made the first observation of the rare $B_s\to \mu^+\mu^-$ decay (8th November 2012). The LHC has also achieved important results in testing the SM and many BSM scenarios as was already shown in Fig. \ref{fig:SM-test}.

\begin{figure}[!ht]
\centering
\hspace*{-1.5cm}\epsfig{file=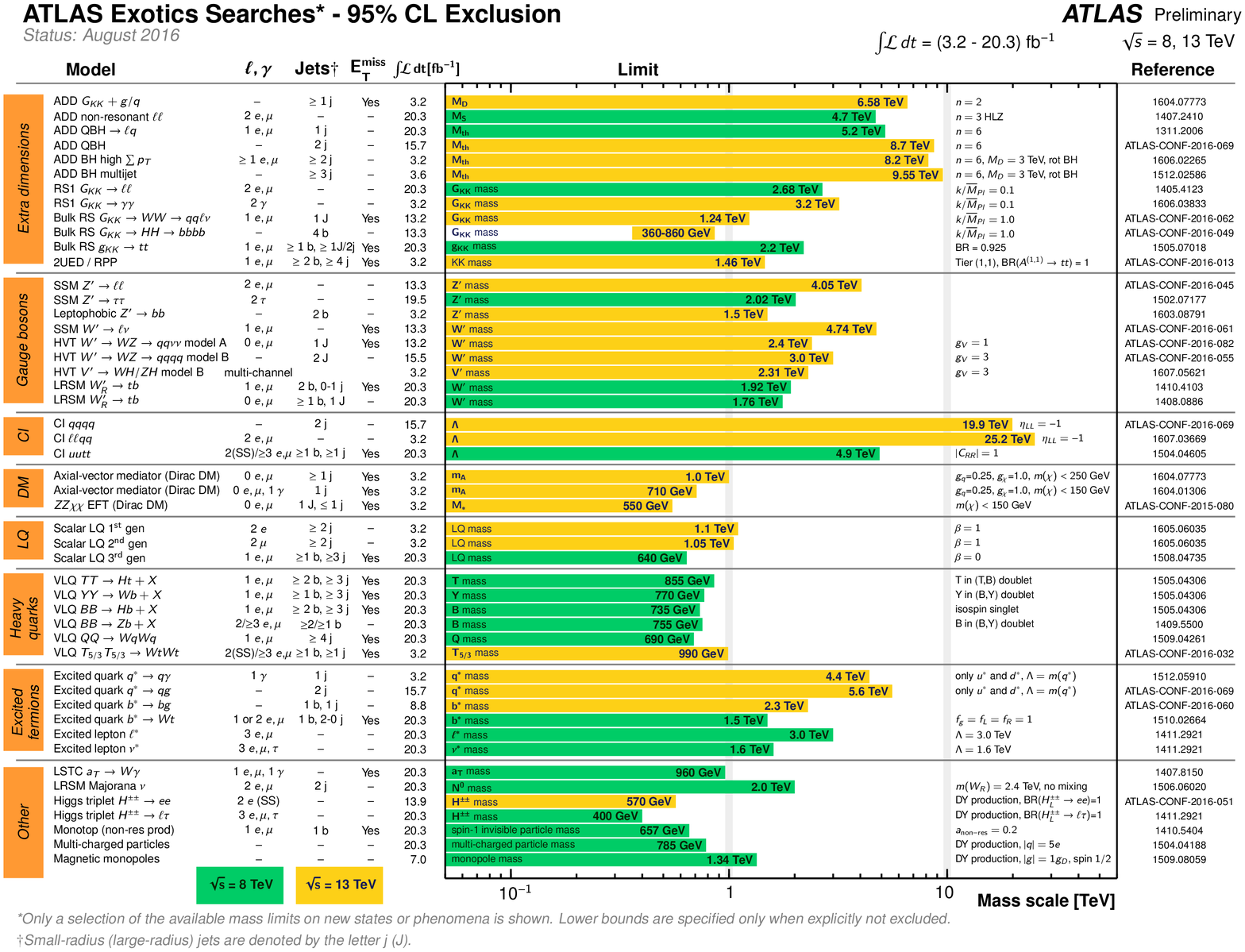, width=1.1\textwidth}
\vspace{-10mm} \caption[Representative selection of the reach of ATLAS searches for new phenomena other than SUSY.]{Representative selection of the reach of ATLAS searches for new phenomena other than SUSY. Yellow (green) bands indicate 13 TeV (8 TeV) data results.}
\label{fig:BSM-test}
\end{figure}

From the point of view of BSM theories the LHC has so far found no evidence of new particles belonging to any new physics theory and bounds on the masses of these new states are being set higher and higher, as shown for example in Fig.~\ref{fig:BSM-test} which was taken from the ALTAS twiki \cite{AtlasTwiki}. The table shows the reach of some representative ATLAS searches for new phenomena other than SUSY. We can already see that the bounds for heavy quarks are between 690 GeV and 990 GeV, and we will analyse these results in more details in Chapter \ref{Chapter:IntroXQ}.
Though it has to be stressed that these bounds strongly depend on the underlying model's assumptions, it is however clear that the first runs of the LHC has already ruled out a consequent part of the accessible parameter space of these BSM theories and the forthcoming run at 13 TeV and 14 TeV of centre of mass energy will allow to test a large part of the remaining accessible parameter space.

%The physics community must then make an important decision regarding a new generation of colliders, and at present both hadron colliders, like the already mentioned HL-LHC and a 100 TeV LHC, and electron-positron colliders, such as the Compact Linear Collider (CLIC), the International Linear Collider (ILC) and new circular electron positron collider (TLEP), are being considered.

In this thesis we will focus on a specific kind of new particles predicted by several BSM models called extra quarks (XQs), and we will study them in a model independent way, i.e. without assuming their specific properties by considering their different characteristics as free parameters.

%Another possibility is to assume that the new quarks are charged under some parity, so that they can only decay into an ordinary quark via a new boson which can eventually play the role of Dark Matter~\cite{Meade:2006dw,Alwall:2010jc}: this happens in extra dimensional models with a K-parity~\cite{Appelquist:2000nn,Servant:2002aq} or in Little Higgs models with T-parity~\cite{Low:2004xc,Matsumoto:2008fq}. The Dark Matter candidate can be either a spin one (like in 5D models, or in Little Higgs ones) or a spin zero particle (like in 6D models~\cite{Cacciapaglia:2009pa}) state.

%% ------------- --------------------- --------------------- ---------------------- --------------------- ---------------------- ----------------------

\section{Plan of the Thesis}
The plan of the Thesis is the following.

In the first chapter we will describe the XQs we are interested in, present their general properties and see how we can study them in a model independent way. We will then consider two different cases: in the second chapter we will focus on XQs decaying to SM particles (visible decay), while in the third chapter we will study XQs decaying to DM (invisible decay). In both cases we will evaluate the effects of large width in the determination of the cross section and in the reinterpretation of bounds from experimental searches. In the case of visible decay we will also 
%describe the recasting code XQCAT and 
study the interference effects taking place in a model containing several XQs, while in the case of invisible decay we will compare our XQ model to a SUSY model leading to the same final state $t \bar{t} + \MET$. Finally, we will conclude in the fifth chapter.

%% ------------- --------------------- --------------------- ---------------------- --------------------- ---------------------- ----------------------
%% Chapter II
%% ------------- --------------------- --------------------- ---------------------- --------------------- ---------------------- ---------------------- 

\chapter{Introduction to extra quarks} \label{Chapter:IntroXQ}

\section{The Standard Model quarks}
\label{sec:SMflavour}

In this section we present in more detail the quark sector of the SM following the approach of \cite{Kuhr:2013hd}.

\subsection{SM quarks and interactions}

In particle Physics the term \emph{flavour} was introduced by Gell-Mann in 1953 \cite{GellMann:1953zza} and is used to describe several copies of the same gauge representation, namely several fields that are assigned the same quantum charges. Within the SM, when thinking of its unbroken $SU(3)_c \times U(1)_Y$ gauge group, there are four different types of particles, each coming in three flavours:
\begin{itemize}
\item up-type quarks in the (3,+2/3) representation: $u$, $c$, $t$;
\item down-type quarks in the (3,1/3) representation: $d$, $s$, $b$;
\item charged leptons in the (1,1) representation: $e$, $\mu$, $\tau$;
\item neutrinos in the (1,0) representation: $\nu_1$, $\nu_2$, $\nu_3$.
\end{itemize}

Here we are only interested in the quark sector, the three generations of quark flavour pairs can be written as:
\begin{equation}
\left(\begin{array}{c} u \\ d \end{array}\right) \quad \left(\begin{array}{c} c \\ s \end{array}\right) \quad \left(\begin{array}{c} t \\ b \end{array}\right).
\label{eq:SMquarks}
\end{equation}

Each pair consists of an up-type quark with electric charge $+2/3$ and a down-type quark with charge $-1/3$. The generations are distinguished by the different masses, increasing from the first to the third.

These quarks are charged under $SU(3)_c$, $SU(2)_L$ and $U(1)_Y$ which means that they can interact strongly, weakly and electromagnetically. Moreover, they can interact with the Higgs boson $H$ because of how they acquire their masses. Indeed, in order to write a mass term for a quark $q$ we need to use a Yukawa term in the Lagrangian that will become after the electroweak symmetry breaking
\begin{equation}
\L_{\text{Yuk}}^q = - \frac{y_q}{\sqrt{2}} \bar q q H - m_q \bar q q
\end{equation}
where $y_q$ is the Yukawa coupling of this quark with the Higgs boson and $m_q = y_q v / \sqrt{2}$ is its mass, with $v$ the Vacuum Expectation Value (VEV) of the Higgs.

\subsection{Quark flavour Physics}

In the SM the flavour quantum number is conserved in strong and electromagnetic interactions. It can only be changed by charged current weak processes, described by the exchange of a $W^\pm$ boson. The neutral current weak interaction ($Z$ boson exchange) is flavour-conserving.

This is a direct consequence of how the electroweak Lagrangian is built combined to the effect of the quarks mixing.

Indeed the quark mass terms in Lagrangian (after spontaneous symmetry breaking) takes the following form using the weak eigenstates
\begin{equation}
\L_Y^q = - \frac{v}{\sqrt{2}} \sum_{j,k} ( \bar d_L^j y_d^{jk} d_R^k + \bar u_L^j y_u^{jk} u_R^k + \hc)
\end{equation}
where the subscript $L/R$ denotes the left/right-handed component of the quark fields which will be defined more precisely in Sec. \ref{sec:chirality}. By defining the mass matrix $\tilde{M}_{u,d} = v \; y_{u,d} / \sqrt{2}$ we can rewrite this Lagrangian
\begin{equation}
\L_Y^q = - \sum_{j,k} ( \bar d_L^j M_d^{jk} d_R^k + \bar u_L^j M_u^{jk} u_R^k + \hc)
\end{equation}
Note that the Yukawa matrices and thus the mass matrices are in general not diagonal in the "interaction space", and especially for the SM they are not.
To obtain the mass eigenstates we diagonalize them using unitary transformations  
\begin{equation}
\tilde{q}_A = V_{A,q} q_A
\end{equation}
with $q \in \{u, d\}$ and $A \in \{R , L\}$, we define a set of 4 unitary matrices such that $V_{A,q} V_{A,q}^\dag = 1$. These matrices are determined by 
\begin{eqnarray}
M_u & = \left(\begin{array}{ccc}
m_u & 0 & 0 \\
0 & m_c & 0 \\
0 & 0 & m_t \\
\end{array}\right) = \frac{v}{\sqrt{2}} V_{L,u} \; y_u \; V_{R,u}^\dag \\
M_d & = \left(\begin{array}{ccc}
m_d & 0 & 0 \\
0 & m_s & 0 \\
0 & 0 & m_b \\
\end{array}\right) = \frac{v}{\sqrt{2}} V_{L,d} \; y_d \; V_{R,d}^\dag
\end{eqnarray}
with the usual Dirac masses $m_q$ giving us the final mass Lagrangian in the "mass space"
\begin{equation}
\L_Y^q = - \bar{\tilde{d}}_L M_d \tilde d_R - \bar{\tilde u}_L M_u u_R + \hc
\end{equation}

If up-type and down-type Yukawa matrices cannot be diagonalised simultaneously, there is an net effect of the basis change on the charged current interaction (which connects up- and down-type). The charged-current interaction gets a flavour structure which is encoded in the Cabibbo-Kobayashi-Maskawa matrix (CKM)
\begin{equation}
V_{CKM} = V_{L,u} V_{L,d}^\dag .
\end{equation} 
The Lagragian for the charged-current interaction can be written as
\begin{eqnarray}
\L_{CC} &\propto & \bar u_L \gamma^\mu W^+_\mu d_L + \bar d_L \gamma^\mu W^-_\mu u_L \\
&\propto & \bar{\tilde u}_L V_{L,u} \gamma^\mu W^+_\mu V_{L,d}^\dag \tilde d_L + \bar{\tilde d}_L V_{L,d}^\dag \gamma^\mu W^-_\mu V_{L,u} \tilde u_L \\
&\propto & \bar{\tilde u}_L \gamma^\mu W^+_\mu V_{CKM} \tilde d_L + \bar{\tilde d}_L \gamma^\mu W^-_\mu V_{CKM}^\dag \tilde u_L
\end{eqnarray}
where $W^\pm_\mu$ is the gauge field related to the $W^\pm$ boson. The element $(V_{CKM})_{ij}$ connects the left-handed $u$-type quark of the $i$th generation with the left-handed $d$-type quark of the $j$th generation. We label the matrix element according to quark flavour instead to the generation index.

On the other hand looking at neutral-current interaction, the Lagrangian has the following form
%\begin{eqnarray}
%\L_{NC} &\propto & \bar u_L \gamma^\mu Z_\mu (\alpha - \gamma^5) u_L + \bar d_L \gamma^\mu (\alpha + \gamma^5) Z_\mu d_L \\
%&\propto & \bar{\tilde u}_L V_{L,u} \gamma^\mu (\alpha - \gamma^5) Z_\mu V_{L,u}^\dag \tilde u_L + \bar{\tilde d}_L V_{L,d}^\dag \gamma^\mu (\alpha + \gamma^5) Z_\mu V_{L,d} \tilde d_L \\
%&\propto & \bar{\tilde u}_L \gamma^\mu (\alpha - \gamma^5) Z_\mu \tilde u_L + \bar{\tilde d}_L \gamma^\mu (\alpha + \gamma^5) Z_\mu \tilde d_L
%\end{eqnarray}
%where $Z_\mu$ is the gauge field related to the $Z$ boson and $\alpha$ depend on the quarks involved. Here we have used the unitarity relation $V_{A,q} V_{A,q}^\dag = 1$ for the last step of the calculation.
\begin{eqnarray}
\L_{NC} &\propto & \bar u_L \gamma^\mu Z_\mu u_L + \bar d_L \gamma^\mu Z_\mu d_L \\
&\propto & \bar{\tilde u}_L V_{L,u} \gamma^\mu Z_\mu V_{L,u}^\dag \tilde u_L + \bar{\tilde d}_L V_{L,d}^\dag \gamma^\mu Z_\mu V_{L,d} \tilde d_L \\
&\propto & \bar{\tilde u}_L \gamma^\mu Z_\mu \tilde u_L + \bar{\tilde d}_L \gamma^\mu Z_\mu \tilde d_L
\end{eqnarray}
where $Z_\mu$ is the gauge field related to the $Z$ boson and where we have used the unitarity relation $V_{A,q} V_{A,q}^\dag = 1$ for the last step of the calculation.

This shows why flavour changing neutral currents (FCNC) do not occur in the SM at tree level while flavour changing charged currents are allowed. 

This makes FCNC processes a good candidate to search for deviations from the SM because new particles or new interactions may introduce flavour changing tree level amplitudes, that are of comparable size or larger than the amplitude of the higher order SM loop processes.

The existence of new XQs besides the six SM ones is among the open problems of particle physics to which the LHC may soon provide an answer. Searches for new quarks are actively undertaken by both ATLAS and CMS experiments, though no signals have been found so far.

%% ------------- --------------------- --------------------- ---------------------- --------------------- ---------------------- ----------------------

\section{Chirality, chiral and vector-like quarks} \label{sec:chirality}

\begin{mydef}
Every particle represented by a spinor $\psi$ has two different components of chirality, a right-handed one and a left-handed one\footnote{The neutrinos seems to be an exception to this rule since only left-handed neutrinos and right-handed antineutrinos have been observed so far.} 
\begin{equation}
\psi = \psi_R + \psi_L \\
= P_R \psi + P_L \psi
\end{equation}
where $P_R = \frac{1 + \gamma^5}{2}$ and $P_L = \frac{1 - \gamma^5}{2}$ are respectively the right-handed and left-handed projection operators. In the massless limit, those two components have respectively an helicity $\lambda_R = +1$ and $\lambda_L = -1$ where $\lambda = \vec{\sigma} \cdot \hat{p}$ represents the projection of the spin on the direction of propagation. 
\end{mydef}

\subparagraph*{Remark}

Using the fact that $P_L$ and $P_R$ are orthogonal projectors, that $\bar{\psi} P_{L/R} = \bar{\psi}_{R/L}$ and that $\{ \gamma^\mu , \gamma^5 \} = 0$ we show that $\bar{\psi} \gmu \psi = (\bar{\psi}_L + \bar{\psi}_R) \gmu (\psi_L + \psi_R) = \bar{\psi}_L \gmu \psi_L + \bar{\psi}_R \gmu \psi_R$ and $\bar{\psi} \psi = \bar{\psi}_L \psi_R + \bar{\psi}_R \psi_L$.

For quarks, we note $q = q_R + q_L$. In the SM $q_L$ has to belong to a doublet of $SU(2)_L$ while the $q_R$ has to belong to a singlet of the same group to explain the experimentally observed maximal violation of the parity for the weak interaction, meaning that in the charged current Lagrangian $\L_W = \frac{g}{\sqrt{2}} \left( J^{\mu +} W_\mu^+ + J^{\mu -} W_\mu^- \right)$ where $J^{\mu +} = J^{\mu +}_L + J^{\mu +}_R$, we only have left-handed charged currents:
\begin{equation}
\left\{ \begin{array}{l}
J^{\mu +}_L = \bar{u}_L \gamma^\mu d_L = \bar{u} \gamma^\mu (1 - \gamma^5) d \rightarrow V - A \\
J^{\mu +}_R = 0 \end{array} \right.
\end{equation}

We say that these weak currents have a \emph{vector-axial} ($V-A$) structure and we call such quarks \emph{chiral}. A fourth generation of SM-like chiral quarks (ChQs) has been excluded by the Higgs discovery because such heavy fermion are expected to contribute significantly to the properties of the Higgs boson, leading to measurable deviations in Higgs production cross sections and Branching Ratios (BRs) that are in contradiction with the SM nature of the scalar boson observed at the LHC\footnote{This constraint can be relaxed if the Higgs sector is extended, for this reason we will consider the possibility of new ChQs in some specific cases in the following.}. Yet this constraint does not apply to ChQs decaying to DM so we will sometimes consider them while studying such scenarios.

Another type of quarks called vector-like (VL) can also be defined:

\begin{mydef}
A \emph{vector-like quark} (VLQ) is a quark whose left- and right-handed chiralities belong to the same representation of the symmetry group $G$ of the underlying theory. For the SM, $G = SU(3)_C \otimes SU(2)_L \otimes U(1)_Y$.
\end{mydef}

\subparagraph*{Remark}

Note that the SM quarks are VL under $SU(3)_C$ and $U(1)_Y$, but not under $SU(2)_L \otimes U(1)_Y$.

These VLQs takes their name from the fact that their electroweak coupling structure allows both left- and right-handed charged currents: the structure of these current is therefore \emph{vector}.
\begin{equation}
J^{\mu +} = J^{\mu +}_L + J^{\mu +}_R = \bar{u}_L \gamma^\mu d_L + \bar{u}_R \gamma^\mu d_R = \bar{u} \gamma^\mu d = V
\end{equation}

\paragraph*{Models predicting VLQs}

These kind of quarks have actually never been observed, but are predicted by many models beyond the SM. Note that even if the structure of the coupling is different between VL and ChQs, their phenomenology is similar so we can study them the same way. 
From a theoretical point of view, VLQs have been introduced in many models; the most studied scenarios which predict the presence of VLQs can be divided into broad categories\footnote{A description of the various models as well as their consistency against the observations of the 125 GeV Higgs-like resonance is beyond the scopes of this thesis; details can be found in the original works and references therein. Here, it is sufficient to note how the emergence of VLQs is a recurrent consequence in many models of BSM physics.}:
\begin{itemize}
\item{\it Composite Higgs Models:} the electroweak symmetry breaking is driven by a condensate of the top quark and a VL singlet involving a see-saw mechanism between the two states \cite{Dobrescu:1997nm,Chivukula:1998wd,Collins:1999rz,He:2001fz,Hill:2002ap,Contino:2006qr,Anastasiou:2009rv,Kong:2011aa,Carmona:2012jk,Gillioz:2012se,Cacciapaglia:2017gzh,Vignaroli:2016opj,Cacciapaglia:2015vrx,Nevzorov:2015sha};
\item{\it Extra Dimensions:} excited partners of SM quarks belonging to heavier tiers of universal extra-dimensional scenarios are VL \cite{Antoniadis:2001cv,Hosotani:2004wv,Agashe:2004rs,Couture:2017mbd};
\item{\it Gauging of the flavour group:} VL fermions are required for anomaly cancellation and can play a role in the mechanisms of quark mass generation \cite{Davidson:1987tr,Babu:1989rb,Grinstein:2010ve,Guadagnoli:2011id,Bobeth:2016llm};
\item{\it Little Higgs Models:} VL states appear as partners of SM fermions in larger representations of the symmetry group \cite{ArkaniHamed:2002qy,Han:2003wu,Perelstein:2003wd,Schmaltz:2005ky,Carena:2006jx,Matsumoto:2008fq,Graham:2009gy}.
\item{\it Supersymmetric non-minimal extensions of the SM:} VL matter can be introduced in non minimal supersymmetric models to increase corrections the Higgs mass without affecting too much electroweak precision observables \cite{Moroi:1991mg,Moroi:1992zk,Babu:2008ge,Martin:2009bg,Graham:2009gy,Martin:2010dc,Hirsch:2015fvq}, and it appears also in non-minimal, GUT-inspired, supersymmetric scenarios \cite{Kang:2007ib}.
\end{itemize}
VLQs can also appear in models which try to explain measured asymmetries in different processes.
\begin{itemize}
\item in \cite{Choudhury:2001hs,Kumar:2010vx} VLQs are introduced to explain the observed $A_{FB}^b$ asymmetry: bottom partners can mix with the bottom quark and induce modifications of its coupling with the Z boson. 
\item the forward-backward asymmetry $A_{FB}$ in top pair production, measured at Tevatron, can be explained with the existence of a colour octet with a large decay width; this condition can be obtained if the colour octet is allowed to decay to a heavy VL state and a SM fermion \cite{Barcelo:2011vk,Barcelo:2011wu}. 
\end{itemize}

Here we will only work with simplified models featuring XQs which can be used to build more complex scenarios. These XQ models can be split in two different types:
\begin{itemize}
\item the ones where XQs decay into visible particles (SM quarks and boson), in which case the possible decays for an XQ $T$ with charge $q_T = +2/3$ are $T \rightarrow W^+ \, d_i, Z \, u_i$ or $H \, u_i$, where $u_i \in \{ u, c, t \}$ and $d_i \in \{ d, s, b \}$,
\item the ones where XQs decay into invisible particles (DM) and a SM quark, in which case the only possible decay for a $T$ is $T \rightarrow {\rm DM} \, u_i$, where ``DM'' is the Dark Matter {\em candidate} (which can be scalar or vector), i.e.\ a neutral massive particle that escapes detection as $\MET$ but whose astrophysical properties remain open. In these models, we impose a specific $\mathcal{Z}_2$ symmetry to the Lagrangian under which the new particles are odd while all the SM ones are even in order to make the DM particle stable\footnote{This is similar to the $R$ parity in SUSY.}.
\end{itemize} %\vspace{\baselineskip}

%% ------------- --------------------- --------------------- ---------------------- --------------------- ---------------------- ----------------------

\section{Model framework}
\label{sec:model}

\subsection{Representations of XQs} \label{sec:representation}

\subsubsection{Interaction terms} \label{sec:interaction terms}

We want to add a new quark to the SM in a model independent way, i.e. we are not interested which theory predicts the new state, we only want to understand the observable consequence of this SM extension. We then consider all the SM particles and an XQ, the chiralities of which eventually belong to different representations of the SM gauge group:
\begin{equation}
(SU(3)_C, SU(2)_L, U(1)_Y) \rightarrow  \left\{ \begin{array}{l}
\psi_L \simeq (3, S_L , Y_L ) = (\boldsymbol{S_L} , Y_L ) \\
\psi_R \simeq (3, S_R , Y_R ) = (\boldsymbol{S_R} , Y_R ) \end{array} \right.
\end{equation}
For the SM ChQs we have 
$\left\{ \begin{array}{l}
\text{left doublet:  } S_L = 2, \ Y_L = 1/6 \\
\text{right singlet: } S_R = 1, \ Y_R^u = 2/3, \ Y_R^d = -1/3 \end{array} \right.$

while for VLQs we have $S_L = S_R = S$ and $Y_L = Y_R = Y$. 

Without requiring interactions with SM quarks, it is possible to add every combination of representations. Moreover, for representations bigger than singlets, weak currents are non-trivial, since the representation product gives $m \otimes m = \otimes_{i=0}^{m-1}(2i + 1)$ which always contains a triplet and can be combined with the $SU(2)_L$ gauge bosons $W_i$ in the kinetic term.

We consider a new quark $\psi$ interacting with a SM quark $q$ and a neutral scalar or vector boson\footnote{This is needed to conserve the spin.} belonging to a singlet or a doublet of $SU(2)_L$ (this choice will be justified later). We call the boson $\beta^S_i$ when scalar and $\beta^V_i = \beta^V_{i \mu} \gamma^\mu$ when vector, with $i \in \{ 1,2 \}$ its dimension of representation under $SU(2)_L$. In the case of XQ decaying to SM particle this boson will be the Higgs doublet while in the case of XQ decaying to DM it will be the DM candidate. The interaction term will therefore be $\bar\psi \beta q$ and we see that only a limited subset of representations is allowed. Note that for a scalar boson the coupling links quarks of opposite chirality while for a vector boson it links quarks of the same chirality. We give here the details of the calculation for a scalar boson, for a vector boson the chiralities of the XQ would be inverted ($L \leftrightarrow R$).
\begin{eqnarray}
\psi_R \otimes \beta^S_1 \otimes q_L \rightarrow & (\boldsymbol{1},0) \otimes (\boldsymbol{2},\frac{1}{6}) = (\boldsymbol{2},\frac{1}{6}) & = S_R \\
\psi_R \otimes \beta^S_2 \otimes q_L \rightarrow & (\boldsymbol{2},\pm \frac{1}{2}) \otimes (\boldsymbol{2},\frac{1}{6}) = (\boldsymbol{1},\frac{1}{6} \pm \frac{1}{2}) \oplus (\boldsymbol{3},\frac{1}{6} \pm \frac{1}{2}) & = S_R \\
\psi_L \otimes \beta^S_1 \otimes u_R \rightarrow & (\boldsymbol{1},0) \otimes (\boldsymbol{1},\frac{2}{3}) = (\boldsymbol{1},\frac{2}{3}) & = S_L \\
\psi_L \otimes \beta^S_1 \otimes d_R \rightarrow & (\boldsymbol{1},0) \otimes (\boldsymbol{1},-\frac{1}{3}) = (\boldsymbol{1},-\frac{1}{3}) & = S_L \\
\psi_L \otimes \beta^S_2 \otimes u_R \rightarrow & (\boldsymbol{2},\pm \frac{1}{2}) \otimes (\boldsymbol{1},\frac{2}{3}) = (\boldsymbol{2},\frac{2}{3} \pm \frac{1}{2}) & = S_L \\
\psi_L \otimes \beta^S_2 \otimes d_R \rightarrow & (\boldsymbol{2},\pm \frac{1}{2}) \otimes (\boldsymbol{1},-\frac{1}{3}) = (\boldsymbol{2},-\frac{1}{3} \pm \frac{1}{2}) & = S_L 
\end{eqnarray}
Therefore, the only allowed representations are:
\begin{eqnarray}
\beta_1^S \rightarrow & \left\{ \begin{array}{l}
\psi_L \simeq (\boldsymbol{S_L} , Y_L ) \\
\psi_R \simeq (\boldsymbol{2},\frac{1}{6}) \end{array} \right.
\quad \left\{ \begin{array}{l}
\psi_L \simeq (\boldsymbol{1},\frac{1}{6} \pm \frac{1}{2}) \\
\psi_R \simeq (\boldsymbol{S_R} , Y_R ) \end{array} \right. \\
\beta_2^S \rightarrow & \left\{ \begin{array}{l}
\psi_L \simeq (\boldsymbol{S_L} , Y_L ) \\
\psi_R \simeq (\boldsymbol{1},\frac{1}{6} \pm \frac{1}{2}) \end{array} \right.
\quad \left\{ \begin{array}{l}
\psi_L \simeq (\boldsymbol{S_L} , Y_L ) \\
\psi_R \simeq (\boldsymbol{3},\frac{1}{6} \pm \frac{1}{2}) \end{array} \right.
\quad \left\{ \begin{array}{l}
\psi_L \simeq (\boldsymbol{2},\frac{1}{6} \pm \frac{1}{2} \pm \frac{1}{2}) \\
\psi_R \simeq (\boldsymbol{S_R} , Y_R ) \end{array} \right.
\end{eqnarray}

So if $\psi$ is a VLQ it can belong to a \emph{singlet}, a \emph{doublet} or a \emph{triplet} under $SU(2)_L$, all other representations being excluded, while if it is a ChQ its left-handed chirality can belong to a \emph{singlet} or a \emph{doublet} and its right-handed chirality can belong to a \emph{singlet}, a \emph{doublet} or a \emph{triplet} in the case of a coupling with a scalar boson.

\subsubsection{Mass terms} \label{sec:mass terms}

The mass terms for the new quarks can be written in a gauge-invariant way without requiring the Higgs mechanism only in the VL scenario because under $SU(2)$
\begin{equation}
M \bar{Q} Q = M \bar{Q}_L Q_R \simeq S_L \otimes S_R = S \otimes S = 1 \oplus ...
\end{equation}
which is always allowed. A mass term arising from Higgs mechanism can be obtained by finding the representations that give a singlet when contracted with the Higgs boson, i.e. $\bar\psi H \psi \simeq S_L \otimes 2 \otimes S_R = 1 \oplus ...$, and the corresponding hypercharge is directly obtained by conservation. The different possibilities are the following
\begin{eqnarray}
\psi_R \simeq (\boldsymbol{2},\frac{1}{6}) & \rightarrow & \psi_L \simeq (\boldsymbol{1},\frac{1}{6} \pm \frac{1}{2}) \\
\psi_L \simeq (\boldsymbol{1},\frac{1}{6} \pm \frac{1}{2}) & \rightarrow & \psi_R \simeq (\boldsymbol{2},\frac{1}{6} \pm \frac{1}{2} \pm \frac{1}{2}) \\
\psi_R \simeq (\boldsymbol{1},\frac{1}{6} \pm \frac{1}{2}) & \rightarrow & \psi_L \simeq (\boldsymbol{2},\frac{1}{6} \pm \frac{1}{2} \pm \frac{1}{2}) \\
\psi_R \simeq (\boldsymbol{3},\frac{1}{6} \pm \frac{1}{2}) & \rightarrow & \psi_L \simeq (\boldsymbol{2},\frac{1}{6} \pm \frac{1}{2} \pm \frac{1}{2}) \oplus (\boldsymbol{4},\frac{1}{6} \pm \frac{1}{2} \pm \frac{1}{2}) \\
\psi_L \simeq (\boldsymbol{2},\frac{1}{6} \pm \frac{1}{2} \pm \frac{1}{2}) & \rightarrow & \psi_R \simeq (\boldsymbol{1},\frac{1}{6} \pm \frac{1}{2} \pm \frac{1}{2} \pm \frac{1}{2}) \oplus (\boldsymbol{3},\frac{1}{6} \pm \frac{1}{2} \pm \frac{1}{2} \pm \frac{1}{2})
\end{eqnarray}
from which the particle content can be obtained
\begin{eqnarray*}
& \text{-- for a coupling with a scalar singlet $\beta_1^S$} \\
& \psi_R = (T,B)_R \rightarrow \left\{ \begin{array}{l}
\psi_L = T_L  \\
\psi_L = B_L  \end{array} \right. & \\
& \psi_L = T_L \rightarrow \left\{ \begin{array}{l}
\psi_R = (X,T)_R  \\
\psi_R = (T,B)_R  \end{array} \right. & \qquad
\psi_L = B_L \rightarrow \left\{ \begin{array}{l}
\psi_R = (T,B)_R  \\
\psi_R = (B,Y)_R  \end{array} \right. \\
& \text{ } \\
& \text{-- for a coupling with a scalar doublet $\beta_2^S$} \\
& \psi_R = T_R \rightarrow \left\{ \begin{array}{l}
\psi_L = (X,T)_L  \\
\psi_L = (T,B)_L  \end{array} \right. & \qquad
\psi_R = B_R \rightarrow \left\{ \begin{array}{l}
\psi_L = (T,B)_L  \\
\psi_L = (B,Y)_L  \end{array} \right. \\
& \psi_R = (X,T,B)_R \rightarrow \left\{ \begin{array}{l}
\psi_L = (X,T)_L  \\
\psi_L = (T,B)_L \\
\psi_L = (X^\prime,X,T,B)_L  \\
\psi_L = (X,T,B,Y)_L \end{array} \right. & \qquad
\psi_R = (T,B,Y)_R \rightarrow \left\{ \begin{array}{l}
\psi_L = (T,B)_L  \\
\psi_L = (B,Y)_L  \\
\psi_L = (X,T,B,Y)_L  \\
\psi_L = (T,B,Y,Y^\prime)_L  \end{array} \right. \\
& \psi_L = (X,T)_L \rightarrow \left\{ \begin{array}{l}
\psi_R = X_R  \\
\psi_R = T_R  \\
\psi_R = (X^\prime,X,T)_R  \\
\psi_R = (X,T,B)_R  \end{array} \right. & \qquad
\psi_L = (T,B)_L \rightarrow \left\{ \begin{array}{l}
\psi_R = T_R  \\
\psi_R = B_R  \\
\psi_R = (X,T,B)_R  \\
\psi_R = (T,B,Y)_R  \end{array} \right. \\
& \psi_L = (B,Y)_L \rightarrow \left\{ \begin{array}{l}
\psi_R = B_R  \\
\psi_R = Y_R  \\
\psi_R = (T,B,Y)_R  \\
\psi_R = (B,Y,Y^\prime)_R  \end{array} \right. 
\end{eqnarray*}
%Here the two first lines correspond to a coupling with a scalar singlet $\beta_1^S$ and the four last ones correspond to a coupling with a scalar doublet $\beta_2^S$, for a coupling with a vector boson the chiralities have to be inverted. 
Here we have respectively called $X^\prime$, $X$, $T$, $B$, $Y$ and $Y^\prime$ quarks with charge $+8/3$, $+5/3$, $+2/3$, $-1/3$, $-4/3$ and $-7/3$.

At the end\footnote{Remember that these results are valid for a coupling with a scalar boson, in the case of a coupling with a vector boson the chiralities have to be inverted.} we are left with the SM-like and VL scenarios
\begin{equation}
\left\{ \begin{array}{l}
\psi_L \simeq \bf 2 \\
\psi_R \simeq \bf 1 \end{array} \right.
\qquad \left\{ \begin{array}{l}
\psi_L \simeq \bf 1 \\
\psi_R \simeq \bf 1 \end{array} \right.
\qquad \left\{ \begin{array}{l}
\psi_L \simeq \bf 2 \\
\psi_R \simeq \bf 2 \end{array} \right.
\qquad \left\{ \begin{array}{l}
\psi_L \simeq \bf 3 \\
\psi_R \simeq \bf 3 \end{array} \right.
\end{equation}
as well as the following exotic chiral scenarios that we will not study in the rest of this thesis (apart in Appendix \ref{app:massless} where we check if we check if they predict massless quarks):
\begin{equation}
\left\{ \begin{array}{l}
\psi_L \simeq \bf 1 \\
\psi_R \simeq \bf 2 \end{array} \right.
\quad \left\{ \begin{array}{l}
\psi_L \simeq \bf 2 \\
\psi_R \simeq \bf 3 \end{array} \right.
\quad \left\{ \begin{array}{l}
\psi_L \simeq \bf 3 \\
\psi_R \simeq \bf 2 \end{array} \right.
\quad \left\{ \begin{array}{l}
\psi_L \simeq \bf 4 \\
\psi_R \simeq \bf 3 \end{array} \right.
\end{equation}
In the following we will only be interested in VLQs and in SM-like ChQs that we will only call ChQs for simplicity.

\subsection{XQs coupling to SM particles}

In this case we have seen previously that extra ChQs have been excluded by the Higgs discovery, we will therefore focus on VLQ in this section. The results presented here are inspired by \cite{Okada:2012gy}.

\subsubsection{Interactions and representations}

The minimal scenarios with the presence of VLQs coupling to SM particles are those in which the new states interact with SM quarks and the Higgs boson through Yukawa couplings, which means that we have $B^S_2 = H$. This coupling generates the mixing of the new heavy fermion with the SM quarks. In the general case this coupling takes the following structure
\begin{equation}
-y_{ij} \bar{q}_i H q_j + \hc
\end{equation}
where the $i$ and $j$ indices run through the flavour. 
%As we experimentally do not observe FCNC at tree level we have $i=j$ in the SM but with XQ we can have $i \neq j$ because we cannot assume a priori that in scenarios of new physics neutral currents conserve flavour. 
Note that we cannot assume a priori that in scenarios of new physics neutral currents conserve flavour so we can have $i \neq j$.

If we make the minimal hypothesis with only one new family of VLQs $T$ or $B$ with respective charge $2/3$ and $-1/3$ we have different possibilities of multiplets of VLQs that are summarised in Table \ref{tab:VLrepresentations}.

\begin{table}[htb]
\begin{eqnarray*}
\setlength{\arraycolsep}{2pt}
\begin{array}{ccccccccccc}
&\multicolumn{3}{c}{\mbox{SM quarks $q$}}&\multicolumn{2}{c}{\mbox{Singlets $Q_R$}}&\multicolumn{3}{c}{\mbox{Doublets $\psi_L$}}&\multicolumn{2}{c}{\mbox{Triplets $\psi_R$}}\\
& 
\begin{array}{c} ~ \\ \left(\begin{array}{c} u \\ d \end{array}\right) \\ ~ \end{array} &
\begin{array}{c} ~ \\ \left(\begin{array}{c} c \\ s \end{array}\right) \\ ~ \end{array} &
\begin{array}{c} ~ \\ \left(\begin{array}{c} t \\ b \end{array}\right) \\ ~ \end{array} &
\begin{array}{c} ~ \\ T \\ ~ \\ ~ \end{array} &
\begin{array}{c} ~ \\ ~ \\ B \\ ~ \end{array} & 
\begin{array}{c} \left(\begin{array}{c} X \\ T \end{array}\right) \\ ~ \\ ~ \end{array} &
\begin{array}{c} ~ \\ \left(\begin{array}{c} T \\ B \end{array}\right) \\ ~ \end{array} &
\begin{array}{c} ~ \\ ~ \\ \left(\begin{array}{c} B \\ Y \end{array}\right) \end{array} & 
\begin{array}{c} \left(\begin{array}{c} X \\ T \\ B \end{array}\right) \\ ~ \end{array} &
\begin{array}{c} ~ \\ \left(\begin{array}{c} T \\ B \\ Y \end{array}\right) \end{array} \\
\midrule
SU(2)_L & 
\multicolumn{3}{c}{\begin{array}{c}q_L=2\\q_R=1\end{array}} & 
\multicolumn{2}{c}{1} & 
\multicolumn{3}{c}{2} & 
\multicolumn{2}{c}{3} \\
\midrule
U(1)_Y  & 
\multicolumn{3}{c}{\begin{array}{c}q_L=1/6\\u_R=2/3\\d_R=-1/3\end{array}} & 2/3 & -1/3 & 7/6 & 1/6 & -5/6 & 2/3 & -1/3 \\
\midrule
\mathcal{L}_Y &
\multicolumn{3}{c}{\begin{array}{c}-y_u^i \bar q_L^i H^c u_R^i \\- y_d^i \bar q_L^i V_{CKM}^{i,j} H d_R^j\end{array}} &
\multicolumn{2}{c}{\begin{array}{c}-\lambda_u^i \bar q_L^i H^c T_R \\-\lambda_d^i \bar q_L^i H B_R \end{array}} &
\multicolumn{3}{c}{\begin{array}{c}-\lambda_u^i \bar \psi_L H^{(c)} u_R^i \\-\lambda_d^i \bar \psi_L H^{(c)} d_R^i \end{array}} &
\multicolumn{2}{c}{\begin{array}{c}-\lambda_i \bar q_L^i \sigma^a H^{(c)} \psi_R^a \end{array}} \\
\midrule
\mathcal{L}_m & \multicolumn{3}{c}{\mbox{not allowed\footnotemark[5]}} & \multicolumn{7}{c}{-M_{VLQ} \, \bar\psi \, \psi}
\end{array} 
\end{eqnarray*}
\caption[Allowed representations for VLQs, with quantum numbers under $SU(2)_L$ and $U(1)_Y$ and Yukawa mixing terms in the Lagrangian.]{Allowed representations for VLQs, with quantum numbers under $SU(2)_L$ and $U(1)_Y$ and Yukawa mixing terms in the Lagrangian. Depending on the chosen representation, the Higgs boson may be $H$ or $H^c$, therefore it has been noted as $H^{(c)}$ when necessary. The gauge invariant mass term common to all representations is a peculiar feature of VLQs.}
\label{tab:VLrepresentations}
\end{table}

\footnotetext[5]{The Higgs Mechanism is necessary for gauge invariance.}
\setcounter{footnote}{5}

\subparagraph*{Remark}

To couple with SM quarks through Yukawa-coupling the $X$ and $Y$ must be part of a doublet or of a triplet, otherwise it is not possible to conserve the hypercharge with a coupling of the form $\bar{q} H X$ as shown in Table \ref{tab:quantum numbers}. Yet, we can have a singlet $X$ or $Y$ if we add two different families of VLQs to the SM, including a doublet $(X, T)$ or $(B, Y)$ for example. 

\begin{table}[ht!]
{\centering
\begin{tabular}{c|ccccccc}
 & $\bar{u}_L$ & $\bar{d}_L$ & $\bar{u}_R$ & $\bar{d}_R$  & $H$ & $X$ & $Y$ \\
\hline
$Y$ & $-1/6$ & $-1/6$ & $-2/3$ & $1/3$ & $\pm 1/2$ & $5/3$ & $-4/3$ \\
\hline
$T_{3L}$ & $-1/2$ & $1/2$ & $0$ & $0$ & $\mp 1/2$ & $0$ & $0$ \\
\hline
$Q$ & $-2/3$ & $1/3$ & $-2/3$ & $1/3$ & $0$ & $5/3$ & $-4/3$ \\
\end{tabular}
\caption{Quantum numbers of SM quarks, VLQs $X$ and $Y$ (as part of singlet) and Higgs boson.}
\label{tab:quantum numbers}
}
\end{table}

\subsubsection{Mixing matrices and coupling parametrisation}

\paragraph*{Mixing matrices}

%While ChQs are always allowed to mix with SM quarks, VLQs can only do it after the Higgs develops its VEV: the mixing occurs in the left-handed sector for the singlet and triplet representations and in the right-handed sector for the doublet representation. We label the mass eigenstates as $\{X_{5/3},t^\prime,b^\prime,Y_{-4/3}\}$. The mass matrices for the SM-partners $t^\prime$ and $b^\prime$ can be diagonalized by unitary $4\times4$ matrices $V^{t,b}_L$ and $V^{t,b}_R$:
While ChQs mix with SM quarks in both the left-handed and right-handed, the mixing of VLQs only occurs in the left-handed sector for the singlet and triplet representations and in the right-handed sector for the doublet representation. We label the mass eigenstates as $\{X_{5/3},t^\prime,b^\prime,Y_{-4/3}\}$. The mass matrices for the SM-partners $t^\prime$ and $b^\prime$ can be diagonalized by unitary $4\times4$ matrices $V^{t,b}_L$ and $V^{t,b}_R$:
\begin{eqnarray}
\left(\begin{array}{cccc} m_u & & & \\ & m_c & & \\ & & m_t & \\ & & & M_{t^\prime} \end{array}\right) &=& (V^t_L)^\dagger \cdot \mathcal{M}_t \cdot (V^t_R) \\
\left(\begin{array}{cccc} m_d & & & \\ & m_s & & \\ & & m_b & \\ & & & M_{b^\prime} \end{array}\right) &=& (V^b_L)^\dagger \cdot \mathcal{M}_b \cdot (V^b_R)
\end{eqnarray}
where the actual expressions of $\mathcal{M}_t$ and $\mathcal{M}_b$ depend on the chosen representations and on the assumptions on the mixing parameters.

\paragraph*{Neutral currents}

The couplings with gauge bosons also depend on the chosen representations, but a common feature of every VLQ scenario is that tree-level FCNCs are developed through the mixing with SM quarks. The general form of $Z q q$ couplings with the presence of VLQs is:
\begin{eqnarray}
\setlength{\arraycolsep}{0pt}
\begin{array}{rcll}
g_{ZL}^{IJ} &=& \frac{g}{c_W} \Big[ \left( T_3 - Q s_W^2 \right) \delta^{IJ} & + f_L (V^{t,b}_L)^{*,q^\prime I} (V^{t,b}_L)^{q^\prime J} \Big] \\
g_{ZR}^{IJ} &=& \frac{g}{c_W} \Big[ \left( - Q s_W^2 \right) \delta^{IJ}              & + f_R (V^{t,b}_R)^{*,q^\prime I} (V^{t,b}_R)^{q^\prime J} \Big]
\end{array}
\label{eq:Zcoupling}
\end{eqnarray}
where $I,J$ run on all quarks, including VLQs, $T_3=\pm1/2$ is the weak isospin of the top or bottom SM quark, and $f_{L,R} \in \{0,\pm1/2,\pm1\}$ are parameters which depend on the VLQ representation and satisfy the relation $T_3^{q^\prime}=T_3+f_L=f_R$; they are listed in Tab.~\ref{fLfR} for each representation.

\begin{table}[htb]
\begin{eqnarray*}
\begin{array}{ccccccccc}
& & \multicolumn{2}{c}{\mbox{Singlets}} & \multicolumn{3}{c}{\mbox{Doublets}} & \multicolumn{2}{c}{\mbox{Triplets}} \vspace*{1mm} \\
& & T & B & \left(\begin{array}{c} X \\ T \end{array}\right) & \left(\begin{array}{c} T \\ B \end{array}\right) & \left(\begin{array}{c} B \\ Y \end{array}\right) & \left(\begin{array}{c} X \\ T \\ B \end{array}\right) & \left(\begin{array}{c} T \\   \\ Y \end{array}\right) \vspace*{1mm} \\
\toprule
%U(1)_Y & & 2/3 & -1/3 & 7/6 & 1/6 & -5/6 & 2/3 & -1/3 \\
%\midrule
T & \begin{array}{c} f_L \\ f_R \end{array} & 
\begin{array}{c} -1/2 \\  0   \end{array} & & 
\begin{array}{c} -1   \\ -1/2 \end{array} & 
\begin{array}{c}  0   \\ +1/2 \end{array} & &
\begin{array}{c} -1/2 \\  0   \end{array} & 
\begin{array}{c} +1/2 \\ +1   \end{array} \\
\midrule
B & \begin{array}{c} f_L \\ f_R \end{array} & & 
\begin{array}{c} +1/2 \\ 0    \end{array} & &
\begin{array}{c}  0   \\ -1/2 \end{array} & 
\begin{array}{c} +1   \\  1/2 \end{array} & 
\begin{array}{c} -1/2 \\ -1   \end{array} & 
\begin{array}{c} +1/2 \\  0   \end{array} 
\end{array}
\end{eqnarray*}
\caption{Neutral current parameters $f_L$ and $f_R$.}
\label{fLfR}
\end{table}

The new form of the coupling from Eq. (\ref{eq:Zcoupling}) has two implications. First of all we see that we have FCNCs between the new state and SM quarks, but also between SM quarks themselves, if the VLQs are allowed to mix with at least two families. Secondly we notice that even flavour conserving neutral currents ($I=J$) are modified by the presence of VLQs. Constraints on FCNCs coming from a large number of observations can therefore provide strong bounds on mixing parameters.

\paragraph*{Charged currents}

Furthermore, charged currents are modified too. The general form of $W q_1 q_2$ couplings with the presence of VLQs is:
\begin{eqnarray}
\setlength{\arraycolsep}{0pt}
g_{WL}^{IJ} = \frac{g}{\sqrt{2}} (V_{CKM}^L)^{IJ} = \frac{g}{\sqrt{2}} (V^t_L)^\dagger \cdot \hat\delta_L \cdot \tilde V_{CKM}^L \cdot V^b_L \\
g_{WR}^{IJ} = \frac{g}{\sqrt{2}} (V_{CKM}^R)^{IJ} = \frac{g}{\sqrt{2}} (V^t_R)^\dagger \cdot \hat\delta_R \cdot \tilde V_{CKM}^R \cdot V^b_R
\label{VCKM}
\end{eqnarray}
where $I,J=1,2,3(,4)$ and the matrices $V^{t,b}$ may or may not be present depending on the scenario considered. The matrices $\hat\delta_{L,R}$ are defined as:
\begin{eqnarray}
\hat\delta_L=\left(\begin{array}{ccc|c} 1 & ~ & ~ & ~ \\ ~ & 1 & ~ & ~ \\ ~ & ~ & 1 & ~ \\ \hline ~ & ~ & ~ & 1 \end{array} \right) \qquad
\hat\delta_R=\left(\begin{array}{ccc|c} 0 & ~ & ~ & ~ \\ ~ & 0 & ~ & ~ \\ ~ & ~ & 0 & ~ \\ \hline ~ & ~ & ~ & 1 \end{array} \right) \qquad
\label{dLdR}
\end{eqnarray}
where the lines mean that the size of the matrices depend on the chosen scenario; in particular, $g_{WR}$ is non-zero only if both an up- and down-type VLQ are present simultaneously, because as we have seen in Sec. \ref{sec:chirality} $J^{\mu +}_R = 0$ for SM ChQs. The matrices $\tilde V_{CKM}^{L,R}$ represent the misalignment between SM quarks in the left- and right-handed sector; $\tilde V_{CKM}^L$ corresponding to the measured $CKM$ matrix in the absence of VLQs presented in Sec. \ref{sec:SMflavour}. Two $CKM$ matrices can thus be defined in the presence of VLQs, for the left- and right-handed sectors. In the case of existence of VLQs, the usual SM $CKM$ matrix we measured experimently corresponds to the $3\times3$ block $(V_{CKM}^L)^{ij}$, with $i,j \in \{ 1,2,3 \}$. Since the full new $CKM$ matrices $\tilde V_{CKM}^{L,R}$ have to be unitary, this would also mean that the measured $3\times 3$ $CKM$ submatrix is not unitary, and it is possible to check that deviations from unitarity are proportional to the mixing between SM quarks and VL states. 

Charged currents may also be present between the exotic states $\{X_{5/3},Y_{-4/3}\}$ and up- or down-type quark respectively. The couplings are:
\begin{eqnarray}
\begin{array}{rcl} 
g_W^{XI} &=& \frac{g}{\sqrt{2}} \left( (V^t_L)^{4I} + (V^t_R)^{4I} \right) \\
g_W^{YI} &=& \frac{g}{\sqrt{2}} \left( (V^b_L)^{4I} + (V^b_R)^{4I} \right) 
\end{array}
\end{eqnarray}

\paragraph*{Coupling to the Higgs boson}

Finally, the couplings to the Higgs bosons can be written as:
\begin{eqnarray}
\setlength{\arraycolsep}{0pt}
\begin{array}{rcl}
C_u^{IJ} &=& \frac{1}{v} \left(\setlength{\arraycolsep}{2pt}\begin{array}{cccc} m_u & & & \\ & m_c & & \\ & & m_t & \\ & & & M_{t^\prime} \end{array}\right) - \frac{M}{v} (V^t_L)^{*,4I} (V^t_R)^{4J} \\
C_d^{IJ} &=& \frac{1}{v} \left(\setlength{\arraycolsep}{2pt}\begin{array}{cccc} m_d & & & \\ & m_s & & \\ & & m_b & \\ & & & M_b^{\prime} \end{array}\right) - \frac{M}{v} (V^b_L)^{*,4I} (V^b_R)^{4J}
\end{array}
\end{eqnarray}
From these expressions it can be inferred that the presence of VLQ can modify the mechanism of production and decay of the Higgs boson with respect to SM predictions.

\subsubsection{The effective Lagrangian}

Finally, we can write the Lagrangian for our effective model describing the phenomenology of the 4 different type of VLQs\footnote{This Lagrandian is only valid for leading left-handed coupling and it suffices to exchange the chiralities L $\leftrightarrow$ R to obtain the Lagrangian for leading right-handed coupling.}:
\begin{multline}
\mathcal{L}_L =  \kappa_T \left\{ \sqrt{\frac{\zeta_i \xi_W^T}{\Gamma_W^0}} \frac{g}{\sqrt{2}}\; [\bar{T}_{L} W_\mu^+ \gamma^\mu d^i_{L} ]  +  \sqrt{\frac{\zeta_i \xi_Z^T}{\Gamma_Z^0}} \frac{g}{2 c_W} \; [\bar{T}_{L} Z_\mu \gamma^\mu u^i_{L} ]  \right.\\
\left. -  \sqrt{\frac{\zeta_i \xi_H^T}{\Gamma_H^0}} \frac{M_T}{v}\; [\bar{T}_{R} H u^i_{L}]  -  \sqrt{\frac{\zeta_3 \xi_H^T}{\Gamma_H^0}} \frac{m_t}{v}\; [\bar{T}_{L} H t_{R}] \right\}\\
 + \kappa_B \left\{ \sqrt{\frac{\zeta_i \xi_W^B}{\Gamma_W^0}} \frac{g}{\sqrt{2}}\; [\bar{B}_{L} W_\mu^- \gamma^\mu u^i_{L} ]  +  \sqrt{\frac{\zeta_i \xi_Z^B}{\Gamma_Z^0}} \frac{g}{2 c_W} \; [\bar{B}_{L} Z_\mu \gamma^\mu d^i_{L} ] -  \sqrt{\frac{\zeta_i \xi_H^B}{\Gamma_H^0}} \frac{M_B}{v}\; [\bar{B}_{R} H d^i_{L} ] \right\} \\
 + \kappa_X \left\{ \sqrt{\frac{\zeta_i}{\Gamma_W^0}} \frac{g}{\sqrt{2}}\; [\bar{X}_{L} W_\mu^+ \gamma^\mu u^i_{L} ]  \right\} + \kappa_Y \left\{ \sqrt{\frac{\zeta_i}{\Gamma_W^0}} \frac{g}{\sqrt{2}}\; [\bar{Y}_{L} W_\mu^- \gamma^\mu d^i_{L} ]  \right\} + h.c.\, \label{eq:param}
\end{multline}

where we have used the same definition for $\zeta_i$, $\zeta^Q_V$ and $\Gamma_V^0$ as in \cite{Buchkremer:2013bha}. The curious reader can refer to this paper for a more detailed presentation of this Lagrangian and of its features.

One interesting feature of this parametrisation is that we have a simple expression of the BRs $BR(Q \to V q_i) = \zeta_i \xi_V^Q$, and therefore we also have $\sum_{i=1}^3 \zeta_i = 1$, $\sum_{V=W,Z,H} \xi_V^Q = 1$.

The mass of the VLQ will determine its production rates, especially for pair production which is dominated by QCD processes. The coupling strength factors $\kappa_Q$ will drive the electroweak pair and single production cross sections, which are therefore sensitive to the overall strength of the coupling, similarly to the single top production processes in the SM.

This effective Lagrangian has been implemented in FeynRules~\cite{Christensen:2008py}, and is described in more detail in the Appendix C of \cite{Buchkremer:2013bha}. The complete FeynRules files, together with the CalcHEP and MadGraph outputs, are available on the FeynRules website for the general model \cite{wwwFeynRules} and on the website of the HEP model database project~\cite{hepmdb}.

\subsection{XQs coupling to Dark Matter} \label{sec:XQDM Lagrangian}

We now consider the case of XQs coupling to DM (invisible decay) and we use the same parametrisation as in \cite{Kraml:2016eti, Moretti:2016gkr}. 

We consider a minimal extension of the SM with one XQ state and one DM state, assuming that the XQ mediates the interaction between the DM and the SM quarks. In order to have a stable DM candidate we impose a extra $\mathcal{Z}_2$ symmetry on the Lagrangian under which all the SM states are even while the new states (XQs and SM) are odd. This symmetry is similar to the $R$ parity from SUSY for example. One of the consequence of this parity is that the XQ do not mix with the SM quarks which means that the $CKM$ matrix is not modified in this case.

The most general Lagrangian terms depend on the representation of the DM and of the XQ. We consider a singlet and doublet DM so as we have seen in Section \ref{sec:interaction terms} the XQ can belong to a singlet, a doublet or a triplet under $SU(2)_L$. %For simplicity we will not consider the case of a triplet here.

We label XQ singlet states as $T$ or $B$, XQ doublet states as $\Psi_Y$ and XQ triplets as $\psi_Y$, where $Y$ corresponds to the weak hypercharge of the multiplet in the convention $Q=T_3+Y$, with $Q$ the electric charge and $T_3$ the weak isospin. 
The doublets can then be $\Psi_{1/6} = (T \ B)^T$ or states which contain exotic components $\Psi_{7/6} = (X_{5/3} \ T)^T$ and $\Psi_{-5/6} = (B \ Y_{-4/3})^T$ and the triplets are $\psi_{2/3} = (X_{5/3} \ T \ B)^T$ and $\psi_{-1/3} = (T \ B \ Y_{-4/3})^T$. 
The DM states are labelled as $S^0_{\rm DM}$ if scalar singlets or $V^{0\mu}_{\rm DM}$ if vector singlets; if the DM belongs to a doublet representation, the multiplet is labelled as $\Sigma_{\rm DM} = (S^+ \ S^0_{\rm DM})^T$ (with the charge conjugate $\Sigma^c_{\rm DM} = (S^{0}_{\rm DM} \ -S^-)^T$) if scalar or $\mathcal{V}_{\rm DM} = (V^+ \ V^0_{\rm DM})^T$ (with the charge conjugate $\mathcal{V}^c_{\rm DM} = (V^0_{\rm DM} \ V^-_{\rm DM})^T$) if vector. 
The couplings between the XQ, the DM and the SM quarks are denoted as $\lambda_{jk}^q$ if the DM is scalar, or $g_{jk}^q$ if the DM is vector: the labels $j,k \in \{ 1,2,3 \}$ indicate the representations of the XQ and DM respectively (1 for singlet, 2 for doublet, 3 for triplet), while $q \in \{ u,d,c,s,b,t \}$ identifies which SM quark the new states are coupled with, in case of ambiguity. In the following $i$ is a flavour index running over the 3 SM generations.
We classify below the Lagrangian terms for the minimal SM extensions with one XQ and one DM representation (singlets and doublets) but we anticipate that in the following, for simplicity, we will only consider scenarios with a DM singlet. 

\paragraph*{Lagrangian terms for a DM singlet} A DM singlet can couple either with an XQ singlet or with an XQ doublet $\Psi_{1/6}={T \choose B}$.
\begin{eqnarray}
\Lag^S_1 &=& 
\left[
\lambda_{11}^{u^i} \bar{T}_L \, u^i_R + 
\lambda_{11}^{d^i} \bar{B}_L \, d^i_R +
\lambda_{21}^i \, \overline\Psi_{1/6,R} {u^i \choose d^i}_L 
\right] 
S^0_{\rm DM} + {\rm h.c.} 
\label{eq:LagSingletDMS}
\\
\Lag^V_1 &=& 
\left[
g_{11}^{u^i} \bar{T}_R \gamma_\mu \, u^i_R + 
g_{11}^{d^i} \bar{B}_R \gamma_\mu \, d^i_R + 
g_{21}^i  \, \overline\Psi_{1/6,L} \gamma_\mu {u^i \choose d^i}_L 
\right] 
V^{0\mu}_{\rm DM} + {\rm h.c.},
\label{eq:LagSingletDMV}
\end{eqnarray}

\paragraph*{Lagrangian terms for a DM doublet} A DM doublet can couple with XQ singlets, doublets or triplets with different hypercharges.
\begin{eqnarray}
\Lag^S_2 &=& 
\Bigg[
\left( \lambda_{12}^{d^i} \bar B_R + \lambda_{32}^{d^i} \bar\psi_{2/3,R}^a \, \tau^a \right) {u^i \choose d^i}_L + \lambda_{22}^{d^i} \overline\Psi_{1/6,L} d^i_R +
(\lambda_{22}^{u^i})^\prime \, \overline\Psi_{5/6,L} u^i_R
\Bigg] \Sigma_{\rm DM}^T \\
&+& \Bigg[
\left( \lambda_{12}^{u^i} \bar T_R + \lambda_{32}^{u^i} \bar\psi_{-1/3,R}^a \, \tau^a \right) {u^i \choose d^i}_L + \lambda_{22}^{u^i} \overline\Psi_{1/6,L} u^i_R +
(\lambda_{22}^{d^i})^\prime \, \overline\Psi_{-1/6,L} d^i_R 
\Bigg] \Sigma^{c,T}_{\rm DM} + {\rm h.c.} \nonumber \\
\Lag^V_2 &=& 
\Bigg[
\left( g_{12}^{d^i} \bar B_L \gamma_\mu + g_{32}^{d^i} \bar\psi_{2/3,L}^a \, \gamma_\mu \tau^a \right) {u^i \choose d^i}_L + g_{22}^{d^i} \overline\Psi_{1/6,R} \gamma_\mu d^i_R +
(g_{22}^{u^i})^\prime \, \overline\Psi_{5/6,R} \gamma_\mu u^i_R
\Bigg] \mathcal{V}_{\rm DM}^{\mu,T} \\
&+& \Bigg[
\left( g_{12}^{u^i} \bar T_L \gamma_\mu + g_{32}^{u^i} \bar\psi_{-1/3,L}^a \, \gamma_\mu \tau^a \right) {u^i \choose d^i}_L + g_{22}^{u^i} \overline\Psi_{1/6,R} \gamma_\mu u^i_R + (g_{22}^{d^i})^\prime \, \overline\Psi_{-1/6,R} \gamma_\mu d^i_R 
\Bigg] \mathcal{V}_{\rm DM}^{c,\mu ,T} + {\rm h.c.} \nonumber 
\label{eq:LagDoubletDM}
\end{eqnarray}
where $\tau^1 = \sigma^- = \frac{\sigma_1 - i \sigma_2}{2}$, $\tau^2 = \sigma_3$ and $\tau^3 = \sigma^+ = \frac{\sigma_1 + i \sigma_2}{2}$.

However, in scenarios with a DM doublet, there are always additional exotic states besides the XQ partners of the SM quarks and the DM state, namely charged scalars or vectors and quarks with charges $+5/3$ or $-4/3$. As mentioned above, in order to stick to a minimal extension of the SM containing a partner of the top quark and the DM candidate as the only new states, in the following we consider only the Lagrangian terms of Eqs.~\eqref{eq:LagSingletDMS} or \eqref{eq:LagSingletDMV}, depending on the spin of the DM. %It is also worth noticing that in the considered scenarios  the XQs do not mix with SM states because they have a different quantum number under the $\mathcal Z_2$ symmetry. 
Depending on the representation of the XQ, one can then identify some limiting cases: 
\begin{itemize}
\item \textit{Vector-like XQ}. If the VLQ is a singlet, only couplings with SM singlets are allowed, and $\lambda_{21}=0$ or $g_{21}=0$. On the other hand, if the VLQ is a doublet, $\lambda_{11}=0$ or $g_{11}=0$.
Unlike cases where VLQs mix with the SM quarks through Yukawa couplings via the Higgs boson, couplings for the opposite chiralities are not just suppressed, they are identically zero.
The mass term for a VLQ can be written in a gauge-invariant way as:
\begin{equation}
\Lag_{\rm VLQ} = - M_{T_{\rm VLQ}} \bar T T
\label{eq:VLQmass}
\end{equation}
where $M_{T_{\rm VLQ}}$ is a new physics mass scale not necessarily related to a Higgs-like mechanism for mass generation.

\item \textit{Chiral XQ}. If the XQ is chiral, all the couplings of Eqs.~\eqref{eq:LagSingletDMS} or \eqref{eq:LagSingletDMV} can be allowed at the same time. ChQs can acquire mass in a gauge invariant way via the Higgs mechanism, analogously to SM quarks:
\begin{align}
\Lag_{\rm ChQ} = & - y_{\rm XQ}^B \bar \Psi_{1/6} H B - y_{\rm XQ}^T \bar \Psi_{1/6} H^c T + {\rm h.c.} \nonumber \\  
                              & \Longrightarrow\; - M_{T_{\rm ChQ}} \bar T T - M_{B_{\rm ChQ}} \bar B B
\label{eq:ChQmass}
\end{align}
where $M_{\{T,B\}_{\rm ChQ}} = y_{\rm XQ}^{\{T,B\}} v / \sqrt{2}$ and $v$ is the Higgs VEV. At this point it has to be mentioned that the contribution of the new ChQ to Higgs production and decay processes can be used to pose constraints on the coupling between the XQ and the Higgs boson, and as a consequence, on the maximum mass the ChQ can acquire through the Higgs mechanism. Yet this is beyond the scope ot these study so we will simply consider the ChQ mass as a free parameter in the following analysis.
\end{itemize}

%% ------------- --------------------- --------------------- ---------------------- --------------------- ---------------------- ----------------------

\section{Production and decay of XQs}

\subsection{Production}

There are two main ways of producing an XQ $Q$: pair-production\footnote{We will not consider the case of pair production of two different kinds of XQ $p p \to Q \bar Q^\prime$ because we will only consider one type of quark at a time in our studies.} $p p \to Q \bar Q$ and single production $p p \to Q \bar q$ or $p p \to Q B$ where $q$ is a SM quark and $B \in \{ H, W^\pm , Z\}$ for visible decay and $B \in \{ S^0_{\rm DM} , V^{0\mu}_{\rm DM} \}$ for invisible decay. These different production modes are described in detail in \cite{Buchkremer:2013bha} for VLQs decaying to SM particles, and they can be generalized to XQs without loss of generality.

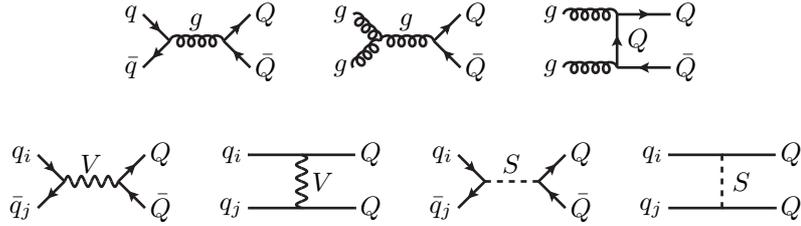
\begin{figure}[ht]
\begin{center}
 \begin{picture}(75,40)(0,-10)
  \SetWidth{1}
  \Line[arrow,arrowpos=0.5,arrowlength=2,arrowwidth=1,arrowinset=0.2](10,20)(20,10)
  \Text(8,20)[rc]{\small{$q$}}
  \Line[arrow,arrowpos=0.5,arrowlength=2,arrowwidth=1,arrowinset=0.2](20,10)(10,0)
  \Text(8,0)[rc]{\small{$\bar q$}}
  \Gluon(20,10)(40,10){2}{4}
  \Text(30,20)[tc]{\small{$g$}}
  \Line[arrow,arrowpos=0.5,arrowlength=2,arrowwidth=1,arrowinset=0.2](40,10)(50,20)
  \Text(52,20)[lc]{\small{$Q$}}
  \Line[arrow,arrowpos=0.5,arrowlength=2,arrowwidth=1,arrowinset=0.2](50,0)(40,10)
  \Text(52,0)[lc]{\small{$\bar{Q}$}}
 \end{picture}
 \begin{picture}(75,40)(0,-10)
  \SetWidth{1}
  \Gluon(10,20)(20,10){2}{3}
  \Text(8,20)[rc]{\small{$g$}}
  \Gluon(10,0)(20,10){2}{3}
  \Text(8,0)[rc]{\small{$g$}}
  \Gluon(20,10)(40,10){2}{4}
  \Text(30,20)[tc]{\small{$g$}}
  \Line[arrow,arrowpos=0.5,arrowlength=2,arrowwidth=1,arrowinset=0.2](40,10)(50,20)
  \Text(52,20)[lc]{\small{$Q$}}
  \Line[arrow,arrowpos=0.5,arrowlength=2,arrowwidth=1,arrowinset=0.2](50,0)(40,10)
  \Text(52,0)[lc]{\small{$\bar{Q}$}}
 \end{picture}
 \begin{picture}(75,40)(0,-10)
  \SetWidth{1}
  \Gluon(10,20)(30,20){2}{4}
  \Text(8,20)[rc]{\small{$g$}}
  \Gluon(10,0)(30,00){2}{4}
  \Text(8,0)[rc]{\small{$g$}}
  \Line[arrow,arrowpos=0.5,arrowlength=2,arrowwidth=1,arrowinset=0.2](30,0)(30,20)
  \Text(34,10)[cl]{\small{$Q$}}
  \Line[arrow,arrowpos=0.5,arrowlength=2,arrowwidth=1,arrowinset=0.2](30,20)(50,20)
  \Text(52,20)[lc]{\small{$Q$}}
  \Line[arrow,arrowpos=0.5,arrowlength=2,arrowwidth=1,arrowinset=0.2](50,0)(30,0)
  \Text(52,0)[lc]{\small{$\bar{Q}$}}
 \end{picture}
\end{center}

\begin{center}
 \begin{picture}(75,40)(0,-10)
  \SetWidth{1}
  \Line[arrow,arrowpos=0.5,arrowlength=2,arrowwidth=1,arrowinset=0.2](10,20)(20,10)
  \Text(8,20)[rc]{\small{$q_i$}}
  \Line[arrow,arrowpos=0.5,arrowlength=2,arrowwidth=1,arrowinset=0.2](20,10)(10,0)
  \Text(8,0)[rc]{\small{$\bar q_j$}}
  \Photon(20,10)(40,10){2}{4}
  \Text(30,20)[tc]{\small{$V$}}
  \Line[arrow,arrowpos=0.5,arrowlength=2,arrowwidth=1,arrowinset=0.2](40,10)(50,20)
  \Text(52,20)[lc]{\small{$Q$}}
  \Line[arrow,arrowpos=0.5,arrowlength=2,arrowwidth=1,arrowinset=0.2](50,0)(40,10)
  \Text(52,0)[lc]{\small{$\bar{Q}$}}
 \end{picture}
 \begin{picture}(75,40)(0,-10)
  \SetWidth{1}
  \Line[](10,20)(30,20)
  \Text(8,20)[rc]{\small{$q_i$}}
  \Line[](30,0)(10,0)
  \Text(8,0)[rc]{\small{$q_j$}}
  \Photon(30,20)(30,0){2}{4}
  \Text(34,10)[cl]{\small{$V$}}
  \Line[](30,20)(50,20)
  \Text(52,20)[lc]{\small{$Q$}}
  \Line[](50,0)(30,0)
  \Text(52,0)[lc]{\small{$Q$}}
 \end{picture}
 \begin{picture}(75,40)(0,-10)
  \SetWidth{1}
  \Line[arrow,arrowpos=0.5,arrowlength=2,arrowwidth=1,arrowinset=0.2](10,20)(20,10)
  \Text(8,20)[rc]{\small{$q_i$}}
  \Line[arrow,arrowpos=0.5,arrowlength=2,arrowwidth=1,arrowinset=0.2](20,10)(10,0)
  \Text(8,0)[rc]{\small{$\bar q_j$}}
  \Line[dash,dashsize=2.5](20,10)(40,10)
  \Text(30,20)[tc]{\small{$S$}}
  \Line[arrow,arrowpos=0.5,arrowlength=2,arrowwidth=1,arrowinset=0.2](40,10)(50,20)
  \Text(52,20)[lc]{\small{$Q$}}
  \Line[arrow,arrowpos=0.5,arrowlength=2,arrowwidth=1,arrowinset=0.2](50,0)(40,10)
  \Text(52,0)[lc]{\small{$\bar{Q}$}}
 \end{picture}
 \begin{picture}(75,40)(0,-10)
  \SetWidth{1}
  \Line[](10,20)(30,20)
  \Text(8,20)[rc]{\small{$q_i$}}
  \Line[](30,0)(10,0)
  \Text(8,0)[rc]{\small{$q_j$}}
  \Line[dash,dashsize=2.5](30,20)(30,0)
  \Text(34,10)[cl]{\small{$S$}}
  \Line[](30,20)(50,20)
  \Text(52,20)[lc]{\small{$Q$}}
  \Line[](50,0)(30,0)
  \Text(52,0)[lc]{\small{$Q$}}
 \end{picture}
\end{center}
\caption[Feynman diagrams for pair production of a generic XQ.]{Feynman diagrams for pair production of a generic XQ. Above the dominant and model independent QCD contributions, below the subdominant and model dependent electroweak contributions. Arrows on fermion lines have been removed to account for both particles and antiparticles, when necessary. Notice the possibility to have FCNCs between SM quarks in the $V$ and $S$ s-channel diagram, which is peculiar to VL scenarios.}
\label{fig:PairVL}
\end{figure}

XQs can in principle be pair-produced by electromagnetic, strong or weak interaction. The probability of production is proportional to $e^2$ for the electromagnetic interaction, to $g_S^2$ for the strong interaction and to $g_W^2$ for the weak interaction, but since $\abs{e} \ll g_S$ and $g_W \ll g_S$, the production by electromagnetic and weak interaction are suppressed. Furthermore, the production through weak currents receives a further -- but light -- suppression from the masses of the propagating $W$ and $Z$ bosons. It is also possible to have pair production of $T\bar{T}$ through the propagation of a Higgs boson in the $t$-channel, but this diagram is strongly suppressed by both the Higgs mass in the propagator and by the small Yukawa couplings between the VLQs and the light SM quarks (the only SM quark for which the Yukawa coupling can be sizeable is the top quark, nevertheless the top is not a parton of the proton). Therefore, we will only consider QCD production pair production in the rest of our study. The cross section for such processes only depend on the XQ mass which means it is \emph{model independent}. The Feynman diagrams for pair production of XQs are shown in Fig.~\ref{fig:PairVL}, the dominant QCD ones being the ones of the first row. 

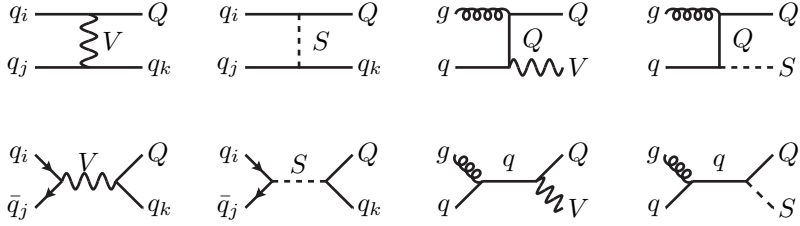
\begin{figure}[ht]
\begin{center}
 \begin{picture}(75,40)(0,-10)
  \SetWidth{1}
  \Line[](10,20)(30,20)
  \Text(8,20)[rc]{\small{$q_i$}}
  \Line[](10,0)(30,0)
  \Text(8,0)[rc]{\small{$q_j$}}
  \Photon(30,20)(30,0){3}{3}
  \Text(35,10)[lc]{\small{$V$}}
  \Line[](30,20)(50,20)
  \Text(52,20)[lc]{\small{$Q$}}
  \Line[](30,0)(50,0)
  \Text(52,0)[lc]{\small{$q_k$}}
 \end{picture}
 \begin{picture}(75,40)(0,-10)
  \SetWidth{1}
  \Line[](10,20)(30,20)
  \Text(8,20)[rc]{\small{$q_i$}}
  \Line[](10,0)(30,0)
  \Text(8,0)[rc]{\small{$q_j$}}
  \Line[dash,dashsize=2.5](30,20)(30,0)
  \Text(35,10)[lc]{\small{$S$}}
  \Line[](30,20)(50,20)
  \Text(52,20)[lc]{\small{$Q$}}
  \Line[](30,0)(50,0)
  \Text(52,0)[lc]{\small{$q_k$}}
 \end{picture}
 \begin{picture}(75,40)(0,-10)
  \SetWidth{1}
  \Line[](10,0)(30,0)
  \Text(8,0)[rc]{\small{$q$}}
  \Gluon(10,20)(30,20){2}{4}
  \Text(8,20)[rc]{\small{$g$}}
  \Line[](30,0)(30,20)
  \Text(35,10)[lc]{\small{$Q$}}
  \Photon(30,0)(50,0){3}{3}
  \Text(52,0)[lc]{\small{$V$}}
  \Line[](30,20)(50,20)
  \Text(52,20)[lc]{\small{$Q$}}
 \end{picture}
 \begin{picture}(75,40)(0,-10)
  \SetWidth{1}
  \Line[](10,0)(30,0)
  \Text(8,0)[rc]{\small{$q$}}
  \Gluon(10,20)(30,20){2}{4}
  \Text(8,20)[rc]{\small{$g$}}
  \Line[](30,0)(30,20)
  \Text(35,10)[lc]{\small{$Q$}}
  \Line[dash,dashsize=2.5](30,0)(50,0)
  \Text(52,0)[lc]{\small{$S$}}
  \Line[](30,20)(50,20)
  \Text(52,20)[lc]{\small{$Q$}}
 \end{picture}
\end{center}

\begin{center}
 \begin{picture}(75,40)(0,-10)
  \SetWidth{1}
  \Line[arrow,arrowpos=0.5,arrowlength=2,arrowwidth=1,arrowinset=0.2](10,20)(20,10)
  \Text(8,20)[rc]{\small{$q_i$}}
  \Line[arrow,arrowpos=0.5,arrowlength=2,arrowwidth=1,arrowinset=0.2](20,10)(10,0)
  \Text(8,0)[rc]{\small{$\bar q_j$}}
  \Photon(20,10)(40,10){3}{3}
  \Text(30,21)[tc]{\small{$V$}}
  \Line[](40,10)(50,20)
  \Text(52,20)[lc]{\small{$Q$}}
  \Line[](50,0)(40,10)
  \Text(52,0)[lc]{\small{$q_k$}}
 \end{picture}
 \begin{picture}(75,40)(0,-10)
  \SetWidth{1}
  \Line[arrow,arrowpos=0.5,arrowlength=2,arrowwidth=1,arrowinset=0.2](10,20)(20,10)
  \Text(8,20)[rc]{\small{$q_i$}}
  \Line[arrow,arrowpos=0.5,arrowlength=2,arrowwidth=1,arrowinset=0.2](20,10)(10,0)
  \Text(8,0)[rc]{\small{$\bar q_j$}}
  \Line[dash,dashsize=2.5](20,10)(40,10)
  \Text(30,20)[tc]{\small{$S$}}
  \Line[](40,10)(50,20)
  \Text(52,20)[lc]{\small{$Q$}}
  \Line[](50,0)(40,10)
  \Text(52,0)[lc]{\small{$q_k$}}
 \end{picture}
 \begin{picture}(75,40)(0,-10)
  \SetWidth{1}
  \Gluon(20,10)(10,20){2}{3}
  \Text(8,20)[rc]{\small{$g$}}
  \Line[](10,0)(20,10)
  \Text(8,0)[rc]{\small{$q$}}
  \Line[](20,10)(40,10)
  \Text(30,20)[tc]{\small{$q$}}
  \Line[](40,10)(50,20)
  \Text(52,20)[lc]{\small{$Q$}}
  \Photon(50,0)(40,10){3}{3}
  \Text(52,0)[lc]{\small{$V$}}
 \end{picture}
 \begin{picture}(75,40)(0,-10)
  \SetWidth{1}
  \Gluon(20,10)(10,20){2}{3}
  \Text(8,20)[rc]{\small{$g$}}
  \Line[](10,0)(20,10)
  \Text(8,0)[rc]{\small{$q$}}
  \Line[](20,10)(40,10)
  \Text(30,20)[tc]{\small{$q$}}
  \Line[](40,10)(50,20)
  \Text(52,20)[lc]{\small{$Q$}}
  \Line[dash,dashsize=2.5](50,0)(40,10)
  \Text(52,0)[lc]{\small{$S$}}
 \end{picture}
\end{center}
\caption[Feynman diagrams for single production of a generic XQ.]{Feynman diagrams for single production of a generic XQ. XQs can interact with SM quarks both through charged currents and neutral currents, allowing FCNCs also within SM states in diagrams with $q_i-q_j-\{V,S\}$ interactions. Arrows on fermion lines have been removed to account for both particles and antiparticles, when necessary. Notice that not all diagrams are allowed for a specific XQ (e.g. neutral currents are not allowed for quarks with exotic electric charges).}
\label{fig:SingleVL}
\end{figure}

On the other hand the single production processes, that we show in Fig.~\ref{fig:SingleVL}, always involve electroweak couplings and depend on the value of the coupling between the XQ and the particles it decays. It is therefore weaker than pair production as well as \emph{model dependent}. Yet it has to be mentioned that the pair production cross section decreases faster than single production when the XQ mass increases due to different PDF scaling, meaning that single production becomes eventually the dominant process when the mass is large enough. The XQ mass corresponding to the equivalence between pair and single production cross sections depends on the specific model and value of the coupling.
  
In consequence we will only consider QCD pair production in the following because it is the dominant process in the mass region of interest and that the cross section only depends on the XQ mass making it model independent.

\subsection{Decay and Narrow-Width Approximation}

The decay channels of XQs are model dependent too, and this is the most relevant problem when trying to interpret experimental bounds on new heavy quarks, due to the fact that these bounds are generally obtained under strong assumptions on the BRs of the new states.

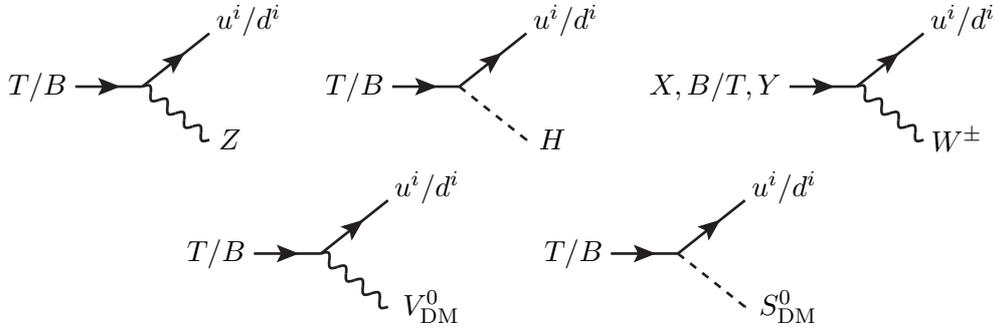
\begin{figure}[ht]
\begin{center}
\begin{picture}(115,50)(0,0)
    \SetWidth{1.0}
    \SetColor{Black}
    \Line[arrow](5,30)(30,30)
    \Text(2,30)[rc]{$T / B$}
    \Line[arrow](30,30)(55,50)
    \Text(82,55)[rc]{$u^i / d^i$}
    \Photon(55,10)(30,30){2}{4}
    \Text(67,10)[rc]{$Z$}
\end{picture} 
\begin{picture}(145,50)(0,0)
    \SetWidth{1.0}
    \SetColor{Black}
    \Line[arrow](5,30)(30,30)
    \Text(2,30)[rc]{$T / B$}
    \Line[arrow](30,30)(55,50)
    \Text(82,55)[rc]{$u^i / d^i$}
    \Line[dash](55,10)(30,30)
    \Text(70,10)[rc]{$H$}
\end{picture} 
\begin{picture}(95,70)(0,0)
    \SetWidth{1.0}
    \SetColor{Black}
    \Line[arrow](5,30)(30,30)
    \Text(2,30)[rc]{$X,B / T,Y$}
    \Line[arrow](30,30)(55,50)
    \Text(82,55)[rc]{$u^i / d^i$}
    \Photon(55,10)(30,30){2}{4}
    \Text(77,10)[rc]{$W^\pm$}
\end{picture}
\end{center}

\begin{center} 
\begin{picture}(130,50)(0,0)
    \SetWidth{1.0}
    \SetColor{Black}
    \Line[arrow](5,30)(30,30)
    \Text(2,30)[rc]{$T / B$}
    \Line[arrow](30,30)(55,50)
    \Text(82,55)[rc]{$u^i / d^i$}
    \Photon(55,10)(30,30){2}{4}
    \Text(82,10)[rc]{$V^0_{\rm DM}$}
\end{picture} 
\begin{picture}(95,50)(0,0)
    \SetWidth{1.0}
    \SetColor{Black}
    \Line[arrow](5,30)(30,30)
    \Text(2,30)[rc]{$T / B$}
    \Line[arrow](30,30)(55,50)
    \Text(82,55)[rc]{$u^i / d^i$}
    \Line[dash](55,10)(30,30)
    \Text(82,10)[rc]{$S^0_{\rm DM}$}
\end{picture} 
\end{center}
\caption[Feynman diagrams for decay of a generic XQ decaying to SM particles or DM.]{Feynman diagrams for decay of a generic XQ. Above the decays into SM particles (visible decay), below the decays into DM (invisible decay).}
\label{fig:DecayVL}
\end{figure}

\begin{figure}[!ht]
{\centering
\includegraphics[height=6cm,angle=270]{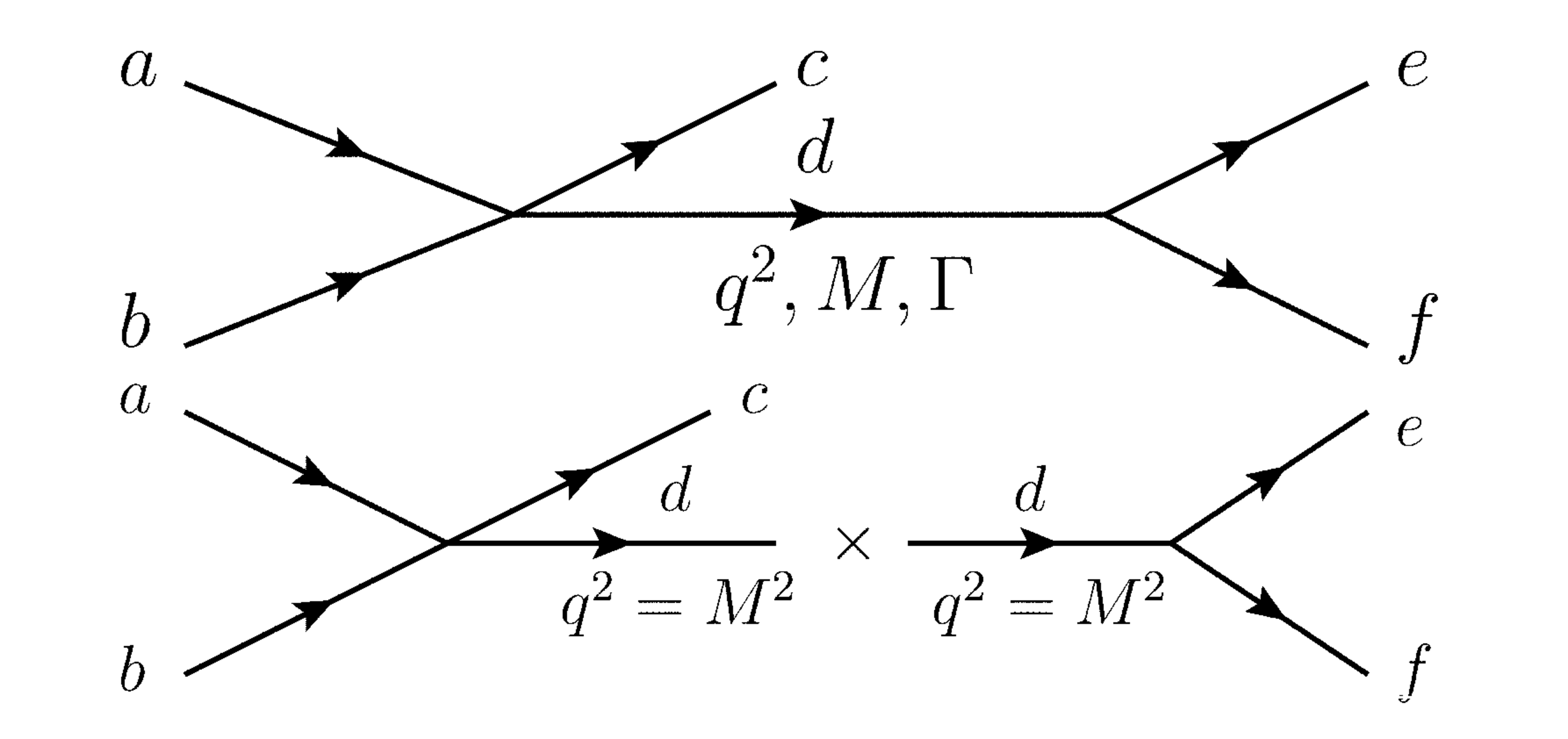}
\caption[Splitting of a process $a \  b \rightarrow c \  e \  f$ into production $a \  b \rightarrow c \  d$ and decay $d \rightarrow e \  f$.]{Splitting of a process $a \  b \rightarrow c \  e \  f$ into production $a \  b \rightarrow c \  d$ and decay $d \rightarrow e \  f$. In the NWA, the particle $d$ is massive and on-shell.}
\label{fig:splitting}
}
\end{figure}

A simple way to remove some of the model-dependency when considering the decay of XQ is to use the Narrow-Width Approximation (NWA). It is a widely applied and useful way to simplify the calculation of complicated processes involving the resonant production of an unstable particle and its decay. The basic idea is to factorise the whole process into the on-shell production and the subsequent decay, as show in Fig. \ref{fig:splitting} for an arbitrary process $a \  b \rightarrow c \  e \  f$.

The particle $d$ is intern to the diagram, its propagator is 
\begin{eqnarray}
\frac{i}{q^2 - M^2 - i M \Gamma} & \qquad \text{if } d \text{ is scalar} \\
\frac{i(\cancel{q} + M)}{q^2 - M^2 - i M \Gamma} & \qquad \text{if } d \text{ has a spin of } 1/2  \\
\frac{-i(g_{\mu\nu} - \frac{q_\mu q_\nu}{M^2})}{q^2 - M^2 - i M \Gamma} & \qquad \text{if } d \text{ has a spin of } 1
\end{eqnarray}
In the first case with a scalar propagator, the total matrix element of the event is
\begin{equation}
\mathcal{M} = \mathcal{M}_P \frac{1}{q^2 - M^2 - i M \Gamma} \mathcal{M}_D
\end{equation}
and the squared matrix element is\footnote{Note that the squared propagator become a Breit-Wigner distribution.}
\begin{equation}
\abs{\bar{\mathcal{M}}}^2 = \abs{\mathcal{M}_P}^2 \frac{1}{(q^2 - M^2)^2 + (M \Gamma)^2} \abs{\mathcal{M}_D}^2
\end{equation}
If the width $\Gamma$ of the particle $d$ is much smaller than its mass $M$ ($\Gamma \ll M$), we have 
\begin{equation}
\frac{1}{(q^2 - M^2)^2 + (M \Gamma)^2} \rightarrow \frac{\pi}{M \Gamma} \cdot \delta(q^2 - M^2)
\end{equation}
which means that off-shell effects are suppressed and that the particle $d$ can be considered as on-shell. Thanks to this result and under some other conditions (see. \cite{Fuchs:2013} for more details) we can show that the total cross section of this event verify, with an error of $\mathcal{O} ( \Gamma/M )$,
\begin{equation}
\sigma \simeq \sigma_P \cdot BR \qquad \text{where} \qquad BR = \frac{\Gamma_D}{\Gamma} \quad \text{is the BR}
\end{equation}
where $\sigma_P$ is the production cross section, $\Gamma_D$ the partial decay width into the particles in the final state of the considered process, and $\Gamma$ the total decay width of the unstable particle. This result can also be proved for spin $1/2$ or $1$ propagator (see \cite{Uhlemann:2007}, Sect. 2.2 for more details). For the example of the Fig. \ref{fig:splitting}, this result can be rewritten as $\sigma_{ab \rightarrow cef} \simeq \sigma_{ab \rightarrow cd} \cdot BR_{d \rightarrow ef}$. Similarly, in the NWA we can write the total cross section for a process such as $pp \rightarrow T \bar{T} \rightarrow W^+ b Z \bar{t}$ as $\sigma_{\rm total} \simeq \sigma_{pp \rightarrow T \bar{T}} \cdot BR_{T \rightarrow W^+ b} \cdot BR_{\bar{T} \rightarrow Z \bar{t}}$.

\paragraph*{Use of this approximation}

Usually the experimental searches focus on scenarios where the widths of the XQs are small (with respect to their masses), such that it is possible to use the NWA to factorize the production and decay parts of the scattering amplitudes, thus neglecting terms of $\mathcal O(\Gamma/M)^n$ (the power $n$ depending on the observable). This approximation is particularly useful in processes where the XQs are produced in pairs via QCD interactions, such that the production cross section depends only on the XQ mass and the assumptions about the XQ interactions with the SM quarks are encoded in their BRs. However, the width of the XQs may not always be small enough for the above approximation to hold: if the XQ couplings are numerically large or if the XQ has many decay channels, the total width may increase to sizeable values, so that it is not possible to factorize production from decay. In this case, only the analysis of the full process, from the initial state to the XQ decay products, can provide a good description of the kinematics of the final states and thus of the determination of the limits on the XQ and DM masses from experimental searches. These large width effects will be considered in detail in Sec. \ref{sec:VLQ_width} and \ref{sec:VLQDM_width}.

%% ------------- --------------------- --------------------- ---------------------- --------------------- ---------------------- ----------------------

\section{Past and current searches}
\label{sec:searches}

Various searches of new heavy states have been undertaken both at Tevatron and at the LHC, though no evidence for the existence of other quarks, beside those of the SM, has been obtained. Direct bounds on ChQs can be interpreted as bound on VLQs, but it must be stressed that decay channels of VLQs are different from decay channels of ChQs. For VLQs charged and neutral currents can have similar BRs, therefore searches performed with specific assumptions on the heavy state decay channel can give a rough idea of the bounds on VLQ mass, once rescaled with the actual BR in the specific channel. Note that all these searches only considered XQs with a relatively small width so they could use the NWA.

\begin{figure}[!ht]
{\centering
\includegraphics[width=.49\textwidth]{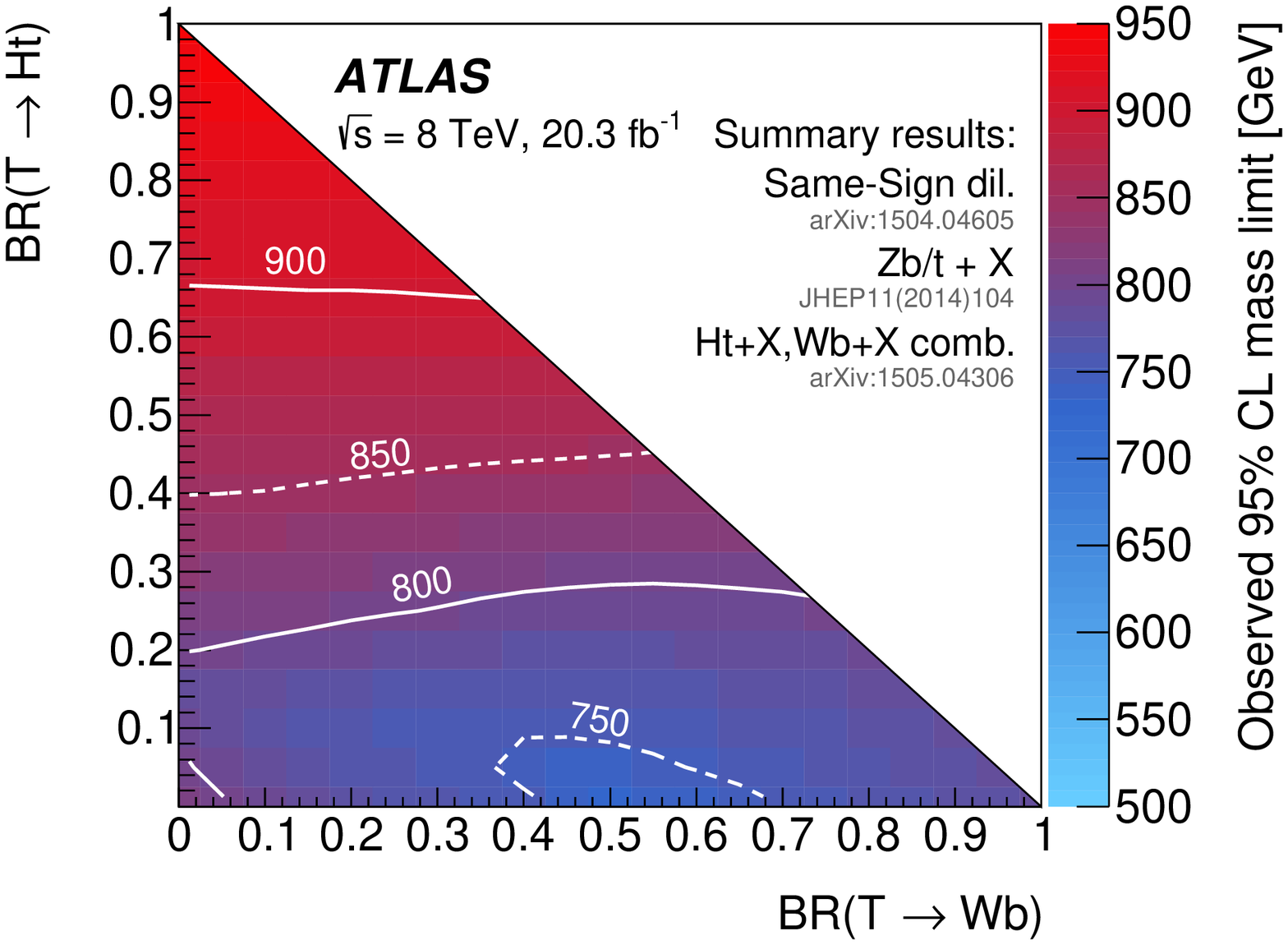}
\includegraphics[width=.49\textwidth]{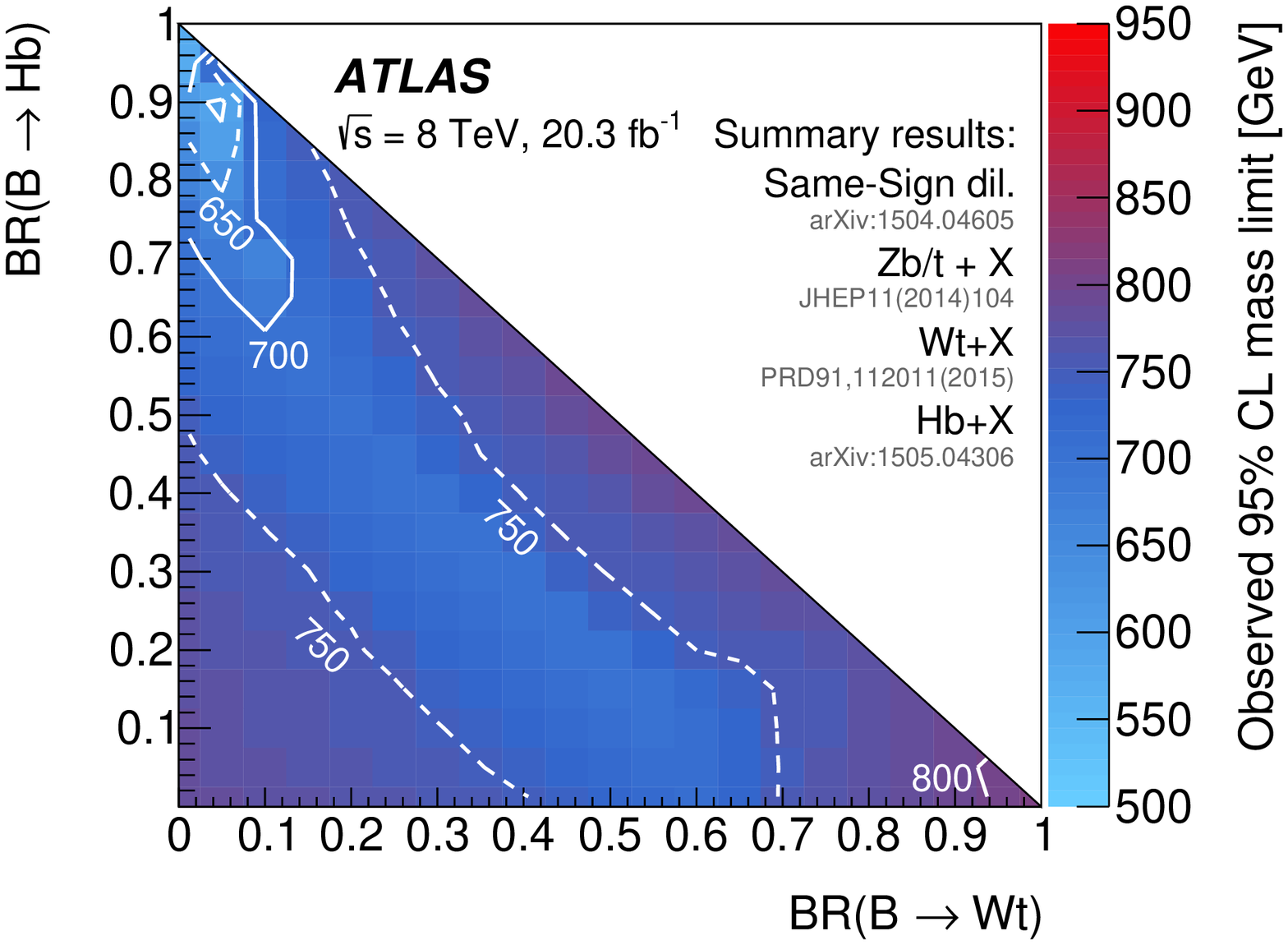}
\caption[Observed lower limits at 95\% C.L. on the mass of VLQs $T$ and $B$ for ATLAS searches with 20/fb of 8 TeV data.]{Observed lower limits at 95\% C.L. on the mass of VLQs $T$ (left) and $B$ (right) for ATLAS searches with 20/fb of 8 TeV data\footnote{Some 13 TeV results from the Run II of the LHC excluding VLQ up to 1300 GeV are starting to be published for this kind of VLQ searches while I am writting this thesis but not for all possible decay channels so we will not present them here.}. Mass exclusions are drawn sequentially for the different analyses. For a given bin in the BR plane, the strongest of all limits considered is shown (i.e. no combination is made of the different analyses, except for the Ht+X and Wb+X analyses which are combined).}
\label{fig:VLQBounds}
}
\end{figure}

An overview of all available searches of XQs (VLQs and ChQs) at Tevatron and at the LHC done before 2012 can be found in \cite{Okada:2012gy}. We show on Fig. \ref{fig:VLQBounds} the ATLAS combined bounds obtained with different 8 TeV searches for VLQs $T$ and $B$ coupling to third generation SM quarks only. We see that for any value of the BR a $T$ ($B$) lighter than 700 (600) GeV is excluded.

In the following we present an overview of the most recent searches for VLQs at the run 2 of the LHC for $p p$ collisions at $\sqrt{s}$ = 13 TeV recorded in 2015 and early 2016, focusing on the assumptions that have been made to obtain the bounds on the heavy quark masses. More details on single searches (kinematic cuts, detector parameters...) can be found in the original publications. Note that we do not present searches for ChQs at the run 2 of the LHC since none were performed at the energy of 13 TeV.

\subsection{Searches for VLQ decaying to SM particles at 13 TeV}

\subsubsection{Searches from the ATLAS collaboration}

\paragraph*{ATLAS @ 3.2 fb$^{-1}$}

In \cite{ATLAS:2016ovj} a search for a singly-produced VLQ $Q \in \{T,Y \}$ decaying to $W b$ and where the $W$ boson decays leptonically is performed. For a $QWb$ coupling strength of $\sqrt{(\kappa^{Wb}_L)^2+(\kappa^{Wb}_R)^2} = 1/\sqrt{2}$, the observed 95 \% CL lower limit on the $Q$ mass is 1.44 TeV. The results are also interpreted as limits on the $QWb$ coupling strength and the mixing with the SM sector for a singlet $T$ quark or a $Y$ quark from a $(B,Y)$ doublet. The smallest excluded coupling-strength values are obtained for VLQ masses around 1000 GeV; they are as small as $|\kappa^{Wb}_L| = 0.45$ for a $T$ quark and $\sqrt{(\kappa^{Wb}_L)^2 + (\kappa^{Wb}_R)^2} = 0.33$ for a $Y$ quark.

In \cite{TheATLAScollaboration:2016gxs} a search for pair production of VLQs $T$ with significant BR to a Higgs boson and a top quark, $T \bar{T} \to Ht+X$, is presented. Data are analysed in the lepton-plus-jets final state, characterised by an isolated electron or muon with high transverse momentum, large missing transverse momentum and multiple jets. 95\% CL lower limits are derived on the $T$ mass under several BR hypotheses assuming contributions only from $T \to Wb, Zt, Ht$. The observed lower limits on the $T$ mass range between 700 GeV and 900 GeV for all possible values of the BRs into the three decay modes. Under the assumption of $BR(T \to Ht) = 1$, a limit of $M_T > 900$ GeV is obtained. The limits for a weak-isospin doublet and singlet are $M_T > 800$ GeV and $M_T > 750$ GeV respectively.

\paragraph*{ATLAS @ 14.7 fb$^{-1}$}

In \cite{ATLAS:2016cuv} a search for the pair production of a heavy VLQ $T$ decaying to $W b$ is performed. Data are analysed in the lepton plus jets final state, characterised by a high-transverse-momentum isolated electron or muon, large missing transverse momentum, multiple jets, of which at least one is b-tagged, and a single large-radius jet or system of two small-radius jets identified as originating from the hadronic decay of a $W$ boson. Under two different assumptions, that of a BR $BR(T \to Wb) = 1$ or an $SU(2)$ singlet, an observed 95\% CL lower limit on the VLQ mass of 1090 GeV and 810 GeV is obtained, respectively. For a VLQ $T$, under the assumption that only the $T \to Wb$, $T \to Zt$ and $T \to Ht$ decay modes contribute, 95\% CL upper limits are derived for various masses in the two-dimensional plane of $BR(T \to Ht)$ versus $BR(T \to Wb)$, ranging between 500 GeV and 1100 GeV.

% In \cite{ATLAS:2016qlg} a search for pair produced VLQs $T$ in the $Zt+X$ channel using final states with exactly one lepton, at least four jets, and large missing transverse momentum is performed. The observed 95\% CL lower limits on the $T$ mass are 810 GeV for the weak-isospin singlet model, 1.03 TeV for the weak-isospin doublet model and 1.13 TeV for the pure $Zt$ decay mode. 

\paragraph*{ATLAS @ 36.1 fb$^{-1}$}

In \cite{Aaboud:2017qpr} a search for pair produced VLQs $T$ using events with exactly one lepton, at least four jets, and large missing transverse momentum is performed. The search is optimised for the $Z(\to \nu \nu)t+X$ decay channel. The observed 95\% CL lower limits on the $T$ mass are 870 GeV for the weak-isospin singlet model, 1.05 TeV for the weak-isospin doublet model and 1.16 TeV for the pure $Z t$ decay mode. Limits are also set on the mass as a function of the decay BRs, excluding large parts of the parameter space for masses below 1 TeV.

\subsubsection{Searches from the CMS collaboration}

\paragraph*{CMS @ 2.3 fb$^{-1}$}

In \cite{Sirunyan:2017ezy} a search for single production of VLQs $T$ and $B$ decaying into a $Z$ boson and a top or a bottom quark, respectively, is presented. An exotic $T$ quark production mode through the decay of a heavy $Z^\prime$ resonance is also considered. The search is performed in events with a $Z$ boson decaying leptonically, accompanied by a bottom or a top quark decaying hadronically. Products of production cross section and branching fraction for $T$ and $B$ quarks from 1.26 and 0.13 pb are excluded at 95\% confidence level for the range of resonance mass considered, which is between 0.7 and 1.7 TeV. 

In \cite{Sirunyan:2016ipo} a search is performed for single electroweak production of a VLQ $T$ in association with a top or bottom quark. The search targets $T$ quarks decaying to a top quark and a Higgs boson in fully hadronic final states. For a $T$ quark with mass above 1 TeV the daughter top quark and Higgs boson are highly Lorentz-boosted and can each appear as a single hadronic jet. Upper limits at 95\% confidence level are set on the product of the single $T$ quark production cross sections and the branching fraction $BR(T \to tH)$, and these vary between 0.31 and 0.93 pb for $T$ quark masses in the range 1000-1800 GeV. 

In \cite{Sirunyan:2017tfc} a search is presented for VLQs $T$ and $Y$, decaying into a $b$ quark and a $W$ boson, which is produced singly in association with a light flavour quark and a $b$ quark. The search is carried out using events containing one electron or muon, at least one $b$-tagged jet with large transverse momentum, at least one jet in the forward region of the detector, and missing transverse momentum. Upper limits at 95\% CL are set on the cross sections for single production of $Y$ and $T$ quarks in the mass range from 0.70 to 1.80 TeV. For $Y$ quarks with coupling of 0.5 and $BR(Y \to bW)= 1$, the observed lower mass limits are 1.40 TeV.

\paragraph*{CMS @ 2.6 fb$^{-1}$}

In \cite{Sirunyan:2017usq} a search for pair production of $T$ and $B$ quarks is presented. The $T$ and $B$ are assumed to decay into $W, Z$ or $H$ and a third generation quark. This search is performed in final states with one charged lepton and several jets, exploiting techniques to identify $W$ or Higgs bosons decaying hadronically with large transverse momenta. Upper limits at 95\% confidence level on the $T$ pair production cross section are set that exclude $T$ masses below 860 GeV in the singlet, and below 830 GeV in the doublet branching fraction scenario. For other branching fraction combinations with $BR(tH) + BR(bW) \geqslant 0.4$, lower limits on the $T$ quark range from 790 to 940 GeV. Limits are also set on pair production of singlet VL $B$ quarks, which can be excluded up to a mass of 730 GeV.

\subsection{Searches for XQ decaying to DM}

No searches for VLQ decaying to DM have been performed at the LHC yet, the only ones exploring such scenarios were done at Tevatron \cite{Aaltonen:2011rr,Aaltonen:2011tq} and excluded the presence of a $T$ of mass smaller than 360 (400) GeV for a DM particle lighter than 100 (70) GeV. 

Yet some LHC searches for SUSY have been recasted to draw limits on VLQ scenarios. Indeed, a large number of searches for final states containing jets and/or leptons plus $\MET$ have been designed by the ATLAS and CMS SUSY groups~\cite{ATLAS:2012hpa}, and the interpretations of the results are typically limits in some SUSY simplified model. Examples are multi-jet + $\MET$ searches being interpreted as limits in the the gluino--neutralino mass plane, or searches for the $t\bar t+\MET$ final state being interpreted in terms of stops decaying to top+neutralino. The same searches can be used to put constraints on scenarios leading to final states with $\MET$ generated by the production of XQs decaying to a bosonic DM candidate and we present here some of the limits obtained with such a recasting.

In \cite{Cacciapaglia:2013wha}, a re-interpretation of a few ATLAS and CMS SUSY searches using 5 fb$^{-1}$ of data at 7~TeV in terms of UED signatures is done, using among others a simplified scenario with top partners decaying to DM and light quarks. A recurrence scale of 600 GeV is excluded at a CL above 99.9\%, whereas a recurrence scale of 700 GeV is disfavoured at the 72\% confidence level.

In \cite{Edelhauser:2015ksa} the applicability of SUSY simplified model results to new physics scenarios with same spin SM partners was analysed also in the context of UED, focussing on the so-called T2 topology which corresponds to squark-antisquark production in the limit of a heavy gluino. Despite sizeable differences in the detection efficiencies due to the spin of the new particles, the limits on particle masses are found to be rather similar, meaning that the supersymmetric simplified models employed in current experimental analyses also provide a reliable tool to constrain same spin BSM scenarios.

In \cite{Baek:2016lnv} a study of constraints and LHC signatures of a scenario with a VLQ $T$ decaying to a top quark and scalar DM $S$ has been performed. $T\bar{T}$ pair produced at the LHC will decay $100\%$ into $t\bar{t} + \MET$ signal when kinematically open. The latest ATLAS 13 TeV 13.2 fb$^{-1}$ data can excluded $M_T$ between 300
(650) and 1150 (1100) GeV for $M_S$ = 40 (400) GeV and the exclusion region can reach up to $M_S\sim 500$ GeV.

Finally, we showed in \cite{Kraml:2016eti} that limits on scenarios featuring a VLQ $T$ decaying to scalar or vector can be obtained by rescaling the limits obtained for SUSY with the VLQ cross section in the NWA. This work is presented in detail in Sec. \ref{sec:VLQ-SUSY}

%% ------------- --------------------- --------------------- ---------------------- --------------------- ---------------------- ----------------------
%% Chapter III
%% ------------- --------------------- --------------------- ---------------------- --------------------- ---------------------- ---------------------- 

\chapter{New quark decaying to Standard Model particles} \label{Chapter:XQ}

In this Chapter we focus on models featuring XQs decaying to SM particles. 
%and we will only consider top partners $T_i$. The possibilities of decay for these quarks will therefore be $T_i \to \{ W^+ b, Z t, H t \}$.

%% ------------- --------------------- --------------------- ---------------------- --------------------- ---------------------- ----------------------

%\section{The XQCAT project} \label{sec:XQCAT}

%% ------------- --------------------- --------------------- ---------------------- --------------------- ---------------------- ----------------------

\section{Interference effects in pair production of XQs} \label{sec:VLQ_inteference}

Experimental searches for VLQs usually adopt a phenomenological approach, assuming that only one new state $Q_V$ is present beyond the SM and, in order to be as model independent as possible, searches usually consider QCD pair production, although single production has also been explored \cite{ATLAS:2016ovj,Sirunyan:2017ezy,Sirunyan:2016ipo,Sirunyan:2017tfc}. Most models, however, predict in general the existence of a new \textit{quark sector}, which implies the presence of more than one new coloured state, some of which being possibly degenerate or nearly degenerate. If two or more quarks of a given model can decay to the same final state, interference effects should be considered in order to correctly evaluate the total cross section and the kinematical distributions of the signal. Current bounds on the masses of new states obtained assuming the presence of only one new particle cannot be easily reinterpreted in more complex scenarios containing more than one new quark, unless interference effects in the total cross section and kinematical distributions are taken into account.

Following the same approach as in \cite{Barducci:2013zaa}, we show that this can be done through a simple formula, which enables one to correctly model such interference effects at both inclusive and exclusive levels\footnote{Most of this work was not done during my PhD, but during a three months master internship I did in Summer 2013 at the University of Southampton, working with the same supervisors on a similar project.}. Note that even though we present the results for visible decays here, such interference effects can also occur in production and invisible decays, their study is one of our next project and we expect to get similar results.

\subsection{Analytical estimation of the interference effects for pair vector-like quarks production}

\subsubsection{Analytical ``master formula" for the interference}

We will assume throughout the analysis that the new heavy quarks undergo two-body decays to SM particles and we will not consider chain decays of heavy quarks into other new states. This approach is generally valid for  models in which the new quarks interact with the SM ones only through Yukawa couplings.
Therefore, the new heavy quarks $Q_V^i$ can decay into either SM gauge bosons or the Higgs boson and ordinary quarks. We will assume that flavour changing neutral currents are present and therefore decays such as $T\to Zt$ and $T\to Ht$ are allowed, alongside $T\to W^+ b$. This is consistent with the embedding of new VLQs in extensions of the SM as we have seen in the previous Chapter. 
If more than one VLQ species is present in the model, then there are two ways to obtain a given final state:
\begin{itemize}
 \item[A.]  The VLQs have the same charge, so a $Q_V^i \bar{Q}_V^i$ pair  decays into the same final state, e.g., \\
 $T_{1,2}\bar{T}_{1,2} \to  W^+W^-b \bar b (W^+Zb\bar t)$;
 \item[B.]  $Q_V^i$ quarks have  different charges  but after decay  their  pair leads to the same final state, e.g.,\\
 $B \bar B \to (t W^-)(\bar t W^+)$ and $X_{5/3} \bar X_{5/3} \to (t W^+)(\bar t W^-)$.
\end{itemize}

We have verified that, while the interference in case B can be safely neglected when the masses of the VLQs are much larger than the masses of the decay products (which is usually the case), because of the largely different kinematics of the final states, case A has to be considered carefully. 
It is worth mentioning that, for the classes of models under consideration, we have quarks of identical charge and with couplings to the same particles, so that the effects of  the mixing between such quarks at loop level could be important and should (eventually) be taken into account. These effects are model dependent though and involve computation of loops that may contain states belonging to new sectors (e.g., new gauge bosons). 
We assume in the following that these effects can be computed and that particle wave-function as well as Feynman rules are already formulated for mass-eigenstates, i.e., the masses and widths that we will be using are those obtained after computing the rotations of the states due to the one-loop mixing terms, so that interference effects can then be explored in a model independent way.

The measure of the interference between $Q_V^i$ and $Q_V^j$ pairs of species $i$ and $j$ decaying into the same final state can be defined by the following simple expression
\begin{equation}
F_{ij}= 
\frac{\sigma^{\rm int}_{ij}}{\sigma_i+\sigma_j}
=
\frac{\sigma^{\rm tot}_{ij}-(\sigma_i+\sigma_j)}{\sigma_i+\sigma_j}
=
\frac{\sigma^{\rm tot}_{ij}}{\sigma_i+\sigma_j}-1
\label{orderparameter}
\end{equation}
where
$\sigma^{\rm tot}_{ij}$ is the total cross section of $Q_V^i$ and $Q_V^j$ pair production including their interference, the $\sigma_{i,j}$'s are their individual production rates while $\sigma^{\rm int}_{ij}$ represents the value of the interference.

The interference term $F_{ij}$ ranges from $-1$ to 1. Completely constructive interference is obviously achieved when $\sigma^{\rm int}_{ij} = \sigma_i + \sigma_j$, while completely destructive interference is obtained when $\sigma^{\rm int}_{ij} = -(\sigma_i + \sigma_j)$.

We have seen in Chapter \ref{Chapter:IntroXQ} that, under very general hypotheses, the couplings of VLQs with SM quarks are dominantly chiral and that the chirality of the coupling depends on the VLQ representation under $SU(2)$. If the VLQ has a half-integer isospin (doublets, quadruplets, ...) couplings are dominantly right-handed while, if the VLQ has to an integer isospin (singlets, triplets, ...) couplings are mostly left-handed. This feature is valid for a wide range of hypotheses about the mixing between VLQs and SM quarks and between VLQs themselves. However, if Yukawa couplings between VLQs and the Higgs boson are large, it is possible to achieve couplings with non-dominant chiralities.

Our results about the analysis of interference effects can be applied in both cases, therefore, we divide our study in two parts. Firstly, we show the results for the interference of two $T$s with the same chiral couplings. Then we generalise the analysis to the case where the couplings of the heavy quarks do not exhibit a dominant chirality.

We would now like to make the ansatz that, in case of chiral new quarks $i$ and assuming small $\Gamma_i/m_i$ values, the interference is proportional to the couplings of the new quarks to the final state particles and to the integral of the {\it scalar} part of the propagator. The range of validity of the ansatz in terms of the $\Gamma_i/m_i$ ratio is explored in a subsequent section.

If the couplings are chiral for both heavy quarks and the chirality is the same we have
\begin{equation}
 \label{eq:intpart}
 \sigma^{\rm int}_{ij} \propto 2 Re\left[g_{i1} g_{j1}^* g_{i2}^* g_{j2} \left(\int_{-\infty}^{+\infty} d q^2 \mathcal{P}_i\mathcal{P}_j^*\right)^2\right]
\end{equation}
where 1 and 2 refer to the two decay branches (1 corresponding to the quark branch and 2 to the antiquark branch) while the scalar part of the propagator for any new quark $i$ is given by
\begin{equation}
\mathcal{P}_i=\frac{1}{q^2-m_i^2+im_i\Gamma_i}.
\end{equation}
The cross section for pair production of species $i$ only is 
\begin{equation}
  \label{eq:prodpart}
 \sigma_{i} \propto |g_{i1}|^2 |g_{i2}|^2 \left(\int d q^2 \mathcal{P}_i\mathcal{P}_i^*\right)^2
\end{equation}
and an analogous expression can be written for species $j$.

Therefore, the analytical  expression which should describe the interference in the case of chiral XQ pair production of species $i$ and $j$ followed by their decay into the same final state, is given by
\begin{equation}
\label{eq:int1}
\kappa_{ij}=\frac{2 Re\left[g_{i1} g_{j1}^* g_{i2}^* g_{j2} \left(\int \mathcal{P}_i\mathcal{P}_j^*\right)^2\right]}{|g_{i1}|^2 |g_{i2}|^2 \left(\int \mathcal{P}_i\mathcal{P}_i^*\right)^2+|g_{j1}|^2 |g_{j2}|^2 \left(\int \mathcal{P}_j\mathcal{P}_j^*\right)^2}.
\end{equation}
Ultimately, $\kappa_{ij}$ should closely describe the true value of the interference term $F_{ij}$ from Eq.~(\ref{orderparameter}) if the ansatz is correct.

After integration $\kappa_{ij}$ takes the following form:
\begin{equation}
\label{eq:int2}
\kappa_{ij}=\frac{8 Re[g_{i1} g_{j1}^* g_{i2}^* g_{j2}] m_i^2 m_j^2 \Gamma_i^2 \Gamma_j^2}{|g_{j1}|^2|g_{j2}|^2m_i^2\Gamma_i^2+|g_{i1}|^2|g_{i2}|^2m_j^2\Gamma_j^2} \frac{(m_i\Gamma_i+m_j\Gamma_j)^2-(m_i^2 - m_j^2)^2}{\left((m_i\Gamma_i+m_j\Gamma_j)^2+(m_i^2 - m_j^2)^2\right)^2}.
\end{equation}

The previous expression can be generalised when the chirality of the coupling is not predominantly left or right. In the approximation in which the final states are massless (in practice, neglecting the top mass) only four sub-diagrams give a non-zero contribution, the ones corresponding to considering the following combinations of chiralities: $Q_1,Q_2,\bar Q_1,\bar Q_2$=$L,L,L,L$ or $L,L,R,R$ or $ R,R,L,L$ or $R,R,R,R$. If the masses of the final state objects cannot be neglected, the non-zero combinations would be 16 because any combination of $Q_1$ would interfere with any combination of $Q_2$, though interferences involving $LR$ or $RL$ flipping are suppressed by the mass of the quarks in the final state. Analogously to the previous case, we have numerically proven that neglecting the masses of the final states is a reasonable assumption in the range of XQ masses still allowed by experimental data, hence we will consider the final state quarks as massless.

The expression in Eq. (\ref{eq:int1}) can therefore be rewritten in the following way:
\begin{equation}
\kappa^{ab}_{ij}=\frac{2 Re\left[g^a_{i1} g_{j1}^{a*} g^{b*}_{i2} g_{j2}^{b} \left(\int \mathcal{P}_i\mathcal{P}_j^*\right)^2\right]}{|g^a_{i1}|^2 |g^b_{i2}|^2 \left(\int \mathcal{P}_i\mathcal{P}_i^*\right)^2+|g^a_{j1}|^2 |g^b_{j2}|^2 \left(\int \mathcal{P}_j\mathcal{P}_j^*\right)^2}=\frac{\mathcal{N}_{ij}^{ab}}{\mathcal{D}_{ij}^{ab}}, \qquad ab=LL,LR,RL,RR.
\end{equation}
After summing over all allowed topologies, we obtain the generalisation of Eq.(\ref{eq:int2}) as:
\begin{equation}
\label{eq:gen1}
\kappa^{gen}_{ij}=
\frac{\sum_{a,b=L,R}2 Re\left[g^a_{i1} g_{j1}^{a*} g^{b*}_{i2} g_{j2}^{b} \left(\int \mathcal{P}_i\mathcal{P}_j^*\right)^2\right]}{\sum_{a,b=L,R}|g^a_{i1}|^2 |g^b_{i2}|^2 \left(\int \mathcal{P}_i\mathcal{P}_i^*\right)^2+|g^a_{j1}|^2 |g^b_{j2}|^2 \left(\int \mathcal{P}_j\mathcal{P}_j^*\right)^2}=\frac{\sum_{ab}\kappa_{ij}^{ab}\mathcal{D}_{ij}^{ab}}{\sum_{ab}\mathcal{D}_{ij}^{ab}},
\end{equation}
which, after integration, becomes
\begin{eqnarray}
\label{eq:gen2}
\kappa^{gen}_{ij}&=&
\frac{8 Re[(g_{i1}^L g_{j1}^{L*} + g_{i1}^R g_{j1}^{R*})(g_{i2}^{L*}g_{j2}^L + g_{i2}^{R*}g_{j2}^R)] m_i^2 m_j^2 \Gamma_i^2 \Gamma_j^2}{\left((|g_{j1}^L|^2+|g_{j1}^R|^2)(|g_{j2}^L|^2+|g_{j2}^R|^2\right)m_i^2\Gamma_i^2+ \left( (|g_{i1}^L|^2+|g_{i1}^R|^2) (|g_{i2}^L|^2+|g_{i2}^R|^2) \right) m_j^2\Gamma_j^2}\cdot\nonumber \\ &&\frac{(m_i\Gamma_i+m_j\Gamma_j)^2-(m_i^2 - m_j^2)^2}{\left((m_i\Gamma_i+m_j\Gamma_j)^2+(m_i^2 - m_j^2)^2\right)^2}.
\end{eqnarray}

%%%%%%%%%%%%%%%%%%%%%%%%%%%%%%%%%%%%%%
%%%%%%%%%%%%%%%%%%%%%%%%%%%%%%%%%%%%%%

\subsubsection{Region of validity of the approximation}

When considering the production and decay of different heavy quarks which couple to the same SM  particles, interference at tree level is not the only one which should potentially be taken into account. Quarks with same quantum numbers can mix at loop level too, which results into the respective mixing matrix of the one-loop corrected propagators and their corresponding interference.
Mass and width eigenstates can be obtained by diagonalising the respective matrices, but the rotations are in general different for these two matrices, therefore mass and width eigenstates may be misaligned. A careful treatment of all such mixing effects is beyond the scope of this analysis but, in order to be able to apply our results, it is crucial to understand when the mixing effect can be neglected.

\begin{figure}[htb]
\begin{center}
 \begin{picture}(250,44)(0,0)
    \SetColor{Black}
    \Gluon(10,22)(50,22){4}{4.5}
    \Text(8,22)[rc]{$g$}
    \Line[arrow,arrowpos=0.5,arrowwidth=2](90,32)(50,22)
    \Text(60,29)[cb]{$\overline{Q}_J$}
    \Line[arrow,arrowpos=0.5,arrowwidth=2](50,22)(90,12)
    \Text(60,15)[ct]{$Q_J$}
    \Vertex(90,32){5}
    \Text(90,40)[cb]{$\delta_{JK}+\Sigma_{JK}$}
    \Vertex(90,12){5}
    \Text(90,4)[ct]{$\delta_{IJ}+\Sigma_{IJ}$}
    \Line[arrow,arrowpos=0.5,arrowwidth=2](130,42)(90,32)
    \Text(132,44)[lc]{$\overline{Q}_K$}
    \Line[arrow,arrowpos=0.5,arrowwidth=2](90,12)(130,2)
    \Text(132,0)[lc]{$Q_I$}
    \Text(250,22)[rc]{$I,J,K=1,2$}
  \end{picture}
\end{center}
\caption{Pair production of two heavy quarks $Q_1$ and $Q_2$, including loop mixing.}
\label{fig:pairprodwithloops}
\end{figure}
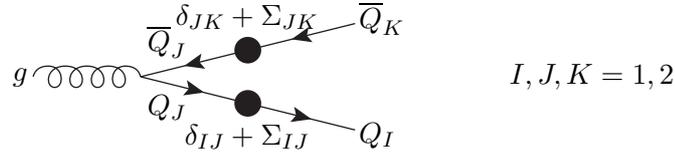

\begin{figure}[htb]
\begin{minipage}{.45\textwidth}
\begin{center}
\begin{center}
 \begin{picture}(150,120)(0,-10)
    \SetColor{Black}
    \Line[arrow,arrowpos=0.5,arrowwidth=2](10,70)(50,70)
    \Text(8,70)[rc]{$Q_I$}
    \Vertex(50,70){2}
    \Text(50,75)[rb]{$A^S$}
    \Arc[arrow,arrowpos=0.5,arrowlength=5,arrowwidth=2,arrowinset=0.2,clock](70,70)(20,-180,-360)
    \Text(70,94)[cb]{$m_f$}
    \Vertex(90,70){2}
    \Text(92,75)[lb]{$B^S$}
    \Arc[dash,dashsize=5,clock](70,70)(20,-0,-180)
    \Text(70,44)[ct]{$m_S$}
    \Line[arrow,arrowpos=0.5,arrowwidth=2](90,70)(130,70)
    \Text(132,70)[lc]{$Q_J$}
    \Text(72,15)[cc]{$A^{S}=(g_L^S)^I P_L + (g_R^S)^I P_R$}
    \Text(72,0)[cc]{$B^{S}=(g_L^S)^J P_L + (g_R^S)^J P_R$}
  \end{picture}
\end{center}
\end{center}
\end{minipage}
\begin{minipage}{.45\textwidth}
\begin{center}
 \begin{picture}(150,120)(0,-10)
    \SetColor{Black}
    \Line[arrow,arrowpos=0.5,arrowwidth=2](10,70)(50,70)
    \Text(8,70)[rc]{$Q_I$}
    \Vertex(50,70){2}
    \Text(50,75)[rb]{$A^V$}
    \Arc[arrow,arrowpos=0.5,arrowlength=5,arrowwidth=2,arrowinset=0.2,clock](70,70)(20,-180,-360)
    \Text(70,94)[cb]{$m_f$}
    \Vertex(90,70){2}
    \Text(92,75)[lb]{$B^V$}
    \PhotonArc[clock](70,70)(20,-0,-180){3}{7}
    \Text(70,44)[ct]{$m_V$}
    \Line[arrow,arrowpos=0.5,arrowwidth=2](90,70)(130,70)
    \Text(132,70)[lc]{$Q_J$}
    \Text(72,15)[cc]{$A^{V}=(g_L^V)^I \gamma^\mu P_L + (g_R^V)^I \gamma^\mu P_R$}
    \Text(72,0)[cc]{$B^{V}=(g_L^V)^J \gamma^\nu P_L + (g_R^V)^J \gamma^\nu P_R$}
  \end{picture}
\end{center}
\end{minipage}
\caption[Loop topologies for corrections to quark propagators.]{Loop topologies for corrections to quark propagators. The particles  in the loop can be any fermion, vector or scalar which are present in the model under consideration.}
\label{fig:loopnotation}
\end{figure}
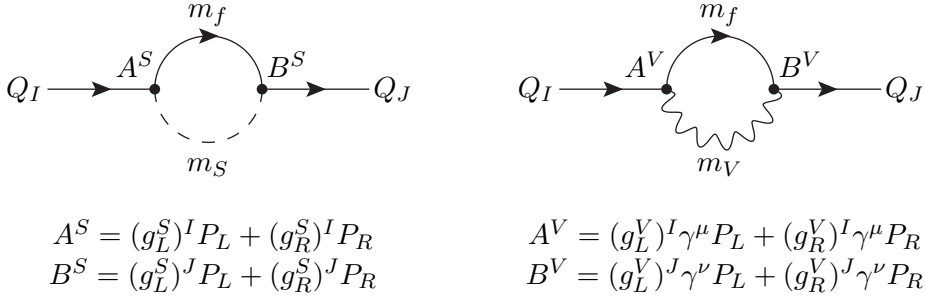

Let us consider the structure of the interference terms for the process of QCD pair production of two heavy quarks, $Q_1$ and $Q_2$, including only the one-loop corrections to the quark propagators and neglecting the vertices unmodified for simplicity. From now on we will consider only the imaginary part of the quark self-energies, that give the corrections to the quark widths, and we will assume real couplings for simplicity. A more detailed treatment of mixing effects under general assumptions in heavy quark pair production will be performed in a dedicated analysis. Considering only the case of s-channel exchange of the gluon for simplicity, and still not including the decays of the heavy quarks, the amplitude of the process depicted in Fig. \ref{fig:pairprodwithloops} is:
\begin{equation}
 \mathcal{M} = \bar{u}_I (\delta_{IJ}+\Sigma_{IJ}) P^+_J V^\sigma P^-_J (\delta_{JK}+\Sigma_{JK}) v_K \mathcal{M}^P_\sigma \quad\mbox{with}\quad I,J,K=1,2
\end{equation}
where the QCD amplitude terms and colour structure have been factorised into the vertex $V^\sigma$ and the term $\mathcal{M}^P_\sigma$, the propagators of the quark and antiquarks are $P^+$ and $P^-$, respectively, and $\Sigma$ represents the loop insertions. The loop contributions depend on the particle content of the model and therefore cannot be evaluated in a model independent way. However, it is straightforward to determine the structure of the loops by noticing that the only allowed topologies are fermion-scalar (fS) and fermion-vector (fV), see Fig. \ref{fig:loopnotation}.

These topologies can be evaluated for general masses and couplings of the particles  in the loops, and therefore the most general structure of the loop insertion is:
\begin{equation}
 \Sigma_{IJ} = \sum_{\mbox{fS loops}} \Sigma_{IJ}^{fS} + \sum_{\mbox{fV loops}} \Sigma_{IJ}^{fV}
\end{equation}
where, in Feynman gauge and adopting the Passarino-Veltman functions $B_0$ and $B_1$:
\begin{small}
\begin{eqnarray}
 \Sigma_{IJ}^{fS} \hspace*{-3mm} &=& \hspace*{-3mm} \left((g^S_L)^I (g^S_L)^J m_f B_0(p^2,m_f^2,m_S^2) + (g^S_R)^I (g^S_L)^J \slash p B_1(p^2,m_f^2,m_S^2) \right) P_L + L \hspace*{-0.5mm}  \leftrightarrow \hspace*{-0.5mm} R \\
 \Sigma_{IJ}^{fV} \hspace*{-3mm} &=& \hspace*{-3mm} \left(4(g^V_R)^I (g^V_L)^J m_f B_0(p^2,m_f^2,m_V^2) \hspace*{-0.5mm} - \hspace*{-0.5mm} 2 (g^V_L)^I (g^V_L)^J \slash p B_1(p^2,m_f^2,m_V^2) \right) P_L \hspace*{-0.5mm} + \hspace*{-0.5mm} L \hspace*{-0.5mm} \leftrightarrow \hspace*{-0.5mm} R
\end{eqnarray}
\end{small}
When $I=J$, the loop contributions correspond to a correction to the diagonal quark propagators while, when $I\neq J$, the loops correspond to the off-diagonal mixing between the quarks.
Without loosing generality, let us consider  the $I,K=1,2$ case, for which we can  define two amplitude matrices, corresponding to production of the quarks $J=1$ and $J=2$ that, through the loop-corrected propagators, become quarks $I,K=1,2$. 

The amplitude matrices are:
\begin{small}
\begin{eqnarray}
\mathcal{M}_{J=1} \hspace*{-2mm} &=& \hspace*{-2mm} \left( \hspace*{-2mm} 
\begin{array}{cc}
 \bar{u}_1 (1+\Sigma_{11}) P^+_1 V^\sigma P^-_1 (1+\Sigma_{11}) v_1 \mathcal{M}^P_\sigma & \bar{u}_1 (1+\Sigma_{11}) P^+_1 V^\sigma P^-_1 \Sigma_{12} v_2 \mathcal{M}^P_\sigma \\
 \bar{u}_2 \Sigma_{21} P^+_1 V^\sigma P^-_1 (1+\Sigma_{11}) v_1 \mathcal{M}^P_\sigma & \bar{u}_2 \Sigma_{21} P^+_1 V^\sigma P^-_1 \Sigma_{12} v_2 \mathcal{M}^P_\sigma
\end{array}
\hspace*{-2mm} \right) \\
\mathcal{M}_{J=2} \hspace*{-2mm} &=& \hspace*{-2mm} \left( \hspace*{-2mm} 
\begin{array}{cc}
 \bar{u}_1 \Sigma_{12} P^+_2 V^\sigma P^-_2 \Sigma_{21} v_1 \mathcal{M}^P_\sigma & \bar{u}_1 \Sigma_{12} P^+_2 V^\sigma P^-_2 (1+\Sigma_{22}) v_2 \mathcal{M}^P_\sigma \\
 \bar{u}_2 (1+\Sigma_{22}) P^+_2 V^\sigma P^-_2 \Sigma_{21} v_1 \mathcal{M}^P_\sigma & \bar{u}_2 (1+\Sigma_{22}) P^+_2 V^\sigma P^-_2 (1+\Sigma_{22}) v_2 \mathcal{M}^P_\sigma
\end{array}
\hspace*{-2mm} \right)
\end{eqnarray}
\end{small}
The interference contribution of the cross section can be obtained by contracting elements of one matrix with elements of the other matrix. Some interesting consequences can be derived from the structure of these matrices.
\begin{enumerate}
 \item It is possible to construct four interference terms by contracting elements with same indices (e.g. $\mathcal{M}_{J=1}|_{(1,1)}$ with $\mathcal{M}_{J=2}|_{(1,1)}$) due to the fact that the quarks in the final state are the same. At lowest order these interference terms will always contain two off-diagonal loop corrections.
 \item Any element of one matrix can be contracted with any element of the other matrix only when considering also the decays of the quarks, there fixing specific decay channels for the quark and antiquark branches. This way it is possible to obtain 16 interference combinations. The order of the interference term and the number of off-diagonal mixing contributions, however, will not always be the same, depending on the contraction. In particular, when contracting the element (1,1) of the $\mathcal{M}_{J=1}$ matrix with the element (2,2) of the $\mathcal{M}_{J=2}$ matrix, there are no off-diagonal loop mixings involved and the contraction after the quark decays will be given by a pure tree level contribution plus diagonal loop corrections while, when contracting the element (2,2) of the $\mathcal{M}_{J=1}$ matrix with the element (1,1) of the $\mathcal{M}_{J=2}$ matrix, there are 4 off-diagonal loop mixings involved,  so that this process, which has mixing terms to a higher power,  is expected to be suppressed.
\end{enumerate}
It is interesting to notice that, in the case of same element contractions before quark decays (case 1), the order of the process is the same as in the case of contractions after quark decays of the element (1,1) of the $\mathcal{M}_{J=1}$ matrix with the element (2,2) of the $\mathcal{M}_{J=2}$ matrix (case 2). Therefore, the 4 interference contributions of case 1 can be competitive with the tree-level interference term after quark decay. However, if the off-diagonal contributions to the mixing matrix are negligible with respect to the diagonal elements, the two amplitude matrices reduce to:
\begin{eqnarray}
\mathcal{M}_{J=1} &\simeq& \left(
\begin{array}{cc}
 \bar{u}_1 (1+\Sigma_{11}) P^+_1 V^\sigma P^-_1 (1+\Sigma_{11}) v_1 \mathcal{M}^P_\sigma & 0 \\
 0 & 0
\end{array}
\right), \\ 
\mathcal{M}_{J=2} &\simeq& \left(
\begin{array}{cc}
 0 & 0 \\
 0 & \bar{u}_2 (1+\Sigma_{22}) P^+_2 V^\sigma P^-_2 (1+\Sigma_{22}) v_2 \mathcal{M}^P_\sigma
\end{array}
\right).
\end{eqnarray}
In this case the same element contraction of case 1 do not enter the determination of the interference terms and the lowest order contribution is given by contracting the only non-zero elements of the matrices at tree level after the decays of the quarks. In other words, the analytical description of the interference developed in the previous section can only be applied in the case of suppressed or negligible mixing between the heavy quarks. One should note that the requirement of suppression of off-diagonal mixing can be potentially quite restrictive, since it will take place in case of cancellation of loop contributions in the kinematic $p^2\simeq M_Q^2$ region where the couplings of the heavy quarks are chosen to compensate the different values of the loop integrals. The verification of such a case is eventually model dependent and requires computing the mixing matrix structure, which in turn depends on the particle content of the model. 
For example in case of the off-diagonal contributions to the propagators of two top partners $T_1$ and $T_2$ that only couple to the third family of SM quarks and with all SM gauge bosons and the Higgs boson, and requiring their sum to be suppressed with respect to the sum of the diagonal contributions, we obtain the following relation:
\begin{equation}\label{eq:interference condition}
  \Sigma_{IJ}=\Sigma_{IJ}^{tH}+\Sigma_{IJ}^{tZ}+\Sigma_{IJ}^{bW}+\Sigma_{IJ}^{tG^0}+\Sigma_{IJ}^{bG^+}\ll\{\Sigma_{II},\Sigma_{JJ}\}
\end{equation}
with $I,J=1,2$, $I\neq J$ and where the two last terms of the sum account for Goldstone bosons. The suppression of the off-diagonal contribution depends on all the masses and couplings involved, plus it also depends on the $p^2$ of the external heavy quarks. However, if it is possible to find coupling configurations which satisfy the relation for a large $p^2$ region, our approach can be safely adopted. A detailed numerical treatment of this relation for different particle contents and coupling values is beyond the scope of this preliminary analysis, but it will be developed in a future one. 
% What we want to stress now is that in this region the treatment developed in the previous section can indeed be applied, but bearing in mind that it is only part of a more general approach, that will be . 
It is also interesting to notice that, if the mass and width eigenvalues are not misaligned, it is possible to diagonalise the matrix of the propagators and define new states with definite mass and eigenstates. In this case it is possible to consider the exact amplitude matrix,
\begin{eqnarray}
\mathcal{M}_{J={1^\prime}} &=& \left(
\begin{array}{cc}
 \bar{u}_{1^\prime} P^+_{1^\prime} V^\sigma P^-_{1^\prime} v_{1^\prime} \mathcal{M}^P_\sigma & 0 \\
 0 & 0
\end{array}
\right), \\ 
\mathcal{M}_{J={2^\prime}} &=& \left(
\begin{array}{cc}
 0 & 0 \\
 0 & \bar{u}_{2^\prime} P^+_{2^\prime} V^\sigma P^-_{2^\prime} v_{2^\prime} \mathcal{M}^P_\sigma
\end{array}
\right),
\end{eqnarray}
then compute the tree-level interference after the decays of the quarks with the method developed in the previous section, but considering quarks with loop-corrected masses and widths. Again, this is a specific situation, but it is a further case when the relations studied in this section can be applied.

\subsection{Numerical results}

\subsubsection{Total cross section}

\begin{figure}[htb]
\begin{center}
 \begin{picture}(160,100)(0,0)
    \SetWidth{2.0}
    \SetColor{Black}
    \Line[arrow,arrowpos=0.5,arrowwidth=2](10,60)(40,50)
    \Text(8,60)[rc]{\small{$p$}}
    \Line[arrow,arrowpos=0.5,arrowwidth=2](10,20)(40,30)
    \Text(8,20)[rc]{\small{$p$}}
    \SetWidth{1.0}
    \Line[arrow,arrowpos=0.5,arrowwidth=2](60,50)(90,60)
    \Text(75,60)[cb]{\small{$T$}}
    \Line[arrow,arrowpos=0.5,arrowwidth=2](60,30)(90,20)
    \Text(75,20)[ct]{\small{$\bar T$}}
    \Line[arrow,arrowpos=0.5,arrowwidth=2](90,60)(120,70)
    \Text(122,70)[lc]{\small{$b$}}
    \Line[arrow,arrowpos=0.5,arrowwidth=2](90,20)(120,10)
    \Text(122,10)[lc]{\small{$\bar t$}}
    \Photon(120,50)(90,60){2}{4.5}
    \Text(122,50)[lc]{\small{$W^+$}}
    \Photon(90,20)(120,30){2}{4.5}
    \Text(122,30)[lb]{\small{$Z$}}
    \GOval(50,40)(15,15)(0){0.882}
  \end{picture}
\end{center}
\caption{Pair production of VLQs $T$ and subsequent decay into a $bW^+\bar t Z$ final state.}
\label{fig:prod-decay}
\end{figure}
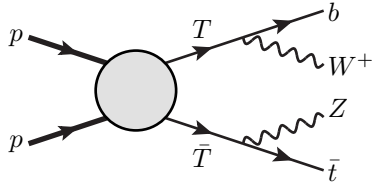

We first consider the production and decay rates of two $T$s pairs decaying into $W^+b$ and $Z \bar t$, see Fig.~\ref{fig:prod-decay}, i.e., we consider the $2\to4$ process
\begin{equation}
\label{eq:proc}
p p \rightarrow T_i \bar T_i \rightarrow W^+ b Z \bar t, \qquad i=1,2,
\end{equation}
with the chirality of the couplings being the same for the two states.
This process has been chosen to provide a concrete example; in general, VLQs can also decay into the Higgs boson, but we have fixed a specific final state to perform the simulations. Selecting different final states involving decays into Higgs would give analogous results.

\begin{figure}[htb]
\begin{minipage}{.48\textwidth}
\hskip 20pt \bf Chiral couplings
\end{minipage}\hfill
\begin{minipage}{.48\textwidth}
\hskip 20pt \bf General couplings
\end{minipage}
\vskip 10pt
\epsfig{file=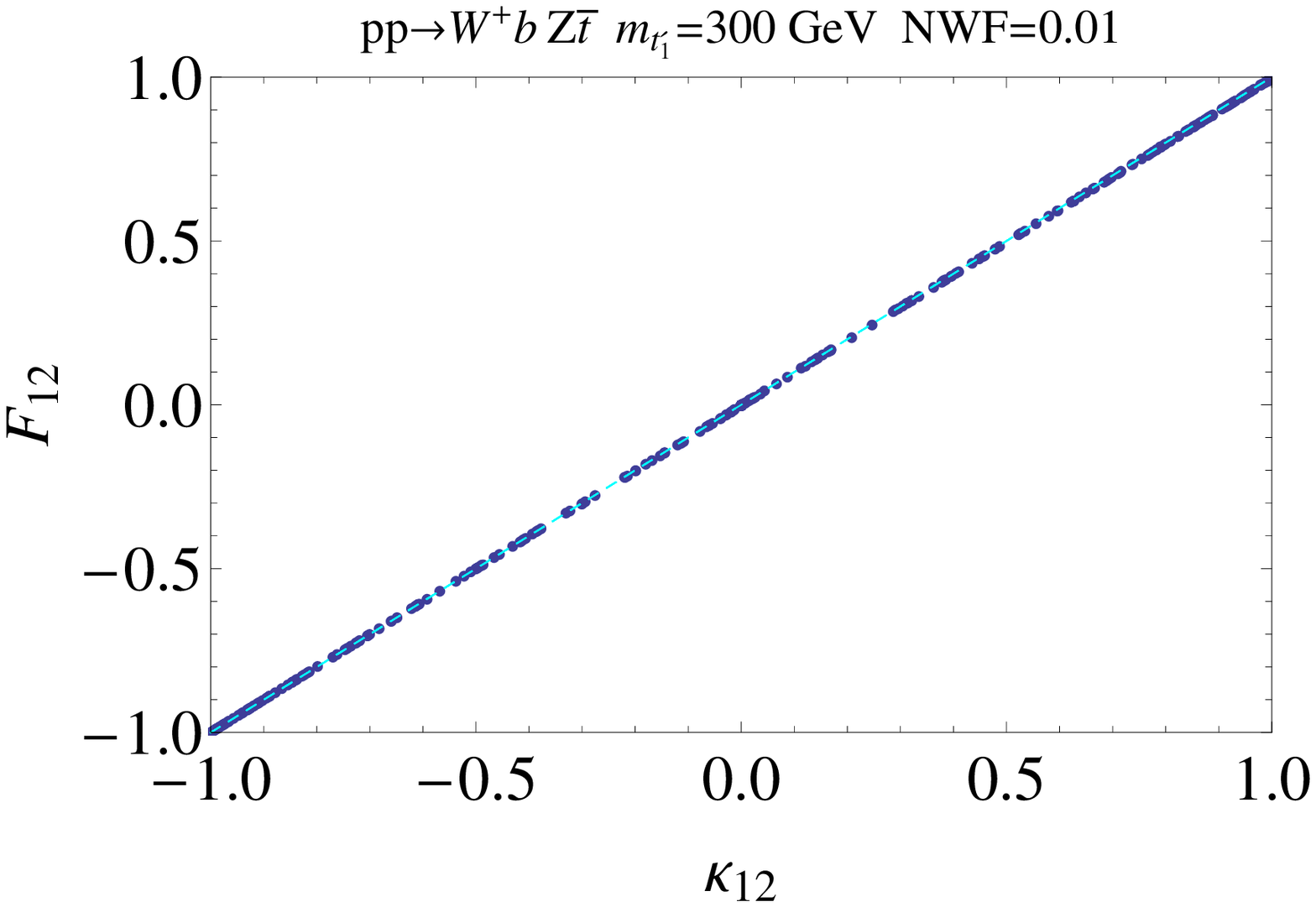, width=0.5\textwidth}%
\epsfig{file=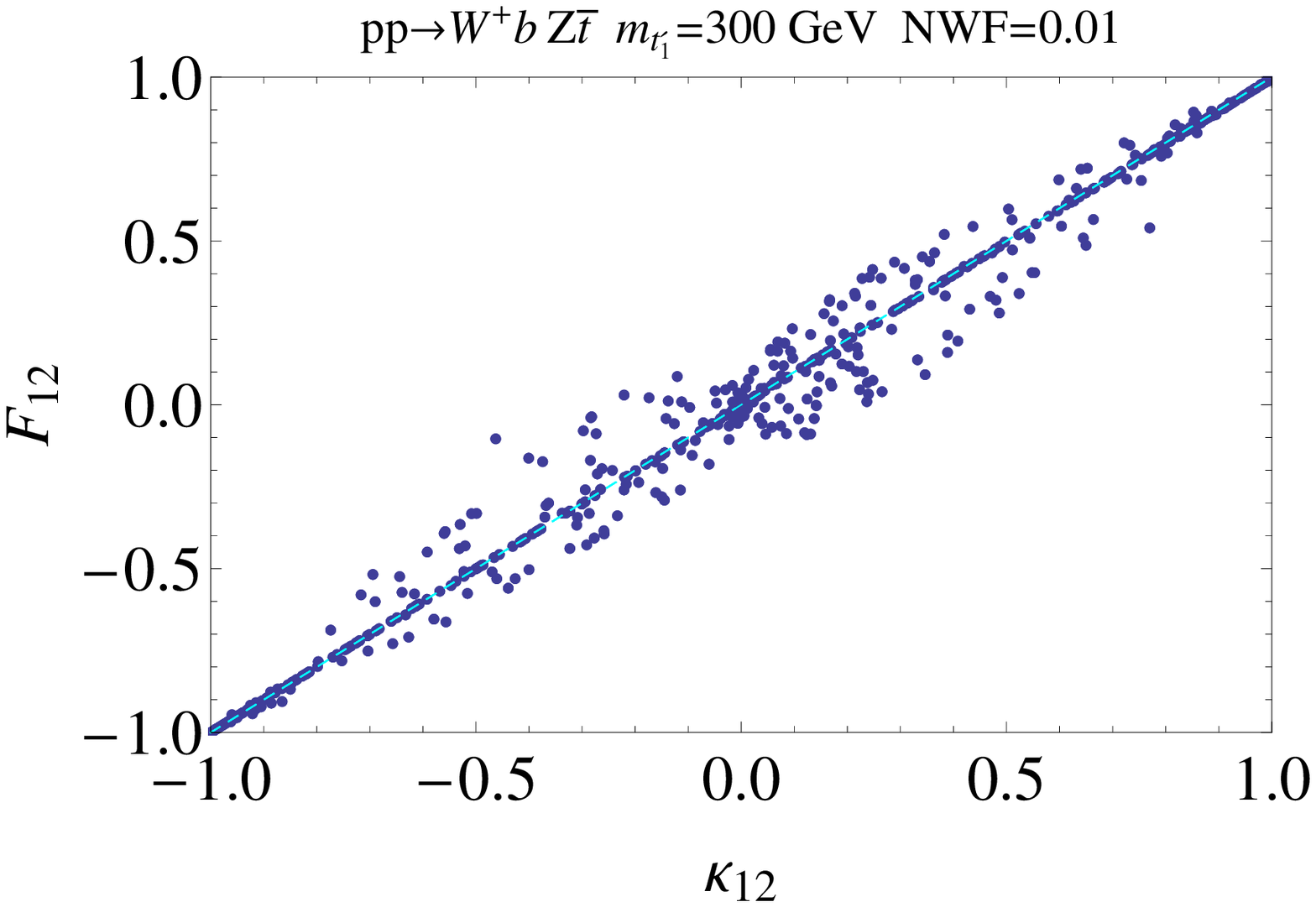, width=0.5\textwidth}\\
\epsfig{file=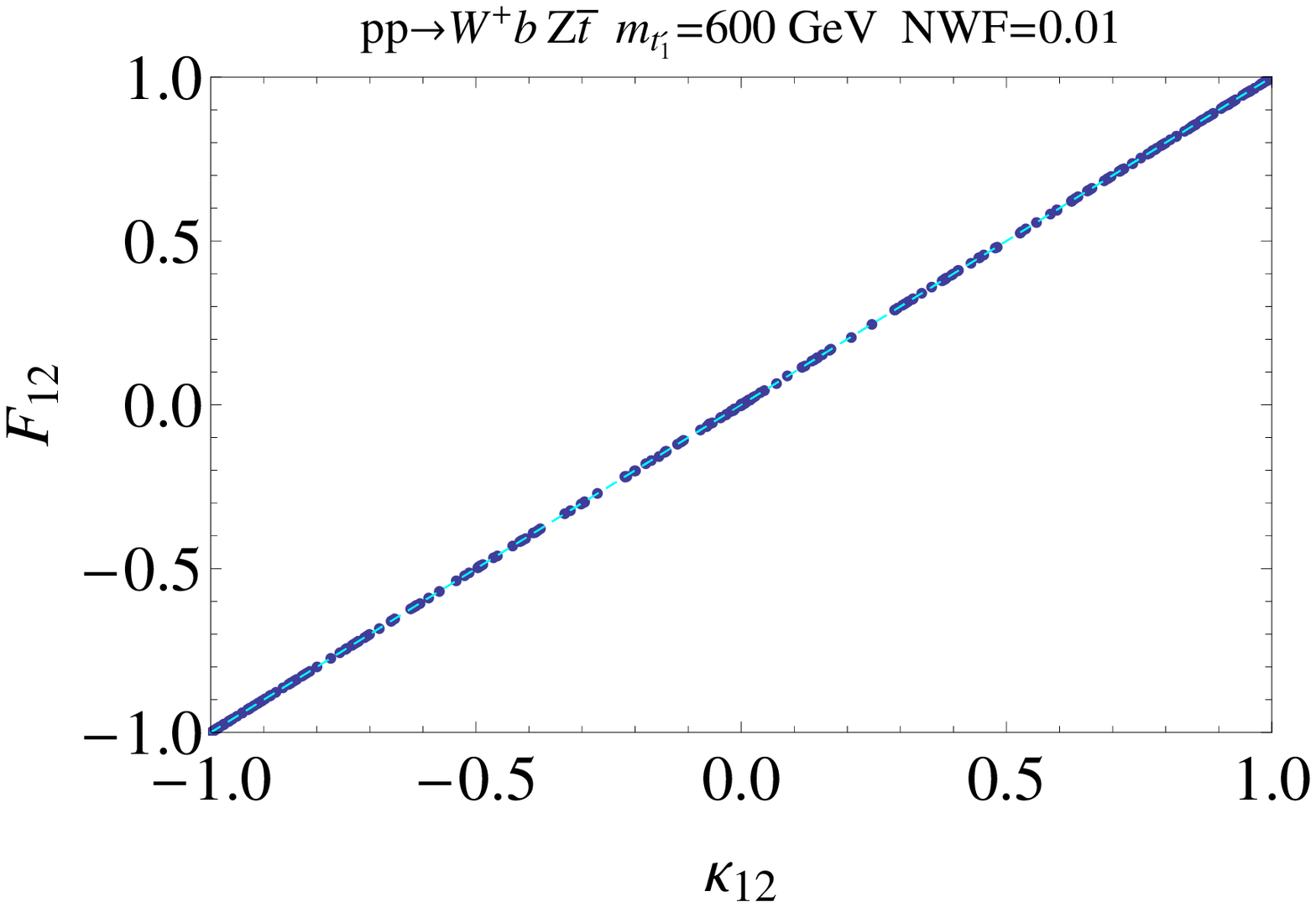, width=0.5\textwidth}%
\epsfig{file=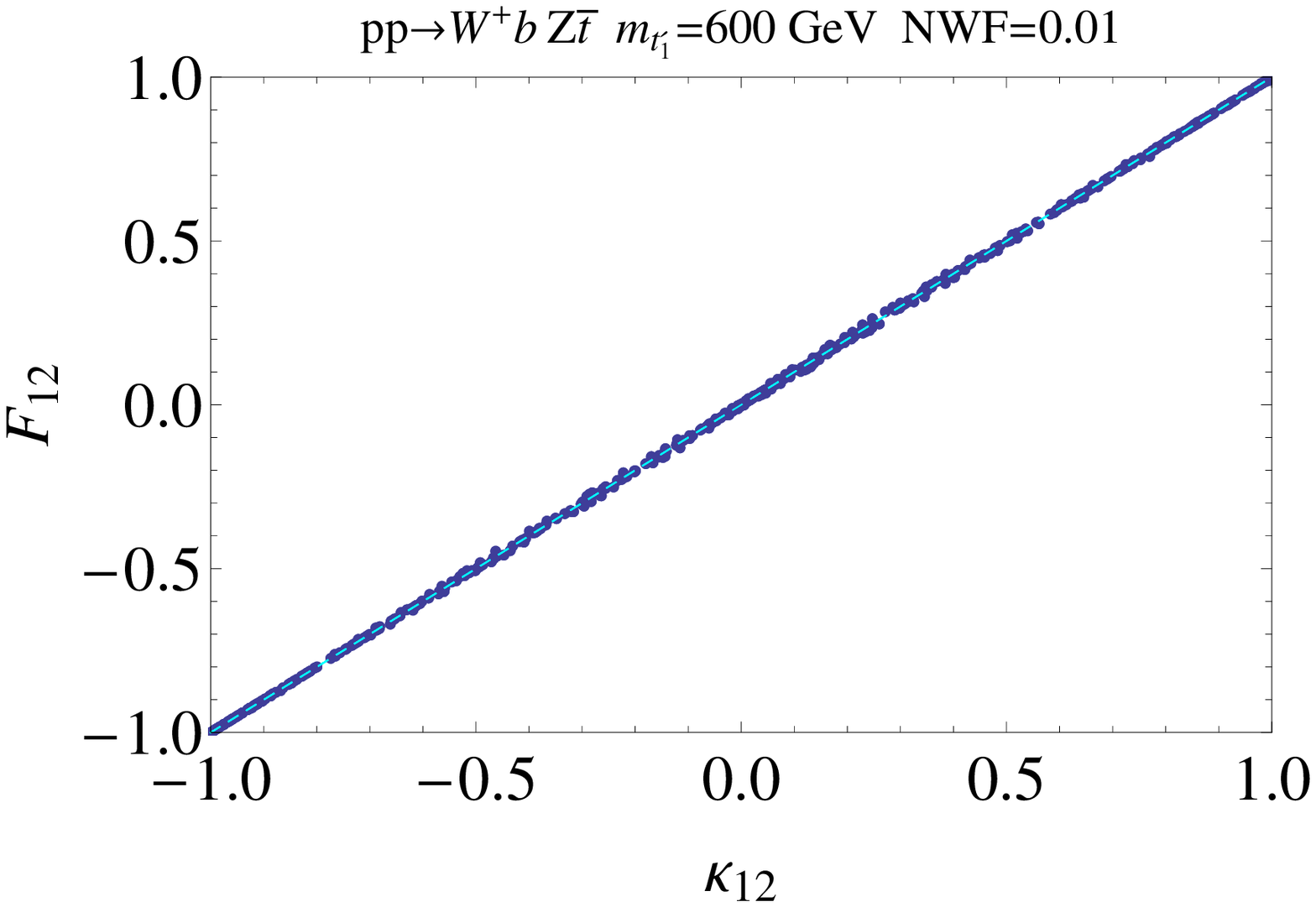, width=0.5\textwidth}\\
\epsfig{file=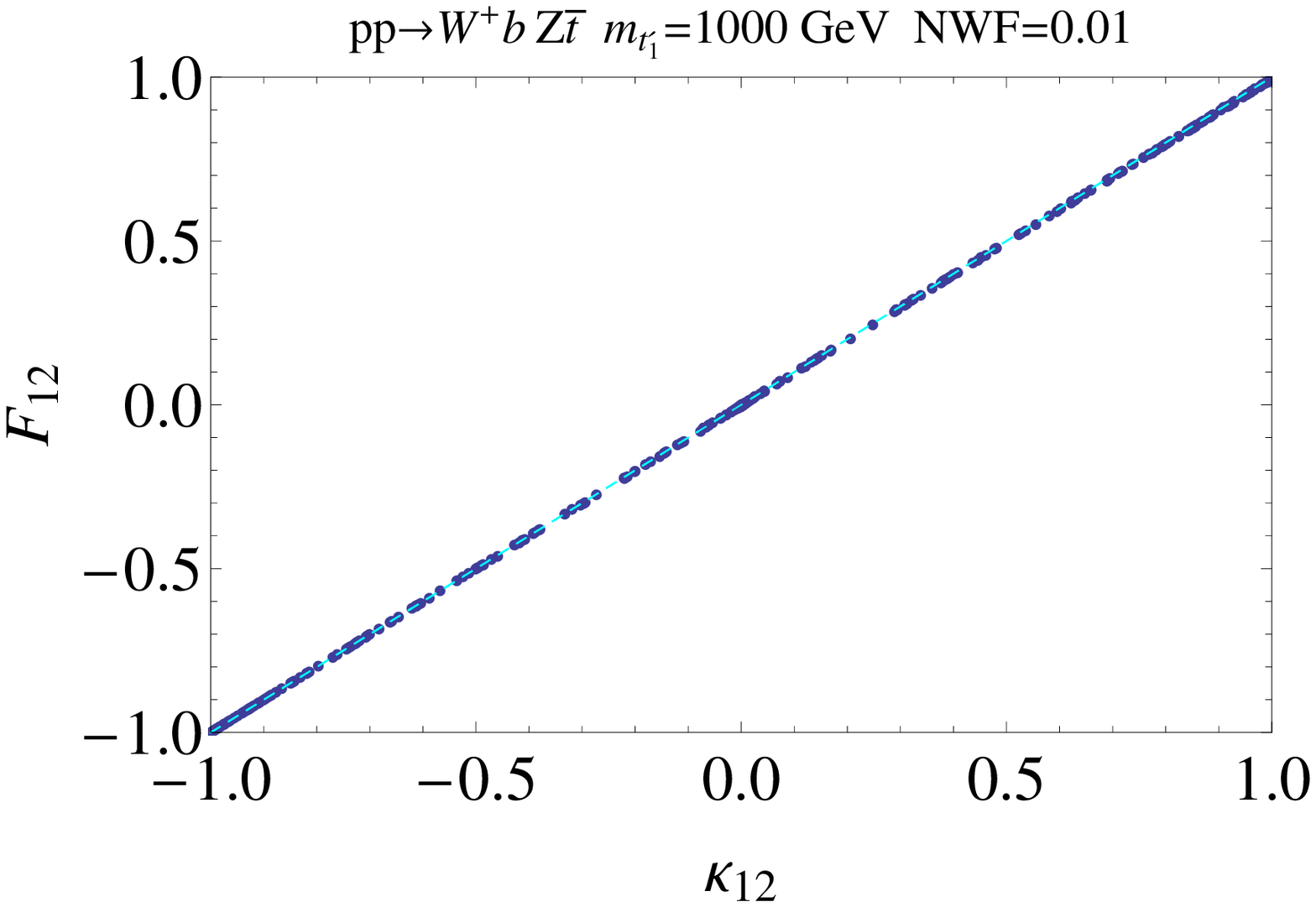, width=0.5\textwidth}%
\epsfig{file=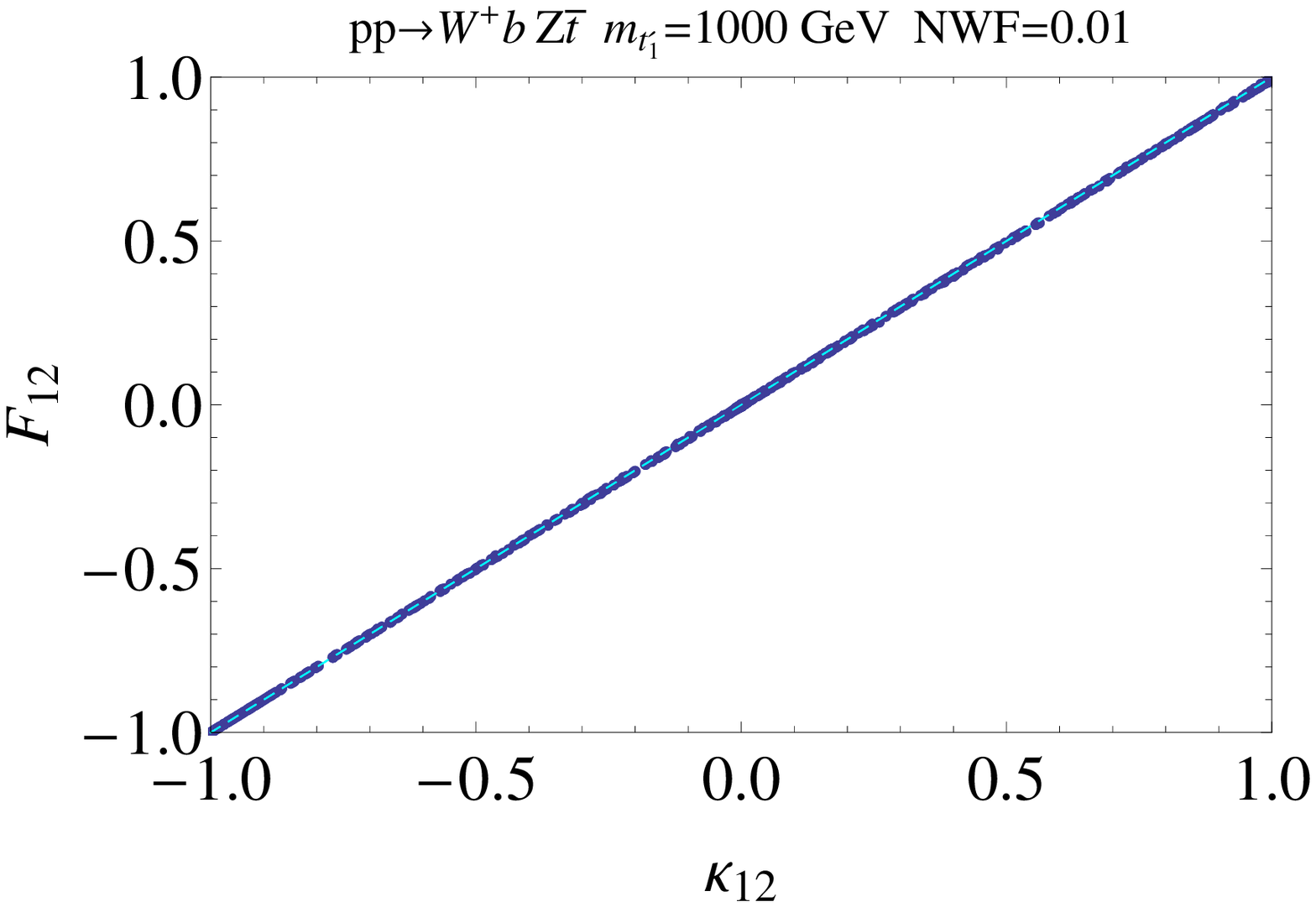, width=0.5\textwidth}
\caption[Interference term $F_{ij}$ as a function of $\kappa_{ij}$.]{Interference term $F_{ij}$ as a function of $\kappa_{ij}$. In the left frame the couplings are chiral while in the right one they are general. The cyan-dashed line is the bisector in the $\kappa_{ij}-F_{ij}$ plane. Blue points are the results of the scan on the couplings for $m_{T_1}=300,600,1000~$GeV, with different values of the mass splitting between $t_1$ and $t_2$. The Narrow Width Factor (NWF) is the upper limit on max$(\Gamma_{T_1}$/$m_{T_1},$ $\Gamma_{T_2}/$ $m_{T_2})$ for each point of the scan.}
\label{fig:wbzt_int}
\end{figure}

We have performed a scan on the VLQs couplings for different values of masses and splitting between the two $T$s and we have obtained the value of the interference term (\ref{orderparameter}) through numerical simulation with MadGraph5~\cite{Alwall:2011uj} and alternatively cross-checked via CalcHEP3.4~\cite{Belyaev:2012qa}.
The results are shown in Fig. \ref{fig:wbzt_int} (left frame), where it is possible to notice a remarkable linear correlation between $F_{ij}$ and the expression in Eq.(\ref{eq:int2}).

If the chirality of the couplings of $T_1$ and $T_2$ with respect to the SM quarks is opposite, interference effects can arise when the masses of the quarks in the final state are not negligible, as is in the case of decay to top quarks. Considering a scenario where $T_1$ decays predominantly to $Zt_L$ and $T_2$ does so in $Zt_R$, then the interference between $t_L$ and $t_R$ may in principle become relevant. We have numerically verified, however, that in case the chirality of the two VLQs is opposite, the interference effect between massive final states is always negligible, unless the XQs masses approach the threshold of the final state. This case implies, however, very light VLQs, with masses of the order of 300 GeV, and this range is already excluded by experimental searches.

We show in Fig. \ref{fig:wbzt_int} (right frame) the results for the analogous process (\ref{eq:proc}) where both chiralities are now present in the couplings of XQs: this process is described by the generalised Eq.(\ref{eq:gen2}). Interference effects between final state quarks of different chiralities become relevant when the masses of the heavy quarks are close to the top mass, but, as already stressed, this scenario has been tested only to show the appearance of  chirality flipping interference effects, since such a low value for the mass of the heavy quarks is already experimentally excluded.

\subsubsection{Differential distributions}

The results of the previous sections only apply to the total cross section of the process of pair production and decay of the heavy quarks. However, it is necessary to evaluate how kinematic distributions are affected by the presence of interference terms, as experimental efficiencies of a given search may be largely different if the kinematics of the final state is not similar to the case without interference. To evaluate the contribution of interference we have considered the process $p p \to W^+ b Z \bar t$, with subsequent semileptonic decay of the top, mediated by two heavy top-like partners $T_1$ and $T_2$ in three limiting cases:
\begin{itemize}
 \item degenerate masses ($m_{T_{1,2}}=600$ GeV) and couplings with same chirality (both left-handed);
 \item degenerate masses ($m_{T_{1,2}}=600$ GeV) and couplings with opposite chirality;
 \item non-degenerate masses ($m_{T_1}=600$ GeV, $m_{T_2}=1.1m_{T_1}=660$ GeV) and couplings with same chirality (both 
left-handed).
\end{itemize}
\begin{figure}[htb]
\begin{minipage}{.48\textwidth}
\textbf{Scalar sum of transverse momentum}
\end{minipage}\hfill
\begin{minipage}{.48\textwidth}
\textbf{Missing transverse energy}
\end{minipage}
\vskip 10pt
\epsfig{file=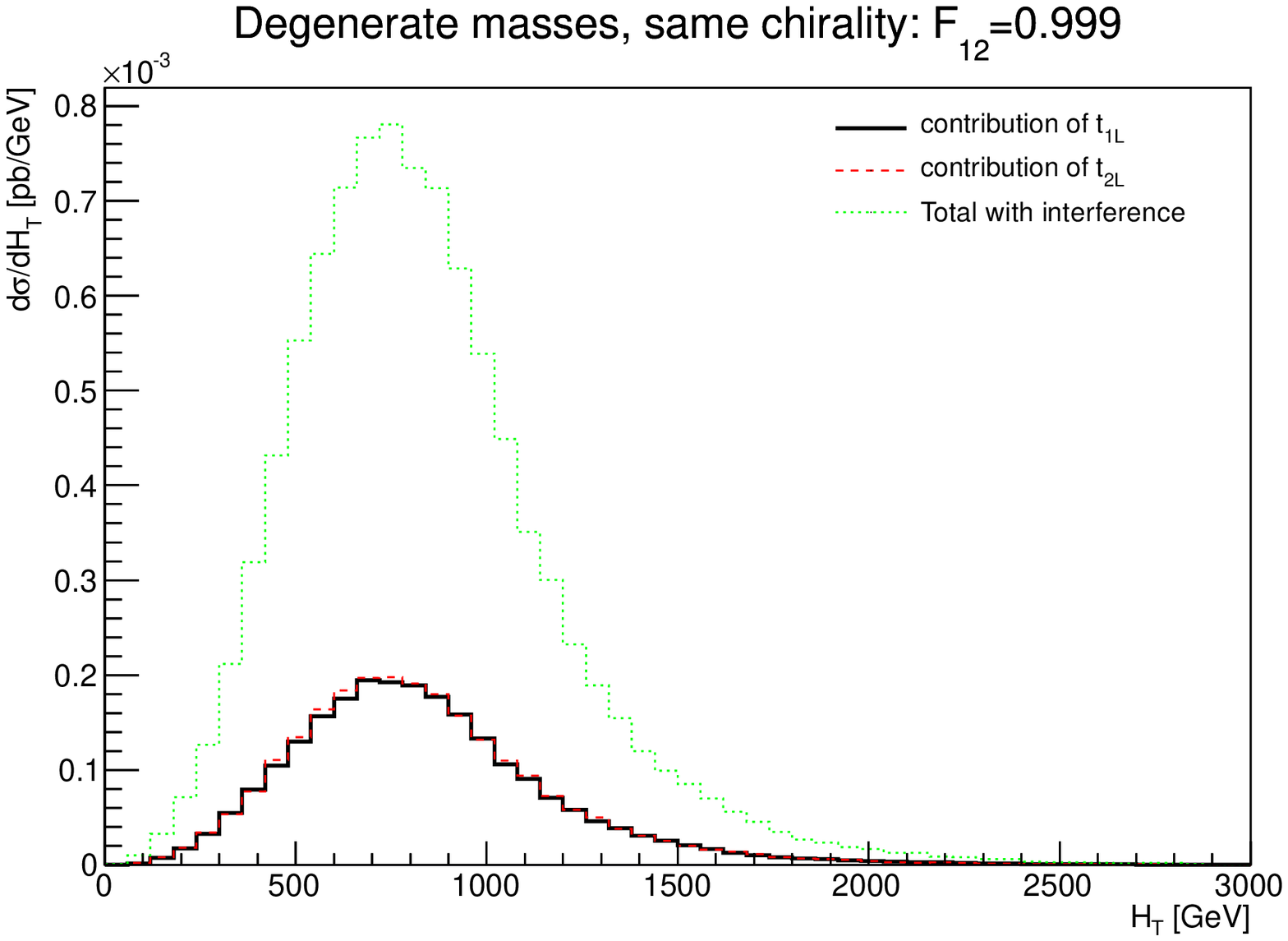, width=0.5\textwidth}%
\epsfig{file=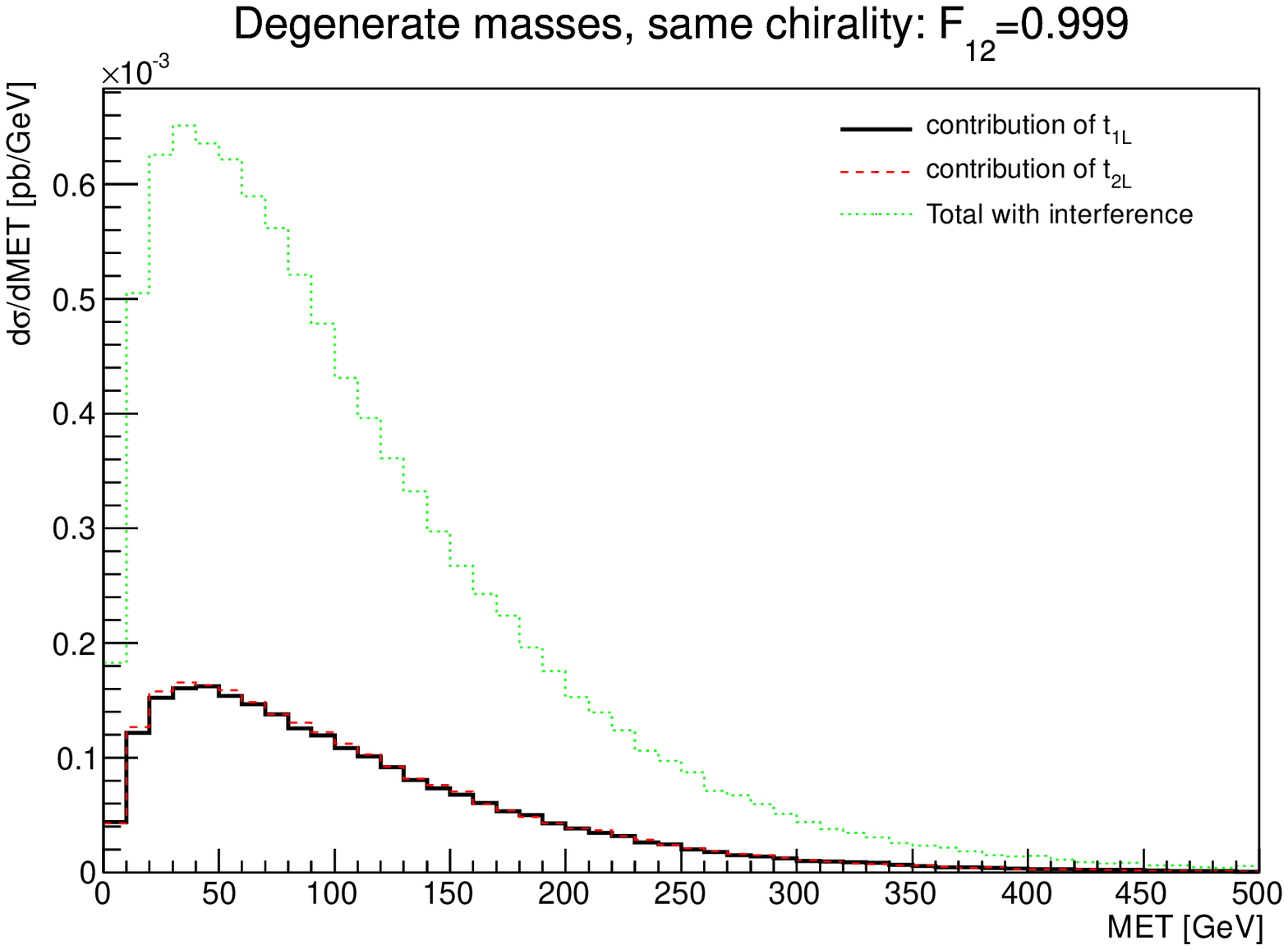, width=0.5\textwidth}\\
\epsfig{file=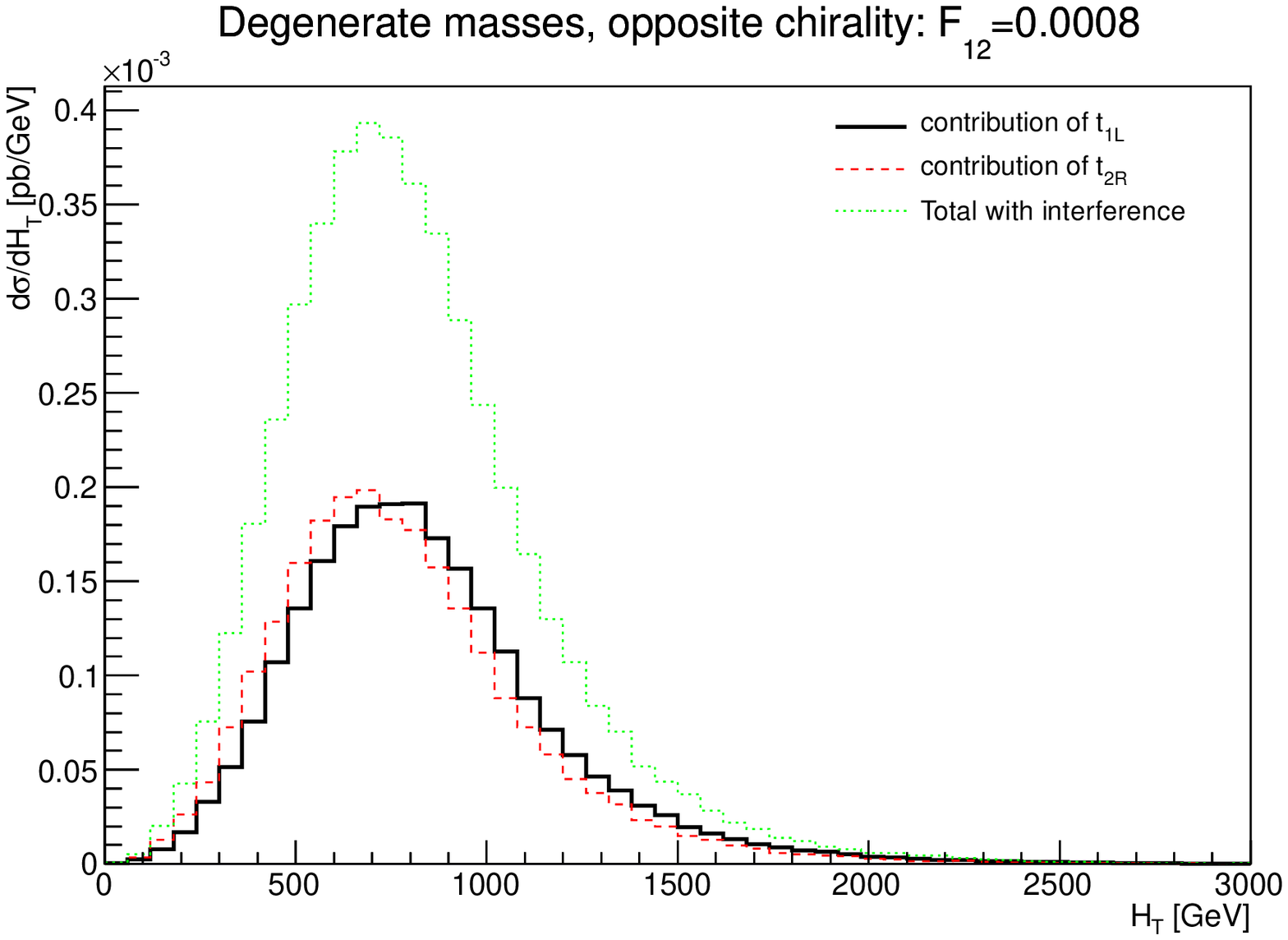, width=0.5\textwidth}%
\epsfig{file=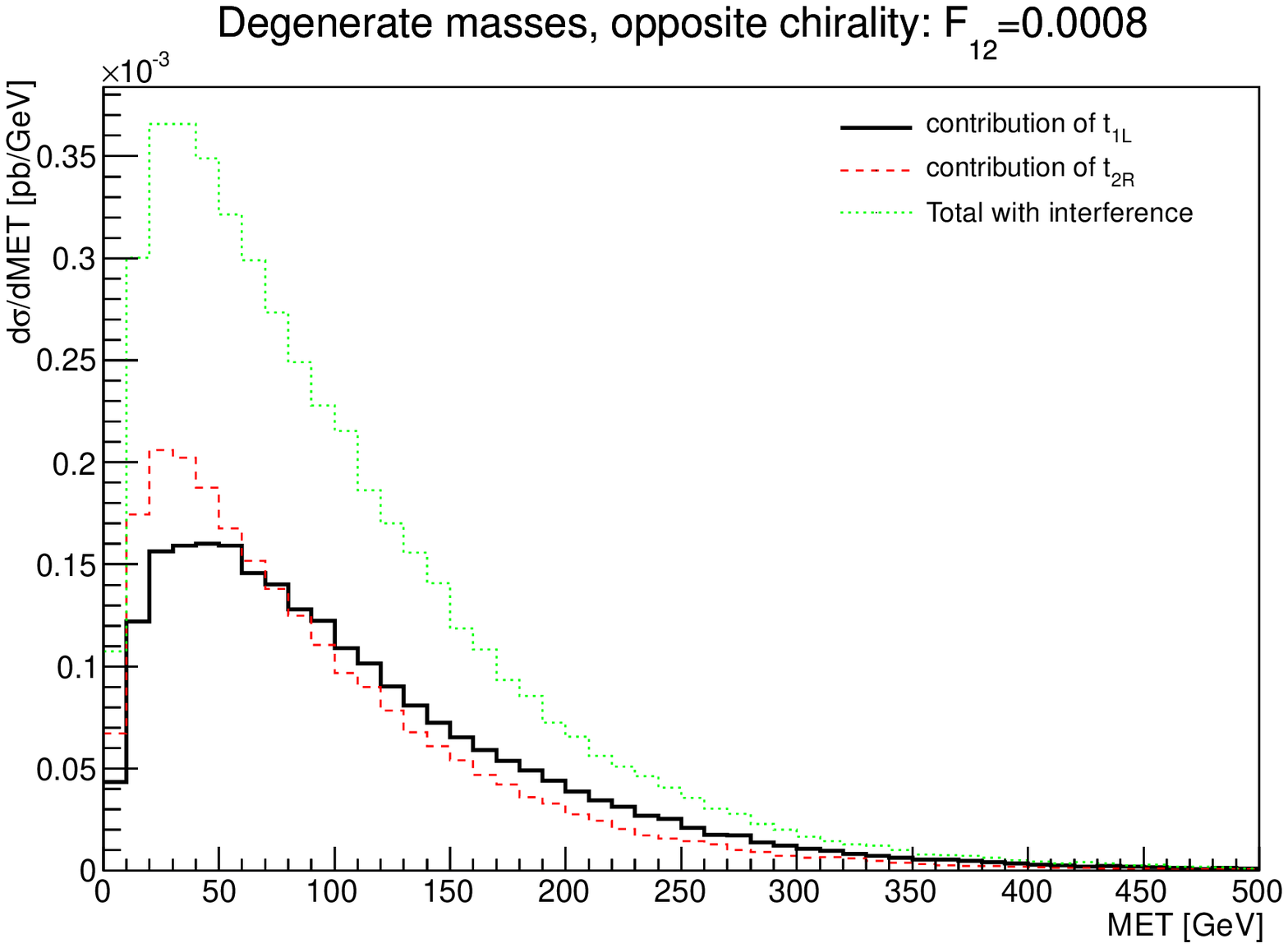, width=0.5\textwidth}\\
\epsfig{file=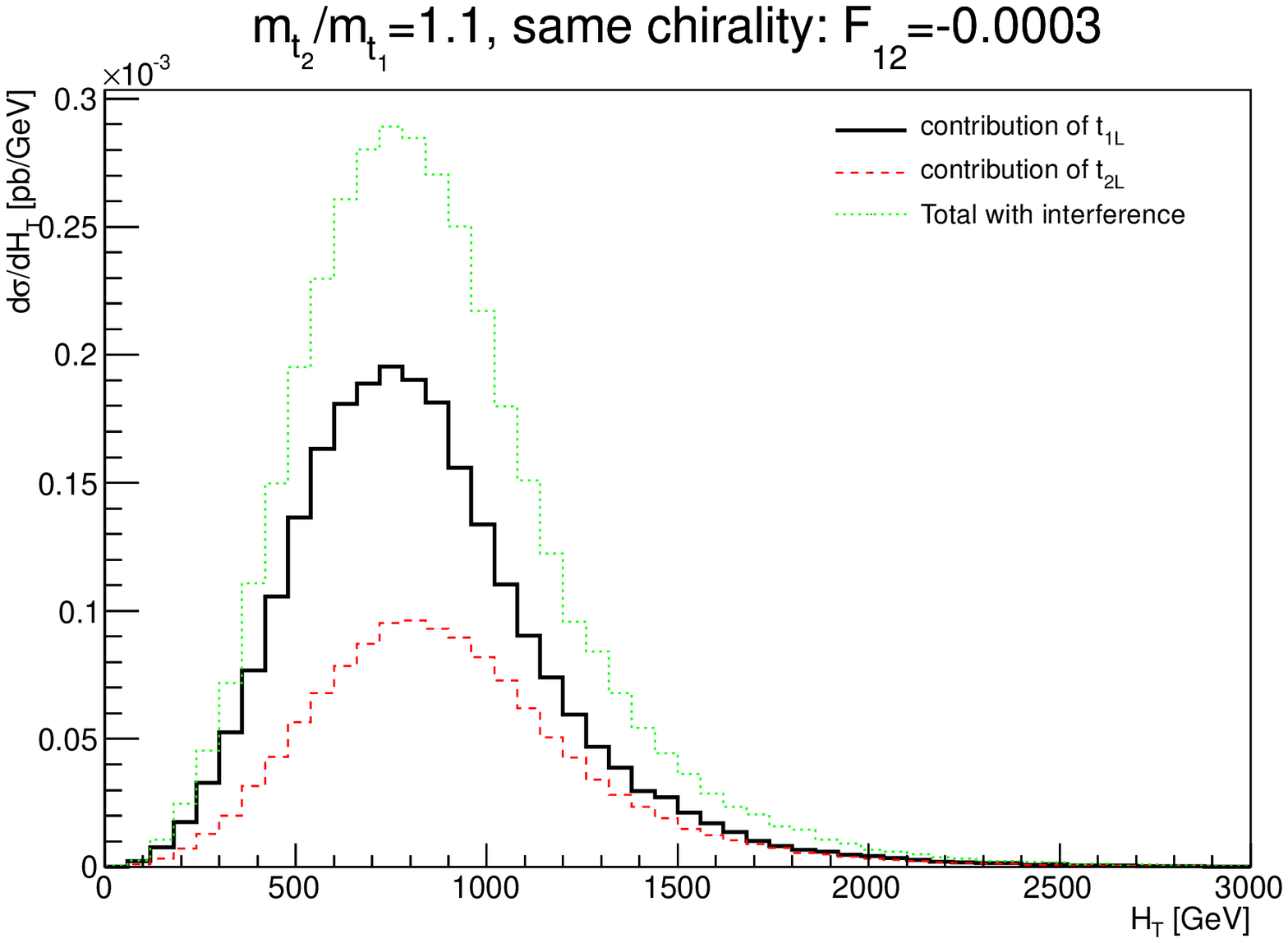, width=0.5\textwidth}%
\epsfig{file=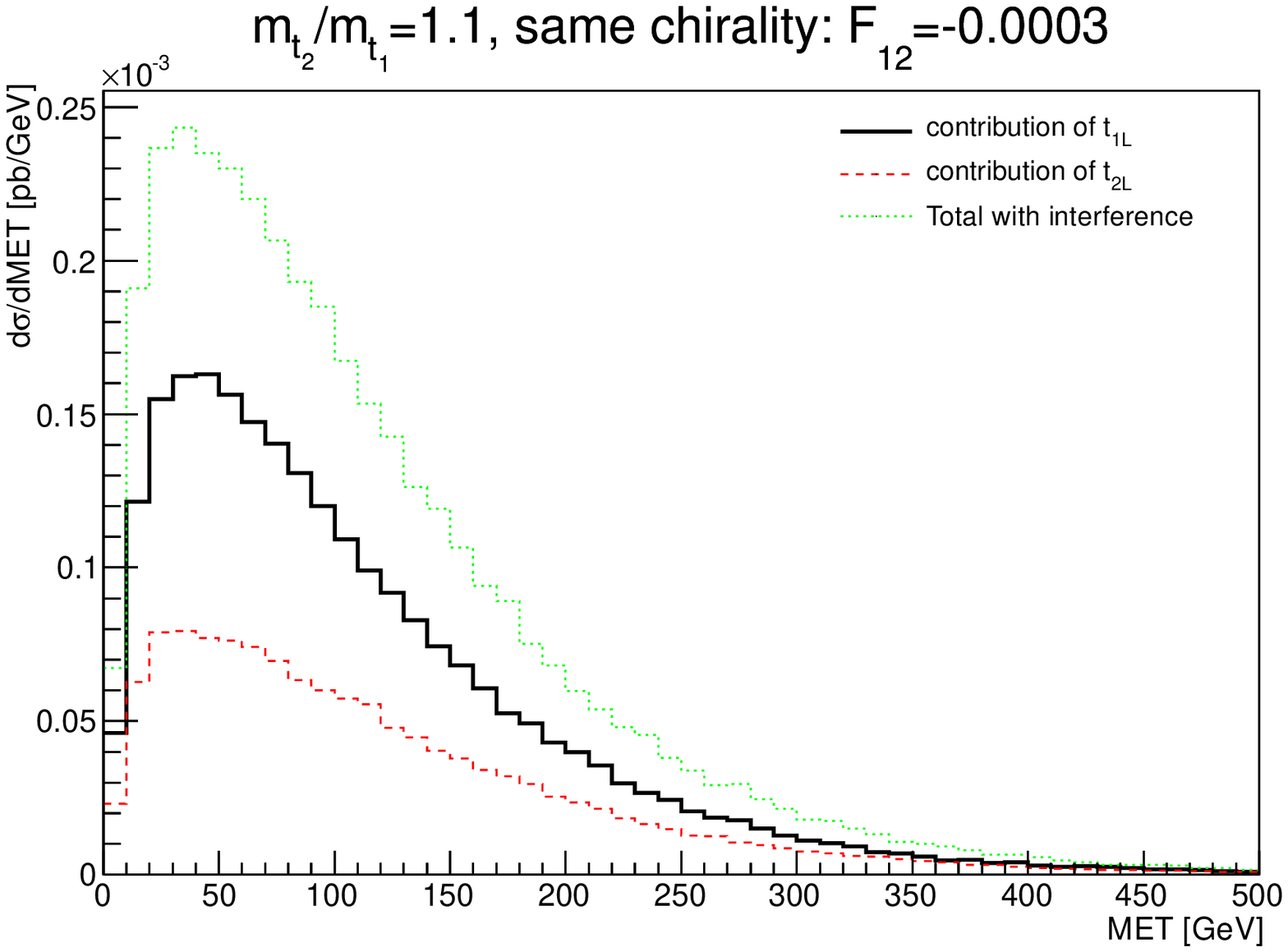, width=0.5\textwidth}
\caption[Differential distributions for $H_T$ and $\MET$ for the process $pp\to W^+bZ\bar t\to W^+bZ \bar b e^- \bar \nu_e$ in three different scenarios: degenerate masses and couplings with same chirality; degenerate masses and couplings with opposite chirality; non-degenerate masses ($m_{T_2}=1.1m_{T_1}$) and couplings with same chirality.]{Differential distributions for $H_T$ and $\MET$ for the process $pp\to W^+bZ\bar t\to W^+bZ \bar b e^- \bar \nu_e$ in three different scenarios: degenerate masses and couplings with same chirality (top); degenerate masses and couplings with opposite chirality (middle); non-degenerate masses ($m_{T_2}=1.1m_{T_1}$) and couplings with same chirality (bottom). Here, $m_{T_1}$ has been fixed to 600~GeV. The values of the interference term $F_{12}$ are shown for each scenario.}
\label{fig:wbzt_dist}
\end{figure}

The results are shown in Fig. \ref{fig:wbzt_dist}, where we display the $H_T$ (scalar sum of the transverse momenta of jets) and $\MET$ (missing transverse energy) differential distributions. When the interference is maximal, all distributions have exactly the same features, that is, the distributions including interference can be obtained by a rescaling of the distributions for production of the two heavy quarks using $(1+\kappa_{ij})$ for the rescaling factor: this relation comes from considering Eq.~(\ref{orderparameter}) and the linear correlation between $F_{ij}$ and $\kappa_{ij}$ verified in the previous section. Therefore, our results for the total cross section can also be applied at differential level and, specifically, it is possible to apply the same experimental efficiencies to the case of a single heavy quark or to the case with degenerate quarks with couplings of identical chirality. 
In contrast, in the two other scenarios we have considered, where interference is negligible, the distributions for production of either $T_1$ or $T_2$ exhibit different features and the distribution of the total process is, for each bin, simply the sum of the distributions of the two heavy quarks (i.e. the rescaling factor is 1 because $k_{ij}\sim0$). Same patterns are seen for all other differential distributions that we have investigated: (pseudo)rapidity, cone separation, etc.

\begin{figure}[htb]
\epsfig{file=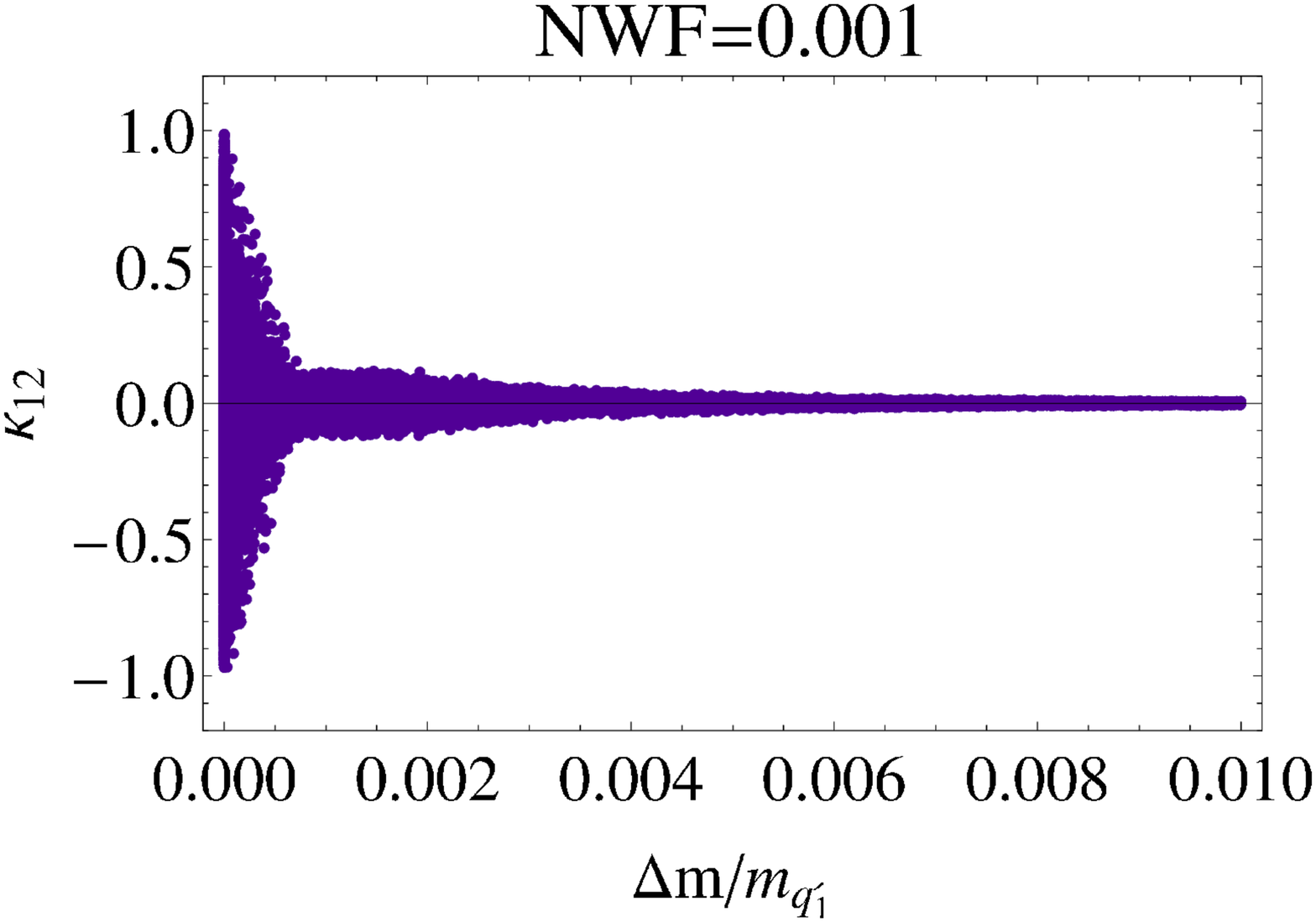, angle=-90,width=.48\textwidth}\hfill
\epsfig{file=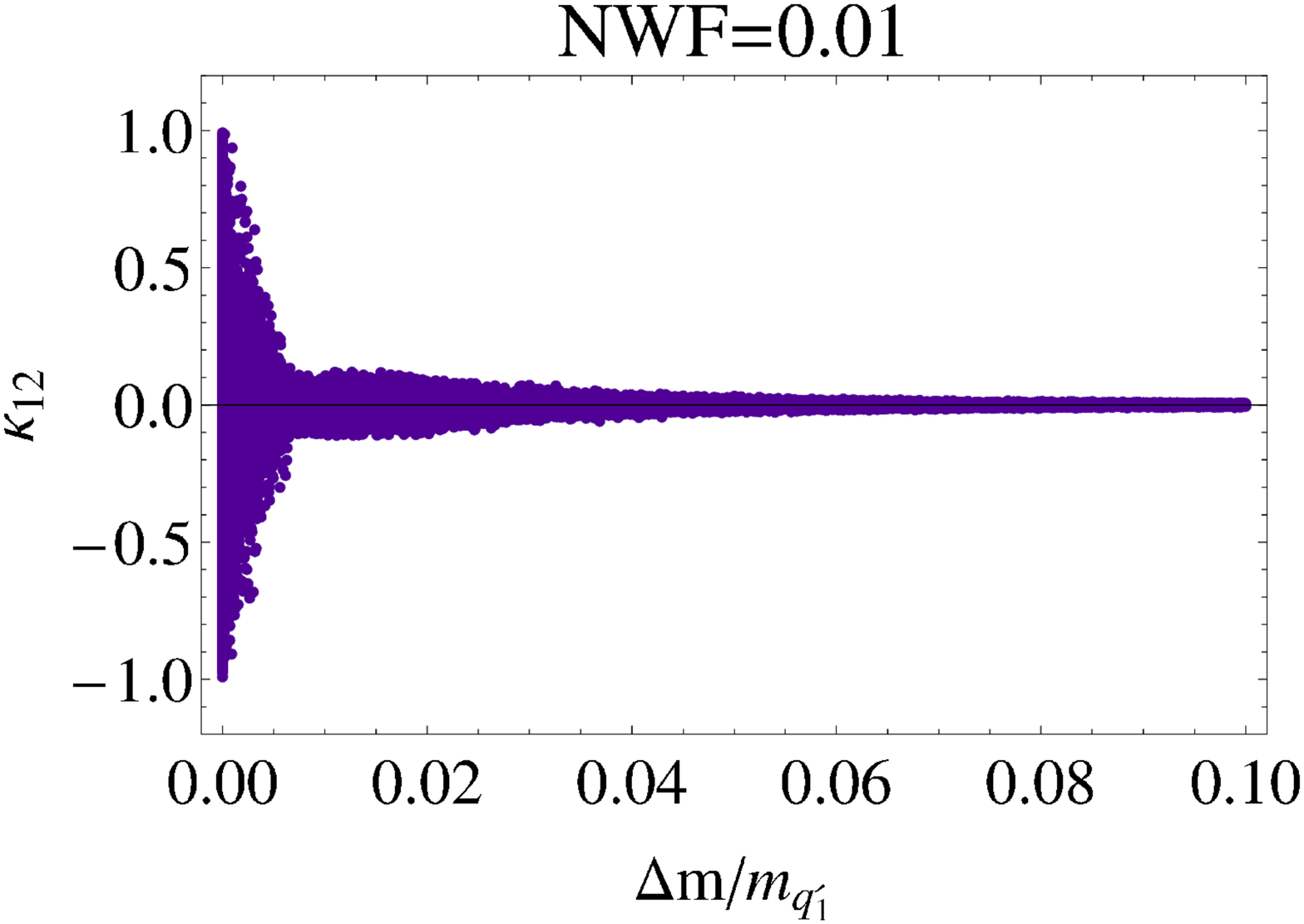, angle=-90,width=.48\textwidth}
\caption[The range of the interference contributions with respect to the mass splitting between the heavy quarks for different values of the NWF.]{The range of the interference contributions with respect to the mass splitting between the heavy quarks for different values of the NWF. Notice the different scales of the $x$ axis.}
\label{fig:intsplit}
\end{figure}

As a final remark, we may ask how much the range of the possible values for the interference term drops by increasing the mass splitting between the heavy quarks and, therefore, when should we consider the interference as always negligible. In Fig. \ref{fig:intsplit} it is possible to notice that the range of values for the parameter $\kappa_{12}$ drops extremely fast with the mass splitting and depends on the value of the NWF. The range of the interference contributions, however, becomes smaller than 10\% in a region of mass splitting where the shapes of the distributions can be safely considered as equivalent.

\subsubsection{Validity range of the model independent approach and ``master formula" for the interference}

In this subsection we discuss the range of validity of the analytical formula for $\kappa_{ij}$ describing the interference effect. Our ansatz was made under the assumption of small $\Gamma/m$ ratios, which, in terms of probability (e.g. amplitude square), means that the QCD production part of the XQs and their subsequent decay can be factorised.
We then took advantage of this consideration by making this factorisation already at amplitude level and writing therefore the interference, Eq.(\ref{eq:intpart}), and pair production, Eq.(\ref{eq:prodpart}), contribution to the total cross section as a modulus squared of quantities that do not involve the QCD production part, then using these two relations to define our $\kappa_{ij}$ parameter in Eq.(\ref{eq:int1}).
This concept of factorisation is valid just in the limit $\Gamma/m \to 0$, for which, however, there will be no decay of the XQ and therefore no interference at all. It is nonetheless clear that this approximation of factorisation of production and decay will be the more accurate the more this ratio is closer to zero. In fact, in the previous subsections we have shown that the formula for $\kappa_{ij}$ reproduces the true interference $F_{ij}$ very accurately in the case of NWF=$\Gamma/m=0.01$. It is however very informative to explore the range of validity of our ansatz in function of the NWF parameter, especially in view of practical applications of our method.

\begin{figure}[htb]
\epsfig{file=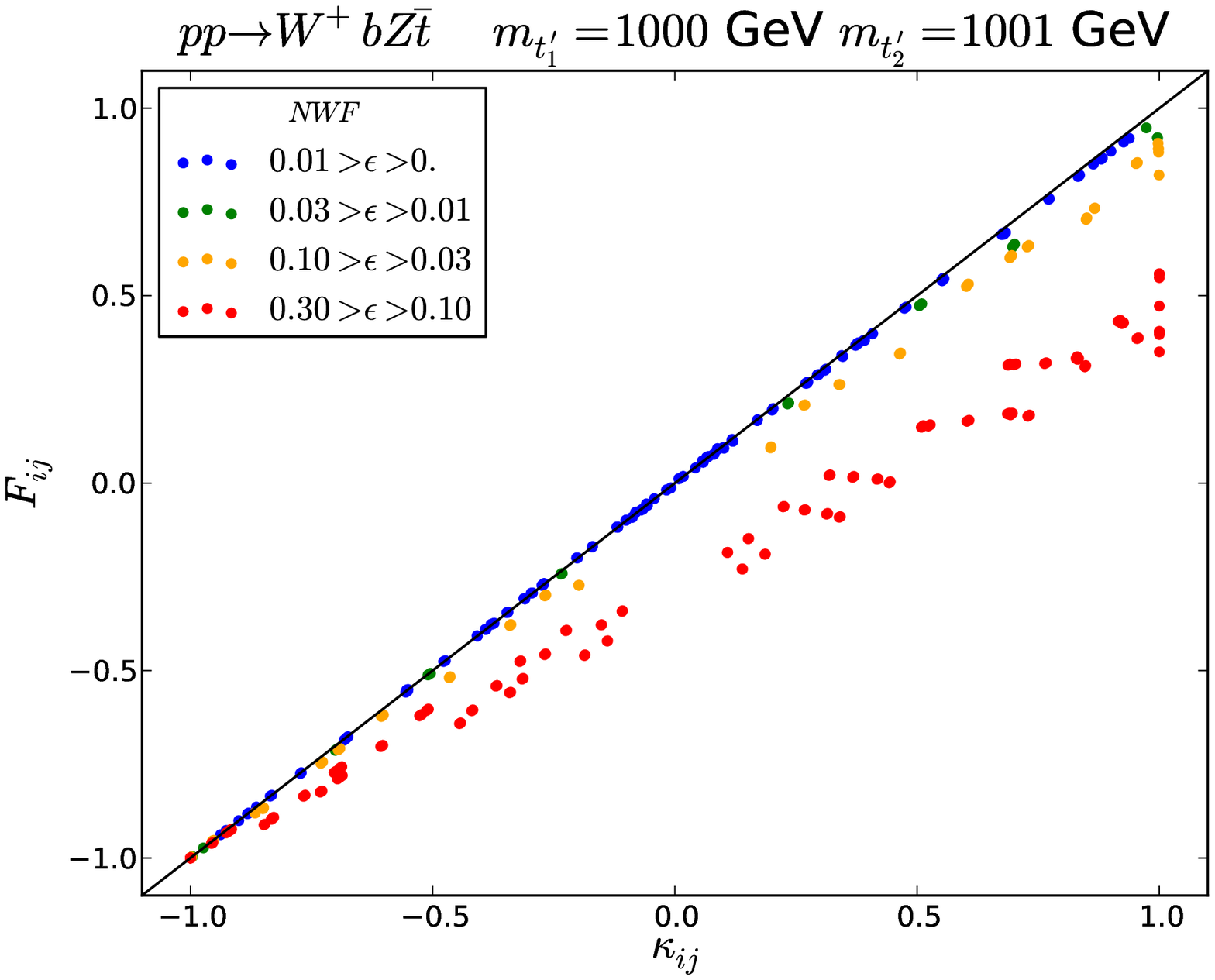,width=0.5\textwidth}%
\epsfig{file=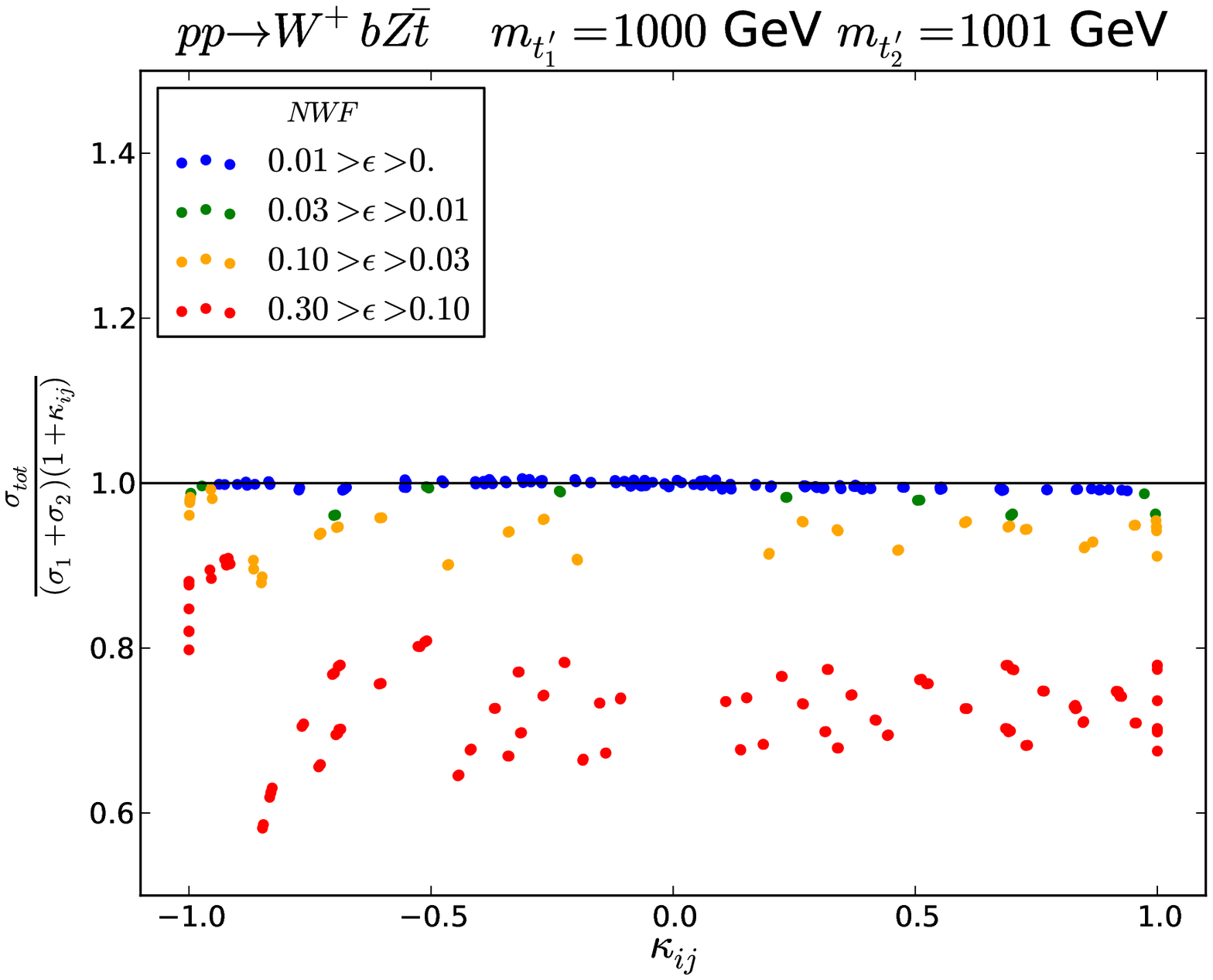,width=0.5\textwidth}%
\caption[$F_{ij}$ versus $\kappa_{ij}$ and $\frac{\sigma_{tot}}{(\sigma_1+\sigma_2)(1+\kappa_{ij})}$ versus $\kappa_{ij}$ for various values of the NWF for the $pp\to W^+bZ\bar{t}$ process.]{$F_{ij}$ versus $\kappa_{ij}$ (left) and $\frac{\sigma_{tot}}{(\sigma_1+\sigma_2)(1+\kappa_{ij})}$ versus $\kappa_{ij}$ (right) for various values of the NWF for the $pp\to W^+bZ\bar{t}$ process.}
\label{fig:range-of-validity}
\end{figure}

In Fig.~\ref{fig:range-of-validity} (left) we present results for $F_{ij}$ versus $\kappa_{ij}$ for values of the NWF in the  0.0--0.3  range for the $pp\to W^+bZ\bar{t}$ process. One can see that our description of the interference remains at a quite accurate level for NWF below about $10\%$ while already in the range 10\%--30\% one can see non-negligible deviations from the analytic formula predictions, i.e., $\kappa_{ij}$, as compared to the true value of the interference, $F_{ij}$.
The ``triangle" shape of the pattern of the left frame of Fig.~\ref{fig:range-of-validity} is simply related to the fact that, in case of large negative interference, the ${\sigma^{\rm tot}_{ij}}$ value is close to zero. Therefore, even in case of large $relative$ deviations, the $predicted$ value of   ${\sigma^{\rm tot}_{ij}}$ will be still close to zero, forcing $F_{ij}$ to be around $-1$, according to Eq.~(\ref{orderparameter}), even in case of large values of the NWF parameter.
Therefore, it is important to look at the complementary plot presenting $\frac{\sigma_{tot}}{(\sigma_1+\sigma_2)(1+\kappa_{ij})}$ versus $\kappa_{ij}$ shown in Fig.~\ref{fig:range-of-validity} (right).
One can see that deviations of the  cross section predicted by the ``master formula", $(\sigma_1+\sigma_2)(1+\kappa_{ij})$, from the real one, $\sigma_{tot}$, depends only on the value of NWF. For large values of NWF one can also see that $\sigma_{\rm tot}$ is below $(\sigma_1+\sigma_2)(1+\kappa_{ij})$, which is related to the fact that in case of  $\sigma_{\rm tot}$ the pure Breit-Wigner shape of the $T_i$ resonances is actually distorted and suppressed on the upper end due  to steeply falling parton distribution functions.
Furthermore, one should note that the quite accurate description of the interference found at the integrated level for NWF $<0.1$ remains true at differential level too.
Finally, we remark that the multi-parametric scan was done using CalcHEP3.4 on the HEPMDB database~\cite{hepmdb}, where the model studied here can be found under the \url{http://hepmdb.soton.ac.uk/hepmdb:1113.0149} link.

\subsection{Conclusions}

We have studied the role of interference in the process of pair production of XQs.
Considering such interference effects is crucial for the reinterpretation of the results of experimental searches of new quarks decaying to the same final state in the context of models with a new quark sector, which is usually not limited to the presence of only one heavy quark.
We have shown that, if the small $\Gamma/m$ approximation holds, and therefore it is possible to factorise the production and decay of the new quarks, the interference contribution can be described by considering a parameter which contains only the relevant couplings and the scalar part of the propagators of the new quarks. 

We have obtained a remarkably accurate description of the exact interference (described by the term $F_{12}$ defined in Eq.~(\ref{orderparameter})) using a simple analytical formula for the parameter $\kappa_{ij}$ defined in Eq.(\ref{eq:int2}).
This description  holds regardless of the chiralities of the couplings between the new and SM quarks, Eq.(\ref{eq:gen2}).
This means that it is possible to analytically estimate, with very good accuracy, the interference contribution to the pair production of two (and possibly more) quarks pairs decaying into the same final state, once couplings, total widths and masses are known, without performing a dedicated simulation or a full analytical computation.
We have also discussed the region of validity of this approximation in connection to the mixing effects at the loop-level contribution to a heavy quark self-energy which could potentially lead to a non-negligible interference. Therefore, in order to use the analytical formula for the interference we have derived, one should verify that  the off-diagonal contributions to the propagators  are suppressed and  check that the relation analogous to Eq.(\ref{eq:interference condition}) takes place for the particular model under study.

We have verified that also at the level of differential distributions it is possible to obtain the distributions including interference by a simple rescaling of those of the heavy quarks decaying to the given final state. Finally, we have checked that the linear correlation does not hold anymore for large values of the $\Gamma/m$ ratio, while it has been verified that for a NWF less than 10\% (which is very typical for all classes of models with XQs), the expressions for $\kappa_{ij}$ do indeed provide an accurate description of the interference term.
When interference effects are relevant and in the range of validity of our expressions, it is therefore possible to apply the same experimental efficiencies used for individual quark pairs to the full process of production and decay of two pairs of XQs.

\newpage

%% ------------- --------------------- --------------------- ---------------------- --------------------- ---------------------- ----------------------

\section{Large width effect on production and decay of XQs} \label{sec:VLQ_width}

We said previously that in order to be as model independent as possible, experimental searches for VLQs exploit an economical approach, assuming that only one new VLQ is present beyond the SM, consider QCD processes alone and parametrise the production and decay dynamics using the NWA. In this section we will follow the approach of \cite{Moretti:2016gkr} and study whether the use of the NWA is justified.

It is well known that, in the case of the top quark, effects induced onto the inclusive cross section by its Finite Width (FW) are of ${\cal O}(\Gamma_t/m_t)^2$, hence generally negligible, as $m_t\approx173$  GeV and $\Gamma_t\approx1.5$  GeV. A study of FW effects in final states corresponding to top pair production has been performed  in Ref.\cite{Kauer:2001sp}. One would naively expect that similar effects in the case of VLQs would be of the same size, i.e., of ${\cal O}(\Gamma_{\rm VLQ}/M_{\rm VLQ})^2$. However, it should be noted that, as
$M_{\rm VLQ}$ is unknown, also $\Gamma_{\rm VLQ}$  is, so that the aforementioned corrections may not be negligible,
if $\Gamma_{\rm VLQ}/M_{\rm VLQ}$ is not very small. In fact, also  differences between the case of the top quark and a VLQ  due to the different structure of their couplings in the charged decay currents would play a role\footnote{Notice that  {VLQs may also decay through flavour changing neutral currents}, involving both the Higgs and $Z$ bosons.}. In this connection, one should recall that, in taking the NWA, as generally done in most Monte Carlo (MC) programs used {in phenomenological and experimental analyses}, one neglects off-diagonal spin effects which stem from the quark (top or VL) being massive and whose size is intimately related to the vector/axial (or left/right) composition of the fermionic state entering the charged decay currents and, of course, to the value of the ratio $\Gamma_{\rm VLQ}/M_{\rm VLQ}$. Furthermore, these 
very same two aspects also enter the interfering terms between the heavy quark (top or VL) signal (whichever way this is defined in terms of Feynman diagrams) and the background (which would then be
represented by all the other graphs leading to the same final state). Needless to say, one should then not assume
that what is valid for the treatment of off-shellness effects of the top quark (and consequent interferences) remains so for VLQs as well. 

Very recently experimental searches for VLQs have started to explore the large width regime, considering single production of top and bottom VLQ partners~\cite{CMS:2017oef,CMS:2017zhw}. However, to our knowledge, no experimental limit has been set for topologies compatible with the pair production channels.
It is the purpose of this study to assess the regions of validity of the NWA for final states compatible with pair production and decay of a VLQ with charge 2/3 but where, due to its FW, the VLQ is produced, via both QCD and EW interactions, in pairs or even singly. Interference effects of various nature will also be considered. We will do so under the assumption that all the decay products of the heavy quark are visible SM states.

%%%%%%%%%%%%%%%%%%%%%%%%%%%%%%%%%%%%%%%%%%%%%%%%%%%%%%%%%%%%%%%%%%%%%%%%%
%%%%%%%%%%%%%%%%%%%%%%%%%%%%%%%%%%%%%%%%%%%%%%%%%%%%%%%%%%%%%%%%%%%%%%%%%
\subsection{Setup}
\label{sec:setup}

%%%%%%%%%%%%%%%%%%%%%%%%%%%%%%%%%%%%%%%%%%%%%%%%%%%%%%%%%%%%%%%%%%%%%%%%%
\subsubsection{Definitions}
To understand the effects of large widths on the signal, we will consider different processes, all leading to the same four-particle final state:

\begin{itemize}

\item\textit{QCD pair production and decay of on-shell VLQs}

This process is usually considered in experimental searches of VLQs. In the NWA it is possible to separate and factorize production and decay of the heavy quarks, thus allowing for a model independent analysis of the results. The cross section for this process is given by (hereafter, in our formulae, $Q$ denotes a VLQ):
\begin{equation}
 \sigma_X \equiv \sigma_{2 \to 2}~{\rm BR}(Q)~{\rm BR}(\bar Q)
\end{equation}
where, obviously, $\sigma_{2 \to 2}$ only takes into account pure QCD topologies.

% \item\textit{QCD pair production and decay of VLQs, but without imposing the on-shell condition}
% 
% The topologies for this process are the same as the previous one, but in this case the VLQs are not required to be on-shell. Production and decays are not factorized. This process is equivalent to the previous one in the narrow width limit, but describes more accurately scenarios where the widths of the VLQs are large. Furthermore, this process takes into account the spin correlations between the quark and antiquark branches, which are lost if production and decays are factorized. The cross section of this process will be labelled as $\sigma_P$. However, it must be noticed that by considering this subset of topologies, the gauge-invariance of the process is lost; the gauge dependence can only be assured by considering the full signal (see below). The quantitative effects of the gauge-dependence of this selection of topologies will be discussed in the following sections.

\item\textit{Full signal}

In this process all the topologies which contain \textit{at least one} VLQ propagator are taken into account. The only assumption is that the QCD and EW order of the processes are the same as in the processes above, for consistency. The full signal includes the pair production process without the on-shell condition described above. The cross section of this process will be labelled as $\sigma_S$. Some example topologies for this process which are not included in the previous ones are in Fig.~\ref{fig:fullsignaltopologies1}. The full signal contains topologies which are generally subleading in the NWA, but that become more and more relevant as the width of the VLQ increases.

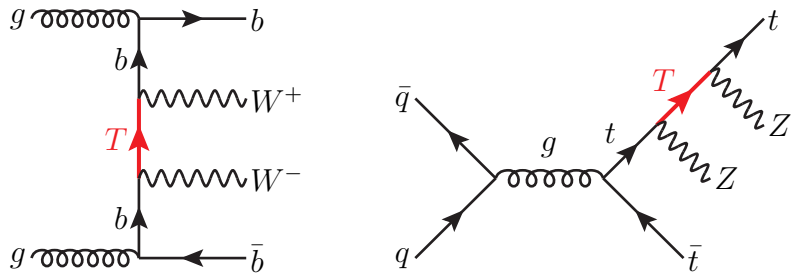
\begin{figure}[H]
\begin{center}
\begin{picture}(120,100)(0,-10)
\SetWidth{1}
\Gluon(10,0)(50,0){3}{6}
\Text(8,0)[rc]{\large $g$}
\Gluon(10,90)(50,90){3}{6}
\Text(8,90)[rc]{\large $g$}
\Line[arrow](90,0)(50,0)
\Text(92,0)[lc]{\large $\bar b$}
\Line[arrow](50,0)(50,30)
\Text(46,15)[rc]{\large $b$}
\SetColor{Red}\SetWidth{1.5}
\Line[arrow](50,30)(50,60)
\Text(46,45)[rc]{\large\Red{$T$}}
\SetColor{Black}\SetWidth{1}
\Line[arrow](50,60)(50,90)
\Text(46,75)[rc]{\large $b$}
\Line[arrow](50,90)(90,90)
\Text(92,90)[lc]{\large $b$}
\Photon(50,30)(90,30){3}{6}
\Text(92,30)[lc]{\large $W^-$}
\Photon(50,60)(90,60){3}{6}
\Text(92,60)[lc]{\large $W^+$}
\end{picture}
\hskip 20pt
\begin{picture}(120,100)(0,-10)
\SetWidth{1}
\Line[arrow](10,0)(40,30)
\Text(8,0)[rc]{\large $q$}
\Line[arrow](40,30)(10,60)
\Text(8,60)[rc]{\large $\bar q$}
\Gluon(40,30)(80,30){3}{5}
\Text(60,38)[cb]{\large $g$}
\Line[arrow](80,30)(100,50)
\Text(83,47)[cc]{\large $t$}
\Line[arrow](110,0)(80,30)
\Text(112,0)[lc]{\large $\bar t$}
\SetColor{Red}\SetWidth{1.5}
\Line[arrow](100,50)(120,70)
\Text(103,67)[cc]{\large\Red{$T$}}
\SetColor{Black}\SetWidth{1}
\Photon(100,50)(120,30){3}{5}
\Text(122,30)[lc]{\large $Z$}
\Line[arrow](120,70)(140,90)
\Text(142,90)[lc]{\large $t$}
\Photon(120,70)(140,50){3}{5}
\Text(142,50)[lc]{\large $Z$}
\end{picture}
\end{center}
\caption{Examples of topologies containing only one VLQ propagator for the $P P \to W^+ b W^- \bar b$ and $P P \to Z t Z \bar t$ processes.}
\label{fig:fullsignaltopologies1}
\end{figure}

\item\textit{SM irreducible background}

This process trivially corresponds to all the $2 \to 4$ topologies which do not involve any VLQ propagators. The cross section will be labelled as $\sigma_B$.

\item\textit{Total process}

This process includes the full signal, the SM background and the interference terms. The cross section will be labelled as $\sigma_T$ and is related to the previous cross sections by the following relation:

\begin{equation}
 \sigma_T = \sigma_S + \sigma_B + \sigma_\text{interference}
\end{equation}

\end{itemize}

In order to determine the effect of large widths on the cross section, we will consider a number of variables:

\begin{itemize}
% \item $\frac{\sigma_P - \sigma_X}{\sigma_X}$: this ratio quantifies the pure off-shell contributions to the pair-production and decay of VLQs
\item $\frac{\sigma_S - \sigma_X}{\sigma_X}$: this ratio takes into account both the off-shell and the subleading contributions given by topologies which contain at least one VLQ propagator. It measures in practice how much the full signal differs from the approximate pair-production-plus-decay signal in the NWA.
\item $\frac{\sigma_T - (\sigma_X + \sigma_B)}{\sigma_X + \sigma_B}$: this ratio measures the correction factor to apply to obtain the full cross section starting with the pair-production in the NWA and the SM background considered independently.
\item $\frac{\sigma_T - (\sigma_S + \sigma_B)}{\sigma_S + \sigma_B}$: this ratio measures the size of the interference effects between signal and SM background.
\end{itemize}

%%%%%%%%%%%%%%%%%%%%%%%%%%%%%%%%%%%%%%%%%%%%%%%%%%%%%%%%%%%%%%%%%%%%%%%%%
\subsubsection{Tools and validation }

Our numerical results at partonic level have been obtained using {\sc MadGraph 5}~\cite{Alwall:2011uj,Alwall:2014hca} with the public VLQ model~\cite{feynrulesVLQ} implemented in {\sc FeynRules}~\cite{Alloul:2013bka}. We have produced events in the five-flavour scheme, using the {\sc cteq6l1}~\cite{Pumplin:2002vw} PDF set. Hadronisation and parton showering have been obtained through the {\sc Pythia\,8} code~\cite{Sjostrand:2014zea}. To obtain the width dependent bounds on the VLQ mass we have considered a combination of searches at 8 TeV and an ATLAS search~\cite{TheATLAScollaboration:2016gxs} at 13 TeV. All the searches we considered are present in the database of the code {\sc CheckMATE\,2}~\cite{Dercks:2016npn}, which exploits the {\sc Delphes\,3} framework~\cite{deFavereau:2013fsa}. We stress here that the purpose of our recasting is not to obtain bounds for large width VLQs but to study the performance of sets of cuts currently adopted in searches for pair production of VLQs or optimised for different final states. Determining an optimised set of selection and kinematics cuts to enhance the sensitivity to the kinematics of a $T$ with large width (and therefore determine a reliable bound in the mass-width plane) will be the scope of a future dedicated study.  \vspace{\baselineskip}

Furthermore, to fully validate our analysis of the NWA results versus the off-shell ones, we developed a separate code where the
Dirac function is obtained as the appropriate limit of the Breit-Wigner distribution,
we have also prepared a dedicated $2\to 6$ program (hence also including the fermionic decays of the bosons stemming from the two $T$ decays, which are SM-like), wherein we have adopted  a suitable mapping of the integrand function, via the standard change of variable
\begin{equation}
p^2-M^2=M\Gamma\tan\theta,
\end{equation}
where $p^2$ is the (squared) moment flowing through a resonance with mass $M$ and width $\Gamma$\footnote{Here the width $\Gamma$ is taken as a constant, meaning that we do not take the $p^2$ dependence into account.}. This factorises the Jacobian
\begin{equation}
\d p^2=\frac{1}{M\Gamma}[(p^2-M^2)^2+M^2\Gamma^2]\d \theta,
\end{equation}
which thus incorporates the resonant behaviour in the sampling of the phase space itself, thereby rendering the multi-dimensional numerical 
integration (done via importance sampling) very efficient. Finally, upon multiplying the integrand function by $\Gamma/\Gamma_{\rm tot}$,
where $\Gamma_{\rm tot}$ is the decaying particle's intrinsic total width,  and taking the limit $\Gamma\to0$, we obtain self-consistently the above transition from the
off-shell to the NWA results. The results obtained this way closely match those obtained through MadGraph 5 for the aforementioned
$2\to2$ (on-shell, times ${\rm BR}$) and   $2\to4$ (off-shell) processes.

\begin{figure}[H]
\centering
\epsfig{file=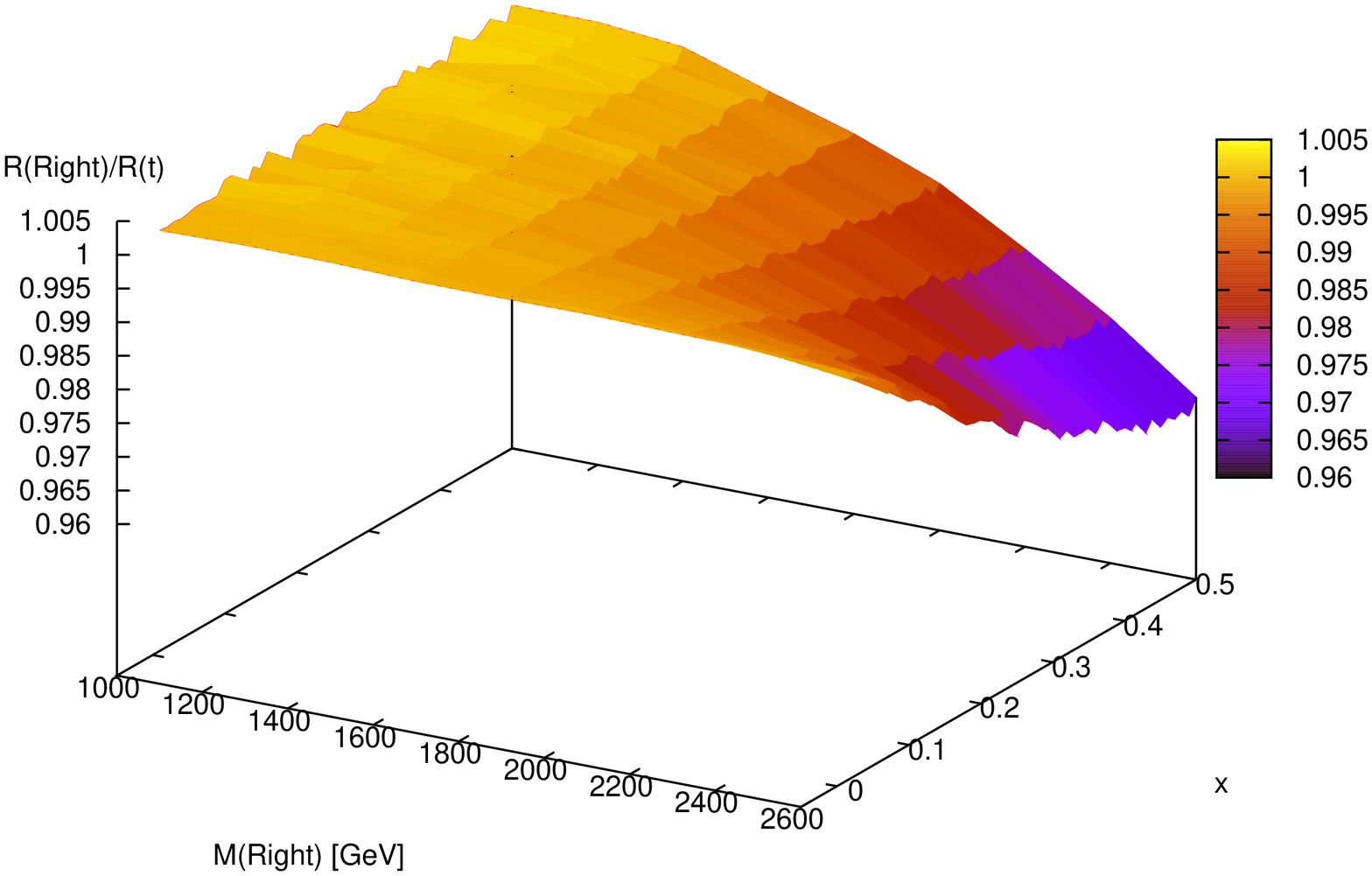,width=0.8\textwidth}
\vspace{-1cm}
\caption{Ratio of FW corrections with respect to the NWA relative to the $V-A$ case of a $V+A$ charged decay current.}
\label{fig:3D}
\end{figure}

As the SM top quark, $t$, and the heavy quark with same electro-magnetic charge, $T$, have a common decay channel, i.e., $bW^+$, as a preliminary exercise meant to address the impact of a potentially very different chiral structures in the transitions $t\to bW^+$ and $T\to bW^+$, we have defined the following quantity
\begin{equation}
R(X)=\frac{\sigma(pp\to X\to bW^+\bar bW^-\to 6~{\rm fermions})_{\rm FW}}
                 {\sigma(pp\to X\to bW^+\bar bW^-\to 6~{\rm fermions})_{\rm NWA}},
\end{equation} 
which measures inclusively the effect of a FW for the cases $X=t$ (a heavy quark with pure $V-A$ couplings,
i.e., top-like) and $X=$~Right (heavy quark with pure $V+A$ couplings). Clearly, these are extreme coupling choices, as 
an interaction eigenstate of a VLQ would have an admixture of $V-A$ and $V+A$ couplings.
However, it should be recalled that VLQ  couplings have always a dominant chirality: this has been demonstrated in  Refs.\cite{delAguila:2000rc,Buchkremer:2013bha}.
In Fig.~\ref{fig:3D} we plot the ratio $R({\rm Right})/R(t)$ mapped as a function of the heavy quark mass $M_{\rm VLQ}$ and relative width $x=\Gamma_{\rm VLQ}/M_{\rm VLQ}$ over the ranges [1000 GeV, 2500 GeV] (i.e., up to the typical mass reach of the LHC for pair production)
and [0, 0.5] (i.e., up to the width limit beyond which the VLQ can no longer be considered a resonance), respectively. One can see that differences are phenomenologically irrelevant.

\vspace{-1cm}

%%%%%%%%%%%%%%%%%%%%%%%%%%%%%%%%%%%%%%%%%%%%%%%%%%%%%%%%%%%%%%%%%%%%%%%%%
%%%%%%%%%%%%%%%%%%%%%%%%%%%%%%%%%%%%%%%%%%%%%%%%%%%%%%%%%%%%%%%%%%%%%%%%%
\subsection{Benchmarks and constraints\label{sec:Benchmarks}} 

In the present analysis we will consider the processes of production of a heavy top-like quark $T$. In principle, from a model independent point of view, the $T$ quark is allowed to interact with all SM quark generations, but to evaluate the effects of large widths in different scenarios, only specific interactions will be switched on in the different examples we will consider. 

Since the purpose of this analysis is to evaluate the effects of large widths on channels commonly explored by experimental analysis, we will consider only final states allowed by $T$ pair production and decay. The full set of channels in which a pair-produced $T$ quark can decay is given by the following matrix:
\begin{eqnarray}
\footnotesize
T\bar T \vspace*{-0.5mm} \to \vspace*{-0.5mm} \left( \vspace*{-1mm} 
\begin{array}{ccc|ccc|ccc}
Wd W\bar d & Wd Z\bar u & Wd H\bar u & Wd W\bar s & Wd Z\bar c & Wd H\bar c & Wd W\bar b & Wd Z\bar t & Wd H\bar t \\
Zu W\bar d & Zu Z\bar u & Zu H\bar u & Zu W\bar s & Zu Z\bar c & Wd H\bar c & Zu W\bar b & Zu Z\bar t & Zu H\bar t \\ 
Hu W\bar d & Hu Z\bar u & Hu H\bar u & Hu W\bar s & Hu Z\bar c & Wd H\bar c & Hu W\bar b & Hu Z\bar t & Hu H\bar t \\ 
\hline                                                                                                     
Ws W\bar d & Ws Z\bar u & Ws H\bar u & Ws W\bar s & Ws Z\bar c & Wd H\bar c & Ws W\bar b & Ws Z\bar t & Ws H\bar t \\ 
Zc W\bar d & Zc Z\bar u & Zc H\bar u & Zc W\bar s & Zc Z\bar c & Wd H\bar c & Zc W\bar b & Zc Z\bar t & Zc H\bar t \\ 
Hc W\bar d & Hc Z\bar u & Hc H\bar u & Hc W\bar s & Hc Z\bar c & Wd H\bar c & Hc W\bar b & Hc Z\bar t & Hc H\bar t \\ 
\hline                                                                                                     
Wb W\bar d & Wb Z\bar u & Wb H\bar u & Wb W\bar s & Wb Z\bar c & Wd H\bar c & Wb W\bar b & Wb Z\bar t & Wb H\bar t \\ 
Zt W\bar d & Zt Z\bar u & Zt H\bar u & Zt W\bar s & Zt Z\bar c & Wd H\bar c & Zt W\bar b & Zt Z\bar t & Zt H\bar t \\ 
Ht W\bar d & Ht Z\bar u & Ht H\bar u & Ht W\bar s & Ht Z\bar c & Wd H\bar c & Ht W\bar b & Ht Z\bar t & Ht H\bar t 
\end{array}
\vspace*{-1mm} \right) 
\label{eq:finalstates1}
\end{eqnarray}
We will focus on two blocks of this matrix, the top-left (corresponding to a $T$ interacting with the first SM generation) and the bottom-right ($T$ interacting with the third SM generation). As we are interested in the width dependence of ratios of cross sections and of mass bounds, we expect that the scenario of mixing with the second generation will not give sizeably different results with respect to the mixing with first generation, so we will not consider it in this analysis. Performing the analysis by selecting specific final states doesn't mean that we are assuming that the $T$ quark only interacts with first or third generation. Effects of large width are different depending on the kinematics of the process and by selecting representative scenarios it is possible to reconstruct intermediate configurations (VLQs interacting partly with heavy and partly with light SM generations).

\begin{figure}[H]
\centering\epsfig{file=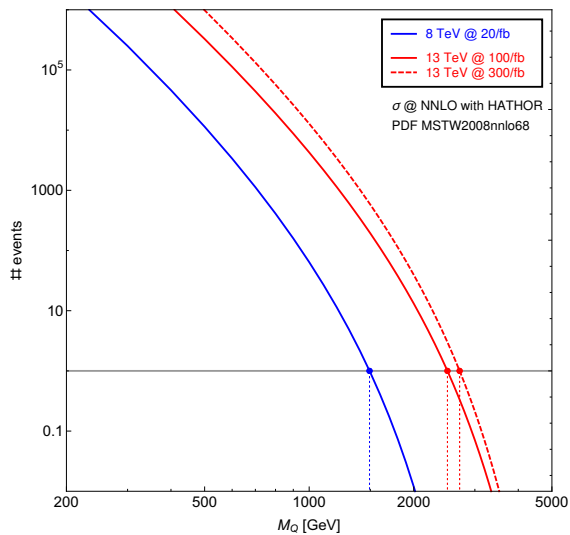, width=.5\textwidth} 
\caption[Number of events at partonic level for $Q \bar Q$ pair production and for different LHC energies and luminosities.]{Number of events at partonic level for $Q \bar Q$ pair production and for different LHC energies and luminosities. The corresponding cross sections have been computed using HATHOR\cite{Aliev:2010zk} with MSTW2008nnlo68 PDFs\cite{Martin:2009bu}.}
\label{fig:Xsigma}
\end{figure}

This analysis is of phenomenological interest only for mass values for which the number of final events is (ideally) larger than 1. In Fig.~\ref{fig:Xsigma} we show the number of events for different LHC luminosities for the 2 to 2 so-called $X$ channel, which is common to all scenarios. The number of events in Fig.~\ref{fig:Xsigma} has been computed considering a NNLO cross-section, however the results in the next sections will correspond to LO cross sections, as we are assuming that for processes of pair production the kinematics won't change appreciably and all the differences can be factorised through a K-factor. From Fig.~\ref{fig:Xsigma} it is possible to see that the ideal practical validity of our results is limited to mass values of around 1500 GeV for LHC@8TeV, 2500 GeV (2700 GeV) for LHC@13TeV with 100/fb (300/fb) integrated luminosity. Of course we are not considering here effects due to experimental acceptances and efficiencies.
% : this study is only meant to assess the role of the complete signal with respect to the common approximations made in theoretical and experimental analyses. 

%%%%%%%%%%%%%%%%%%%%%%%%%%%%%%%%%%%%%%%%%%%%%%%%%%%%%%%%%%%%%%%%%%%%%%%%%
%%%%%%%%%%%%%%%%%%%%%%%%%%%%%%%%%%%%%%%%%%%%%%%%%%%%%%%%%%%%%%%%%%%%%%%%%
\subsubsection{How large can the width be?}

\begin{figure}[H]
\centering
\epsfig{file=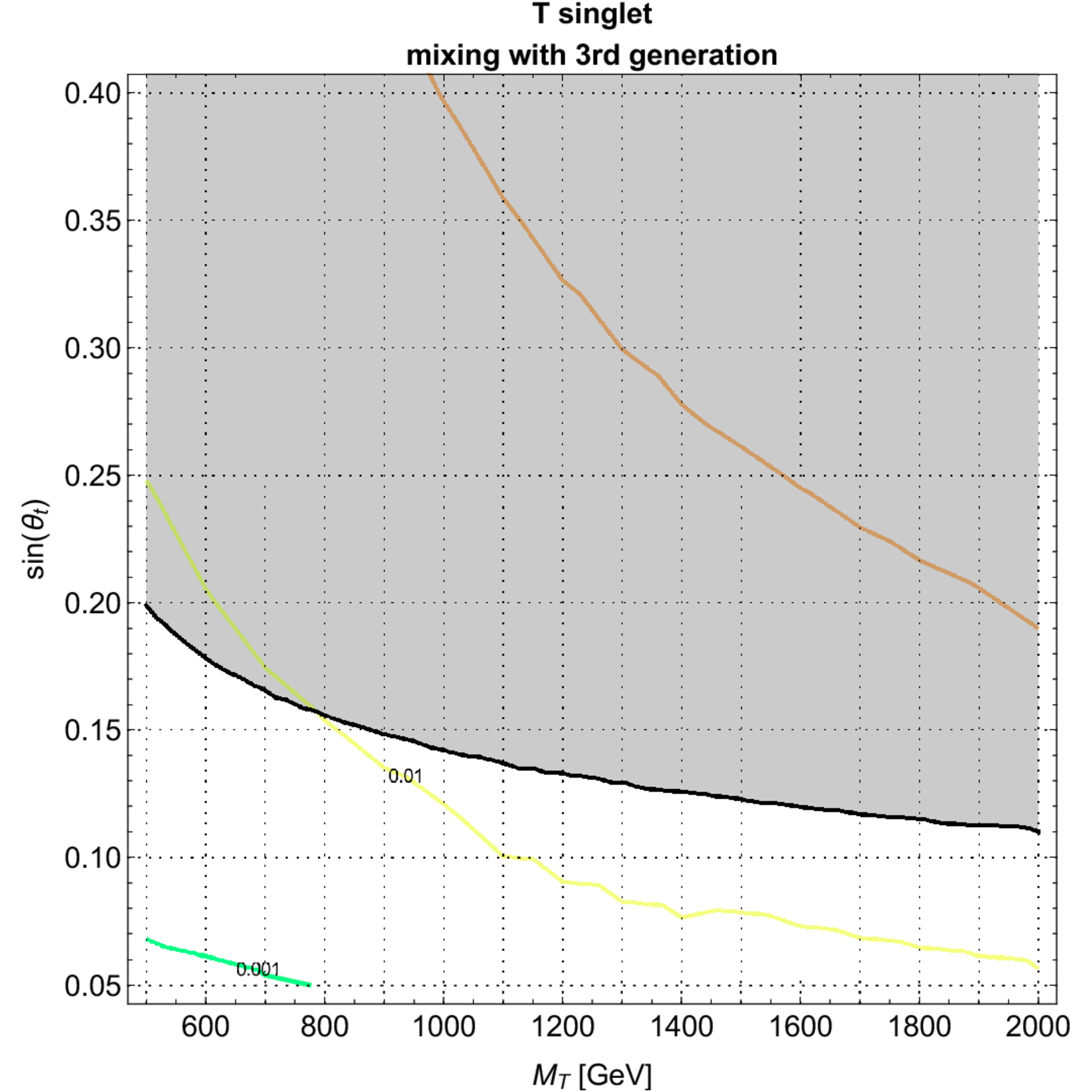, width=.32\textwidth}\hfill
\epsfig{file=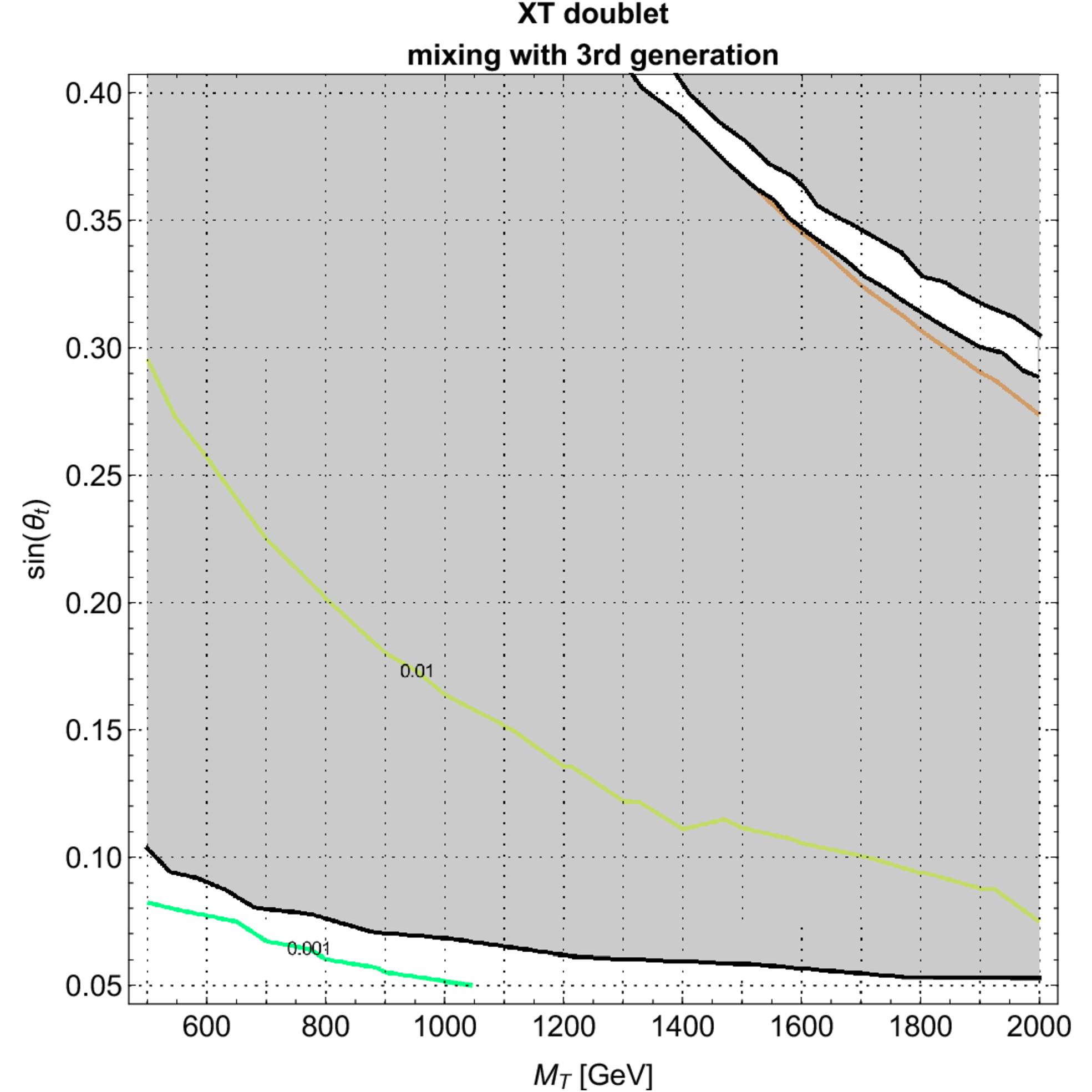, width=.32\textwidth}\hfill
\epsfig{file=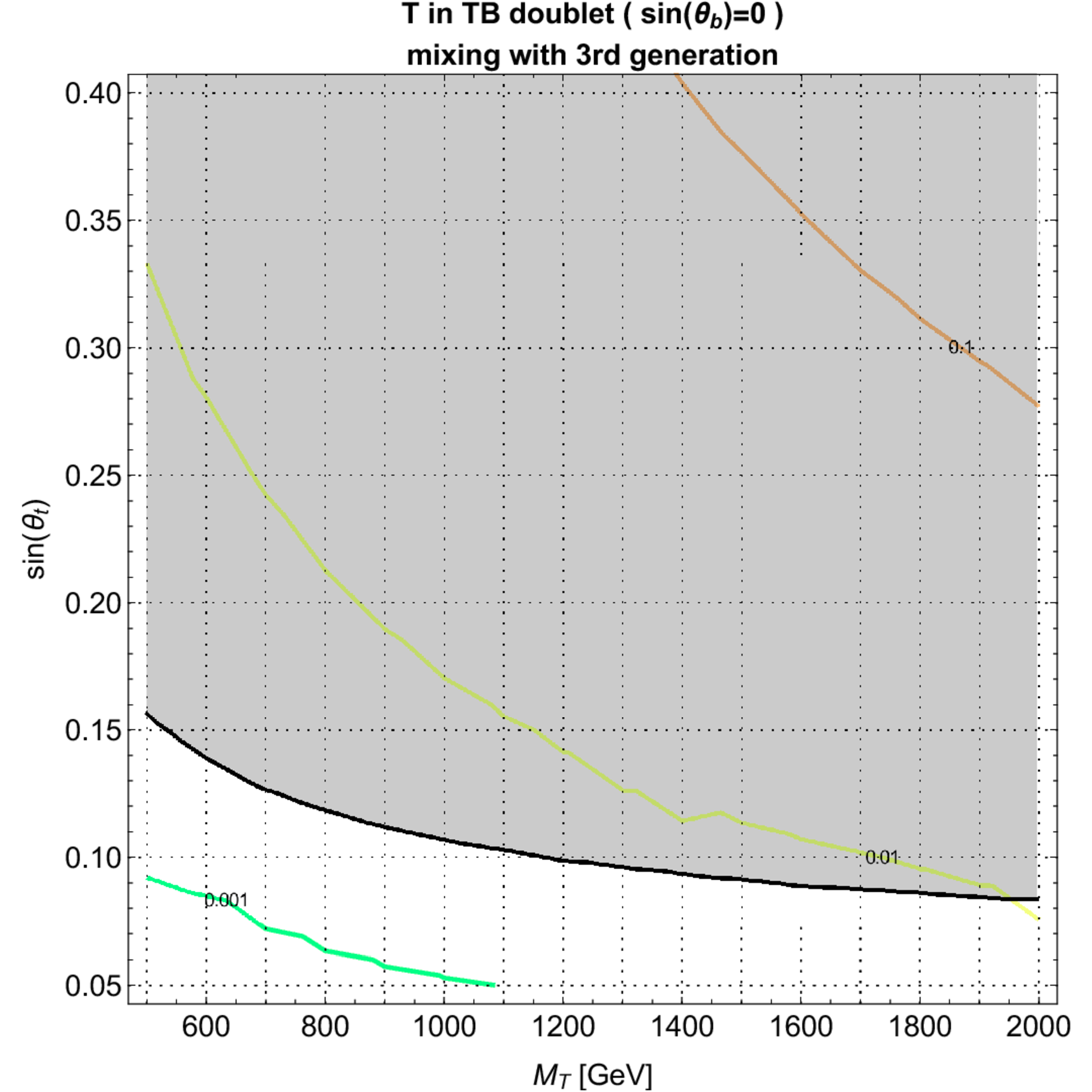, width=.32\textwidth}
\caption[Contours with constant $\Gamma/M$ ratio as function of $T$ mass and mixing angle for $T$ belonging to different representations and with different mixing hypotheses.]{Contours with constant $\Gamma/M$ ratio as function of $T$ mass and mixing angle for $T$ belonging to different representations and with different mixing hypotheses. The excluded (shaded) regions from~\cite{Chen:2017hak} have been superimposed.}
\label{fig:Tsimplifiedbounds}
\end{figure}

In a simplified model where the SM is only augmented by the presence of a VLQ representation containing a $T$ quark the couplings of the VLQ are constrained by different observables \cite{Okada:2012gy}. In contrast, a VLQ $T$ with a large width in such a scenario can only be obtained if its couplings are large. It is therefore important to determine how large the width can be in simplified scenarios if constraints on the $T$ couplings are saturated to the current bounds. Such bounds depend on the specific representation the $T$ state belongs to.
We will consider here as representative scenarios a $T$ singlet and a $T$ as part of a doublet (both $(X,~T)$ and $(T,~B)$). In both cases the BRs depend on both mass and width, but for the singlet the couplings are dominantly left-handed, while for the doublet the couplings are dominantly right-handed.
In Fig. \ref{fig:Tsimplifiedbounds} we show the contours with constant $\Gamma/M$ ratio for different values of the $T$ mass and mixing angle with the SM top quark, to which we have superimposed the excluded regions from EWPTs and $Zbb$ constraints, borrowed from Ref.\cite{Chen:2017hak}. 

Clearly, simplified models where the SM is extended with one VLQ representation containing a $T$ with large mixing are strongly constrained, and therefore the $T$ width cannot become larger than few percent of the mass (at best). The scenarios are even more constrained for $T$ quarks mixing with light generations, for which the bounds are tighter~\cite{Cacciapaglia:2011fx,Okada:2012gy}.
Therefore, to keep a model independent perspective we must assume that the width of the $T$ can become large because of the presence of further (yet undiscovered) new states lighter than the VLQ $T$, which results in a larger number of decay channels into further BSM particles, and/or because of mixing with other VLQs, which may relax constraints from flavour or precision observables because of cancellations of effects~\cite{Cacciapaglia:2009ic}. Hence, for the purposes of this analysis, the {\it total width} of the $T$ will be considered as a free parameter, limited to be less than the extreme value of 50\% of the mass of the VLQ. In practice, we will consider values up to 40\% of the $T$ mass for our numerical evaluations.

%%%%%%%%%%%%%%%%%%%%%%%%%%%%%%%%%%%%%%%%%%%%%%%%%%%%%%%%%%%%%%%%%%%%%%%%%
%%%%%%%%%%%%%%%%%%%%%%%%%%%%%%%%%%%%%%%%%%%%%%%%%%%%%%%%%%%%%%%%%%%%%%%%%
\subsection{Extra $T$ quark mixing with third generation SM quarks}

%%%%%%%%%%%%%%%%%%%%%%%%%%%%%%%%%%%%%%%%%%%%%%%%%%%%%%%%%%%%%%%%%%%%%%%%%
\subsubsection{Large width effects on the signal at parton level}

\begin{figure}[H]
\centering
\epsfig{file=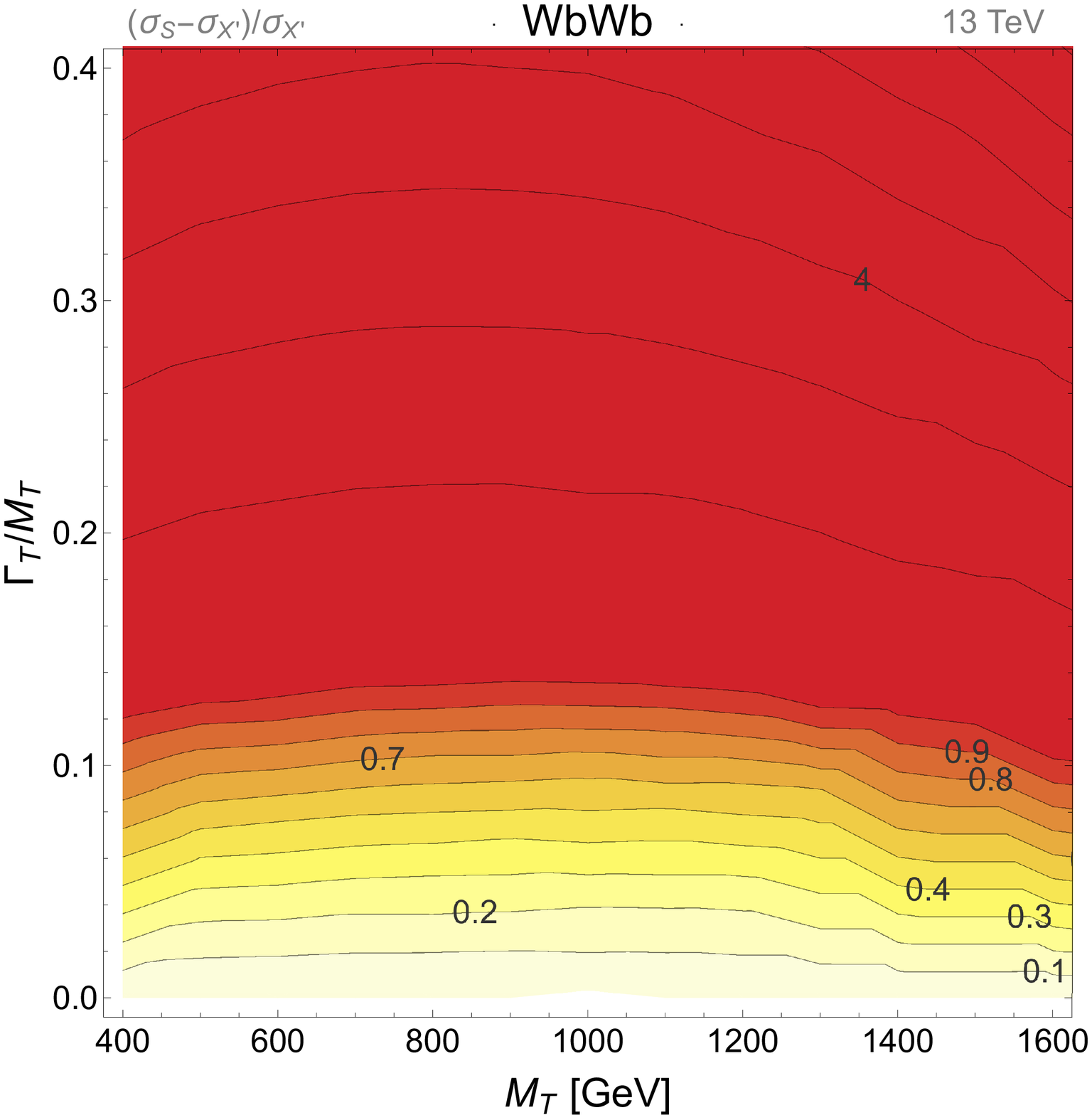, width=.3\textwidth}
\epsfig{file=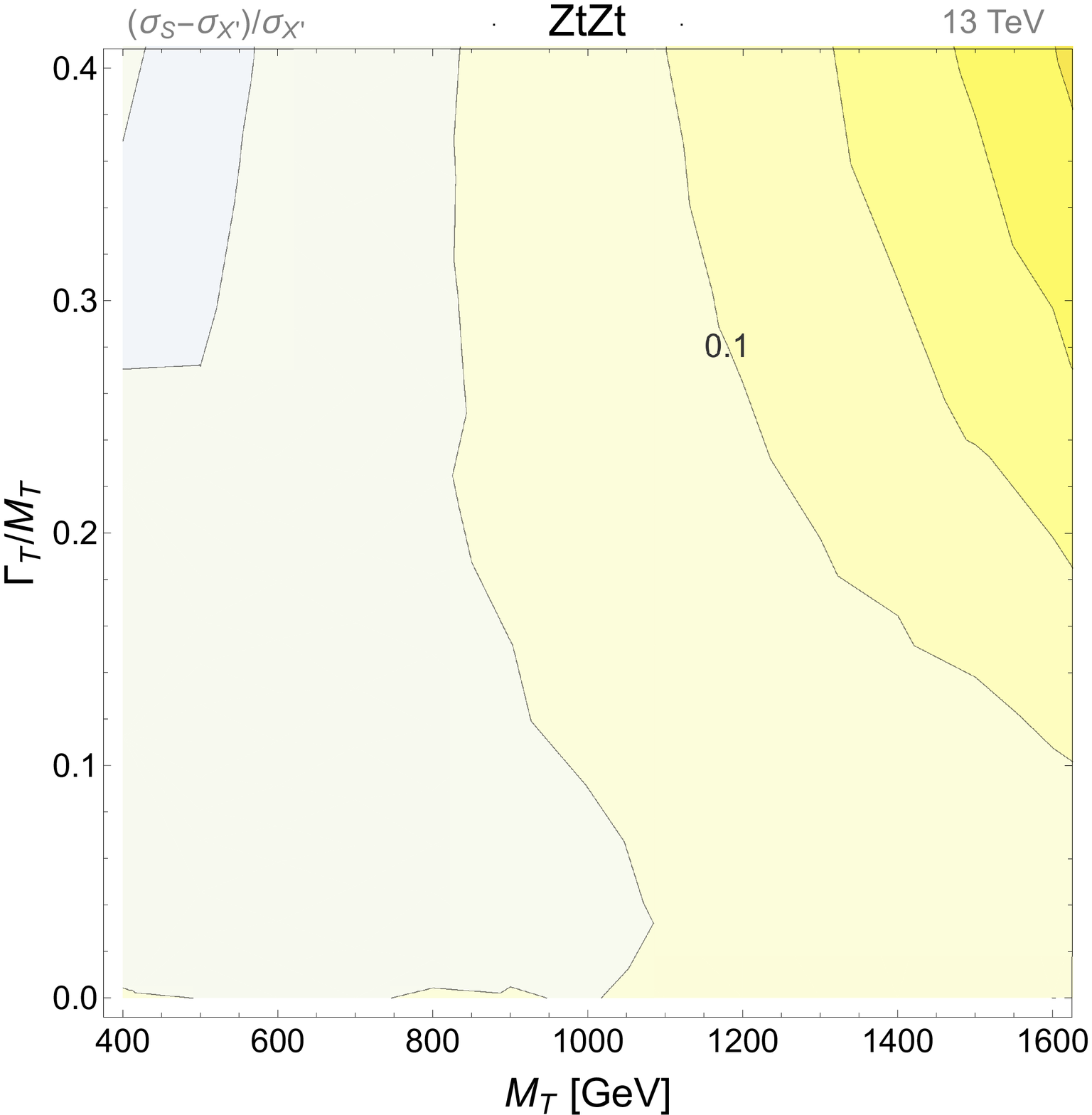, width=.3\textwidth} 
\epsfig{file=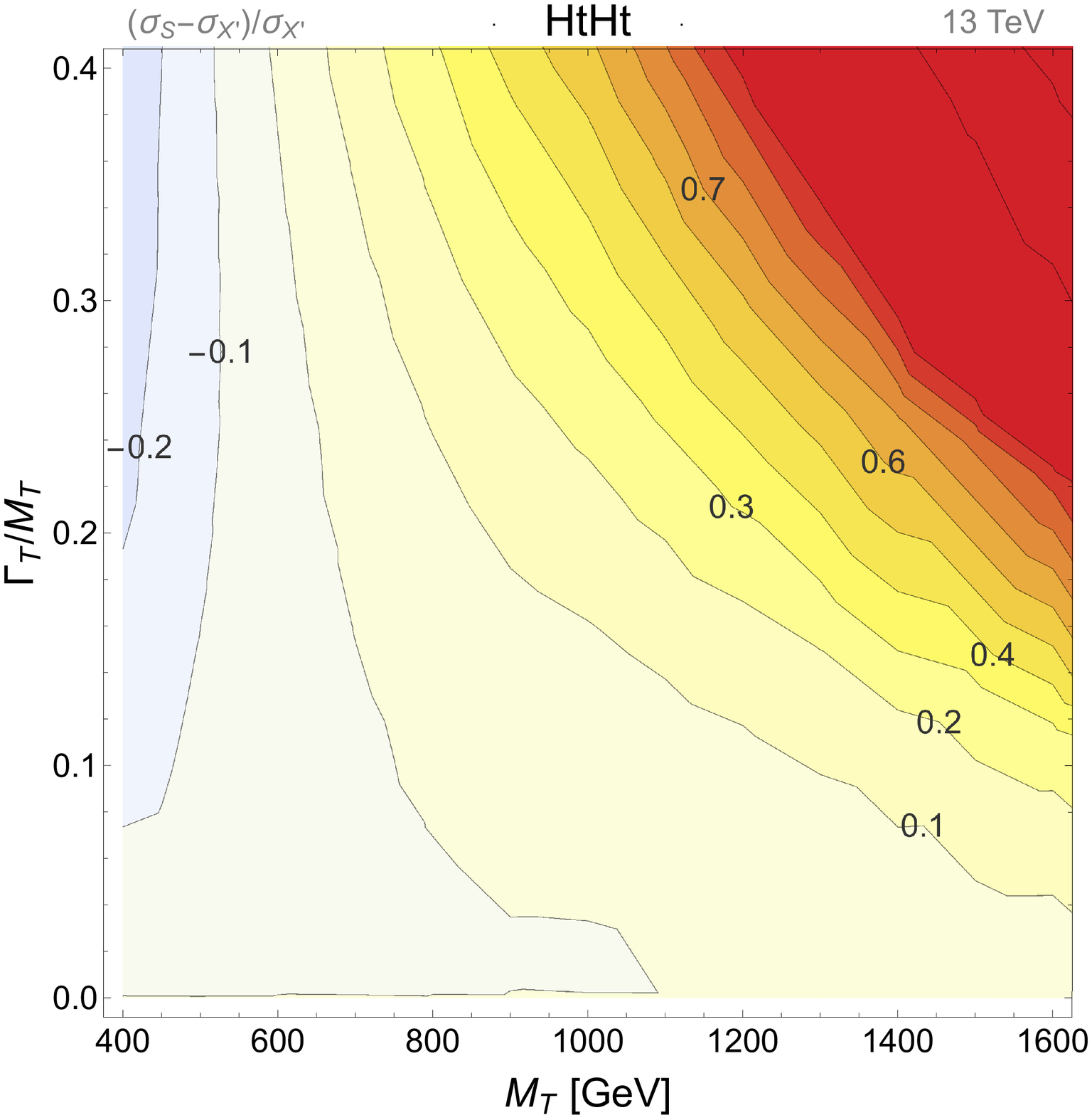, width=.3\textwidth}\\

\epsfig{file=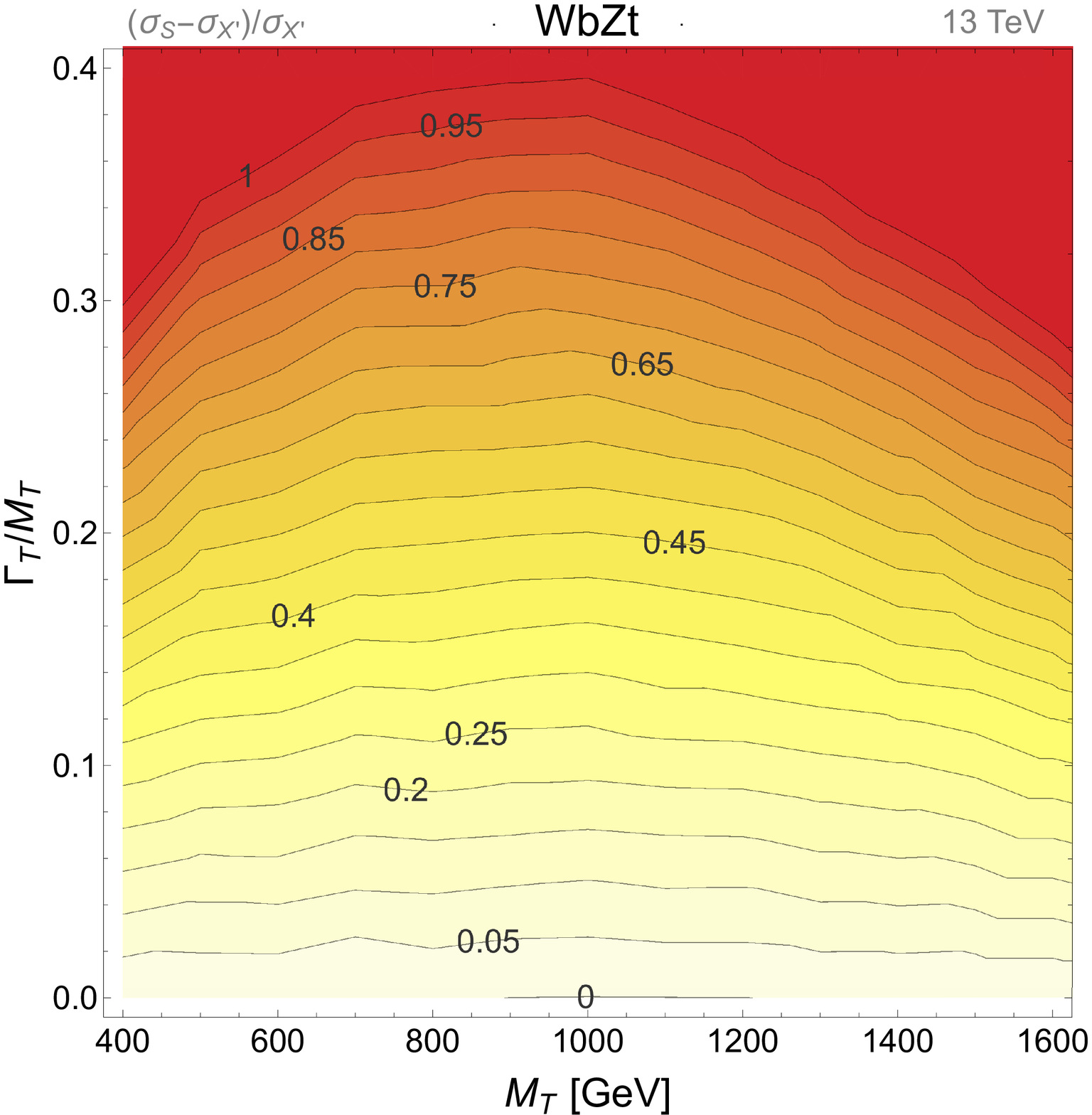, width=.3\textwidth} 
\epsfig{file=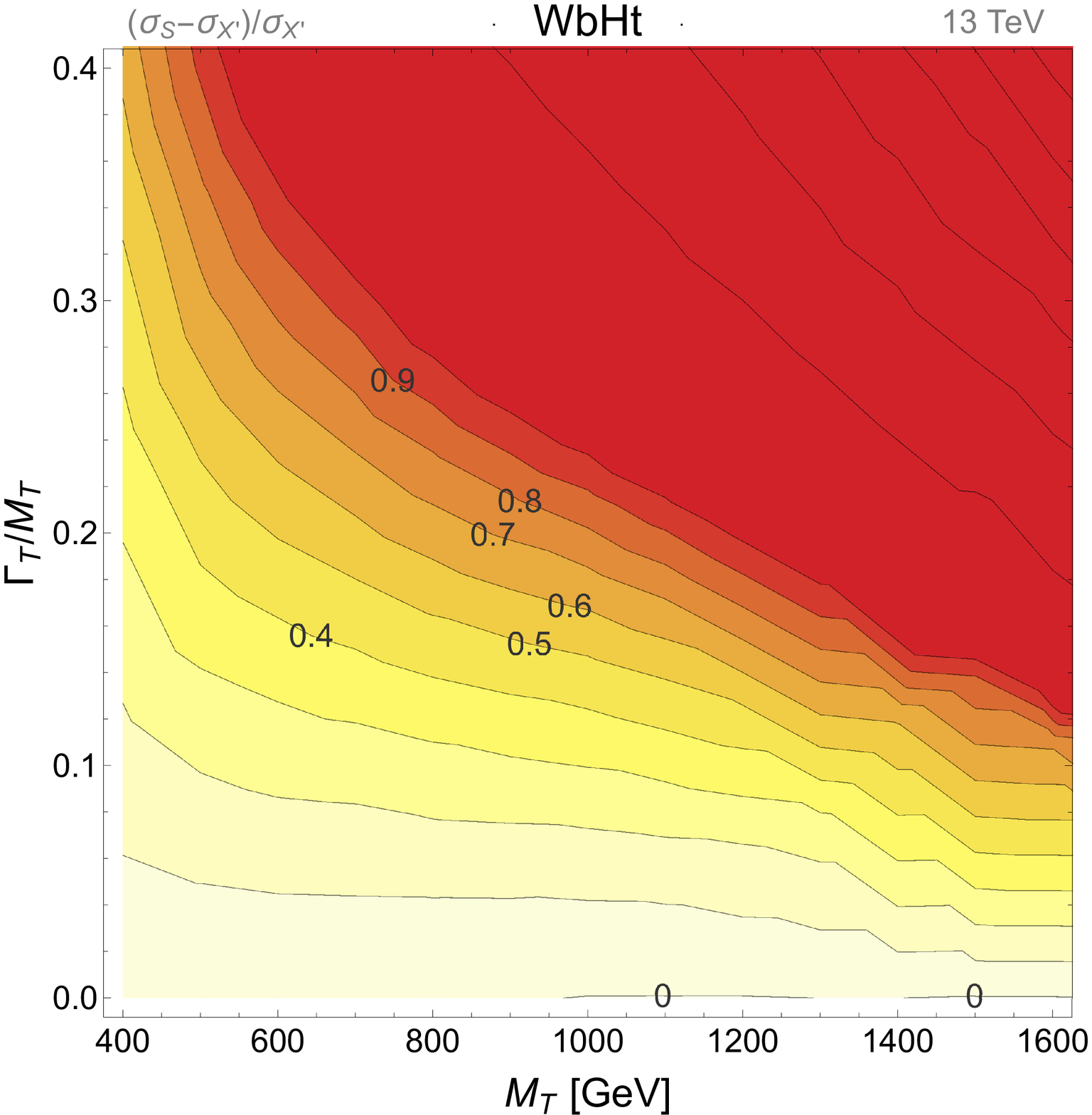, width=.3\textwidth} 
\epsfig{file=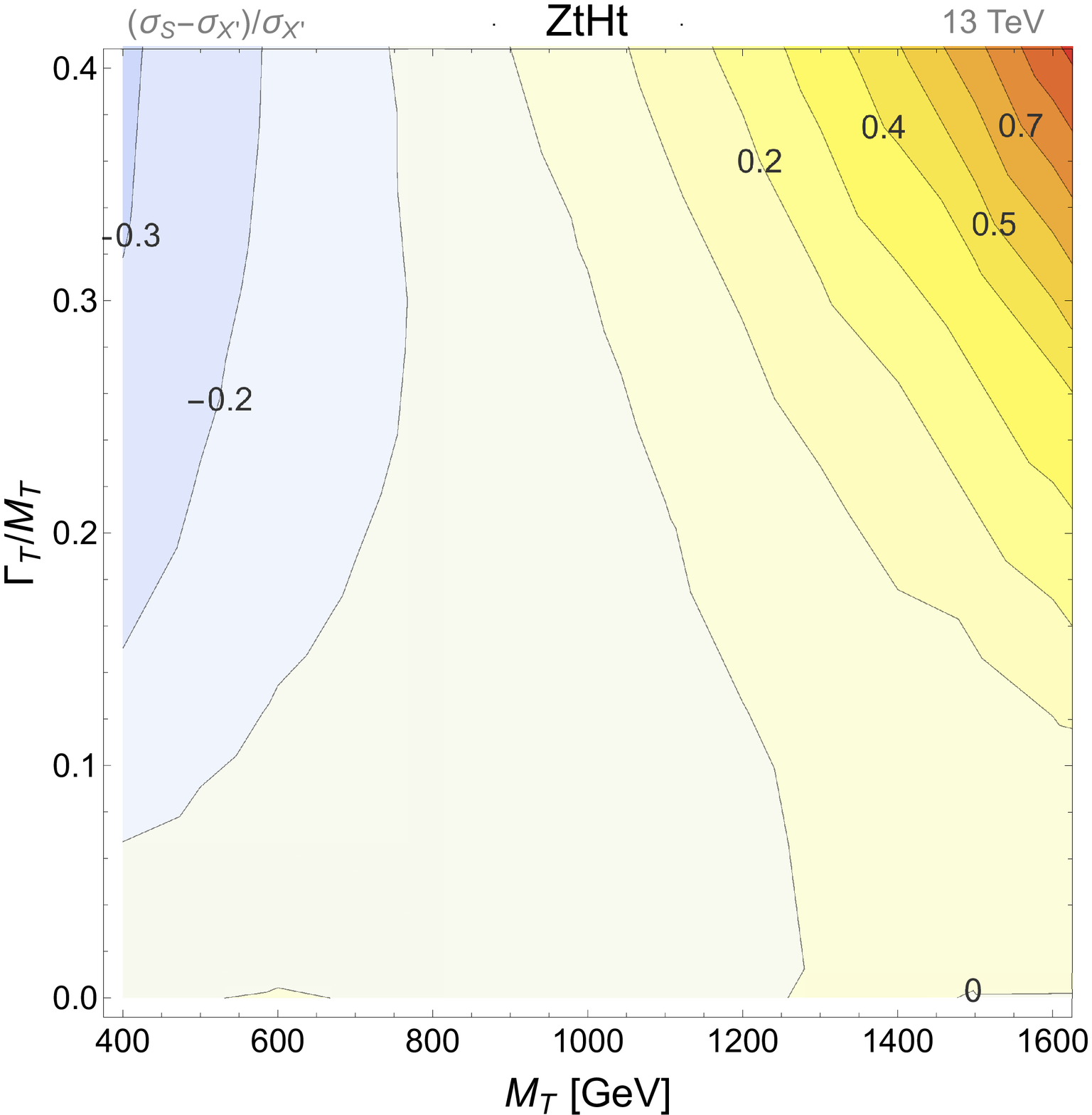, width=.3\textwidth} 

\caption{Relative difference between the full signal cross section $\sigma_S$ and the cross section of QCD pair production $\sigma_X$ for $T$ mixing with the SM top quark.}
\label{fig:SXthird1}
\end{figure}

The effect of a large width in the cross section due to off-shell contributions and to topologies which are absent in the NWA limit is shown in Fig.~\ref{fig:SXthird1}. At parton level we will only show results at 13 TeV. We verified that the results at 8 TeV are qualitatively similar.

As a first sanity check of our calculations we observe that, as expected, in the NWA limit the off-shell contributions are negligible. The contributions of off-shellness and new topologies become more and more relevant as the width of the $T$ increases and the cross section may eventually become several factors larger than in the NWA for some final states. The large increase of the cross section even for small $T$ masses for channels with the bottom in the final state is explained by the presence of diagrams where the $b$-jets are radiated directly from the initial state or generated by gluon splittings:  such topologies are enhanced by collinear divergences. We will not explore this aspect further, as the isolation and kinematics cuts applied at analysis level usually remove such enhanced contributions, independently of the $T$ mass and width as we will show in Sec. \ref{sec:detector3}.

\begin{figure}[H]
\centering
\epsfig{file=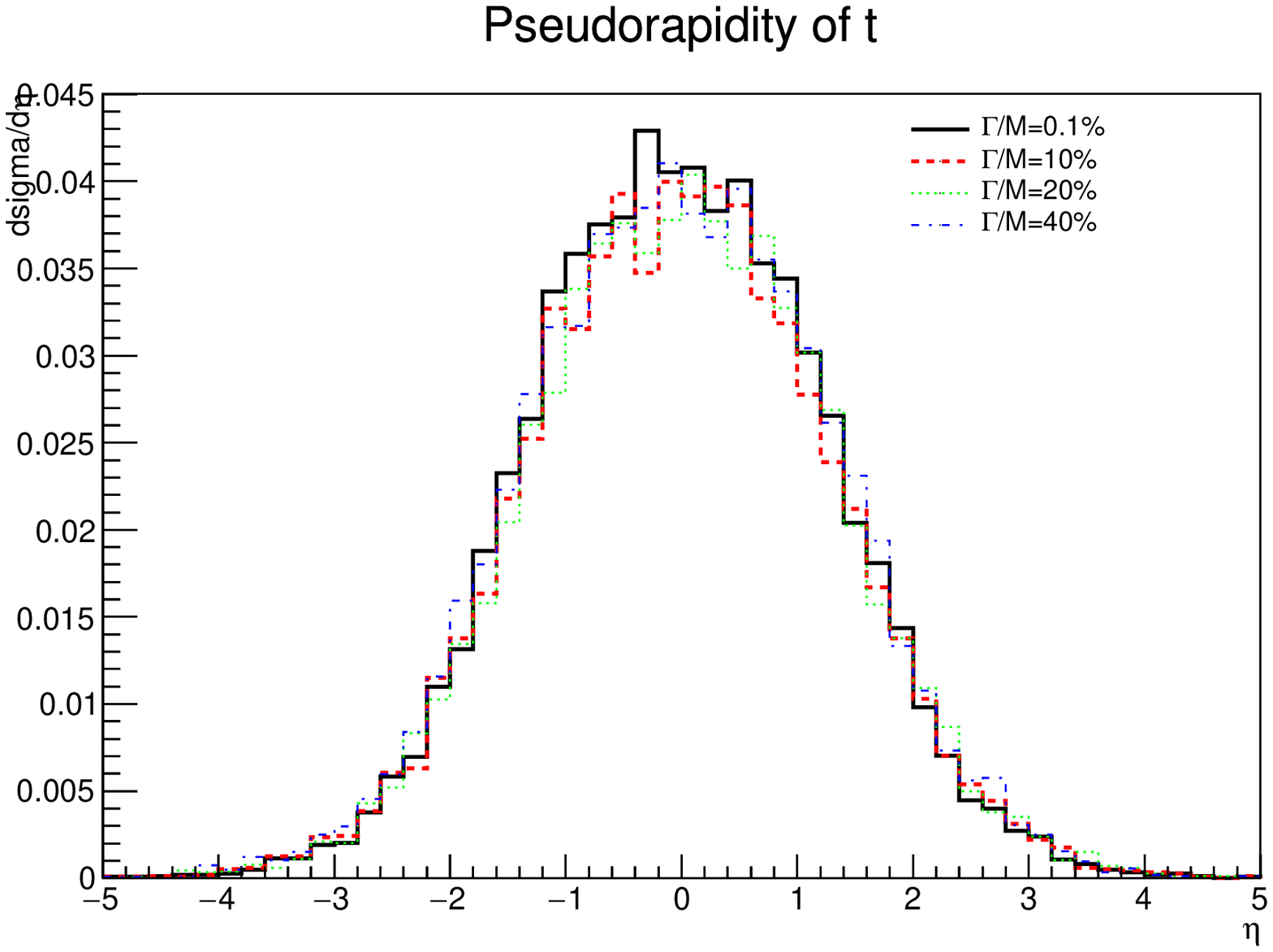, width=.45\textwidth}
\epsfig{file=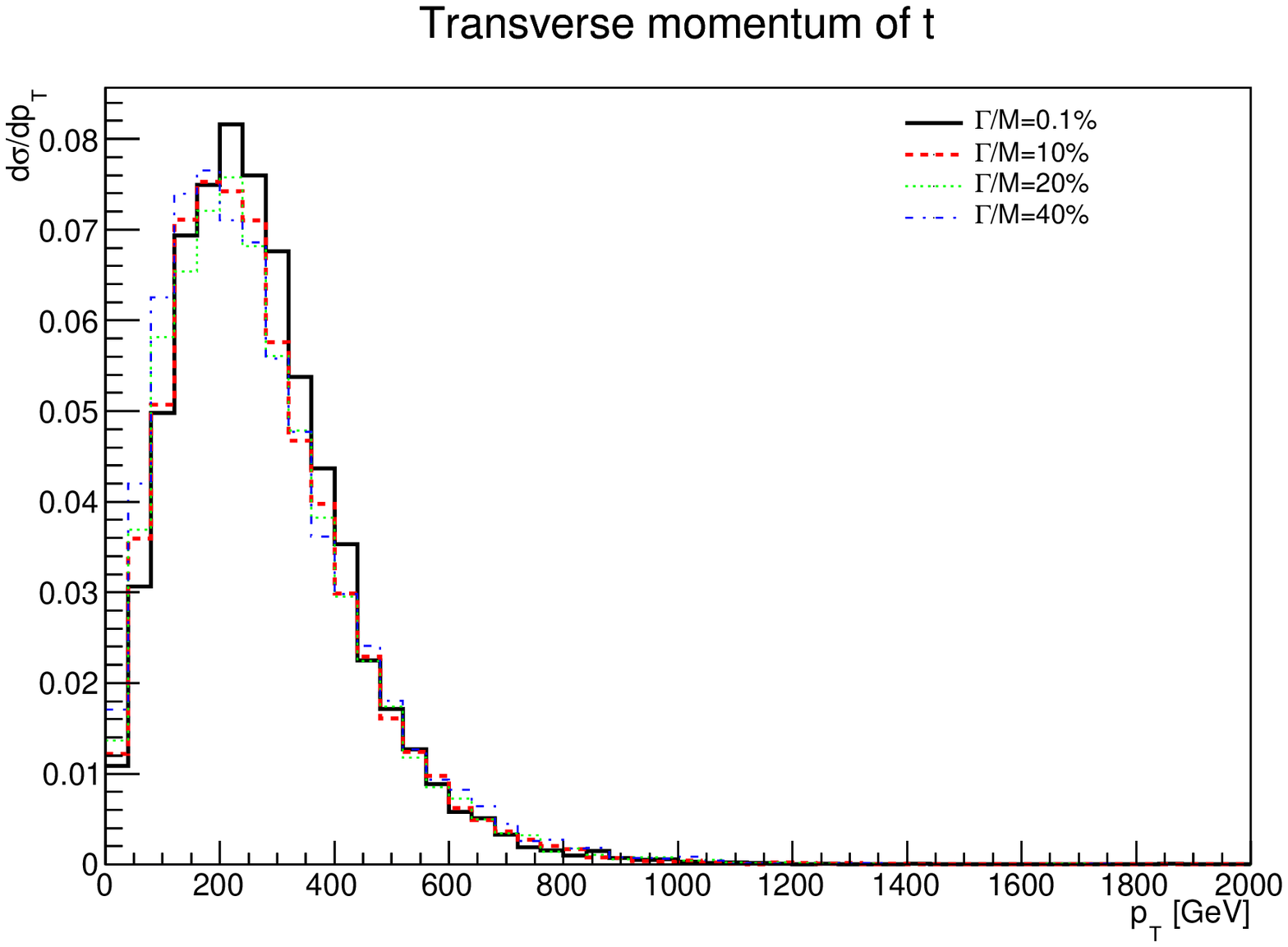, width=.45\textwidth}\\
\epsfig{file=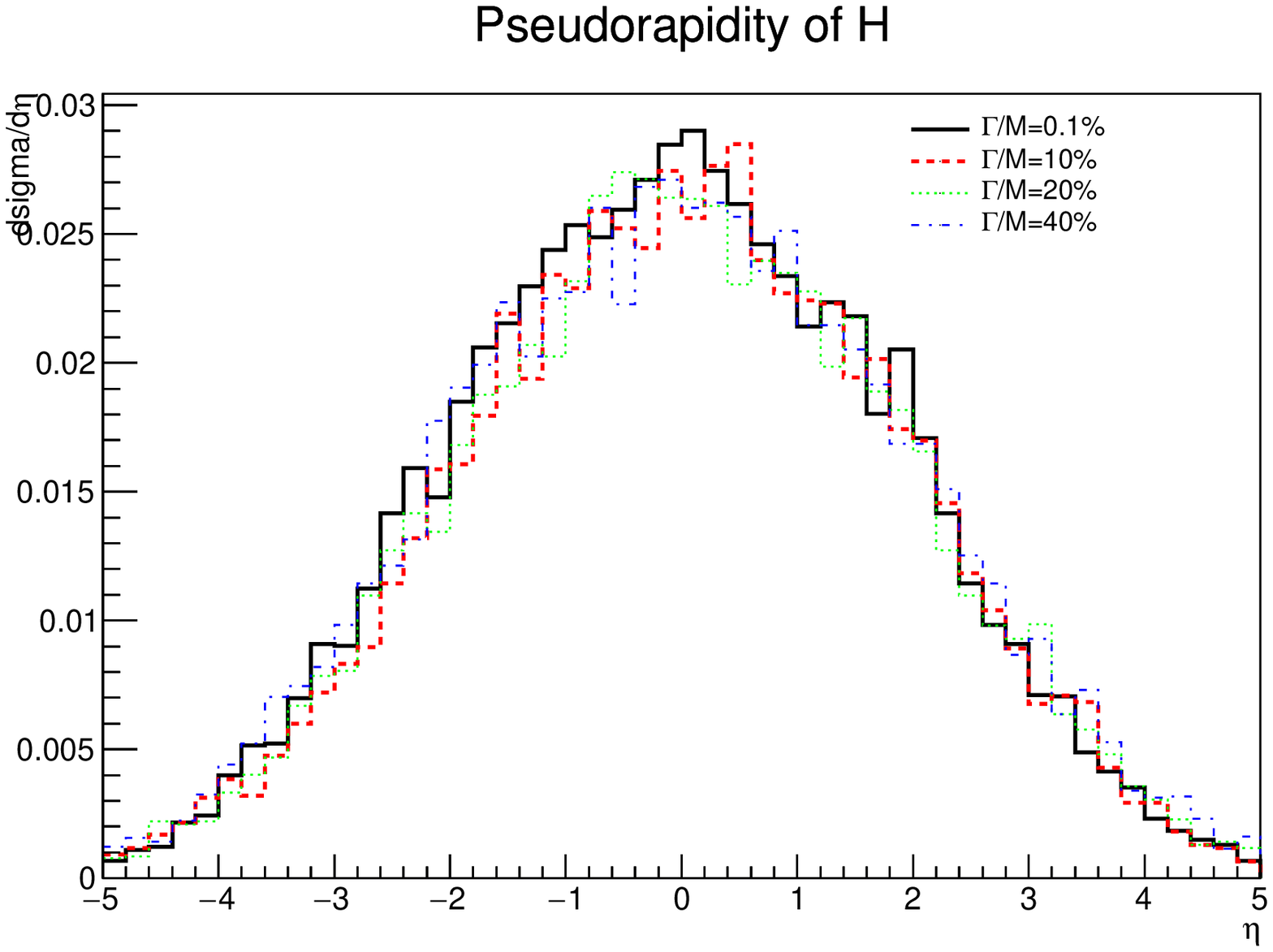, width=.45\textwidth}
\epsfig{file=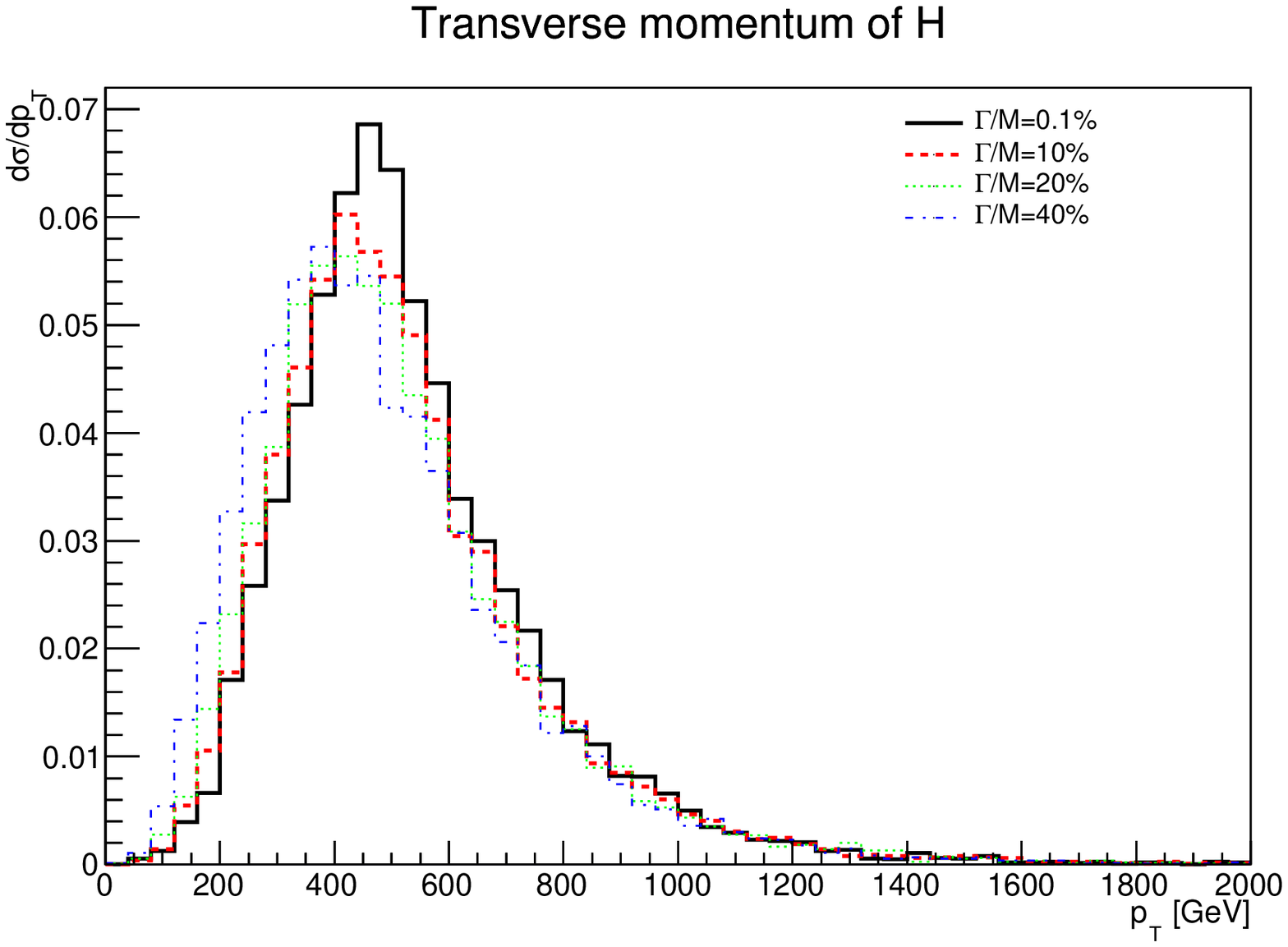, width=.45\textwidth}
\caption[Partonic level differential cross sections for the $HtHt$ channel.]{Partonic level differential cross sections for the $HtHt$ channel. From left to right and top to bottom: $\eta_t$, $p_{Tt}$, $\eta_H$ and $p_{TH}$. All distributions correspond to a $T$ mass of 600 GeV, for which $\sigma_S\sim\sigma_X$ almost independently of the $T$ width.}
\label{fig:disththt}
\end{figure}

\begin{figure}[H]
\centering
\epsfig{file=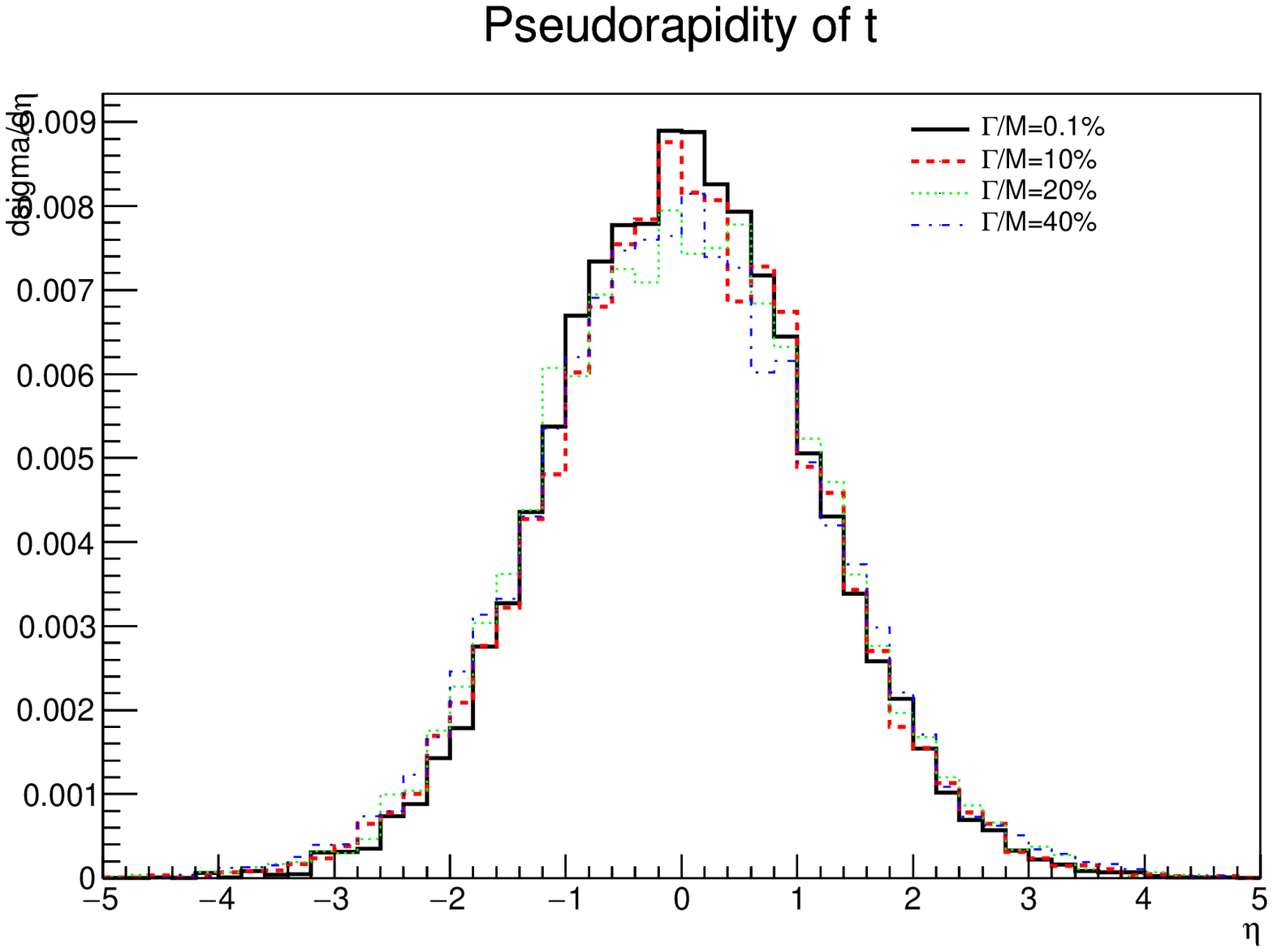, width=.45\textwidth}
\epsfig{file=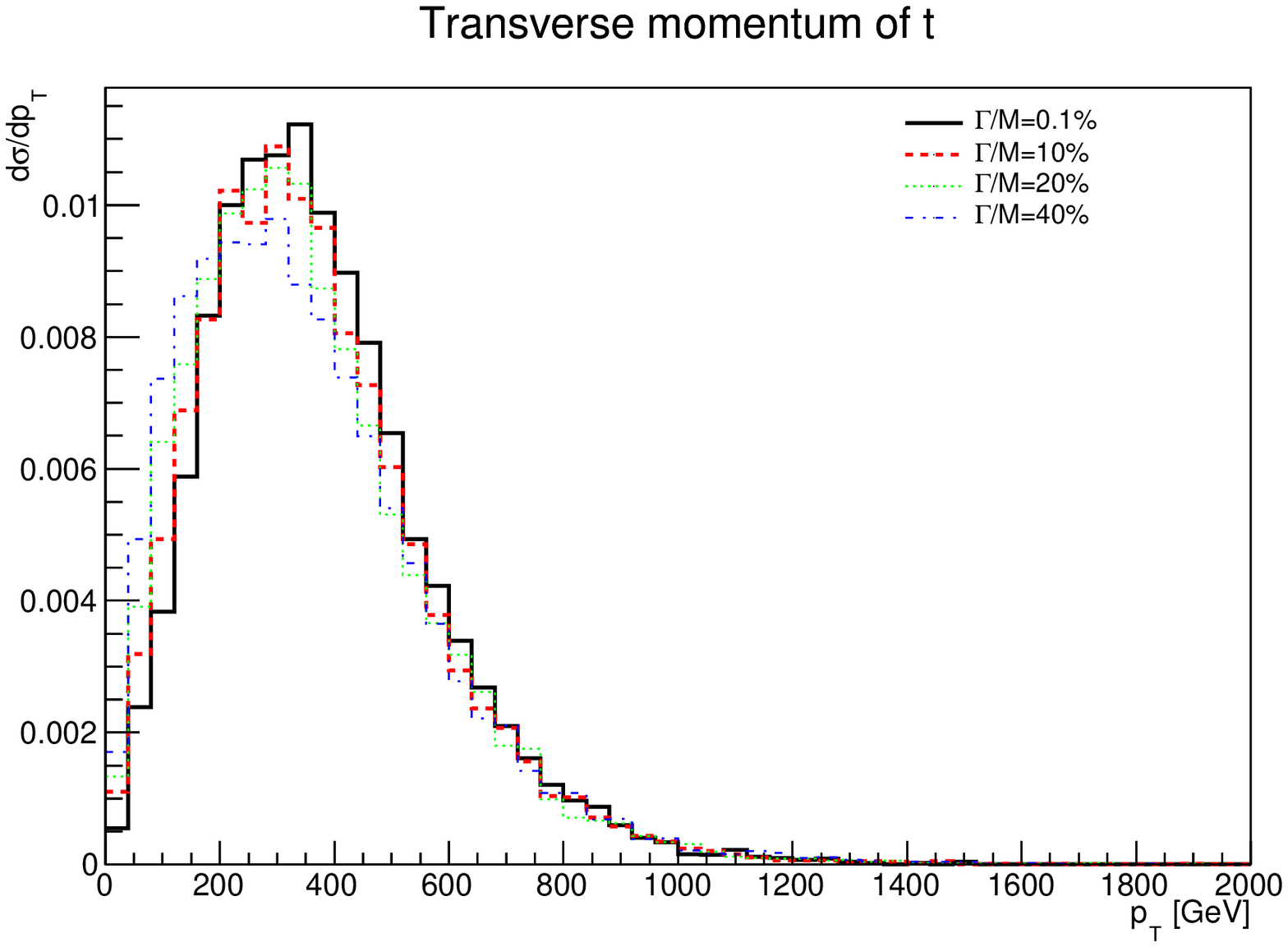, width=.45\textwidth}\\
\epsfig{file=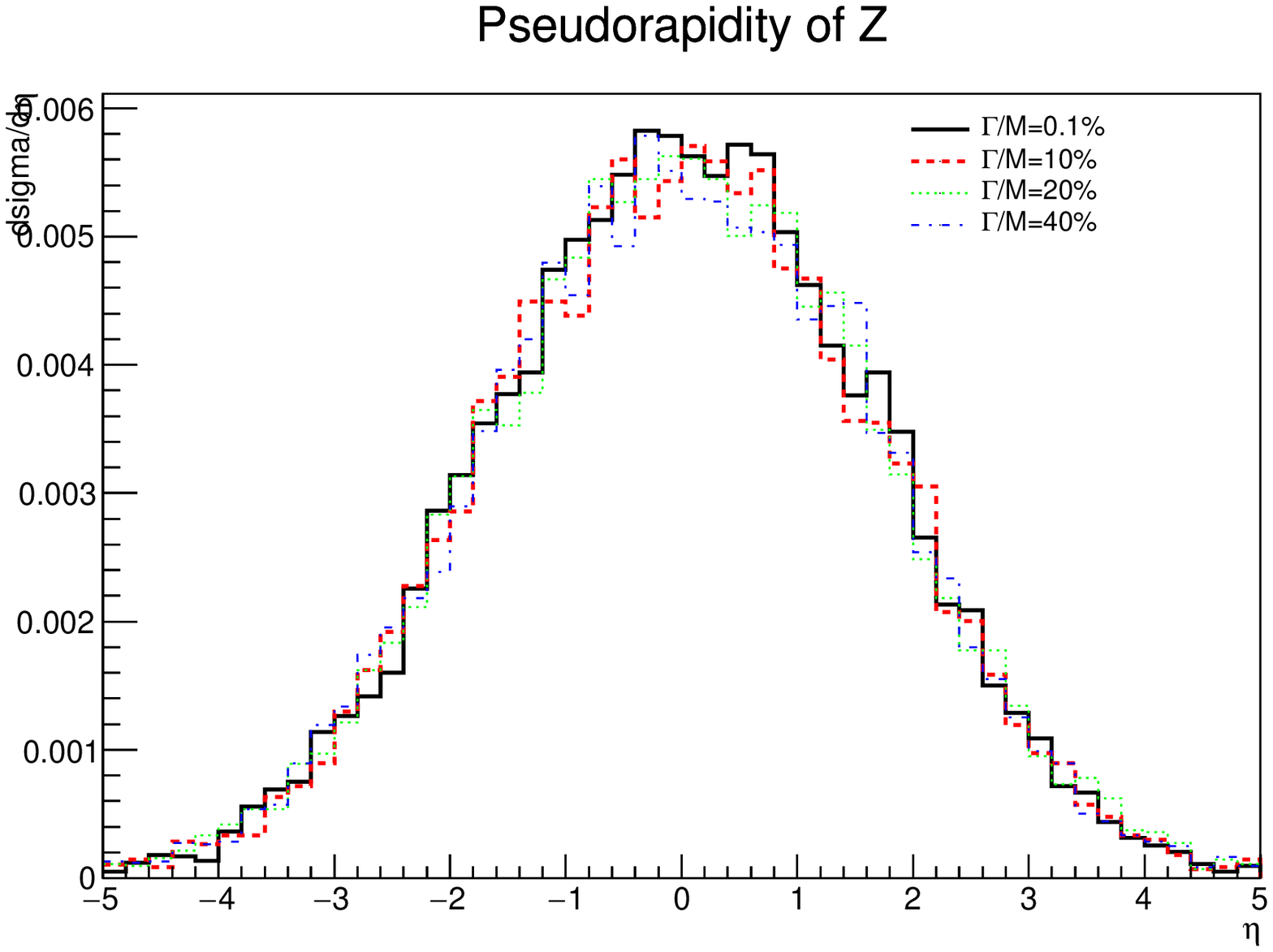, width=.45\textwidth}
\epsfig{file=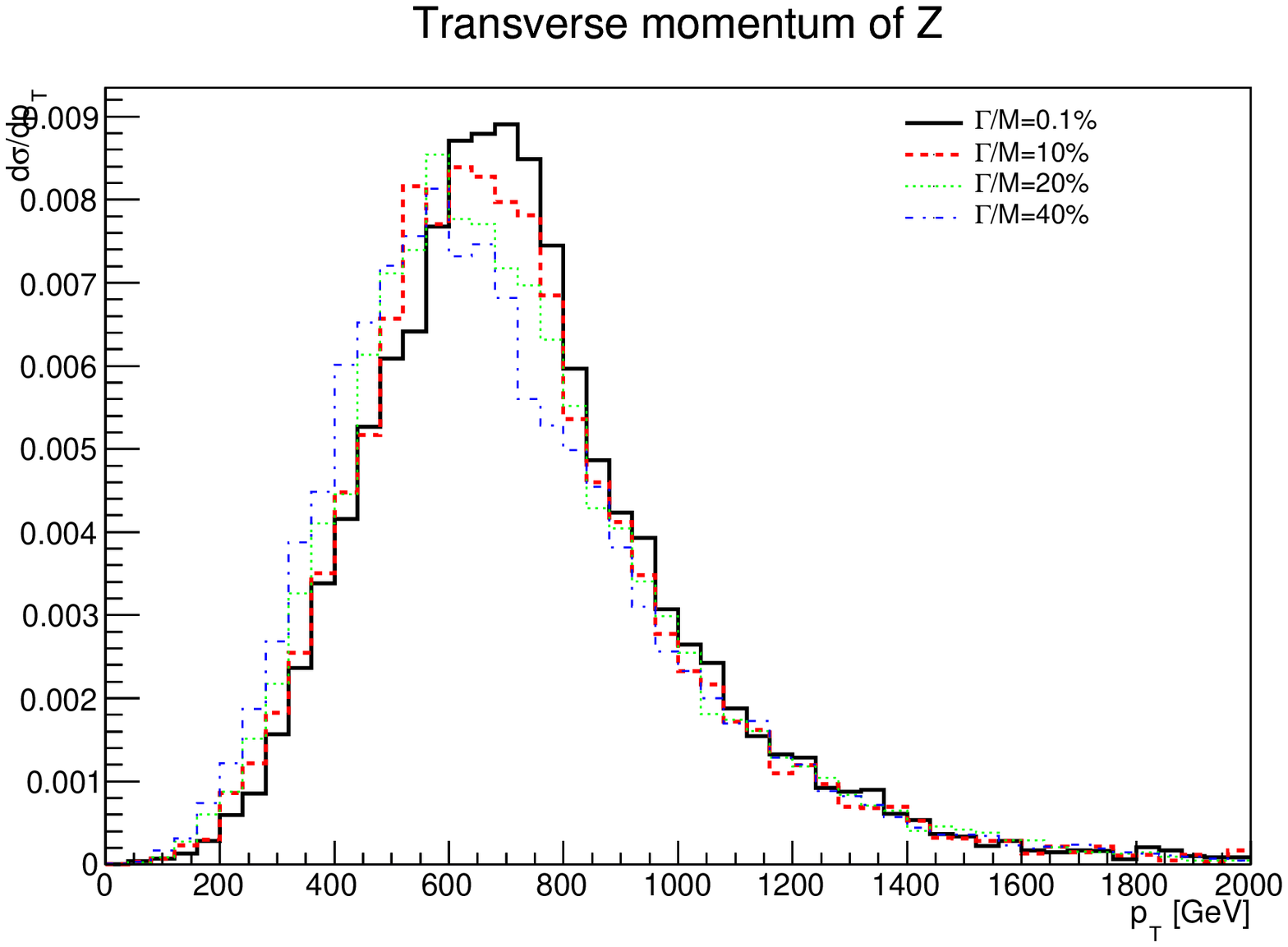, width=.45\textwidth}
\caption[Partonic level differential cross sections for the $ZtZt$ channel.]{Partonic level differential cross sections for the $ZtZt$ channel. From left to right and top to bottom: $\eta_t$, $p_{Tt}$, $\eta_Z$ and $p_{TZ}$. All distributions correspond to a $T$ mass of 800 GeV, for which $\sigma_S\sim\sigma_X$ almost independently of the $T$ width.}
\label{fig:distztzt}
\end{figure}

For some channels it is possible to notice a cancellation of effects which makes the QCD pair production cross section similar to the cross section including off-shell contributions even for large values of the width. The cancellations appear at different values of the $T$ mass, depending on the channel and for processes involving the bottom quark in the final state they are partially masked by the large increase of the cross section due to the collinear divergences caused by topologies where the bottom quarks arise from gluon splitting, as the one shown in Fig.~\ref{fig:fullsignaltopologies1}. Such cancellations are due to the different scaling of phase space between the large and narrow width regimes. Indeed, if the VLQ $T$ has a large width, the transferred momentum of the process can have values in a larger range than in the NWA case, where it is constrained by the resonant production of the $T$ pair: this means in turn that the PDFs are sampled at different scales and therefore the cross section receives a non-trivial mass and width dependent contribution which results in the observed behaviour. Of course, this does not necessarily mean that the NWA approximation can be used along the cancellation regions. Sample kinematical distributions of the decay products of the $T$ in different width regimes are shown in Fig.~\ref{fig:disththt} for the $HtHt$ channel and $M_T=600$ GeV and in Fig.~\ref{fig:distztzt} for the $ZtZt$ channel and $M_T=800$ GeV. In both cases, while the $\eta$ distribution does not change significantly as the width increases, the $p_T$ distributions exhibits a visible shift towards the softer region. 

%%%%%%%%%%%%%%%%%%%%%%%%%%%%%%%%%%%%%%%%%%%%%%%%%%%%%%%%%%%%%%%%%%%%%%%%%
\subsubsection{Interference with SM background}

When considering processes of pair production of heavy quarks in the NWA, interferences with the SM background are zero by construction, but if the width of the heavy quark is large, it is crucial to explore the relevance of interference terms in the determination of the total number of events. Moreover, understanding this contribution for regions which are not usually explored in experimental analyses may be useful in the determination of sets of kinematical cuts for the optimisation of future searches, if any hint of a VLQ with large width  appears in the data.

\begin{figure}[H]
\centering
\epsfig{file=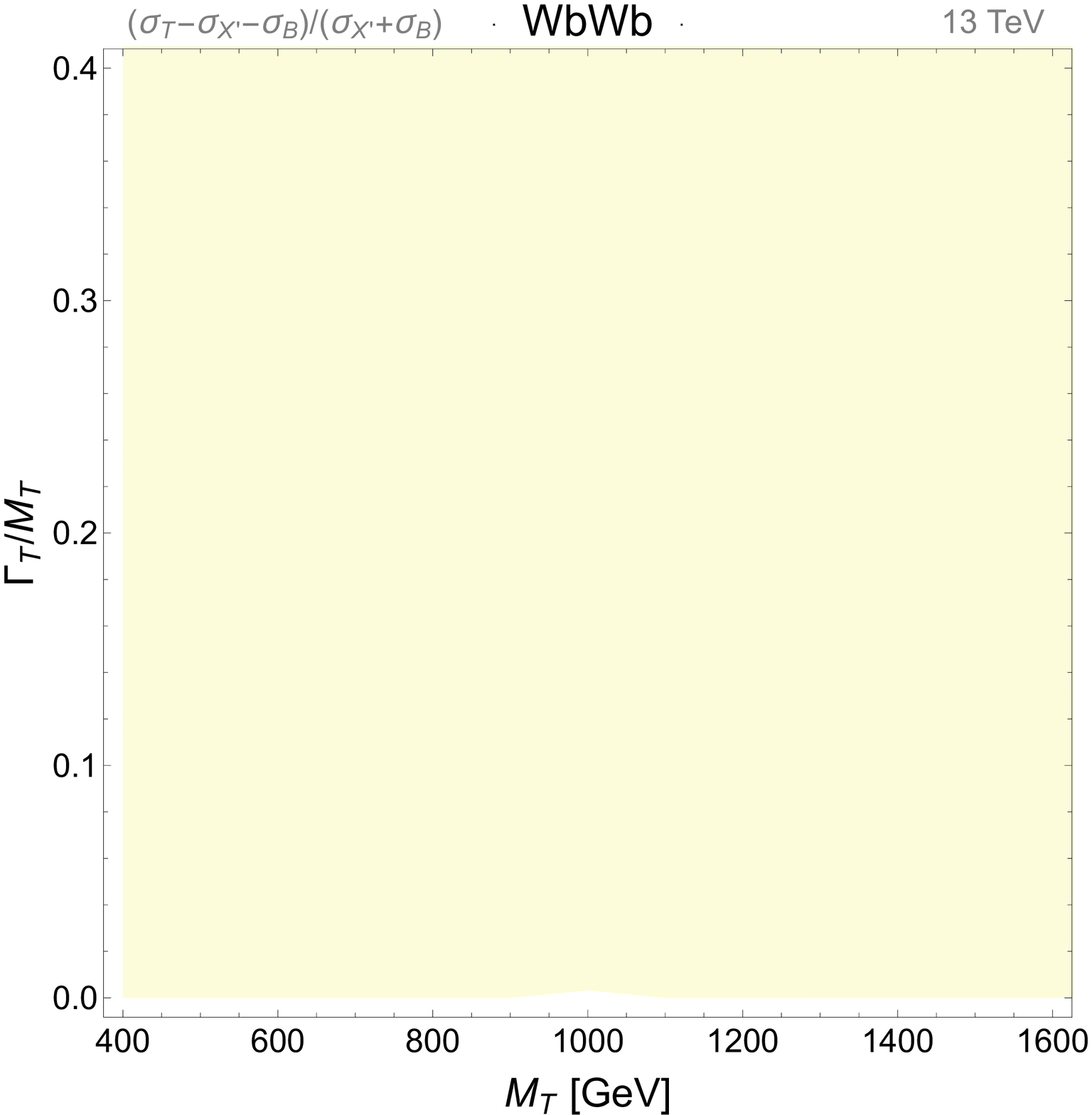, width=.3\textwidth}
\epsfig{file=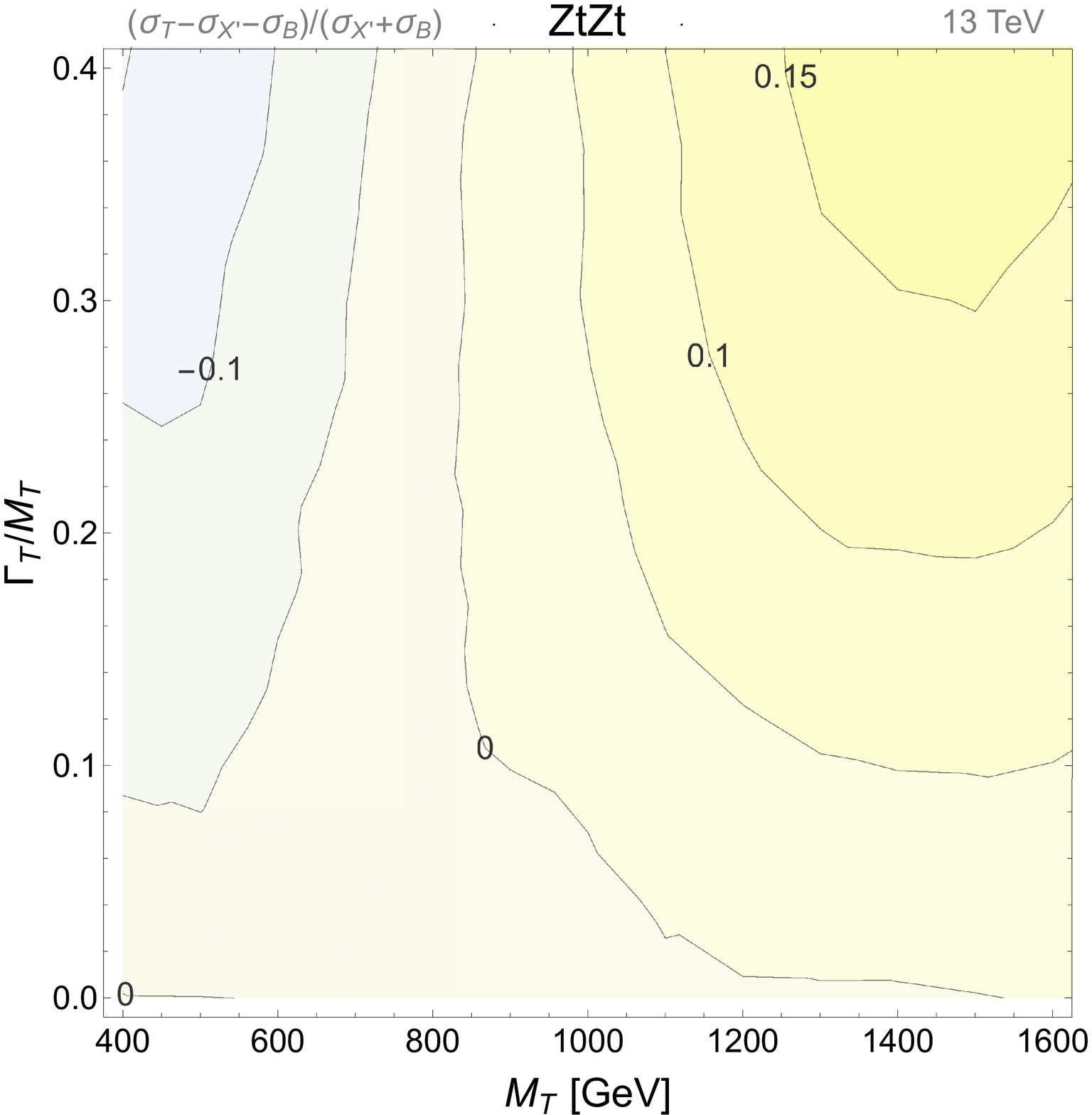, width=.3\textwidth}
\epsfig{file=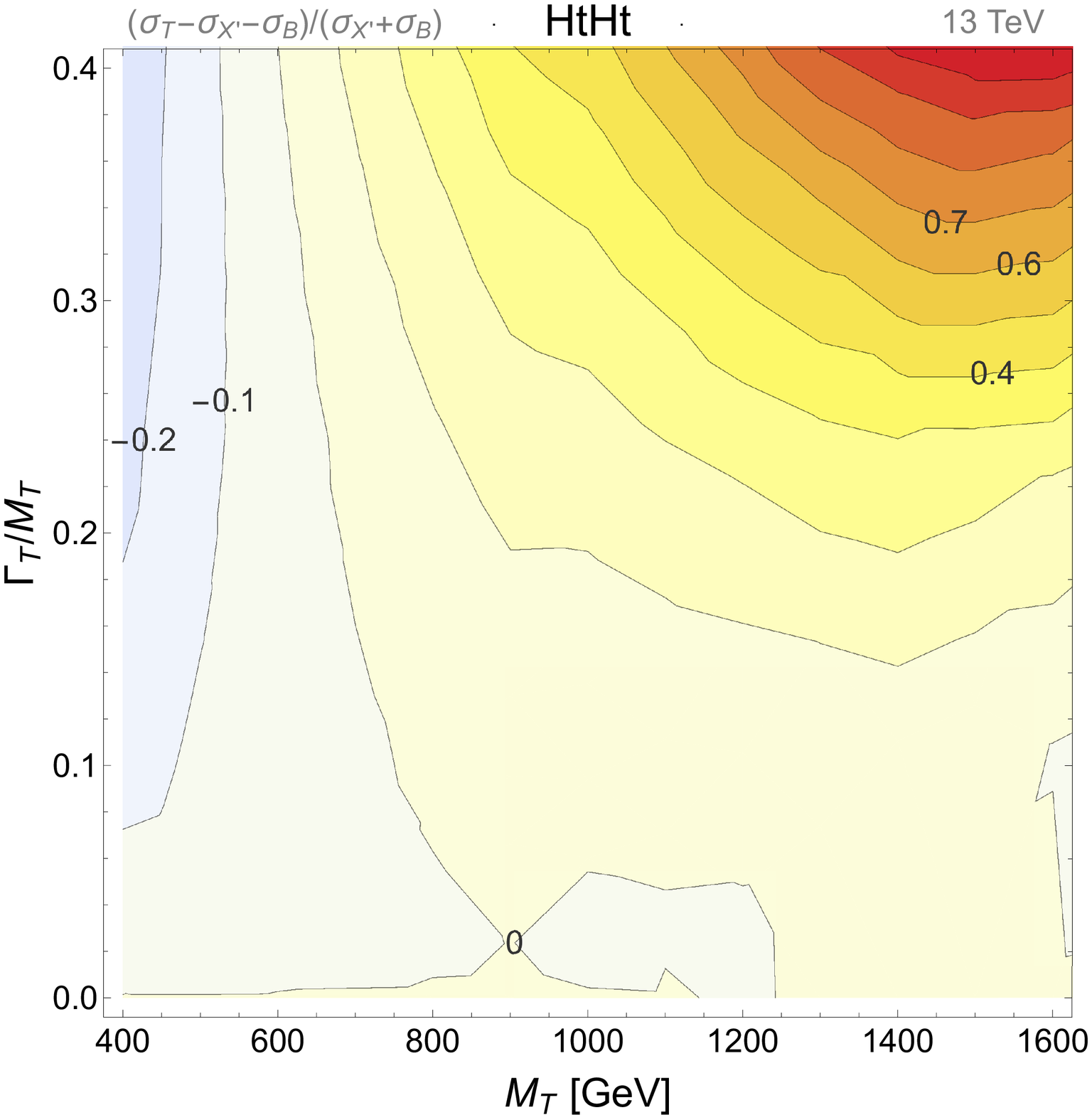, width=.3\textwidth}\\
\epsfig{file=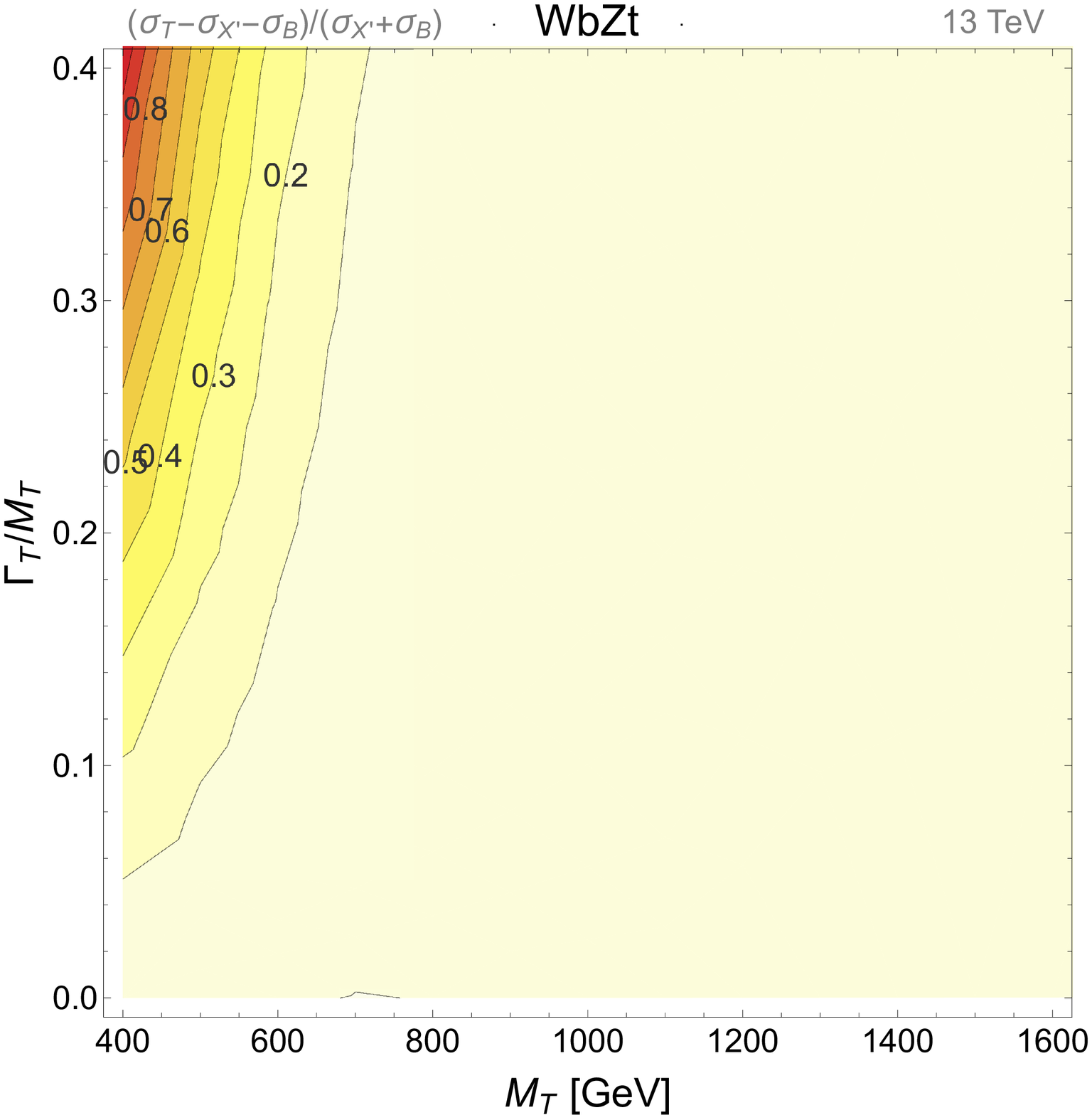, width=.3\textwidth}
\epsfig{file=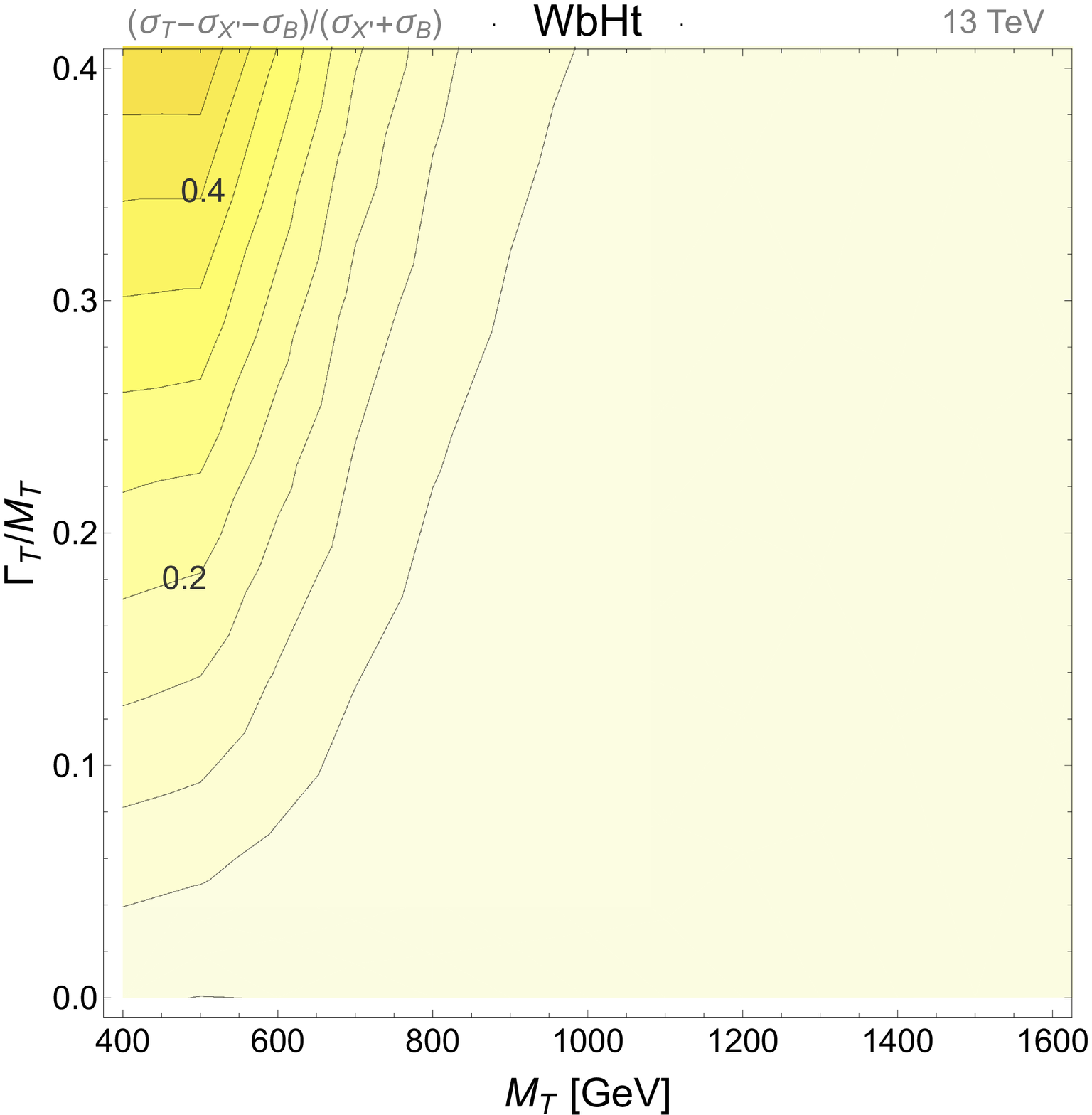, width=.3\textwidth}
\epsfig{file=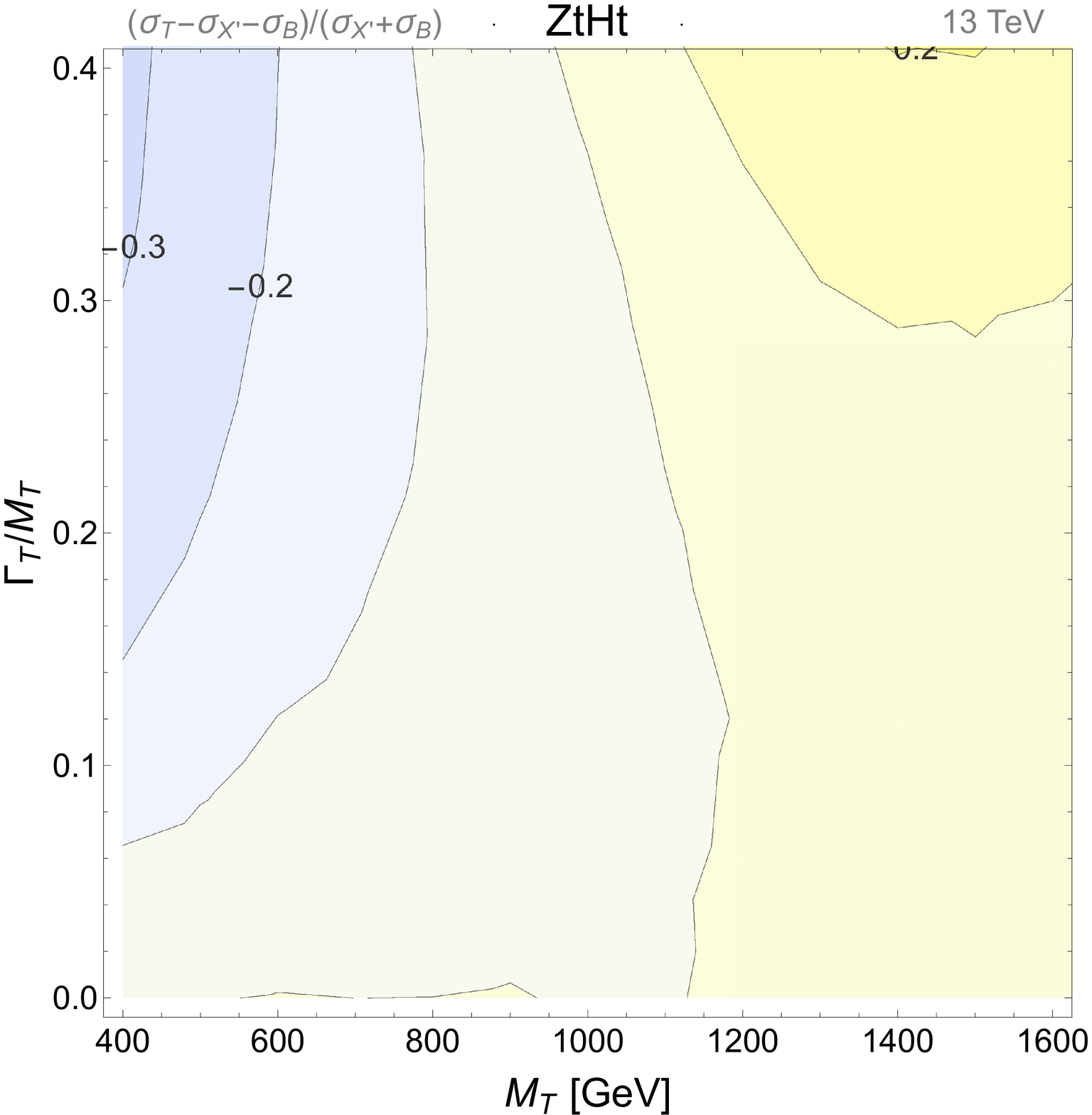, width=.3\textwidth}
\caption[Relative difference in cross section between the total $2 \to 4$ process, including the SM background and the sum of QCD pair production and SM backgrounds.]{Relative difference in cross section between the total $2\to4$ process, including the SM background and the sum of QCD pair production and SM backgrounds. Top row: final states on the diagonal of the matrix in Eq. \ref{eq:finalstates1} (third generation mixing); bottom row: off-diagonal final states (third generation mixing).}
\label{fig:TXBthird}
\end{figure}

The correction factor between the total cross section and the sum of NWA pair production and SM backround cross section is plotted in Fig.~\ref{fig:TXBthird}. Such correction factors depend on the relative weight of the SM background contribution in the determination of the total cross section: they are almost negligible in the whole parameter space where the background is the dominant contribution to the total signal, while they are become larger where the new physics signal has a more relevant role. This can easily be understood by considering what affects the various terms of the ratio. Herein, $\sigma_B$ is a constant term (for fixed final state), $\sigma_X$ only depends on the $T$ mass and $\sigma_T$ is the only term which depends on the both the $T$ mass and width. For the $WbWb$ case, however, $\sigma_T$ is almost entirely dominated by the SM background contribution (mostly by the top pair production process) and therefore the contribution of the $T$ is just a small correction, which does not produce relevant effects in the whole range of masses and widths we have explored. For the $ZtZt$ and $HtHt$ scenarios, on the contrary, the SM background is comparable or negligible with respect to the signal contribution, and therefore the dependence on the $T$ mass and width is much more evident. \footnote{Note that the change of cross section due to the large width certainly depends on the kinematic properties of the final state, i.e. it may be more prominent in some kinematic regions. Plotting these cross section ratios after applying the experimental cuts that define the signal region instead of the total cross sections would provide us with more information. Yet this would have to be done independently for each analysis and would depend on the signal region considered. For these reason we only show here these simpler plots which do not depend on the search considered and study the value of the efficiencies of some specific signal regions later.}

\begin{figure}[H]
\centering
\epsfig{file=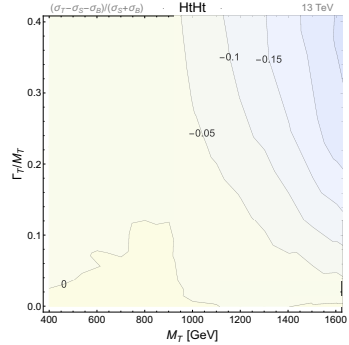, width=.3\textwidth}
\caption[Relative contribution of the interference between the full signal and the SM background.]{Relative contribution of the interference between the full signal and the SM background. $HtHt$ is the only channel for which this contribution can reach values above 10\% in size.}
\label{fig:TSBthird}
\end{figure}

The full contribution of interference terms, considering the full signal instead of the signal in the NWA, is always numerically negligible. In Fig.~\ref{fig:TSBthird} we have shown the only channel for which the contribution can become larger than 10\% in absolute value. The inclusion of single-resonance effects, therefore, changes the picture in a substantial way, showing that interference effects between the full signal and the SM background are always negligible, except for the $HtHt$ channel in the large width and large $M_T$ region. This has to be expected because the kinematical properties of signal and background are usually different. However, this can only be seen by taking into account the full signal contribution. This means that, if searches for VLQs with large width are designed, considering the full signal instead of rescaling the NWA results would almost in any case automatically kill any contribution from interference with the SM background. This suppression of the interference effect is especially important for scenarios where the SM background and the signal are comparable and where the interference effects could therefore be important. 

%%%%%%%%%%%%%%%%%%%%%%%%%%%%%%%%%%%%%%%%%%%%%%%%%%%%%%%%%%%%%%%%%%%%%%%%%
\subsubsection{Results at detector level} \label{sec:detector3}

In this section we will study the performance of 8 TeV and 13 TeV searches from both ATLAS and CMS in determining the excluded region in the $\{M_T,\Gamma_T/M_T\}$ plane. We will consider only final states in the diagonal of the matrix of Eq.~\ref{eq:finalstates1} because non-diagonal final states would not represent, by themselves, physically valid scenarios. Such final states arise only if the VLQ has non-zero {\rm BR}s in different channels, and a consistent treatment would require the combination of diagonal and off-diagonal final states together. As stated above, the purpose of this study is not to set limits, but to study the performance of experimental searches in regions yet unexplored for these scenarios. Indeed, the set of searches we consider are not necessarily optimised for the discovery of VLQs at the LHC, therefore our recast bounds are not likely to be competitive with current bounds for pair production of VLQs in the NWA, and, in this respect, we will not compare our results with other bounds from direct searches for pair production of VLQs.\\

\begin{figure}[H]
\centering
\epsfig{file=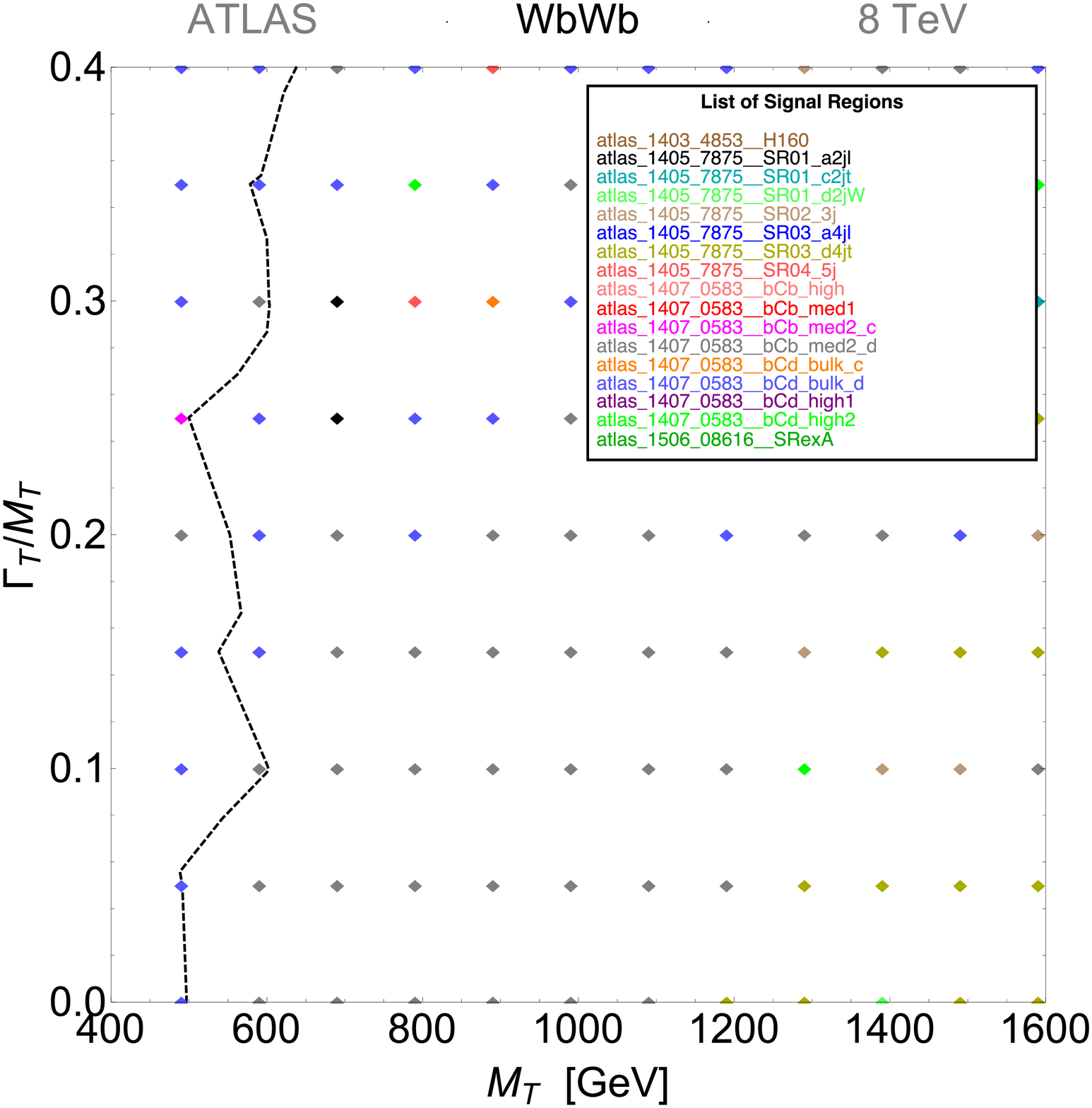, width=.3\textwidth}
\epsfig{file=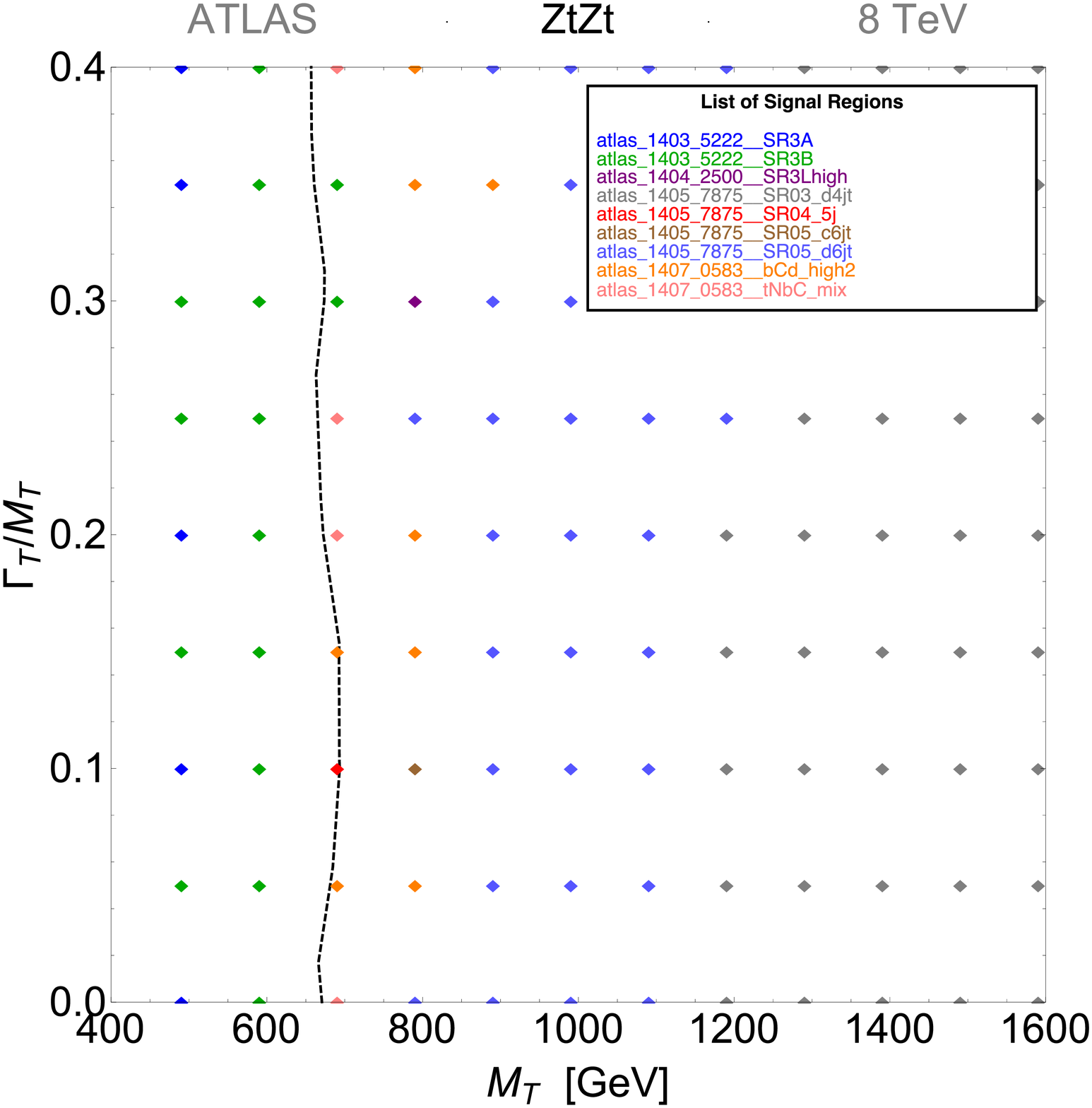, width=.3\textwidth}
\epsfig{file=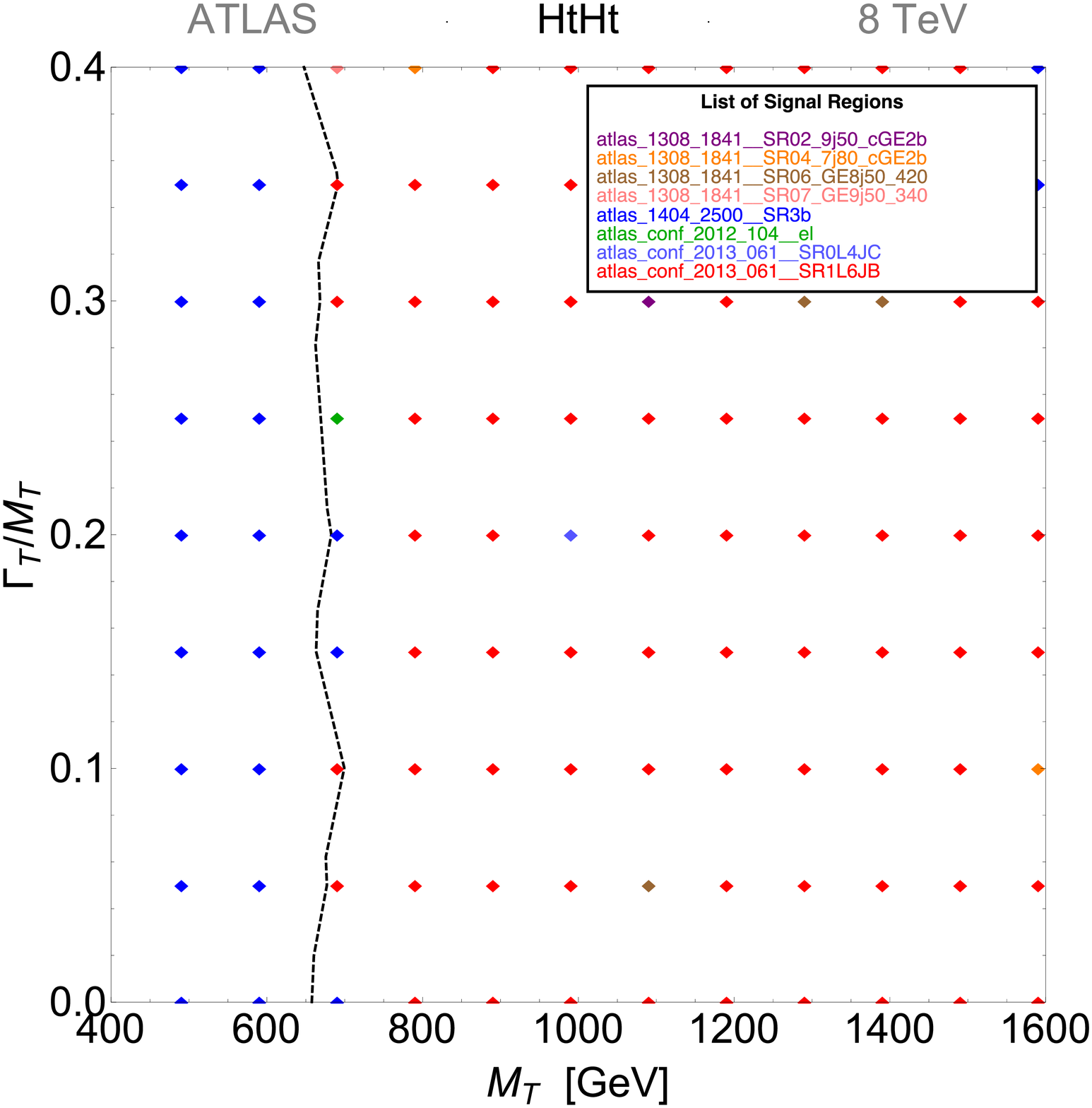, width=.3\textwidth}\\
\epsfig{file=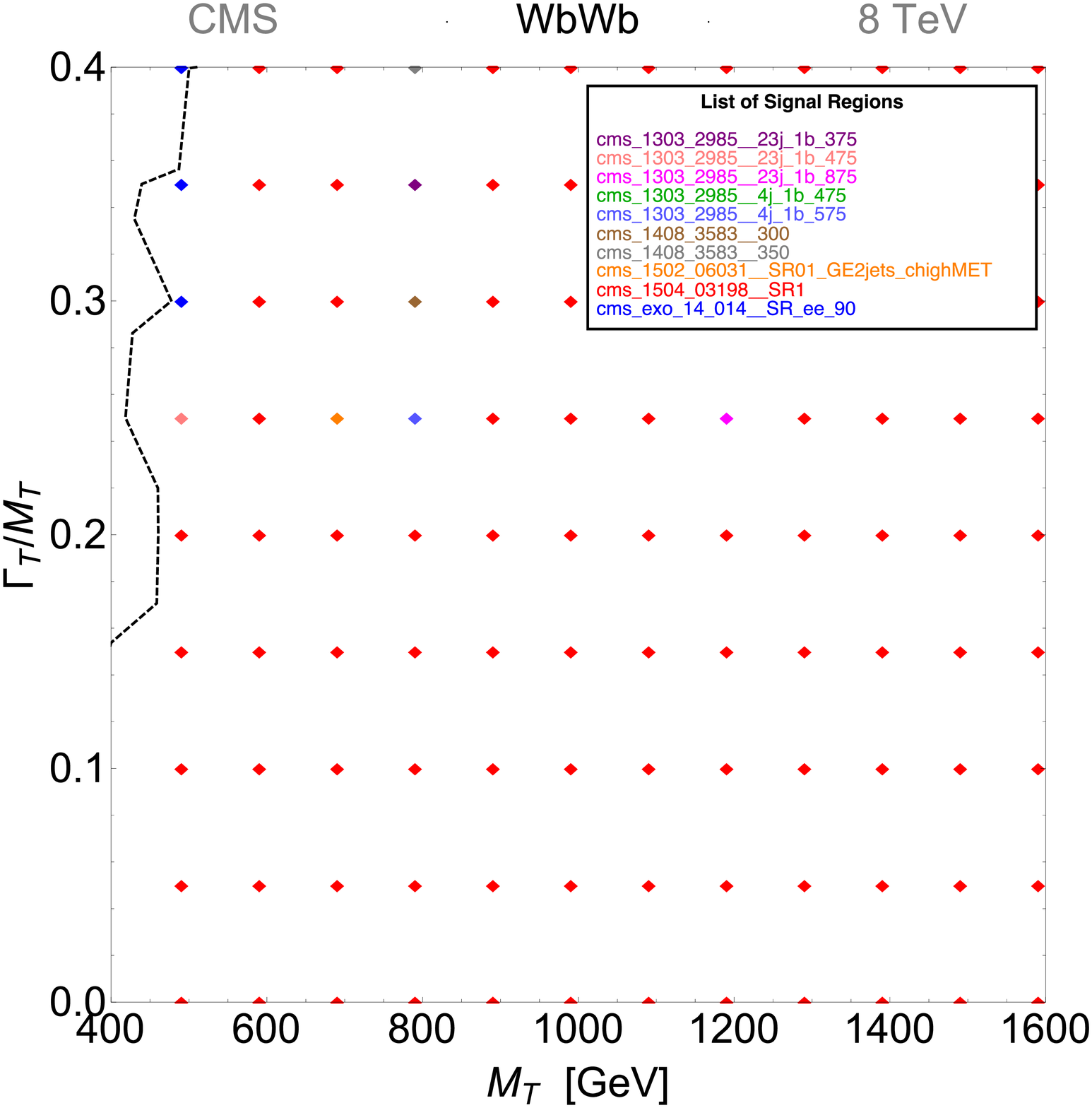, width=.3\textwidth}
\epsfig{file=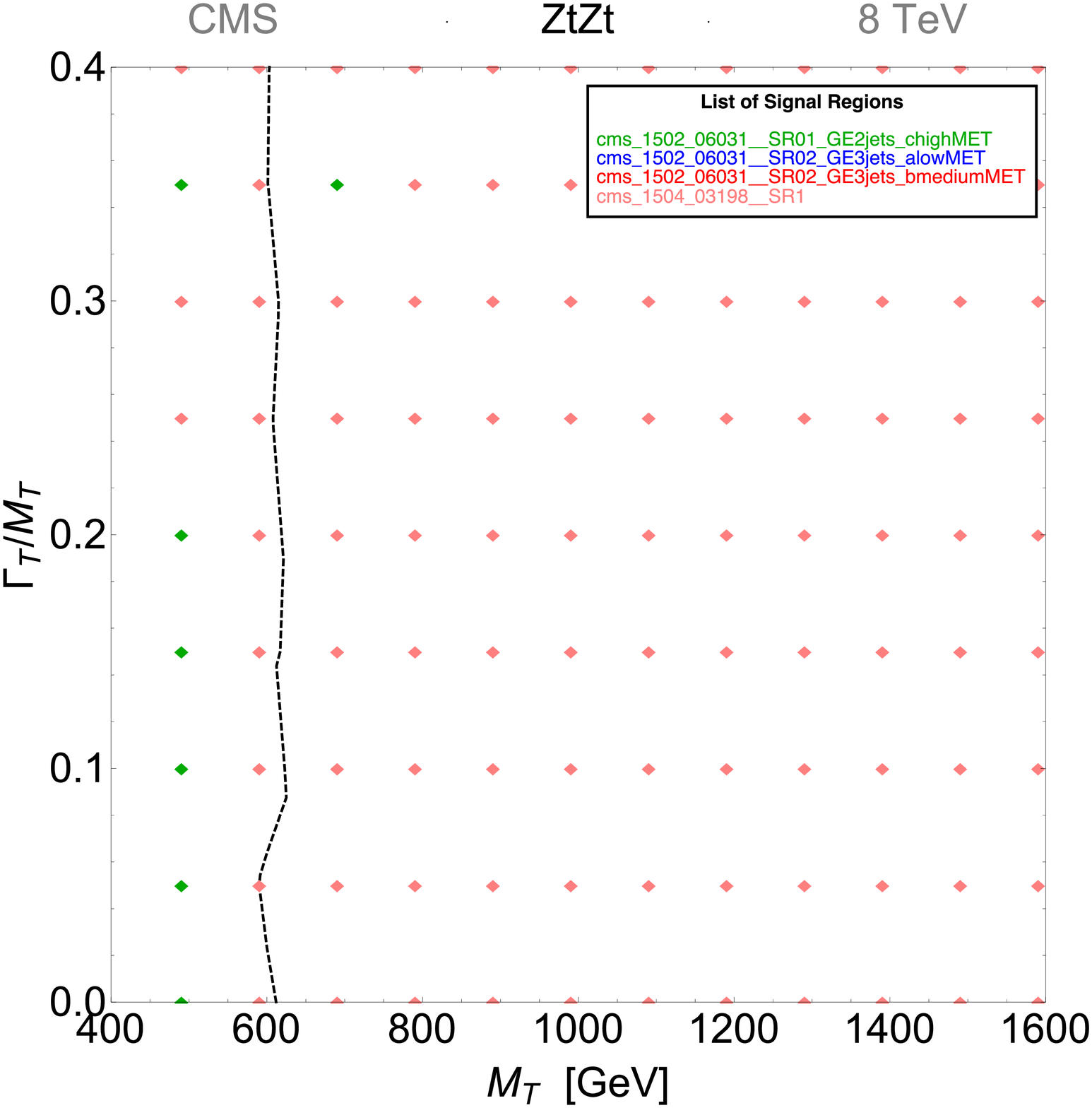, width=.3\textwidth}
\epsfig{file=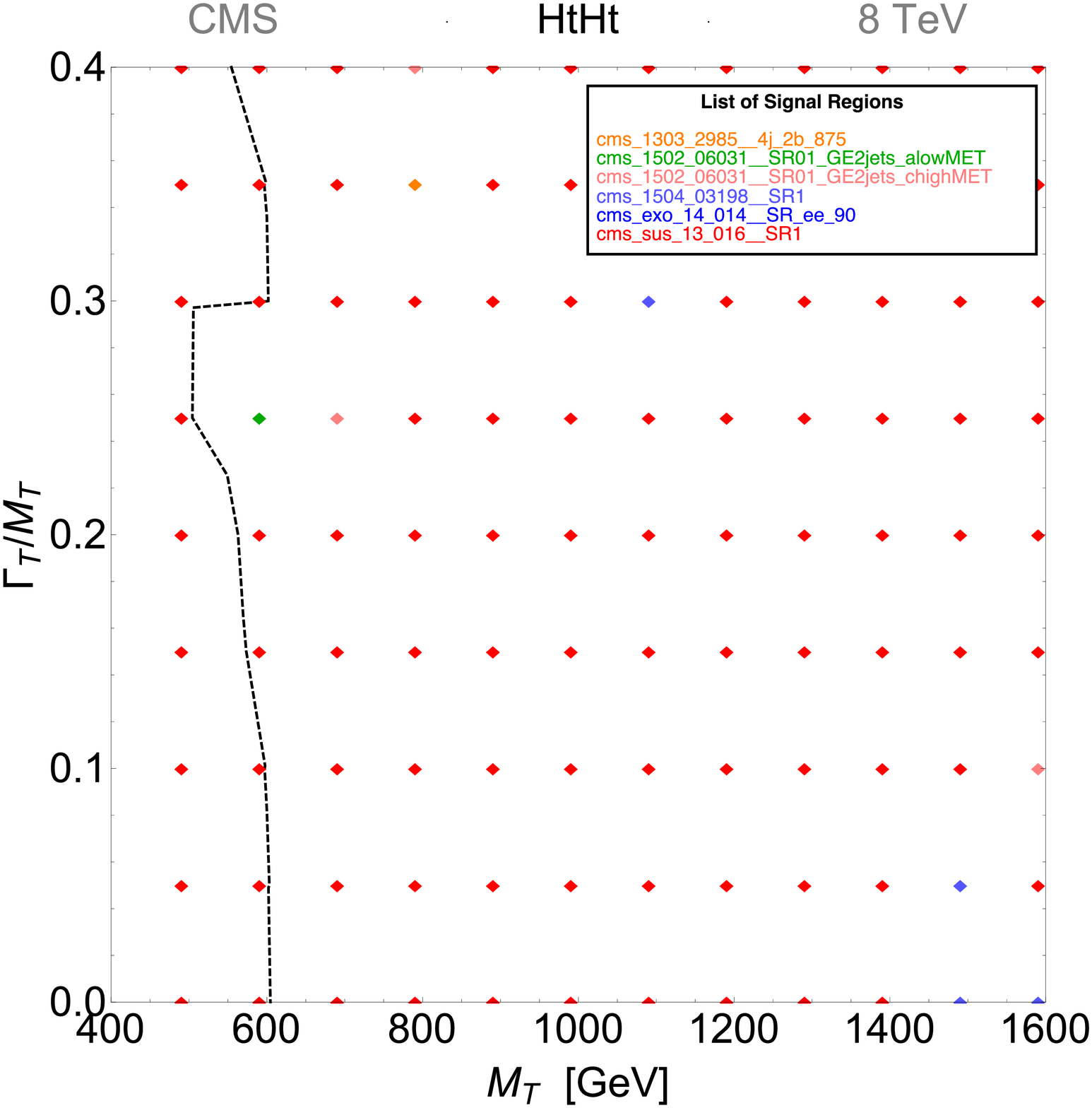, width=.3\textwidth}
\caption[Recast bounds in the ($M_T, \Gamma_T/M_T$) plane with a set of ATLAS and CMS searches at 8 TeV for diagonal final states.]{Recast bounds in the ($M_T, \Gamma_T/M_T$) plane with a set of ATLAS (top row) and CMS (bottom row) searches at 8 TeV for diagonal final states.}
\label{fig:Detector8TeV3rdgen}
\end{figure}

We show in Fig.~\ref{fig:Detector8TeV3rdgen} the exclusion lines for combinations of 8 TeV searches from both ATLAS and CMS for the three diagonal final states compatible with pair production and decay of VLQs $T$. Our results show that none of the Signal Regions (SRs) in the considered searches is sensitive to the large width scenarios: the exclusion bound are, for all final states, analogous to the NWA limit.

\begin{figure}[H]
\centering
\epsfig{file=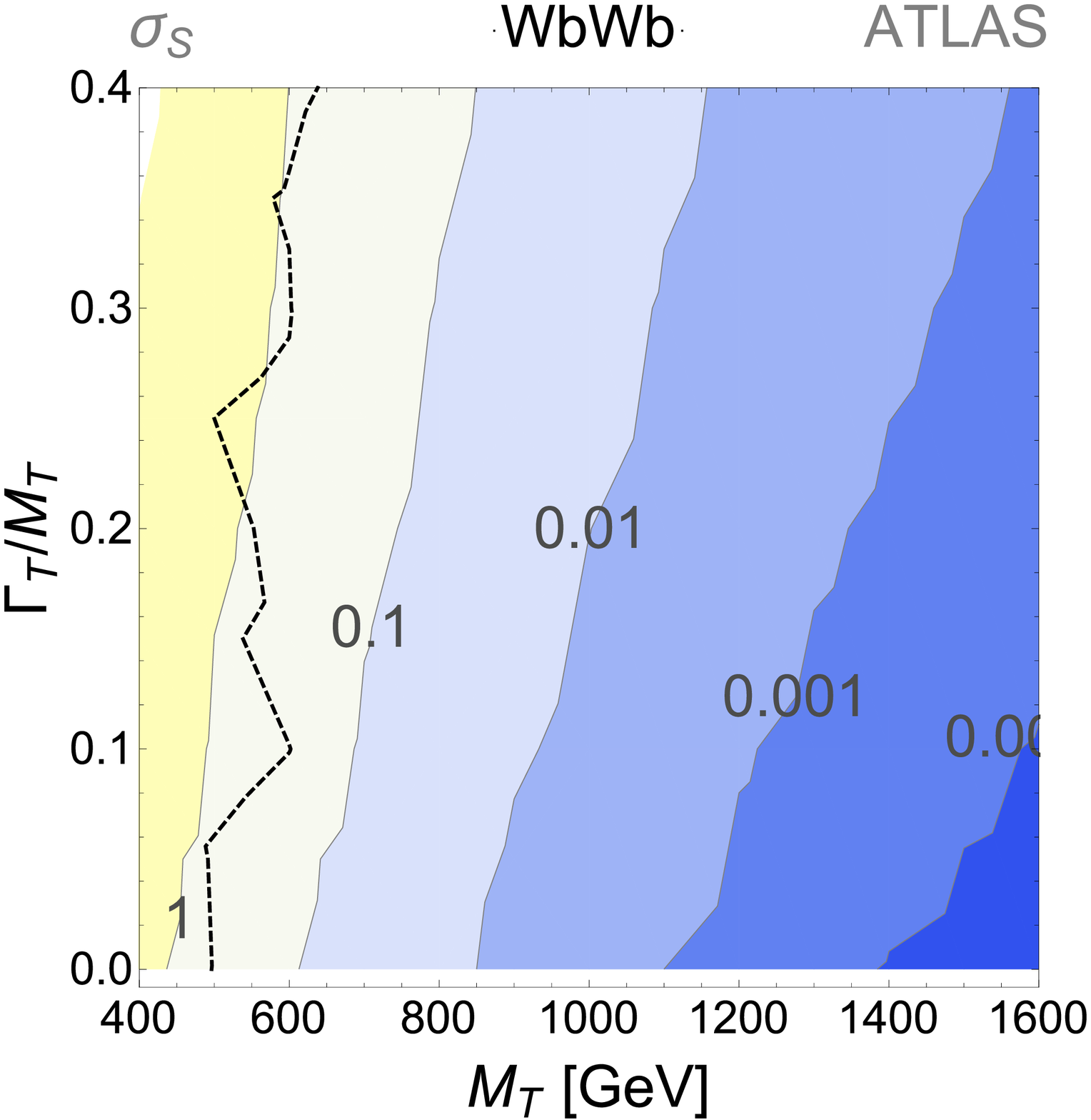, width=.3\textwidth}
\epsfig{file=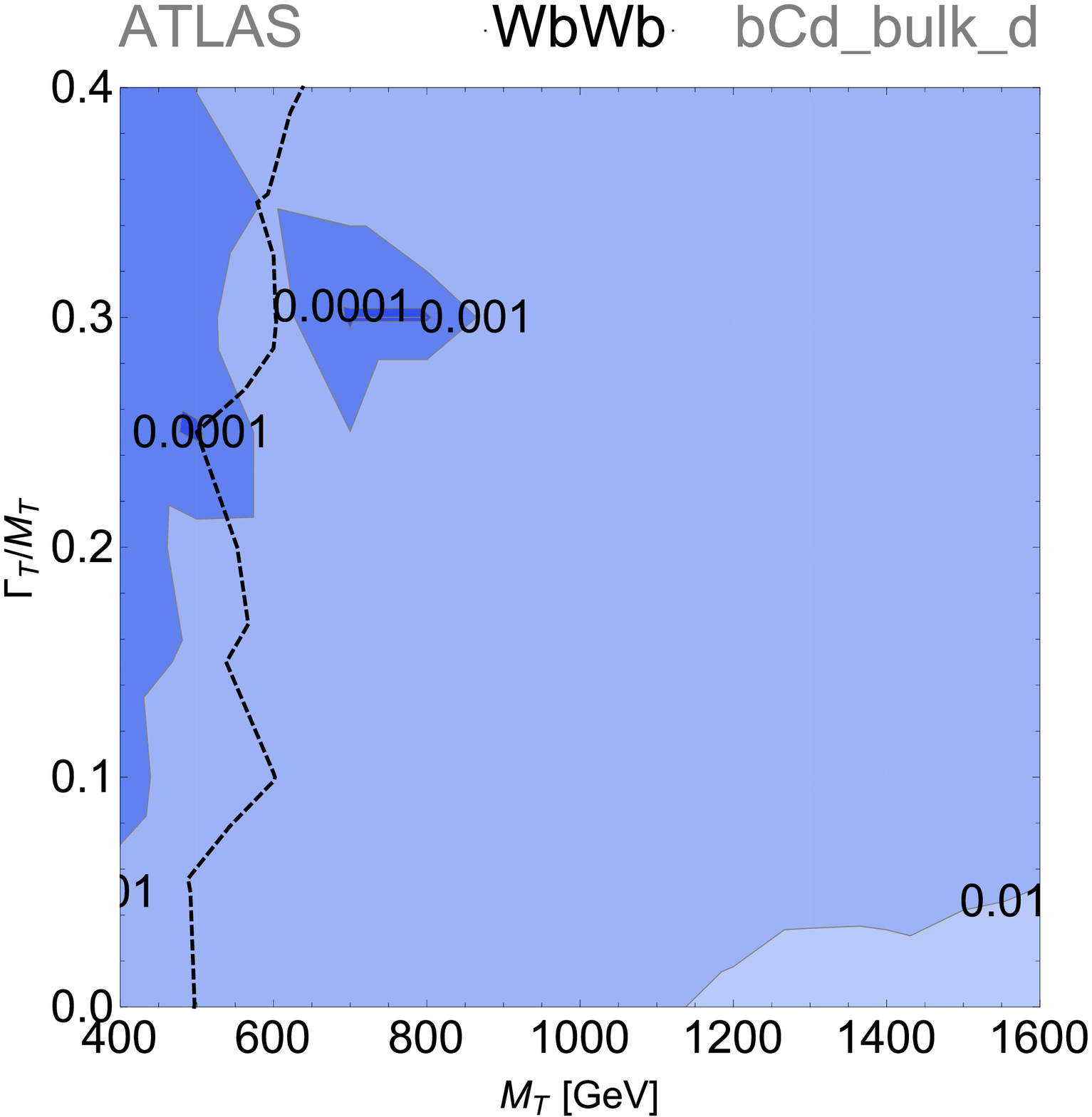, width=.3\textwidth}
\caption[cross section and efficiency of the best ATLAS SR for the $WbWb$ channel, compared with the bound.]{cross section and efficiency of the best ATLAS SR (bCd\_bulk\_d of \cite{Aad:2014kra}) for the $WbWb$ channel, compared with the bound.}
\label{fig:Combined8TeV3genWbWb}
\end{figure}

This can be understood by considering the cross section of the full signal, $\sigma_S$, and the dependence on the $T$ width of the efficiencies of the SRs which is most marked near the bounds. In Fig.~\ref{fig:Combined8TeV3genWbWb} we superimpose the bound from the combination of ATLAS searches at 8 TeV with the cross section of the full signal for the $WbWb$ channel (the others are qualitatively similar): the dependence on the width of the cross section is weak in the region where the searches fix the exclusion limit, and becomes slightly stronger for higher (allowed) masses. Moreover, the variation of the kinematics of the final states is not large enough to increase the sensitivity of the search cuts, as can be seen by looking at the efficiency of the the SR  bCd\_bulk\_d of the ATLAS search~\cite{Aad:2014kra}, which depends rather weakly on the width of the $T$. 
                                                                                                 
\begin{figure}[H]
\centering
\epsfig{file=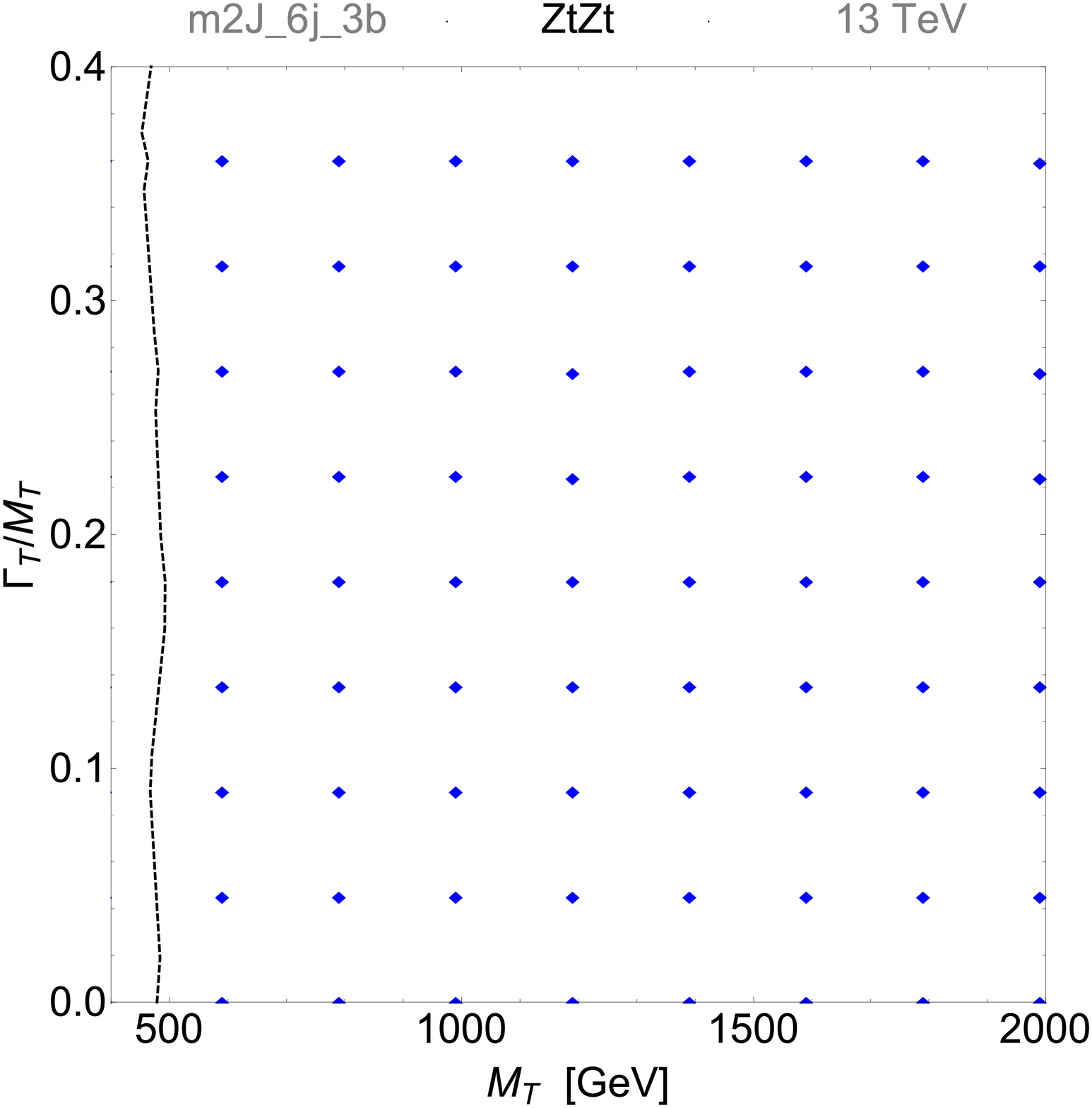, width=.3\textwidth}
\epsfig{file=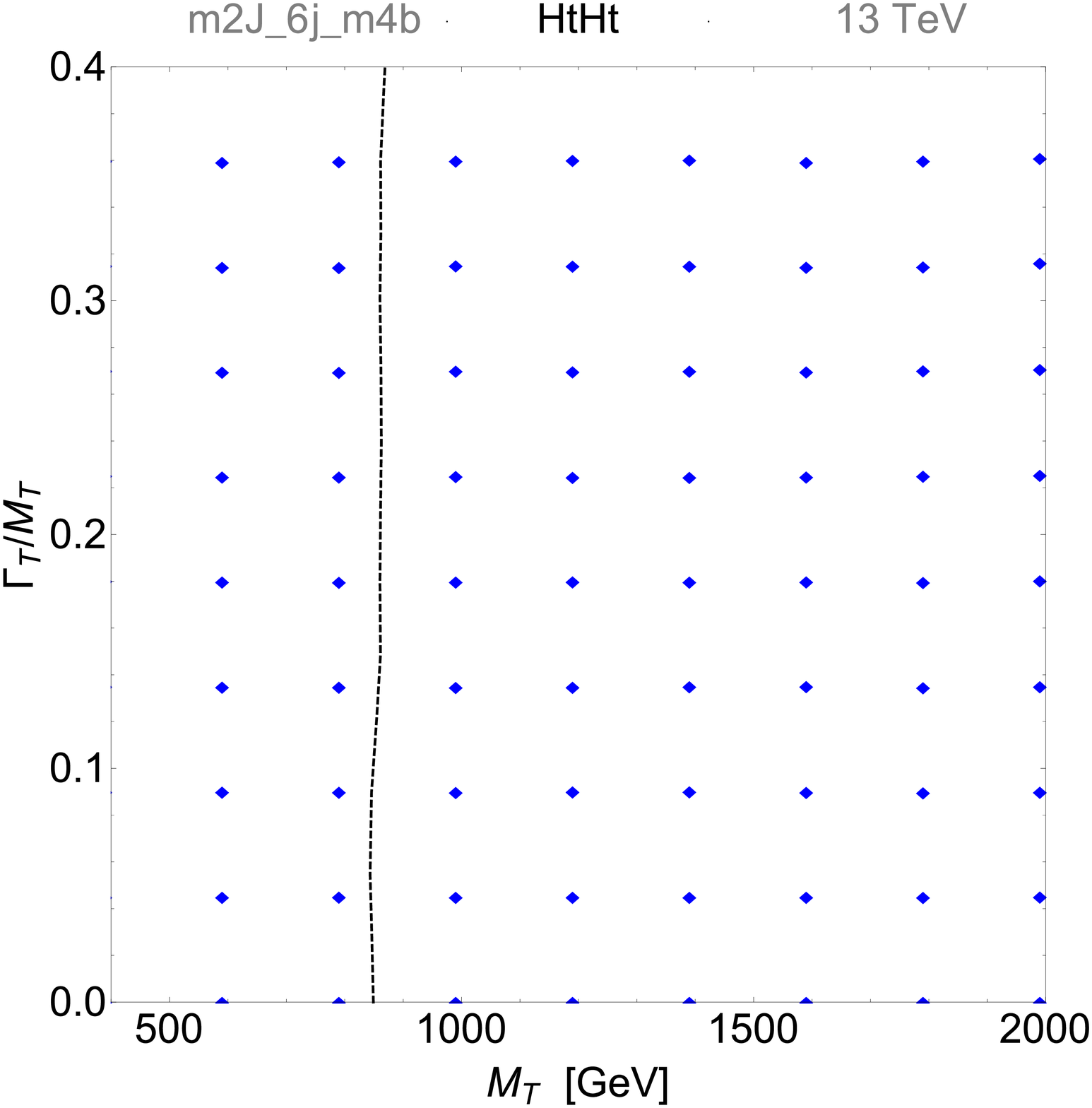, width=.3\textwidth}
 \caption[Same as Fig.~\ref{fig:Detector8TeV3rdgen} for the ATLAS search at 13 TeV implemented in CheckMATE.]{Same as Fig.~\ref{fig:Detector8TeV3rdgen} for the ATLAS search at 13 TeV~\cite{TheATLAScollaboration:2016gxs} implemented in CheckMATE. The plot for the $WbWb$ channel is not shown because within the explored range the recasting does not set any limit.}
\label{fig:Detector13TeV3rdgen}
\end{figure}

Our results at 13 TeV have been obtained considering a dedicated search for pair production of a VLQ $T$~\cite{TheATLAScollaboration:2016gxs} implemented in CheckMATE. The results exhibit a similar behaviour as the set of 8 TeV ones. Our bounds are rather different from those reported in Ref.~\cite{TheATLAScollaboration:2016gxs}. However, we did not rescale the bounds considering different BRs, as we have not factorised the production from decay, and we are mostly interested in the dependence on the width of such bounds. In this respect, the bounds weakly depend on the $T$ width, as can be seen in Fig.~\ref{fig:Detector13TeV3rdgen}. As for the 8 TeV case, the slight increase in cross section, and relative deformation of kinematics distribution of the final state objects is compensated by an increase of the efficiencies of the SRs cuts. This information can be exploited for the design of future dedicated searches if the discovery of VLQs with large width are among the goals of the studies.

% \begin{figure}[H]
% \centering
% \epsfig{file=/Sigma_S_combined_plots/13TeV/Log_13_TeV/S_WbWb_Diag_8, width=.3\textwidth}
% \epsfig{file=/Sigma_S_combined_plots/13TeV/Log_13_TeV/S_ZtZt_Diag_8, width=.3\textwidth}
% \epsfig{file=/Sigma_S_combined_plots/13TeV/Log_13_TeV/S_HtHt_Diag_8, width=.3\textwidth}
% \caption{\label{fig:CombinedSBound13TeV3gen} Recast bounds in the $M_T-\Gamma_T/M_T$ plane with a set of ATLAS (top row) and CMS (bottom row) searches at 8 TeV for diagonal final states.}
% \end{figure}

%%%%%%%%%%%%%%%%%%%%%%%%%%%%%%%%%%%%%%%%%%%%%%%%%%%%%%%%%%%%%%%%%%%%%%%%%
%%%%%%%%%%%%%%%%%%%%%%%%%%%%%%%%%%%%%%%%%%%%%%%%%%%%%%%%%%%%%%%%%%%%%%%%%
\subsection{Extra $T$ quark mixing with first generation SM quarks}

%%%%%%%%%%%%%%%%%%%%%%%%%%%%%%%%%%%%%%%%%%%%%%%%%%%%%%%%%%%%%%%%%%%%%%%%%
\subsubsection{Large width effects on the signal at parton level}

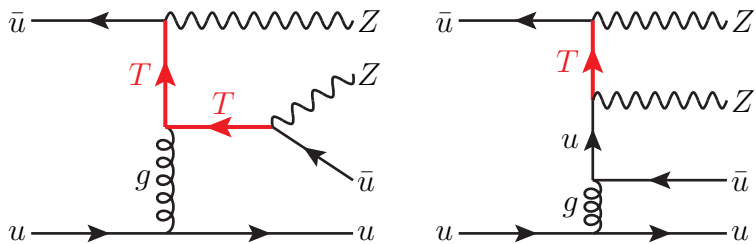
\begin{figure}[H]
\begin{center}
\begin{picture}(140,100)(0,-10)
\SetWidth{1}
\Line[arrow](0,0)(50,0)
\Text(-2,0)[rc]{\large $u$}
\Line[arrow](50,80)(0,80)
\Text(-2,80)[rc]{\large $\bar u$}
\Photon(50,80)(120,80){3}{9}
\Text(122,80)[lc]{\large $Z$}
\Line[arrow](50,0)(120,0)
\Text(122,0)[lc]{\large $u$}
\Gluon(50,40)(50,0){3}{5}
\Text(45,20)[rc]{\large $g$}
\SetColor{Red}\SetWidth{1.5}
\Line[arrow](50,40)(50,80)
\Text(45,60)[rc]{\large \Red{$T$}}
\Line[arrow](90,40)(50,40)
\Text(72,45)[cb]{\large \Red{$T$}}
\SetColor{Black}\SetWidth{1}
\Photon(90,40)(120,60){3}{4}
\Text(122,60)[lc]{\large $Z$}
\Line[arrow](120,20)(90,40)
\Text(122,20)[lc]{\large $\bar u$}
\end{picture}\hskip 20pt
\begin{picture}(140,100)(0,-10)
\SetWidth{1}
\Line[arrow](0,0)(50,0)
\Text(-2,0)[rc]{\large $u$}
\Line[arrow](50,80)(0,80)
\Text(-2,80)[rc]{\large $\bar u$}
\Photon(50,80)(100,80){3}{7}
\Text(102,80)[lc]{\large $Z$}
\Line[arrow](50,0)(100,0)
\Text(102,0)[lc]{\large $u$}
\Gluon(50,20)(50,0){3}{3}
\Text(45,10)[rc]{\large $g$}
\SetColor{Red}\SetWidth{1.5}
\Line[arrow](50,50)(50,80)
\Text(45,65)[rc]{\large \Red{$T$}}
\SetColor{Black}\SetWidth{1}
\Line[arrow](50,20)(50,50)
\Text(45,35)[rc]{\large $u$}
\Photon(100,50)(50,50){3}{7}
\Text(102,50)[lc]{\large $Z$}
\Line[arrow](100,20)(50,20)
\Text(102,20)[lc]{\large $\bar u$}
\end{picture}
\end{center}
\caption{Examples of neutral-current topologies for heavy quarks with large width mixing with first generation.}
\label{fig:firstgentopologies1}
\end{figure}

If the $T$ interacts with first generation SM quarks, topologies where gluons splitting into light quarks increase the cross section due to collinear enhancements are present also for neutral currents, as shown in Fig.~\ref{fig:firstgentopologies1}. In the case of mixing with third generation, such topologies were not present for neutral currents due to the large top mass.

\begin{figure}[H]
\centering
\epsfig{file=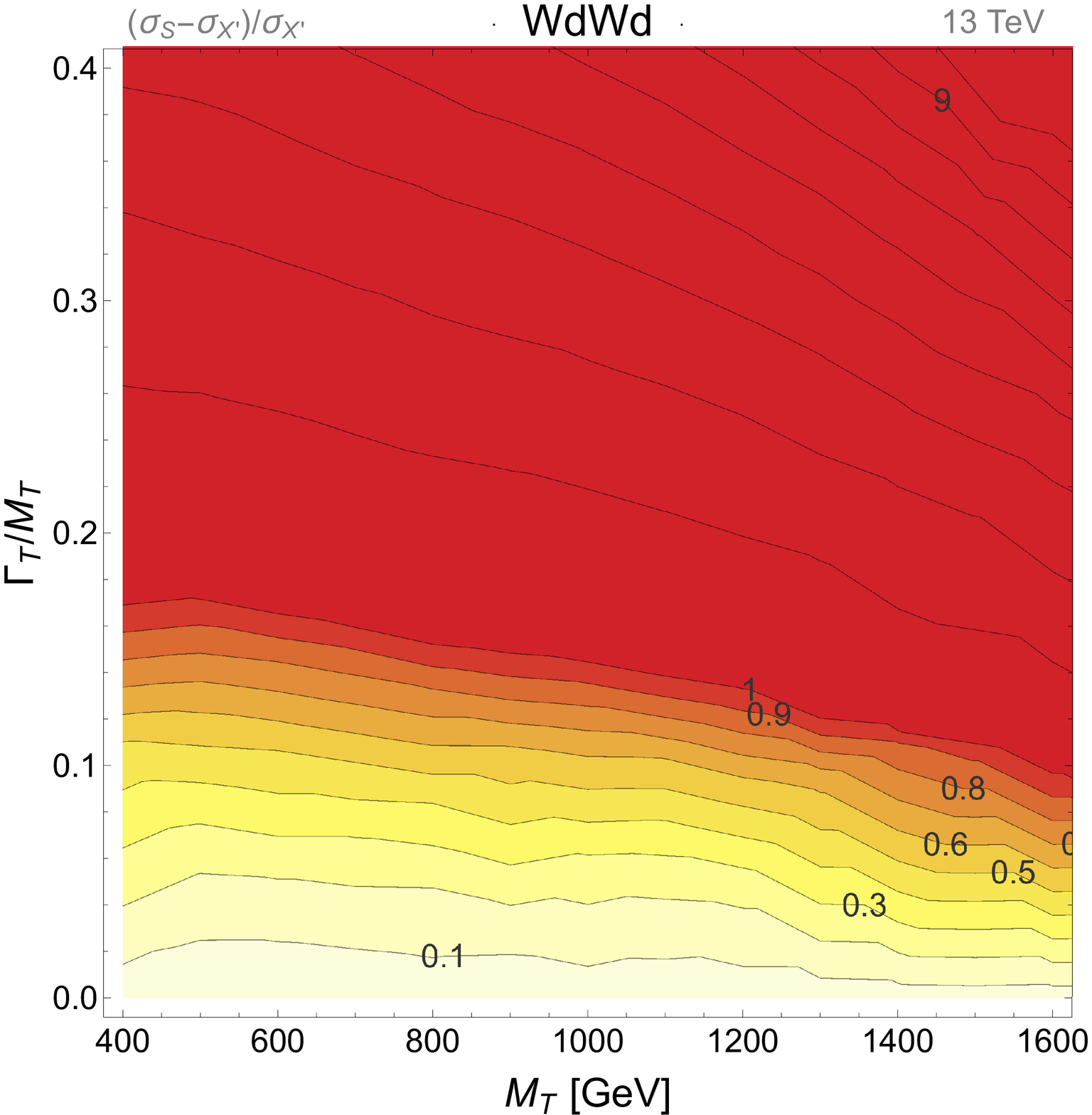, width=.3\textwidth}
\epsfig{file=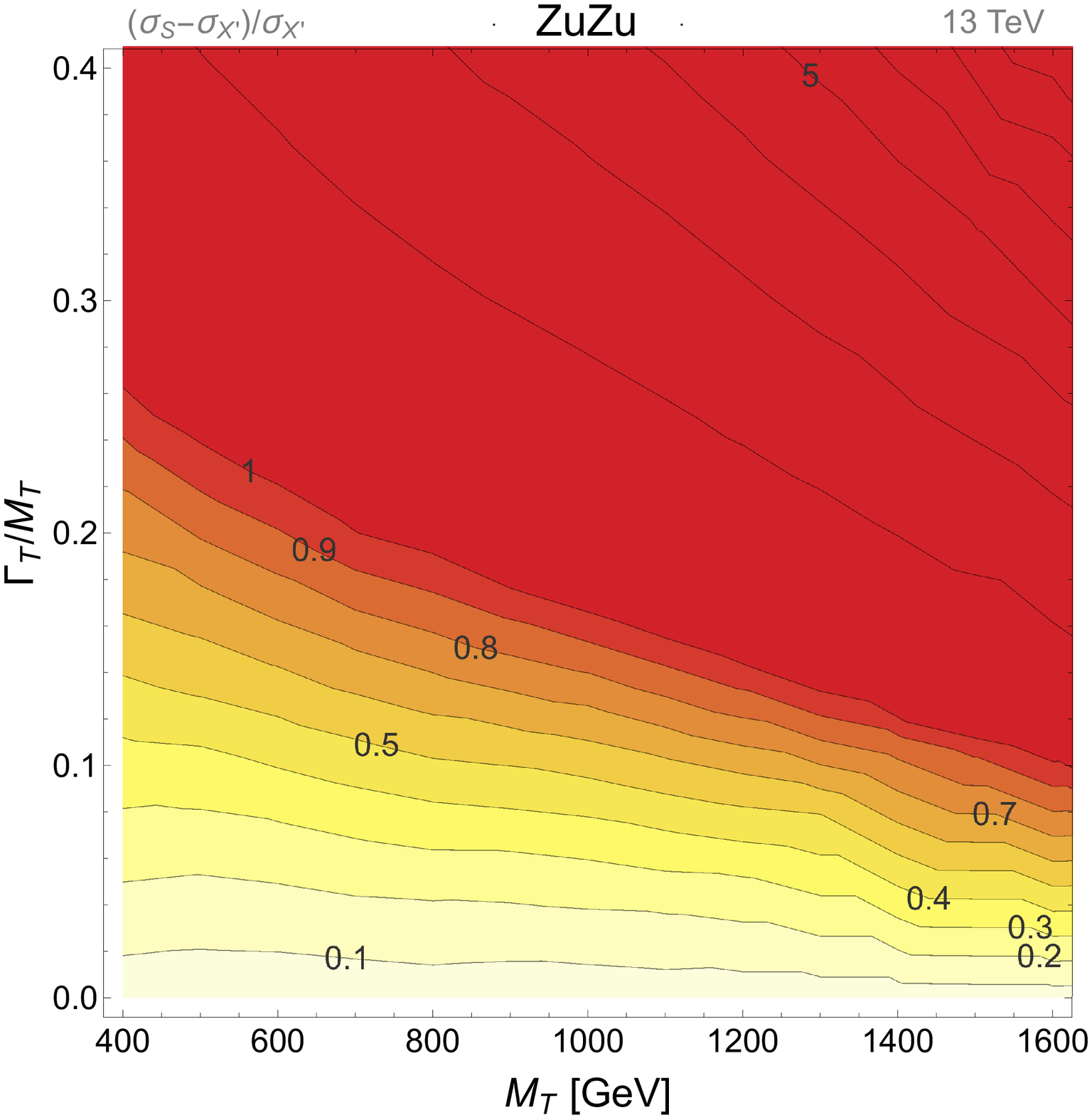, width=.3\textwidth} 
\epsfig{file=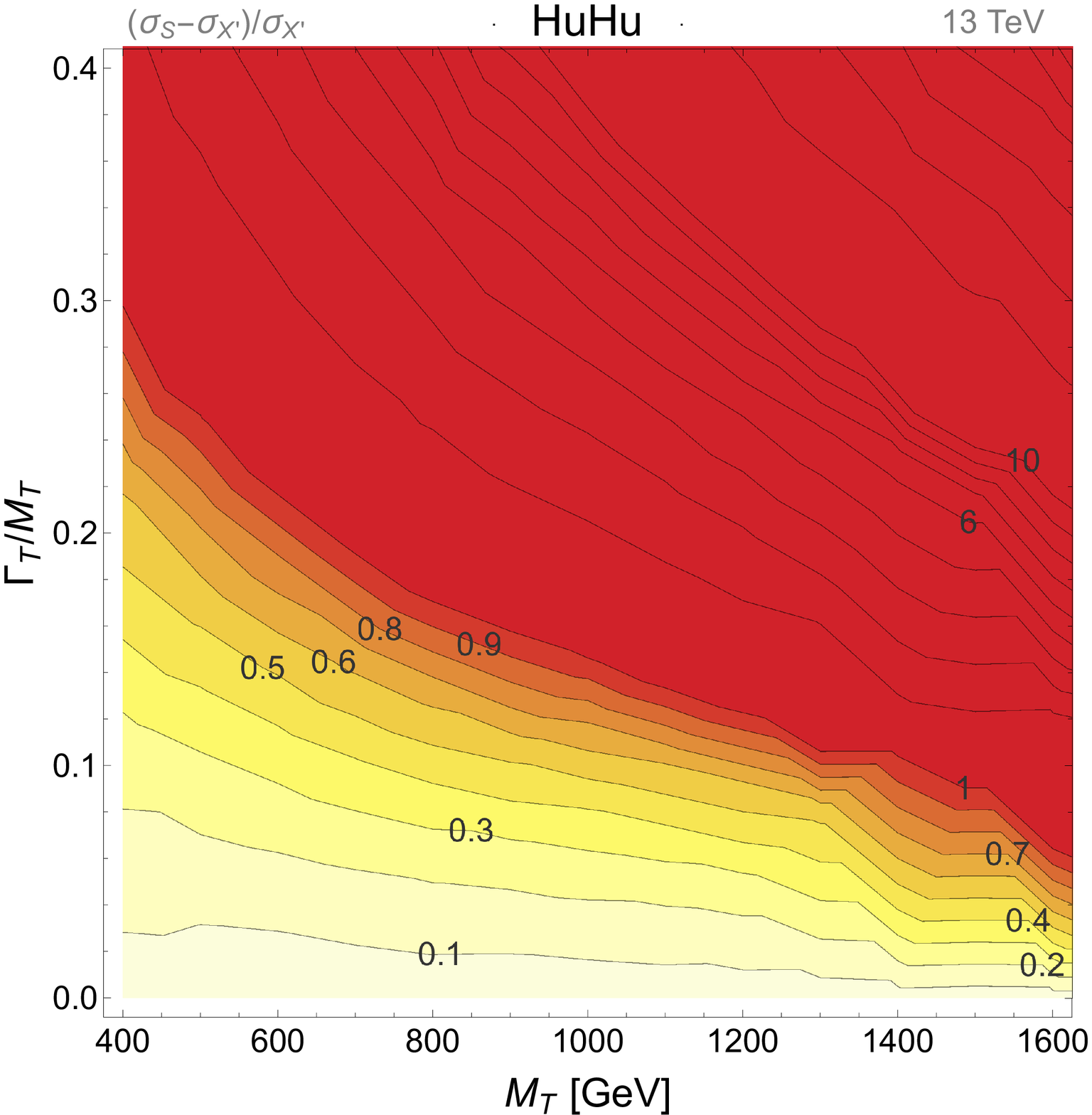, width=.3\textwidth}\\
\epsfig{file=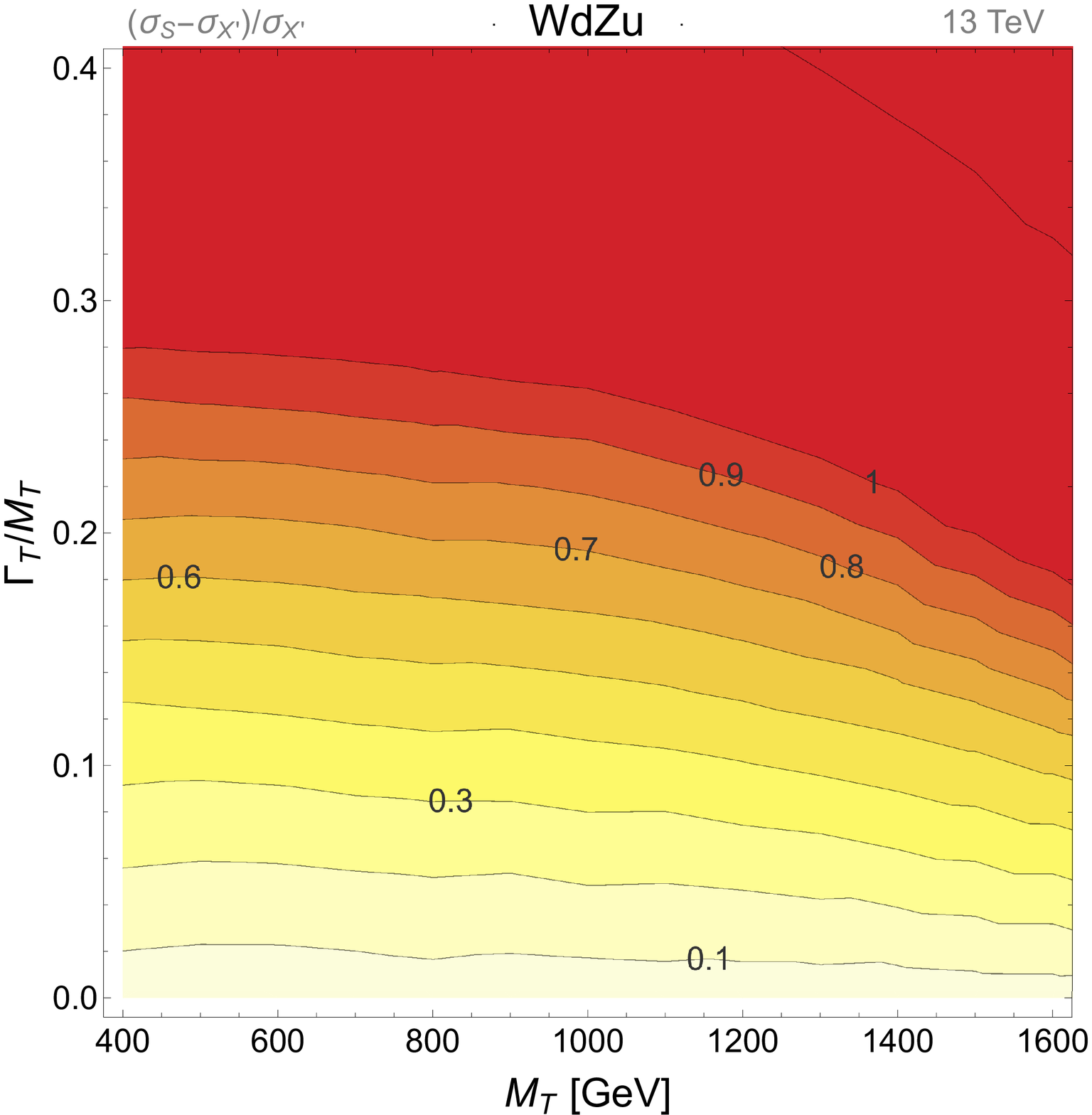, width=.3\textwidth} 
\epsfig{file=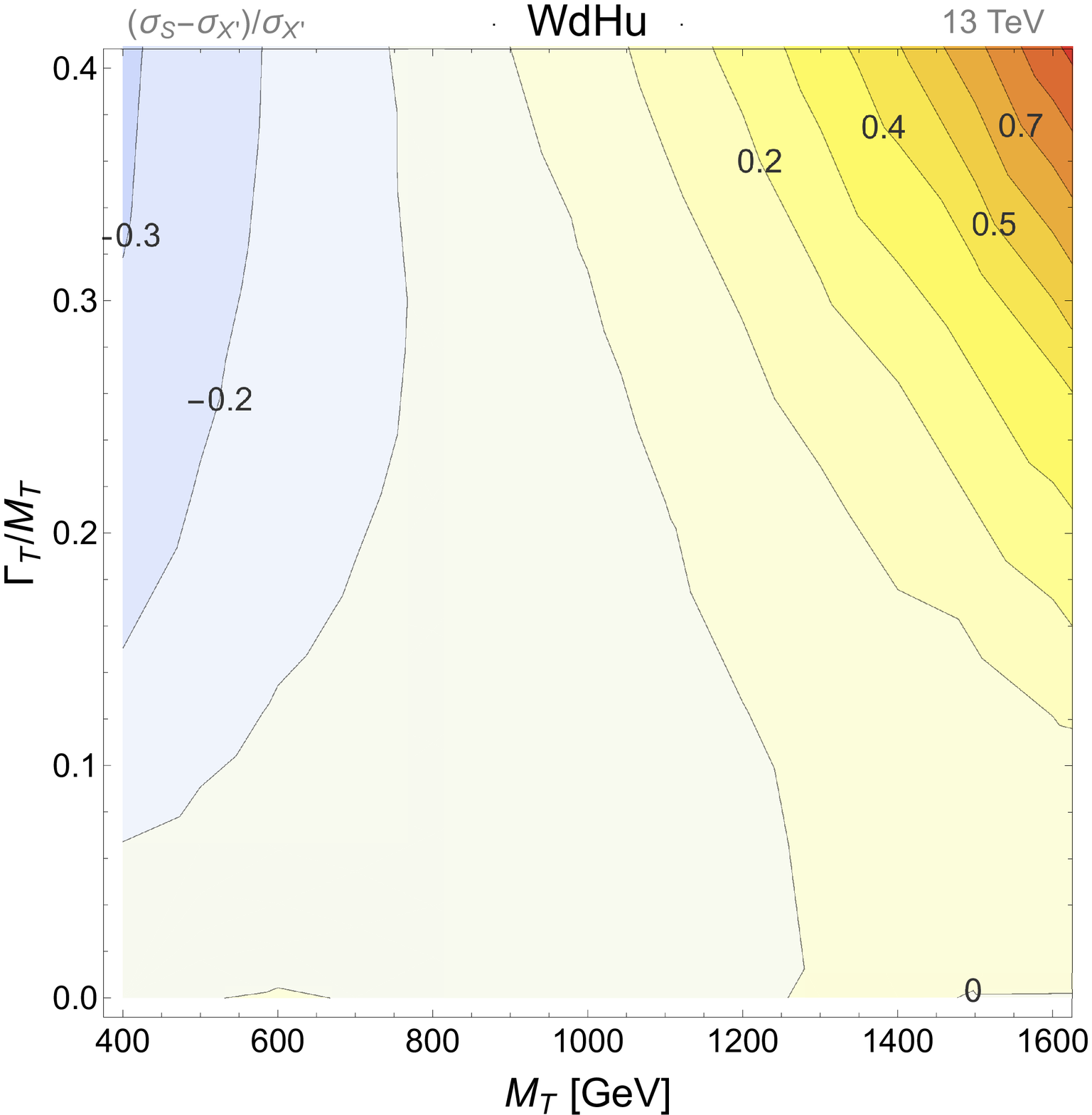, width=.3\textwidth} 
\epsfig{file=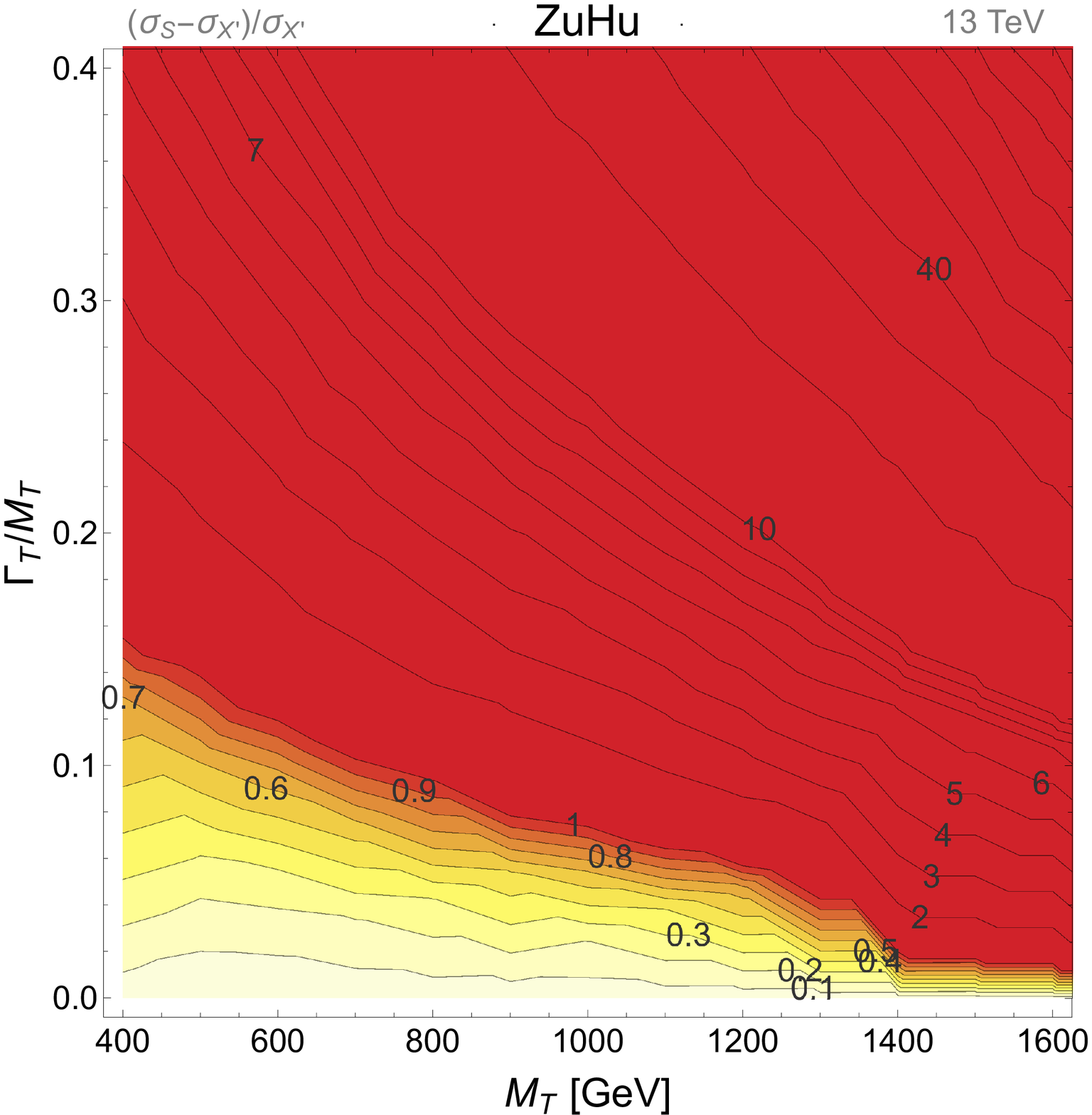, width=.3\textwidth} 
\caption{Same as Fig.~\ref{fig:SXthird1} for $T$ mixing with first generation.}
\label{fig:SXfirst1}
\end{figure}

The relative increase of the cross section with respect to the NWA regime is shown in Fig.~\ref{fig:SXfirst1} for an energy of 13 TeV (we have checked that the results at 8 TeV are analogous), where it is possible to notice the large enhancement due to topologies with collinear divergences for all final states.

%%%%%%%%%%%%%%%%%%%%%%%%%%%%%%%%%%%%%%%%%%%%%%%%%%%%%%%%%%%%%%%%%%%%%%%%%
\subsubsection{Interference with SM background}

\begin{figure}[H]
\centering
\epsfig{file=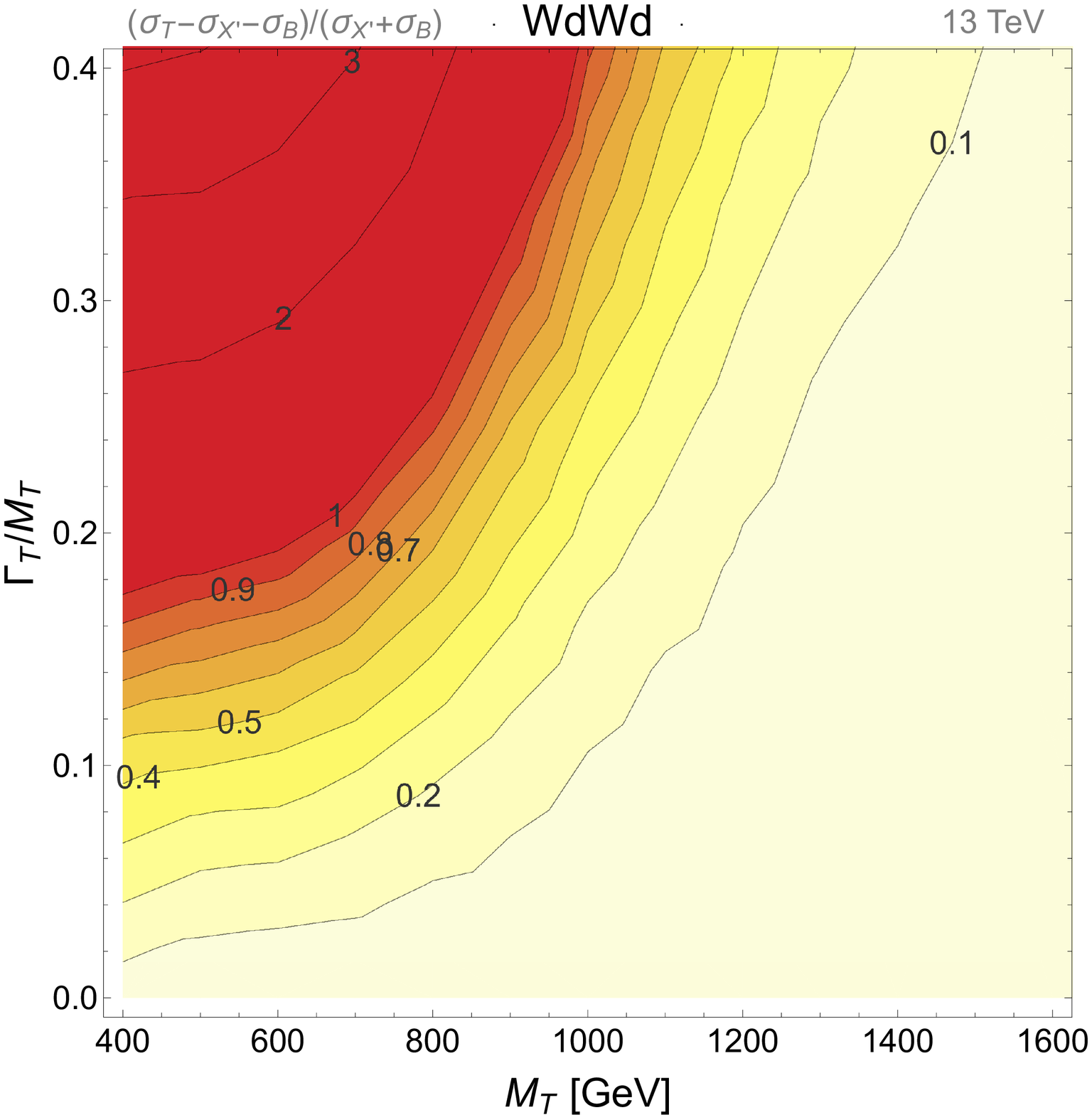, width=.3\textwidth}
\epsfig{file=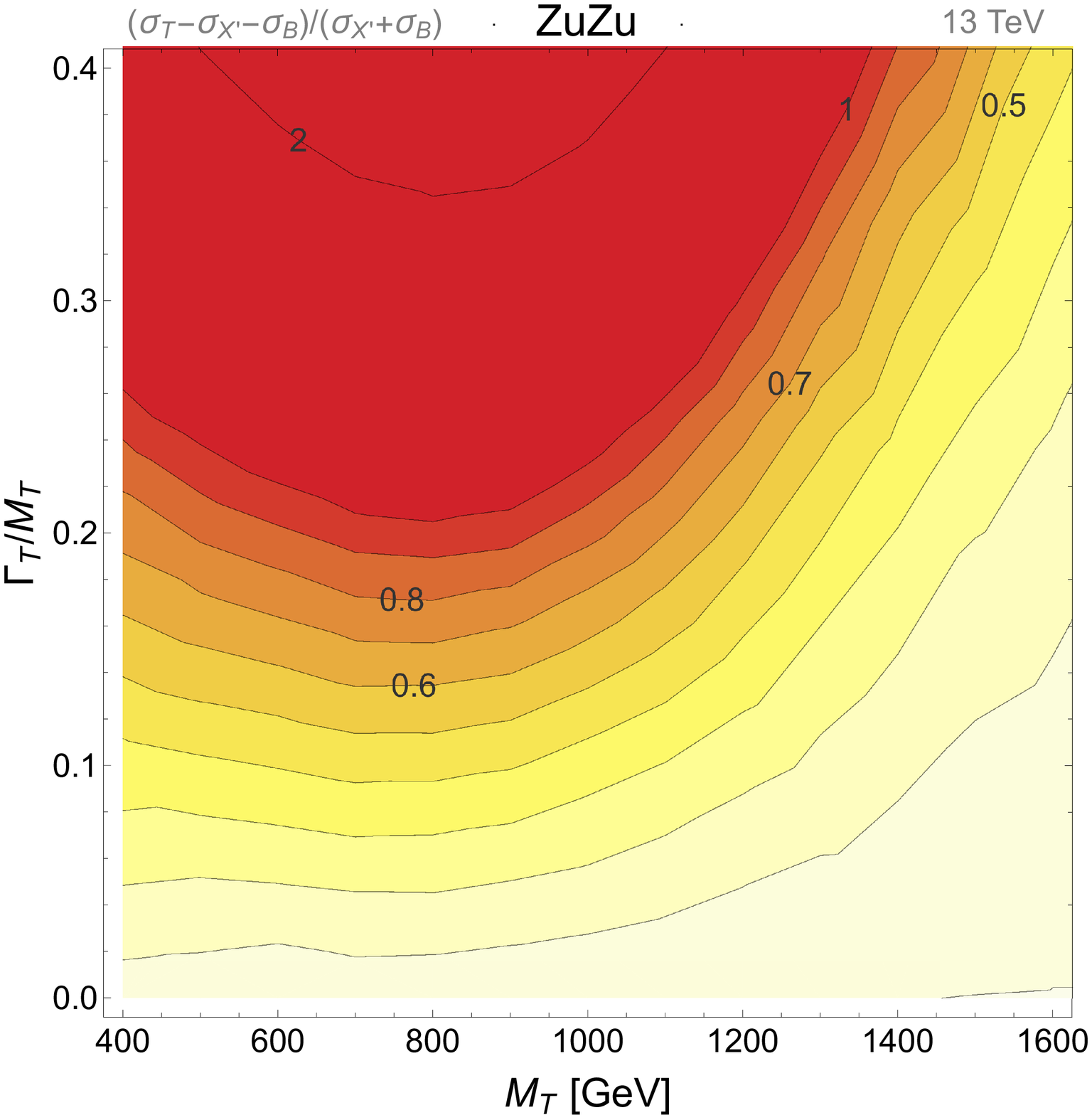, width=.3\textwidth}
\epsfig{file=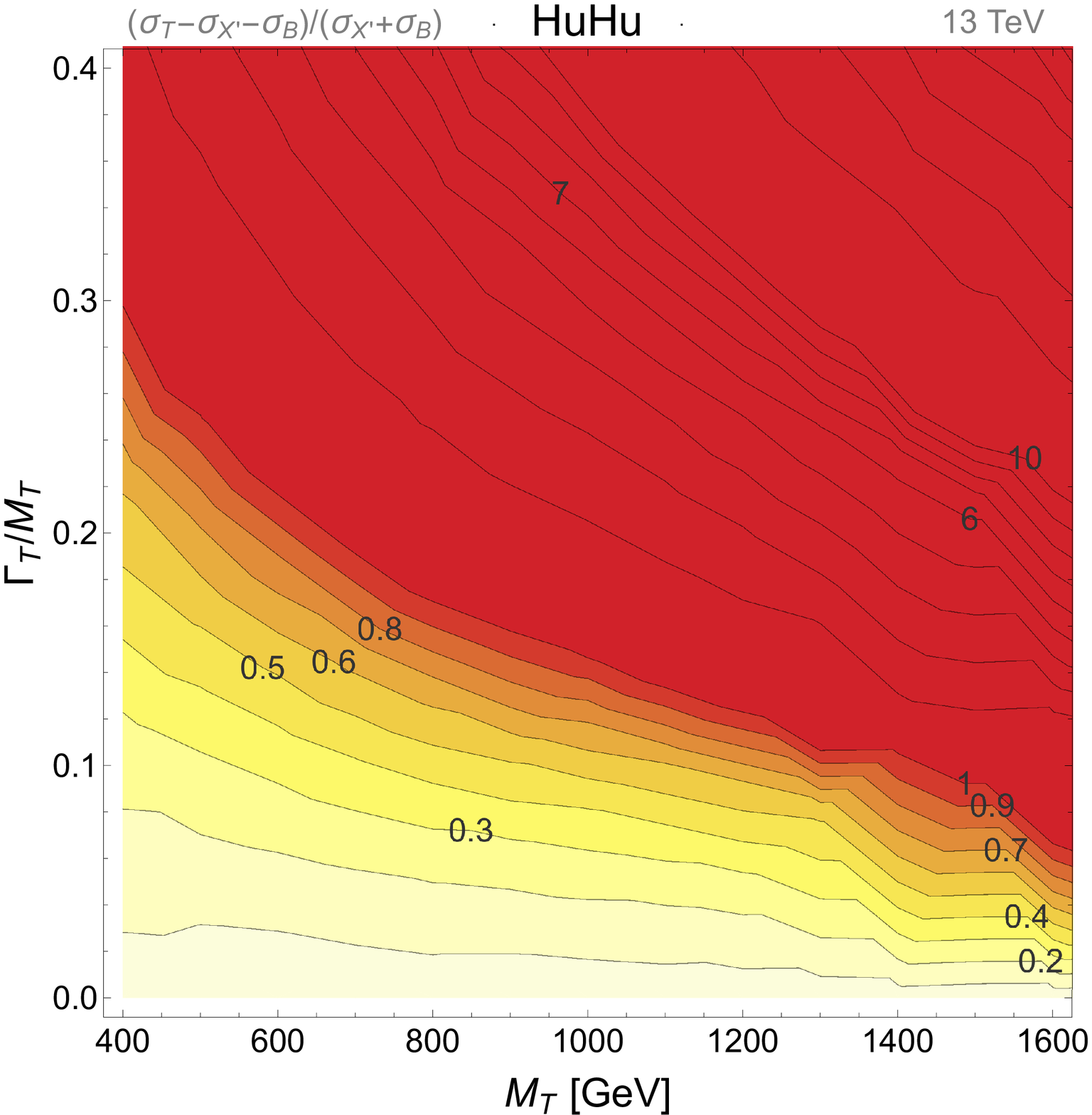, width=.3\textwidth}\\
\epsfig{file=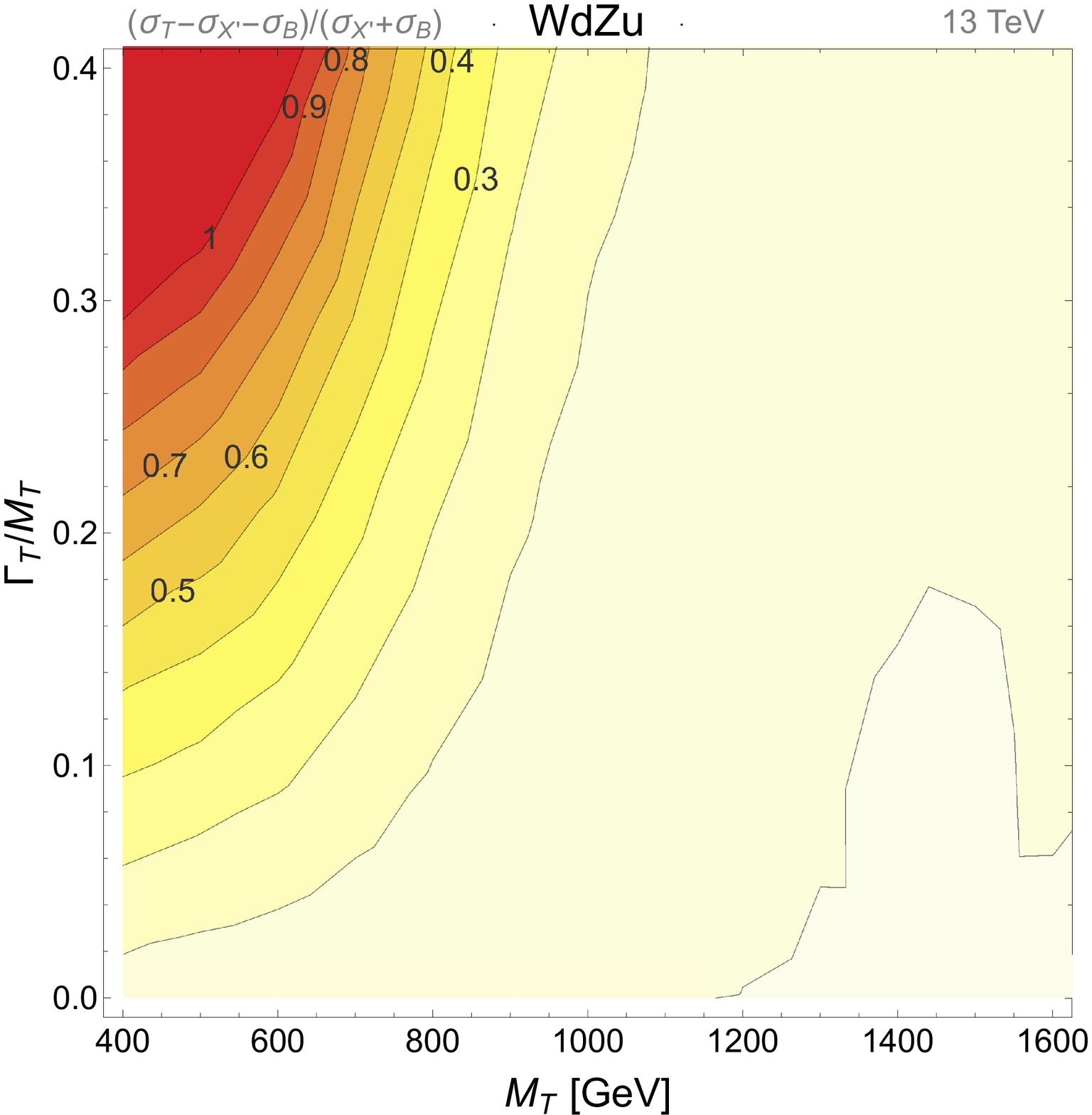, width=.3\textwidth}
\epsfig{file=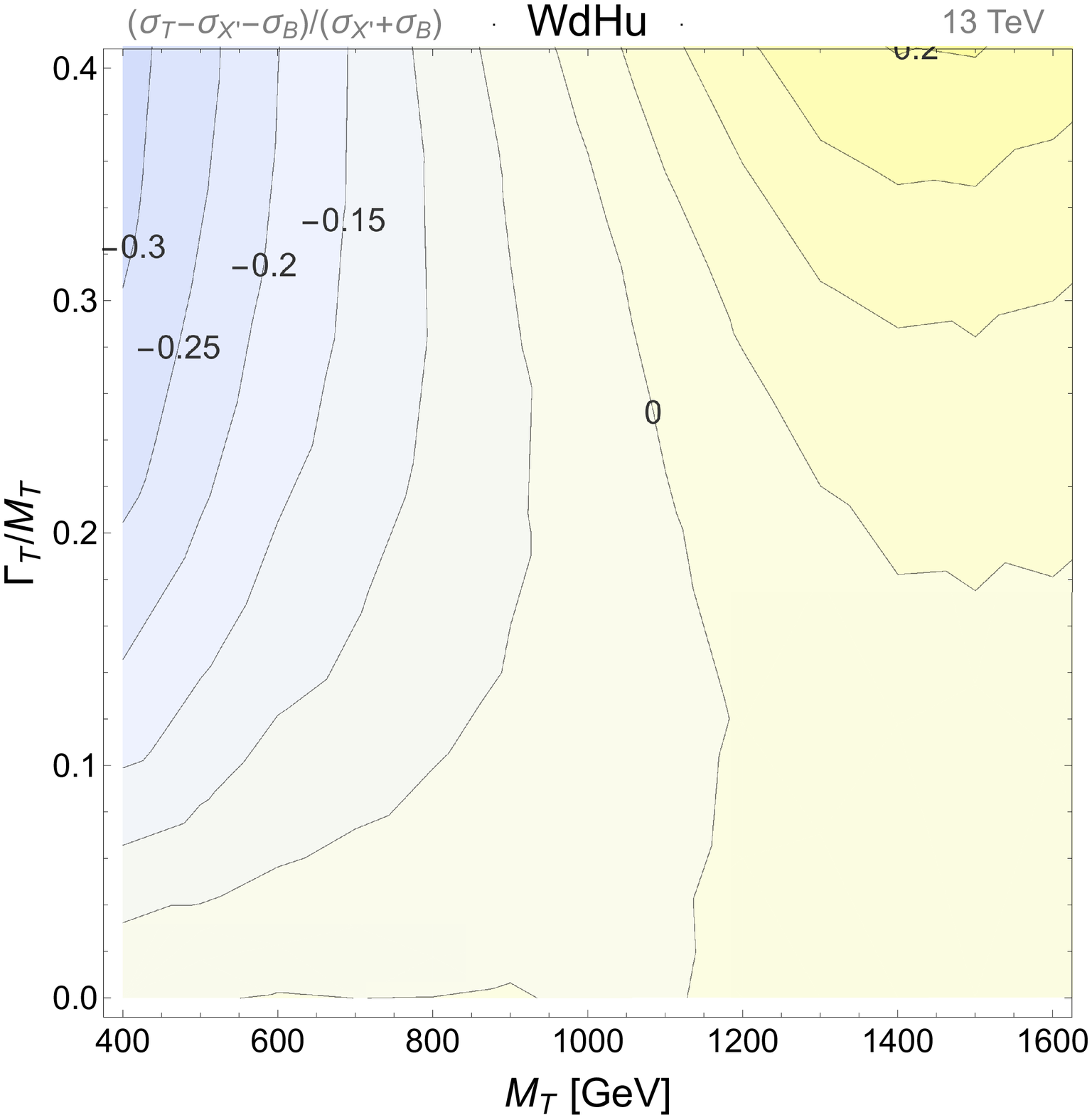, width=.3\textwidth}
\epsfig{file=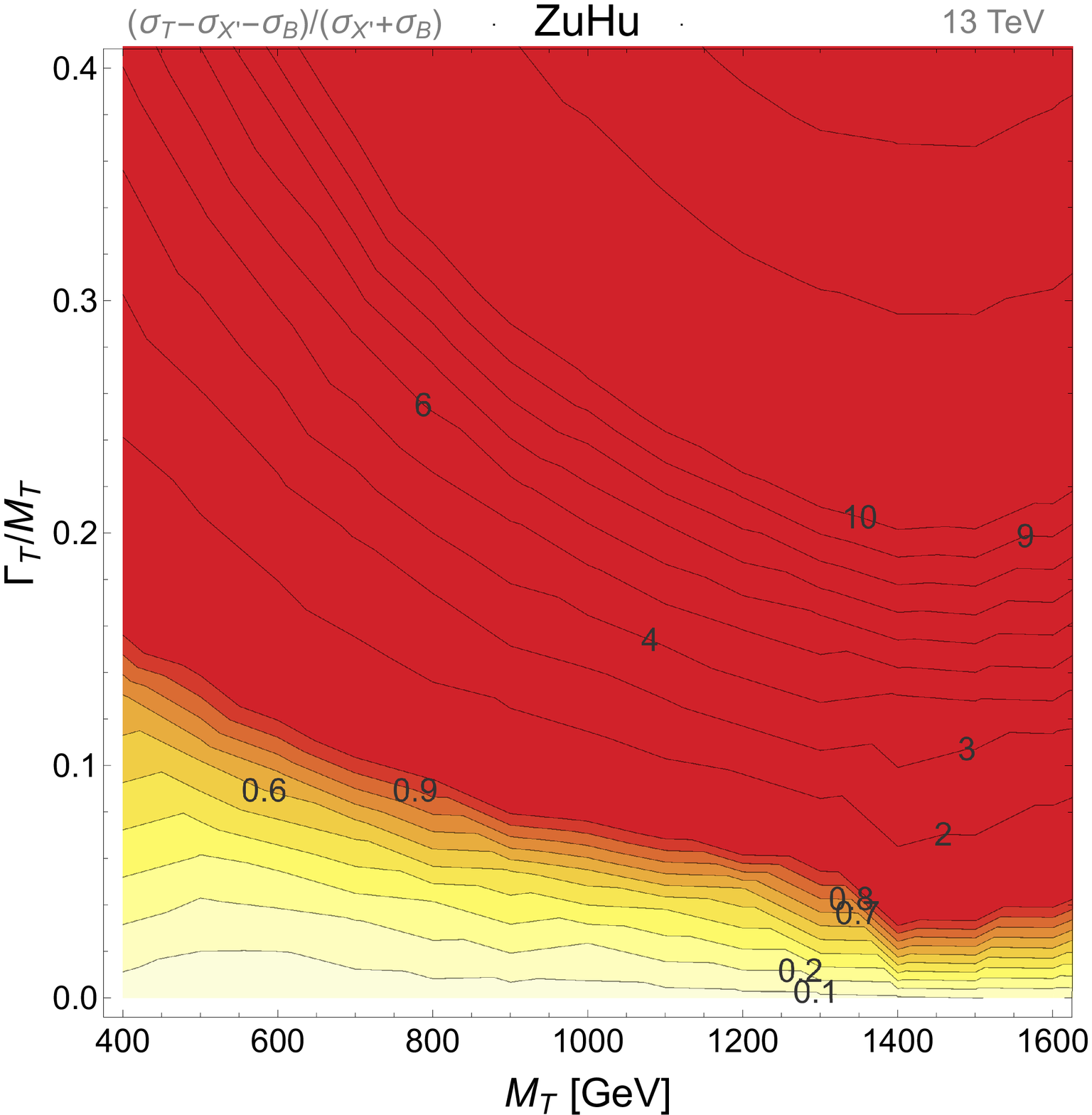, width=.3\textwidth}
\caption{Same as Fig.~\ref{fig:TXBthird} for $T$ mixing with first generation.}
\label{fig:TXBfirst}
\end{figure}

The correction factors to multiply to the sum of NWA cross section and SM background to obtain the interference term are plotted in Fig.~\ref{fig:TXBfirst}. For all channels the correction factor becomes quickly large as the $T$ width increases, even if in different fashions depending on the channel. The relative differences between signal and background are small in this case, such that $\sigma_T$ receives a large contribution from the signal. However, when taking into account the full signal, including the large width effects, the interference effects with the SM background become small or negligible in the whole parameter space with respect to the total cross-section. As in the case of mixing with third generation, these results show that searches for the exploration of scenarios where the VLQs mix with light generations and have a large width would be significantly more accurate by considering the full signal instead than reinterpreting the NWA results.

\begin{figure}[H]
\centering
\epsfig{file=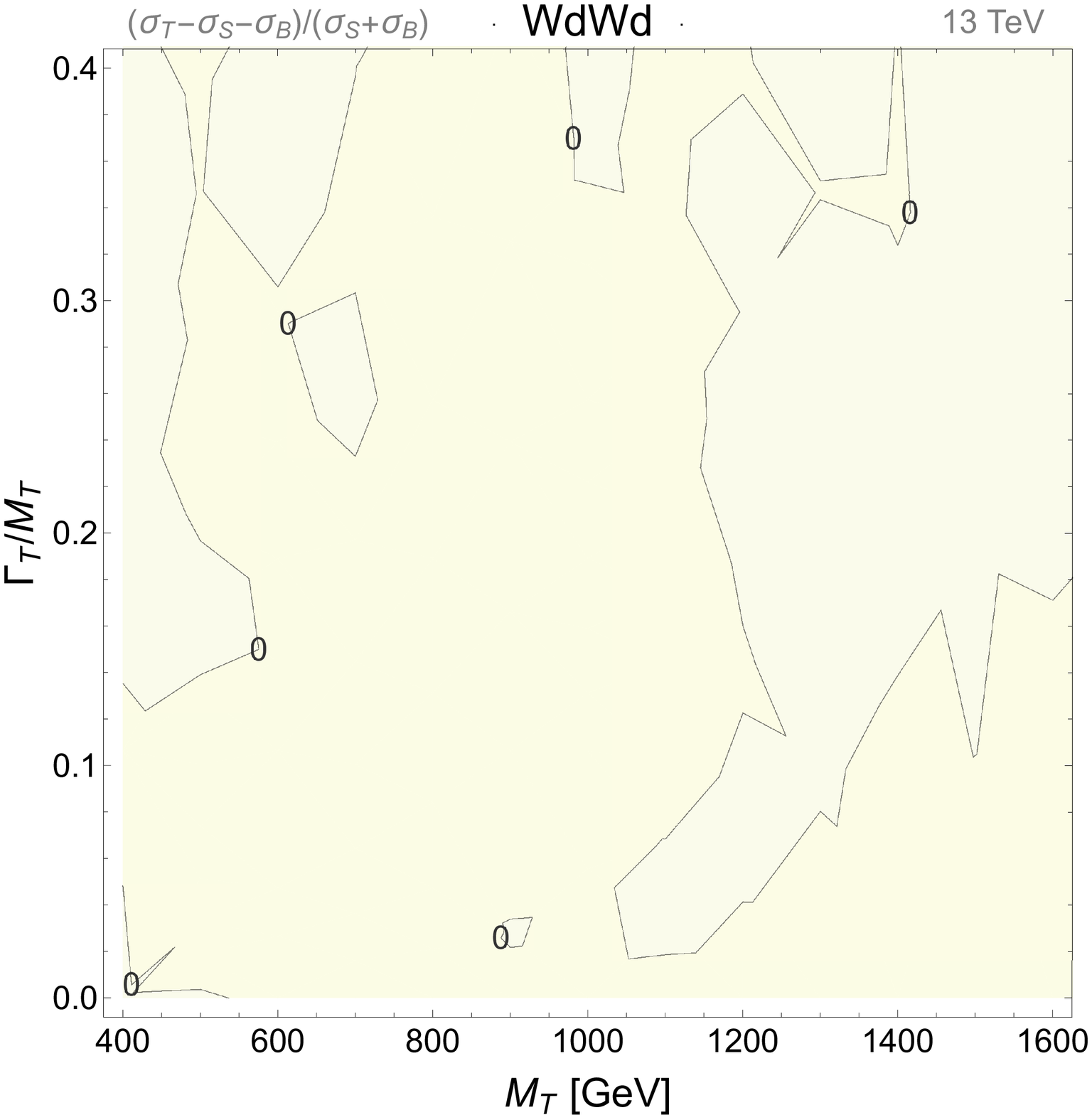, width=.3\textwidth}
\epsfig{file=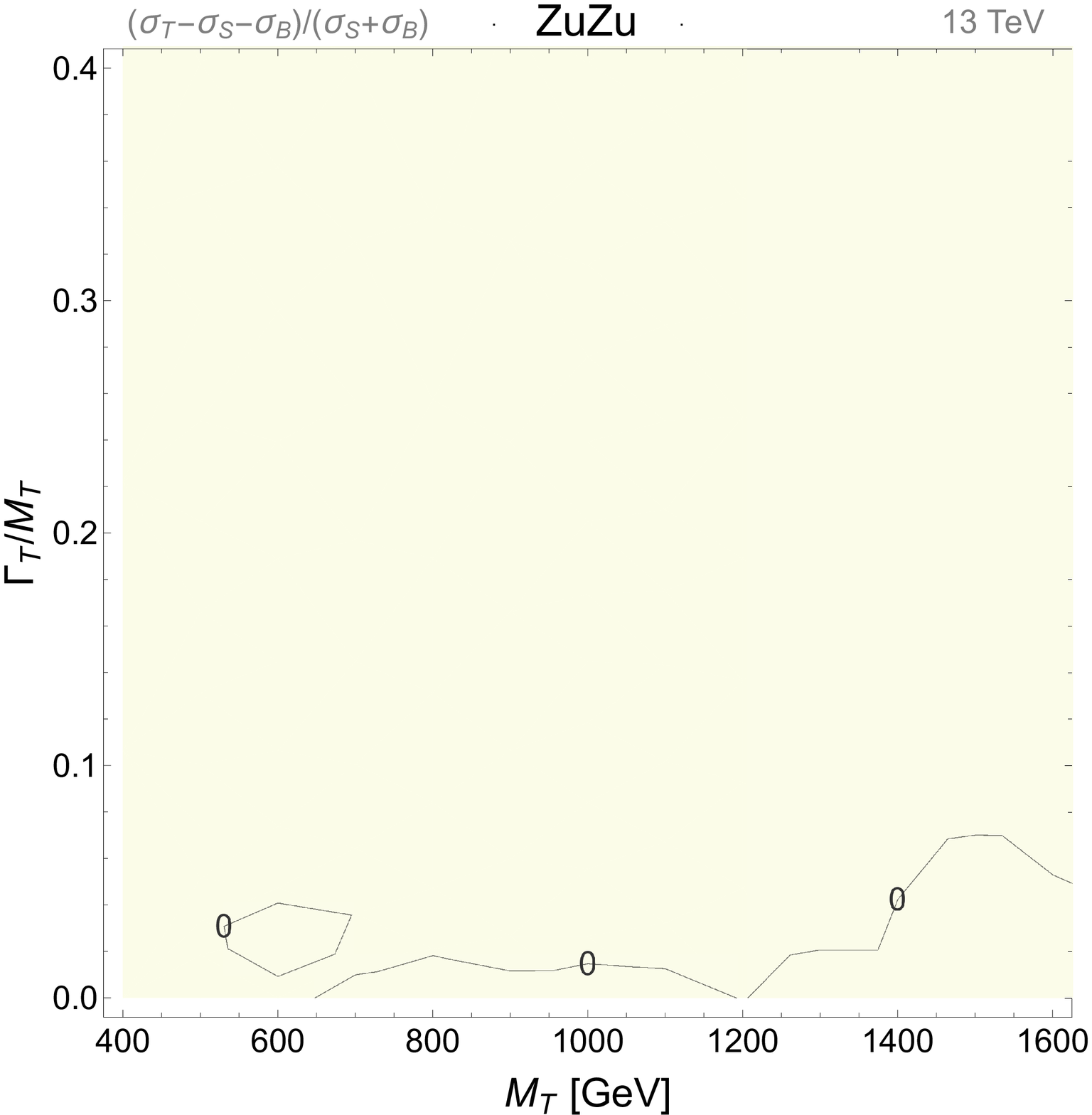, width=.3\textwidth}
\epsfig{file=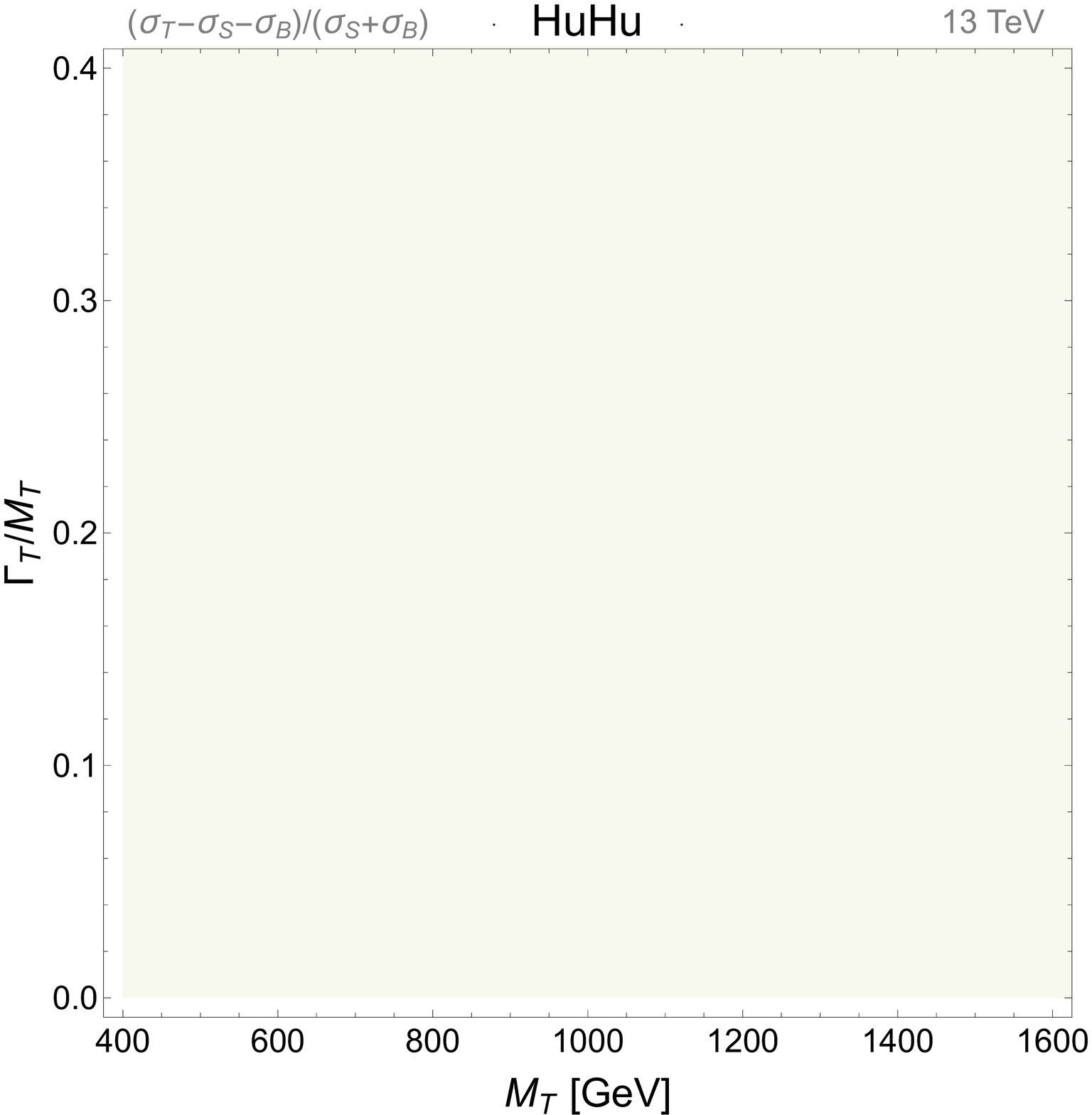, width=.3\textwidth}\\
\epsfig{file=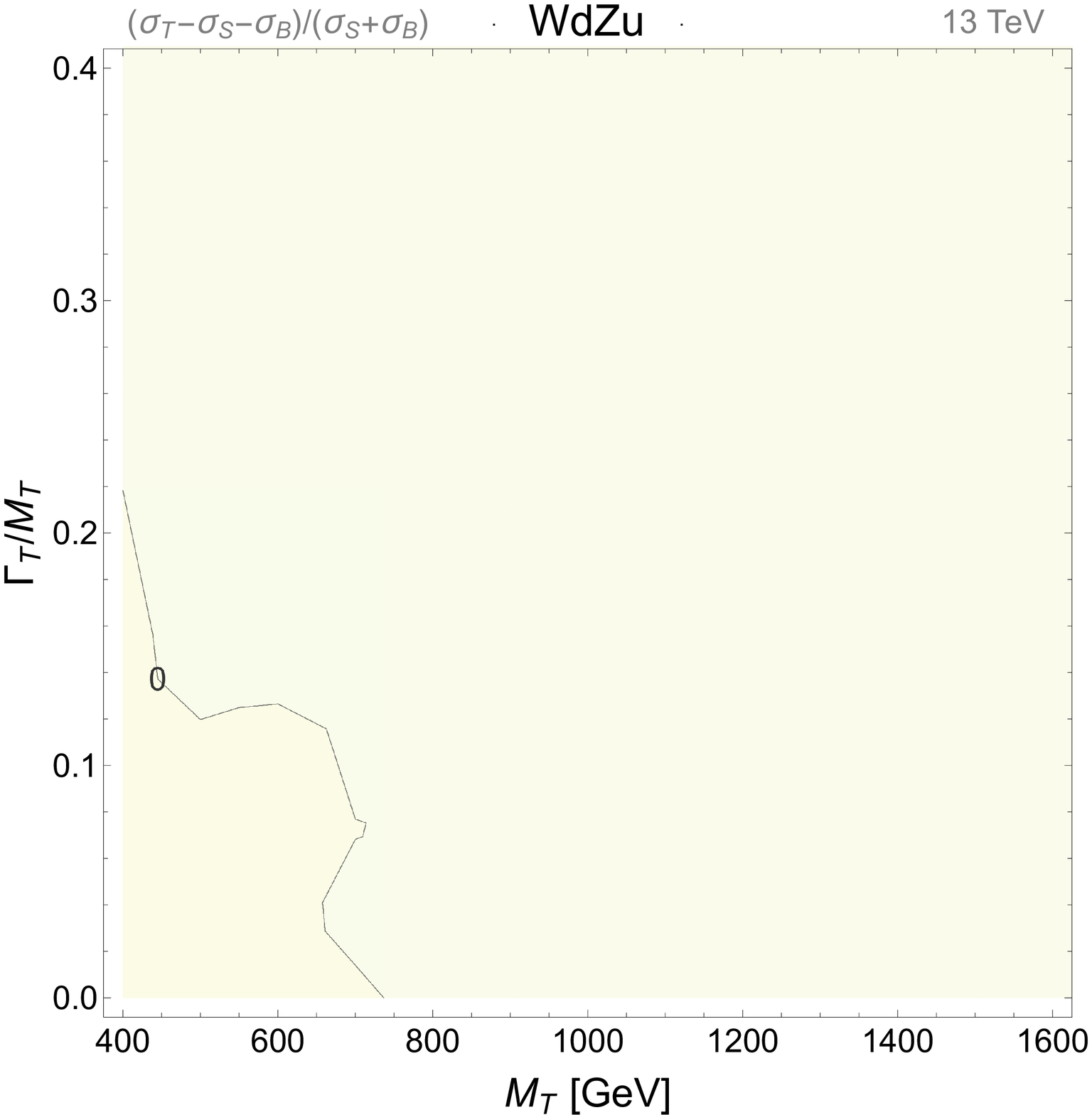, width=.3\textwidth}
\epsfig{file=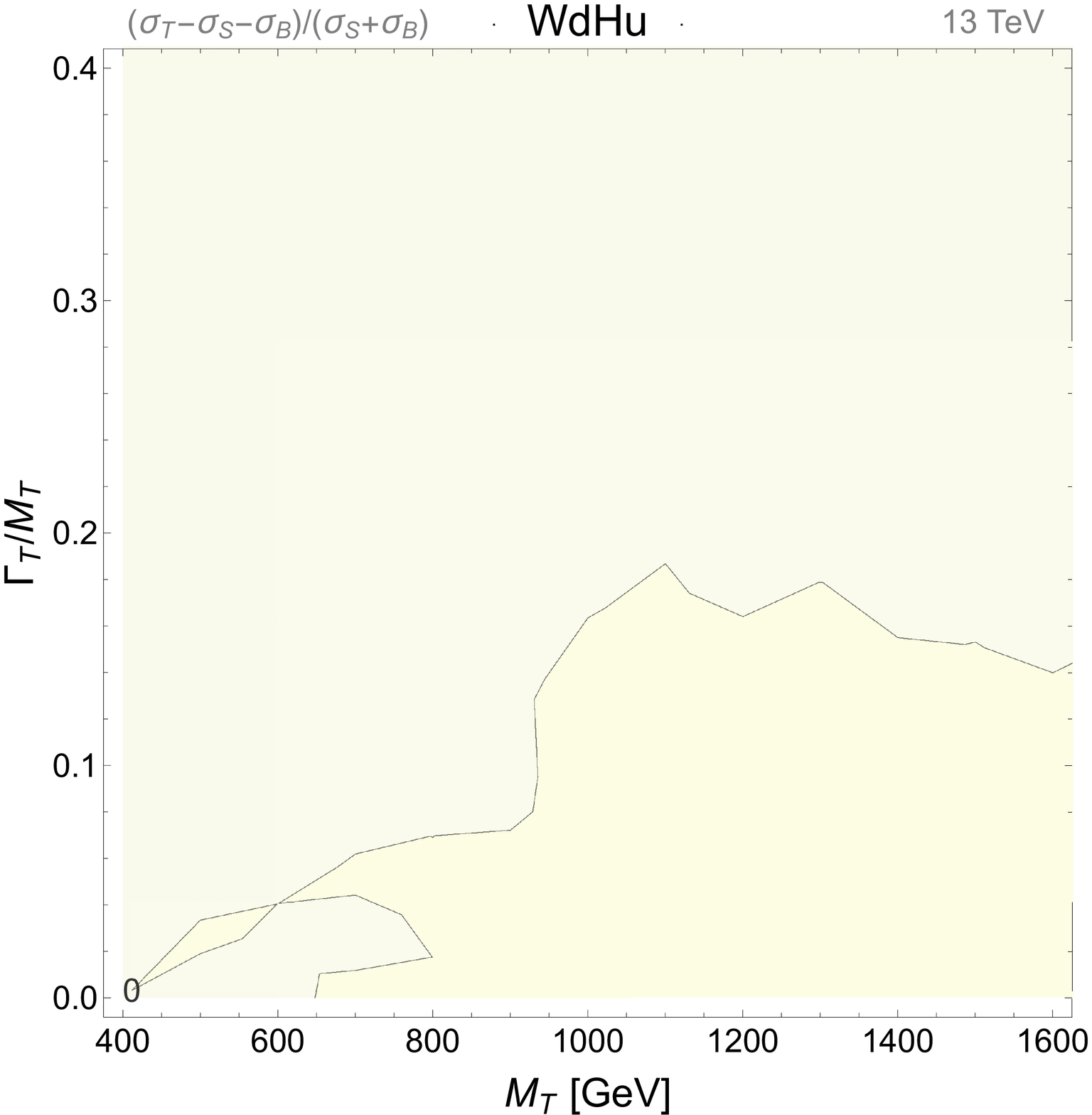, width=.3\textwidth}
\epsfig{file=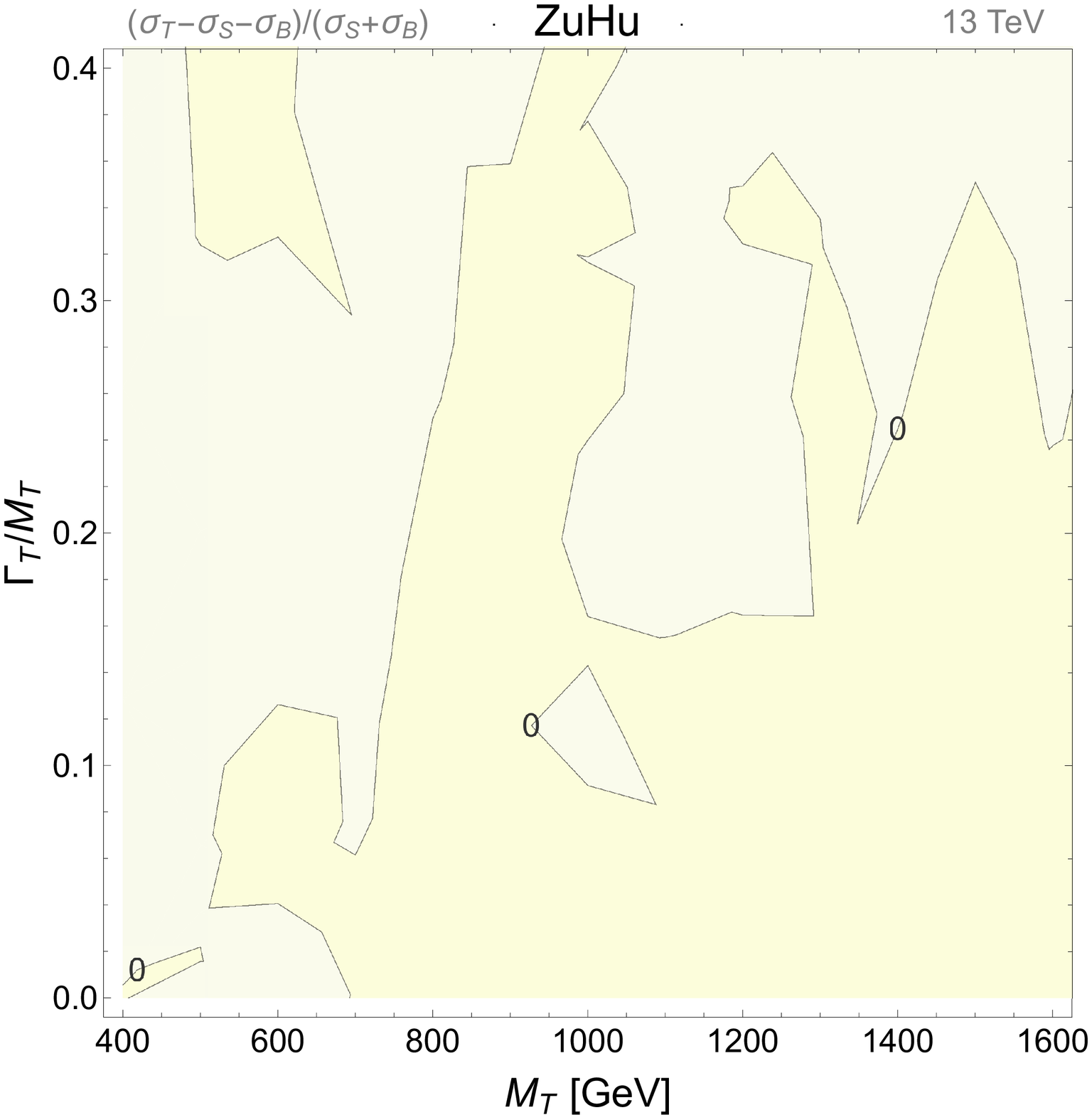, width=.3\textwidth}
\caption{Same as Fig.~\ref{fig:TSBthird} for $T$ mixing with first generation.}
\label{fig:TSBfirst}
\end{figure}

\subsubsection{Results at detector level}

\begin{figure}[H]
\centering
\epsfig{file=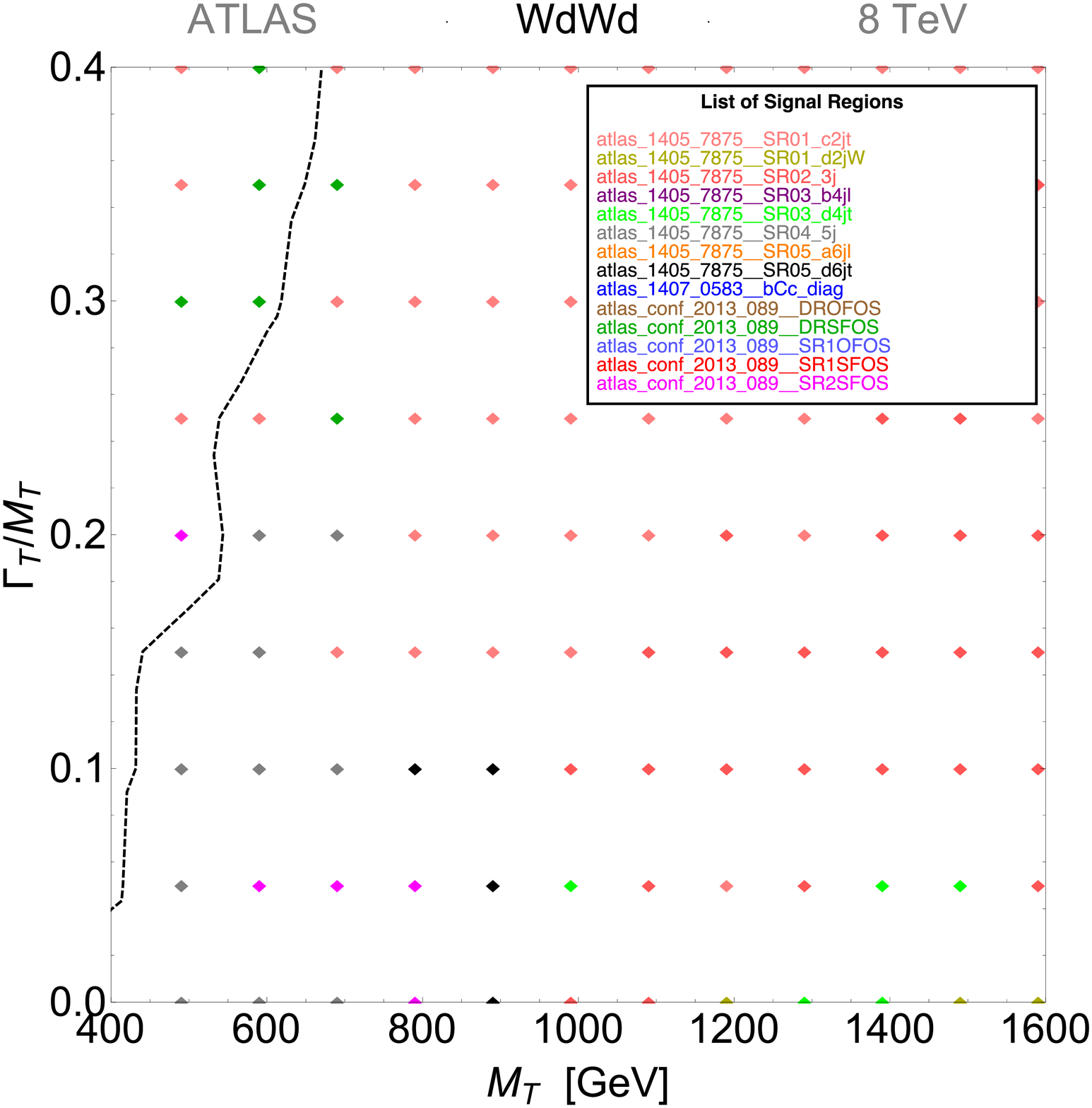, width=.3\textwidth}
\epsfig{file=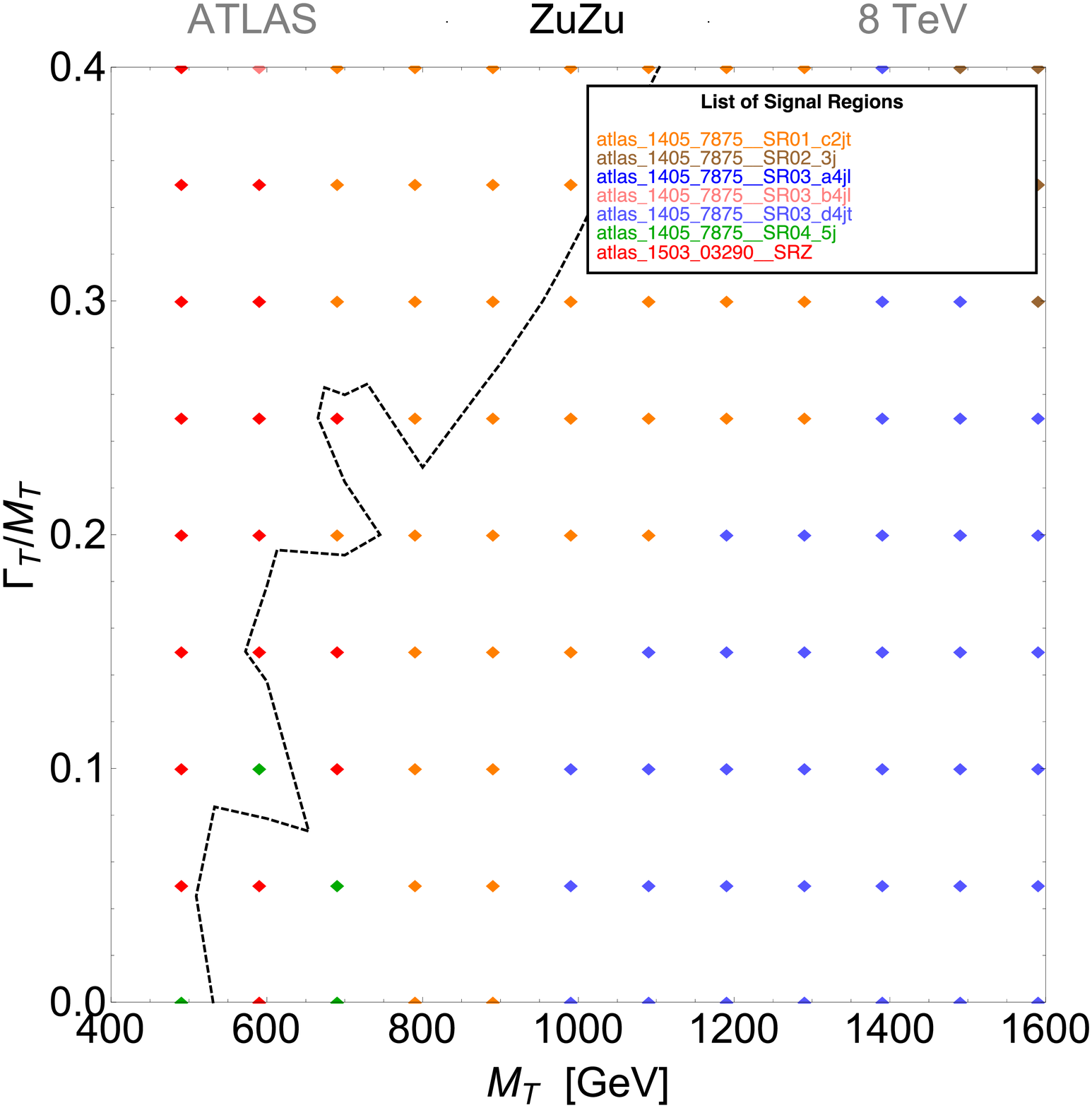, width=.3\textwidth}
\epsfig{file=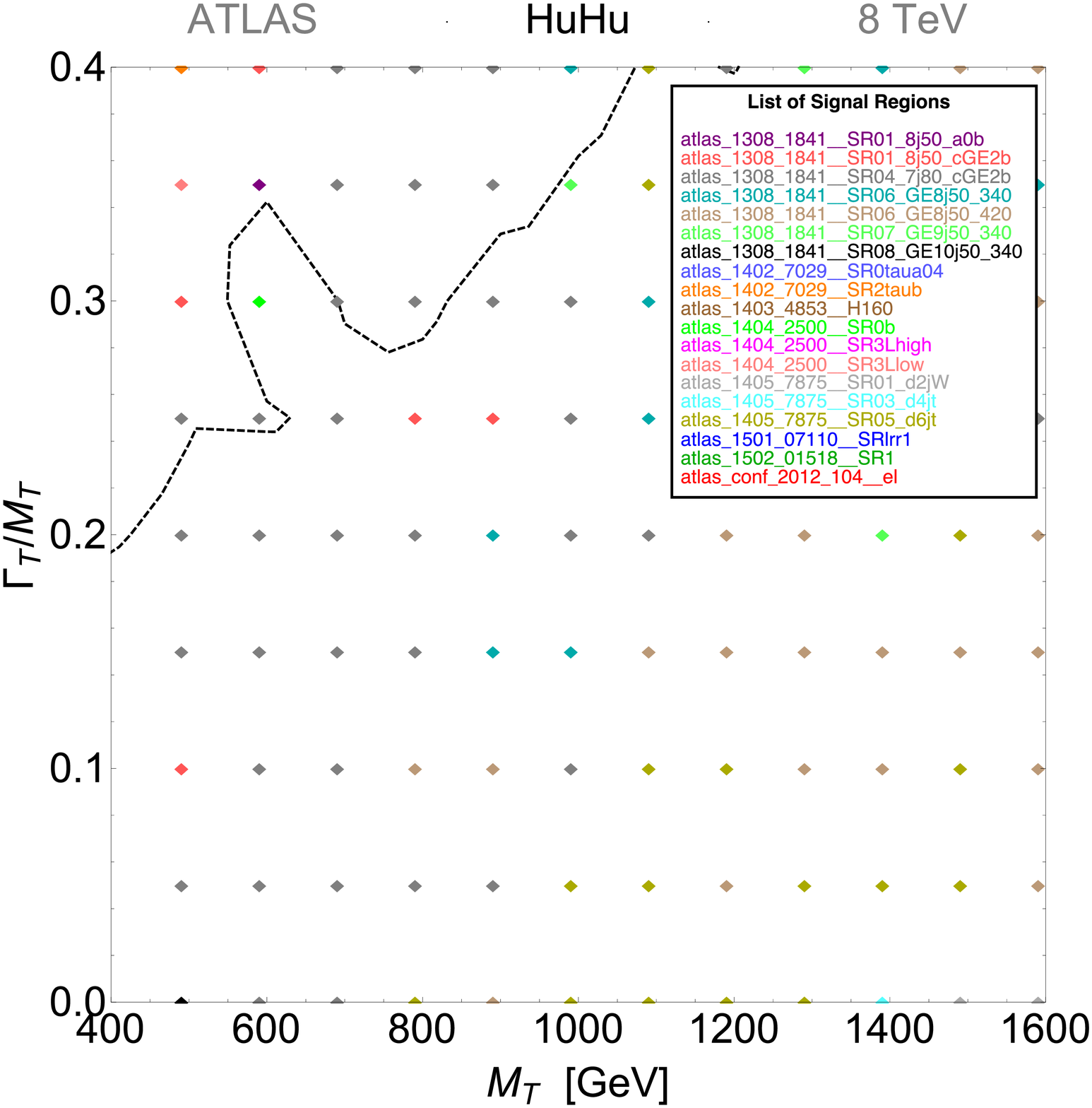, width=.3\textwidth}\\
\epsfig{file=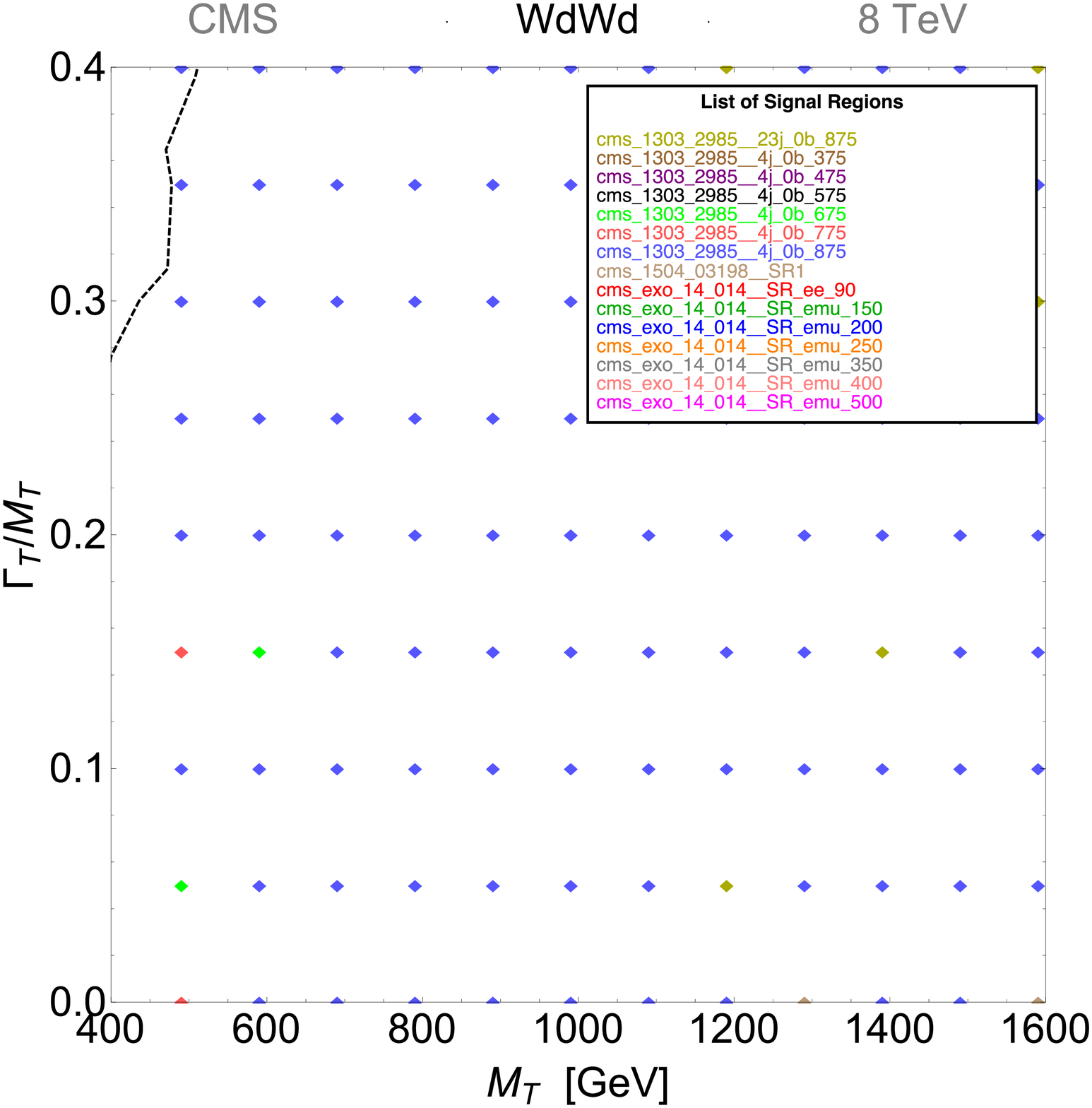, width=.3\textwidth}
\epsfig{file=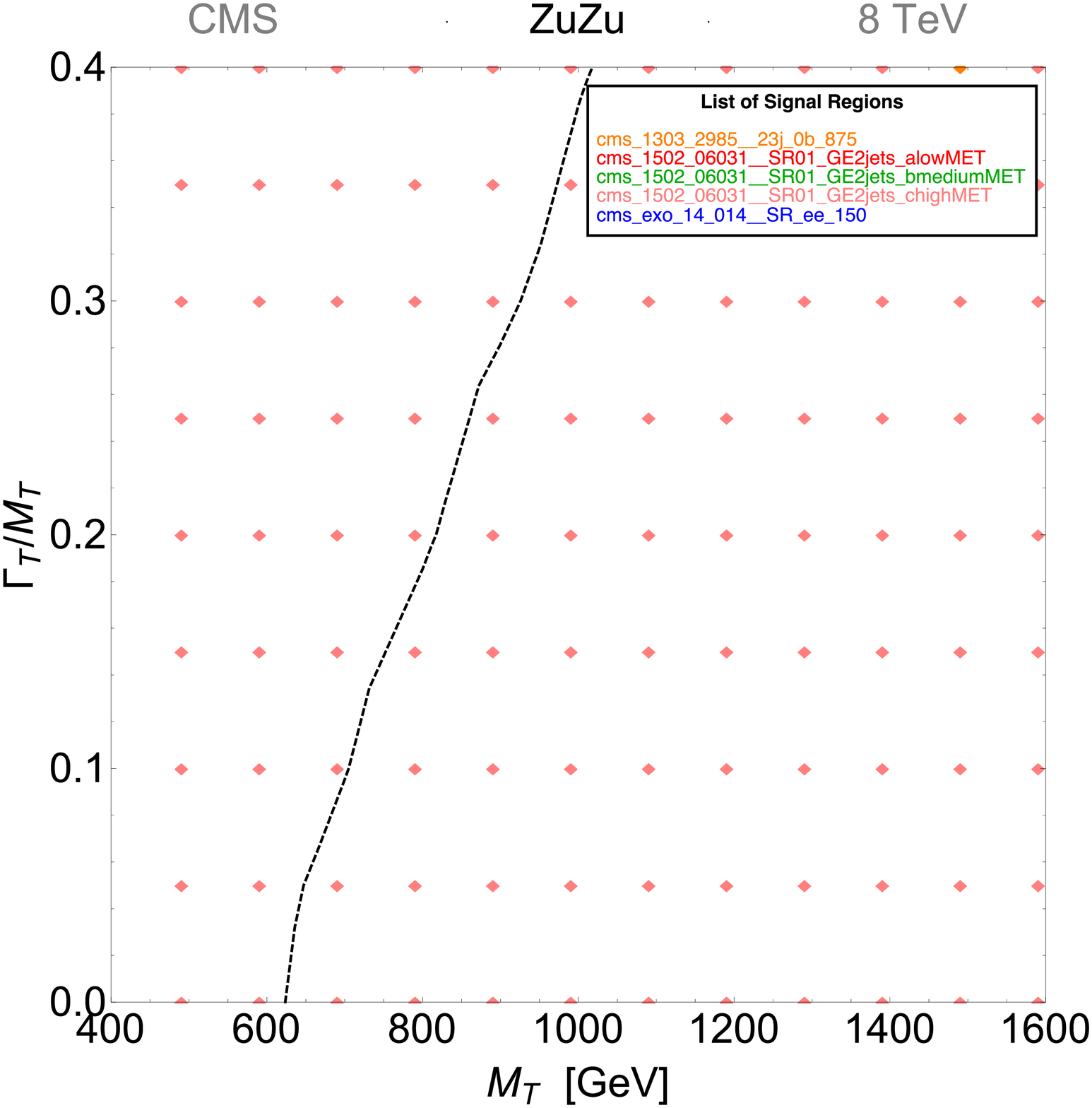, width=.3\textwidth}
\epsfig{file=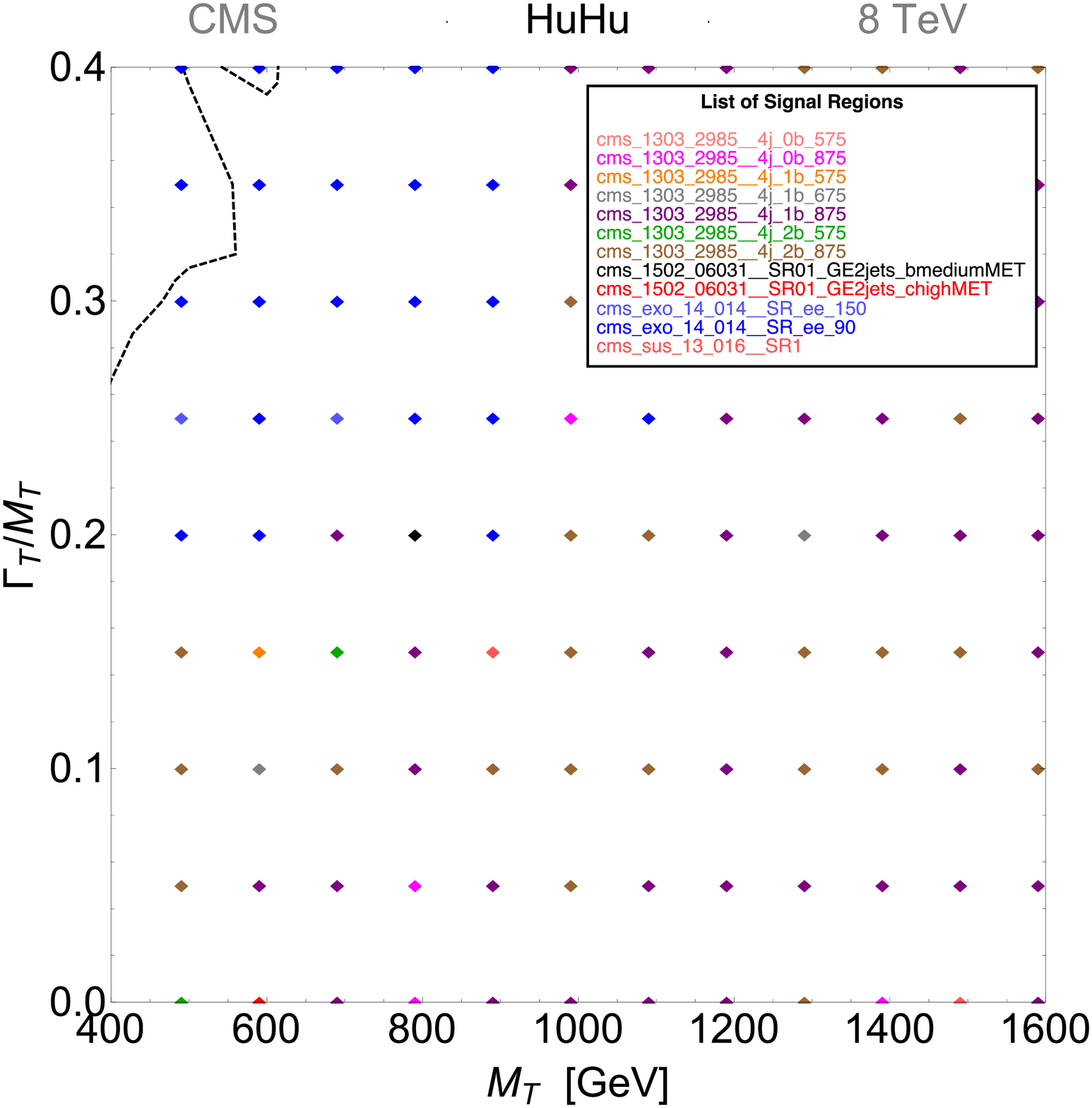, width=.3\textwidth}
\caption{Same as Fig.~\ref{fig:Detector8TeV3rdgen} for $T$ mixing with first generation.}
\label{fig:Detector8TeV1stgen}
\end{figure}

\begin{figure}[H]
\centering
\epsfig{file=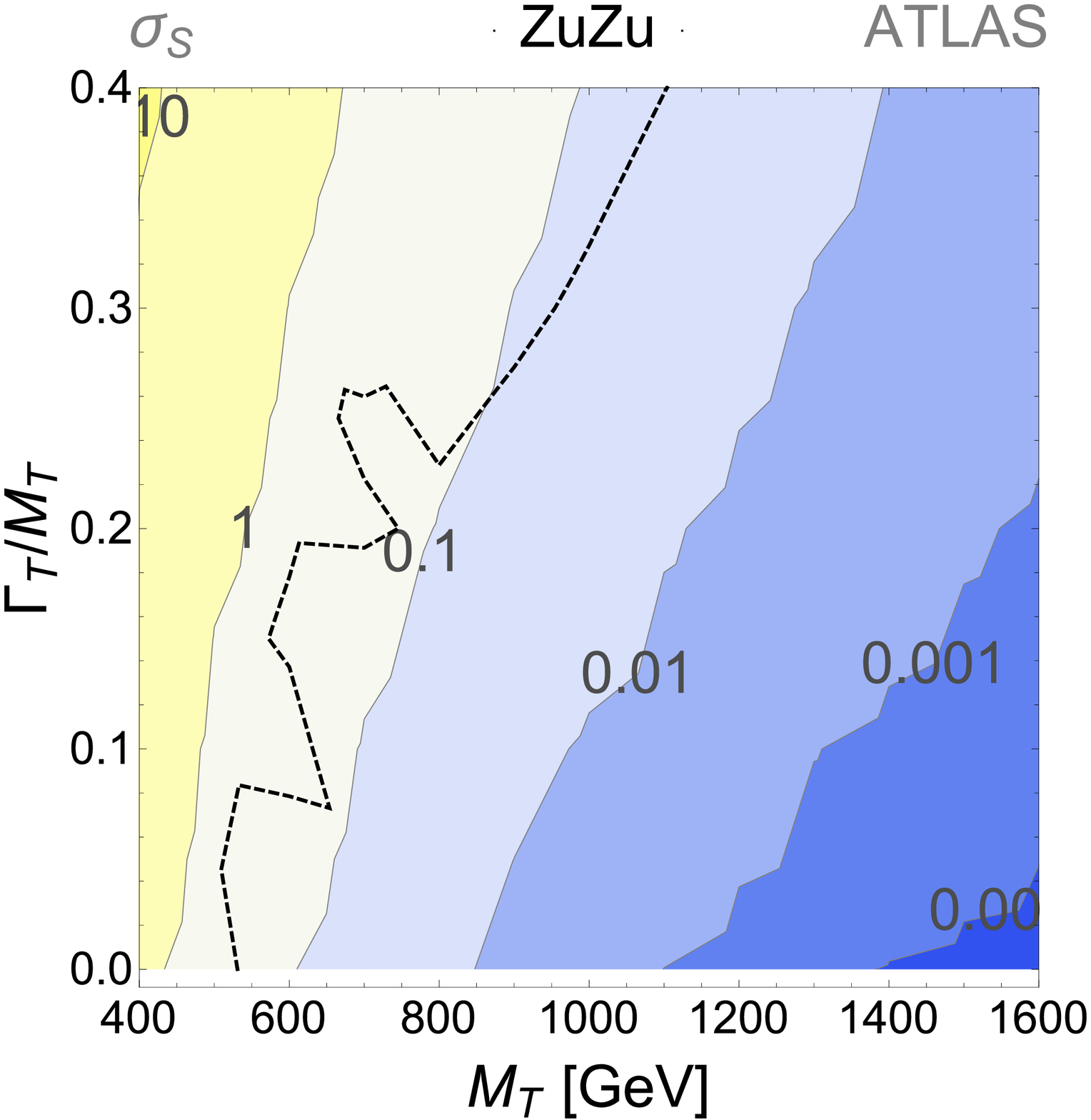, width=.3\textwidth}
\epsfig{file=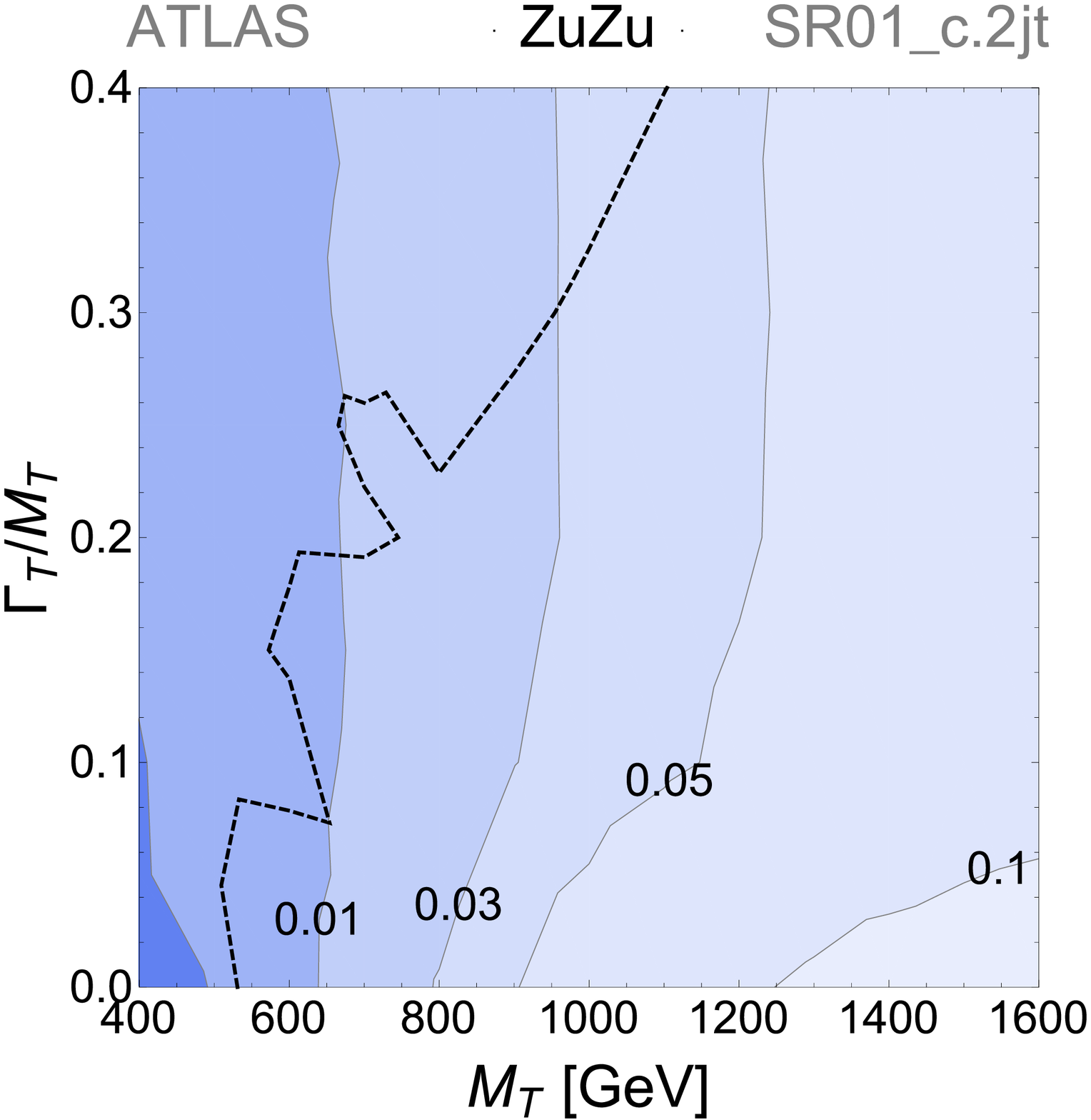, width=.3\textwidth}
\caption[Cross section and efficiency of the best ATLAS SR for the $ZuZu$ channel, compared with the bound.]{Cross section and efficiency of the best ATLAS SR (SR01\_c.2jt of \cite{Aad:2014wea}) for the $ZuZu$ channel, compared with the bound.}
\label{fig:CombinedBound8TeVZuZu}
\end{figure}

Our recast results, obtained considering the same set of ATLAS and CMS searches at 8 TeV as in the case of mixing with third generation, are shown in Fig.~\ref{fig:Detector8TeV1stgen}.
The dependence of the bound on the $T$ width is stronger than in the case of mixing with third generation. For all channels the bound on the $T$ mass becomes stronger as the $T$ width increases. This behaviour has again to be put in relation with the dependence of the signal cross section, $\sigma_S$, on the $T$ mass and width, shown in the example of Fig.~\ref{fig:CombinedBound8TeVZuZu} for the bound on the $ZuZu$ channel from ATLAS searches. It is possible to see that the bound roughly tracks the cross section, which unlike in the case of third generation mixing is much more dependent on the width of the $T$, and that the width dependence of the efficiency on the other hand is weakly increasing with both width and mass of $T$ along the bound. 

%%%%%%%%%%%%%%%%%%%%%%%%%%%%%%%%%%%%%%%%%%%%%%%%%%%%%%%%%%%%%%%%%%%%%%%%%
%%%%%%%%%%%%%%%%%%%%%%%%%%%%%%%%%%%%%%%%%%%%%%%%%%%%%%%%%%%%%%%%%%%%%%%%%
\subsection{Conclusions}

\vspace*{-2mm}

We have performed an analysis of off-shell and interference contributions to the process of pair production of heavy quarks at the LHC in the context of minimal scenarios where the SM is extended by adding only a new quark state. As, according to current experimental limits, the latter cannot have the $V-A$ structure of the top quark (unless the Higgs sector is extended, which is not the case in our analysis), we have first assessed how off-shellness impacts on the heavy quark decay signature common to the one of top quark pairs, i.e., $W^-\bar b W^+b$, showing that a $V+A$ chiral structure would be similarly affected over the LHC kinematical regime for pair production of heavy quarks
which can be profiled through a resonance. In this case then, the implementation of FW effects for heavy quarks can be subsumed under the well established procedures already put in place for the top quark, by simply rescaling the mass of the fermion. Many more decays are however possible for a generic heavy quark pair. Of all the latter, as representative examples, we have chosen to focus on the production and decay of a heavy VL top partner $T$ in the singlet representation and considered two scenarios in which it mixes  with either the first or third generation of SM quarks. 

\vspace*{-0.5mm}

The results of our analysis quantify the relevance of the large width regime in the determination of the cross section and  the importance of  interference effects between signal and SM background. Clearly, the differences in the cross section are ultimately reflected in different kinematical distributions, which result in different experimental efficiencies for specific sets of kinematical cuts on the final state. The effect of interference is also found to be generally relevant if the NWA approximation is adopted, while its role is almost negligible if the full signal is considered. Finally, we have evaluated the performance of a set of ATLAS and CMS searches at both 8 and 13 TeV in the determination of the excluded region in the $(M_T,\Gamma_T/M_T)$ plane. We found that the SRs which are most relevant for the determination of the constraints are weakly sensitive to the $T$ width if the $T$ mixes with the SM top quark, while they can pose higher mass bounds (with respect to the NWA limits) if the $T$ mixes with the up quark. 

\vspace*{-0.5mm}

To summarise, we showed that it is not possible to trivially rescale the mass bounds for VLQs decaying to SM states obtained considering processes of pair production and decay in the NWA to determine constraints for VLQ with large widths. Further, given the weak dependence on the $T$ width of a large set of SRs of 8 TeV ATLAS and CMS analyses, and also of SRs from a dedicated 13 TeV ATLAS analysis~\cite{TheATLAScollaboration:2016gxs} looking at pair production of VLQs $T$, we think that designing different SRs in experimental analyses to explore the large width regime by taking into account the full kinematical properties of the signal is advisable for a more comprehensive search of heavy quarks at the LHC. A prerequisite to this is to dismiss at MC generation level both the NWA (which leads to severe mis-estimates) and a naive generalisation to a FW approach using the same topologies as in the NWA (which is potentially strongly gauge dependent) in favour of a full determination of every contribution (off-shellness and new topologies) to the signal.

%% ------------- --------------------- --------------------- ---------------------- --------------------- ---------------------- ----------------------
%% Chapter IV
%% ------------- --------------------- --------------------- ---------------------- --------------------- ---------------------- ---------------------- 

\chapter{New quark decaying to Dark Matter} \label{Chapter:XQDM}

In this Chapter we will focus on (chiral or VL) XQ decaying to a DM candidate and a SM quark, and we will only consider a heavy top partner $T$. The different possibilities for its decay are therefore $T \to S^0_{\rm DM} t$ and $T \to V^0_{\rm DM} t$.

%% ------------- --------------------- --------------------- ---------------------- --------------------- ---------------------- ----------------------

%\section{XQCATDM} \label{sec:XQCATDM}

%% ------------- --------------------- --------------------- ---------------------- --------------------- ---------------------- ----------------------

\section{Study of $t \bar{t} + \MET$ searches for SUSY and XQ scenarios} \label{sec:VLQ-SUSY}

If this new physics is responsible for the DM of the universe in the form of weakly interacting massive particles, its signatures at the LHC and other future colliders are expected to be characterized by events with an excess of missing transverse energy, $\MET$. An intense experimental effort is thus being made at the LHC to isolate such signatures, though no signal has been observed so far.\footnote{Of course, $\MET$ signatures cannot be univocally associated with the production of DM. Neutral long-lived particles which decay outside the detector would produce the very same signatures without being DM. However, the observation of a signature compatible with DM at the LHC would allow to focus on specific regions of the parameter space to be corroborated by other observations, like DM direct and/or indirect detection.} 

The prototype for a new physics model leading to $\MET$ signatures is R-parity conserving SUSY, in particular the minimal supersymmetric standard model (MSSM) with a neutralino as the lightest supersymmetric particle \cite{Martin:1997ns,Drees:2004jm,Baer:2006rs}.  
Indeed, a large number of searches for final states containing jets and/or leptons plus $\MET$ have been designed by the ATLAS and CMS SUSY groups~\cite{ATLAS:2012hpa}, and the interpretations of the results are typically limits in some SUSY simplified model. 
Examples are multi-jet + $\MET$ searches being interpreted as limits in the the gluino--neutralino mass plane, or searches for the $t\bar t+\MET$ final state being interpreted in terms of stops decaying to top+neutralino. 

The same searches can be used to put constraints on scenarios leading to final states with $\MET$ generated by the production of XQs decaying to a bosonic DM candidate. A common feature of these models is that the new states have the same spin as their SM partners, while in SUSY the spins differ by half a unit.

In these XQ models, the lightest odd particle is a DM candidate which interacts with the SM states through new mediator particles. A crucial property of scenarios where the mediators are odd is that they can only be produced in pairs or in association with other odd particles. This is then followed by (cascade) decays into SM particles and the DM candidate. Since the spins in the decays are all correlated, if it was possible to identify the spin of the mediator, this would give information on the bosonic/fermionic nature of the DM candidate as well.

It is therefore interesting to ask how the current results from SUSY searches constrain other models of new physics that would lead to the same signatures, and how same spin and different spin scenarios could be distinguished should a signal be observed. In this section, we concentrate on the first of these questions, comparing the cases of pair production of scalar (SUSY) and fermionic (XQ) top partners with charge 2/3, which decay into $t+{\rm DM}$, thus leading to a $t\bar t + \MET$ final state. Concretely, we consider the processes 
\begin{eqnarray*}
 \text{Top partner with spin 0: } &\quad& pp \to \tilde t\, \tilde t^* \to t \bar t + \tilde\chi^0 \tilde\chi^0 \\
 \text{Top partner with spin 1/2: } &\quad& pp \to T \bar T \to t \bar t + \{S^0 S^0 \text{ or } V^0 V^0\}
\end{eqnarray*}
where $\tilde\chi^0$, $S^0$ and $V^0$ represent fermionic, scalar, and vectorial DM candidates respectively. 
Recasting a number of ATLAS and CMS searches for stops~\cite{ATLAS:2013cma,Aad:2014kra,Chatrchyan:2013xna,Aad:2014qaa} from Run~1 of the LHC, as well as a generic search for gluinos and squarks~\cite{Aad:2014wea} by means of {\sc CheckMATE}~\cite{Drees:2013wra} and {\sc MadAnalysis}\,5~\cite{Conte:2014zja,Dumont:2014tja}, we compare the efficiencies of these searches for the processes above. 
This allows us to determine whether cross section upper limit maps or efficiency maps derived in the context of stop--neutralino  simplified models can safely be applied to XQ scenarios where the $t\bar t + \MET$ final state arises from the production of heavy $T$ quarks. Such maps are used in public tools like {\sc SModelS}~\cite{Kraml:2013mwa,Kraml:2014sna} and {\sc XQCAT}~\cite{Barducci:2014ila,Barducci:2014gna}, and it is relevant to know how generically they can be applied. 
Moreover, we determine up-to-date bounds in the parameter space of the XQ and DM masses -- such bounds were posed by a few early searches at the Tevatron \cite{Aaltonen:2011rr, Aaltonen:2011na} and the LHC at 7~TeV \cite{ATLAS:2011mda, CMS:2012dwa}, but can be improved by a reinterpretation of the 8~TeV LHC results as we do in this section following the approach of \cite{Kraml:2016eti}.

Related studies exist in the literature and were already mentioned in Sec. \ref{sec:searches}. Here, we extend these works by considering specifically top partners and by applying up-to-date recasting tools.

%==============================================================================
\subsection{Benchmark scenarios}
\label{sec:models}
%==============================================================================

%-------------------------------------------------------------------------------
\subsubsection{The SUSY case: stop--neutralino simplified model}
%-------------------------------------------------------------------------------

The prototype for the $t\bar t + E_T^{\rm miss}$ signature in the SUSY context is a stop--neutralino simplified model. 
This assumes that the lighter stop, $\tilde t_1$, and the lightest neutralino, $\tilde\chi^0_1$, taken to be the lightest SUSY particle and the DM candidate, are the only accessible sparticles --- all other sparticles are assumed to be heavy. 
In this case, direct stop pair production is the only relevant SUSY production mechanism. Moreover, for large enough mass difference, the $\tilde t_1$ decays to 100\% into $t+\tilde\chi^0_1$. The process we consider thus is 
\begin{equation}
  pp\to \tilde t_1^{} \tilde t_1^* \to t\bar t \tilde\chi^0_1\tilde\chi^0_1 \,.
\end{equation}
Following the notation of \cite{Gajdosik:2004ed}, the top--stop--neutralino interaction is given by  ($i=1,2$; $k=1,...,4$)
\begin{align}
  {\cal L}_{t\st\nt}  
   &= g\,\bar t\,( f_{Lk}^{\st}\PR + h_{Lk}^{\st}\PL )\,\nt_k\,\st_L^{} +
      g\,\bar t\,( h_{Rk}^{\st}\PR + f_{Rk}^{\st}\PL )\,\nt_k\,\st_R^{} + {\rm h.c.} \nonumber \\
   & = g\,\bar t\,( a^{\,\st}_{ik}\PR + b^{\,\st}_{ik}\PL )\,\nt_k\,\st_i^{} + {\rm h.c.}
  \label{eq:LagSUSYSMS}
\end{align}
where $P_{R,L}^{}  = \frac{1}{2}(1\pm\gamma_5)$ are the right and left projection operators,  and 
\begin{align}
   a^{\,\st}_{ik} &= f_{Lk}^{\st}\,R_{i1}^{\st} +
                      h_{Rk}^{\st}\,R_{i2}^{\st}\,,  \nonumber \\ % \label{eq:aik}\\
   b^{\,\st}_{ik} &= h_{Lk}^{\st}\,R_{i1}^{\st} +
                      f_{Rk}^{\st}\,R_{i2}^{\st}\,.   \label{eq:bik}
\end{align}
The $f_{L,R}^{\st}$ and $h_{L,R}^{\st}$ couplings are
\begin{align}
  f_{Lk}^{\,\ti t} &= -\sfrac{1}{\sqrt 2}\,(N_{k2} +\sfrac{1}{3}\tan\theta_W N_{k1}) \,, \nonumber\\
  f_{Rk}^{\,\ti t} & = \sfrac{2\sqrt 2}{3}\,\tan\theta_W N_{k1}\,, \hspace{24mm} % \nonumber\\
  h_{Rk}^{\ti t} = -y_t\, N_{k4} = h_{Lk}^{\ti t*}\,, 
\end{align}
with $N$ the neutralino mixing matrix and $y_t=m_t/(\sqrt{2}m_W\sin\beta)$ the top Yukawa coupling in the MSSM. 
Finally, $R$ is the stop mixing matrix,  
\begin{equation}
  {\tilde t_1 \choose \tilde t_2}  = R \, {\tilde t_L \choose \tilde t_R} \,, \quad 
  R =  \left(\begin{array}{rr} \cos\theta_{\tilde t} & \sin\theta_{\tilde t} \\ -\sin\theta_{\tilde t} & \cos\theta_{\tilde t} \end{array}\right) \,. \label{eq:stopMixing}
\end{equation}
All this follows SLHA \cite{Skands:2003cj} conventions. 

Under the above assumption that all other neutralinos besides the $\tilde\chi^0_1$ and the charginos are heavy, the $\tilde\chi^0_1$ is dominantly a bino.  Neglecting the wino and higgsino components $N_{12}$ and $N_{14}$, the $t\st_1\nt_1$ interaction from Eq. \eqref{eq:LagSUSYSMS}  simplifies to 
\begin{align}
  {\cal L}_{t\st_1\nt_1}  
  & \approx - \frac{g}{3\sqrt{2}} \tan\theta_W N_{11} \, \bar t 
   \left( \cos\theta_{\tilde t}\, P_R^{} - 4 \sin\theta_{\tilde t}\, P_L^{} \right) \tilde\chi^0_1\, \tilde t_1^{} + {\rm h.c.}\,.
  \label{eq:LagSUSYSMSsimple}
\end{align}
While in practice one never has a {\em pure} bino,  this approximation shows that the polarisation of the tops originating from the $\tilde t_1\to t\tilde\chi^0_1$ decays will reflect the chirality of the $\tilde t_1$\footnote{Here we are talking about the "chirality" of a scalar which is a language abuse, we actually refer to $\tilde t_{L,R}$ as defined in eq. \eqref{eq:stopMixing}.}.
(The wino interaction also preserves the chirality, while the higgsino one flips it.)
This will be relevant for defining XQ benchmark scenarios analogous to SUSY ones, since the $p_T$ and angular distributions of the top decay products somewhat depend on the top polarisation~\cite{Cao:2006wk,Nojiri:2008ir,Shelton:2008nq,Perelstein:2008zt,Berger:2012an,Chen:2012uw,Bhattacherjee:2012ir,Belanger:2012tm,Low:2013aza,Belanger:2013gha,Wang:2015ola}.

%-------------------------------------------------------------------------------
\subsubsection{The extra quark scenario: conventions and Lagrangian terms}
%-------------------------------------------------------------------------------

As the XQ analogue of the SUSY case above, we consider a minimal extension of the SM with one XQ state and one DM state, assuming that the XQ mediates the interaction between the DM and the SM quarks of the third generation. Interactions between the XQ, DM and lighter quarks are neglected. 
The most general Lagrangian was already presented in Sec. \ref{sec:XQDM Lagrangian}. Here we are only interested in a $T$ coupling to a singlet DM and the top quark for the following study so our Lagrangians takes the following expression:
\begin{eqnarray}
\Lag^S_1 &=& \lambda_{11}^t \bar{T} P_R^{} t S^0_{\rm DM} + {\rm h.c.} 
\label{eq:LagSingletDMST3}
\\
\Lag^V_1 &=& g_{11}^t \bar{T} \gamma_\mu P_R^{} t V^{0\mu}_{\rm DM} + {\rm h.c.} 
\label{eq:LagSingletDMVT3}
\end{eqnarray}

%-------------------------------------------------------------------------------
\subsubsection{Benchmark points}
%-------------------------------------------------------------------------------

In order to compare the XQ and SUSY scenarios, it is useful to consider benchmark points with the same top partner and DM masses as well as the same left and right couplings (leading to $t_L$ or $t_R$ in the final state) for the two models. To this end, we start from the stop--neutralino simplified model and  choose two mass combinations: 
$(m_{\tilde t_1},\,m_{\tilde\chi^0_1})=(600,\,10)$~GeV and $(m_{\tilde t_1},\,m_{\tilde\chi^0_1})=(600,\,300)$~GeV. 
The first one is excluded by the 8~TeV searches, while the second one lies a bit outside the 8~TeV bounds~\cite{Aad:2015pfx,CMS:2014yma,Chatrchyan:2013xna,Chatrchyan:2014lfa,CMS:2014wsa}.\footnote{The $(m_{\tilde t_1},\,m_{\tilde\chi^0_1})=(600,\,300)$~GeV mass combination actually lies just on the edge of the new 13~TeV bounds presented by CMS \cite{CMS:2016nhb} at the Moriond 2016 conference.}
Moreover, since the searches for $\st_1\to t\nt_1$ exhibit a small dependence on the top polarisation~\cite{Aad:2014kra}, 
we consider the two cases $\tilde t_1\sim \tilde t_R$ and $\tilde t_1\sim \tilde t_L$.\footnote{Strictly speaking, because of SU(2), a $\tilde t_1\sim \tilde t_L$ should be accompanied by a $\tilde b_L$ of similar mass; with no other 2-body decay being kinematically open, the sbottom would however decay to 100\% into $b\nt_1$and thus not contribute to the $t\bar t+\MET$signature.} The results for arbitrary stop mixing (or top polarisation) will then always lie  between these two extreme cases. This leads to four benchmark scenarios, which we denote by 
\begin{equation*}
   \rm  (600,\,10)L\,;\quad (600,\,10)R \,;\quad (600,\,300)L \,; \quad (600,\,300)R \,. 
\end{equation*}
The strategy then is to use the same mass combinations $(m_T,\,m_{\rm DM})$ and left/right couplings for the XQ case. 
For XQ$+S^0_{\rm DM}$, we directly use  $\lambda_{11}^t=b_{11}^{\st}$ and $\lambda_{21}^t=a_{11}^{\st}$. 
For XQ$+V^0_{\rm DM}$, however, the width of the XQ would be too large if we were using the same parameters as in the SUSY or scalar DM case; to preserve the NWA, we therefore reduce the couplings by a factor 10, i.e.\ $g_{11}^t=b_{11}^{\st}/10$ and $g_{21}^t=a_{11}^{\st}/10$. The concrete values for the different benchmark scenarios are listed in Table~\ref{tab:BPlist}. 

\begin{table}
\scriptsize
\centering
\begin{tabular}{c| rr | rr}
\toprule
& \multicolumn{2}{c|}{\bf (600,\,10)L} & \multicolumn{2}{c}{\bf (600,\,300)L} \\
$\tilde t_1 \sim \tilde t_L$ & 
$a_{11}^{\st} = -8.3649\;10^{-2}$ & $b_{11}^{\st} = 1.5406\;10^{-3}$ & 
$a_{11}^{\st}  = -8.3638\;10^{-2}$ & $b_{11}^{\st} = 2.5811\;10^{-3}$ \\
XQ + $S^0_{\rm DM}$ & 
$\lambda_{21}^t = -8.3649\;10^{-2}$ & $\lambda_{11}^t = 1.5406\;10^{-3}$ & 
$\lambda_{21}^t = -8.3638\;10^{-2}$ & $\lambda_{11}^t = 2.5811\;10^{-3}$ \\ 
XQ + $V^0_{\rm DM}$ & 
$g_{21}^t = -8.3649\;10^{-3}$ & $g_{11}^t = 1.5406\;10^{-4}$ & 
$g_{21}^t = -8.3638\;10^{-3}$ & $g_{11}^t = 2.5811\;10^{-4}$ \\
\midrule
& \multicolumn{2}{c|}{\bf (600,\,10)R} & \multicolumn{2}{c}{\bf (600,\,300)R} \\
$\tilde t_1 \sim \tilde t_R$ &
$a_{11}^{\st} = 1.1425\;10^{-3}$ & $b_{11}^{\st} = 3.3467\;10^{-1}$ &
$a_{11}^{\st} = 2.1823\;10^{-3}$ & $b_{11}^{\st} = 3.3466\;10^{-1}$ \\
XQ + $S^0_{\rm DM}$ & 
$\lambda_{21}^t = 1.1425\;10^{-3}$ & $\lambda_{11}^t = 3.3467\;10^{-1}$ & 
$\lambda_{21}^t = 2.1823\;10^{-3}$ & $\lambda_{11}^t = 3.3466\;10^{-1}$ \\
XQ + $V^0_{\rm DM}$ & 
$g_{21}^t = 1.1425\;10^{-4}$ & $g_{11}^t = 3.3467\;10^{-2}$ & 
$g_{21}^t = 2.1823\;10^{-4}$ & $g_{11}^t = 3.3466\;10^{-2}$ \\
\bottomrule
\end{tabular}
\caption{Benchmark points for the SUSY and XQ scenarios.}
\label{tab:BPlist}
\end{table}

The alert reader will notice that in Table~\ref{tab:BPlist}, although there is a strong hierarchy between the left and right couplings, both of them are non-zero. Moreover, the couplings for the (600,\,300)L case are not the same as for the (600,\,10)L case;  
the same is true for (600,\,300)R vs.\ (600,\,10)R. 
The reason for this is as follows. The pure left or pure right case, $\tilde t_1\equiv\tilde t_L$ or $\tilde t_R$, would require that the off-diagonal entry in the stop mixing matrix is exactly zero, that is $A_t\equiv\mu/\tan\beta$, where $A_t$ is the trilinear stop-Higgs coupling, $\mu$ is the higgsino mass parameter and $\tan\beta=v_2/v_1$ is the ratio of the Higgs VEVs. To avoid such tuning, and also because the $\tilde\chi^0_1$ will never be a 100\% pure bino even if the winos and higgsinos are very heavy, we refrain from using the approximation of Eq.~\eqref{eq:LagSUSYSMSsimple} with $N_{11}=1$ and $\cos\theta_{\tilde t}=1$ or $0$.  Instead, we choose the masses of the benchmark points as desired by appropriately adjusting the relevant soft terms while setting all other soft masses to 3--5 TeV. From this we then compute the stop and neutralino mixing matrices and the full $\tilde\chi^0_1\tilde t_1t$ couplings $a_{11}^{\st}$ and $b_{11}^{\st}$ of of Eq.~\eqref{eq:LagSUSYSMS}, using 
{\sc{SuSpect}}~v2.41~\cite{Djouadi:2002ze}. 
The resulting values are $N_{11}\simeq 1$, $\cos\theta_{\tilde t}\simeq 1$ (or $\sin\theta_{\tilde t}\simeq 1$) to sub-permil precision, but nonetheless this leads to a small non-zero value of the ``other'' sub-dominant coupling, and to a slight dependence on the $\tilde\chi^0_1$ mass. 
An interesting consequence is that since we started by defining our SUSY benchmark points so that they are consistent and non-excluded by the current searches and we then used the same couplings for our XQ benchmark points, our comparison between SUSY and XQ is effectively between SUSY and ChQ scenarios because both chiralities of the new particles are non-zero. A comparison between SUSY and VLQ scenarios would require $\tilde t_1 \equiv \tilde t_L$ or $\tilde t_1 \equiv \tilde t_R$.
Our conclusions however do not depend on this.

%==============================================================================
\subsection{Monte Carlo event generation}
\label{sec:MC}
%==============================================================================

%-------------------------------------------------------------------------------
\subsubsection{Setup and tools}
%-------------------------------------------------------------------------------

For the MC analysis, we simulate the $2\to 6$ process 
$$pp\to t\,\bar t~{\rm DM~DM}\to (W^+b)(W^-\bar b)~{\rm DM~DM}$$ with {\sc MadGraph}\,5~\cite{Alwall:2011uj,Alwall:2014hca}, where $\rm DM$ is the neutralino in the SUSY scenario or the scalar/vector boson in the XQ scenario.
This preserves the spin correlations in the  $t\to Wb$  decay. 
Events are then passed to {\sc Pythia}\,6~\cite{Sjostrand:2006za}, which takes care of the decay $W \to 2f$ as well as hadronisation and parton showering.\footnote{In \cite{Boughezal:2013pja} it was argued that certain kinematic distributions show sizeable differences between LO and NLO, which can be ameliorated by including initial state radiation of extra jets. We tested this but did not find any relevant differences with and without simulating extra jets for the analyses we consider in this paper.  We therefore conclude that LO matrix element plus parton showering is sufficient for the scope of this study, in particular as it saves a lot of CPU time.}

For the SUSY scenarios we make use of the MSSM model file in {\sc MadGraph}, while for the XQ simulation we implemented the model in {\sc Feynrules}~\cite{Alloul:2013bka} to obtain the UFO model format to be used inside {\sc MadGraph}.  
For the PDFs we employ the cteq6l1 set~\cite{Pumplin:2002vw}. 
To analyse and compare the effects of various ATLAS and CMS 8~TeV analyses, we employ {\sc CheckMATE}~\cite{Drees:2013wra} as well as {\sc MadAnalysis}\,5~\cite{Conte:2014zja}.
Both frameworks use {\sc Delphes\,3}~\cite{deFavereau:2013fsa} for the emulation of detector effects.

\begin{figure}[ht!]
\centering \includegraphics[width=0.95\textwidth]{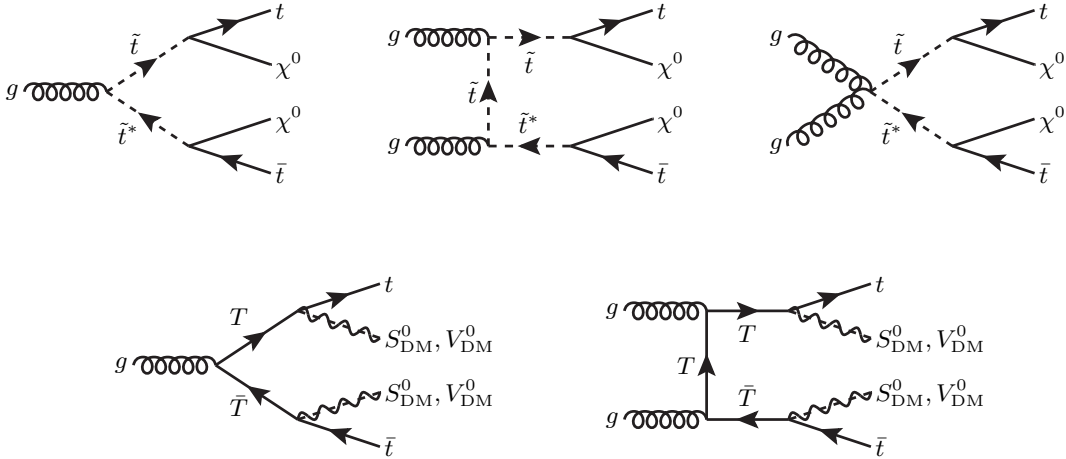}
\caption[Feynman diagrams for the production of $t\bar t + \MET$ in the SUSY and XQ scenarios.]{Feynman diagrams for the production of $t\bar t + \MET$ in the SUSY and XQ scenarios. We have omitted for simplicity the $gg$ and $q \bar q$ initial states which are common for the s-channel gluon topologies.}
\label{fig:topologies}
\end{figure}

\begin{figure}[ht!]
\centering \includegraphics[width=0.7\textwidth ,angle=270]{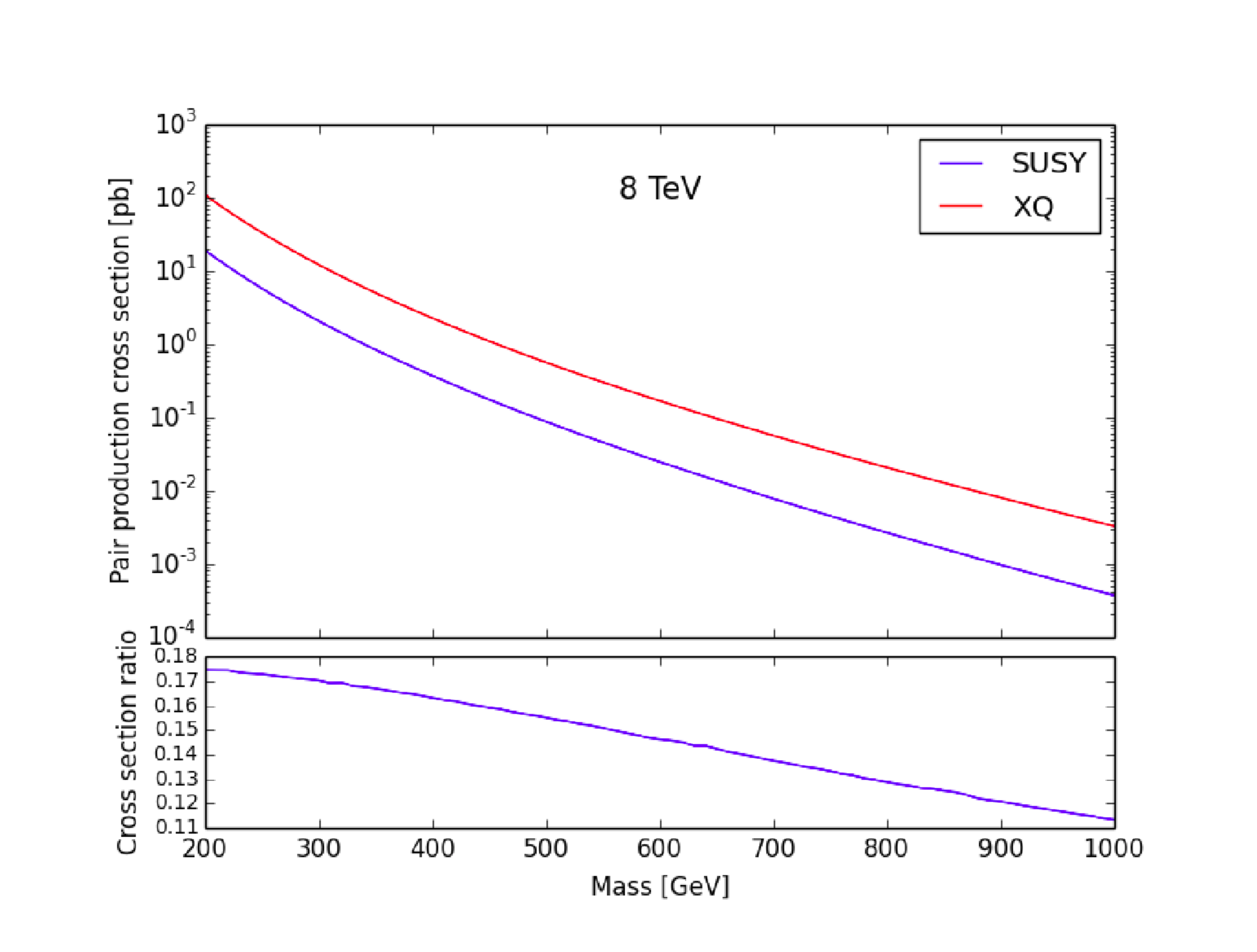} 
\caption{Production cross sections for SUSY and XQ top partners at $\sqrt{s}=8$~TeV.}
\label{fig:XS}
\end{figure}

The Feynman diagrams relevant for the SUSY and XQ processes are shown in Fig.~\ref{fig:topologies}. We observe that besides the difference in the spin of the mediator and DM, in the SUSY case there is a topology which is not present in the XQ case, namely the 4-leg diagram initiated by two gluons. 
The $pp\to \st_1^{}\st_1^*$ and $pp\to T\bar T$ production cross sections at $\sqrt{s}=8$~TeV are compared in Fig.~\ref{fig:XS}. 
The comparison is done at the highest available order for each scenario, i.e.\ at NLO+NLL for SUSY~\cite{nllfast,Beenakker:1996ch,Kulesza:2008jb,Kulesza:2009kq,Beenakker:2009ha,Beenakker:2011fu,Beenakker:1997ut,Beenakker:2010nq} and at NLO+NNLL for XQ~\cite{Cacciari:2011hy}. We see that, for the same mass, the XQ cross section is about a factor 5--10 larger than the SUSY cross section.  The same experimental analysis targeting $t\bar t+\MET$ will therefore have a significantly higher reach in fermionic (XQ)  than in scalar (SUSY) top partner masses. For instance, an excluded cross section of 20\,fb corresponds to $m_{\st_1}\gtrsim 620$~GeV in the SUSY case but  $m_{T}\gtrsim 800$~GeV in the XQ case. The precise reach will, of course,  depend on the specific cut acceptances in the different models.

%-------------------------------------------------------------------------------
\subsubsection{Generator-level distributions}
%-------------------------------------------------------------------------------

\begin{figure}[h!]\vspace*{4mm}\centering
\includegraphics[width=0.34\textwidth]{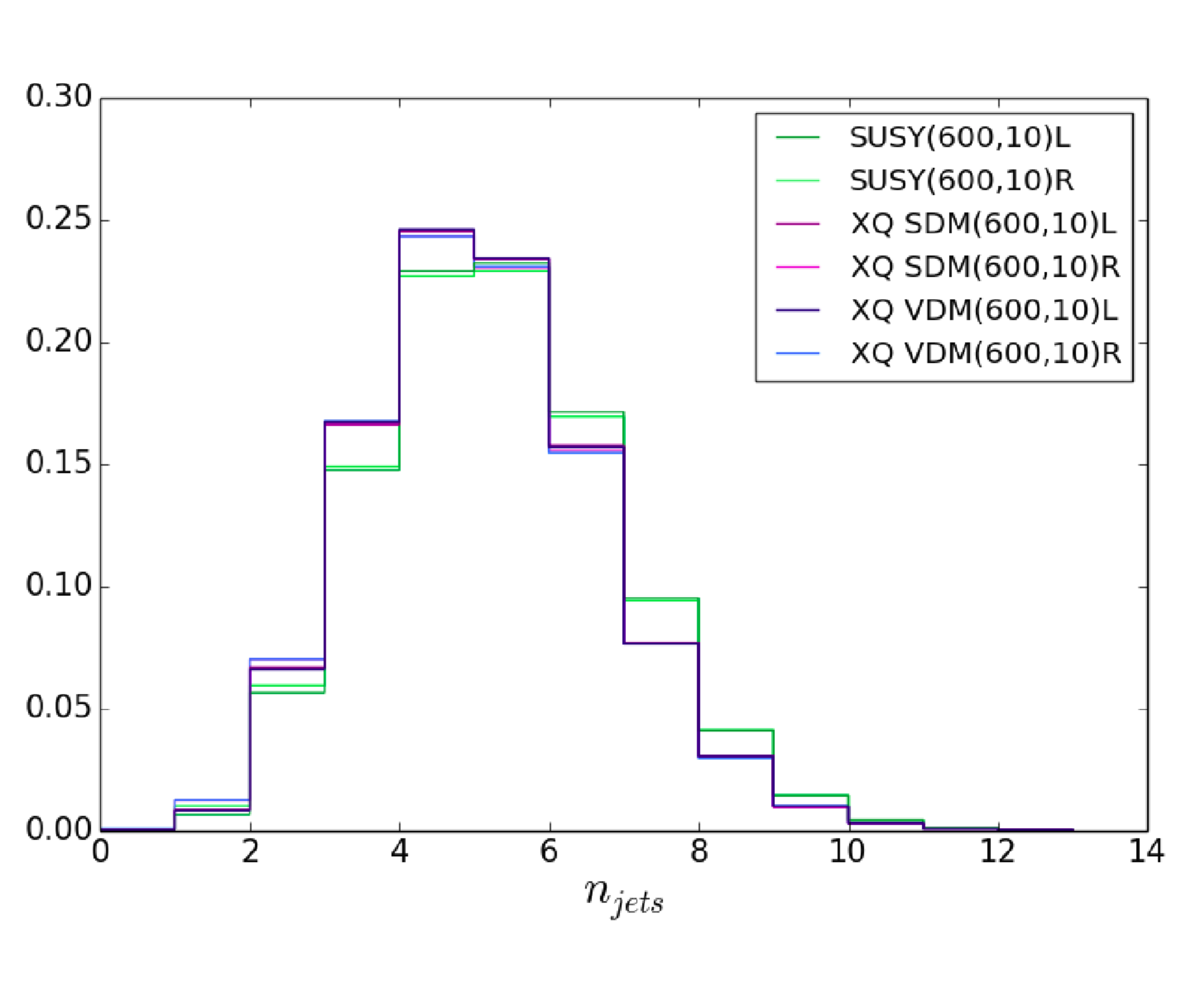}%
\includegraphics[width=0.34\textwidth]{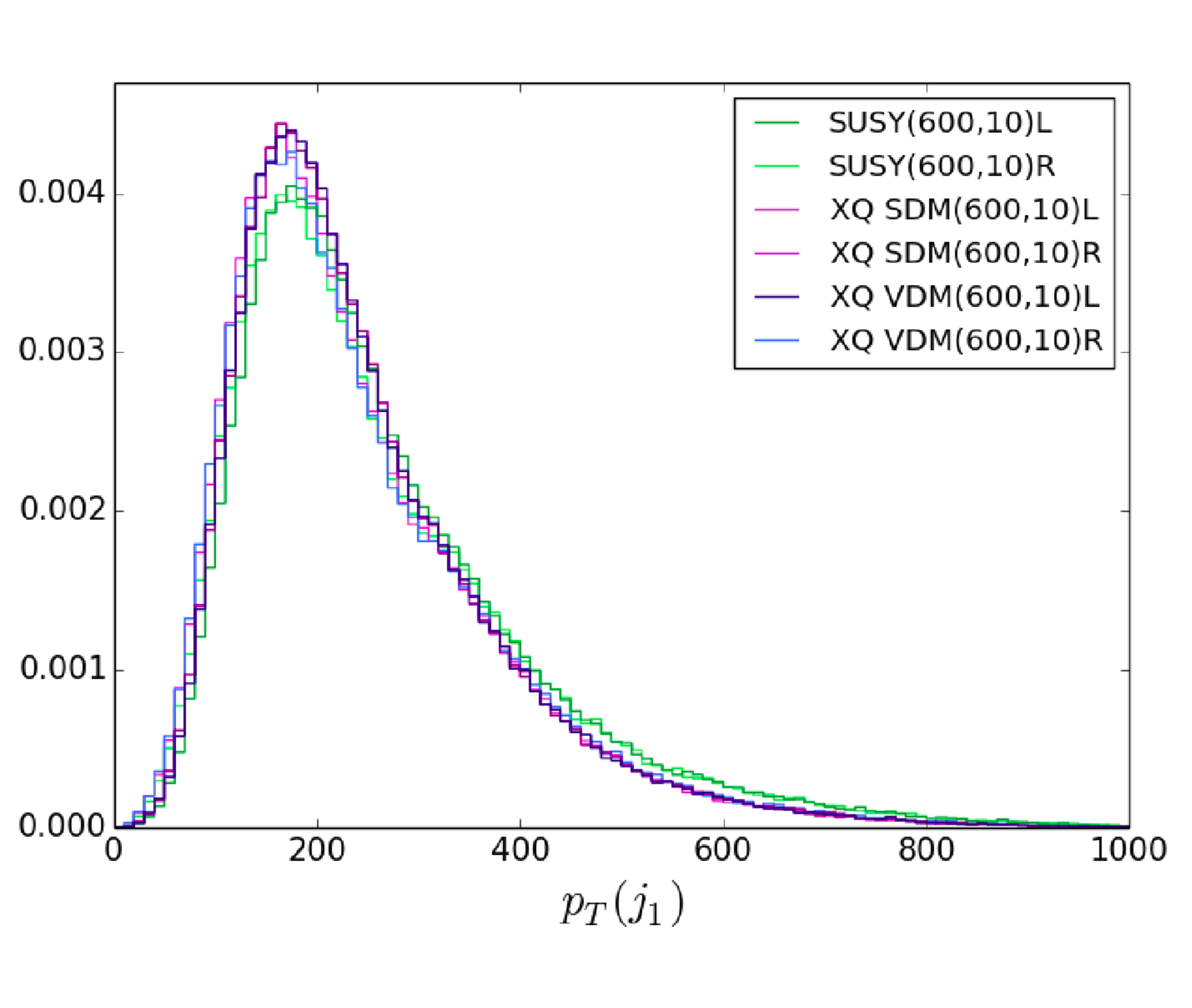}%
\includegraphics[width=0.34\textwidth]{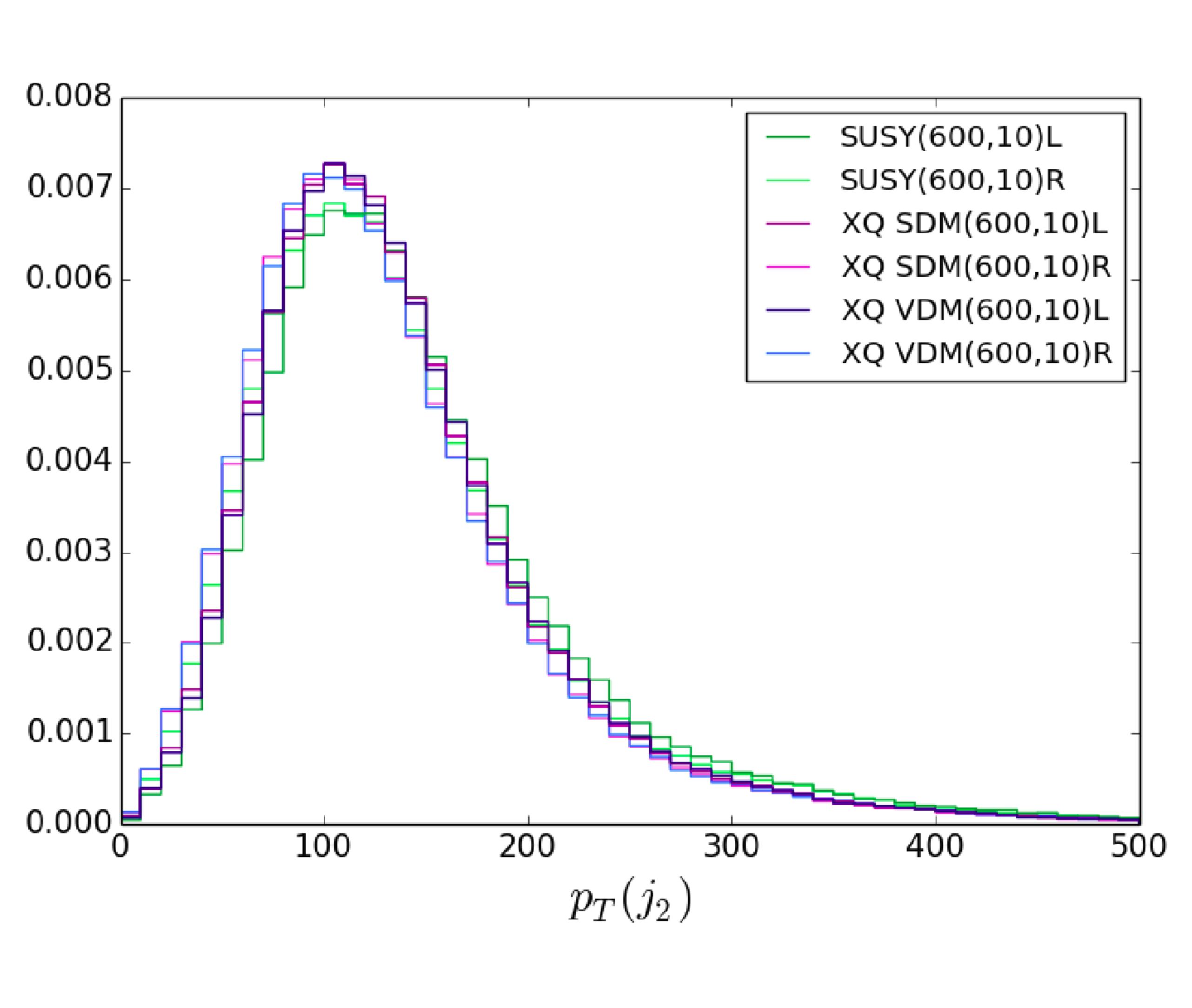}\\
\includegraphics[width=0.34\textwidth]{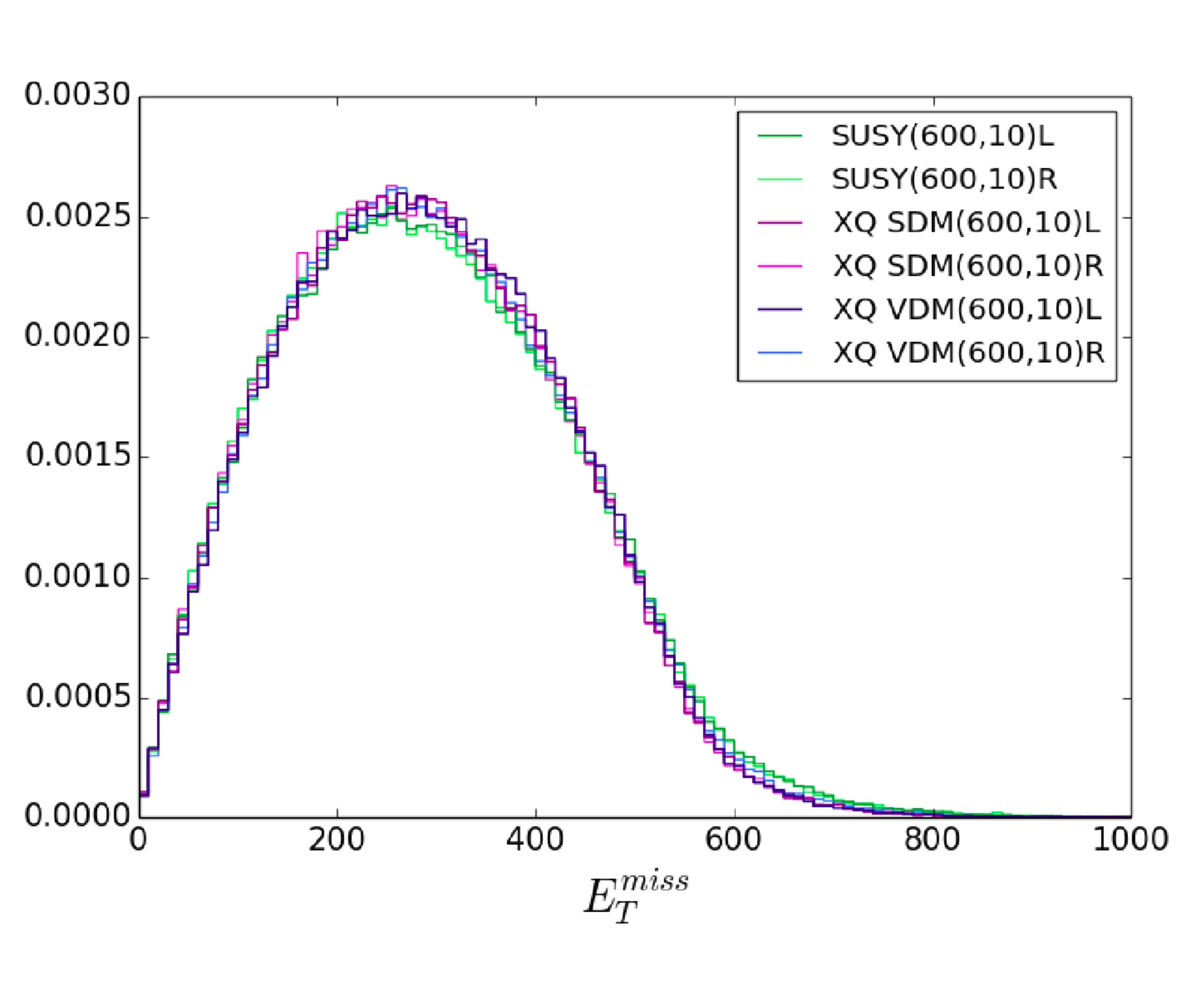}%
\includegraphics[width=0.34\textwidth]{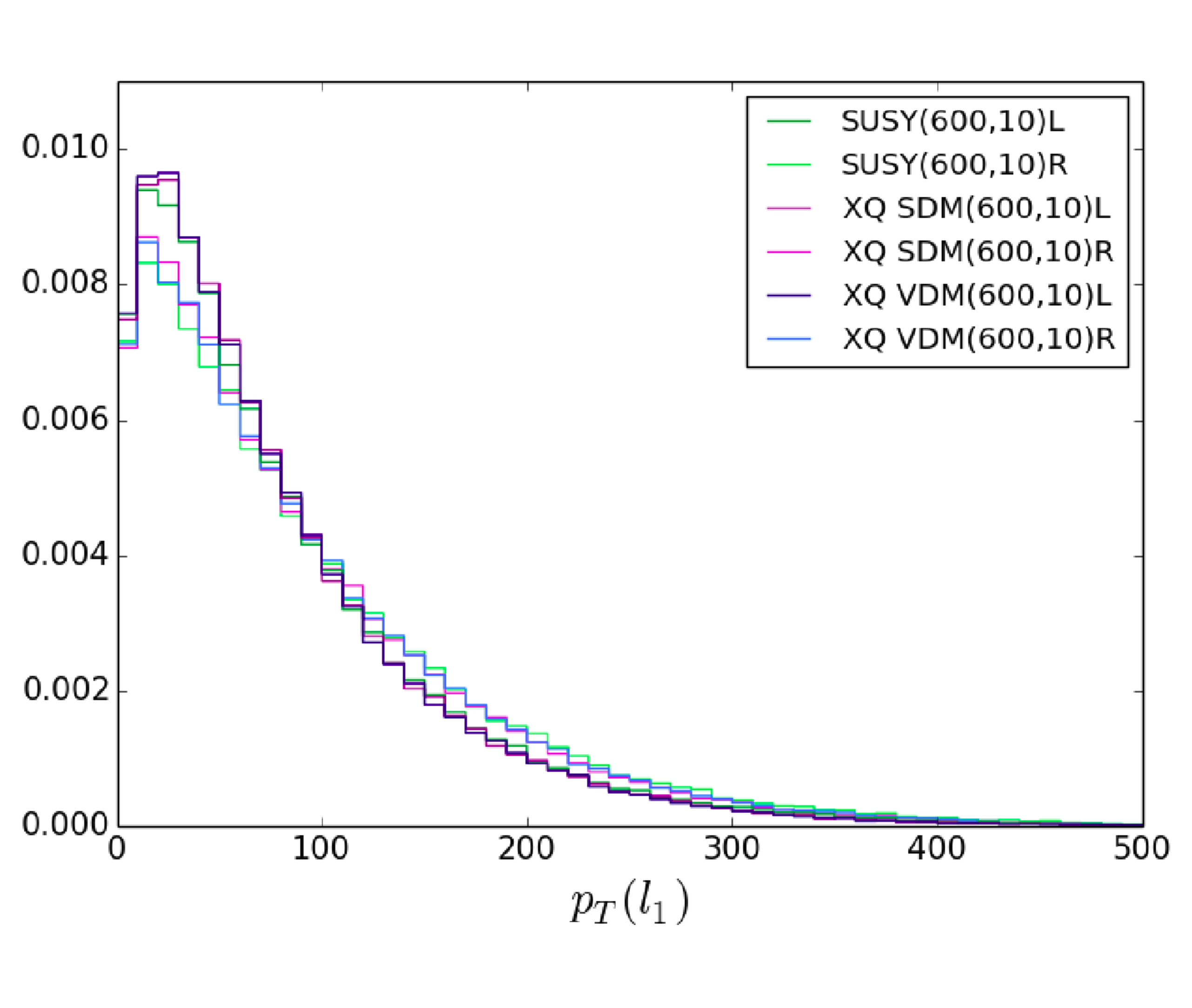}%
\includegraphics[width=0.34\textwidth]{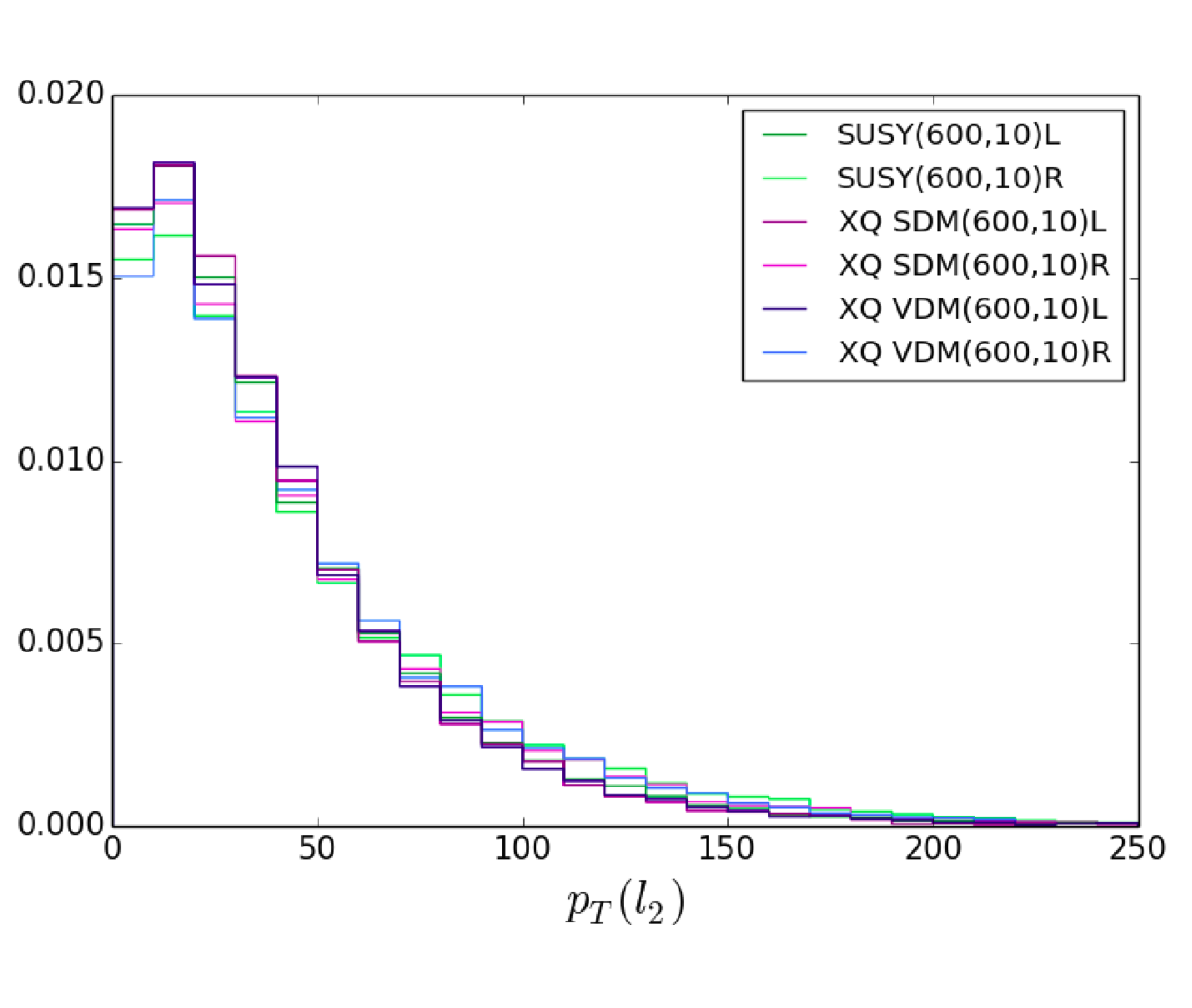}
\caption{Differential distributions (normalized to one) of jet multiplicity $n_{\rm jets}$, transverse momentum of the leading and sub-leading jet $p_T(j_1)$ and $p_T(j_2)$, missing transverse energy $\MET$, and $p_T$ of the leading and sub-leading lepton $p_T(l_1)$ and $p_T(l_2)$ for the mass combination $(600,\,10)$.}
\label{fig:GenLevelDist1}
\end{figure}

As a first check whether we can expect specific differences in the cut efficiencies between the SUSY and XQ models, it is instructive to consider 
some basic parton-level distributions, as shown in Fig.~\ref{fig:GenLevelDist1} for the (600,\,10) mass combination. 
These distributions have been obtained using {\sc MadAnalysis}\,5 and considering the showered and hadronised event files from {\sc Pythia}; jets have been processed through {\sc FastJet}~\cite{Cacciari:2005hq,Cacciari:2011ma} using the anti-kt algorithm with minimum $p_T=5$~GeV and cone radius $R=0.5$. 
We see that the SUSY events tend to have more jets and a slightly harder $\MET$ spectrum. 
Moreover, the leading and sub-leading jets tend to be somewhat harder in the SUSY than in the XQ cases. 
Overall, these differences are however rather small and will likely not lead to any significant differences in the %SUSY and XQ 
cut efficiencies. 

\begin{figure}[h!]
\centering
\includegraphics[width=0.34\textwidth]{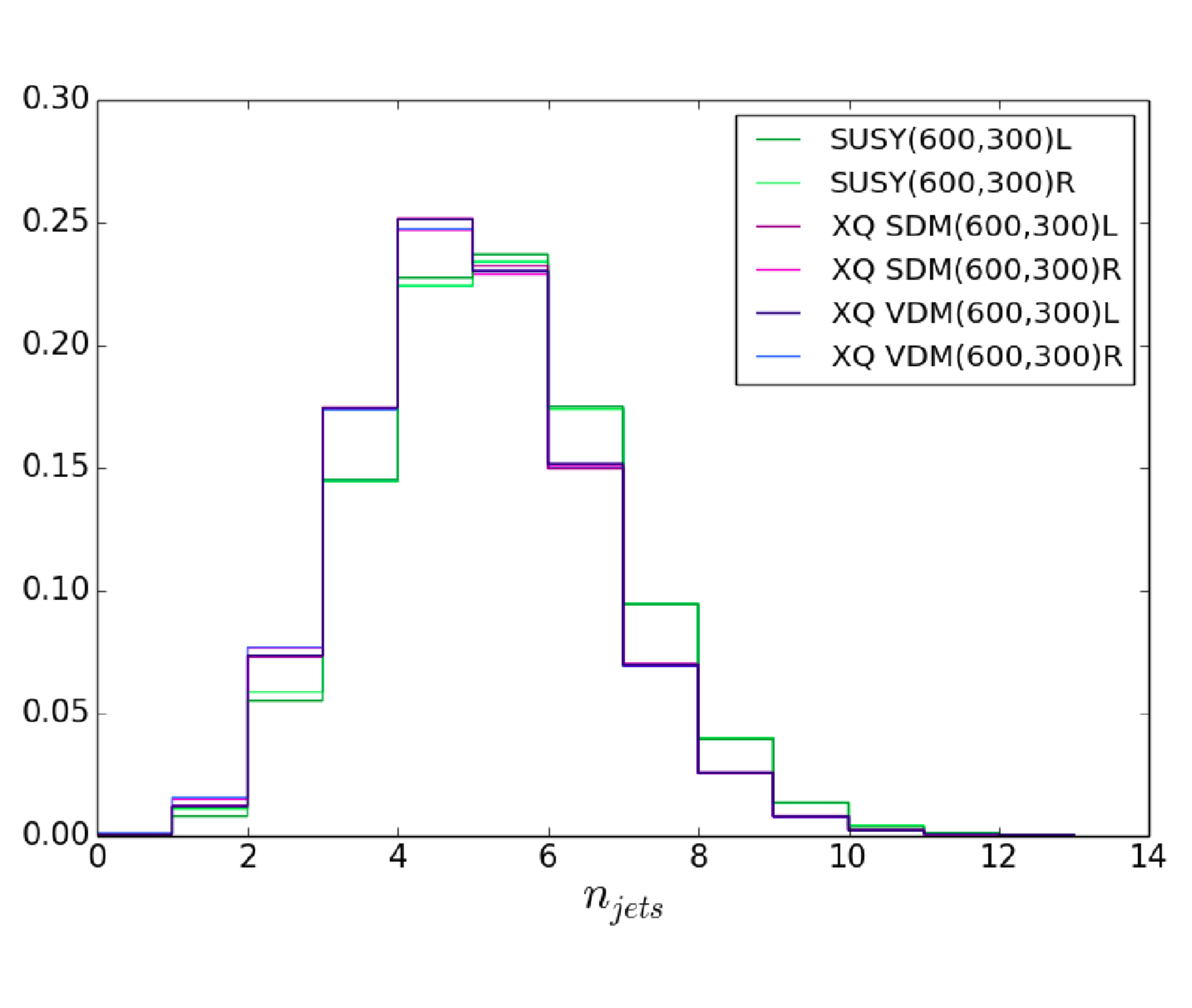}%
\includegraphics[width=0.34\textwidth]{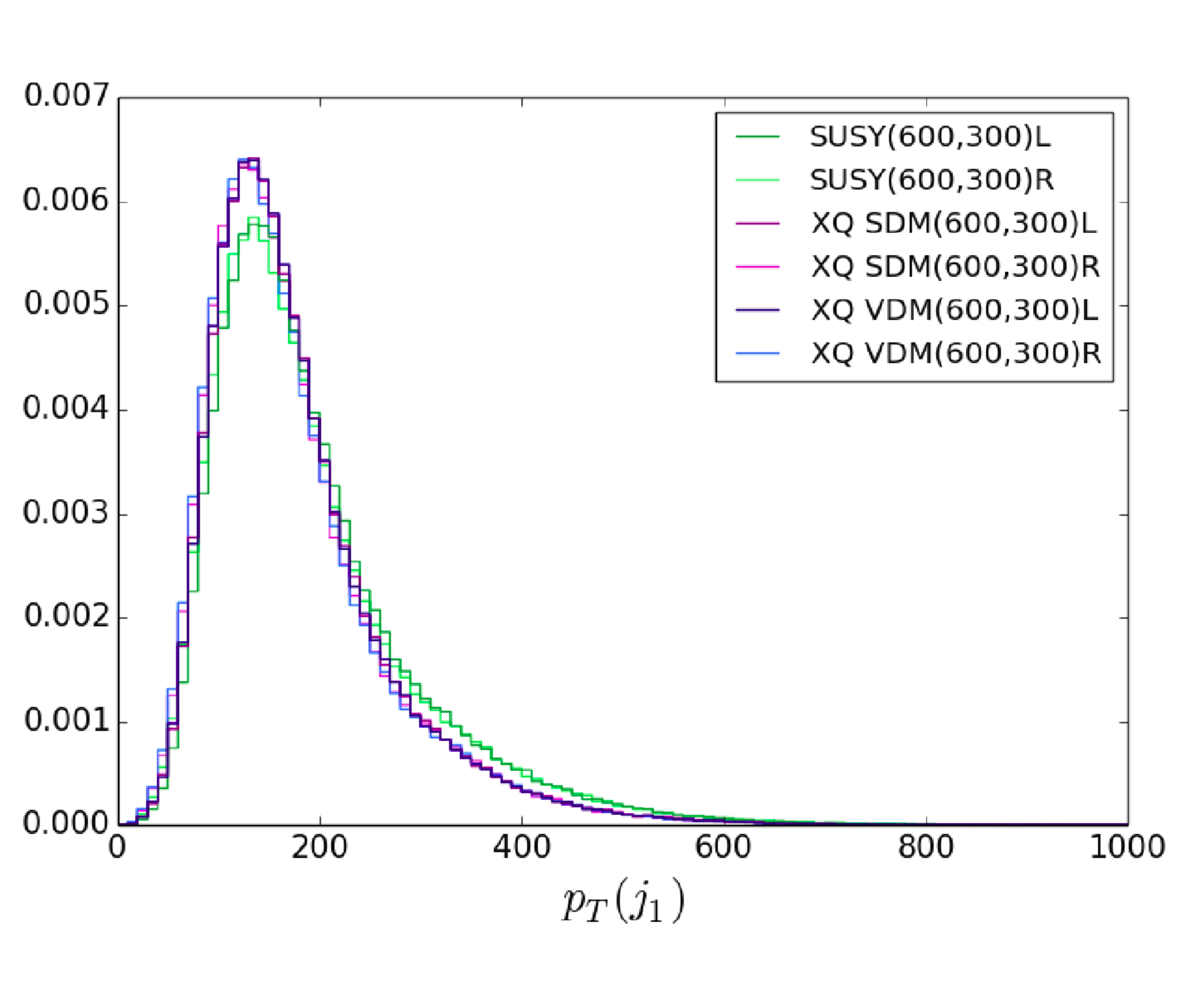}%
\includegraphics[width=0.34\textwidth]{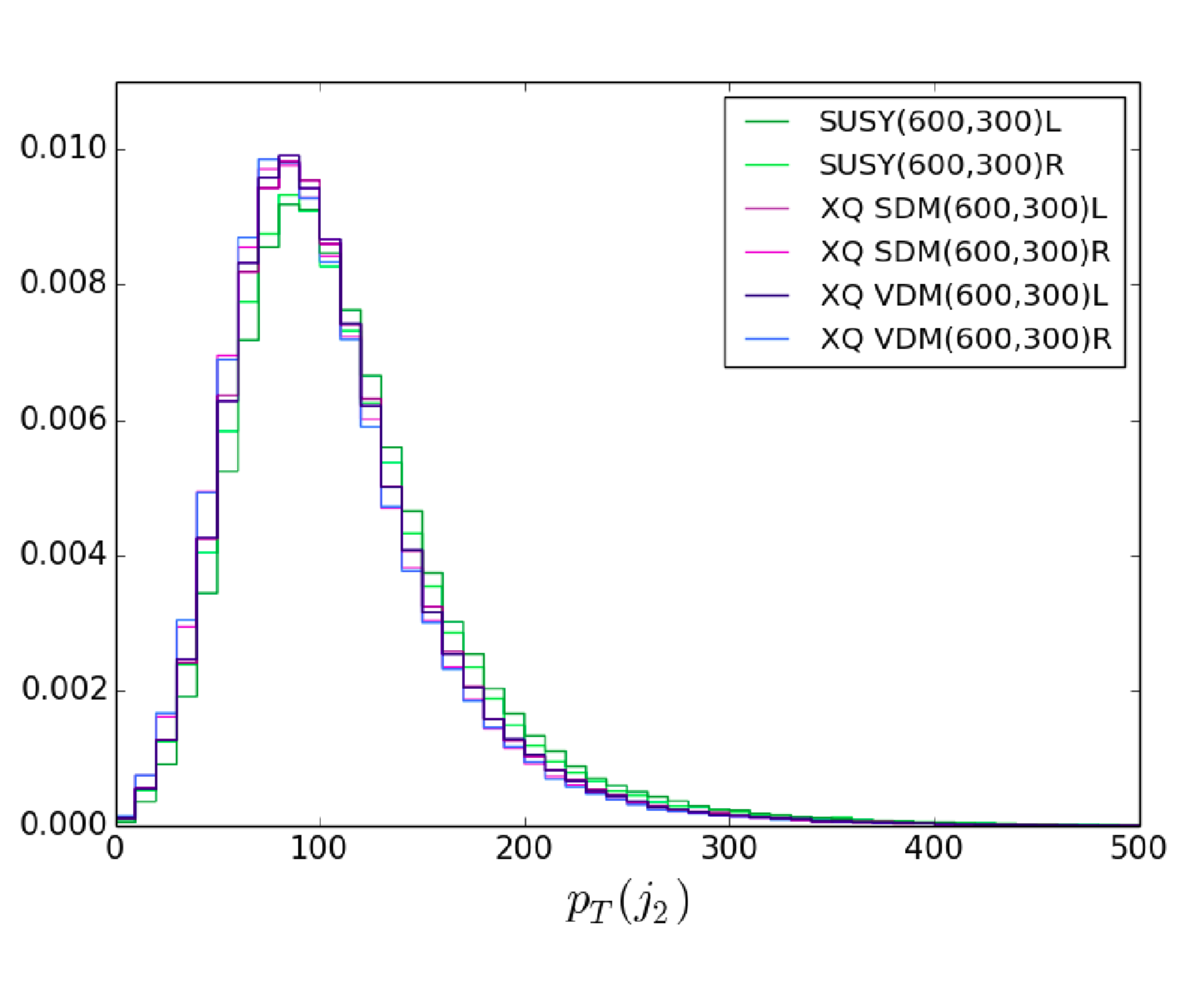}\\
\includegraphics[width=0.34\textwidth]{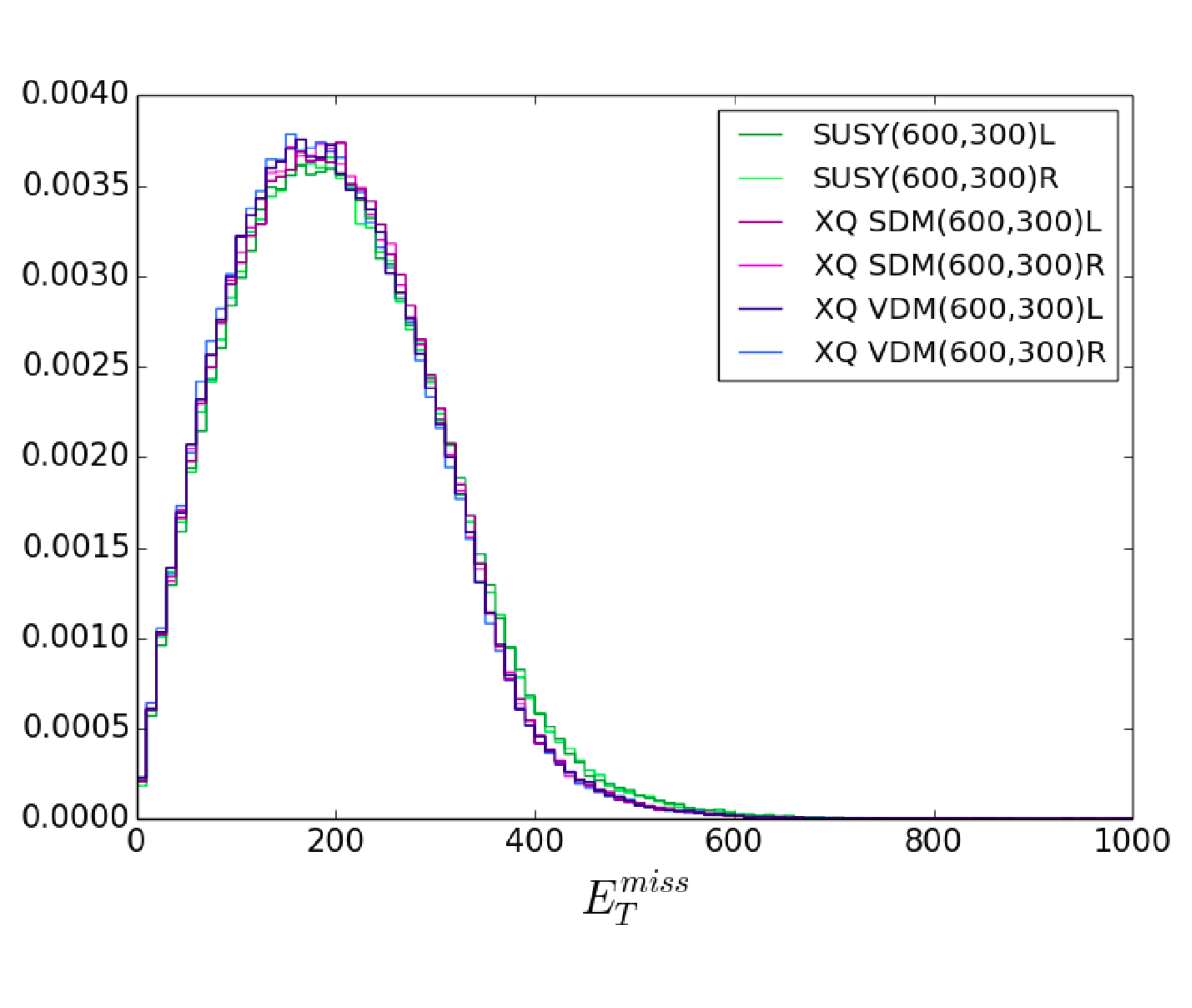}%
\includegraphics[width=0.34\textwidth]{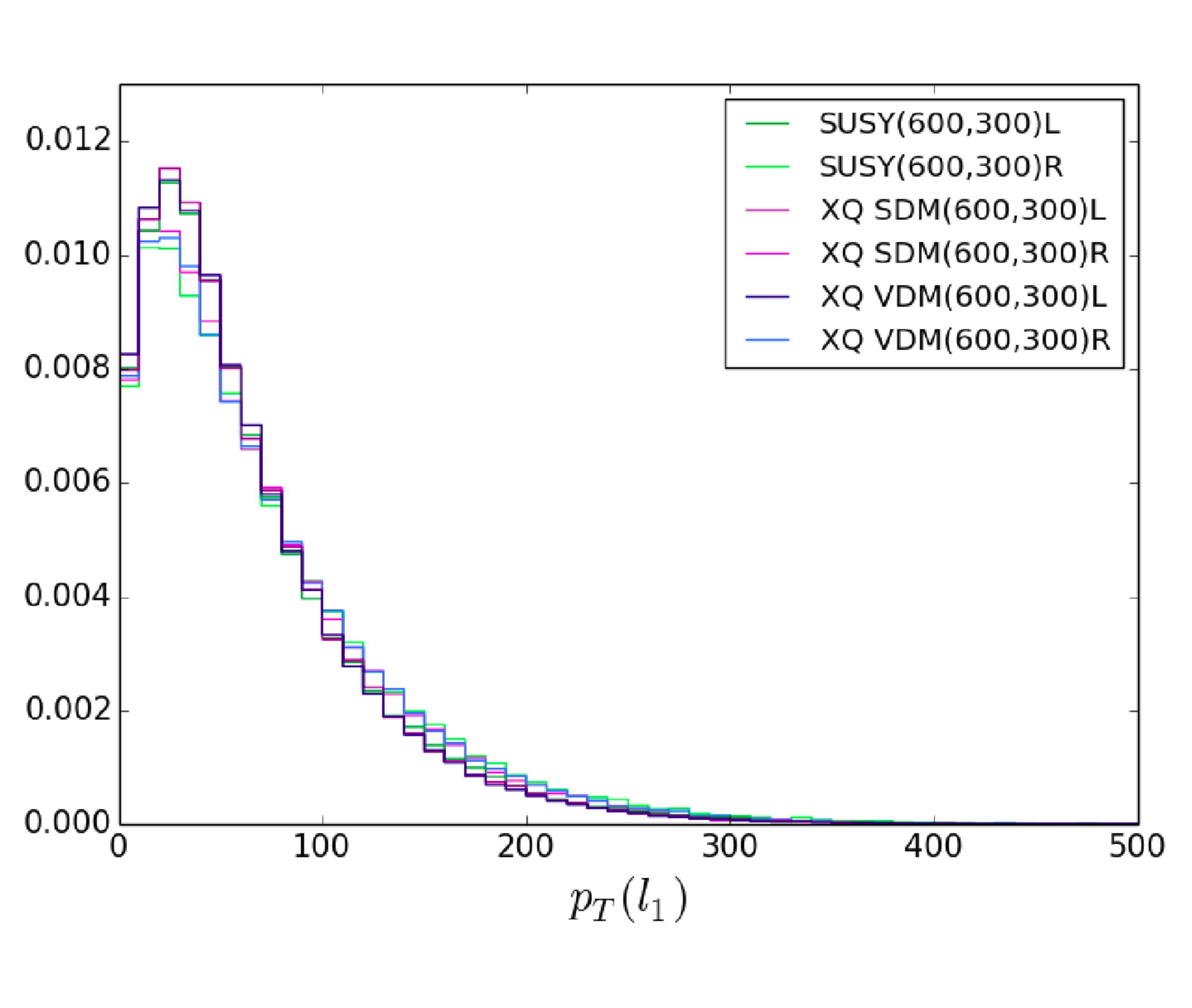}%
\includegraphics[width=0.34\textwidth]{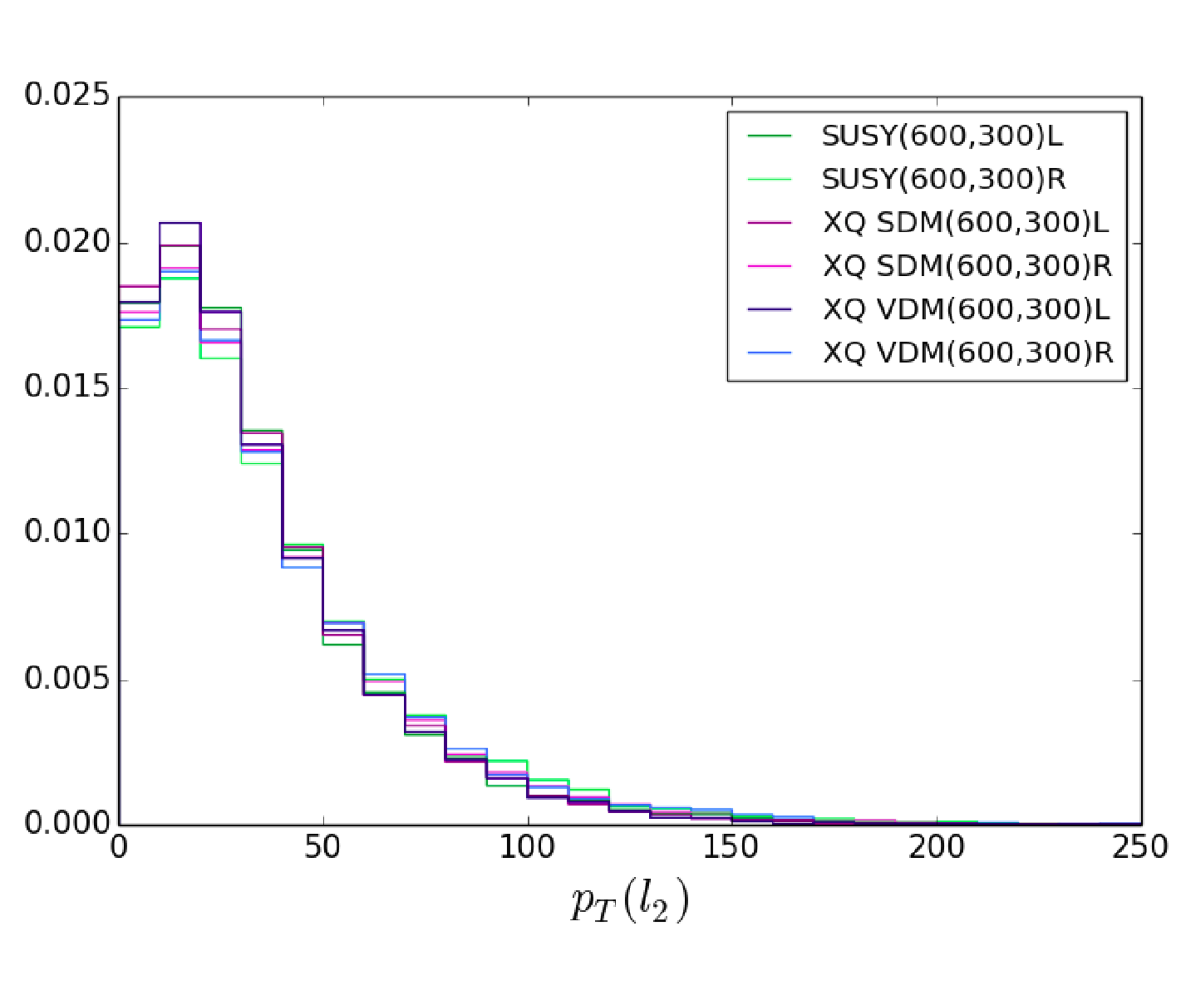}
\caption{Same as Fig.~\ref{fig:GenLevelDist1} but for the $(600,\,300)$ mass combination.}
\label{fig:GenLevelDist2}
\end{figure}

Regarding the lepton $p_T$, the small difference that appears is between the L and R cases 
rather than between SUSY and XQ: all the (600,\,10)R scenarios exhibit somewhat harder $p_T(l)$
than the (600,\,10)L scenarios. This comes from the fact that the top polarisation influences the $p_T$ of the top decay products.  %as discussed in \cite{Belanger:2012tm,Belanger:2013gha} and references therein. 
These features persist for smaller top partner--DM mass difference, see Fig.~\ref{fig:GenLevelDist2}.

Polarisation effects in stop decays were studied in detail in \cite{Cao:2006wk,Nojiri:2008ir,Shelton:2008nq,Perelstein:2008zt,Berger:2012an,Chen:2012uw,Bhattacherjee:2012ir,Belanger:2012tm,Low:2013aza,Belanger:2013gha,Wang:2015ola}. Sizeable effects were found in kinematic distributions of the final-state leptons and $b$-quarks, and in particular in their angular correlations. While this might help to constrain the relevant mixing angles in precision studies of a positive signal~\cite{Perelstein:2008zt,Berger:2012an,Bhattacherjee:2012ir,Belanger:2012tm,Low:2013aza,Belanger:2013gha} and possibly to characterise the spin of the top partner mediators and of the DM states through the structure of their coupling~\cite{Shelton:2008nq,Berger:2012an,Chen:2012uw}, as we will see, the current experimental analyses are not very sensitive to these effects. 

%\clearpage

\clearpage
%==============================================================================
\subsection{Effects in existing 8 TeV analyses}
\label{sec:analyses8tev}
%==============================================================================

Let us now analyse how the cut acceptances of existing 8 TeV analyses compare for the SUSY and XQ  scenarios. 
To this end, we consider the following ATLAS and CMS analyses implemented in {\sc CheckMATE}~\cite{Drees:2013wra} 
or the {\sc MadAnalysis}\,5 Public Analysis Database ({\sc MA5} PAD)~\cite{Dumont:2014tja}: 

\begin{itemize}
\item Fully hadronic stop search: ATLAS-CONF-2013-024~\cite{ATLAS:2013cma}  implemented in {\sc CheckMATE}, see Section~\ref{sec:fullhadstop}
\item Stop searches in the single lepton mode from ATLAS~\cite{Aad:2014kra}  ({\sc CheckMATE}) and CMS~\cite{Chatrchyan:2013xna} ({\sc MA5} PAD, recast code \cite{ma5code:cms-sus-13-011}), see Section~\ref{sec:1lepstop}
\item The stop search with 2 leptons from ATLAS~\cite{Aad:2014qaa} implemented in {\sc CheckMATE}, see Section~\ref{sec:2lepstop}
\item The generic gluino/squark search in the 2--6 jets plus missing energy channel from ATLAS~\cite{Aad:2014wea}  ({\sc MA5} PAD, recast code \cite{MA5-ATLAS-multijet-1405}), see Section~\ref{sec:genglusq}
\end{itemize}

%-----------------------------------------------------------------------------
\subsubsection{Fully hadronic stop search}
\label{sec:fullhadstop}
%-----------------------------------------------------------------------------

The ATLAS analysis~\cite{ATLAS:2013cma} implemented in {\sc CheckMATE} targets stop-pair production followed by stop decays into a top quark and the lightest neutralino, $pp\to \tilde t_1^{}\tilde t_1^*\to t \bar t \tilde\chi^0_1\tilde\chi^0_1$ 
in the fully-hadronic top final state, $t\to bW\to bq\bar q$. 
The search is thus conducted in events with large missing transverse momentum and six or more jets, of which $\ge 2$ must have been $b$-tagged. The two leading jets are required to have $p_T >80$~GeV with the remaining jets having $p_T >35$~GeV. Pre-selected electrons or muons, as well as taus are vetoed. Further requirements are imposed on azimuthal angle ($\Delta\phi$) and transverse mass ($m_T$) variables and on two 3-jet systems. 
Then three overlapping SRs are defined by requirements on $\ETmiss$, 
SR1:  $\ETmiss\ge 200$~GeV,  SR2:  $\ETmiss\ge 300$~GeV and SR3:  $\ETmiss\ge 350$~GeV.%
\footnote{We note that the conference note~\cite{ATLAS:2013cma} was superseded by the paper publication \cite{Aad:2014bva}, which has six SRs targeting the $\tilde t_1\to t\tilde\chi^0_1$ decay instead of three. Four of these, SRA1--4, are for ``fully resolved'' events with $\ge 6$ jets and a stacked $\ETmiss$ cut of 150, 250, 300 and 350~GeV. This is similar to the conference note. Two more SRs, SRB1--2, are for ``partially resolved'' events with 4 or 5 jets and higher $\ETmiss$, designed to target high stop masses.  Moreover, the paper considers three SRs, SRC1--3, optimized for stop decays into charginos. The limit is then set from a combination of SRA+B or SRA+C. Since this cannot be reproduced without a prescription of how to combine the SRs, 
we keep using the {\sc CheckMATE} implementation of the conference note to test the efficiencies of the hadronic stop search for our benchmark points. This is also justified by the fact that we are not primarily interested in the absolute limit but in potential differences in selection efficiencies between scalar and fermionic top partners.}

\begin{table}[ht!]
\scriptsize
\centering %\hspace*{-8mm}
\begin{tabular}{lccc}
\hline
         & SUSY & XQ-SDM & XQ-VDM \\
\hline         
Initial no.\ of events & 200000 & 200000 & 200000 \\
$\ETmiss>80$~GeV (Trigger) & 187834 (-6.08 \%)  &  187872 (-6.06 \%)  & 188358 (-5.82 \%) \\ 
muon veto ($p_T>10$~GeV) & 154643 (-17.67 \%)  & 153946 (-18.06 \%) &  154710 (-17.86 \%) \\
electron veto ($p_T>10$~GeV) &  123420 (-20.19 \%)  & 122439 (-20.47 \%)  & 123247 (-20.34 \%) \\
$\ETmiss > 130$~GeV & 113638 (-7.93 \%)  & 112808 (-7.87 \%)  & 113620 (-7.81 \%) \\
$\ge 6$ jets, $p_T>80,80,35$~GeV & 33044 (-70.92 \%)  & 27987 (-75.19 \%)  & 28285 (-75.11 \%) \\
reconstr.\ $\ETmiss{^{\rm,track}} > 30$~GeV & 32564 (-1.45 \%)  & 27563 (-1.51 \%)  & 27901 (-1.36 \%) \\
$\Delta\phi$($\ETmiss, \ETmiss{^{\rm,track}}) < \pi/3$ & 31200 (-4.19 \%)  & 26583 (-3.56 \%)  & 26939 (-3.45 \%) \\
$\Delta\phi$($\ETmiss$, 3 hdst jets) > 0.2$\pi$ & 26276 (-15.78 \%)  &  22795 (-14.25 \%)  & 23129 (-14.14 \%) \\ 
tau veto & 22880 (-12.92 \%) & 19967 (-12.41 \%)  & 20354 (-12.00 \%) \\
2 $b$ jets & 9668 (-57.74 \%)  & 8510 (-57.38 \%) &  8660 (-57.45 \%) \\
$m_T(b\,{\rm jets}) > 175$~GeV & 7202 (-25.51 \%)  &  6447 (-24.24 \%)  & 6579 (-24.03 \%) \\
3 closest jets 80--270 GeV & 6437 (-10.62 \%)  & 5877 (-8.84 \%)  & 5929 (-9.88 \%) \\
same for second closest jets & 3272 (-49.17 \%) &  3186 (-45.79 \%)  & 3351 (-43.48 \%) \\
\hline
$\ETmiss\ge 150$~GeV & 3230 (-1.28 \%) &  3156 (-0.94 \%) &  3312 (-1.16 \%) \\
$\ETmiss\ge 200$~GeV (SR1) & 3067 (-5.05 \%) &  3000 (-4.94 \%) &  3161 (-4.56 \%) \\
$\ETmiss\ge 250$~GeV & 2795 (-8.87 \%) &  2732 (-8.93 \%) &  2867 (-9.30 \%) \\
$\ETmiss\ge 300$~GeV (SR2) & 2413 (-13.67 \%) & 2373 (-13.14 \%) & 2490 (-13.15 \%)\\
$\ETmiss\ge 350$~GeV (SR3) & 1948 (-19.27 \%) & 1926 (-18.84 \%) & 2010 (-19.28 \%)\\
\hline
\end{tabular}
\caption{Cut-flow of the hadronic stop analysis of ATLAS for Point~(600,\,10)L, derived with {\sc CheckMATE}.}
\label{tab:cutflow-CM-1}
\end{table}

The effect of the various cuts is illustrated in Table~\ref{tab:cutflow-CM-1} for the example of Point (600,\,10)L. 
We observe that most preselection cuts have very similar efficiencies\footnote{Here and in the following, we use the term ``efficiency'' for the percentage of events remaining after one or more cuts. Strictly speaking this is the quantity acceptance$\times$efficiency, $A\epsilon$.}
when comparing SUSY and XQ cases. 
Small differences, of the level of few percent, occur only in the requirement of at least six jets (cf.\ Fig.~\ref{fig:GenLevelDist1}) and 
the condition on ``3 closest jets'' and ``second closest jets'', but these differences tend to compensate  each other. 
Finally, the effect of the $\ETmiss$ cuts that define the three SRs is almost the same for the SUSY and XQ scenarios. 
Consequently, the final numbers of events in each of the SRs agree within $\lesssim 5\%$ for the SUSY and XQ scenarios. 

The total efficiencies in the three SRs, cross section excluded at 95\%~CL and corresponding top partner mass limits in GeV are compared in Table~\ref{tab:hadr-eff-limits} for all four benchmark scenarios.\footnote{Given the upper limit on the cross section together with the cross section prediction as a function of the top partner mass one can estimate the 95\% CL mass limit under the assumption that the efficiency is flat. While this kind of extrapolation is not a substitute for determining the true limit through a scan over the masses, it does give an indication of i) the impact of the differences in the excluded cross section and ii) the higher reach in XQ as compared to SUSY. As we will see, this extrapolation works reasonably well for the stop searches 
but not for analyses that involve cuts which are directly sensitive to the overall mass scale.} 
We see that for a specific mass combination, the total efficiencies and hence the upper limit on the cross section are very similar for the SUSY and XQ hypotheses.
The derived lower limit on the top partner mass of course depends on the input cross section (whether it is assumed SUSY-like or XQ-like), and is thus higher for the XQ interpretation than for the SUSY interpretation. 
However, the differences in the mass limits arising from applying SUSY, XQ-SDM or XQ-VDM efficiencies are generally small. Indeed, for the (600,\,10) scenarios, i.e.\ large mass splitting, they are only 2--4~GeV, which is totally negligible. For smaller mass splittings, represented by the (600,\,300) scenarios, they reach about 10--20~GeV, which is still negligible.      
Finally, note that the effect on the mass limit from considering L vs.\ R polarised tops is of comparable size.

\begin{table}[ht!]\centering
\scriptsize
\begin{tabular}{| l | ccc | ccc |}
\hline
& \multicolumn{3}{c|}{\bf Point~(600,\,10)L} &   \multicolumn{3}{c|}{\bf Point~(600,\,10)R}\\
\hline
         & SUSY & XQ-SDM & XQ-VDM & SUSY & XQ-SDM & XQ-VDM\\
\hline         
eff. SR1 & 0.015 & 0.015 & 0.016 & 0.014 & 0.015 & 0.014 \\
eff. SR2 & 0.012 & 0.012 & 0.012 & 0.011 & 0.012 & 0.011 \\
eff. SR3$^*$ & 0.0097 & 0.0096 & 0.010 & 0.0092 & 0.0095 & 0.0094 \\
\hline
excl. XS [pb] & 0.0196 & 0.0199 & 0.0189 & 0.0209 & 0.0201 & 0.0205 \\
mass limit/SUSY XS & 619 & 618 & 622 & 613 & 617 & 615 \\
mass limit/XQ XS & 805 & 803 & 808 & 798 & 802 & 800 \\
\hline
$1-{\rm CLs}$ & 0.98 & 1 & 1 & 0.97 & 1 & 1 \\
\hline
\end{tabular}\\[4mm]
\begin{tabular}{| l | ccc | ccc |}
\hline
& \multicolumn{3}{c|}{\bf Point~(600,\,300)L} &   \multicolumn{3}{c|}{\bf Point~(600,\,300)R} \\
\hline
         & SUSY & XQ-SDM & XQ-VDM & SUSY & XQ-SDM & XQ-VDM \\
\hline         
eff. SR1$^*$ & 0.0074 & 0.0064 & 0.0062  & 0.0066 & 0.0060 & 0.0053 \\
eff. SR2 & 0.0039 & 0.0032 & 0.0031  & 0.0035 & 0.0032 & 0.0026 \\
eff. SR3 & 0.0022 & 0.0016 & 0.0017  & 0.0018 & 0.0016 & 0.0013 \\
\hline
excl. XS [pb] & 0.0647 & 0.0759 & 0.0772 & 0.0726 & 0.0805 & 0.0910 \\
mass limit/SUSY XS & 522 & 510 & 509 & 514 & 506 & 497 \\
mass limit/XQ XS & 687 & 671 & 670 & 676 & 666 & 655 \\
\hline
$1-{\rm CLs}$ & 0.59  & 1 & 1 & 0.54 & 1 & 1 \\
\hline
\end{tabular}
\caption[Efficiencies in the three SRs, cross section (XS) excluded at 95\%~CL, corresponding extrapolated top partner mass limits in GeV, and CLs exclusion value from the hadronic stop analysis of ATLAS derived with {\sc CheckMATE}.]{Efficiencies in the three SRs, cross section (XS) excluded at 95\%~CL, corresponding extrapolated top partner mass limits in GeV, and CLs exclusion value from the hadronic stop analysis of ATLAS derived with {\sc CheckMATE}. ``mass limit/SUSY XS'' means that the excluded XS is translated to a mass limit using the SUSY production cross section from Fig.~\ref{fig:XS}, while ``mass limit/XQ XS'' means the limit is estimated using the XQ cross section. The exclusion CL is obtained considering the corresponding cross sections at 600 GeV, $\sigma(\st_1\st_1^*)=0.024$~pb for stop production and $\sigma(T\bar T)=0.167$~pb for XQ production. The most sensitive SR used for the limit setting is marked with a star.}
\label{tab:hadr-eff-limits}
\end{table}

%\clearpage
%-----------------------------------------------------------------------------
\subsubsection{Stop search in the single lepton final state}
\label{sec:1lepstop}
%-----------------------------------------------------------------------------

Stops are also searched for in final states with a single lepton, jets and $\ETmiss$, arising from one $W$ decaying leptonically while the other one decays hadronically. The ATLAS analysis \cite{Aad:2014kra} for this channel is implemented in {\sc CheckMATE}, while the (cut-based version of) the corresponding CMS analysis~\cite{Chatrchyan:2013xna} is implemented in the {\sc MA5} PAD. 

In the CMS analysis~\cite{Chatrchyan:2013xna}, events are required to contain one isolated electron (muon) 
with $p_T > 30$ (25) GeV, no additional isolated track or hadronic $\tau$ candidate, at least four jets with 
$p_T > 30$~GeV at least one of which must be $b$-tagged, $\ETmiss > 100$~GeV and $M_T>120$~GeV. 
The analysis further makes use of the quantity $M_{\rm T2}^W$, a hadronic top $\chi^2$ ensuring that  three of the jets in the event be consistent with the $t\to bW\to bq\bar q$ decay, and the topological variable $\Delta\phi(\ETmiss,\, {\rm jet})$. Various SRs are defined targeting $\tilde t_1^{}\to t \tilde\chi^0_1$ or $\tilde t_1^{}\to b \tilde\chi^+_1$ decays with small or large mass differences between the stop and the neutralino or chargino. 

As an illustrative example, we show in Table~\ref{tab:cutflow-MA5-1lep} the cut-flow for the ``$\tilde t_1^{}\to t \tilde\chi^0_1$, high $\Delta M$, $\ETmiss>300$~GeV'' SR for Point~(600,\,10)R, which is the most sensitive SR for this benchmark.  
The only noticeable difference, though hardly of the level of 5\% in the cut efficiency, arises from the requirement of at least four jets. All other cuts have again almost the same effects on the SUSY and XQ models. 
Altogether, starting from the same number of events, we end up with slightly more SUSY than XQ events in this SR, but this difference is only 6--7\%. 

\begin{table}[ht!]
\scriptsize
\centering 
\begin{tabular}{lccc}
\hline
         & SUSY & XQ-SDM & XQ-VDM \\
\hline         
Initial no.\ of events & 200000 & 200000 & 200000 \\
$\ge1$ candidate lepton & 51097 (-74.45 \%) & 50700 (-74.65 \%) & 50417 (-74.79 \%) \\
$\ge4$ central jets & 23737 (-53.55 \%) & 21333 (-57.92 \%) & 20997 (-58.35 \%)  \\
$\ETmiss > 50$~GeV & 23203 (-2.25 \%) & 20848 (-2.27 \%) & 20548 (-2.14 \%)  \\
$\ETmiss > 100$~GeV & 21640 (-6.74 \%) & 19393 (-6.98 \%) & 19206 (-6.53 \%)  \\
$\ge 1~b$-tagged jet & 18339 (-15.25 \%) & 16643 (-14.18 \%) & 16512 (-14.03 \%)  \\
isol lepton and track veto & 17370 (-5.28 \%) & 15892 (-4.51 \%) & 15750 (-4.61 \%)  \\
hadronic tau veto & 17061 (-1.78 \%) & 15646 (-1.55 \%) & 15487 (-1.67 \%)  \\
$M_T > 120$~GeV & 13811 (-19.05 \%) & 12788 (-18.27 \%) & 12691 (-18.05 \%)  \\
$\Delta\phi$($\ETmiss$, j1 or j2) > 0.8 & 12006 (-13.07 \%) & 11251 (-12.02 \%) & 11164 (-12.03 \%)  \\
$\chi^2 < 5$ & 7079 (-41.04 \%) & 6771 (-39.82 \%) & 6750 (-39.54 \%)  \\
$\ETmiss > 300$~GeV & 4138 (-41.55 \%) & 3820 (-43.58 \%) & 3929 (-41.79 \%)  \\
$M_{\rm T2}^W > 200$~GeV & 3030 (-26.78 \%) & 2830 (-25.92 \%) & 2851 (-27.44 \%)  \\
\hline
\end{tabular}
\caption[Cut-flow for the ``$\tilde t_1^{}\to t \tilde\chi^0_1$, high $\Delta M$, $\ETmiss>300$~GeV'' SR (denoted SR-A) of the CMS stop search in the 1-lepton channel for Point~(600,\,10)R, derived with the {\sc MadAnalysis}\,5 recast code.]{Cut-flow for the ``$\tilde t_1^{}\to t \tilde\chi^0_1$, high $\Delta M$, $\ETmiss>300$~GeV'' SR (denoted SR-A) of the CMS stop search in the 1-lepton channel for Point~(600,\,10)R, derived with the {\sc MadAnalysis}\,5 recast code~\cite{ma5code:cms-sus-13-011}. Note that the event weighting to account for trigger and lepton identification efficiencies and for initial-state radiation effects is not included in this cut-flow. More details about these aspects and their implementation of the recast code can be found in the original references~\cite{Chatrchyan:2013xna} and \cite{ma5code:cms-sus-13-011}.} 
\label{tab:cutflow-MA5-1lep}
\end{table}

\begin{table}[ht!]\centering
\scriptsize
\begin{tabular}{| l | ccc | ccc |}
\hline
& \multicolumn{3}{c|}{\bf Point~(600,\,10)L} &   \multicolumn{3}{c|}{\bf Point~(600,\,10)R} \\
\hline
         & SUSY & XQ-SDM & XQ-VDM & SUSY & XQ-SDM & XQ-VDM \\
\hline         
eff. SR-A & 0.0108 & 0.0109 & 0.0111 & 0.0108$^*$ & 0.0106$^*$ & 0.0107$^*$ \\
eff. SR-B & 0.0181$^*$ & 0.0176$^*$ & 0.0184$^*$ & 0.0154 & 0.0152 & 0.0153 \\
\hline
excl. XS [pb] & 0.0169 & 0.0173 & 0.0166 & 0.0210 & 0.0213 & 0.0211 \\
mass limit/SUSY XS & 631 & 629 & 633 & 613 & 611 & 612 \\
mass limit/XQ XS & 820 & 818 & 822 & 798 & 796 & 797 \\
\hline
$1-{\rm CLs}$ &0.99 & 1 & 1 & 0.97 & 1 & 1 \\
\hline
\end{tabular}\\[4mm]
\begin{tabular}{| l | ccc | ccc |}
\hline
& \multicolumn{3}{c|}{\bf Point~(600,\,300)L} &   \multicolumn{3}{c|}{\bf Point~(600,\,300)R} \\
\hline
         & SUSY & XQ-SDM & XQ-VDM & SUSY & XQ-SDM & XQ-VDM \\
\hline         
eff. SR-A & \!0.00360\! & 0.00366 & 0.00346  & \!0.00340\! & 0.00321 & 0.00315 \\
eff. SR-B& \!0.00748\!$^*$ & 0.00685$^*$ & 0.00632$^*$  & \!0.00597\!$^*$ & 0.00570$^*$ & 0.00536$^*$ \\
\hline
excl. XS [pb] & 0.0399 & 0.0448 & 0.0480 & 0.0507 & 0.0530 & 0.0563 \\
mass limit/SUSY XS & 560 & 551 & 546 & 541 & 538 & 533 \\
mass limit/XQ XS & 733 & 722 & 715 & 710 & 706 & 700 \\
\hline
$1-{\rm CLs}$ &0.81 & 1 & 1 & 0.72 & 1 & 1 \\
\hline
\end{tabular}
\caption[Efficiencies for the ``$\tilde t_1^{}\to t \tilde\chi^0_1$, high $\Delta M$, $\ETmiss>300$~GeV'' (denoted SR-A) and ``$\tilde t_1^{}\to b \tilde\chi^+_1$, high $\Delta M$, $\ETmiss>250$~GeV'' (denoted SR-B) SRs, cross sections excluded at 95\%~CL, corresponding extrapolated top partner mass limits in GeV, and CLs exclusion value from the 1-lepton stop analysis of CMS, derived with the {\sc MadAnalysis}\,5 recast code.]{Efficiencies for the ``$\tilde t_1^{}\to t \tilde\chi^0_1$, high $\Delta M$, $\ETmiss>300$~GeV'' (denoted SR-A) and ``$\tilde t_1^{}\to b \tilde\chi^+_1$, high $\Delta M$, $\ETmiss>250$~GeV'' (denoted SR-B) SRs, cross sections excluded at 95\%~CL, corresponding extrapolated top partner mass limits in GeV, and CLs exclusion value from the 1-lepton stop analysis of CMS, derived with the {\sc MadAnalysis}\,5 recast code~\cite{ma5code:cms-sus-13-011}. The most sensitive SR used for the limit setting is indicated by a star.}
\label{tab:lept-eff-limits-cms}
\end{table}

Table~\ref{tab:lept-eff-limits-cms} summarises the total efficiencies in the two most important SRs of this analysis,
the cross sections excluded at 95\%~CL and the corresponding top partner mass limits in GeV for all four benchmark scenarios.
Note that, for large mass splitting, the SRs ``$\tilde t_1^{}\to b \tilde\chi^+_1$, high $\Delta M$, $\ETmiss>250$~GeV'' (here denoted as SR-B) which is optimized for $\tilde t_1^{}\to b \tilde\chi^+_1$ decays and ``$\tilde t_1^{}\to t \tilde\chi^0_1$, high $\Delta M$, $\ETmiss>300$~GeV'' (denoted SR-A) optimized for $\tilde t_1^{}\to t \tilde\chi^0_1$ have very similar sensitivities. 
In fact we observe that the most sensitive SR depends on the top polarisation. 
Events with left polarised tops are more likely to pass the additional requirement of SR-B on the leading $b$-jet, $p_T > 100$~GeV.
Concretely, in the SUSY scenario the expected upper limits are $0.0290$ pb in SR-A versus $0.0251$ pb in SR-B for (600,10)L and $0.0291$ pb vs.\ $0.0295$ pb for (600,10)R.
CMS has observed a small underfluctuation in both these SRs: 2 observed events vs.
 $4.7 \pm 1.4$ expected in SR-A and 5 observed events vs.\ $9.9 \pm 2.7$ expected in SR-B.
Overall the observed cross section limit is somewhat lower in the left-polarised scenario.
An analogous observation holds for the XQ scenarios; the differences between SUSY and XQ scenarios are negligible.

Finally, for smaller mass gaps, SR-B is more sensitive in all considered scenarios and we observe differences at the level of 10--15\% in the total signal selection efficiencies, which translate into up to about 20\% differences in the excluded cross sections, 
or $\lesssim 5\%$ in the estimated mass limits. 
The uncertainty from considering scenarios that lead to left or right polarised tops is of similar magnitude.
The latter is consistent with the observation in \cite{Chatrchyan:2013xna} that the limits on the $\tilde t_1^{}$ and $\tilde\chi^0_1$ masses vary by $\pm10$--$20$~GeV depending on the top-quark polarisation; the polarisation dependence in the $\tilde t_1^{}\to b\tilde\chi^+_1$ channel can be somewhat larger.

\begin{table}[ht!]\centering 
\scriptsize
\begin{tabular}{lccc}
\hline
         & SUSY & XQ-SDM & XQ-VDM \\
\hline        
Initial no.\ of events & 200000 & 200000 & 200000 \\ 
Trigger & 158881 (-20.56 \%) & 158929 (-20.54 \%) & 160073 (-19.96 \%)  \\
DQ & 154759 (-2.59 \%) & 155073 (-2.43 \%) & 156148 (-2.45 \%)  \\
\hline
1 baseline electron &  30142 (-80.52 \%) & 29980 (-80.67 \%) & 30019 (-80.78 \%) \\
1 signal electron &  22342 (-25.88 \%) & 22177 (-26.03 \%) & 22169 (-26.15 \%) \\
$\ge3$ jets $p_T\ge25$~GeV &  19865 (-11.09 \%) & 19241 (-13.24 \%) & 19262 (-13.11 \%) \\
$\ge4$ jets $p_T\ge25$~GeV &  14458 (-27.22 \%) & 13275 (-31.01 \%) & 13355 (-30.67 \%)  \\
\ldots & & & \\
{\tt tN\_med} $e$ & 1892 (-86.91 \%) & 1951 (-85.30 \%) & 1987 (-85.12 \%) \\
{\tt bCd\_high1} $e$ & 1792 (-87.61 \%) & 1651 (-87.56 \%) & 1748 (-86.91 \%) \\
{\tt bCd\_bulk} $e$ & 4359 (-69.85 \%) & 4180 (-68.51 \%) & 4262 (-68.09 \%) \\
\hline 
1 baseline $\mu$  & 27993 (-81.91 \%) & 28381 (-81.70 \%) & 28119 (-81.99 \%)  \\
1 signal $\mu$ &  23123 (-17.40 \%) & 23383 (-17.61 \%) & 23088 (-17.89 \%) \\ 
$\ge3$ jets $p_T\ge25$~GeV &  20695 (-10.50 \%) & 20624 (-11.80 \%) & 20302 (-12.07 \%)  \\
$\ge4$ jets $p_T\ge25$~GeV &  15197 (-26.57 \%) & 14448 (-29.95 \%) & 14163 (-30.24 \%) \\
\ldots & & & \\
{\tt tN\_med} $\mu$ & 2108 (-86.13 \%) & 1970 (-86.36 \%) & 1977 (-86.04 \%) \\
{\tt bCd\_high1} $\mu$ & 1790 (-88.22 \%) & 1821 (-87.40 \%) & 1747 (-87.67 \%) \\
{\tt bCd\_bulk} $\mu$ & 4582 (-69.85 \%) & 4415 (-69.44 \%) & 4340 (-69.36 \%) \\
\hline
\end{tabular}
\caption[Partial cut-flows for the ATLAS stop search in the 1-lepton channel for Point~(600,\,10)R, derived with {\sc CheckMATE}.]{Partial cut-flows for the ATLAS stop search in the 1-lepton channel for Point~(600,\,10)R, derived with {\sc CheckMATE}. Shown are the effects of the preselection cuts and the final numbers of events in specific SRs. The cut-flows are given separately for electrons and muons.}
\label{tab:cutflow-CM-1lep}
\end{table}

The corresponding ATLAS search \cite{Aad:2014kra} for this channel is implemented in {\sc CheckMATE}. 
Here, the signal selection requires a least one ``baseline'' lepton with $p_T>10$~GeV, which is later tightened 
to exactly one isolated lepton with $p_T > 25$~GeV.\footnote{Except for the SR with soft-lepton
selections which employ a $p_T$ > 6(7) GeV requirement for muons (electrons).} Events containing additional
baseline leptons are rejected.  
The analysis comprises 15 non-exclusive SRs, 4 of which target $\tilde t_1^{}\to t \tilde\chi^0_1$ (labelled `{\tt tN\_}'), 9 target $\tilde t_1^{}\to b \tilde\chi^+_1$ (labelled `{\tt bC\_}'), and the last 2 target 3-body and mixed decays. 
A minimum number of jets ranging between 2 and 4 is required depending on the SR, together with $b$-tagging requirements and an $\ETmiss$ cut of at least 100~GeV. 
As for the CMS analysis, a number of kinematic variables 
($m_T$, $am_{T2}$, $\Delta\phi(\ETmiss,\vec p_T({\rm jet}))$, etc.) are exploited for reducing the background. 
The relevant SRs for our benchmark points are {\tt tN\_med}, {\tt bCd\_high} and {\tt bCd\_bulk}.\footnote{Note that the ATLAS search has a dedicated SR to target boosted final states, {\tt tN\_boost}. This SR is not considered here, as the relevant ``topness'' variable is not implemented in {\sc CheckMATE}.} 
Of course, for the limit setting only the most sensitive one is used.  
A partial cut-flow example is given in Table~\ref{tab:cutflow-CM-1lep} for Point (600,\,10)R. 
The results for all four benchmark points are summarised in Table~\ref{tab:lept-eff-limits-atlas-corrected}.

\begin{table}[ht!]\centering
\scriptsize
\begin{tabular}{| l | ccc | ccc |}
\hline
& \multicolumn{3}{c|}{\bf Point~(600,\,10)L} &   \multicolumn{3}{c|}{\bf Point~(600,\,10)R} \\
\hline
         & SUSY & XQ-SDM & XQ-VDM & SUSY & XQ-SDM & XQ-VDM \\
\hline
eff. bCd{\tiny \_}bulk{\tiny \_}d & 0.0298* & 0.0287  & 0.0297  & 0.0278* & 0.0264* & 0.0270*\\
eff. bCd{\tiny \_}high1 & 0.0208  & 0.0204* & 0.0210* & 0.0179  & 0.0174  & 0.0175 \\
\hline
excl. XS [pb] & 0.0250 &  0.0335 & 0.0324 & 0.0267 & 0.0281 & 0.0274\\
mass limit/SUSY XS & 598 &  574 &  577 & 593 & 588 & 590 \\
mass limit/XQ XS & 780 & 750 & 754 & 773 & 768 & 770 \\
\hline
$1-{\rm CLs}$ &0.94 & 1 & 1 & 0.93 & 1 & 1 \\
\hline
\end{tabular}\\[4mm]
\begin{tabular}{| l | ccc | ccc |}
\hline
& \multicolumn{3}{c|}{\bf Point~(600,\,300)L} &   \multicolumn{3}{c|}{\bf Point~(600,\,300)R} \\
\hline
         & SUSY & XQ-SDM & XQ-VDM & SUSY & XQ-SDM & XQ-VDM \\
\hline
eff. bCd{\tiny \_}high1 & 0.00919* & 0.00810* & 0.00761* & 0.00777  & 0.00691  & 0.00638 \\
eff. tN{\tiny \_}med & 0.00927     & 0.00869  & 0.00836  & 0.00877* & 0.00862* & 0.00775*\\
\hline
excl. XS [pb] & 0.0742 & 0.0845& 0.0898 & 0.0509 & 0.0517 & 0.0579\\
mass limit/SUSY XS & 512 & 502 &  498 & 541 & 540 & 531 \\
mass limit/XQ XS & 673 & 661 & 656 & 709 & 708& 697\\
\hline
$1-{\rm CLs}$ &0.35 & 1 & 1 & 0.69 & 1 & 1 \\
\hline
\end{tabular}
\caption[Efficiencies for selected SRs, cross sections excluded at 95\%~CL , corresponding extrapolated top partner mass limits in GeV, and CLs exclusion values for the ATLAS stop search in the 1-lepton channel, derived with {\sc CheckMATE}.]{Efficiencies for selected SRs, cross sections excluded at 95\%~CL , corresponding extrapolated top partner mass limits in GeV, and CLs exclusion values for the ATLAS stop search in the 1-lepton channel, derived with {\sc CheckMATE}. The most sensitive SR used for the limit setting is indicated by a star.}
\label{tab:lept-eff-limits-atlas-corrected}
\end{table}

As in the CMS analysis, we observe very similar sensitivities in several SRs, 
and it depends on details of the scenario which SR turns out as the best one. 
It should be noted here that small differences in selection efficiencies can have a considerable impact on the observed limit if they yield different SRs as the most sensitive one. 
In particular, ATLAS has observed more events than expected in SR {\tt bCd\_high1} (16 observed events vs.\ $11 \pm 1.5$ expected). Consequently, limits obtained from this SR are weaker than those using {\tt tN\_med} (12 observed vs.\ $13 \pm 2.2$ expected) or {\tt bCd\_bulk\_d} (29 observed vs.\ $26.5 \pm 2.6$ expected). 
This is relevant, for example, for Point (600,\,10)L. 
Nonetheless, the differences when comparing SUSY, XQ-SDM and XQ-VDM cases remain small, 
in particular always well below the 20--30\% estimated systematic uncertainty inherent to recasting with fast simulation tools.
It is also worth pointing out that, in contrast to its CMS counterpart, this ATLAS analysis tends to give stronger limits for  R than for L scenarios. The effect is more pronounced for smaller mass differences, in agreement with Fig.~24 in \cite{Aad:2014kra}. Overall, the sensitivity to polarisation effects, while larger than for the CMS analysis, remains small.

%\clearpage
%-----------------------------------------------------------------------------
\subsubsection{Stop search in the 2-leptons final state}
\label{sec:2lepstop}
%-----------------------------------------------------------------------------

Let us next discuss the 2-lepton final state considered in the ATLAS analysis~\cite{Aad:2014qaa}. 
This analysis searches for direct stop-pair production with $\tilde t_1^{}\to b \tilde\chi^+_1\to bW^{(*)}\tilde\chi^0_1$ or 
$\tilde t_1^{}\to t \tilde\chi^0_1\to bW\tilde\chi^0_1$, targeting leptonic $W$ decays. 
Events are required to have exactly two oppositely charged signal leptons
(electrons, muons or one of each, defining same flavour (SF) and different-flavour (DF) selections). 
At least one of these electrons or muons must have
$p_T > 25$~GeV and $m_{\ell\ell} > 20$~GeV.
Events with a third preselected electron or muon are rejected. 
The analysis is subdivided into a ``leptonic mT2'' and ``hadronic mT2''  analysis, as well a 
multivariate analysis, which cannot be reproduced with our simulation frameworks. 
The ``leptonic mT2'' (4 SRs) and ``hadronic mT2'' (1 SR) analyses respectively use $m_{T2}$ and $m_{T2}^{b{\rm -jet}}$ 
as the key discriminating variable. Other kinematic variables used include
$\Delta\phi_j$ ($\Delta\phi_\ell$), the azimuthal angular distance between the $p_T^{\rm miss}$
vector and the direction of the closest jet (highest $p_T$ lepton).

The ``leptonic mT2'' analysis has 4 overlapping SRs defined by $m_{T2}>90$, 100, 110 and 120~GeV. 
From these, seven statistically independent SRs denoted S1--S7 are defined in the (jet selections, $m_{T2})$ plane, 
where `jet selections' refers to the number of jets with a certain minimum $p_T$, see Fig.~13 in \cite{Aad:2014qaa}. 
The most sensitive one for our benchmark points is S5, which has $m_{T2}>120$~GeV and at least two jets with 
$p_T({\rm jet1})>100$~GeV and $p_T({\rm jet2})>50$~GeV.

\begin{table}[ht!]\centering 
\scriptsize
\begin{tabular}{lccc}
\hline
         & SUSY & XQ-SDM & XQ-VDM \\
\hline        
Initial no.\ of events & 200000 & 200000 & 200000 \\ 
2 leptons, $p_T>10$~GeV &  63129 (-68.44 \%) & 63877 (-68.06 \%) & 63604 (-68.20 \%) \\
\hline\hline
same flavour &  31464 (-50.16 \%) & 32040 (-49.84 \%) & 31643 (-50.25 \%) \\
isolation & 28096 (-10.70 \%) & 28538 (-10.93 \%) & 28234 (-10.77 \%) \\
opposite sign &  27961 (-0.48 \%) & 28402 (-0.48 \%) & 28078 (-0.55 \%) \\
$m_{\ell\ell} > 20$~GeV & 27457 (-1.80 \%) & 27874 (-1.86 \%) & 27586 (-1.75 \%) \\
$p_T(\ell) > 25$~GeV & 26505 (-3.47 \%) & 26948 (-3.32 \%) & 26625 (-3.48 \%) \\
$Z$ veto &  21448 (-19.08 \%) & 21682 (-19.54 \%) & 21374 (-19.72 \%) \\
$\Delta\phi_j>1$ & 12664 (-40.95 \%) & 13463 (-37.91 \%) & 13375 (-37.42 \%) \\
$\Delta\phi_b<1.5$ &  11779 (-6.99 \%) & 12638 (-6.13 \%) & 12460 (-6.84 \%) \\
$m_{T2}>120$~GeV &  4824 (-59.05 \%) & 5441 (-56.95 \%) & 5368 (-56.92 \%) \\
\hline
S5 -- SF (2 jets, $p_T>100,50$~GeV) & 2378 (-50.70 \%) & 2621 (-51.83 \%) & 2446 (-54.43 \%)\\
\hline\hline
different flavour & 31665 (-49.84 \%) & 31837 (-50.16 \%) & 31961 (-49.75 \%) \\
... & & & \\
$m_{T2}>120$~GeV &  5955 (-59.74 \%) & 6515 (-58.31 \%) & 6697 (-57.45 \%) \\
S5 -- DF (2 jets, $p_T>100,50$~GeV) & 3032 (-49.08 \%) & 3013 (-53.75 \%) & 3030 (-54.76 \%)\\
\hline\hline
S5 -- SF+DF & 5410 & 5634 & 5476\\
\hline
\end{tabular}
\caption[Cut-flow example for the ATLAS stop search in the 2-lepton channel for Point~(600,\,10)R, derived with {\sc CheckMATE}.]{Cut-flow example for the ATLAS stop search in the 2-lepton channel for Point~(600,\,10)R, derived with {\sc CheckMATE}. Here, the leptonic $W$ decay was enforced to enhance statistics.}
\label{tab:cutflow-CM-2lep}
\end{table}

Table~\ref{tab:cutflow-CM-2lep} shows a cut-flow example for the SF selection for Point~(600,\,10)R, as well as an abbreviated version for the DF selection.
Note that the leptonic $W$ decay was enforced in {\sc Pythia} to increase statistics. 
The SF selection gives fewer events than the DF one because the $Z$ veto removes about 20\% of events in the former but none in the latter. 
The combined count for SR S5 is given as the last line in the table. 
As was already the case for the other analyses, no significant differences occur at any particular step of the cut-flow. 
At the end we are left with the marginal difference of 4\% more XQ than SUSY events in a total selection efficiency 
of barely 3 permil (when considering events where the W is allowed to decay to anything).
% after folding the $W$ decay BRs back in). 

The picture is similar for Point~(600,\,10)L, for which %an abbreviated 
the cut-flow is given in Table~\ref{tab:cutflow-CM-2lep-Left}. 
Noteworthy is the fact that the initial difference in Points (600,\,10)R and (600,\,10)L from the 2 lepton selection (the first cut) is inverted by the last cut, so that in the final SR there remain more events for (600,\,10)L than for (600,\,10)R. 
This is a consequence of the dependence on the top polarisation already noted in the parton-level plots in Figs.~\ref{fig:GenLevelDist1} and \ref{fig:GenLevelDist2}.

\begin{table}[ht!]\centering 
\scriptsize
\begin{tabular}{lccc}
\hline
         & SUSY & XQ-SDM & XQ-VDM \\
\hline        
Initial no.\ of events & 200000 & 200000 & 200000 \\ 
2 leptons, $p_T>10$~GeV &  60379 (-69.81 \%) & 61193 (-69.40 \%) & 60812 (-69.59 \%)  \\
\hline\hline
same flavour &   30109 (-50.13 \%) & 30508 (-50.14 \%) & 30419 (-49.98 \%) \\
isolation & 26759 (-11.13 \%) & 27108 (-11.14 \%) & 27066 (-11.02 \%) \\
opposite sign &  26660 (-0.37 \%) & 26994 (-0.42 \%) & 26987 (-0.29 \%) \\
$m_{\ell\ell} > 20$~GeV & 26043 (-2.31 \%) & 26364 (-2.33 \%) & 26381 (-2.25 \%) \\
$p_T(\ell) > 25$~GeV & 25062 (-3.77 \%) & 25251 (-4.22 \%) & 25345 (-3.93 \%) \\
$Z$ veto &  19570 (-21.91 \%) & 19765 (-21.73 \%) & 19642 (-22.50 \%) \\
$\Delta\phi_j>1$ & 11797 (-39.72 \%) & 12485 (-36.83 \%) & 12522 (-36.25 \%) \\
$\Delta\phi_b<1.5$ &  11270 (-4.47 \%) & 11943 (-4.34 \%) & 12035 (-3.89 \%) \\
$m_{T2}>120$~GeV &  4390 (-61.05 \%) & 4785 (-59.93 \%) & 4815 (-59.99 \%)  \\
\hline
S5 -- SF (2 jets, $p_T>100,50$~GeV) & 2711 (-38.25 \%) & 2803 (-41.42 \%) & 2841 (-41.00 \%) \\
\hline\hline
different flavour &  30270 (-49.87 \%) & 30685 (-49.86 \%) & 30393 (-50.02 \%) \\
... & & & \\
$\Delta\phi_j>1$ & 15273 (-38.59 \%) & 16117 (-36.31 \%) & 15896 (-36.21 \%)  \\
$\Delta\phi_b<1.5$ &  14683 (-3.86 \%) & 15505 (-3.80 \%) & 15260 (-4.00 \%) \\
$m_{T2}>120$~GeV &  5581 (-61.99 \%) & 6149 (-60.34 \%) & 5985 (-60.78 \%) \\
S5 -- DF (2 jets, $p_T>100,50$~GeV) & 3524 (-36.86 \%) & 3562 (-42.07 \%) & 3503 (-41.47 \%) \\
\hline\hline
S5 -- SF+DF & 6235 & 6365 & 6344\\
\hline
\end{tabular}
\caption[Cut-flow example for the ATLAS stop search in the 2-lepton channel for Point~(600,\,10)L, derived with {\sc CheckMATE}.]{Cut-flow example for the ATLAS stop search in the 2-lepton channel for Point~(600,\,10)L, derived with {\sc CheckMATE}. To be compared with Table~\ref{tab:cutflow-CM-2lep}.  $W$s were again forced to decay leptonically to enhance statistics.}
\label{tab:cutflow-CM-2lep-Left}
\end{table}

Either way, as can be seen from Table~\ref{tab:lept-eff-limits-atlas}, there is again no significant difference %whatsoever
in the total efficiencies and excluded cross sections between SUSY, XQ-SDM and XQ-VDM scenarios. 

\begin{table}[ht!]\centering
\scriptsize
\begin{tabular}{| l | ccc | ccc |}
\hline
& \multicolumn{3}{c|}{\bf Point~(600,\,10)L} &   \multicolumn{3}{c|}{\bf Point~(600,\,10)R} \\
\hline
         & SUSY & XQ-SDM & XQ-VDM & SUSY & XQ-SDM & XQ-VDM \\
\hline
efficiency & 0.00314 & 0.00334& 0.00323& 0.00276 & 0.00285 & 0.00286\\
excl. XS [pb] & 0.0470 &  0.0443 & 0.0455 & 0.0535 & 0.0520 & 0.0518\\
mass limit/SUSY XS & 547 & 552 & 550 & 537 & 539 & 540\\
mass limit/XQ XS & 717 & 723 & 720 & 705 & 707 & 708\\
\hline
$1-{\rm CLs}$ &0.79  & 1 & 1 & 0.74 & 1 & 1 \\
\hline
\end{tabular}\\[4mm]
\begin{tabular}{| l | ccc | ccc |}
\hline
& \multicolumn{3}{c|}{\bf Point~(600,\,300)L} &   \multicolumn{3}{c|}{\bf Point~(600,\,300)R} \\
\hline
         & SUSY & XQ-SDM & XQ-VDM & SUSY & XQ-SDM & XQ-VDM \\
\hline
efficiency & 0.00134 & 0.001425 & 0.00138 & 0.00111 & 0.00118 & 0.00100 \\
excl. XS [pb] & 0.109 & 0.104 & 0.108 & 0.133 & 0.125 & 0.148\\
mass limit/SUSY XS & 484 & 487 & 484 & 469 & 473 & 462\\
mass limit/XQ XS & 638 & 642 & 639 & 620 & 626 & 611\\
\hline
$1-{\rm CLs}$ &0.49  & 1 & 1 & 0.43 & 1 & 1 \\
\hline
\end{tabular}
\caption[Efficiencies, cross sections excluded at 95\%~CL, corresponding extrapolated top partner mass limits in GeV, and CLs exclusion value for the ATLAS stop search in the 2-lepton channel, derived with {\sc CheckMATE}.]{Efficiencies, cross sections excluded at 95\%~CL, corresponding extrapolated top partner mass limits in GeV, and CLs exclusion value for the ATLAS stop search in the 2-lepton channel, derived with {\sc CheckMATE}. All numbers correspond to the most sensitive SR, SR5.}
\label{tab:lept-eff-limits-atlas}
\end{table}

%\clearpage
%-----------------------------------------------------------------------------
\subsubsection{Gluino/squark search in the 2--6 jets final state \label{sec:gluinosquark}}
\label{sec:genglusq}
%-----------------------------------------------------------------------------

For completeness, we also include a generic SUSY search (nominally for squarks and gluinos) in final states containing 
high-$p_T$ jets, missing transverse momentum and no electrons or muons in our analysis. Concretely, we here consider the 
ATLAS analysis~\cite{Aad:2014wea} via the {\sc MadAnalysis}\,5 recast code~\cite{MA5-ATLAS-multijet-1405}. (A {\sc CheckMATE} implementation of the same analysis was done in \cite{Cao:2015ara} and will be used in Appendix \ref{app:CM results}). 
Our original purpose was to compare the performance of the hadronic stop analysis  to that of a multi-jet analysis which was not optimized for the $t\bar t+\MET$ signature. But, as we will see, the effective mass $M_{\rm eff}$ variable employed in the generic gluino/squark search offers a useful complementary probe. 

Regarding the signal selection, the ATLAS analysis~\cite{Aad:2014wea} comprises 15 inclusive SRs characterized by increasing minimum jet multiplicity, $N_j$, from two to six jets. 
Hard cuts are placed on missing energy and the $p_T$ of the two leading jets: $\MET>160$~GeV, $p_T(j_1)>130$~GeV and $p_T(j_2)>60$~GeV.  
For the other jets, $p_T>60$~or 40~GeV is required depending on the SR. 
In all cases, events are discarded if they contain electrons or muons with $p_T>10$~GeV. 
Depending on $N_j$, additional requirements are placed on the minimum azimuthal separation between any of the jets and the $\MET$, $\Delta\phi({\rm jet},\MET)$, as well as on $\MET/\sqrt{H_T}$ or $\MET/M_{\rm eff}(N_j)$. 
Finally, a cut is placed on $M_{\rm eff}({\rm incl.})$, which sums over all jets with $p_T>40$~GeV and $\MET$.
A cut-flow example is shown in Table~\ref{tab:cutflow-gluino-squark} for Point (600,10)R for a SR with 4 jets (SR {\tt 4jl}). 
Note that, starting from 200K events, we end up with about 15\% (11\%) more SUSY than XQ-SDM (XQ-VDM) events in this SR. 
The reason for this is that the cuts on $p_T(j)$ and $M_{\rm eff}$ remove somewhat more XQ than SUSY events, as expected from the distributions in Fig.~\ref{fig:GenLevelDist1}.

\begin{table}[ht!]\centering 
\scriptsize
\begin{tabular}{lccc}
\hline
         & SUSY & XQ-SDM & XQ-VDM \\
\hline        
Initial no.\ of events & 200000 & 200000 & 200000 \\ 
$\MET > 160$~GeV & 158489 (-20.76\%) & 158497 (-20.75\%) & 159683 (-20.16\%) \\
$N_j > 1$ & 150908 (-4.78\%) & 150121 (-5.28\%) & 151311 (-5.24\%) \\ 
lepton veto & 100139 (-33.64\%) & 100462 (-33.08\%) & 101404 (-32.98\%) \\ 
$p_T(j_1)>130$~GeV & 62585 (-37.50\%) & 58754 (-41.52\%) & 59482 (-41.34\%) \\ 
$p_T(j_2)>60$~GeV & 62045 (-0.86\%) & 58188 (-0.96\%) & 58886 (-1.00\%) \\ 
$p_T(j_3)>60$~GeV & 56729 (-8.57\%) & 52649 (-9.52\%) & 53312 (-9.47\%) \\ 
$p_T(j_4)>60$~GeV & 39150 (-30.99\%) & 34856 (-33.80\%) & 35258 (-33.86\%) \\ 
$\Delta\phi(j_1),\MET) > 0.4$ & 38811 (-0.87\%) & 34616 (-0.69\%) & 35000 (-0.73\%) \\ 
$\Delta\phi(j_2),\MET) > 0.4$ & 37199 (-4.15\%) & 33304 (-3.79\%) & 33635 (-3.90\%) \\ 
$\Delta\phi(j_3),\MET) > 0.4$ & 35447 (-4.71\%) & 31870 (-4.31\%) & 32211 (-4.23\%) \\ 
$\Delta\phi(j_4),\MET) > 0.2$ & 34535 (-2.57\%) & 31064 (-2.53\%) & 31435 (-2.41\%) \\ 
$\MET/\sqrt{H_T}>10$ & 25451 (-26.30\%) & 23522 (-24.28\%) & 24004 (-23.64\%) \\ 
$M_{\rm eff}({\rm incl.})>1$~TeV & 17695 (-30.47\%) & 15062 (-35.97\%) & 15714 (-34.54\%) \\ 
\hline
\end{tabular}
\caption[Cut-flow for the {\tt 4jl} SR of the ATLAS gluino and squark search in the 2--6 jets channel for Point~(600,\,10)R, derived with the {\sc MadAnalysis}\,5 recast code.]{Cut-flow for the {\tt 4jl} SR of the ATLAS gluino and squark search in the 2--6 jets channel for Point~(600,\,10)R, derived with the {\sc MadAnalysis}\,5 recast code~\cite{MA5-ATLAS-multijet-1405}.}
\label{tab:cutflow-gluino-squark}
\end{table}

Table~\ref{tab:limits-gluino-squark} summarises the total efficiencies in the most important SRs of this analysis together with the cross sections excluded at 95\%~CL and the corresponding estimated top partner mass limits for all four benchmark scenarios. 
We observe about 20\% difference in the excluded cross sections between SUSY and XQ interpretations. 
However, the mass limits derived from the excluded cross sections are not reliable because for this search the total efficiencies strongly depend on the top-partner mass. 
As we will see in the next section, while this analysis does provide a limit on $T\bar T$ production because of the larger cross section, it is not sensitive to $\st_1^{}\st_1^*$ production.

\begin{table}[ht!]\centering
\scriptsize
\begin{tabular}{| l | ccc | ccc |}
\hline
& \multicolumn{3}{c|}{\bf Point~(600,\,10)L} &   \multicolumn{3}{c|}{\bf Point~(600,\,10)R} \\
\hline
         & SUSY & XQ-SDM & XQ-VDM & SUSY & XQ-SDM & XQ-VDM \\
\hline
efficiency &  0.08898 & 0.07454 & 0.07752  &  0.08847 & 0.07531  &  0.07857  \\
excl. XS [pb] & 0.0535 & 0.0639 & 0.0612 &  0.0538 & 0.0631 & 0.0605   \\
mass limit/SUSY XS  &  537 & 523 & 527  &   537 & 524 & 528 \\
mass limit/XQ XS      &  705 & 688 & 692   &   704 & 689 & 693   \\
\hline
$1-{\rm CLs}$ &0.65  & 1 & 1 & 0.66 & 1 & 1 \\
\hline
\end{tabular}\\[4mm]
\begin{tabular}{| l | ccc | ccc |}
\hline
& \multicolumn{3}{c|}{\bf Point~(600,\,300)L} &   \multicolumn{3}{c|}{\bf Point~(600,\,300)R} \\
\hline
         & SUSY & XQ-SDM & XQ-VDM & SUSY & XQ-SDM & XQ-VDM \\
\hline
efficiency &  0.05183 & 0.04242 & 0.04159  &  0.05231 & 0.04281  &  0.04020 \\
excl. XS [pb] &  0.257 & 0.313 & 0.320 &   0.254 & 0.311 & 0.330   \\
mass limit/SUSY XS  &  424 & 410 & 409  &   424 & 411 & 407  \\
mass limit/XQ XS      &  563 & 547 & 545  &  564 & 547 & 542   \\
\hline
$1-{\rm CLs}$ &0.13  & 0.67 & 0.66 & 0.13 & 0.68 & 0.65 \\
\hline
\end{tabular}
\caption[Efficiencies, cross sections excluded at 95\%~CL and corresponding extrapolated top partner mass limits in GeV for the ATLAS gluino and squark search in the 2--6 jets channel, derived with the {\sc MadAnalysis}\,5 recast code]{Efficiencies, cross sections excluded at 95\%~CL and corresponding extrapolated top partner mass limits in GeV for the ATLAS gluino and squark search in the 2--6 jets channel, derived with the {\sc MadAnalysis}\,5 recast code~\cite{MA5-ATLAS-multijet-1405}. The last entry is the CLs exclusion value. The most sensitive SR is 4jl for the (600,\,10) mass combination and 4jlm for the (600,\,300) mass combination. Note that for this search the efficiencies strongly depend on the top partner mass, so the extrapolation of the mass limit is unreliable; this is to large extent due to the cut on $M_{\rm eff}$.}
\label{tab:limits-gluino-squark}
\end{table}

%\clearpage
%==============================================================================
\subsection{Results in the  top partner versus DM mass plane}
\label{sec:bounds8TeV}
%==============================================================================

Having analysed the differences, or lack thereof, in the cut efficiencies of the experimental analyses for our four benchmark points, we next perform a scan in the plane of top partner versus DM mass to derive the 95\% CL exclusion lines. 
For definiteness, we keep the couplings fixed to the same values as for the (600,\,10)L and (600,\,10)R benchmark points.

\begin{figure}[H]\centering 
\includegraphics[width=0.48\textwidth]{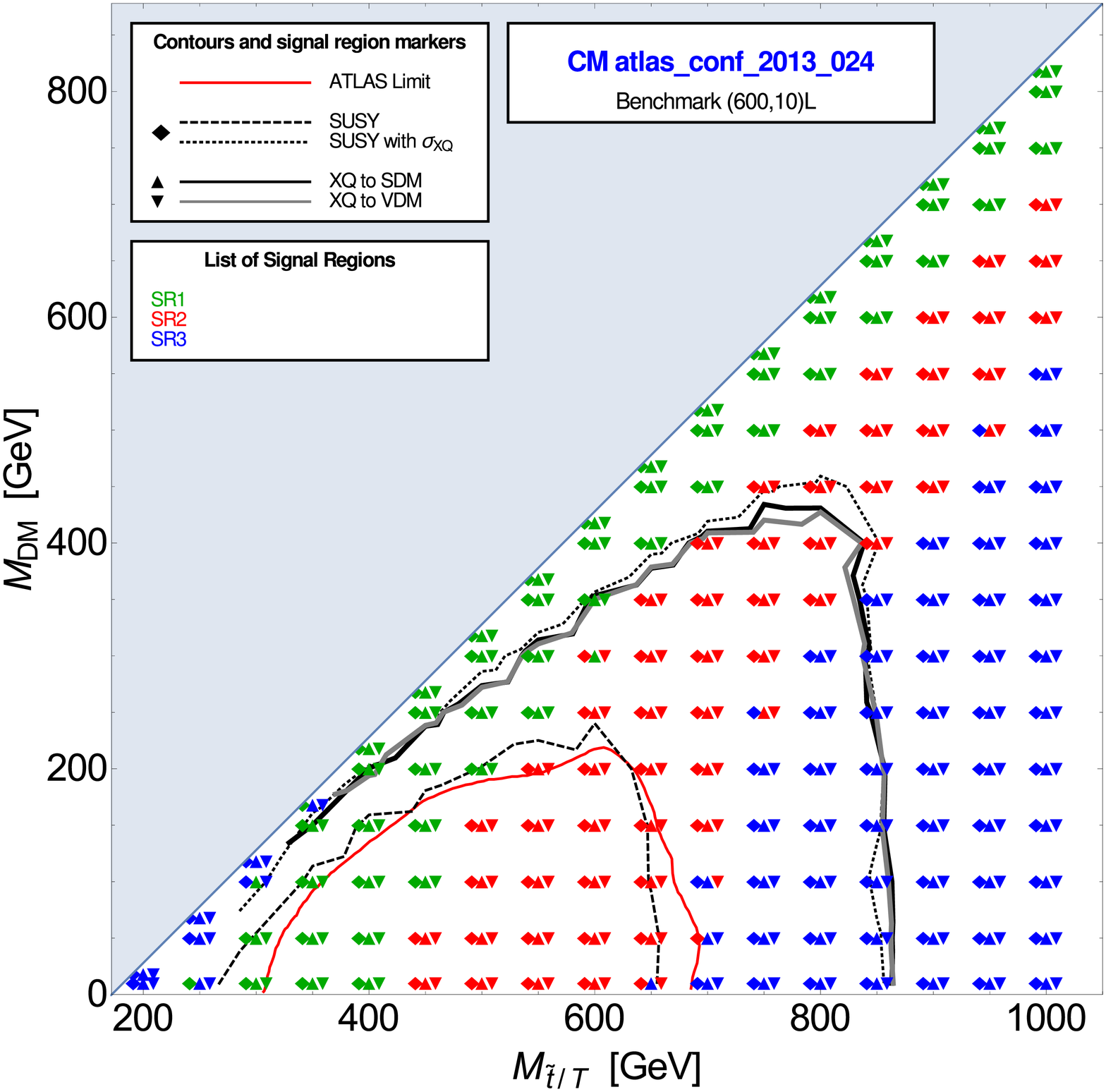}\quad%
\includegraphics[width=0.48\textwidth]{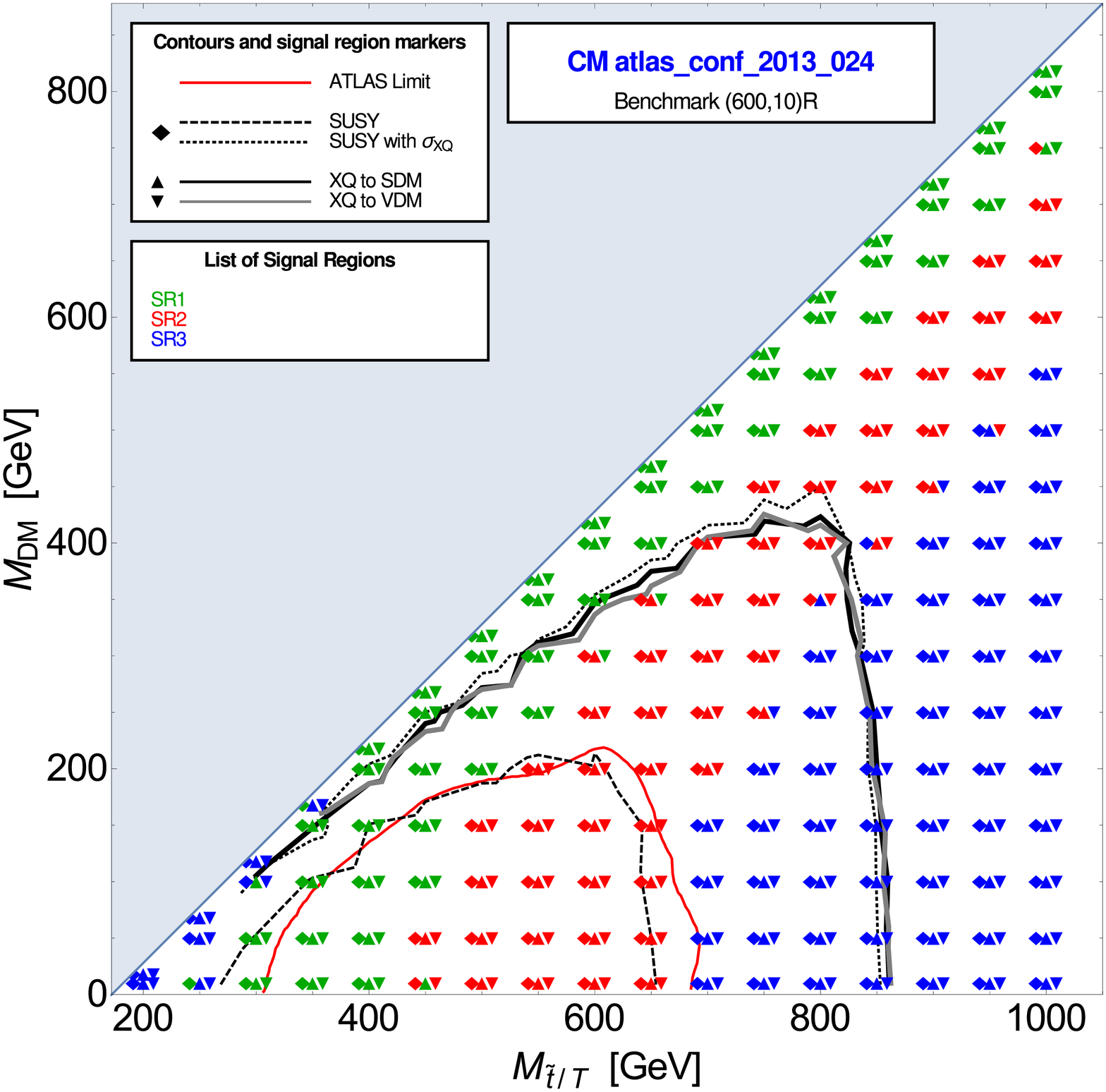}\\[3mm]
\includegraphics[width=0.48\textwidth ,angle=270]{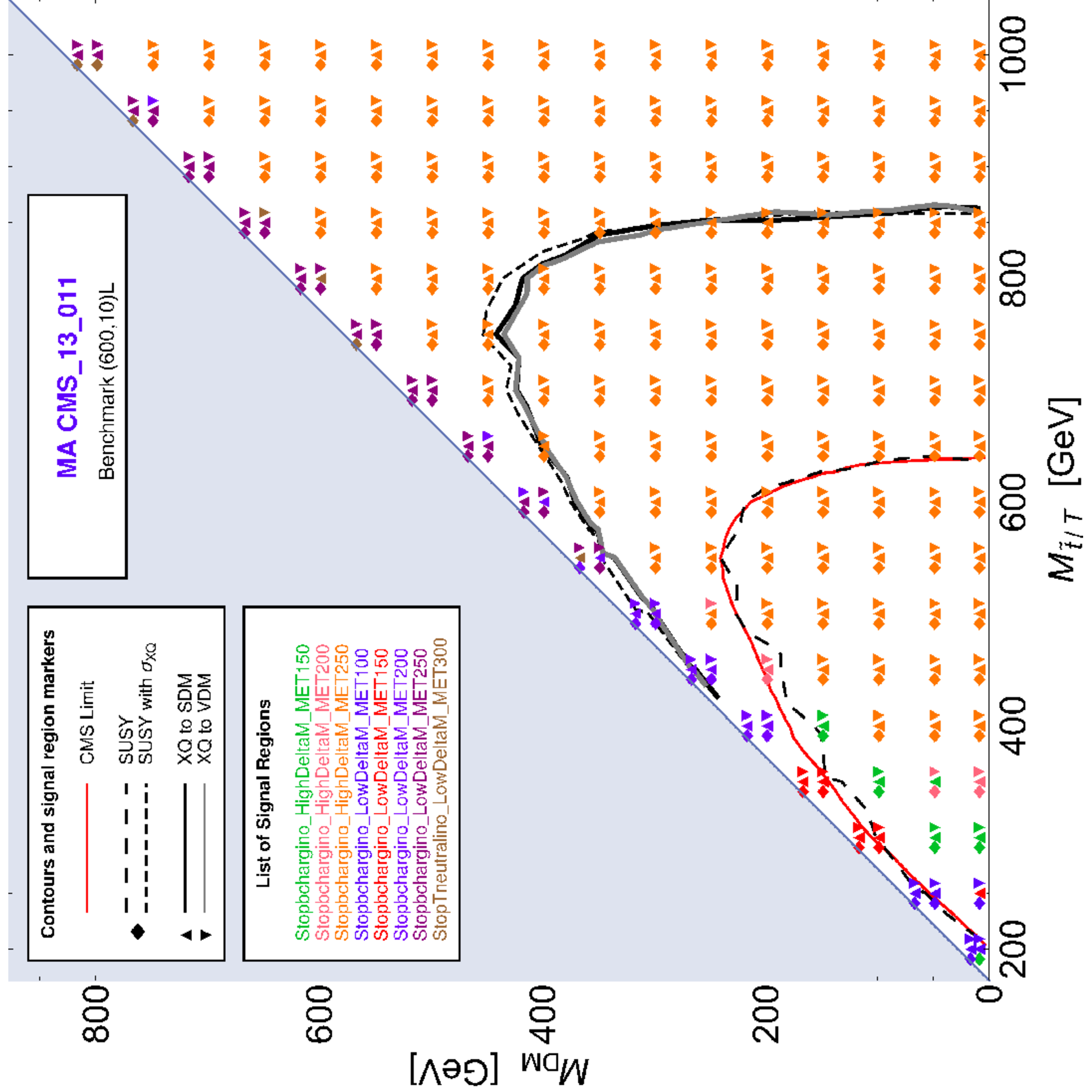}\quad%
\includegraphics[width=0.48\textwidth ,angle=270]{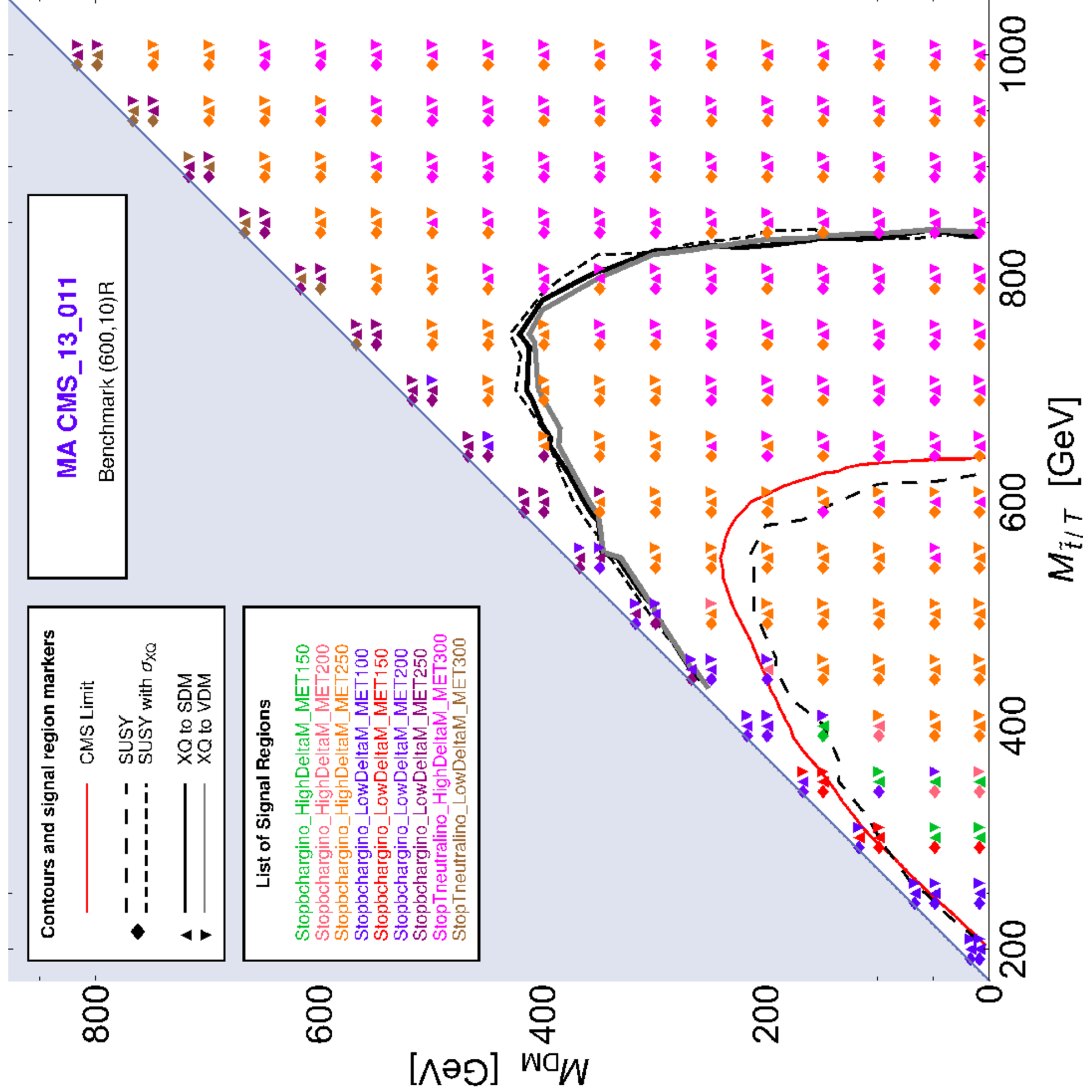}\\[3mm]
\includegraphics[width=0.48\textwidth]{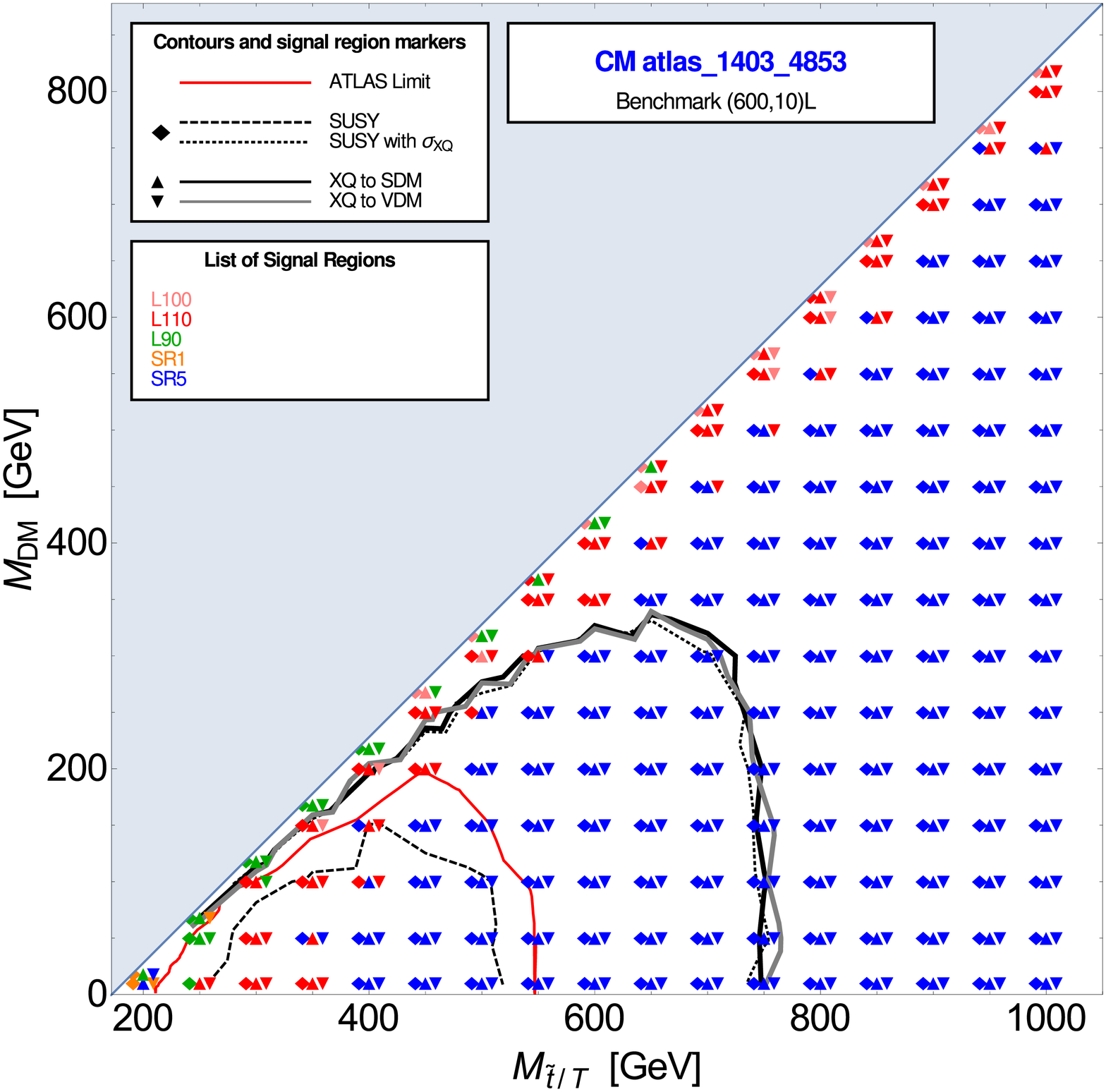}\quad%
\includegraphics[width=0.48\textwidth]{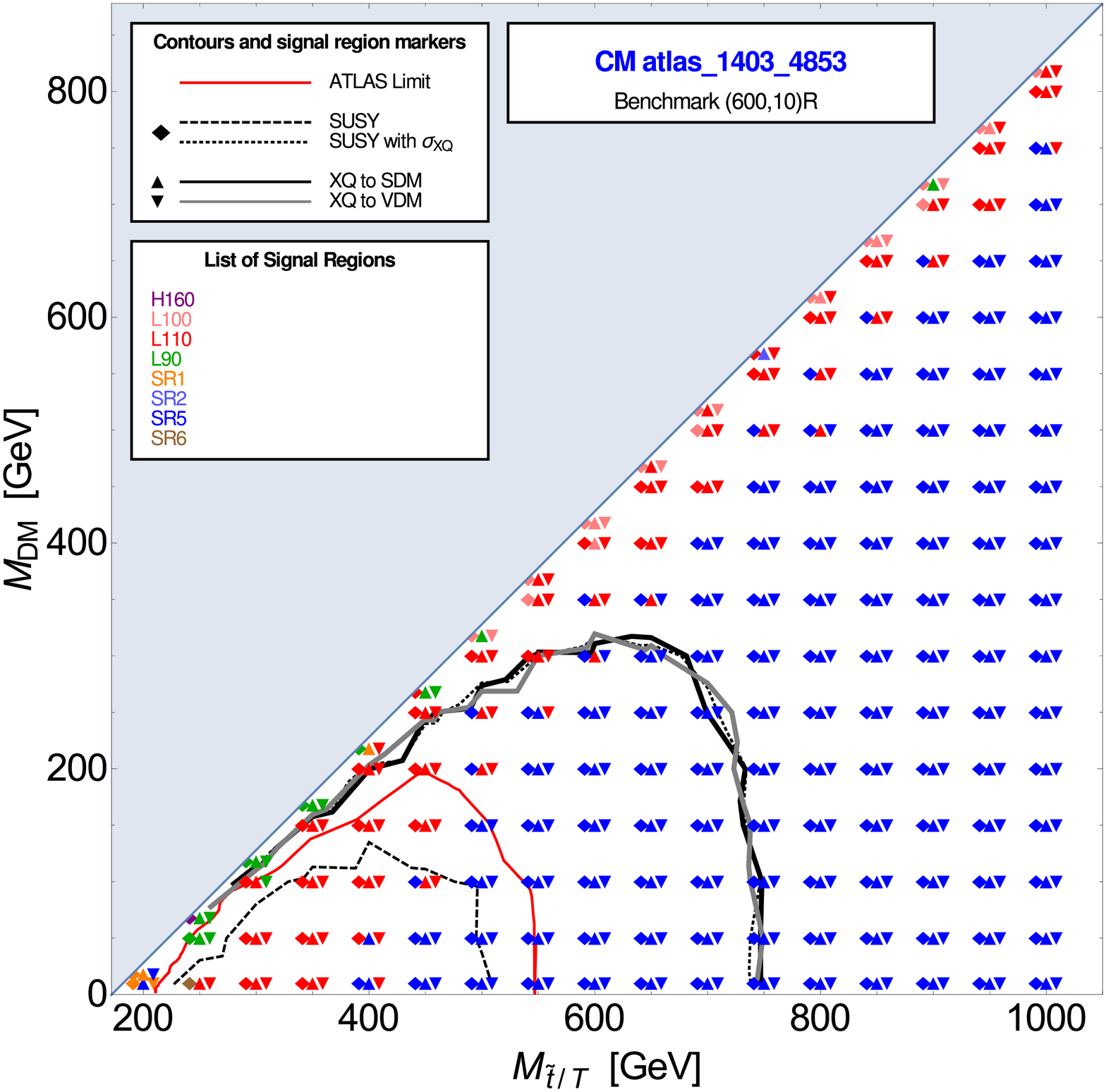}%
\caption[Comparisons of constraints in the top partner versus DM mass plane for the fully hadronic stop search from ATLAS recast with {\sc CheckMATE}, the 1-lepton stop search from CMS recast with {\sc MadAnalysis}\,5, and the 2-lepton stop search from ATLAS recast with {\sc CheckMATE}.]{Comparisons of constraints in the top partner versus DM mass plane for the fully hadronic stop search from ATLAS recast with {\sc CheckMATE} (top), the 1-lepton stop search from CMS recast with {\sc MadAnalysis}\,5 (middle), and the 2-lepton stop search from ATLAS recast with {\sc CheckMATE} (bottom). See text for details. 
%The various lines indicate the regions excluded at 8 TeV for the SUSY and XQ cases, and for the case where the SUSY efficiencies are applied to the XQ cross sections. The plots also contain the information which SRs are the most sensitive ones for each point of the scan. For reference, the official ATLAS/CMS exclusion lines are also shown (full red lines).
}
\label{fig:contours1}
\end{figure}

Figure~\ref{fig:contours1} presents the results for the ATLAS fully hadronic stop search implemented in {\sc CheckMATE} (top row), the CMS 1-lepton stop search recast with {\sc MadAnalysis}\,5 (middle row) and the ATLAS stop search in the 2-lepton final state recast with {\sc CheckMATE} (bottom row). 
The left panels are for the couplings of Point (600,\,10)L, the right panels for the couplings of Point (600,\,10)R, see Table~\ref{tab:BPlist}. 
Shown are the 95\%~CL exclusion lines obtained from SUSY, XQ-SDM and XQ-VDM event simulation (dashed black, full black and full grey lines, respectively), as well as the exclusion lines obtained from rescaling SUSY efficiencies with XQ cross sections (dotted black line). For each bin, the most sensitive SR used for the limit setting in the SUSY, XQ-SDM and XQ-VDM case is indicated by a coloured symbol as shown in the plot legends.
For reference, the official ATLAS/CMS exclusion lines are also shown as full red lines.

For the CMS 1-lepton search, our  exclusion line for left stops agrees remarkably well with the official CMS line (from the cut-based analysis). This is somewhat accidental, as {\it i)} the official CMS limit is for unpolarised stops, and {\it ii)} in our simulation the limit is mostly obtained from a SR optimised for decays to bottom and chargino, not from one optimised for decays to top and neutralino. 
On the other hand, the fairly large discrepancy for the ATLAS 2-lepton search is explained by the fact that the official exclusion curve was obtained using an multivariate analysis not available in {\sc CheckMATE}.

We see that over most of the mass plane, the best SR is the same for SUSY, XQ-SDM and XQ-VDM. 
(For the points where they are different, the sensitivities of the best and 2nd best SRs are actually quite similar.) 
The main conclusions which can be inferred from the plots are the following: 

\begin{enumerate}
\item There are no significant differences between the XQ scenarios where the top partner decays to scalar or vector DM. This is expected because in the NWA the process is largely dominated by the resonant contribution, the cross section of which can be factorised into production cross section times BRs. Since in our framework the BRs are 100\% in the $t+{\rm DM}$ channel, there are no relevant differences between different DM hypotheses. 
\item The contours obtained by rescaling the SUSY efficiencies with the XQ cross sections coincide quite well with the ``true'' XQ exclusion lines obtained by simulating XQ events. 
This means, efficiency maps or cross section upper limit maps for the stop--neutralino simplified model can safely be applied to the XQ case under consideration in this paper. It would thus be of advantage if the official maps by ATLAS and CMS extended to high enough masses to cover the 95\% CL reach for fermionic top partners, which is currently not the case.  
\end{enumerate}

\begin{figure}[ht!]\centering 
\includegraphics[width=0.48\textwidth]{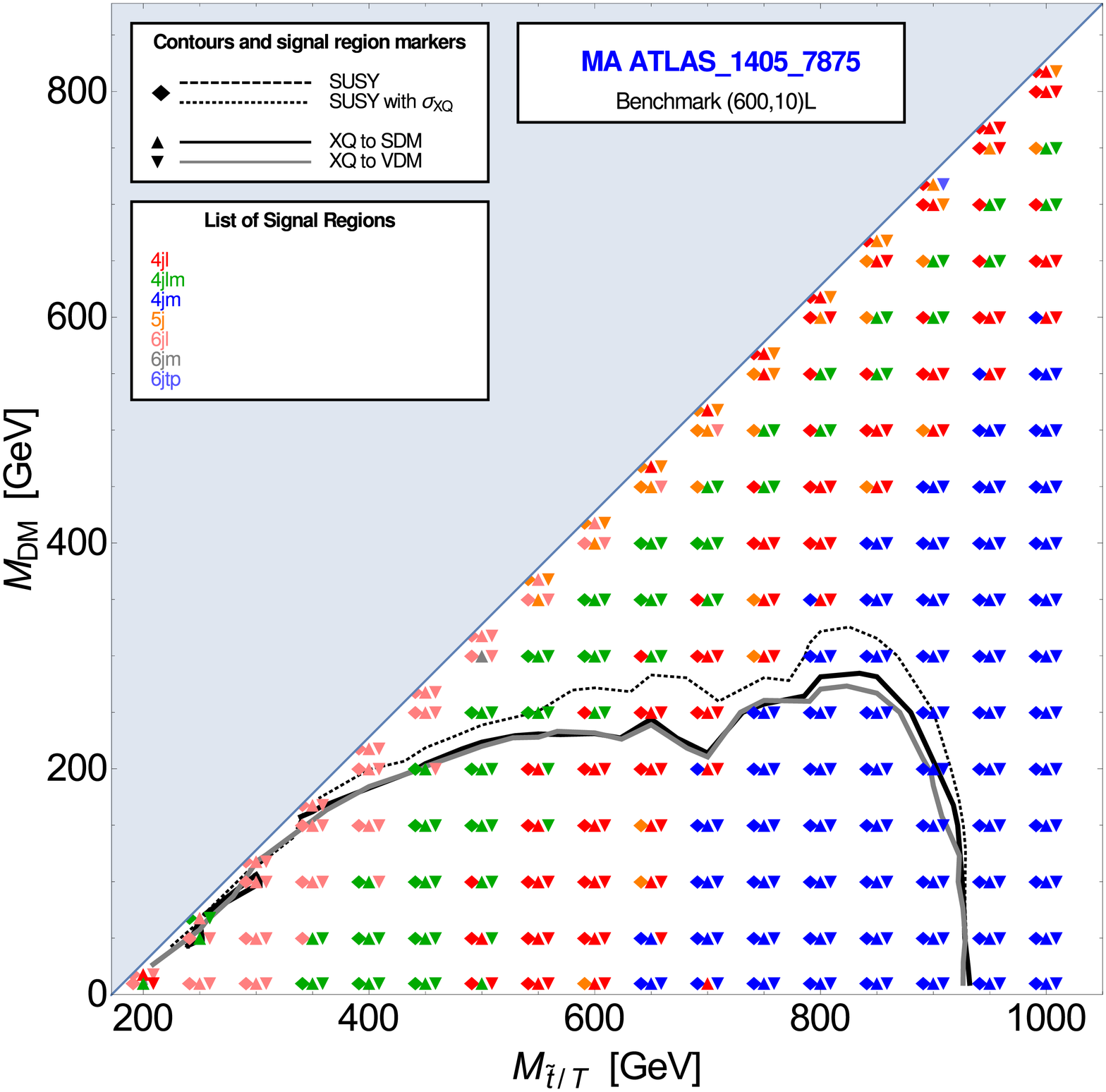}\quad%
\includegraphics[width=0.48\textwidth]{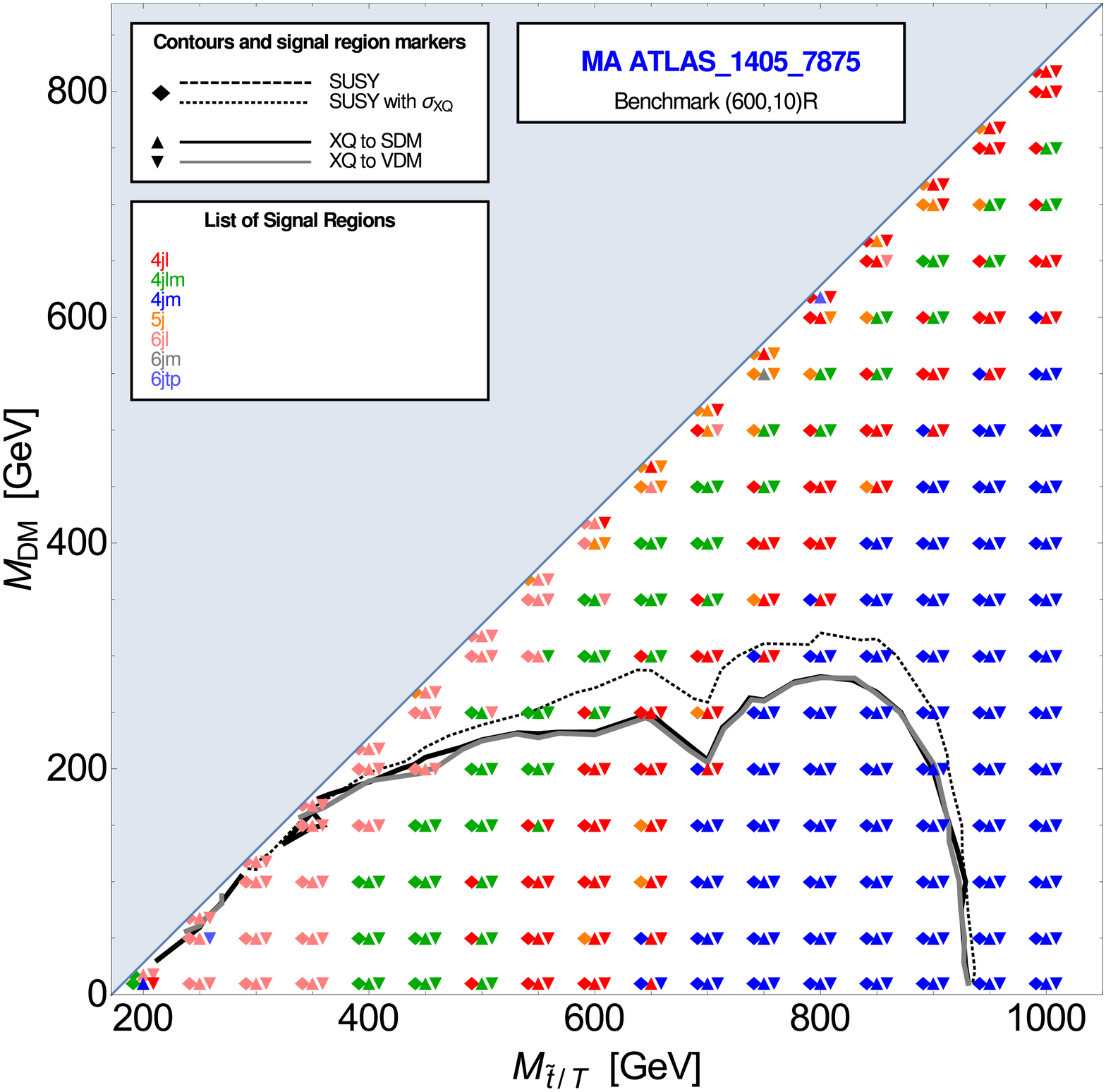}
\caption[Comparison of constraints in the top partner versus DM mass plane based on the {\sc MadAnalysis}\,5 recast code for the ATLAS gluino/squark search with 2--6 jets.]{Comparison of constraints in the top partner versus DM mass plane based on the {\sc MadAnalysis}\,5 recast code for the ATLAS gluino/squark search with 2--6 jets. As in Fig.~\ref{fig:contours1}, the various lines indicate the regions excluded at 8 TeV for the SUSY and XQ cases, and for the case where the SUSY efficiencies are applied to the XQ cross sections. The plots also contain the information which SRs are the most sensitive ones for each point of the scan. Note that no stop--neutralino mass limit is obtained from this analysis.}
\label{fig:contours2}
\end{figure}

\noindent 
The situation is different for the generic gluino/squark search in the multi-jet + $\ETmiss$ channel shown in Fig.~\ref{fig:contours2}.\footnote{To produce this figure, we have extended the {\sc MadAnalysis}\,5 recast code with the SRs {\tt 2jl}, {\tt 4jm} and {\tt 6jm}, which are not present in the PAD version \cite{MA5-ATLAS-multijet-1405}. We note, however, that these SRs could not be validated, as no cut-flows or kinematic distributions are available for them from ATLAS.}  
Contrary to the estimated stop mass limit of about 400--500~GeV in Table~\ref{tab:limits-gluino-squark},    
in the scan we do not obtain any limit on stops from this analysis. As already mentioned in Section~\ref{sec:gluinosquark}, the reason is that the efficiency of the $M_{\rm eff}$ cut strongly depends on the overall mass scale, rendering the extrapolation of the limit unreliable. This can also be seen from the fact that the most sensitive SR changes more rapidly with the top partner mass, see the colour code in Fig.~\ref{fig:contours2}. (The {\sc CheckMATE} implementation of the same analysis gives slightly stronger constraints on the SUSY case, excluding the region $m_{\tilde t}\approx 300-400$~GeV and $m_{\tilde\chi^0_1}\lesssim 50$~GeV, see the Appendix \ref{app:CM results}.)
Likewise, also the limit for the XQ case derived from the scan differs from the estimated one in Table~\ref{tab:limits-gluino-squark}, although here the effect goes in the opposite direction: the actual limit is stronger than the extrapolated one.
In fact, due to the increased efficiencies at high mass scales, this search can give stronger constraints on the XQ case than the stop searches, extending the limit up to $m_T\approx 900$--950~GeV for $m_{\rm DM}\lesssim 300$~GeV. 
The naive rescaling of SUSY efficiencies with XQ cross sections (dashed lines) however somewhat overestimates the reach for the XQ scenario. For this kind of analysis it will thus be interesting to produce efficiency maps specifically for the XQ model.

%\clearpage

%==============================================================================
\subsection{Conclusions}\label{sec:conclusions1}
%==============================================================================

We have studied how various analyses targeting $t\bar t +\MET$ signatures, carried out 
by ATLAS and CMS in the context of SUSY searches, perform for models with fermionic top partners. 
Taking a simplified XQ model with one extra $T$ quark and one DM state and comparing it to a 
simplified stop--neutralino model, we found that given the same kinematical configuration, SUSY and XQ 
efficiencies are very similar.  
The situation is different for generic multi-jet + $\MET$ searches targeting light-flavour squark and gluino production: 
here we found larger efficiencies for the SUSY than for the XQ case.  

Putting everything together, we conclude that cross section upper limit maps and efficiency maps obtained for stop simplified models in stop searches can also be applied to analogous models with fermionic top partners and a DM candidate, provided the NWA applies. An exception may be the region of very small mass differences, where uncertainties in the total cut efficiencies become sizeable, though this does not influence much the actual limit.\footnote{However, this region could become important for scenarios in which multiple degenerate or nearly degenerate top partners occur, as in this case the cross section might be enhanced by interference effects. Separate efficiency maps for the scalar or fermionic top partners would therefore be useful in this regime.} 
To fully exploit the applicability to different top partner models, we encourage the experimental collaborations to present their cross section upper limit and efficiency maps for a wide enough mass range, covering not only the reach for stops but also the reach for fermionic top partners. 
For the generic multi-jet + $\MET$ searches, on the other hand, it would be worthwhile to have efficiency maps specifically for the XQ model. 
As a service to the reader and potential user of our work, we provide the efficiency maps which we derived with {\sc CheckMATE} and {\sc MadAnalysis\,5} as auxiliary material\footnote{The efficiency maps can be downloaded from 
\href{http://lpsc.in2p3.fr/projects-th/recasting/susy-vs-vlq/ttbarMET/}{http://lpsc.in2p3.fr/projects-th/recasting/susy-vs-vlq/ttbarMET/}}. 
The numbers of expected background and observed events from the experimental analyses, needed for the statistical interpretation, are summarized in Appendix \ref{app:exp data}.

\begin{figure}[ht!]\centering 
\includegraphics[width=0.66\textwidth ,angle=270]{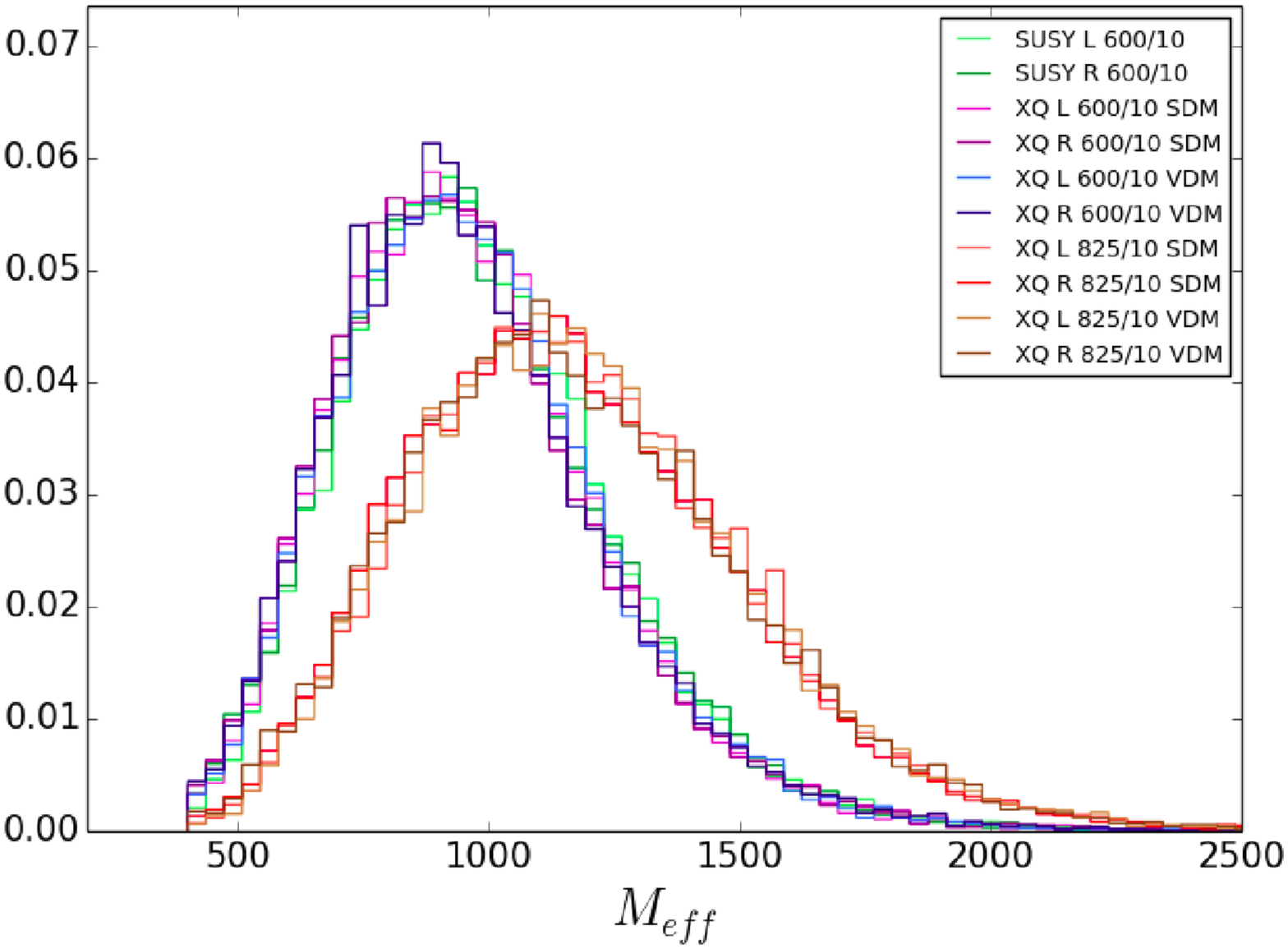}\
\caption[Comparison of the $M_{\rm eff}$ distributions for SUSY and XQ scenarios, after preselection cuts of the CMS 1-lepton stop search.]{Comparison of the $M_{\rm eff}$ distributions for SUSY and XQ scenarios, after preselection cuts of the CMS 1-lepton stop search~\cite{Chatrchyan:2013xna}. Here, $M_{\rm eff}$ is computed as $\sum p_T({\rm jets})+p_T(l)+\MET$. The green, violet and blue histograms are for the default (600,\,10) benchmark points, while the orange and brown histograms show XQ scenarios that would give roughly the same visible cross sections as the (600,\,10) SUSY cases.}
\label{fig:meff}
\end{figure}

The similarity of SUSY and XQ efficiencies also means that, should a signal be observed in $t\bar t +\MET$ events, it is not immediately obvious whether it comes from scalar or fermionic top partners. Since the production cross section (assumed here to be pure QCD) is significantly larger for fermionic than for scalar top partners, one way of discrimination may be to correlate the effective mass scale, $M_{\rm eff}$, or the effective transverse mass \cite{Cabrera:2012cj}, with the observed number of events, see Fig.~\ref{fig:meff} for an illustrative example. (This was also observed in \cite{Datta:2005vx}. However, as pointed out in \cite{Cheng:2005as}, for small XQ--DM mass splittings the decay products become softer and the discrimination from the SUSY case by cross section and $M_{\rm eff}$ is lost.)
Moreover, in the case of fermionic top partners, a corroborating signal may show up in generic gluino/squark searches, which have much less sensitivity to scalar top partners. Finally, the distinction between the two scenarios may be refined by considering special kinematic distributions as discussed in \cite{Han:2008gy,Smillie:2005ar,Datta:2005zs}.

%% ------------- --------------------- --------------------- ---------------------- --------------------- ---------------------- ----------------------

\section{Large width effect on production and decay of XQs decaying to DM} \label{sec:VLQDM_width}

In this section we will present the work done in \cite{Moretti:2017qby} where we study large width effects similar to the ones described in Sec. \ref{sec:VLQ_width} but in the case of XQ decaying to DM instead of SM particules. 

We will focus on a simplified scenario where a top-like XQ interacts with SM quarks and DM candidates and its width is large relatively to its mass (up to $40\%$ of it). We will consider final states compatible with processes of pair production of the $T$ and subsequent decay into a bosonic DM candidate and a SM quark. Then, we will evaluate the effects of large width in the determination of the cross section and in the reinterpretation of bounds from experimental searches. We will distinguish scenarios with a scalar DM from scenarios with a vector DM and we will analyse in detail scenarios where the $T$ state interacts either with the SM up or top quark, such that the final states we will consider are either $2j+\MET$ or $t\bar t + \MET$, respectively. For scenarios where $T$ interacts with the charm quark, leading to a final state analogous to the case of the up quark in terms of reconstructed objects if charm-tagging is not considered, only the main results will be provided. It is important to notice that, unlike in the case of scenarios where the XQs decay only into SM states~\cite{Moretti:2016gkr}, interference terms with the SM background are absent if the XQs decay to DM candidates, as the only (irreducible) source of $\MET$ in the SM is given by final states containing neutrinos.

%%%%%%%%%%%%%%%%%%%%%%%%%%%%%%%%%%%%%%%%%%%%%%%%%%%%%%%%%%%%%%%%%%%%%%%%%
%%%%%%%%%%%%%%%%%%%%%%%%%%%%%%%%%%%%%%%%%%%%%%%%%%%%%%%%%%%%%%%%%%%%%%%%%
\subsection{Model and conventions}
\label{sec:Conventions}

%--------------------------------------------------------------- \subsubsection{Lagrangian terms}
%---------------------------------------------------------------

We concentrate once again on a top partner XQ $T$ and a DM real scalar or real vector singlet respectively called $S^0_{\rm DM}$ and $V^{0\mu}_{\rm DM}$. This time we will consider XQs coupling to different SM quark generation so our Lagrangians takes the following expression:
\begin{eqnarray}
\Lag^S_1 &=& 
\left[
\lambda_{11}^{u^i} \bar{T}_L \, u^i_R +
\lambda_{21}^i \, \overline\Psi_{1/6,R} {u^i \choose d^i}_L 
\right] 
S^0_{\rm DM} + {\rm h.c.} 
\\
\Lag^V_1 &=& 
\left[
g_{11}^{u^i} \bar{T}_R \gamma_\mu \, u^i_R + 
g_{21}^i  \, \overline\Psi_{1/6,L} \gamma_\mu {u^i \choose d^i}_L 
\right] 
V^{0\mu}_{\rm DM} + {\rm h.c.},
\end{eqnarray}
where the different notations were already presented in Sec. \ref{sec:XQDM Lagrangian}.

In the following analysis we will expore in detail scenarios where the $T$ has a purely left-handed coupling ({\it i.e.} it belongs to a VLQ doublet), but we will show (for specific benchmarks) how the experimental limits change in the large width regime when considering alternative hypotheses, such as pure right-handed couplings (VLQ singlet) or couplings where the left- and right-handed components are equal in size with same or opposite sign (ChQ scenarios).

%---------------------------------------------------------------
\subsubsection{Observables and conventions}
%---------------------------------------------------------------

To understand the effects of large widths on the signal, we will consider two different processes, both leading to the same four-particle final state ${\rm DM} \; q \ {\rm DM} \; \bar{q} \equiv q\bar q+\MET$, where $q(\bar q)$ is an ordinary SM (anti)quark. These processes are similar to the ones defined in Section \ref{sec:setup} in the case of visible decay.

\begin{itemize}

\item The \textit{QCD pair production and decay of on-shell XQs} as usually considered in experimental searches. In the NWA, it is possible to separate production and decay of the heavy quarks, thus allowing for a model independent analysis of the results. The cross section for this process is given by (hereafter, in our formulae, $Q$ denotes an XQ):
\begin{equation}
 \sigma_X \equiv \sigma_{2 \to 2}~{\rm BR}(Q)~{\rm BR}(\bar Q)
\end{equation}
where, for simplicity, $\sigma_{2 \to 2}$ only takes into account the dominant (pure) QCD topologies. This factorisation of production and decay only makes sense in  NWA so this process is {\sl dynamically} independent of the width, i.e., $\sigma_X\equiv \sigma_X (M_Q)$, though $\Gamma_Q$ obviously enter in the definition of the BRs of $Q$ and $\bar Q$. 

\item The \textit{full signal} where all the topologies which lead to the same four-particle final state and contain \textit{at least one} XQ propagator are taken into account. The only assumption we make, to allow a consistent comparison with the NWA results, is that the order of the QCD $\alpha_s$ in the full signal topologies is the same as in the NWA case. 
The pair production and decay topologies are included, but for the full signal the XQs are not strictly required to be on-shell. Furthermore, diagrams with only one XQ propagator are also included. We stress that the NWA limit is indeed recovered when the XQ width becomes small with respect to its mass: in this limit, factorisation of production and decay can still be done, as the contribution of all the subleading topologies considered in the full signal becomes negligible and the dominant contribution is given only by pair-production topologies where the XQ is on-shell. If the XQ width is large with respect to its mass, the contribution of other topologies becomes relevant and the factorisation is not possible anymore. 
Hence, this approach, on the one hand, describes  accurately scenarios where the widths of the XQs are large and, on the other hand, is fully gauge invariant (like the NWA approach). Furthermore, it takes into account the spin correlations between the $Q$ quark and antiquark decay branches, which are lost in the NWA. 
The cross section of this process will be labelled as $\sigma_S$ and depends  upon both the mass and width of the XQ: $\sigma_S\equiv\sigma_S (M_Q, \Gamma_Q)$. Some example topologies for this process, which are not included in the previous one, are given in Fig.~\ref{fig:fullsignaltopologies2}.

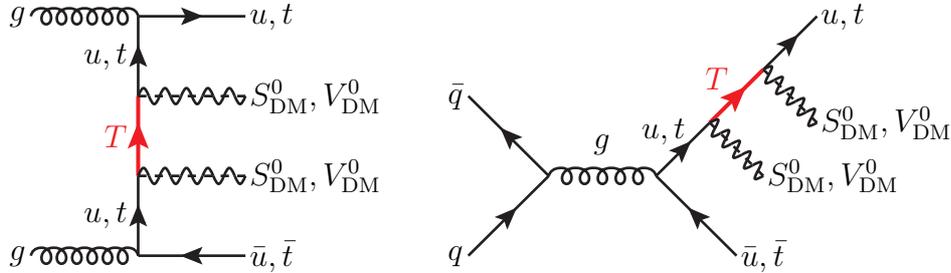
\begin{figure}[ht!]
\begin{center}
\begin{picture}(140,10)(0,0)
\SetWidth{1}
\Gluon(10,0)(50,0){3}{6}
\Text(8,0)[rc]{\large $g$}
\Gluon(10,90)(50,90){3}{6}
\Text(8,90)[rc]{\large $g$}
\Line[arrow](90,0)(50,0)
\Text(92,0)[lc]{\large $\bar u, \bar t$}
\Line[arrow](50,0)(50,30)
\Text(46,15)[rc]{\large $u,t$}
\SetColor{Red}\SetWidth{1.5}
\Line[arrow](50,30)(50,60)
\Text(46,45)[rc]{\large\Red{$T$}}
\SetColor{Black}\SetWidth{1}
\Line[arrow](50,60)(50,90)
\Text(46,75)[rc]{\large $u,t$}
\Line[arrow](50,90)(90,90)
\Text(92,90)[lc]{\large $u,t$}
\Line[dash](50,30)(90,30)
\Photon(50,30)(90,30){3}{5}
\Text(92,30)[lc]{\large $S^0_{\rm DM},V^0_{\rm DM}$}
\Line[dash](50,60)(90,60)
\Photon(50,60)(90,60){3}{5}
\Text(92,60)[lc]{\large $S^0_{\rm DM},V^0_{\rm DM}$}
\end{picture}
\hskip 20pt
\begin{picture}(140,100)(0,0)
\SetWidth{1}
\Line[arrow](10,0)(40,30)
\Text(8,0)[rc]{\large $q$}
\Line[arrow](40,30)(10,60)
\Text(8,60)[rc]{\large $\bar q$}
\Gluon(40,30)(80,30){3}{5}
\Text(60,38)[cb]{\large $g$}
\Line[arrow](80,30)(100,50)
\Text(83,47)[cc]{\large $u,t$}
\Line[arrow](110,0)(80,30)
\Text(112,0)[lc]{\large $\bar u, \bar t$}
\SetColor{Red}\SetWidth{1.5}
\Line[arrow](100,50)(120,70)
\Text(103,67)[cc]{\large\Red{$T$}}
\SetColor{Black}\SetWidth{1}
\Line[dash](100,50)(120,30)
\Photon(100,50)(120,30){3}{5}
\Text(122,30)[lc]{\large $S^0_{\rm DM},V^0_{\rm DM}$}
\Line[arrow](120,70)(140,90)
\Text(142,90)[lc]{\large $u,t$}
\Line[dash](120,70)(140,50)
\Photon(120,70)(140,50){3}{5}
\Text(142,50)[lc]{\large $S^0_{\rm DM},V^0_{\rm DM}$}
\end{picture}
\end{center}
\caption{Examples of topologies containing only one XQ propagator for final states compatible with XQ pair production and decay into scalar or vector DM and SM quarks of first or third generation.}
\label{fig:fullsignaltopologies2}
\end{figure}

\end{itemize}

In order to determine the difference between the two approaches above, we will consider the variable $(\sigma_S - \sigma_X)/\sigma_X$. This ratio takes into account effects of both the off-shellness of $T$ and $\bar T$ in their pair production as well as contributions given by topologies which contain at least one XQ propagator (including interference between the two). It measures in practice how much the full signal differs from the approximate pair-production-plus-decay one computed in the NWA.

%%%%%%%%%%%%%%%%%%%%%%%%%%%%%%%%%%%%%%%%%%%%%%%%%%%%%%%%%%%%%%%%%%%%%%%%%
\subsubsection{Channels \label{sec:Channels}}

In the present analysis we consider the processes of production of a heavy top-like quark $T$. In principle, from a model independent point of view, the $T$ quark is allowed to interact with all SM quark generations, but to evaluate the effects of large widths in different scenarios, only specific interactions will be switched on in the different scenarios we will consider. 

Since the purpose of this analysis is to evaluate the effects of large widths on channels commonly explored by experimental analysis, we will consider only final states allowed by $T$ pair production and decay. The full set of channels in which a pair-produced $T$ quark can decay is given by the following matrix:
\begin{eqnarray}
\footnotesize
T\bar T \to \left(
\begin{array}{ccc|ccc}
S^0_{\rm DM}u\; S^0_{\rm DM}\bar u & \ \; S^0_{\rm DM}u\; S^0_{\rm DM}\bar c & \ \; S^0_{\rm DM}u\; S^0_{\rm DM}\bar t & \ \; S^0_{\rm DM}u\; V^0_{\rm DM}\bar u & \ \; S^0_{\rm DM}u\; V^0_{\rm DM}\bar c & \ \; S^0_{\rm DM}u\; V^0_{\rm DM}\bar t \\
S^0_{\rm DM}c\; S^0_{\rm DM}\bar u & \ \; S^0_{\rm DM}c\; S^0_{\rm DM}\bar c & \ \; S^0_{\rm DM}c\; S^0_{\rm DM}\bar t & \ \; S^0_{\rm DM}c\; V^0_{\rm DM}\bar u & \ \; S^0_{\rm DM}c\; V^0_{\rm DM}\bar c & \ \; S^0_{\rm DM}c\; V^0_{\rm DM}\bar t \\ 
S^0_{\rm DM}t\; S^0_{\rm DM}\bar u & \ \; S^0_{\rm DM}t\; S^0_{\rm DM}\bar c & \ \; S^0_{\rm DM}t\; S^0_{\rm DM}\bar t & \ \; S^0_{\rm DM}t\; V^0_{\rm DM}\bar u & \ \; S^0_{\rm DM}t\; V^0_{\rm DM}\bar c & \ \; S^0_{\rm DM}t\; V^0_{\rm DM}\bar t \\ 
\hline
V^0_{\rm DM}u\; S^0_{\rm DM}\bar u & \ \; V^0_{\rm DM}u\; S^0_{\rm DM}\bar c & \ \; V^0_{\rm DM}u\; S^0_{\rm DM}\bar t & \ \; V^0_{\rm DM}u\; V^0_{\rm DM}\bar u & \ \; V^0_{\rm DM}u\; V^0_{\rm DM}\bar c & \ \; V^0_{\rm DM}u\; V^0_{\rm DM}\bar t \\
V^0_{\rm DM}c\; S^0_{\rm DM}\bar u & \ \; V^0_{\rm DM}c\; S^0_{\rm DM}\bar c & \ \; V^0_{\rm DM}c\; S^0_{\rm DM}\bar t & \ \; V^0_{\rm DM}c\; V^0_{\rm DM}\bar u & \ \; V^0_{\rm DM}c\; V^0_{\rm DM}\bar c & \ \; V^0_{\rm DM}c\; V^0_{\rm DM}\bar t \\ 
V^0_{\rm DM}t\; S^0_{\rm DM}\bar u & \ \; V^0_{\rm DM}t\; S^0_{\rm DM}\bar c & \ \; V^0_{\rm DM}t\; S^0_{\rm DM}\bar t & \ \; V^0_{\rm DM}t\; V^0_{\rm DM}\bar u & \ \; V^0_{\rm DM}t\; V^0_{\rm DM}\bar c & \ \; V^0_{\rm DM}t\; V^0_{\rm DM}\bar t  
\end{array}
\right) \notag
\label{eq:finalstates2}
\end{eqnarray}

To limit ourselves to representative and simple scenarios, we will focus on the diagonal terms of this matrix and analyse in detail XQs coupling either to first or third generation quarks (though the main results for couplings with second generation will also be provided). 
Effects of large width are different depending on the kinematics of the process and by selecting representative scenarios it is always possible to reconstruct intermediate configurations (XQs interacting partly with heavy and partly with light SM generations).

This analysis is of phenomenological interest only for mass values for which the number of final events is (ideally) larger than 1. We have seen in Fig. \ref{fig:Xsigma} from Sec. \ref{sec:Benchmarks} that the ideal practical validity of our results is limited to mass values of around 1500 GeV for LHC@8TeV, 2500 GeV (2700 GeV) for LHC@13TeV with 100/fb (300/fb) integrated luminosity. Of course, we are not considering here effects due to experimental acceptances and efficiencies: this study is only meant to assess the role of the complete signal with respect to the common approximations made in theoretical and experimental analyses.

%%%%%%%%%%%%%%%%%%%%%%%%%%%%%%%%%%%%%%%%%%%%%%%%%%%%%%%%%%%%%%%%%%%%%%%%%
%%%%%%%%%%%%%%%%%%%%%%%%%%%%%%%%%%%%%%%%%%%%%%%%%%%%%%%%%%%%%%%%%%%%%%%%%
\subsection{Analysis tools and experimental searches\label{sec:Monte Carlo}}

As intimated, herein, we want to study the ratio of cross sections $(\sigma_S - \sigma_X)/\sigma_X$ (where we recall that $\sigma_S$ corresponds to the full signal and $\sigma_X$ to the NWA) as well as understand which influence the width of the XQ, in turn triggering the contribution of the forementioned new topologies not present in pair production, can have on its mass bounds. To do so we consider an XQ top partner belonging to the doublet representation $\Psi_{1/6}=(T \ B)^T$ (corresponding to pure left-handed couplings in Eqs. \eqref{eq:LagSingletDMS} and \eqref{eq:LagSingletDMV}) and scan over the parameters $M_T$, $M_{\rm DM}$ and $\Gamma_T$.

For our simulation we analyse in detail scenarios where the DM state has masses $M_{\rm DM}$ = 10 GeV, 500 GeV and 1000 GeV and with an XQ of mass $M_T > M_{\rm DM} + m_q$, with $q \in \{u,c,t\}$ (such that its on-shell decay is kinematically allowed) up to $M_T^{\rm max}$ = 2500 GeV, which is the maximal value of a $T$ mass so that it can be produced for LHC@13TeV with 100/fb integrated luminosity as shown in Fig. \ref{fig:Xsigma}. We also consider values of the $T$ width from $\Gamma_T / M_T$ $\simeq$ 0\% (NWA) to 40\% of the $T$ mass.

Our numerical results at partonic level are obtained using {\sc MadGraph5} \cite{Alwall:2011uj,Alwall:2014hca} and a model we implemented in {\sc Feynrules} \cite{Alloul:2013bka} to obtain the UFO interface format. The model we used is the same as the one in the analysis of Ref.\cite{Kraml:2016eti}. For the MC simulation we use the PDF set {\sc cteq6l1}~\cite{Pumplin:2002vw}. Events are then passed to {\sc Pythia}\,8~\cite{Sjostrand:2007gs,Sjostrand:2006za}, which takes care of the hadronisation and parton showering.

To analyse and compare the effects of a set of 13 TeV analyses considering final states compatible with our scenarios, we employ {\sc CheckMATE 2}~\cite{Dercks:2016npn}, which uses the {\sc Delphes\,3} \cite{deFavereau:2013fsa} framework for the emulation of detector effects. In our simulations we include all the ATLAS and CMS (carried out at 13 TeV) analyses available within the CheckMATE database but we will only list here the most relevant ones for our study. These analysis are the following ATLAS searches: 
\begin{itemize}
\item ATLAS 1604.07773 \cite{Aaboud:2016tnv}, a search for new phenomena in final states with an energetic jet and large missing transverse momentum,
\item ATLAS 1605.03814 \cite{Aaboud:2016zdn}, a search for squarks and gluinos in final states containing hadronic jets, missing transverse momentum but no electrons or muons, 
\item ATLAS-CONF-2016-050 \cite{ATLAS:2016ljb}, a search for the stop in final states with one isolated electron or muon, jets and missing transverse momentum.
\end{itemize}

%%%%%%%%%%%%%%%%%%%%%%%%%%%%%%%%%%%%%%%%%%%%%%%%%%%%%%%%%%%%%%%%%%%%%%%%%
%%%%%%%%%%%%%%%%%%%%%%%%%%%%%%%%%%%%%%%%%%%%%%%%%%%%%%%%%%%%%%%%%%%%%%%%%
\subsection{Extra $T$ quark interacting with Dark Matter and the SM top quark}
\label{sec:Tt}

In this section we will study the case of XQs coupling to third generation SM quarks only. The possible decay channels are therefore $t \bar t + \{S^0_{\rm DM} S^0_{\rm DM}, V^0_{\rm DM} V^0_{\rm DM}\}$, {i.e.}  $t \bar t + \MET$. We start from this channel because, from a theoretical point of view, the top quark is considered the most likely to be affected by new physics phenomena.

%%%%%%%%%%%%%%%%%%%%%%%%%%%%%%%%%%%%%%%%%%%%%%%%%%%%%%%%%%%%%%%%%%%%%%%%%

\subsubsection{Large width effects at parton level}
\label{sec:Parton3}

\begin{figure}[ht!]
\centering
\epsfig{file=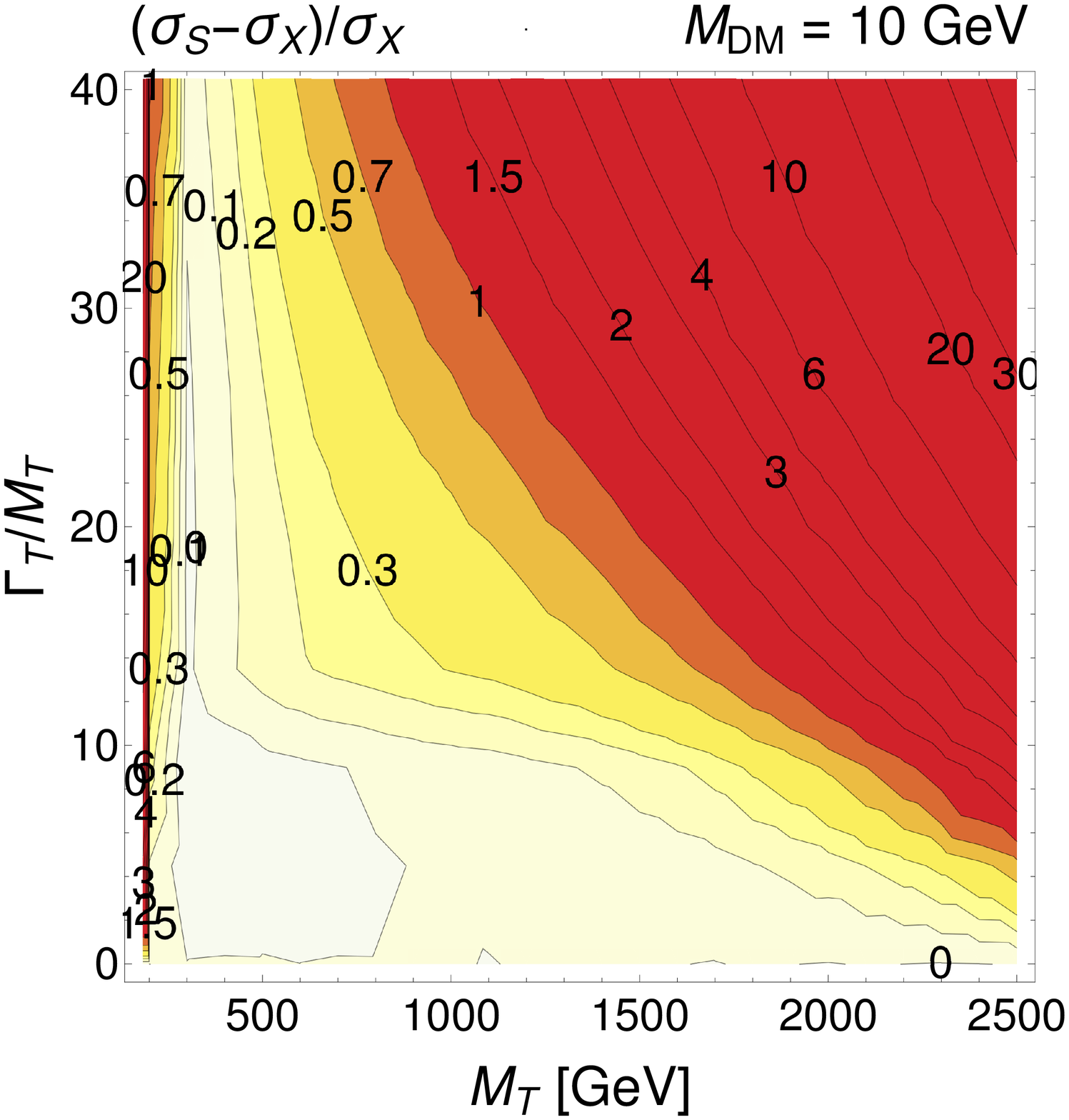,   width=.32\textwidth} 
\epsfig{file=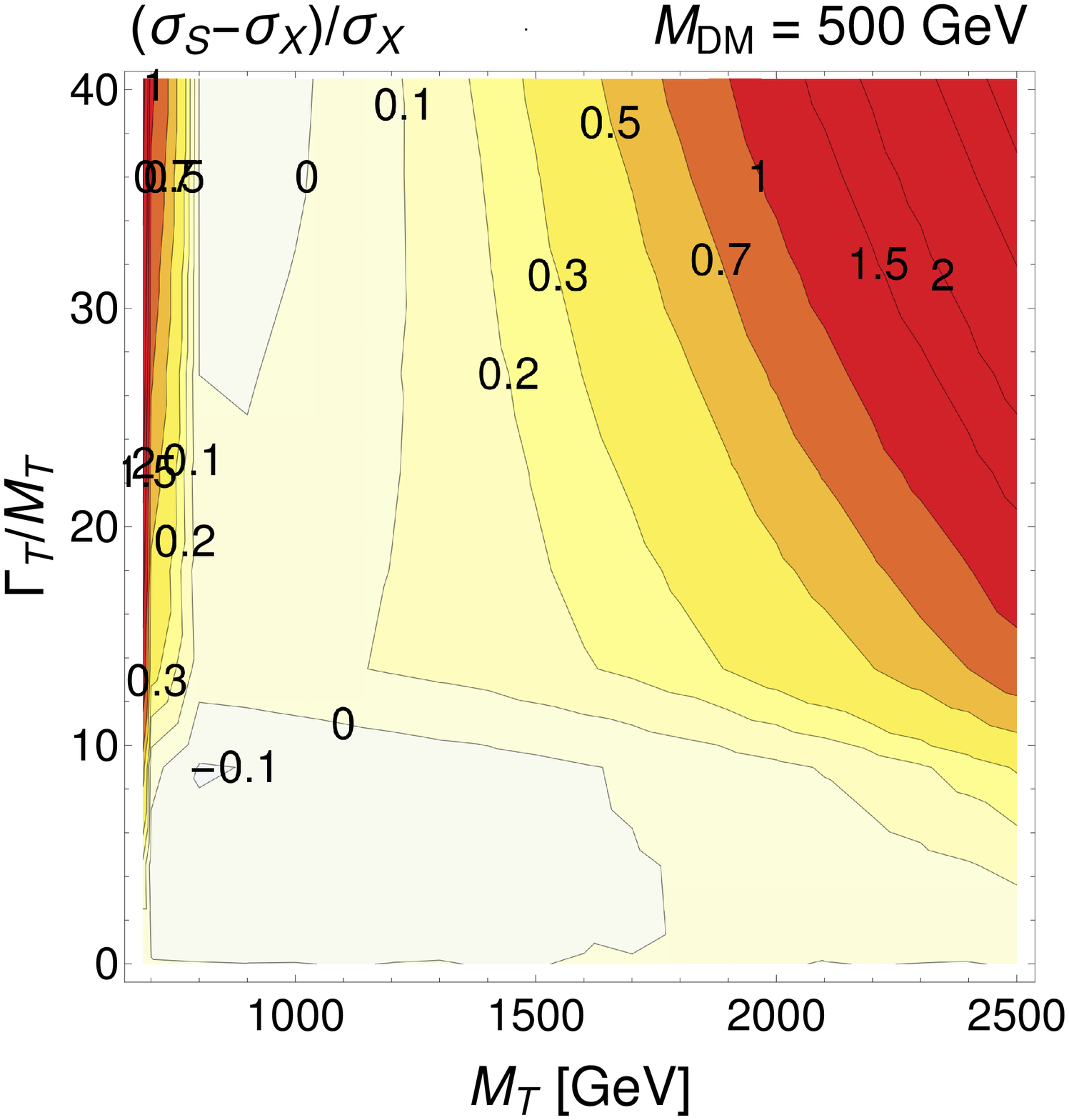,  width=.32\textwidth} 
\epsfig{file=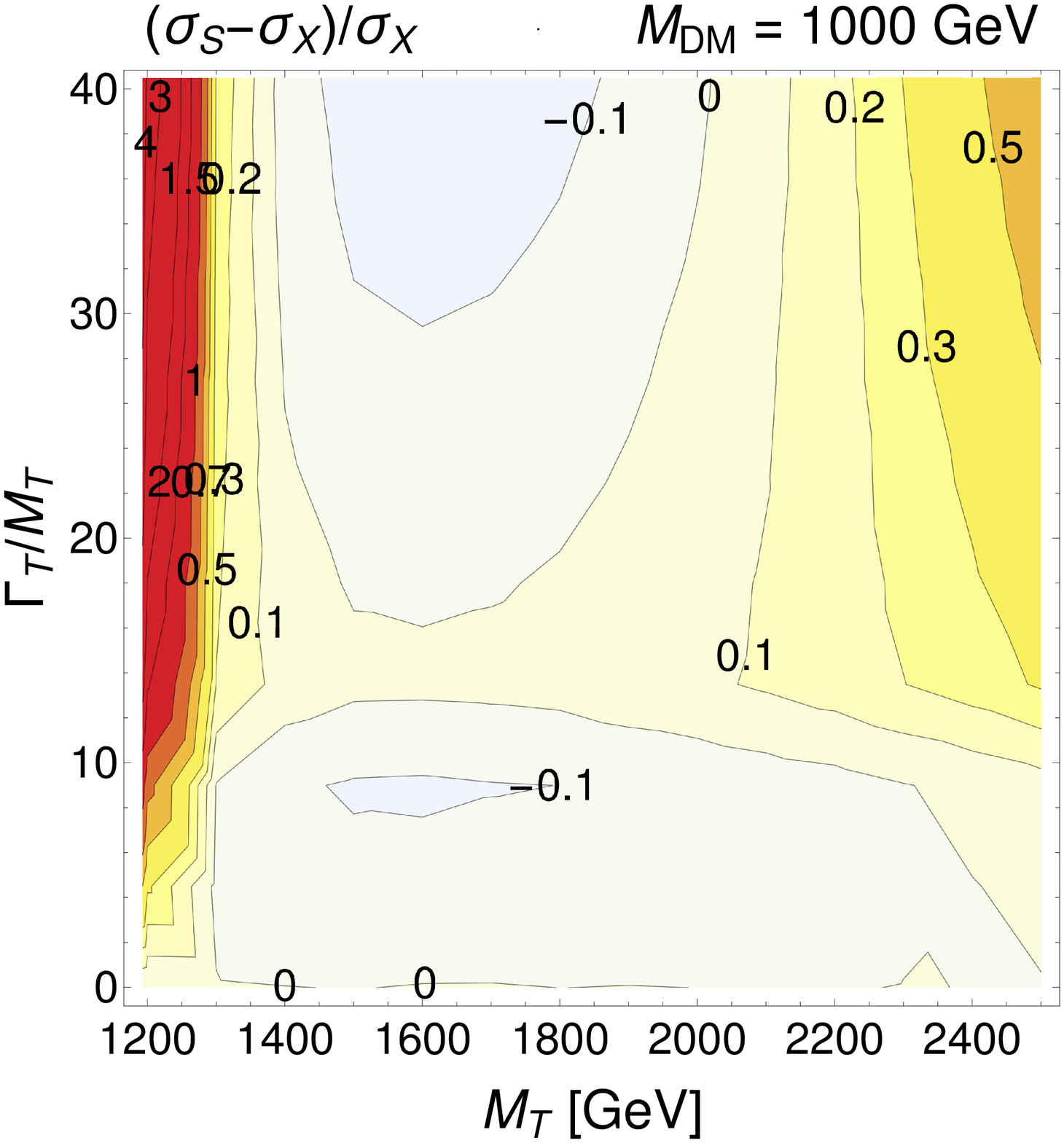, width=.32\textwidth}\\[5pt]  
\epsfig{file=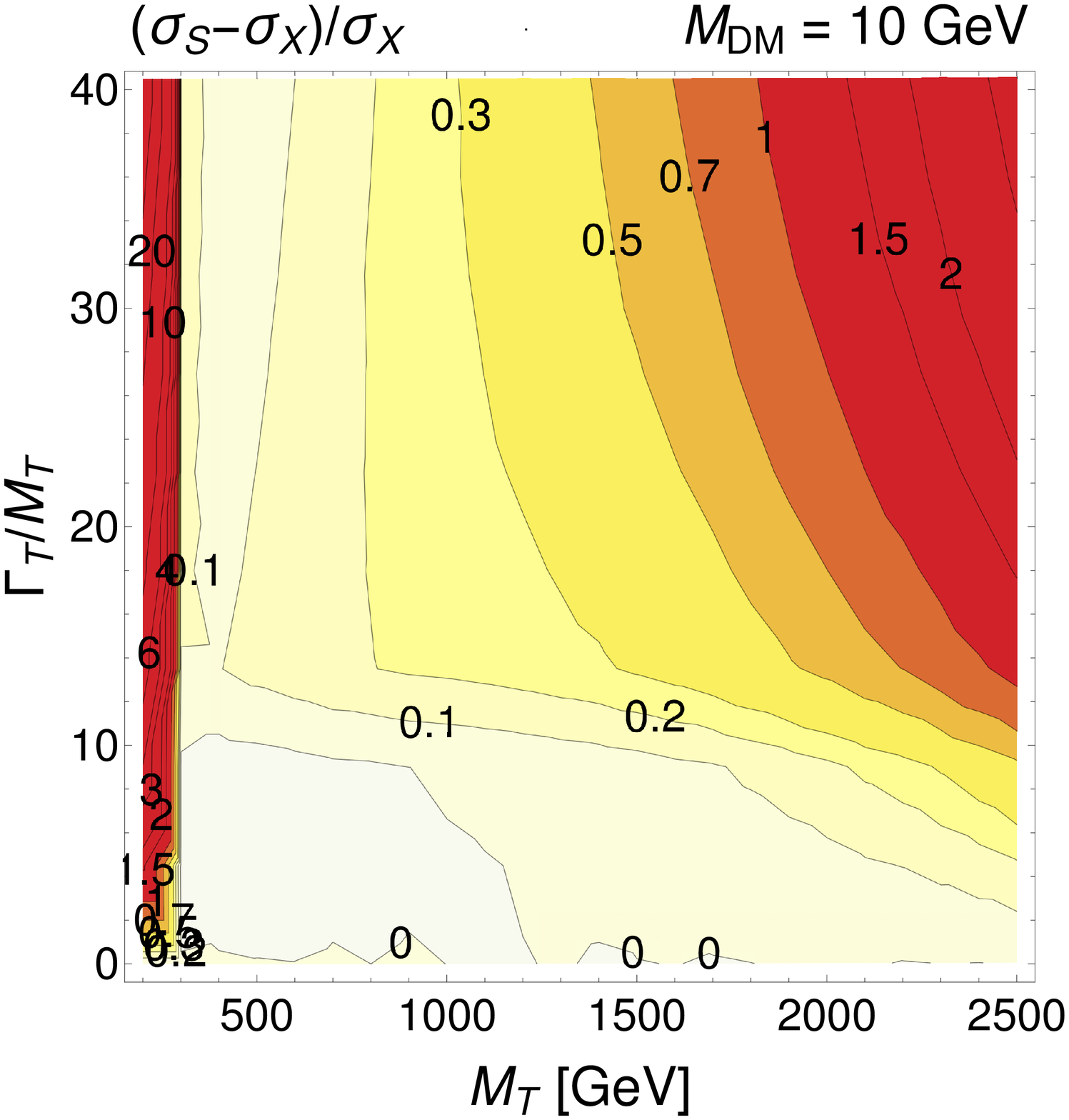,   width=.32\textwidth} 
\epsfig{file=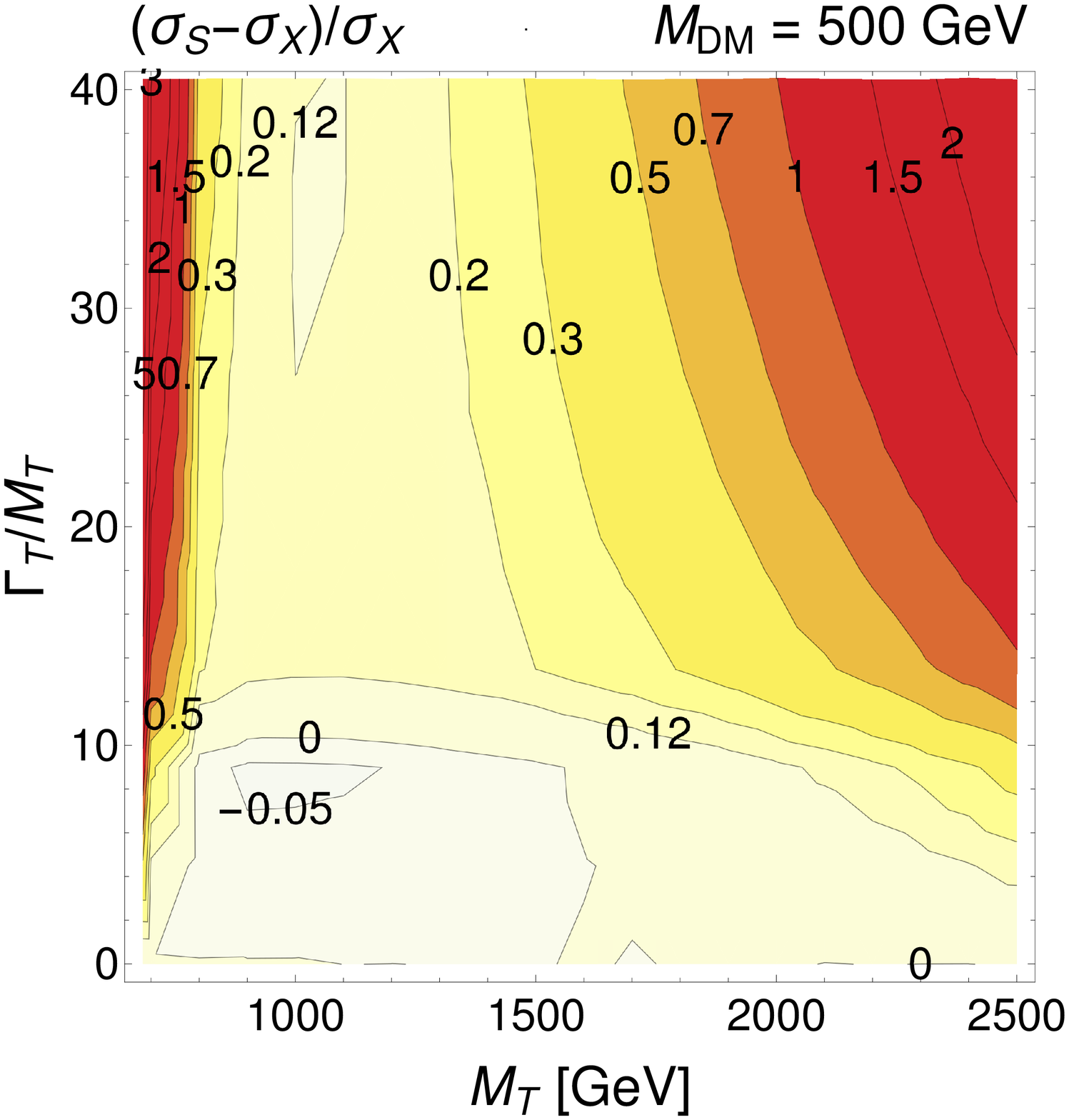,  width=.32\textwidth} 
\epsfig{file=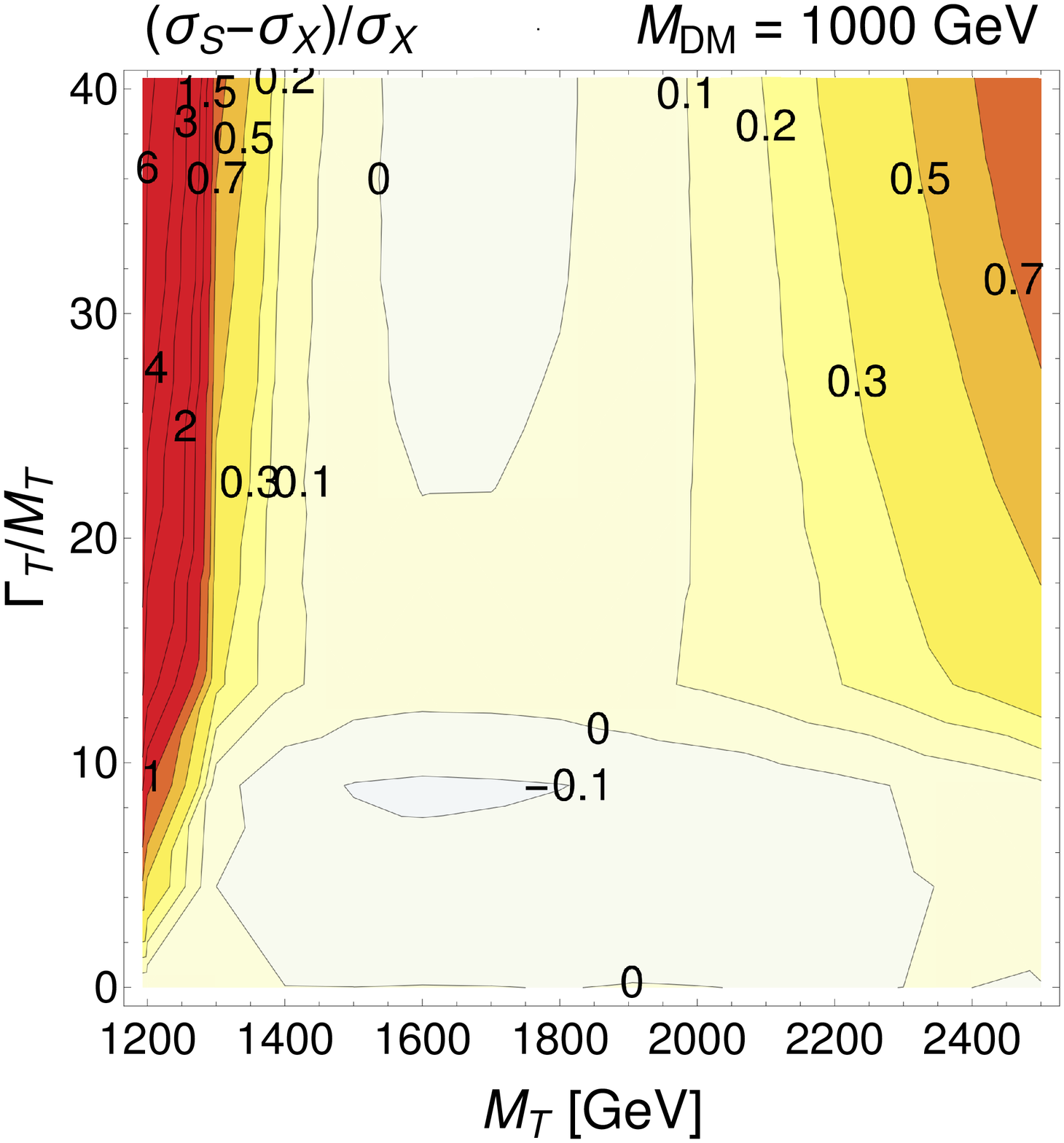, width=.32\textwidth} 
\caption[Relative difference between the full signal and the QCD pair production cross sections for a $T$ coupling to a DM particle (coupling to third generation) of mass 10 GeV, 500 GeV and 1000 GeV.]{Relative difference between the full signal and the QCD pair production cross sections for a $T$ coupling to a DM particle (coupling to third generation) of mass 10 GeV, 500 GeV and 1000 GeV. Top row: scalar DM; bottom row, vector DM.}
\label{fig:SXthird2}
\end{figure}

In Fig.~\ref{fig:SXthird2} the relative differences between the full signal and the QCD pair production cross sections $(\sigma_S - \sigma_X)/\sigma_X$ are plotted for an LHC energy of 13 TeV. Notice that here and in the following we do not apply cuts on $\MET$ at parton level.

\begin{figure}[ht!]
\centering
\subfigure[Scalar DM: $M_{\rm DM}=1$ TeV, $M_T=2$ TeV]{
\epsfig{file=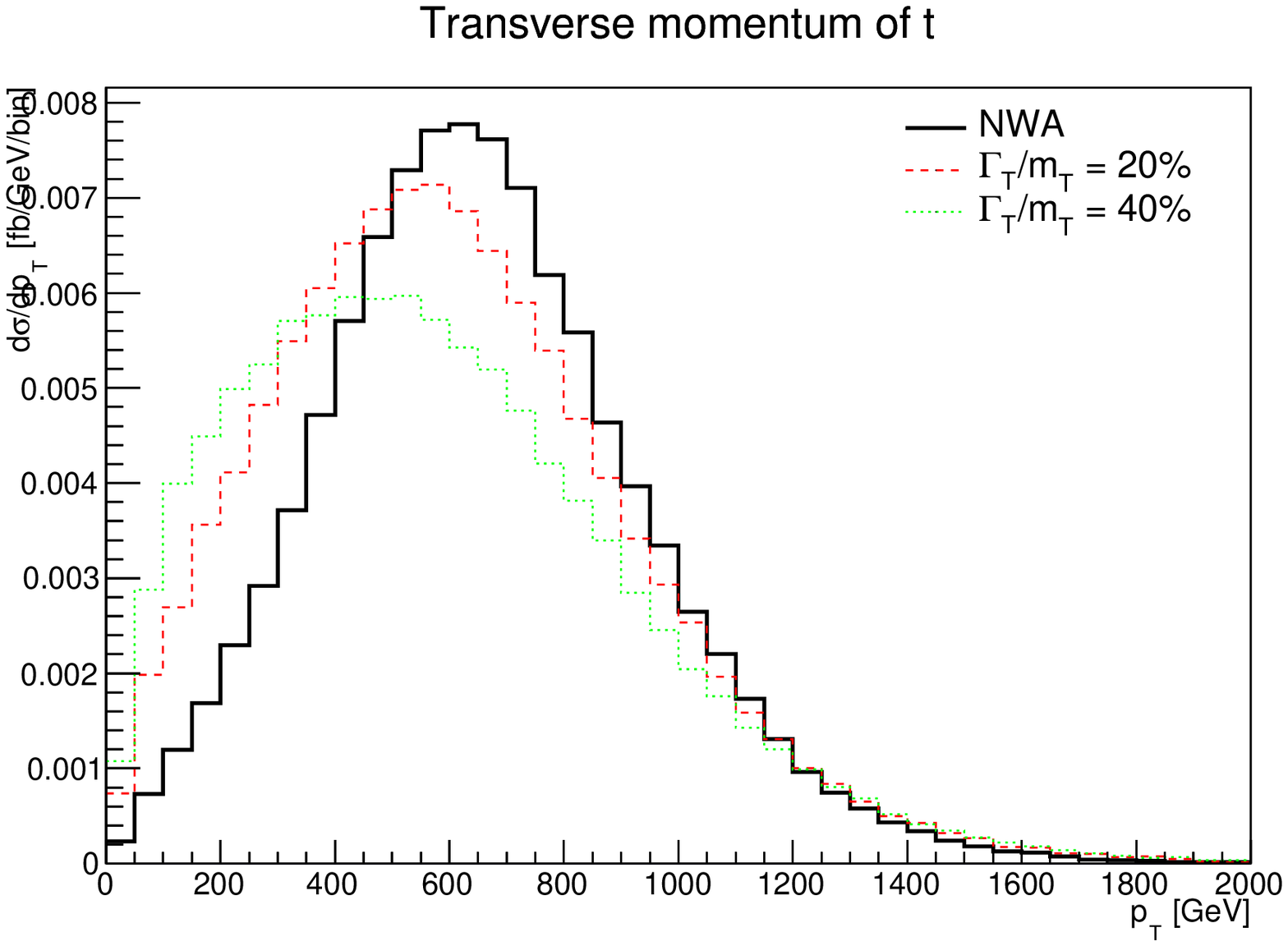, width=.45\textwidth} 
\epsfig{file=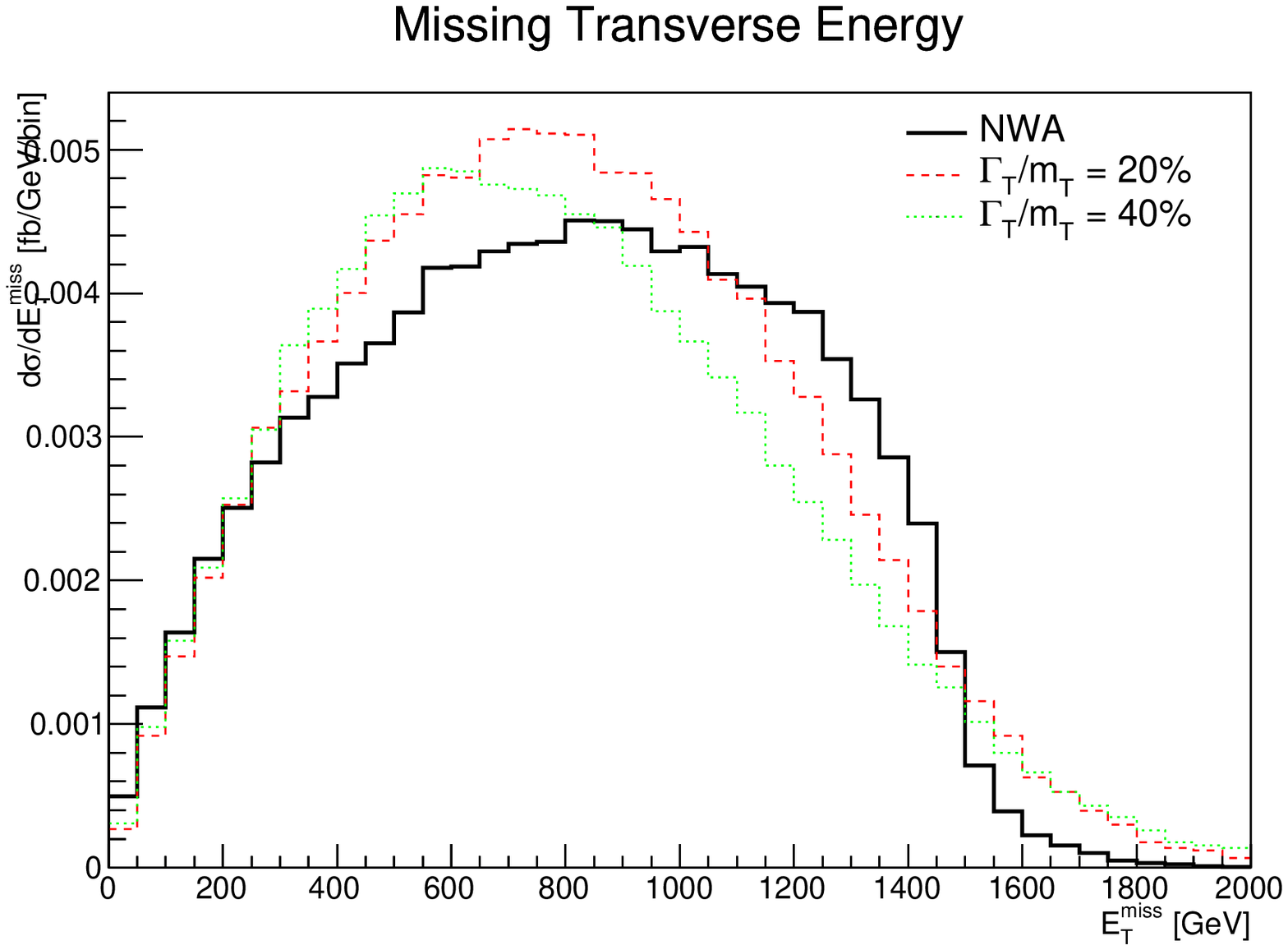, width=.45\textwidth}
}\\
\subfigure[Vector DM: $M_{\rm DM}=10$ GeV, $M_T=400$ GeV]{
\epsfig{file=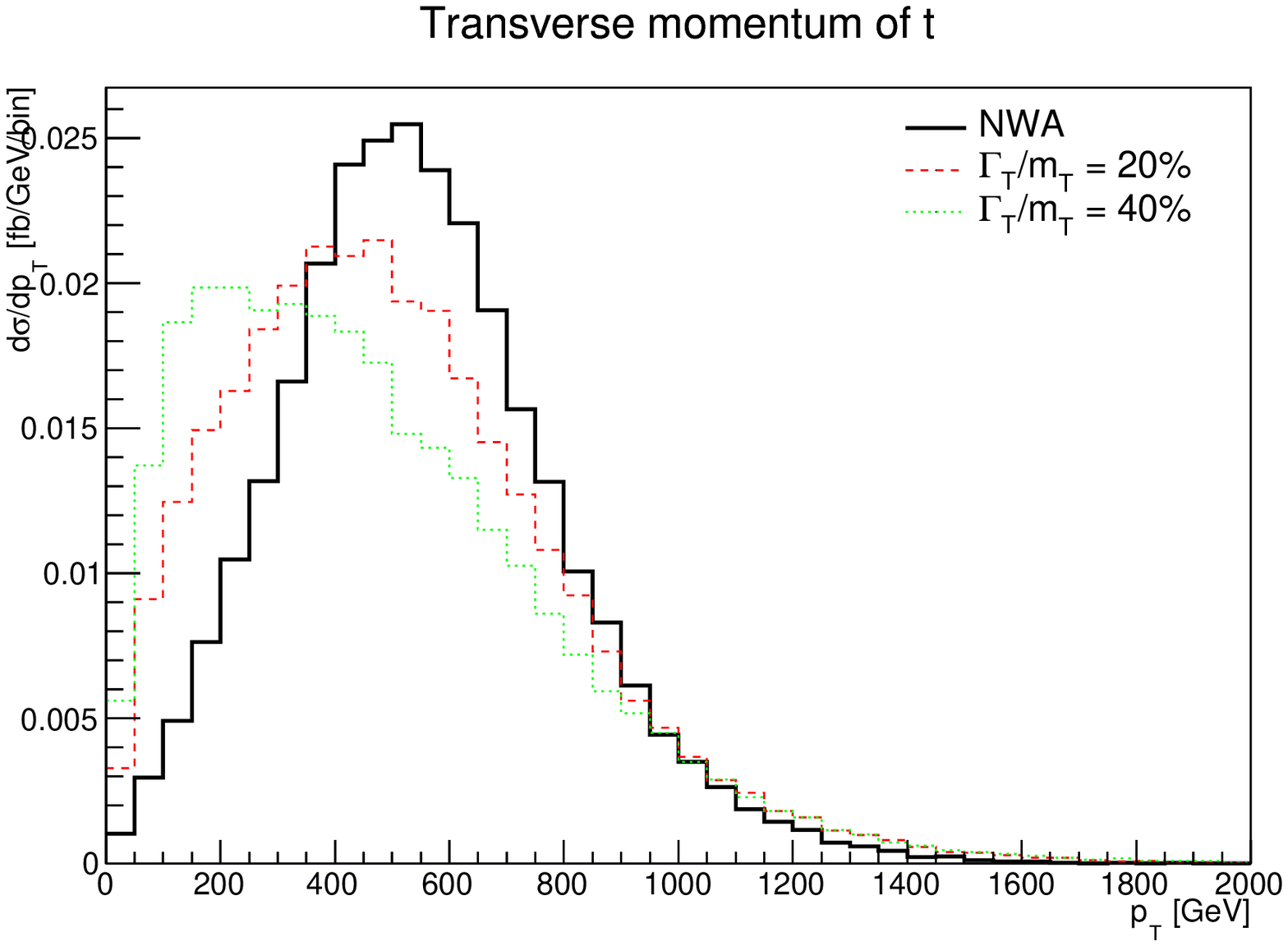, width=.45\textwidth} 
\epsfig{file=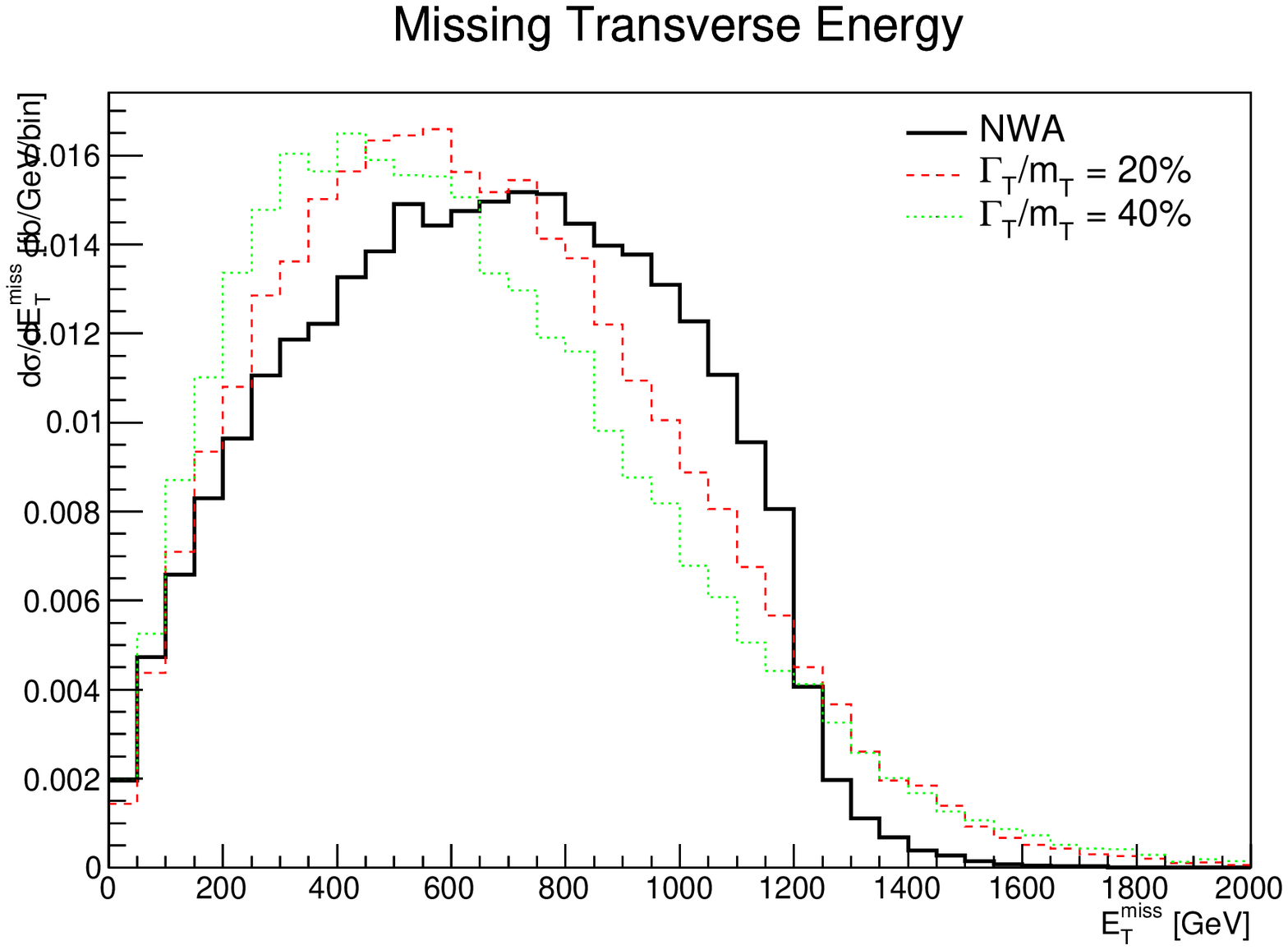, width=.45\textwidth}
}
\caption{Differential distributions of transverse momentum of the top quark and $\MET$ along the cancellation line for scalar and vector DM.}
\label{fig:DistributionCancellation}
\end{figure}

A number of conclusions can be derived from the observation of these results:
\begin{itemize}
\item As expected, and as a health check of our results, in the NWA limit ($\Gamma_T/M_T\to 0$) the QCD pair production channel is always an excellent approximation, as the off-shell and non-doubly-resonant contributions become negligible.
\item The effects of increasing the width becomes quickly relevant, independently of the DM spin, eventually becoming very large near the kinematics limit ($M_T=M_{DM}+m_t$) and for high $T$ masses, where the ratio can reach values above 100\% (represented by red regions in Fig.~\ref{fig:SXthird2}). The increase near the kinematics limit can be explained by a non-trivial combination of factors, the most relevant being the fact that a larger width opens a larger phase space for the decay of the $T$, which is more limited (in the NWA) as the gap between the masses decreases. It is interesting to notice that the cross section for the full signal is large for values of $M_T$ beyond those ideally accessible in the NWA (see Fig. \ref{fig:Xsigma}). Therefore, even if the $T$ mass is too large to produce enough events in the NWA, if its width is sizeable it might still be possible to detect it, unless the experimental acceptances drop with a comparable rate with respect to the NWA values. In this respect, the performance of the aforementioned experimental searches will be discussed in the following section.
\item For all channels, and in specific regions, a cancellation of effects takes place. Such cancellation makes the QCD pair production cross section similar to the cross section of the full signal even for large values of the width. The cancellation appears at different values of the $T$ mass depending on the mass of the DM and of its spin and becomes stronger  when the value of $M_{\rm DM}$ increases. Yet this cancellation does not mean that results in the NWA approximation are valid also for larger widths, as the cancellation is an accidental result due to the different scaling of the cross sections in NWA and large width regime. The differences between NWA and large width results are clearer at differential level. In Fig.~\ref{fig:DistributionCancellation} we show the differential distributions of the missing transverse energy and of the transverse momentum of the top quark along the cancellation line for a scalar DM particle of mass 1000 GeV and for a vector DM particle of mass 10 GeV. A similar effect was already observed in \cite{Moretti:2016gkr}, considering XQ decaying to SM particles instead of DM.
\end{itemize}

%%%%%%%%%%%%%%%%%%%%%%%%%%%%%%%%%%%%%%%%%%%%%%%%%%%%%%%%%%%%%%%%%%%%%%%%%

\subsubsection{Large width effects at detector level \label{sec:Detector3}}

\begin{figure}[ht!]
\centering
\epsfig{file=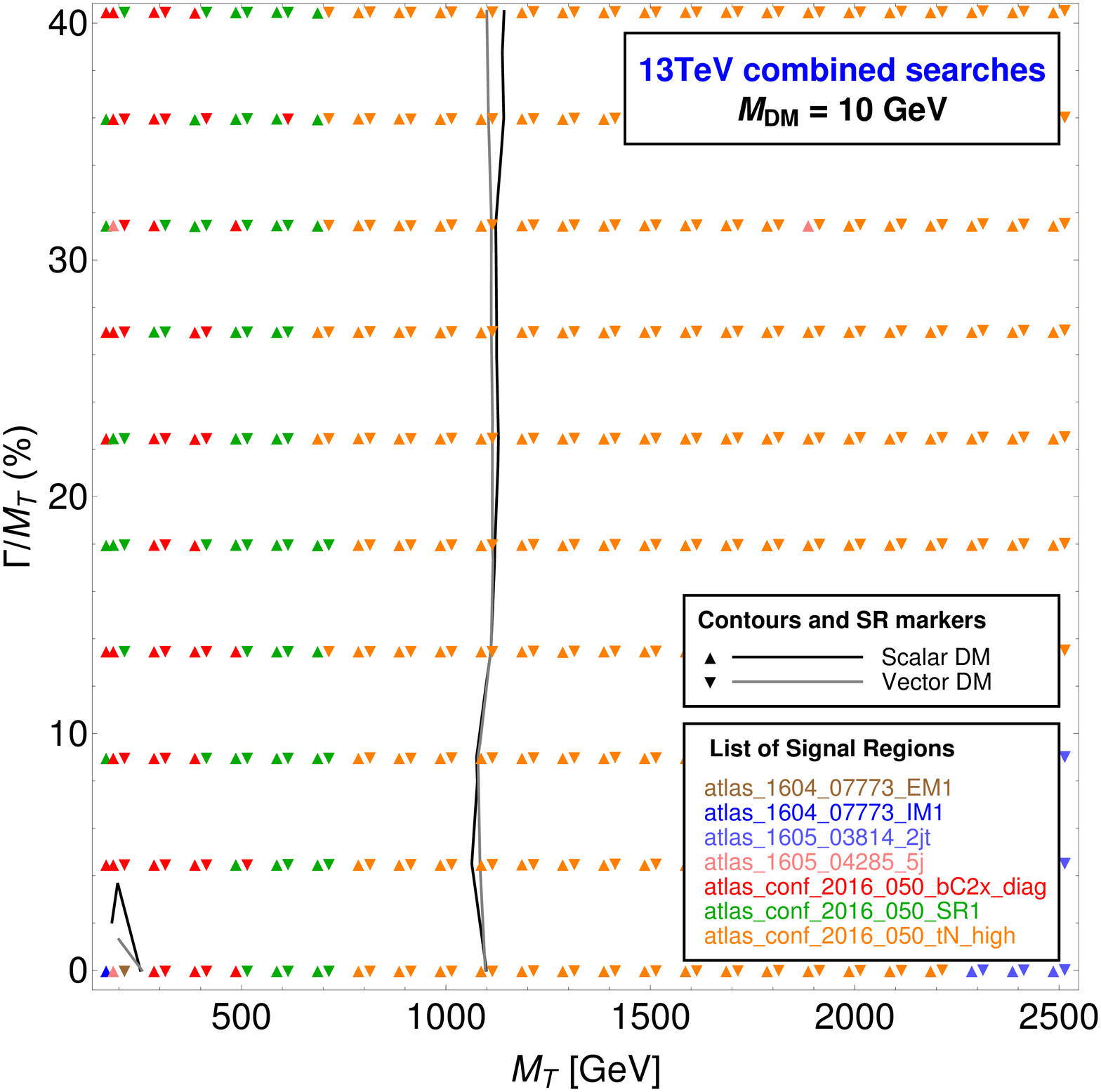, width=.32\textwidth} 
\epsfig{file=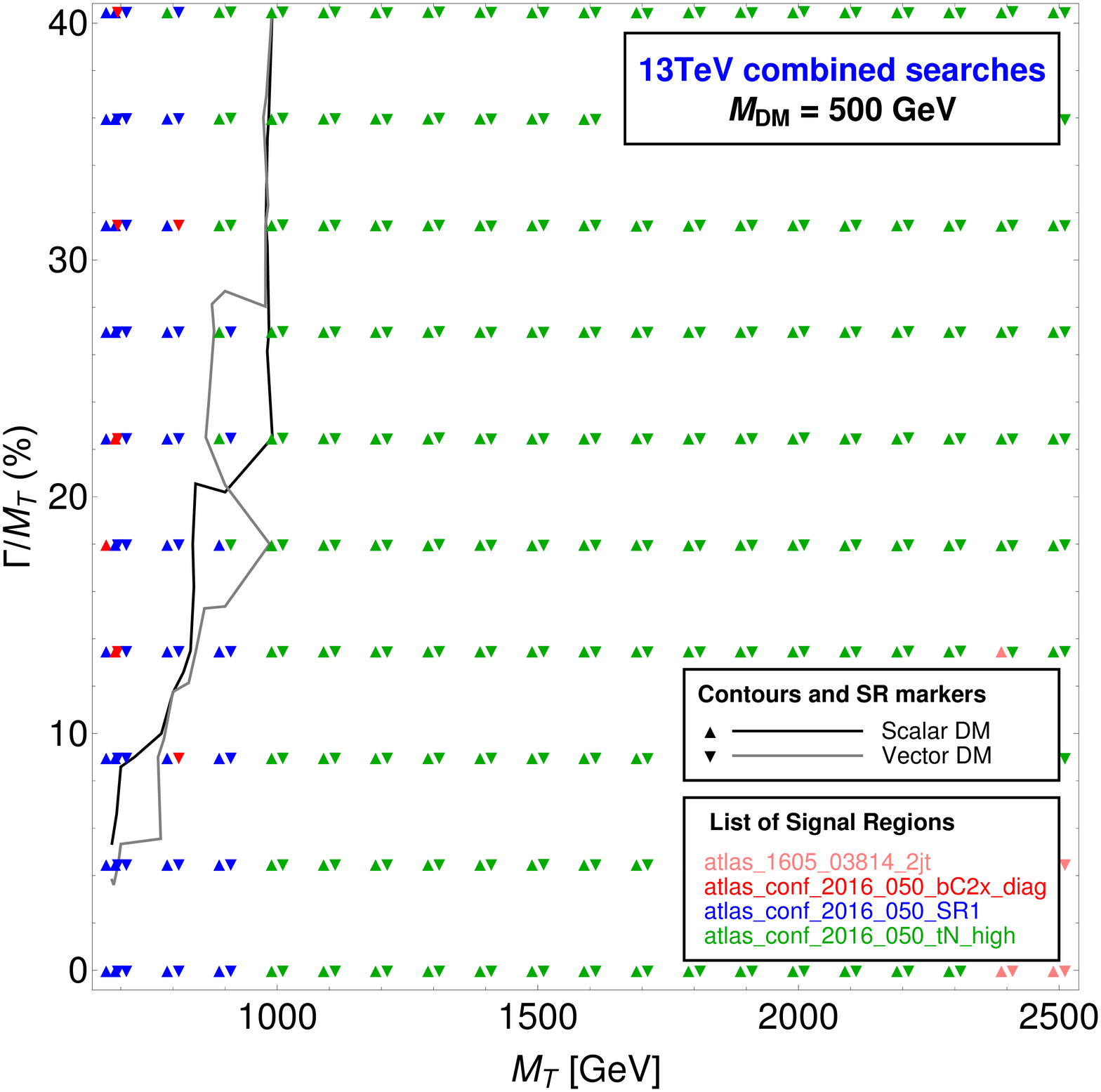, width=.32\textwidth} 
\epsfig{file=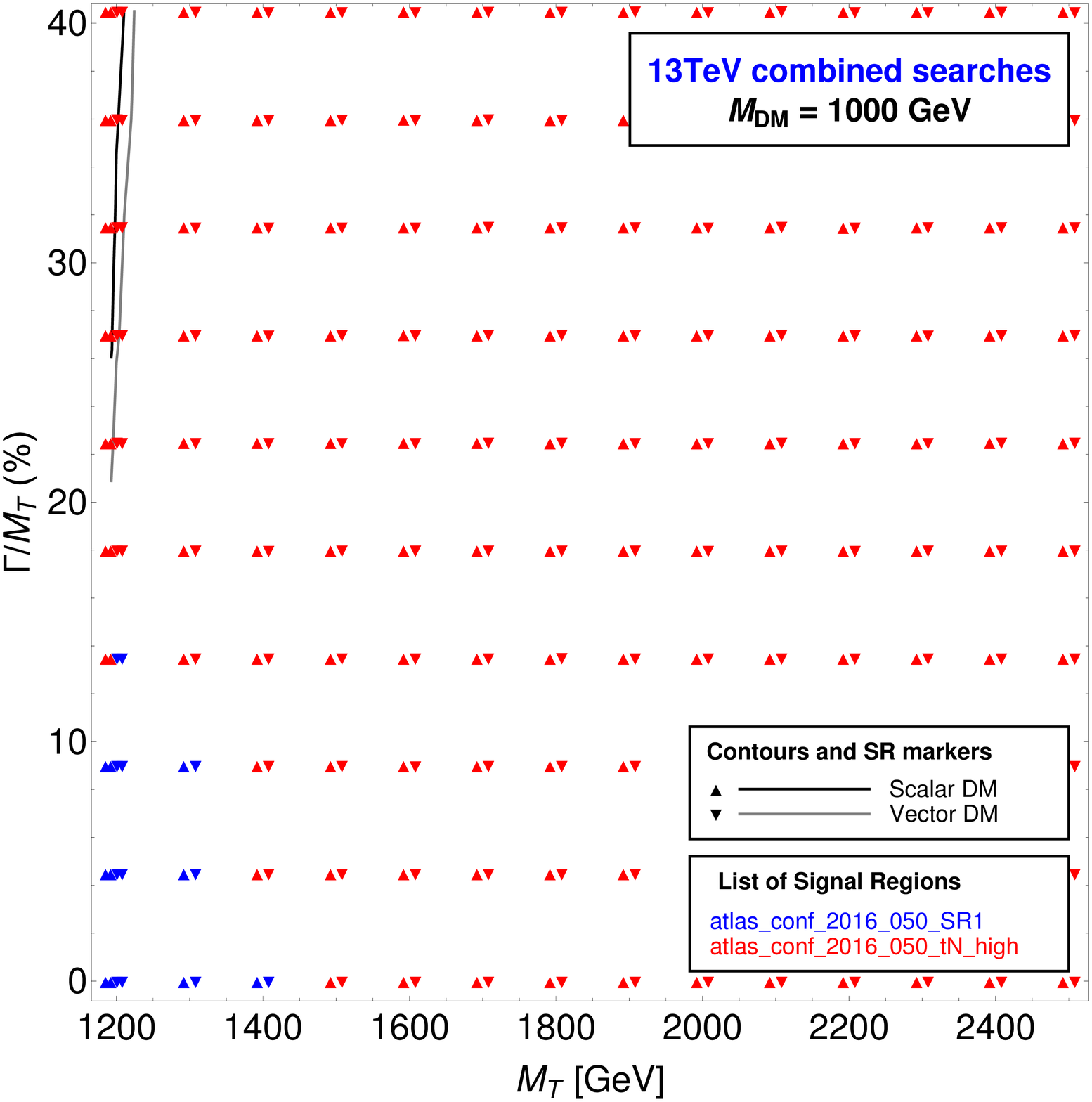, width=.32\textwidth} 
\caption[{\sc CheckMATE} results for a $T$ coupling to a DM particle (coupling to third generation) of mass 10 GeV, 500 GeV and 1500 GeV.]{{\sc CheckMATE} results for a $T$ coupling to a DM particle (coupling to third generation) of mass 10 GeV, 500 GeV and 1500 GeV. The black (grey) line show which part of the parameter space is excluded for the scalar (vector) DM scenario. }
\label{fig:Exclusion3}
\end{figure}

In this section we consider the effects of large widths on the exclusion limits for the $T$ mass. We show in Fig~\ref{fig:Exclusion3} the exclusion limit (corresponding to $r_{\rm max} = 1$ as defined in \cite{Drees:2013wra}) in the $(M_T, \Gamma_T / M_T)$ plane for both scalar and vector DM scenarios and for the same values of the DM mass previously considered.
For each simulated point the best SR is also shown using a colour code. 

\begin{figure}[ht!]
\centering
\epsfig{file=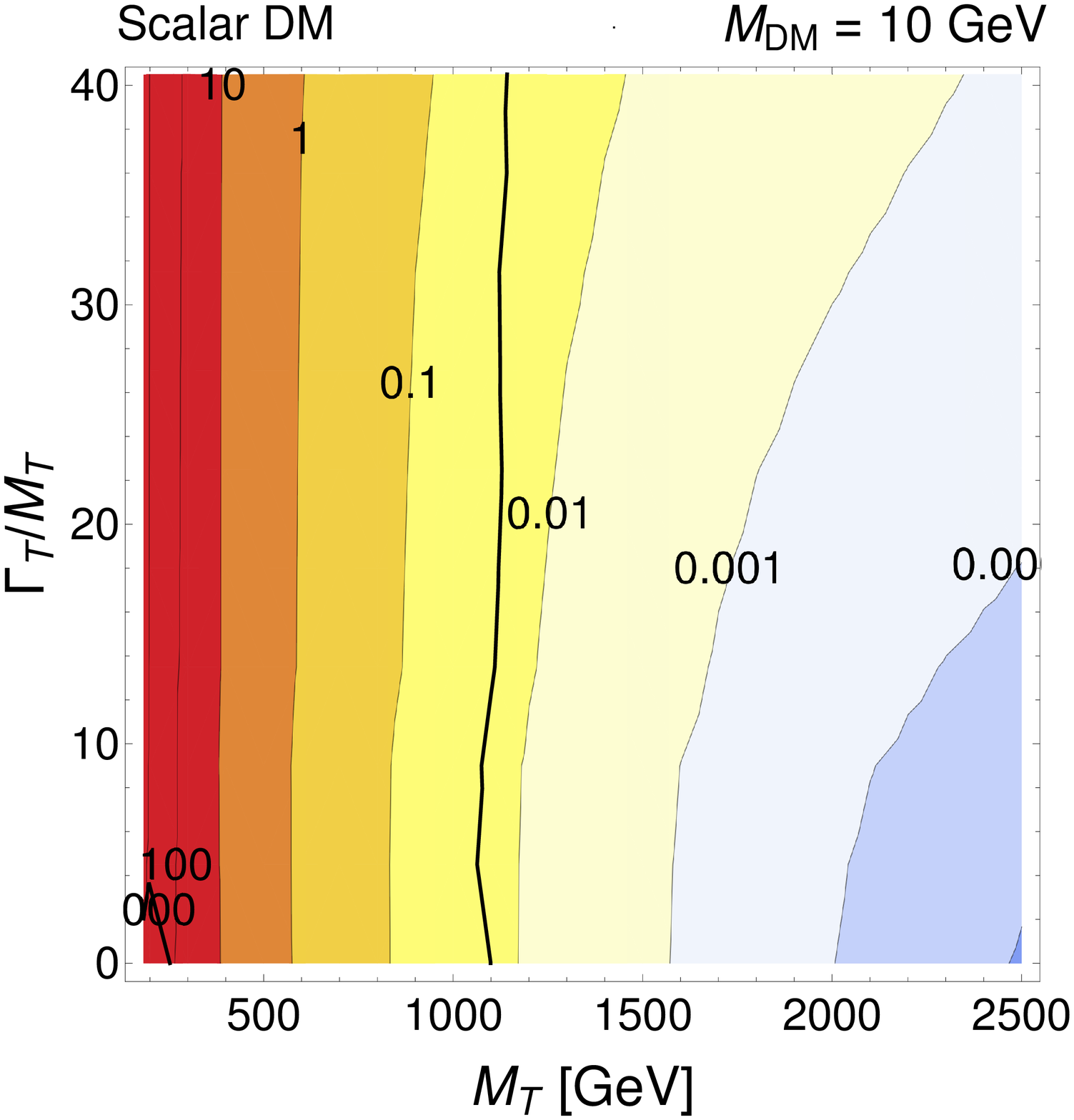,   width=.32\textwidth} 
\epsfig{file=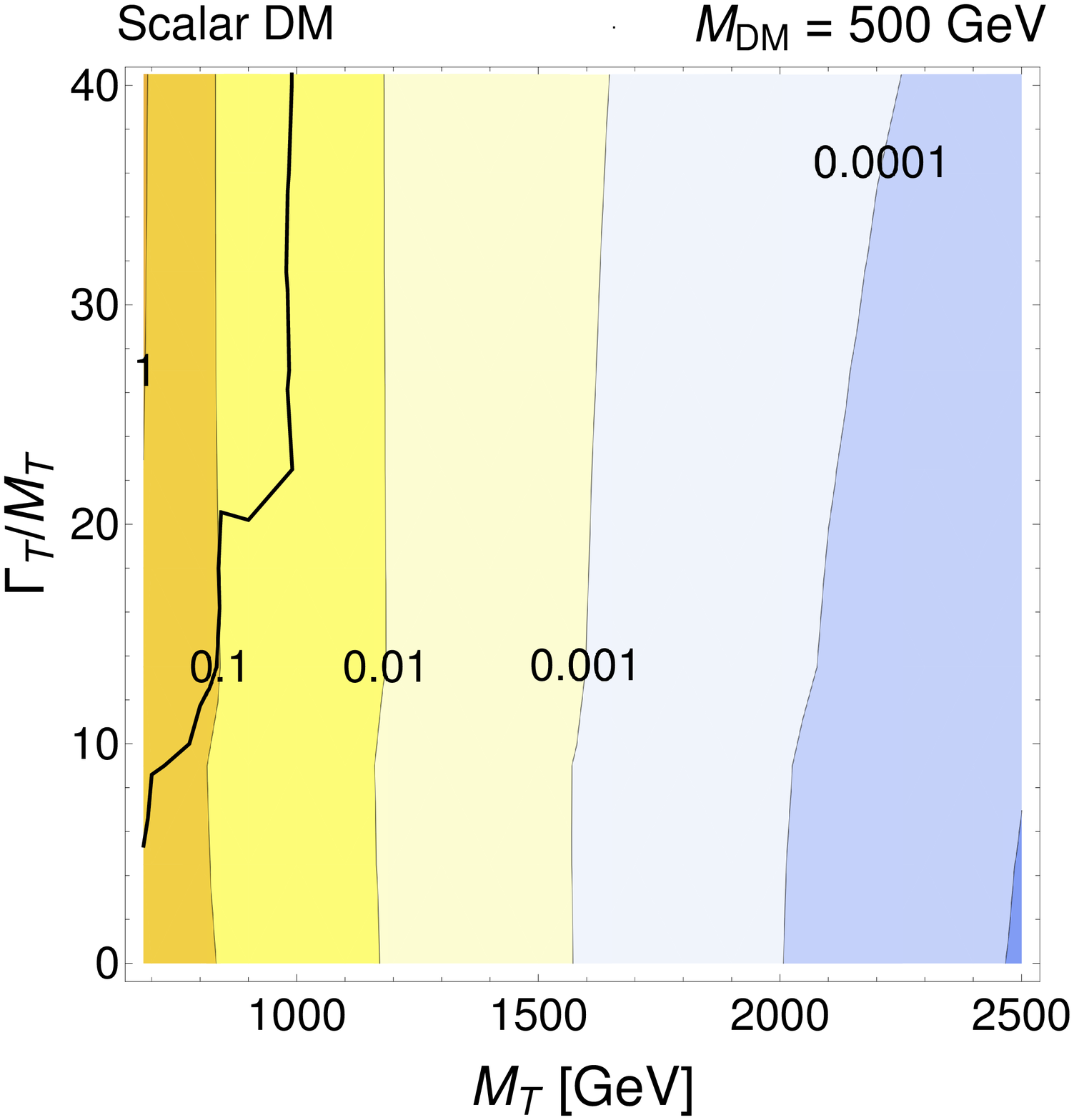,  width=.32\textwidth} 
\epsfig{file=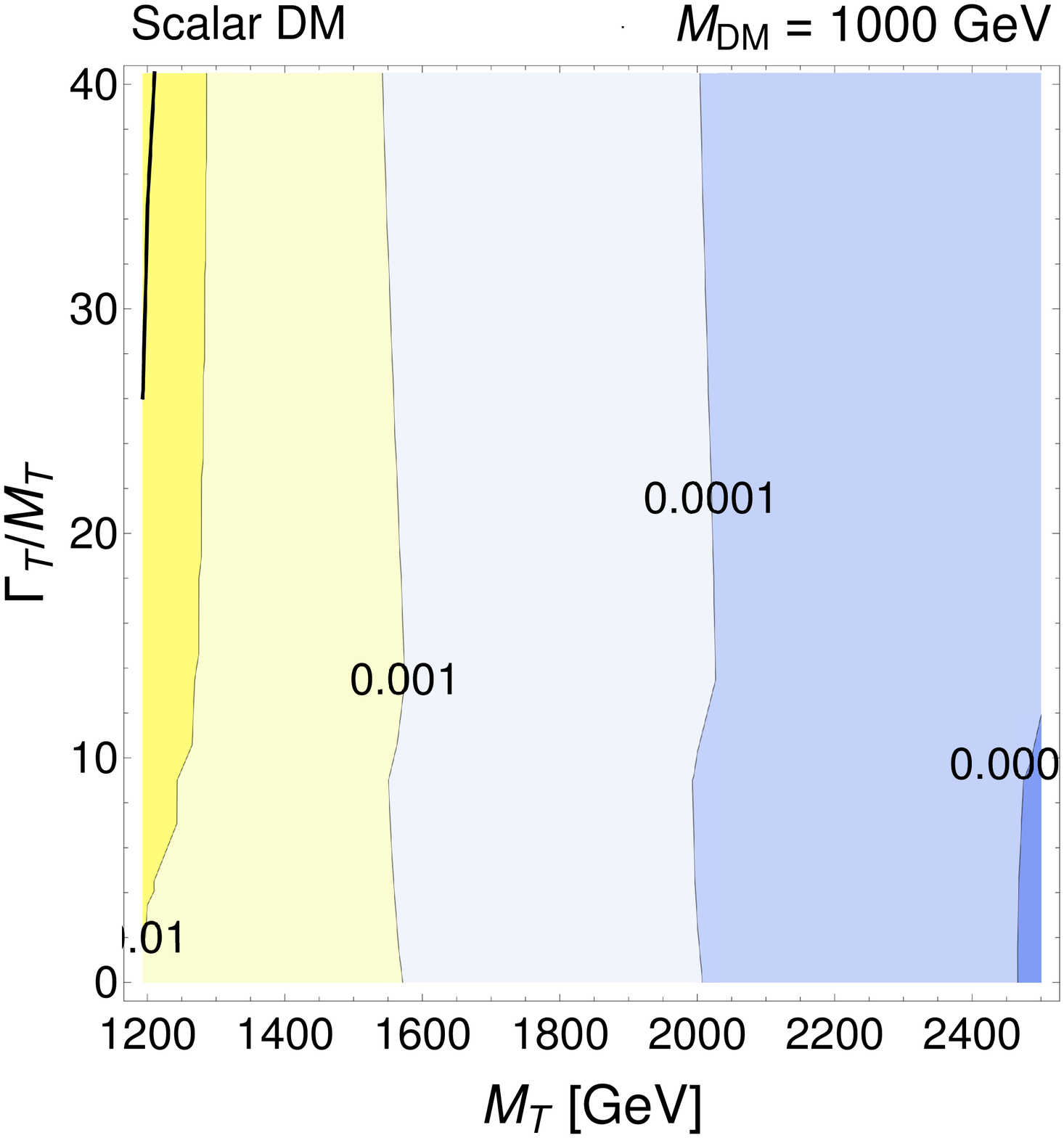, width=.32\textwidth}\\ 
\epsfig{file=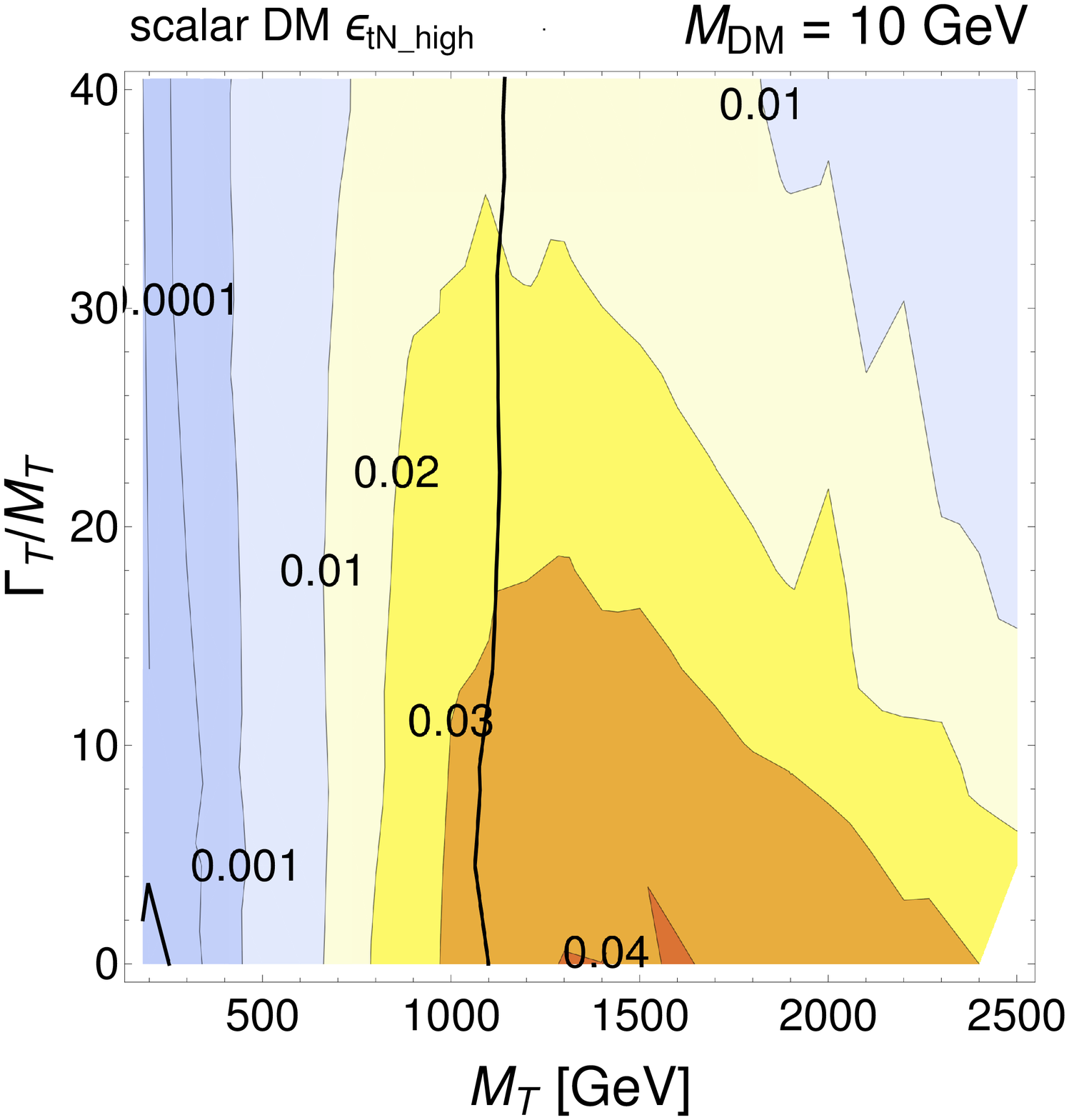,    width=.32\textwidth} 
\epsfig{file=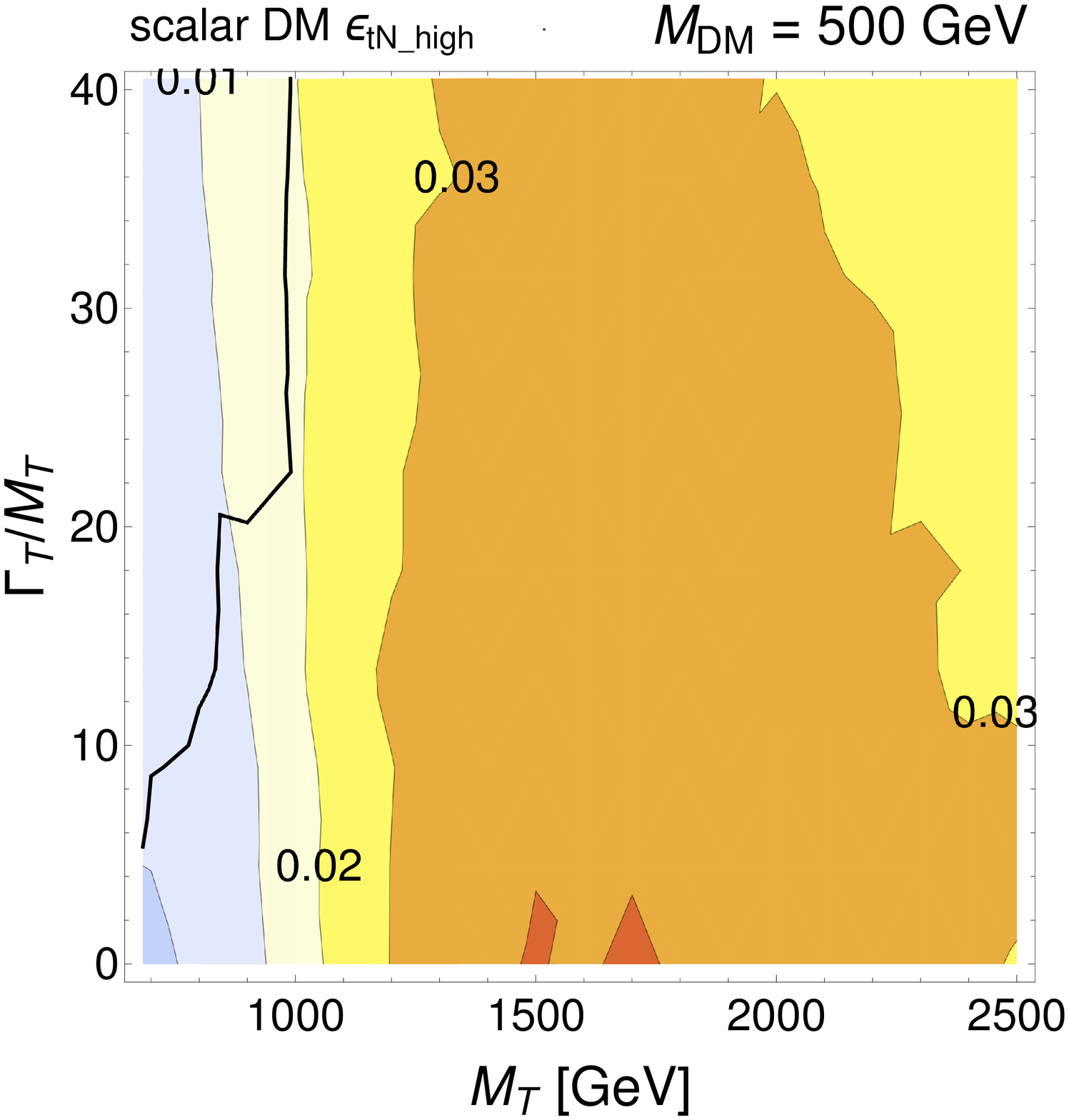,   width=.32\textwidth} 
\epsfig{file=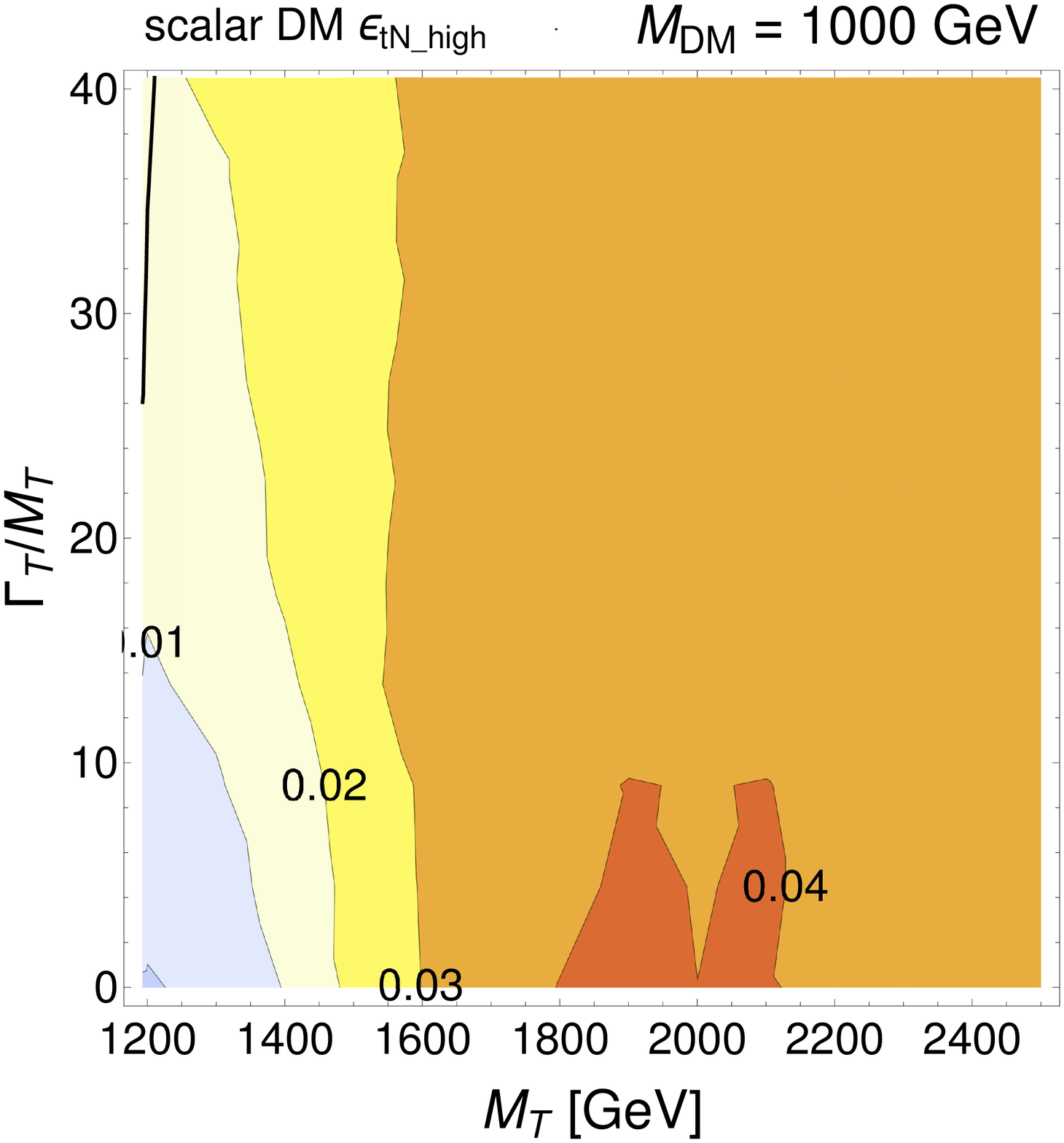,  width=.32\textwidth}
\caption[Full signal cross sections for the scalar DM case and efficiencies of the SR tN\_high from the analysis ATLAS-CONF-2016-050 for different scalar DM masses.]{Top row: full signal cross sections for the scalar DM case. Bottom row: efficiencies of the SR tN\_high from the analysis ATLAS-CONF-2016-050~\cite{ATLAS:2016ljb} for different scalar DM masses.}
\label{fig:sigmaEffs3}
\end{figure}

\begin{figure}[ht!]
\centering
\epsfig{file=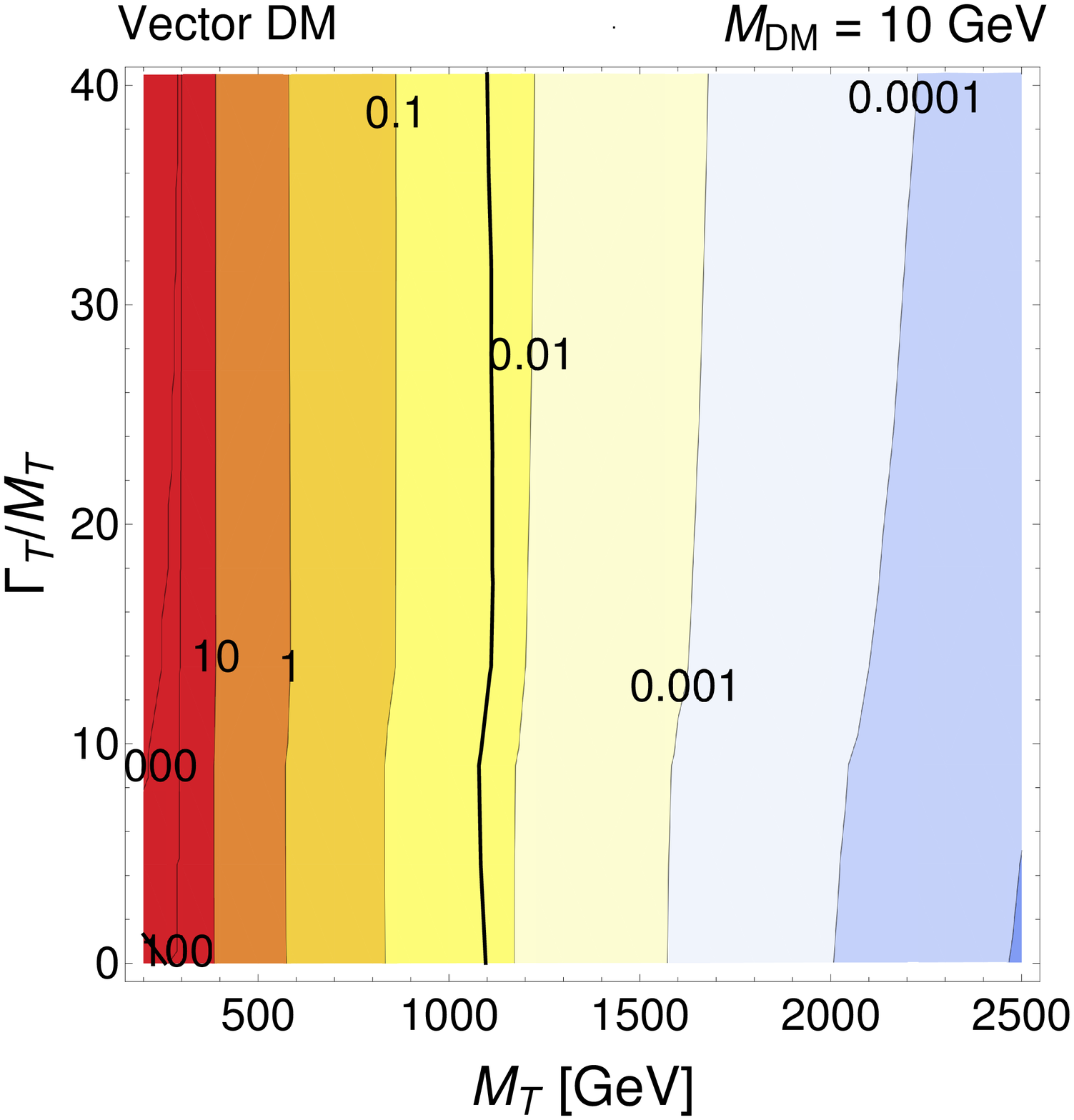,   width=.32\textwidth} 
\epsfig{file=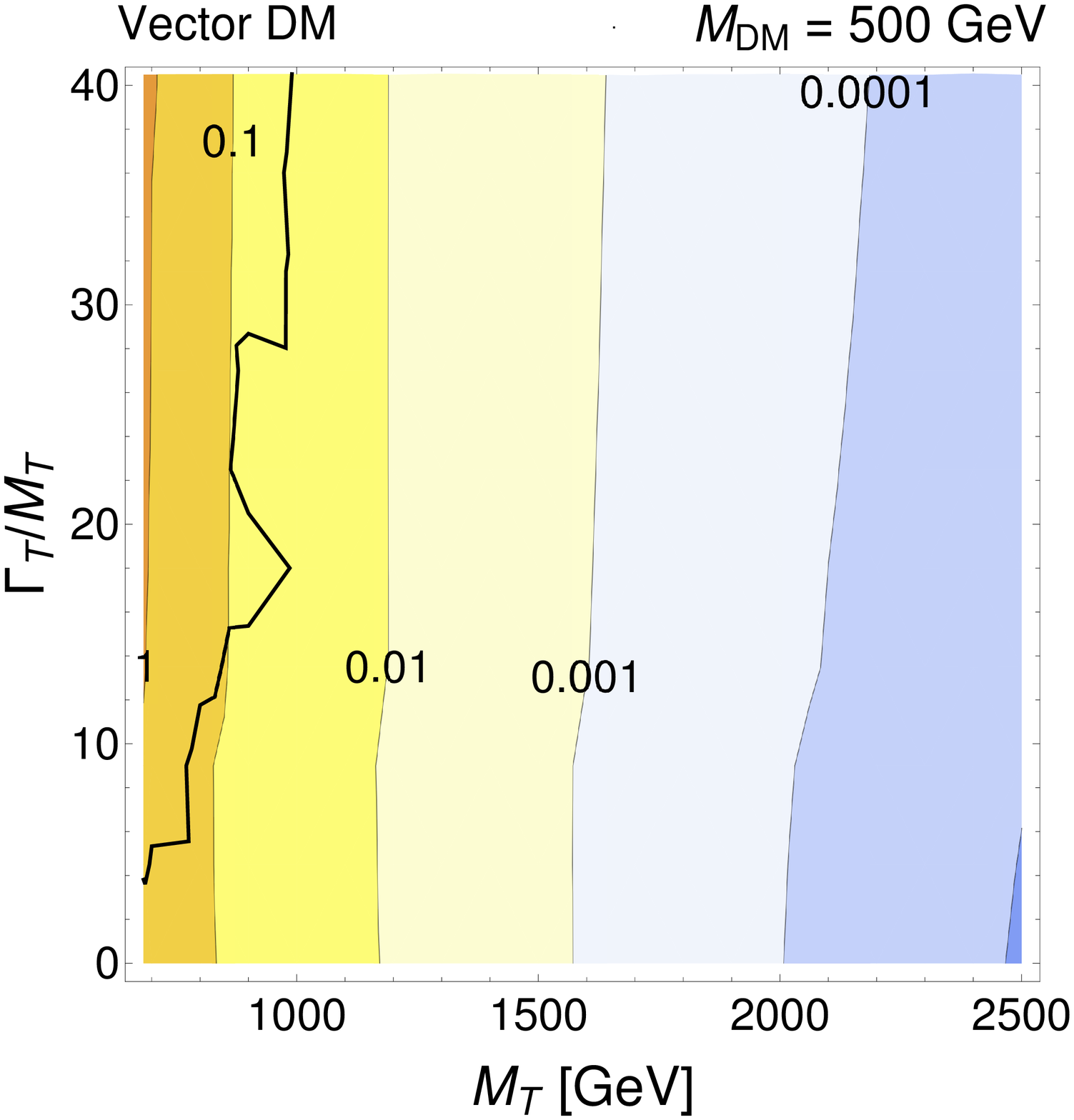,  width=.32\textwidth} 
\epsfig{file=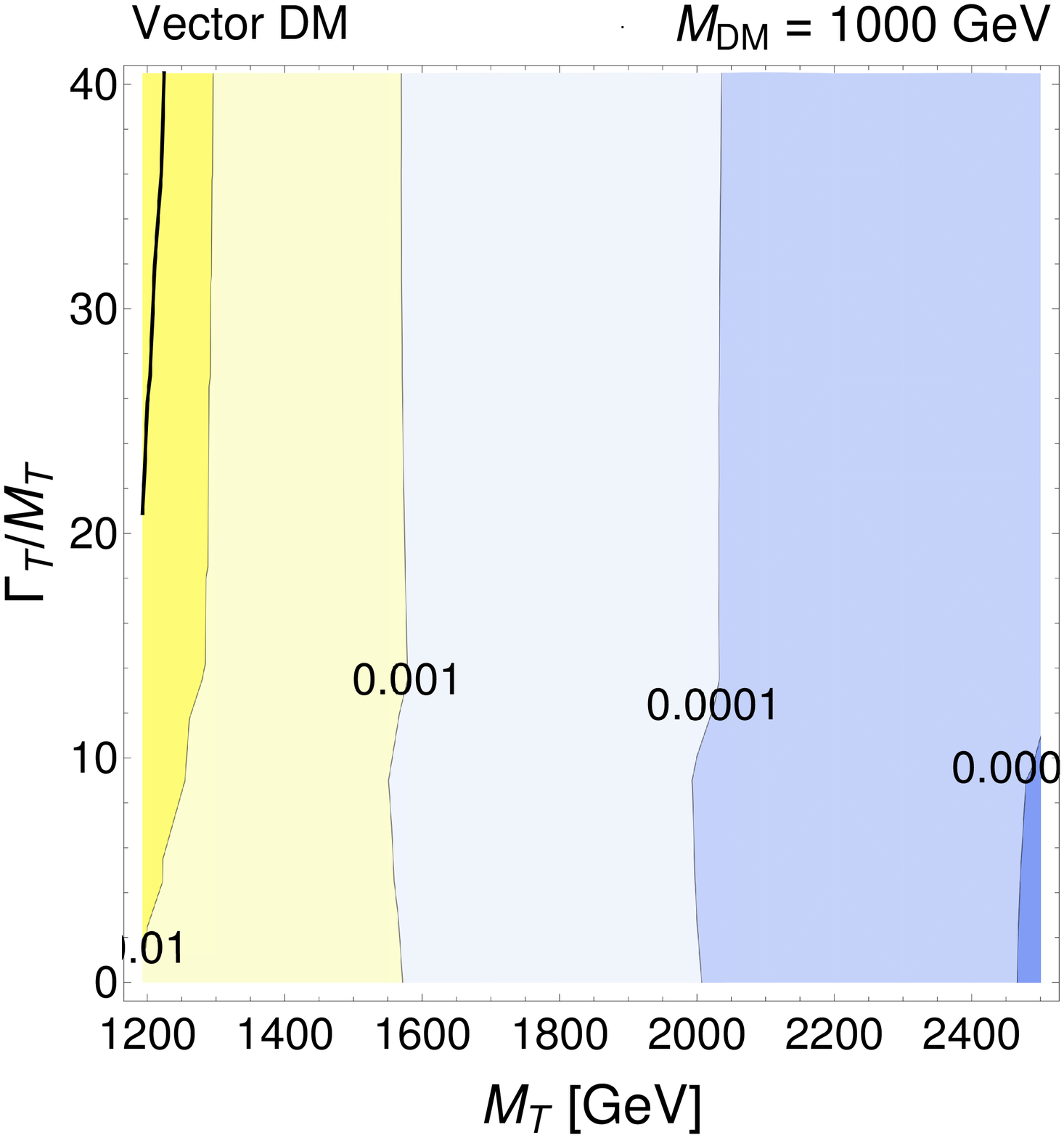, width=.32\textwidth}\\ 
\epsfig{file=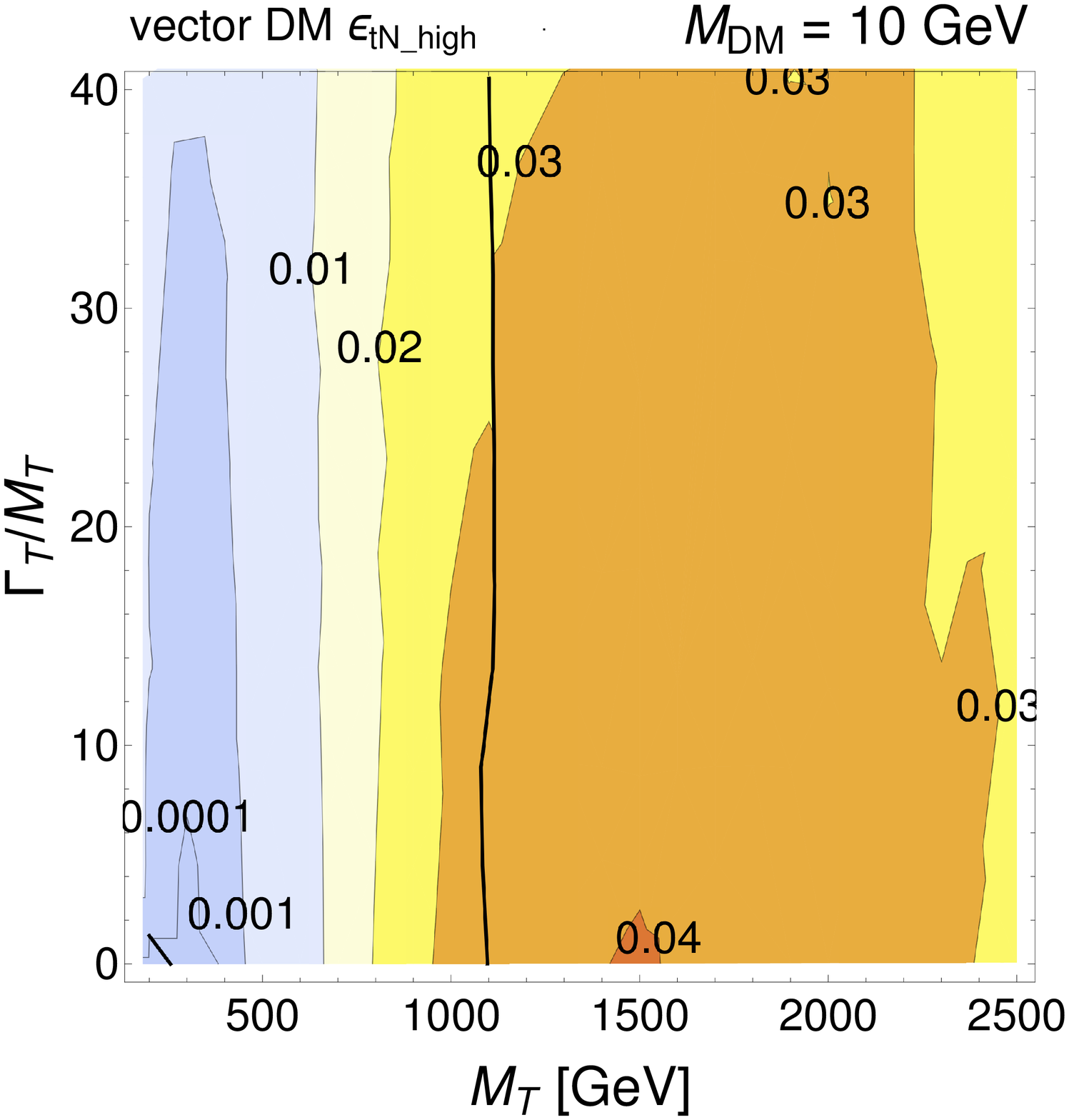,    width=.32\textwidth} 
\epsfig{file=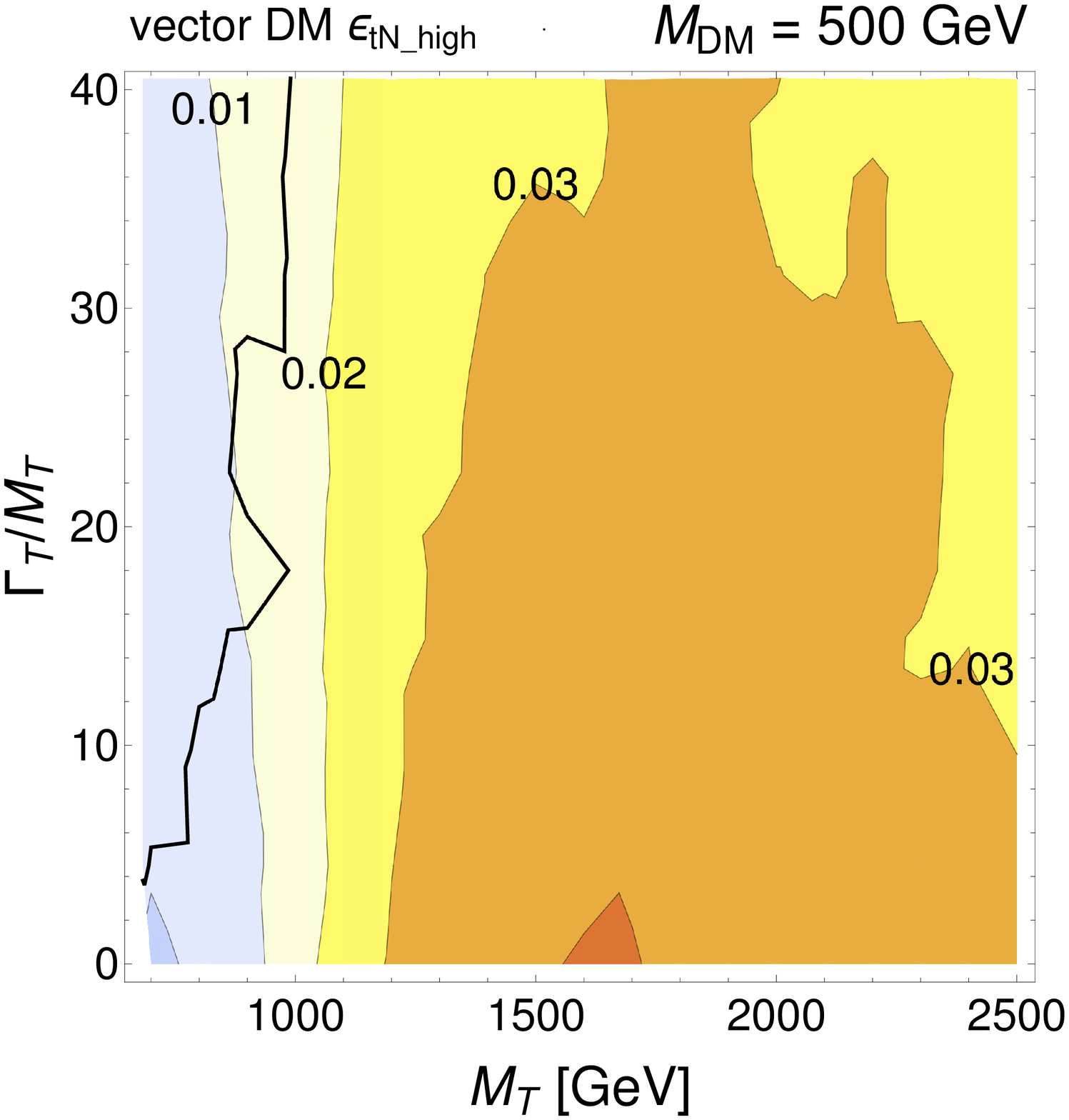,   width=.32\textwidth} 
\epsfig{file=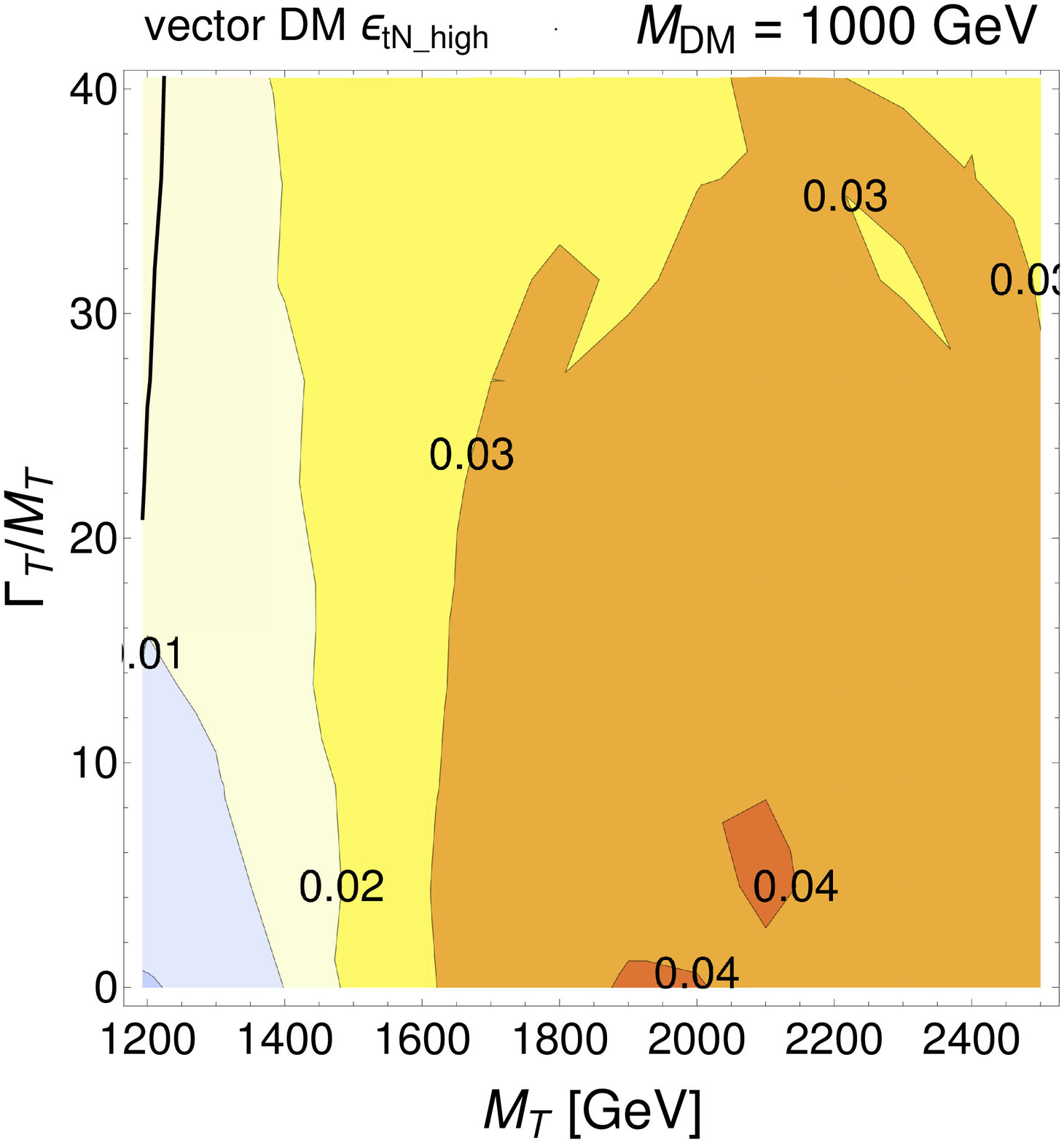,  width=.32\textwidth}
\caption[Full signal cross sections for the vector DM case and efficiencies of the SR tN\_high from the analysis ATLAS-CONF-2016-050 for different vector DM masses.]{Top row: full signal cross sections for the vector DM case. Bottom row: efficiencies of the SR tN\_high from the analysis ATLAS-CONF-2016-050~\cite{ATLAS:2016ljb} for different vector DM masses.}
\label{fig:sigmaEffv3}
\end{figure}

\begin{figure}[ht!]
\centering
\epsfig{file=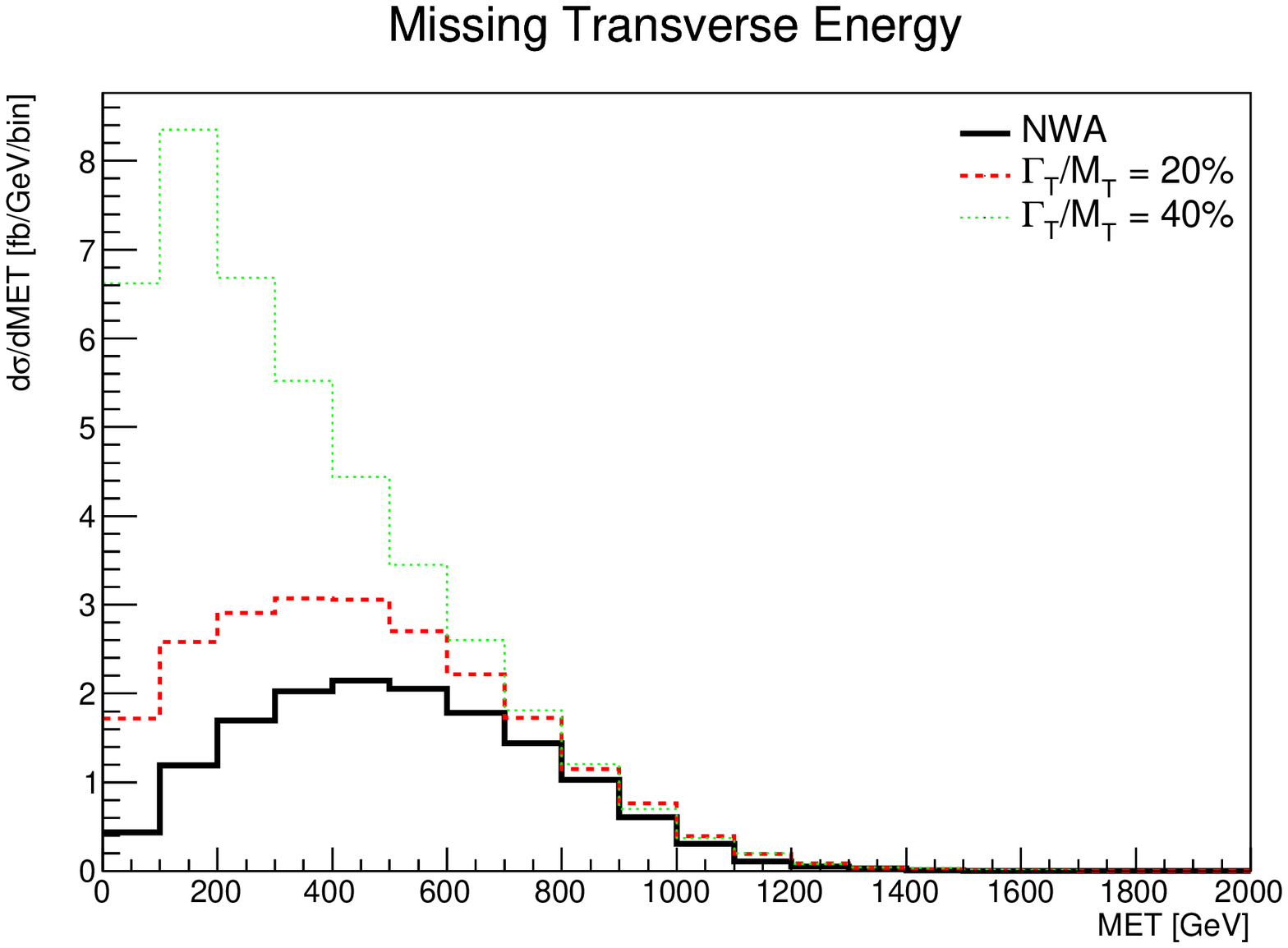, width=.45\textwidth} 
\epsfig{file=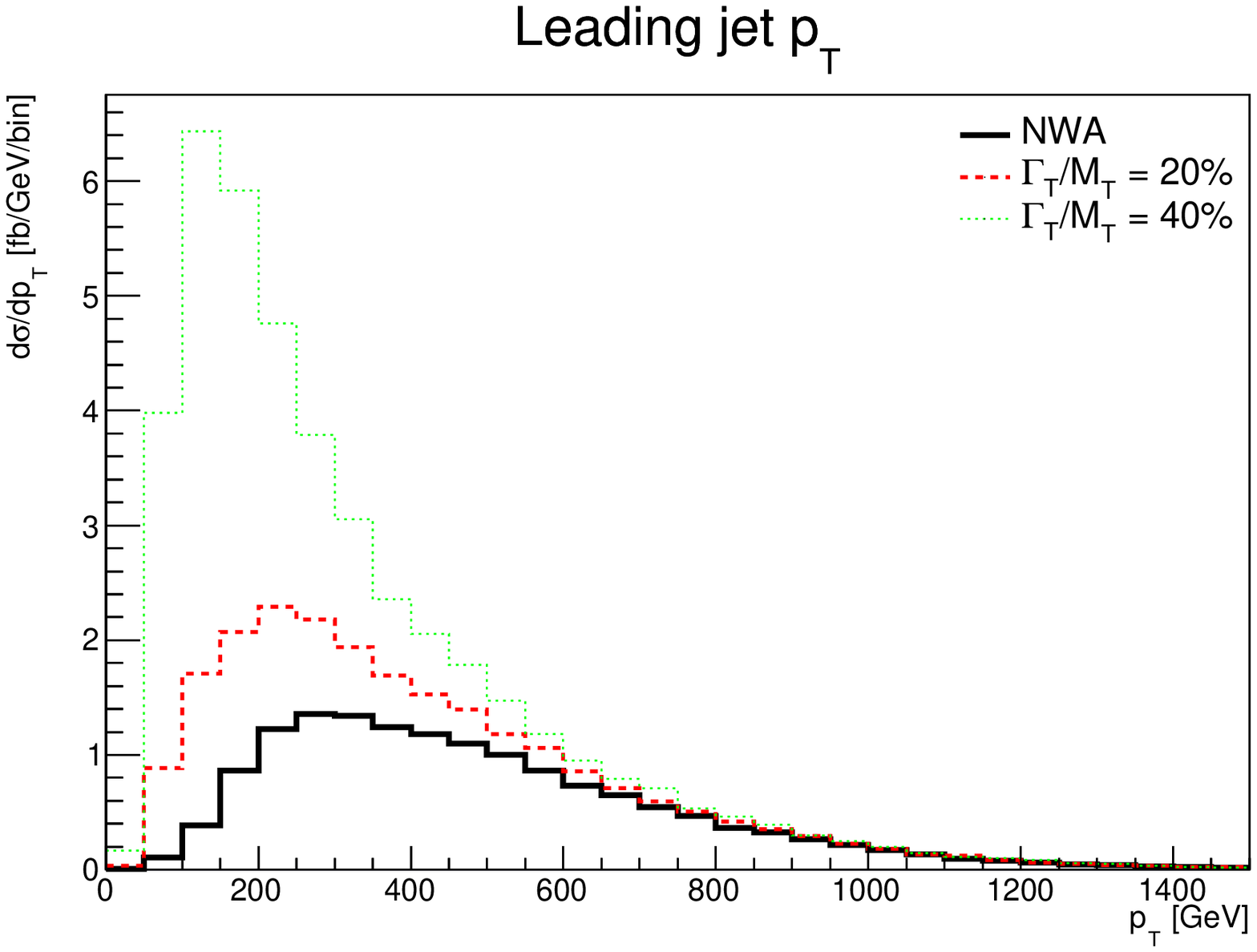, width=.45\textwidth} 
\caption{Differential distributions along the bound for a $T$ with mass $M_T=1100$ GeV coupling to the top quark and scalar DM with mass $M_{DM}=10$ GeV.}
\label{fig:cut3G10}
\end{figure}

The main conclusions which can be derived are the following:
\begin{itemize}
\item For all values of the DM mass the bounds for scalar and vector DM do not show sizeable differences. The most sensitive SR is almost always tN\_high from the analysis ATLAS-CONF-2016-050, which is optimised for ``high mass splitting, leading to very boosted top quarks where the decay products are close-by and can be reconstructed within a single large-R jet''~\cite{ATLAS:2016ljb}. Therefore, this SR is dominantly sensitive to topologies of resonant production, which depend weakly on the spin of the DM particle. 
\item For $M_{\rm DM}$ = 10 GeV the exclusion bound is around $M_T = 1100$ GeV and has basically no width dependence. It is therefore instructive compare the width dependence of the full signal cross section and of the efficiency for the tN\_high SR, shown in Figs. \ref{fig:sigmaEffs3} and \ref{fig:sigmaEffv3}. Clearly, the increase in the cross section is compensated by an analogous decrease in the efficiency of this SR, and this compensation accounts for the fact that the bound is almost independent of the width. The reduction of the efficiency between small and large widths in the bound region is mostly due to the cuts on the $\MET$ and on the $p_T$ of the 4 jets, respectively 450 GeV and \{120,80,50,25\} GeV in this SR~\cite{ATLAS:2016ljb}. In Fig. \ref{fig:cut3G10} we plot the distributions of these observables at detector level, where it is possible to see that cutting on these variables has a stronger effect for the large width scenarios. It is worth noticing that points where the $T$ mass is close to the top mass and its width approaches the NWA are not excluded: in such region the top background hides the XQ signal and makes it undetectable.
\item For $M_{\rm DM}$ = 500 GeV and 1000 GeV the bound shows a slight dependence on the width: the larger the width, the stronger the exclusion. This could be understood looking again at the relation between the efficiencies of the most sensitive SR and the full signal cross section. It's also worth noticing that for these DM masses the NWA region is never excluded, only XQ with a large width can be excluded, and only up to mass of $M_T \sim$ 1000 (1200) GeV for $M_{\rm DM}$ = 500 (1000) GeV. 
\item For higher DM masses the exclusion contour is gradually pushed to the kinematics limit and above the maximum value of the width-over-mass ratio we have tested (40\%), and eventually disappears due to the limited sensitivity of the detector for small mass splittings between $T$ and $DM$. 
\end{itemize} 

%\vspace{\baselineskip}
\paragraph*{Dependence on the chirality of the couplings}

\ \vspace{\baselineskip}

\begin{figure}[ht!]
\centering
\epsfig{file=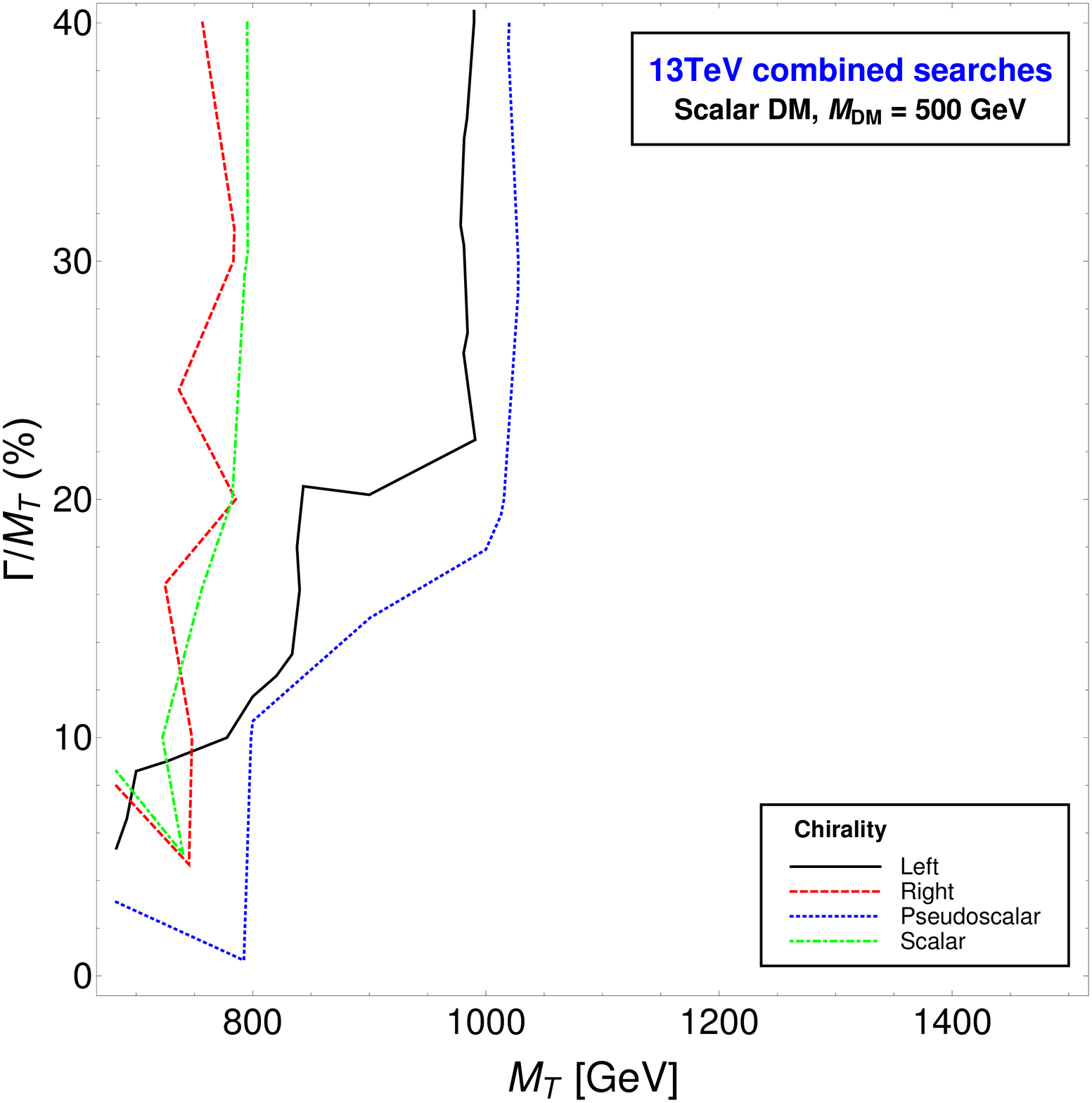, width=.45\textwidth} 
\epsfig{file=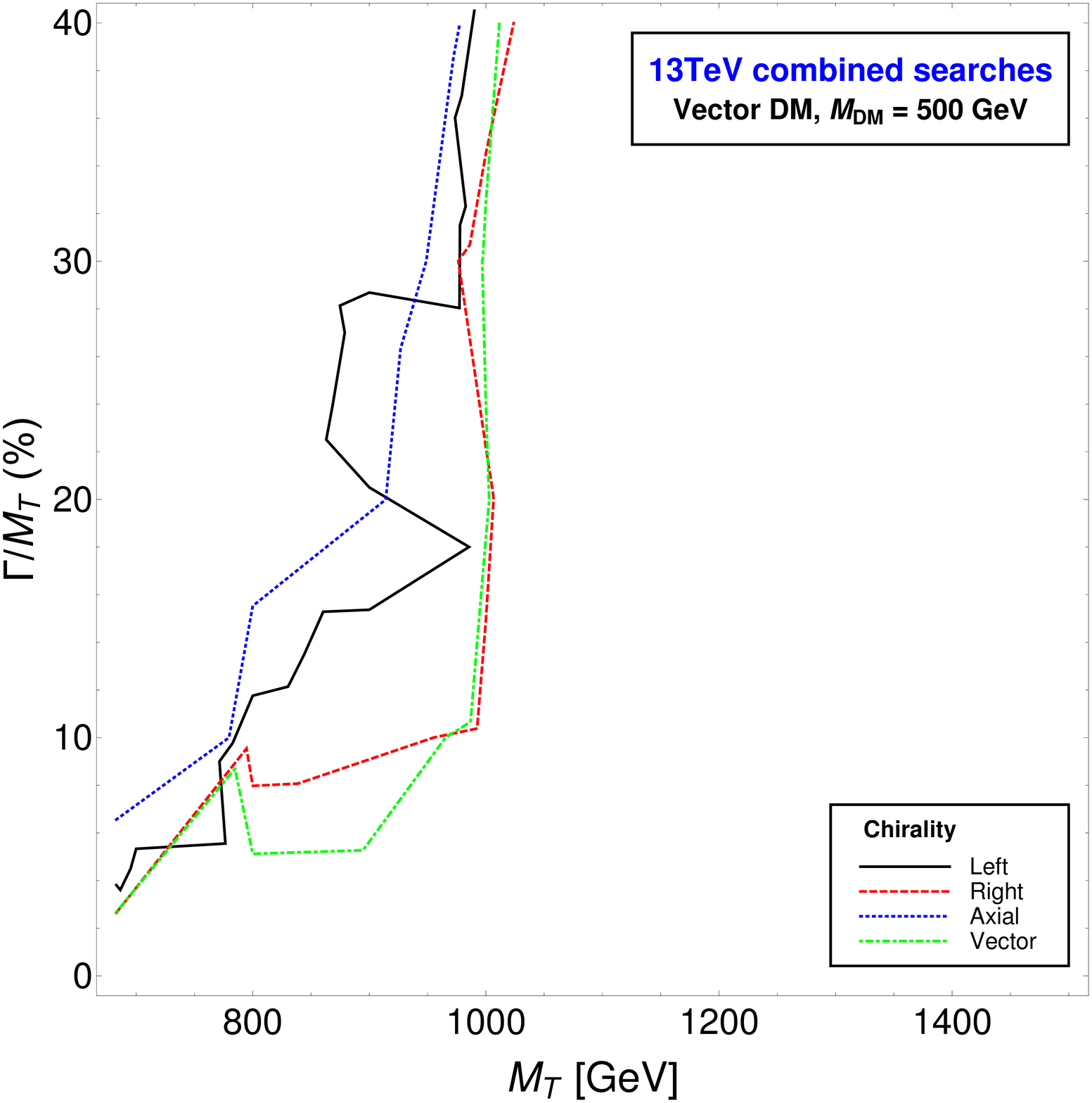, width=.45\textwidth} 
\caption[Exclusion bounds for a $T$ interacting with the SM top quark and DM for different hypotheses on the chirality of the couplings]{Exclusion bounds for a $T$ interacting with the SM top quark and DM for different hypotheses on the chirality of the couplings: for a VLQ $T$ pure left-handed and pure right-handed couplings, and for a ChQ $T$ pure scalar (vector) or pseudoscalar (axial-vector) couplings if $T$ interacts with scalar (vector) DM.}
\label{fig:3Gchirality}
\end{figure}

To conclude the analysis of XQs interacting with DM states and third generation SM quarks, we consider how the bounds change if the $T$ quark is a VLQ singlet (pure right-handed couplings) or a ChQ (where we consider either pure scalar or pseudoscalar couplings if the DM is a scalar or pure vector or axial-vector couplings if the DM is a vector). In Fig.~\ref{fig:3Gchirality} the bounds are shown for all the aforementioned scenarios: keeping in mind that the uncertainty due to the use of a recasting tool is quite large, it is possible to see that with the set of experimental searches considered in this study, the differences between various chiralities are not significant for the vector DM scenario, while there are visible differences if the DM is scalar. Therefore, with the set of cuts currently used to optimise the discovery of new physics in the $t \bar t + \MET$ channels, a characterisation of the couplings of a $T$ interacting with a vector DM and the top quark would be challenging even in the large width regime. If the DM is scalar there could be more room for a characterisation of the properties of the $T$. Designing SRs optimised for the discrimination of different coupling hypotheses and for different $\Gamma/M$ regimes would be advisable in case of discovery of a signal in this channel, but this goes beyond the scope of the present analysis and we defer this to a future study.

%%%%%%%%%%%%%%%%%%%%%%%%%%%%%%%%%%%%%%%%%%%%%%%%%%%%%%%%%%%%%%%%%%%%%%%%%
%%%%%%%%%%%%%%%%%%%%%%%%%%%%%%%%%%%%%%%%%%%%%%%%%%%%%%%%%%%%%%%%%%%%%%%%%

%\newpage

\subsection{Extra $T$ quark interacting with Dark Matter and the SM up quark}
\label{sec:Tu}

In this section we will study the case of XQs coupling to first generation SM quarks and a DM candidate. The possible final states are therefore $S^0_{\rm DM}u \; S^0_{\rm DM}\bar u$ and $\ V^0_{\rm DM}u \; V^0_{\rm DM}\bar u$.

%%%%%%%%%%%%%%%%%%%%%%%%%%%%%%%%%%%%%%%%%%%%%%%%%%%%%%%%%%%%%%%%%%%%%%%%%

\subsubsection{Large width effects at parton level}

\begin{figure}[ht!]
\begin{center}
\begin{picture}(140,100)(0,-10)
\SetWidth{1}
\Line[arrow](0,0)(50,0)
\Text(-2,0)[rc]{\large $u$}
\Line[arrow](50,80)(0,80)
\Text(-2,80)[rc]{\large $\bar u$}
\Photon(50,0)(120,0){3}{9}
\Line[dash](50,0)(120,0)
\Text(122,0)[lc]{\large $S^0_{\rm DM}, V^0_{\rm DM}$}
\SetColor{Red}\SetWidth{1.5}
\Line[arrow](50,0)(50,40)
\Text(45,20)[rc]{\large \Red{$T$}}
\Line[arrow](50,40)(50,80)
\Text(45,60)[rc]{\large \Red{$T$}}
\SetColor{Black}\SetWidth{1}
\Photon(50,80)(120,80){3}{9}
\Line[dash](50,80)(120,80)
\Text(122,80)[lc]{\large $S^0_{\rm DM}, V^0_{\rm DM}$}
\Gluon(50,40)(90,40){3}{6}
\Line[arrow](90,40)(120,60)
\Text(122,60)[lc]{\large $u$}
\Line[arrow](120,20)(90,40)
\Text(122,20)[lc]{\large $\bar u$}
\end{picture}\hskip 60pt
\begin{picture}(140,100)(0,-10)
\SetWidth{1}
\Line[arrow](0,0)(50,0)
\Text(-2,0)[rc]{\large $u$}
\Line[arrow](50,80)(0,80)
\Text(-2,80)[rc]{\large $\bar u$}
\Photon(50,40)(120,40){3}{9}
\Line[dash](50,40)(120,40)
\Text(122,40)[lc]{\large $S^0_{\rm DM}, V^0_{\rm DM}$}
\Line[arrow](50,0)(50,40)
\Text(45,20)[rc]{\large $u$}
\SetColor{Red}\SetWidth{1.5}
\Line[arrow](50,40)(50,80)
\Text(45,60)[rc]{\large \Red{$T$}}
\SetColor{Black}\SetWidth{1}
\Photon(50,80)(120,80){3}{9}
\Line[dash](50,80)(120,80)
\Text(122,80)[lc]{\large $S^0_{\rm DM}, V^0_{\rm DM}$}
\Gluon(50,0)(90,0){3}{6}
\Line[arrow](90,0)(120,10)
\Text(122,10)[lc]{\large $u$}
\Line[arrow](120,-10)(90,0)
\Text(122,-10)[lc]{\large $\bar u$}
\end{picture}
\end{center}
\caption{Examples of topologies which are peculiar to scenarios with heavy quarks coupling to first generation.}
\label{fig:firstgentopologies2}
\end{figure}
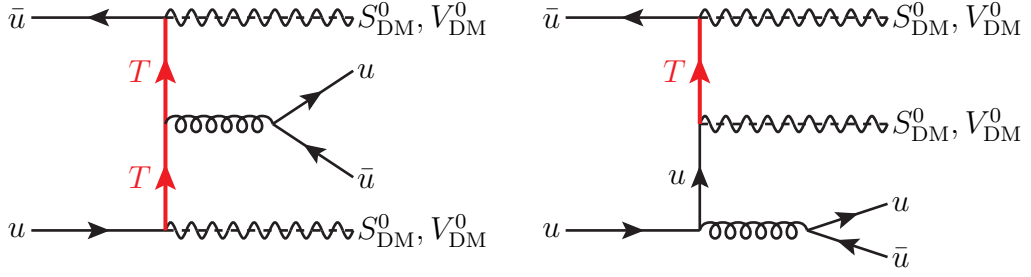

When the $T$ quark couples to quarks of the first generation, the $2\to 4$ process contains topologies where the initial state partons interact directly with the $T$ (examples are shown in Fig.~\ref{fig:firstgentopologies2}) which are absent in the case of coupling to third generation. 

\begin{figure}[ht!]
\centering
\epsfig{file=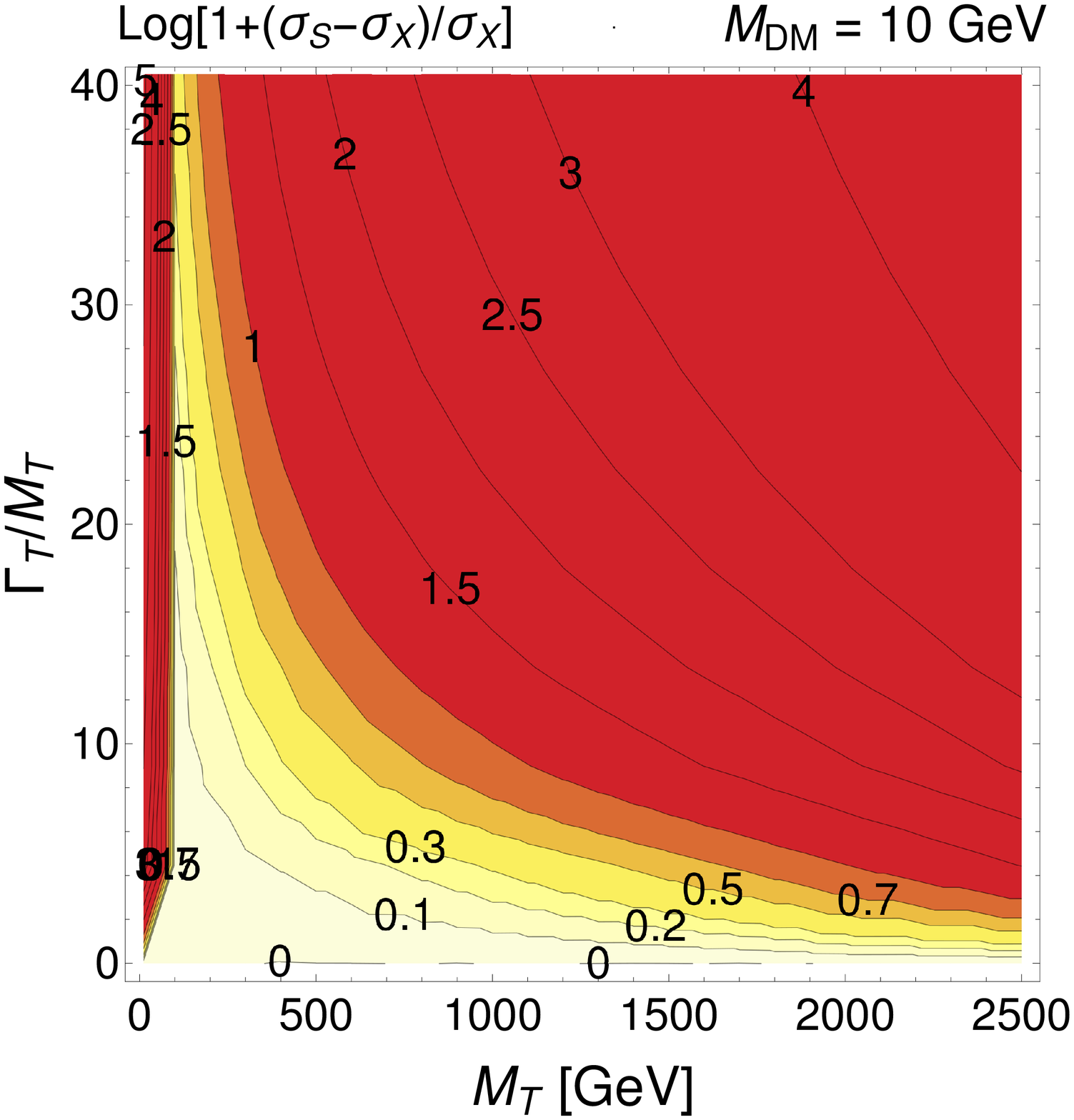,   width=.32\textwidth} 
\epsfig{file=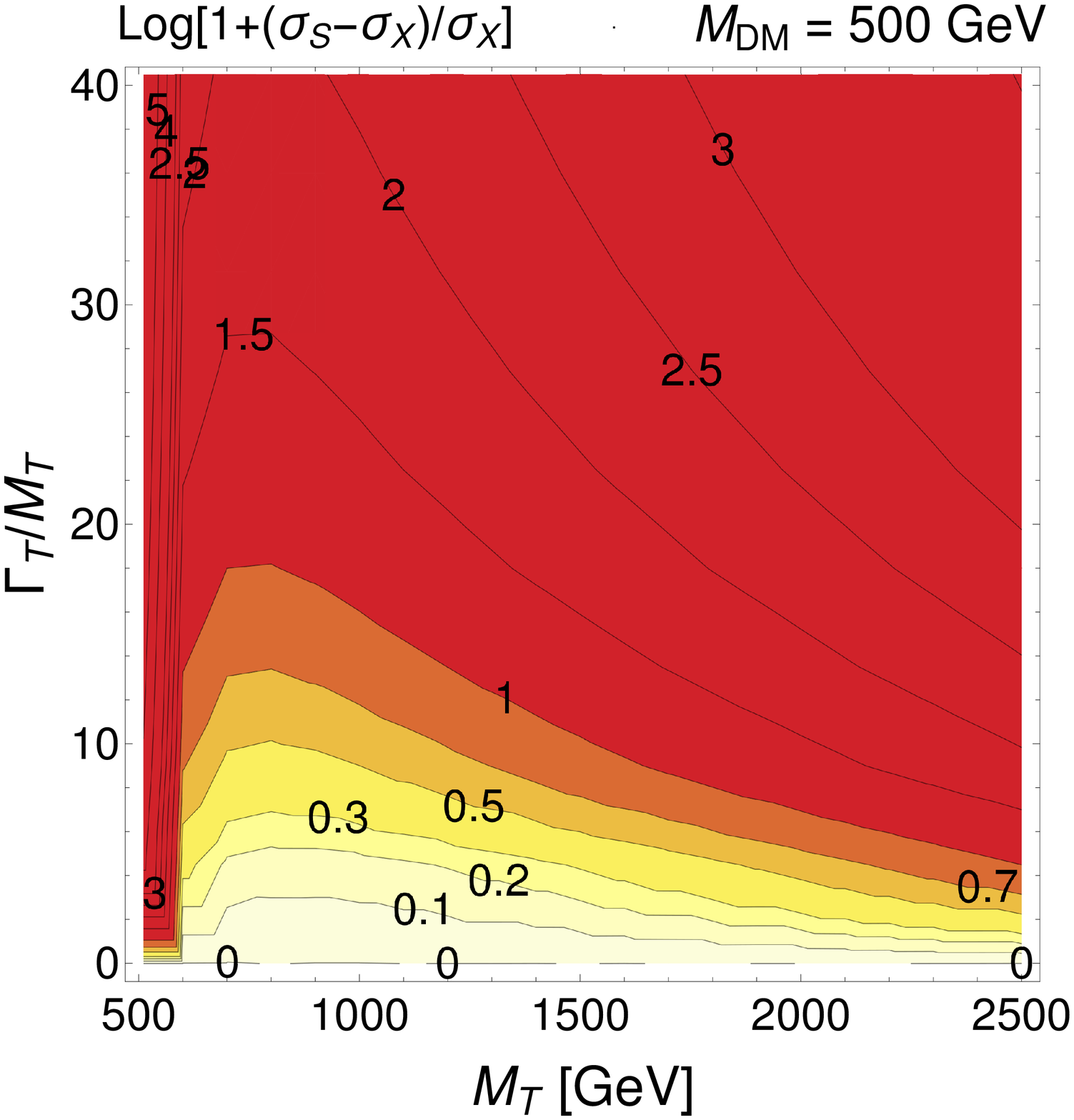,  width=.32\textwidth} 
\epsfig{file=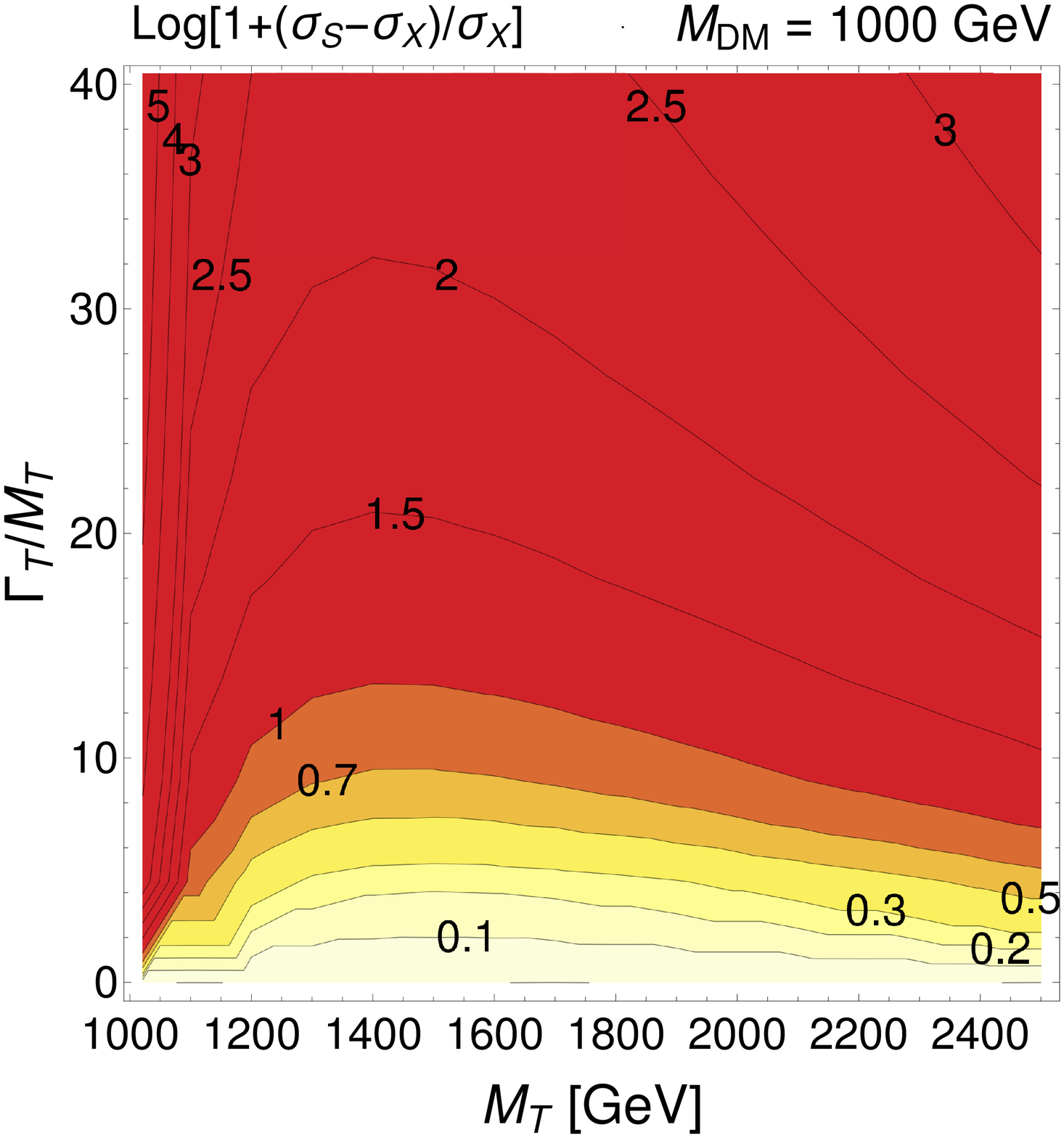, width=.32\textwidth}\\[5pt] 
\epsfig{file=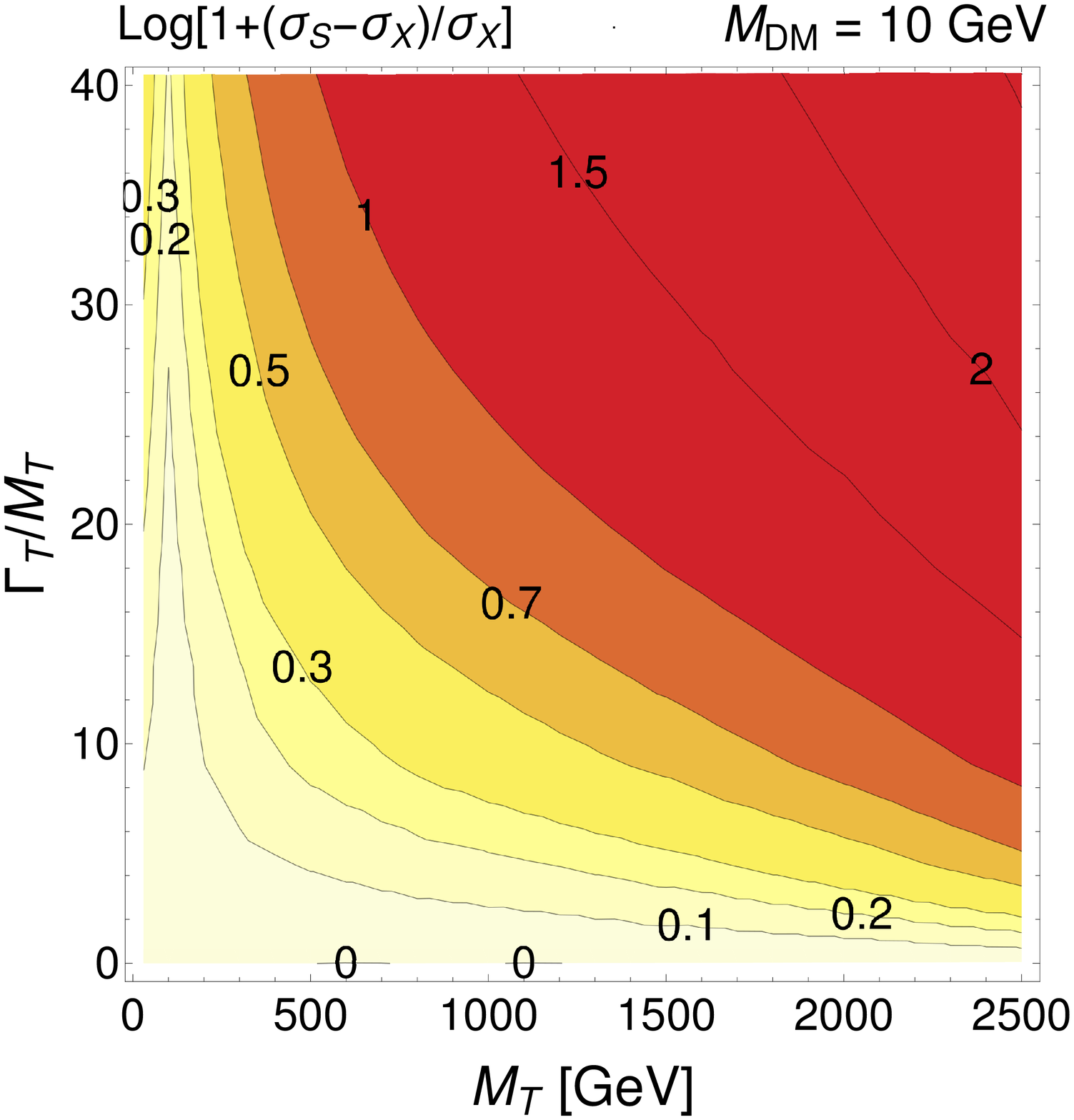,   width=.32\textwidth} 
\epsfig{file=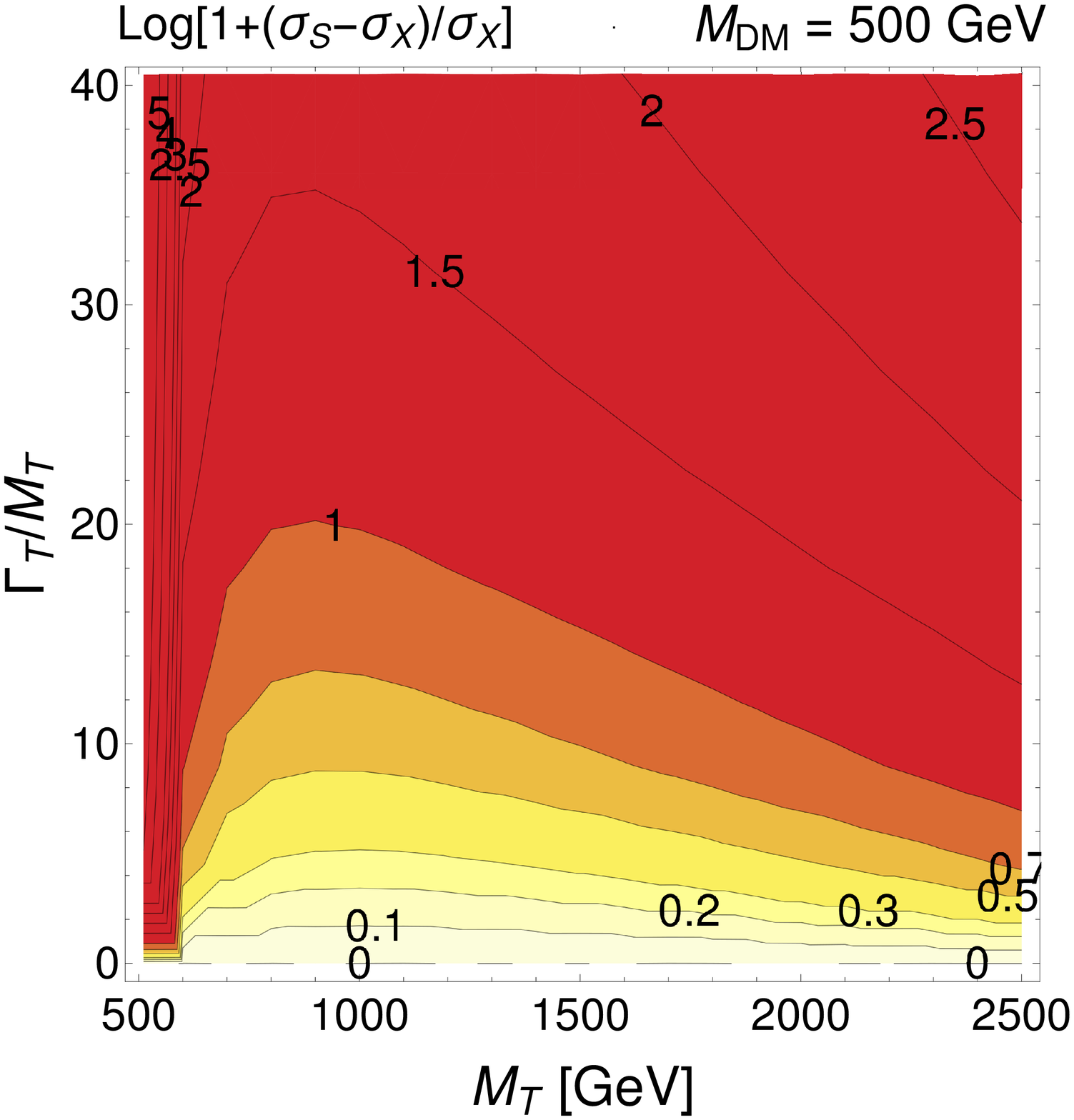,  width=.32\textwidth} 
\epsfig{file=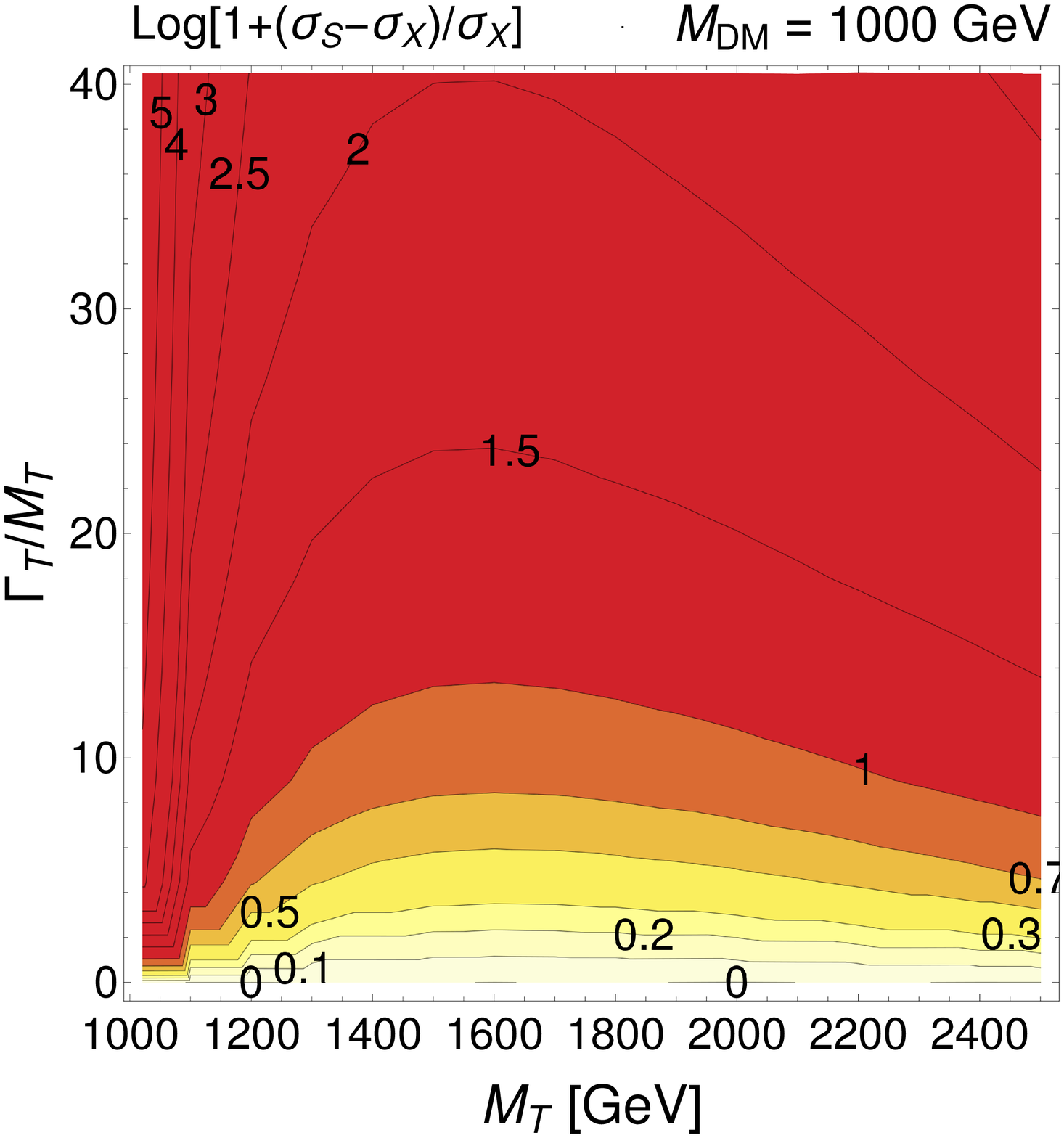, width=.32\textwidth} 
\caption[Relative difference between the full signal and the QCD pair production cross sections for a $T$ coupling to a DM particle (coupling to first generation) of mass 10 GeV, 500 GeV and 1000 GeV.]{Relative difference between the full signal and the QCD pair production cross sections for a $T$ coupling to a DM particle (coupling to first generation) of mass 10 GeV, 500 GeV and 1000 GeV. Due to the large differences between cross sections, the ratio is plotted as $\log [1+(\sigma_S - \sigma_X)/\sigma_X]$ instead of $(\sigma_S - \sigma_X)/\sigma_X$. Notice that in that case the contours at 0.1, 0.2, 0.3, 0.5 and 1 respectively correspond to a value of $(\sigma_S - \sigma_X)/\sigma_X$ equal to 26\%, 58\%, 100\%, 216\% and 900\%. Top row: scalar DM; bottom row, vector DM.}
\label{fig:SXfirst2}
\end{figure}

These topologies contain collinear divergences, due to the gluon splitting, which drastically enhance the full signal cross section with respect to QCD pair-production. In Fig.~\ref{fig:SXfirst2} the logarithm of the relative differences between the full signal cross section and the QCD pair production cross section are plotted for an LHC energy of 13 TeV. Notice that to allow a consistent comparison with the NWA case no cuts have been applied on the light jet at parton level. 

The main conclusions which can be derived from our results are the following:
\begin{itemize}
\item In the NWA the full signal and the QCD pair production topologies become equivalent, as expected. The latter topologies describe the process in an excellent way in the NWA, as subleading topologies and off-shell contributions are indeed negligible.
\item The contributions of new topologies and of off-shell $T$ become more and more relevant as the width of the $T$ increases, quickly becoming extremely relevant for the determination of the cross section, especially when the mass of the XQ and of the DM particle are close. 
\item The cancellation of effects which makes the $\sigma_S$ similar to $\sigma_X$ as in the case of coupling to third generation is not observed in this case. However, a minimum of the cross section ratio (for fixed $\Gamma_T/M_T$) appears for all value of the DM mass and spin in regions that are very similar to the cancellation region observed in section \ref{sec:Parton3}. This decrease is due again to a different scaling of the phase space in the NWA and large width regimes, but due to the additional diagrams in the case of coupling with first generation, the cancellation only lowers the cross section ratio and does not bring it to zero as it was the case for third generation coupling.
\end{itemize}

%%%%%%%%%%%%%%%%%%%%%%%%%%%%%%%%%%%%%%%%%%%%%%%%%%%%%%%%%%%%%%%%%%%%%%%%%

\subsubsection{Large width effects at detector level}

\begin{figure}[ht!]
\centering
\epsfig{file=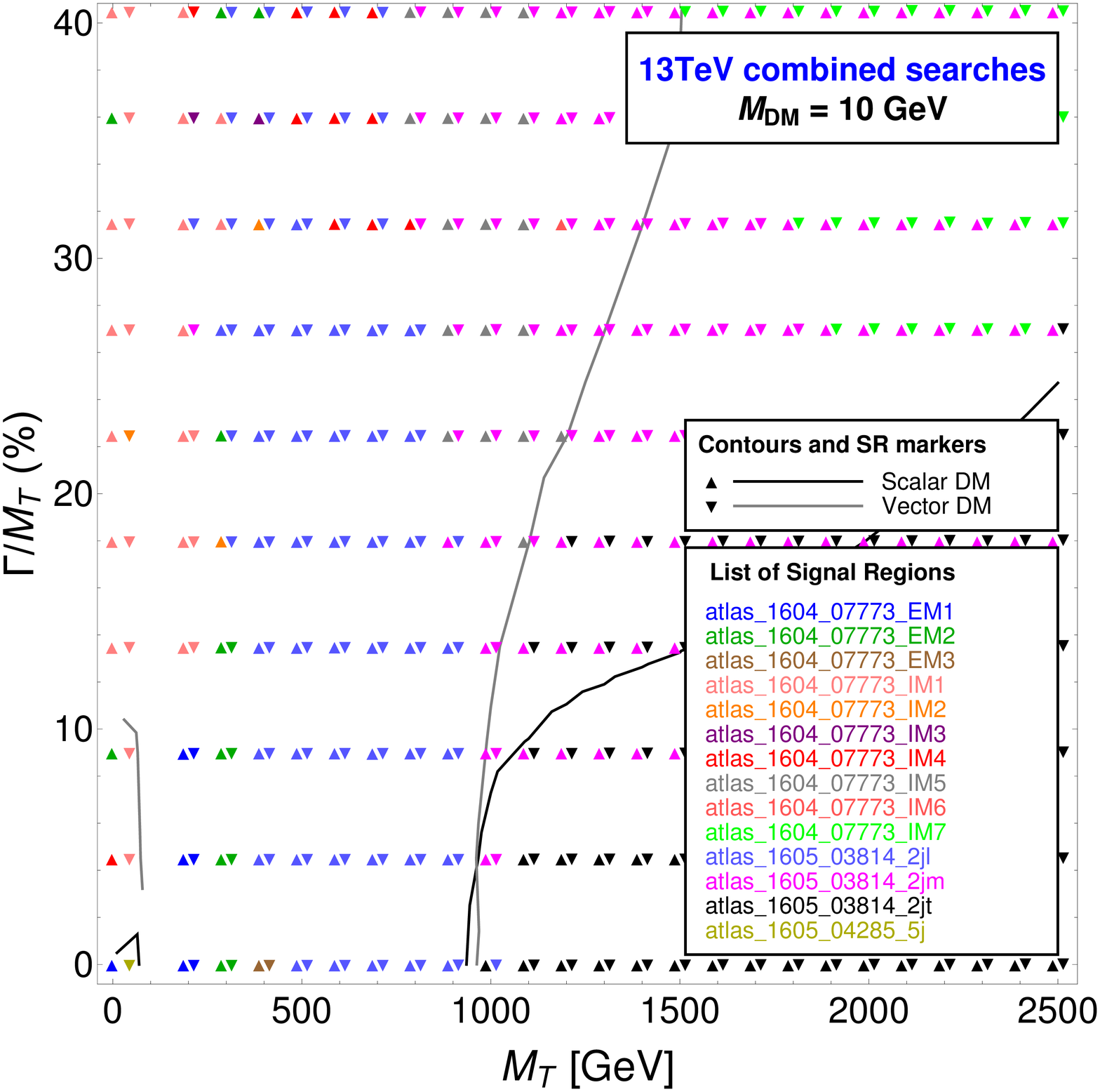, width=.32\textwidth} 
\epsfig{file=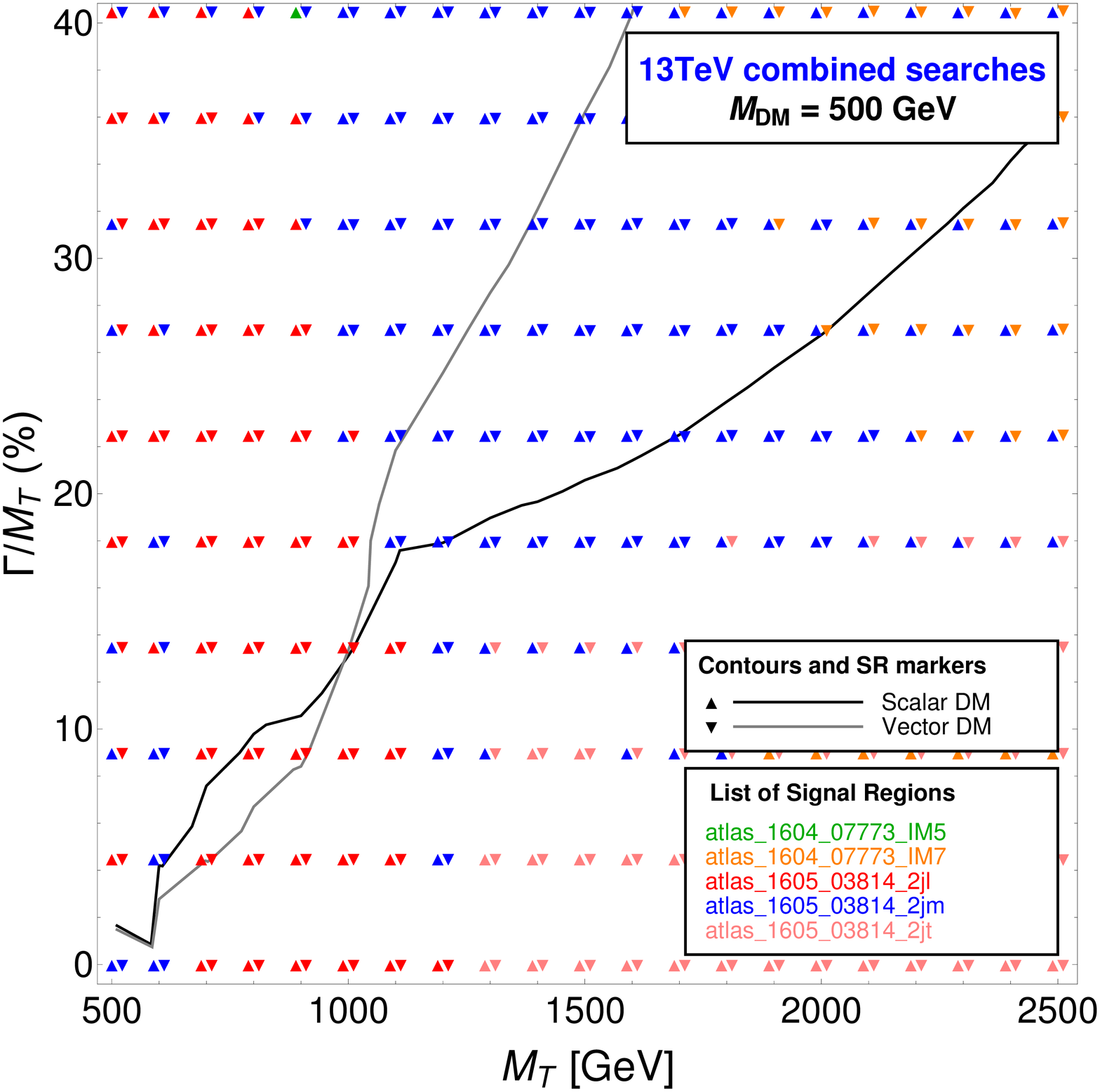, width=.32\textwidth} 
\epsfig{file=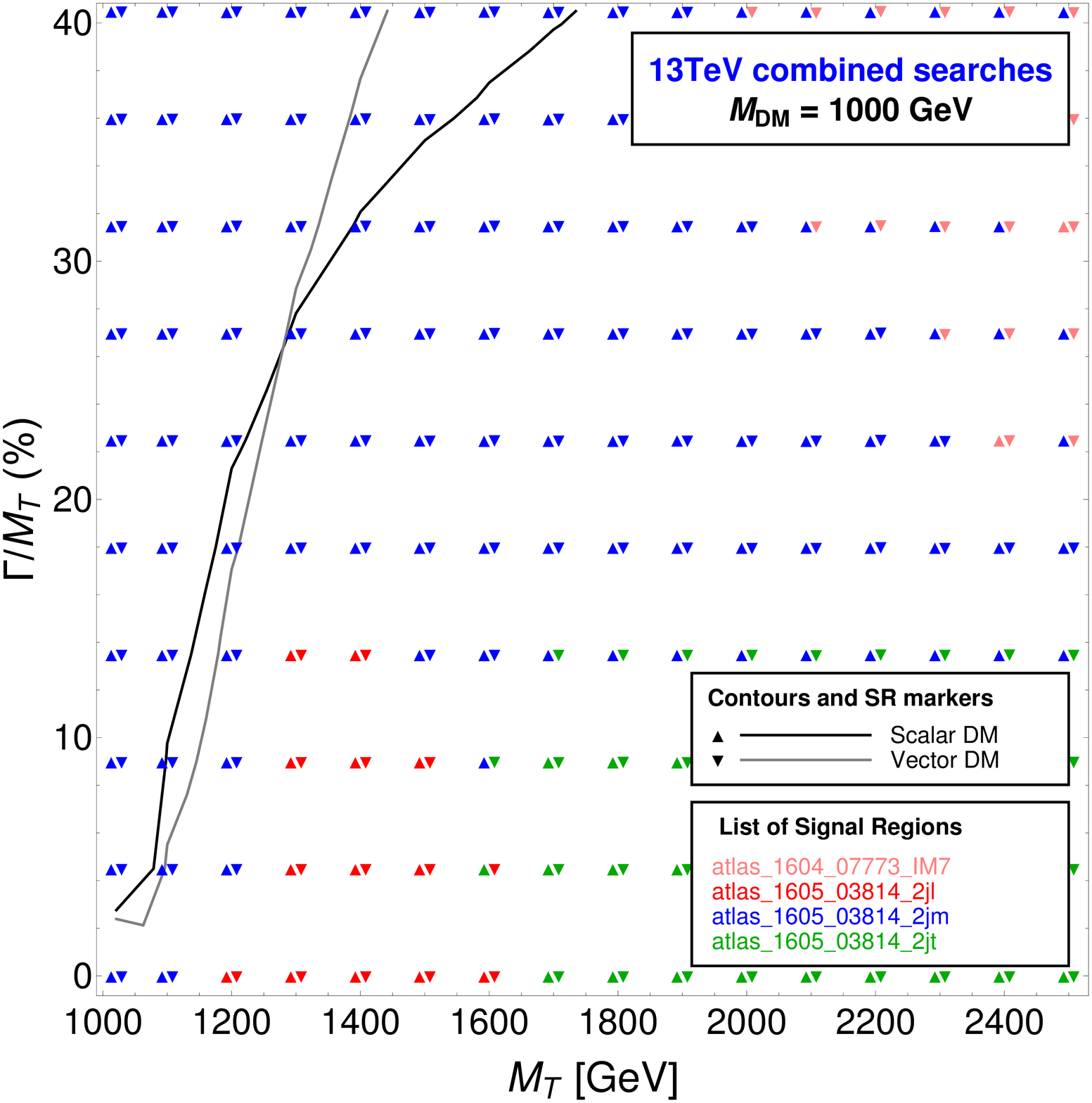, width=.32\textwidth} 
\caption[{\sc CheckMATE} results for a $T$ coupling to a DM particle (coupling to first generation) of mass 10 GeV, 500 GeV and 1000 GeV.]{{\sc CheckMATE} results for a $T$ coupling to a DM particle (coupling to first generation) of mass 10 GeV, 500 GeV and 1000 GeV. The black (grey) line shows which part of the parameter space is excluded in the scalar (vector) DM scenario.}
\label{fig:Exclusion1}
\end{figure}

\begin{figure}[ht!]
\centering
\epsfig{file=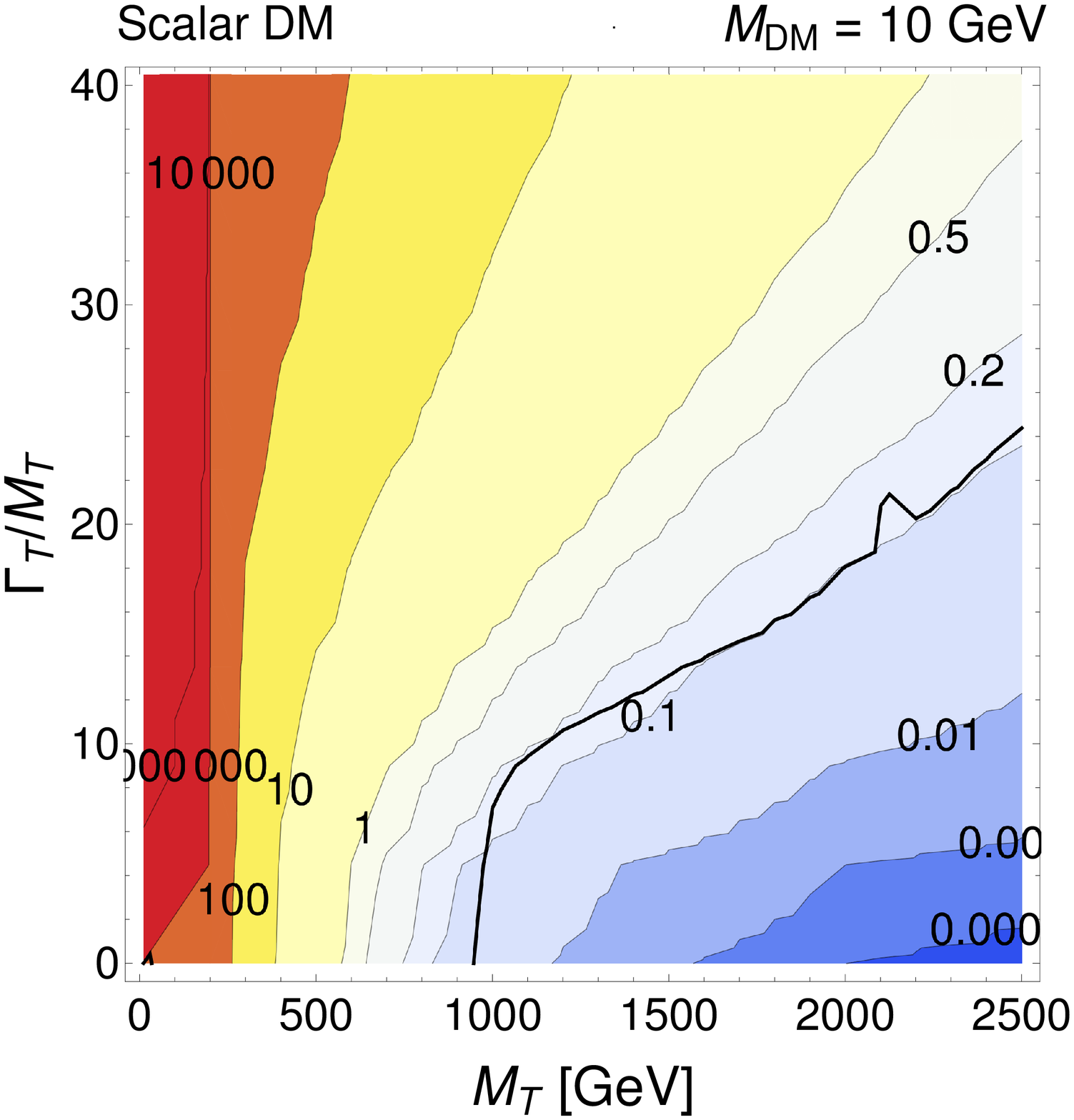,   width=.30\textwidth} 
\epsfig{file=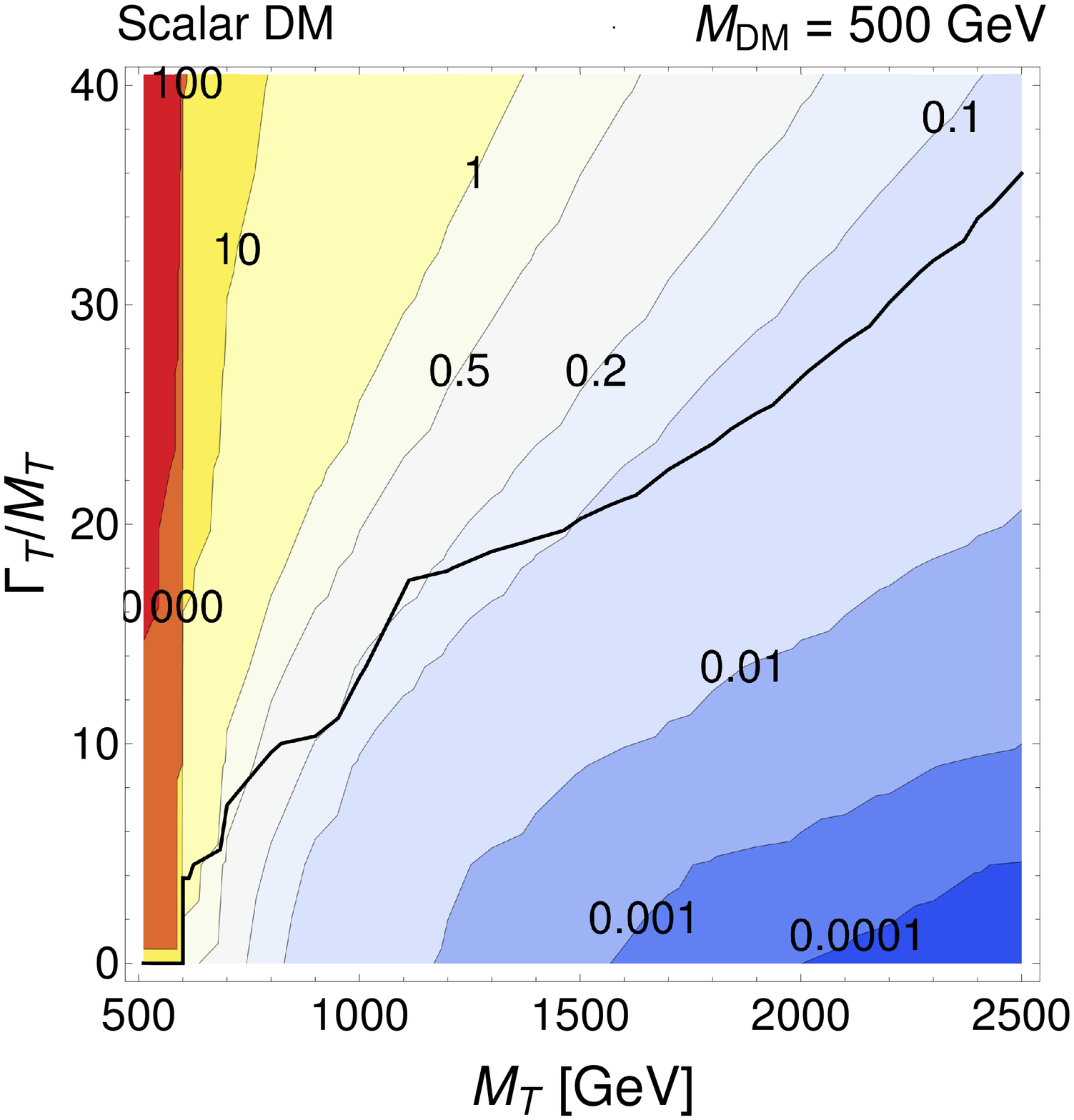,  width=.30\textwidth} 
\epsfig{file=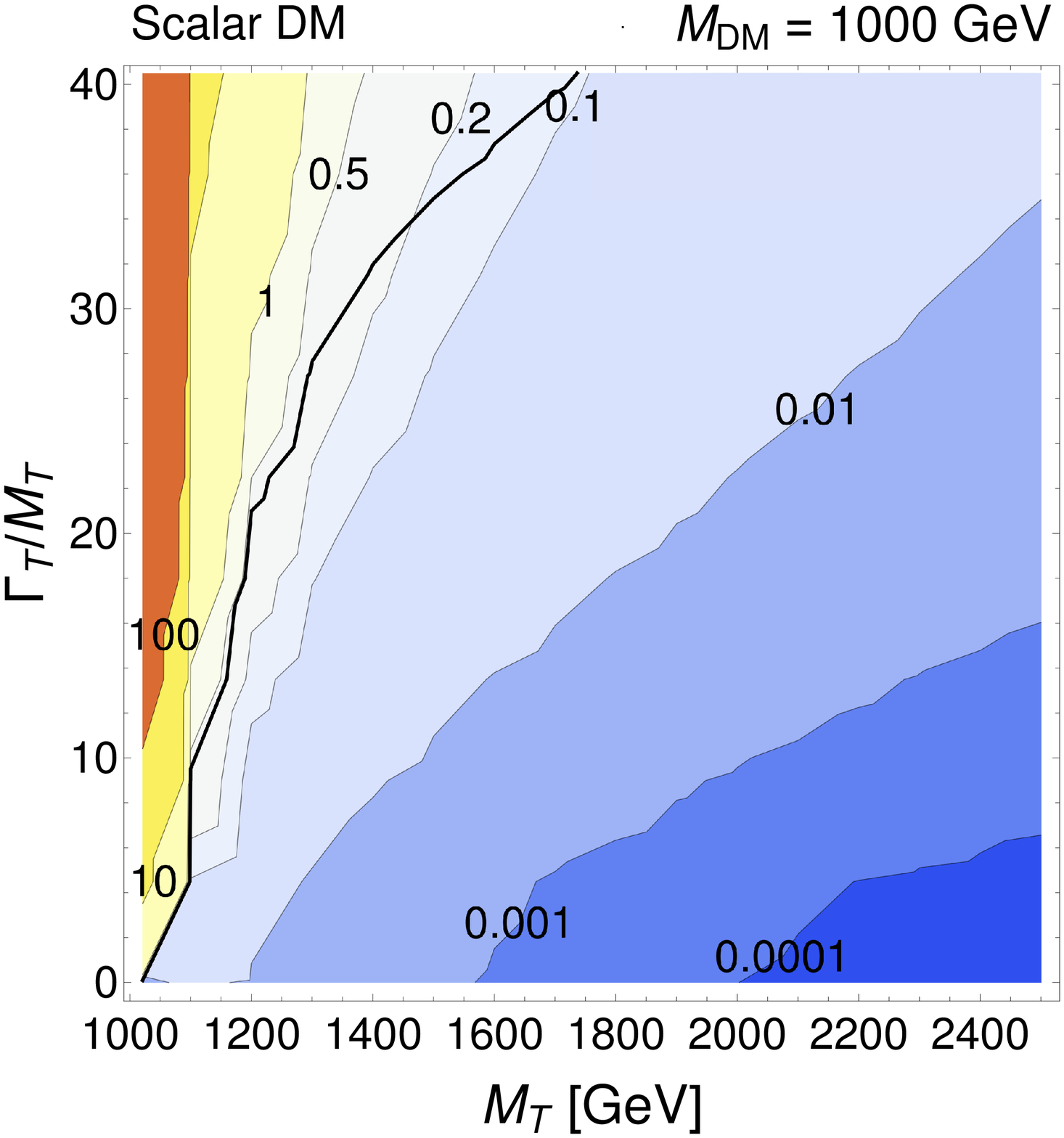, width=.30\textwidth}\\ 
\epsfig{file=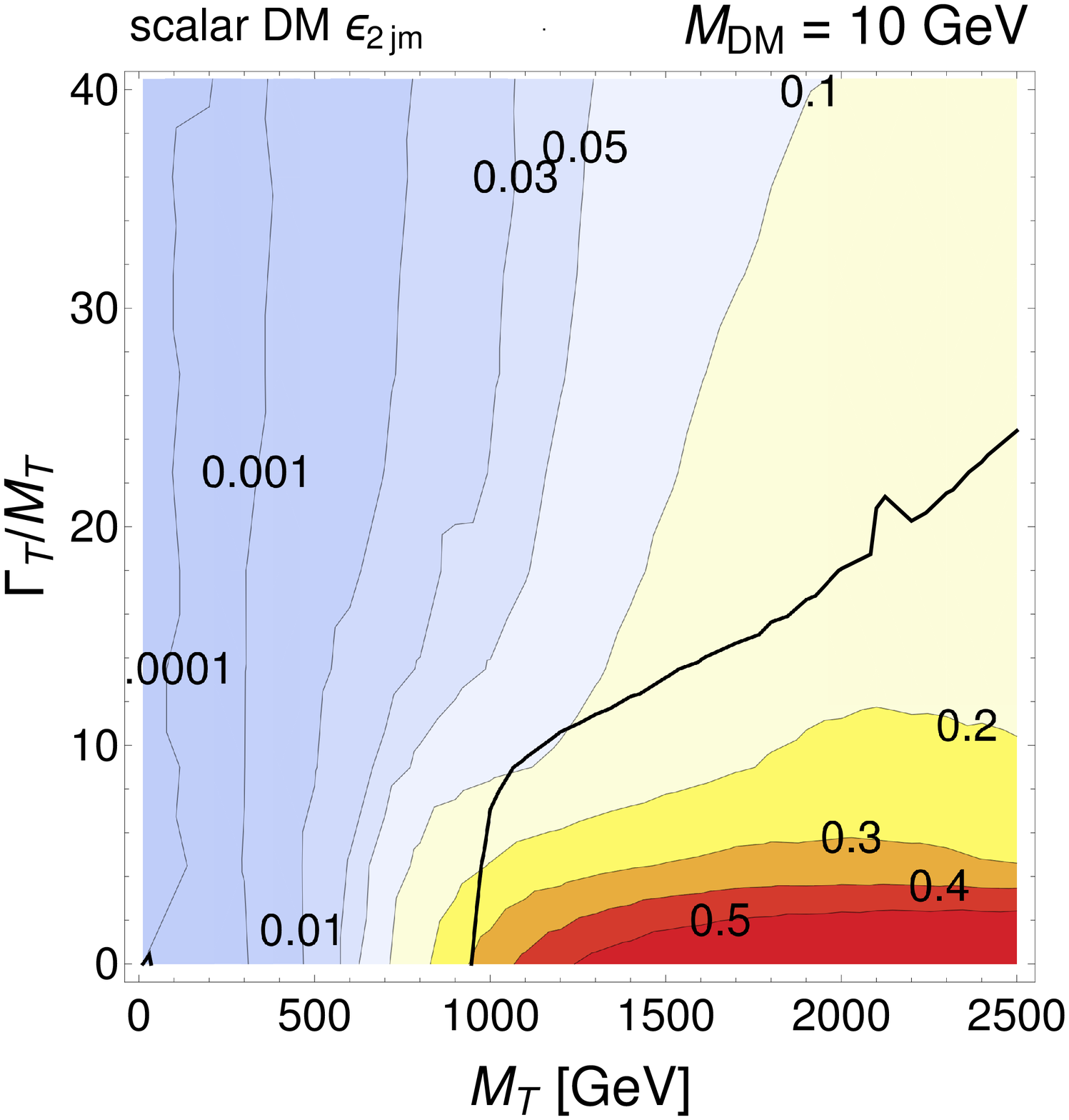,    width=.30\textwidth} 
\epsfig{file=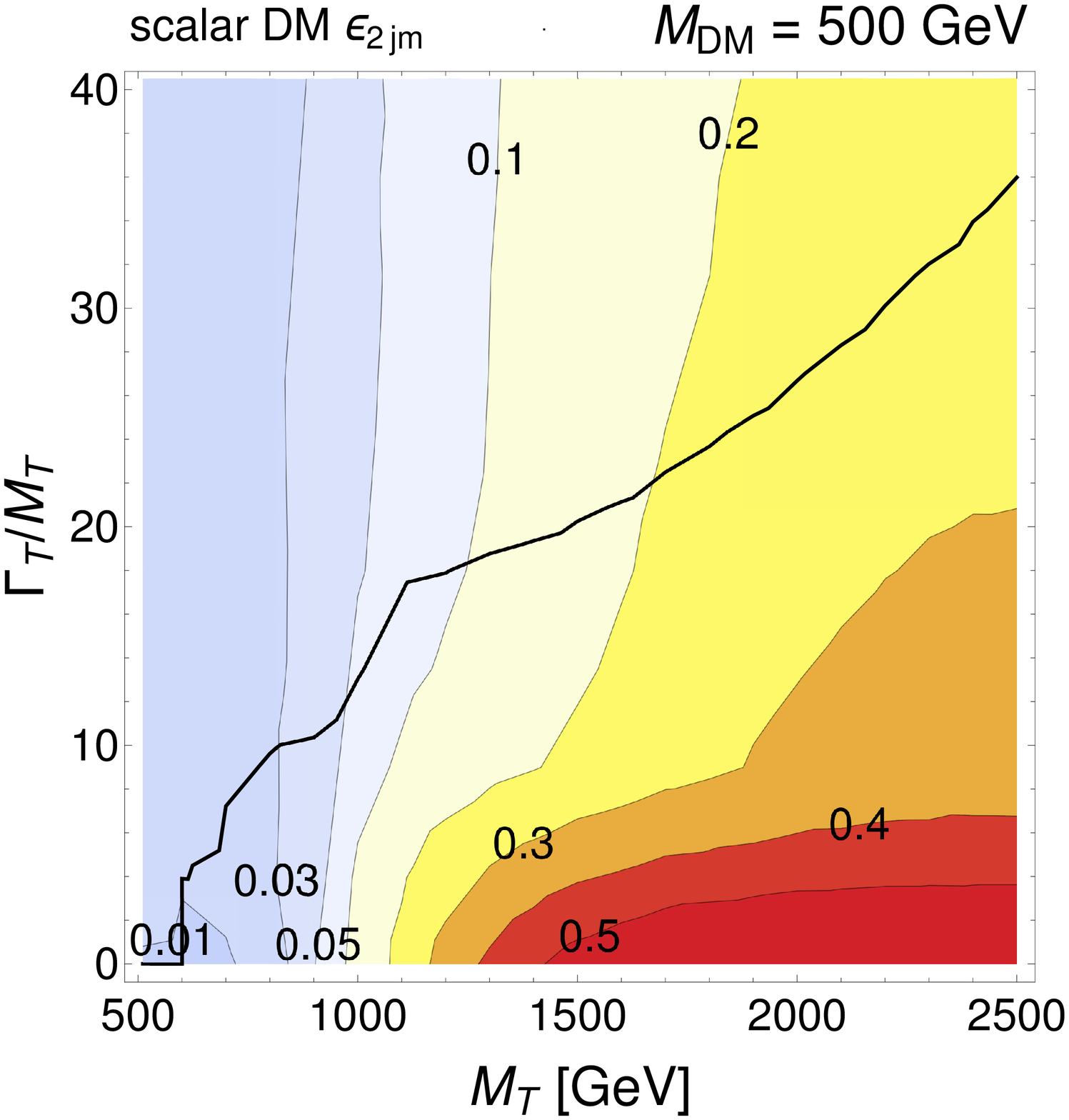,   width=.30\textwidth} 
\epsfig{file=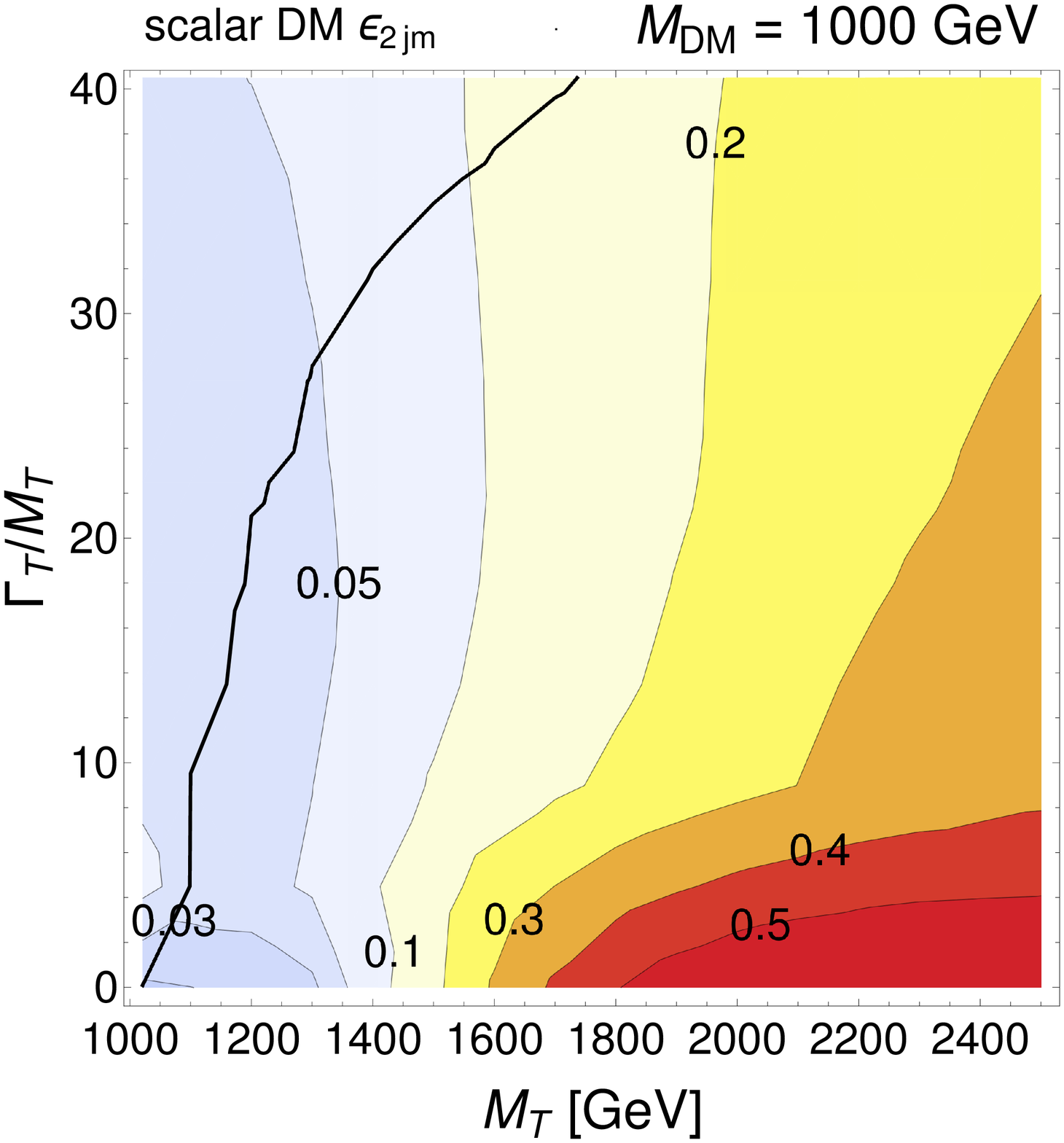,  width=.30\textwidth}
\caption[Full signal cross sections for the scalar DM case and efficiencies of the SR 2jm from the ATLAS search for different scalar DM masses.]{Top row: full signal cross sections for the scalar DM case. Bottom row: efficiencies of the SR 2jm from the ATLAS search~\cite{Aaboud:2016zdn} for different scalar DM masses.}
\label{fig:sigmaEffs1}
\end{figure}

\begin{figure}[ht!]
\centering
\epsfig{file=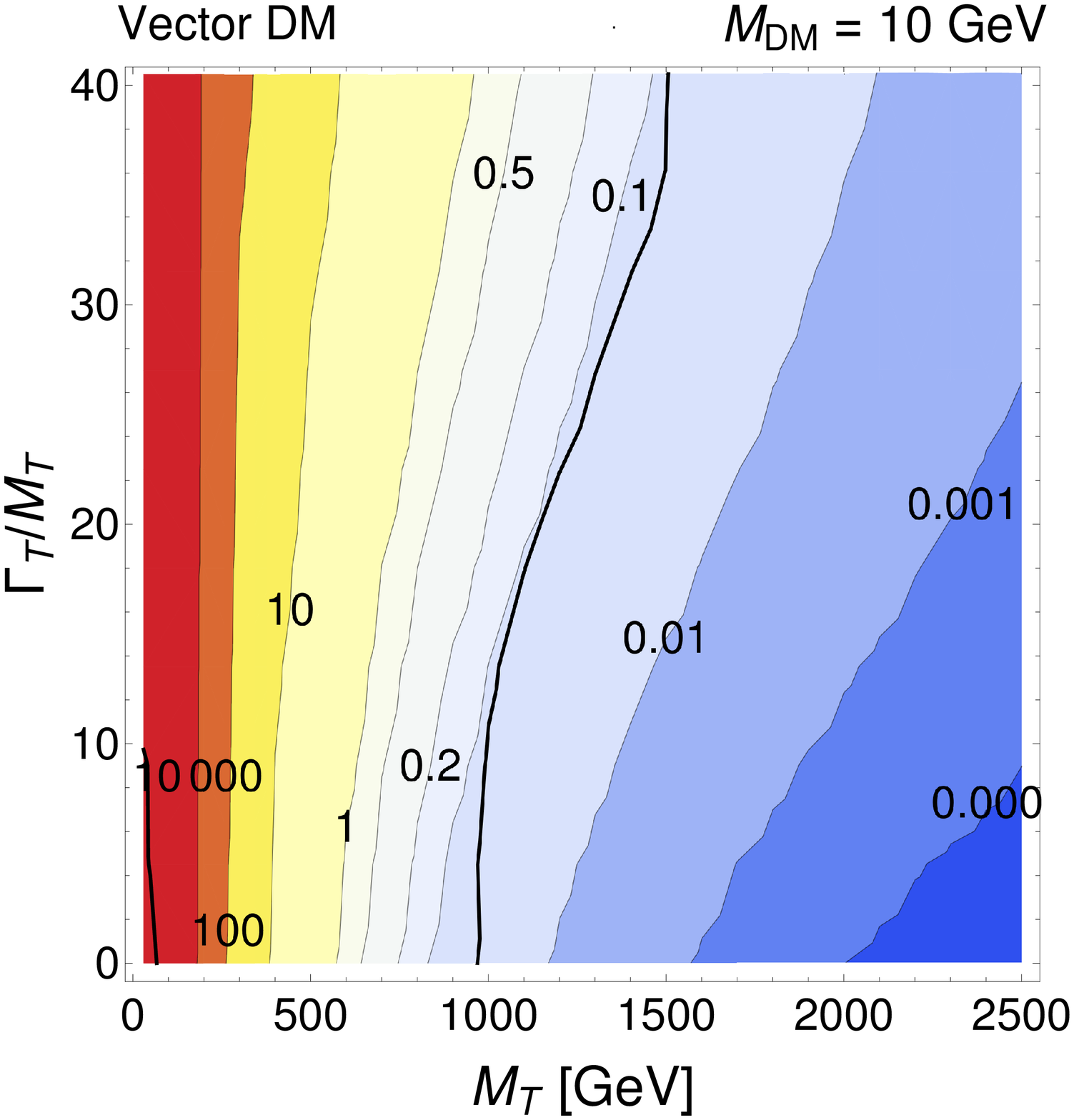,   width=.30\textwidth} 
\epsfig{file=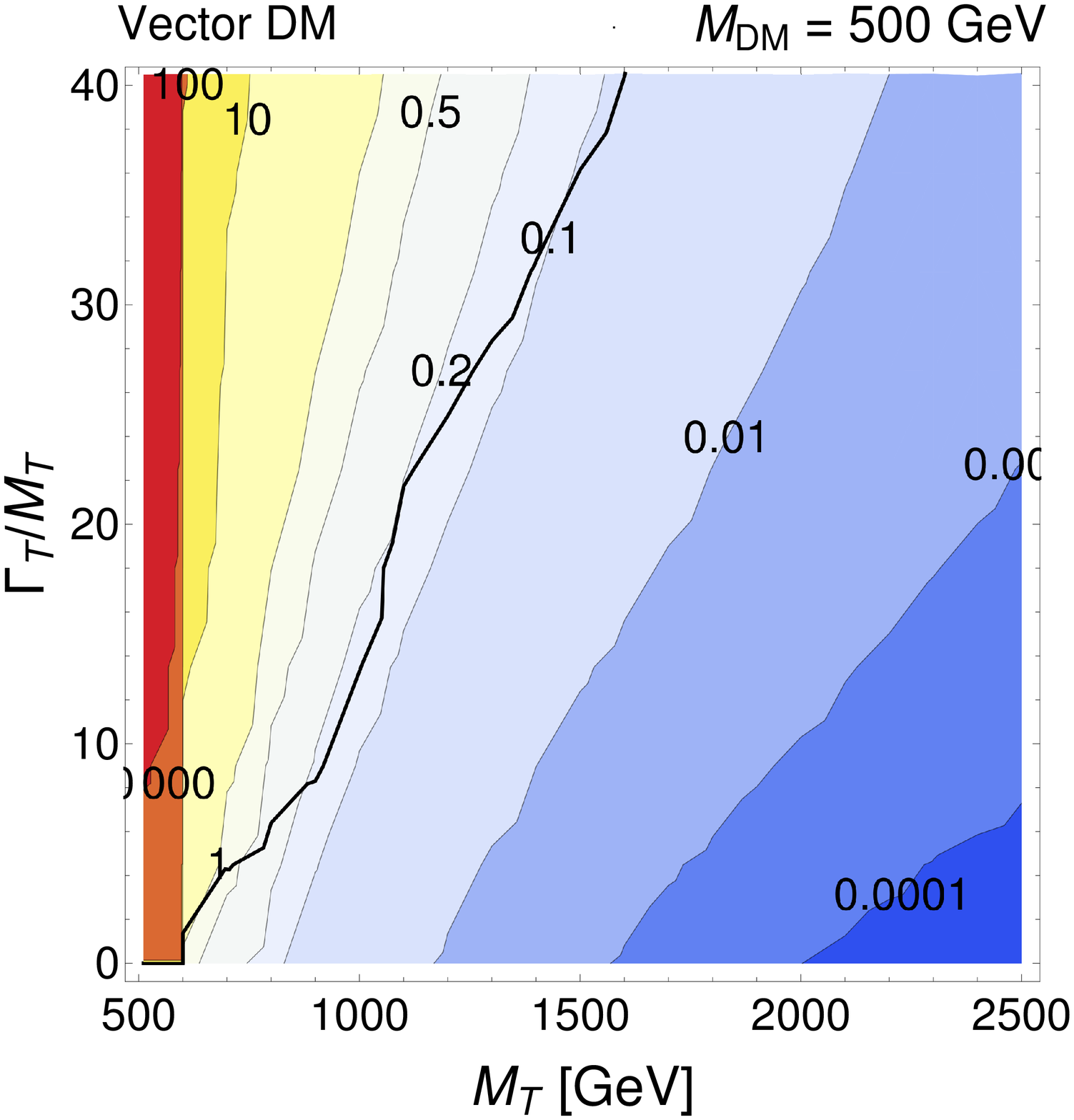,  width=.30\textwidth} 
\epsfig{file=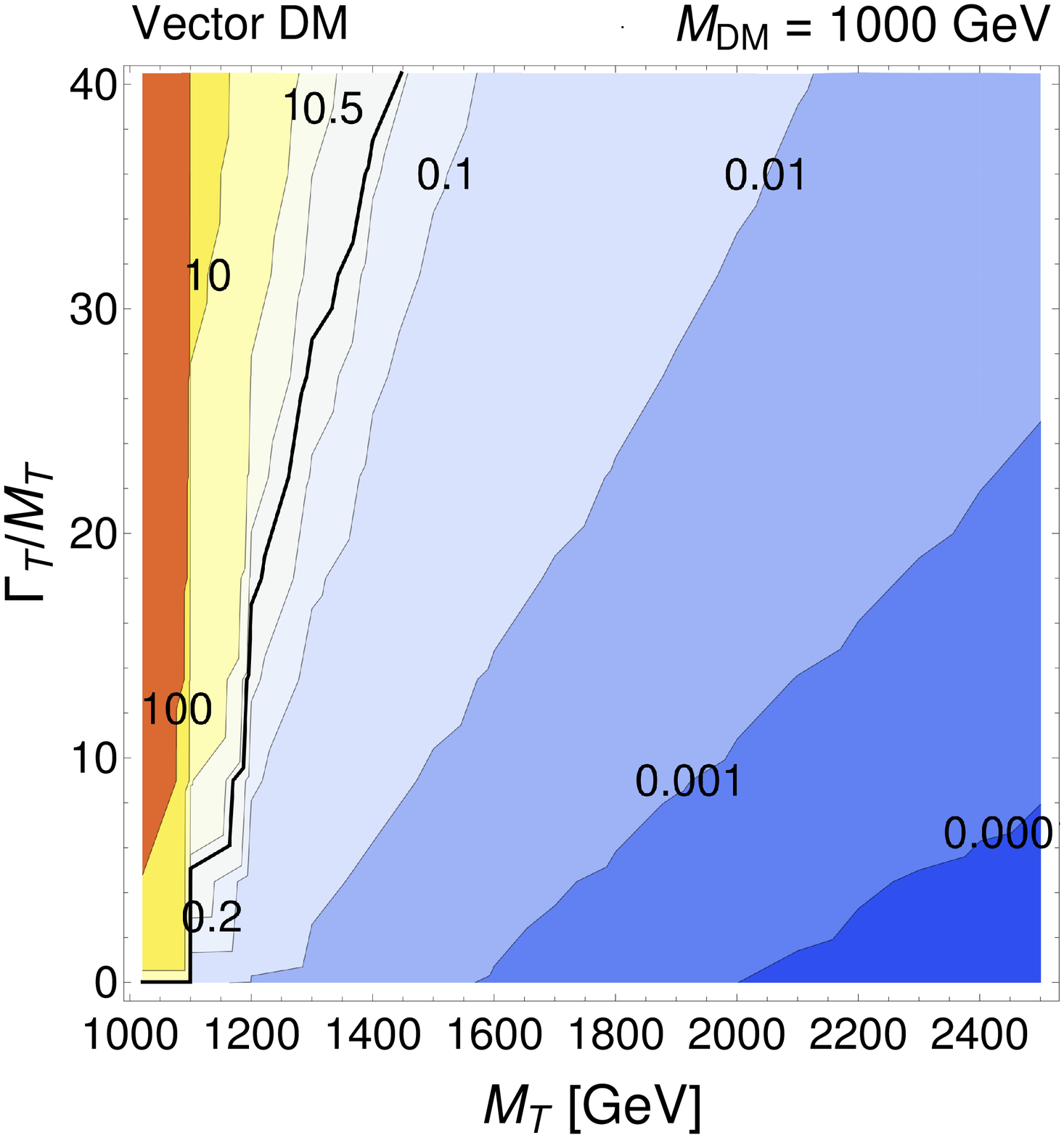, width=.30\textwidth}\\ 
\epsfig{file=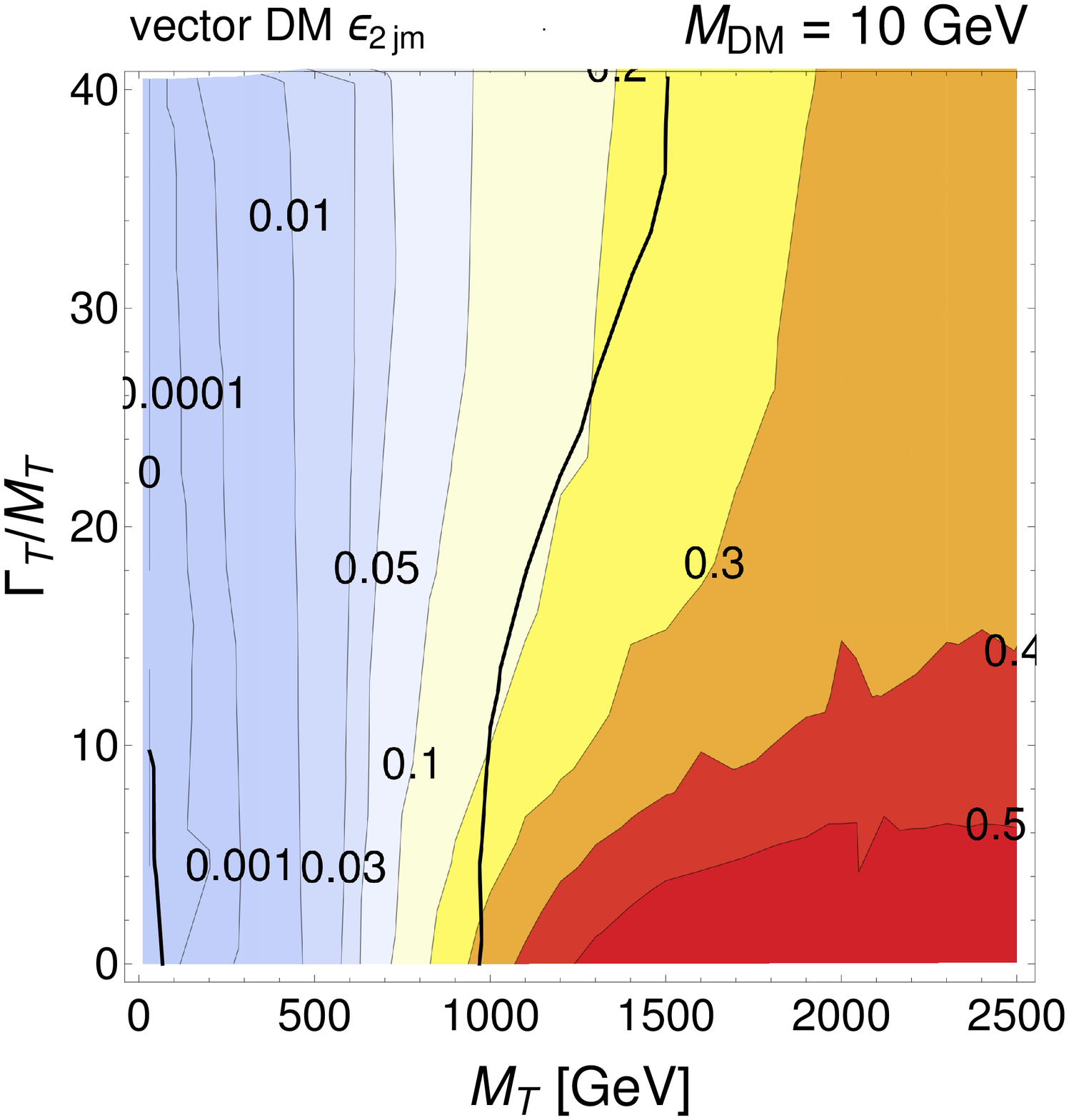,    width=.30\textwidth} 
\epsfig{file=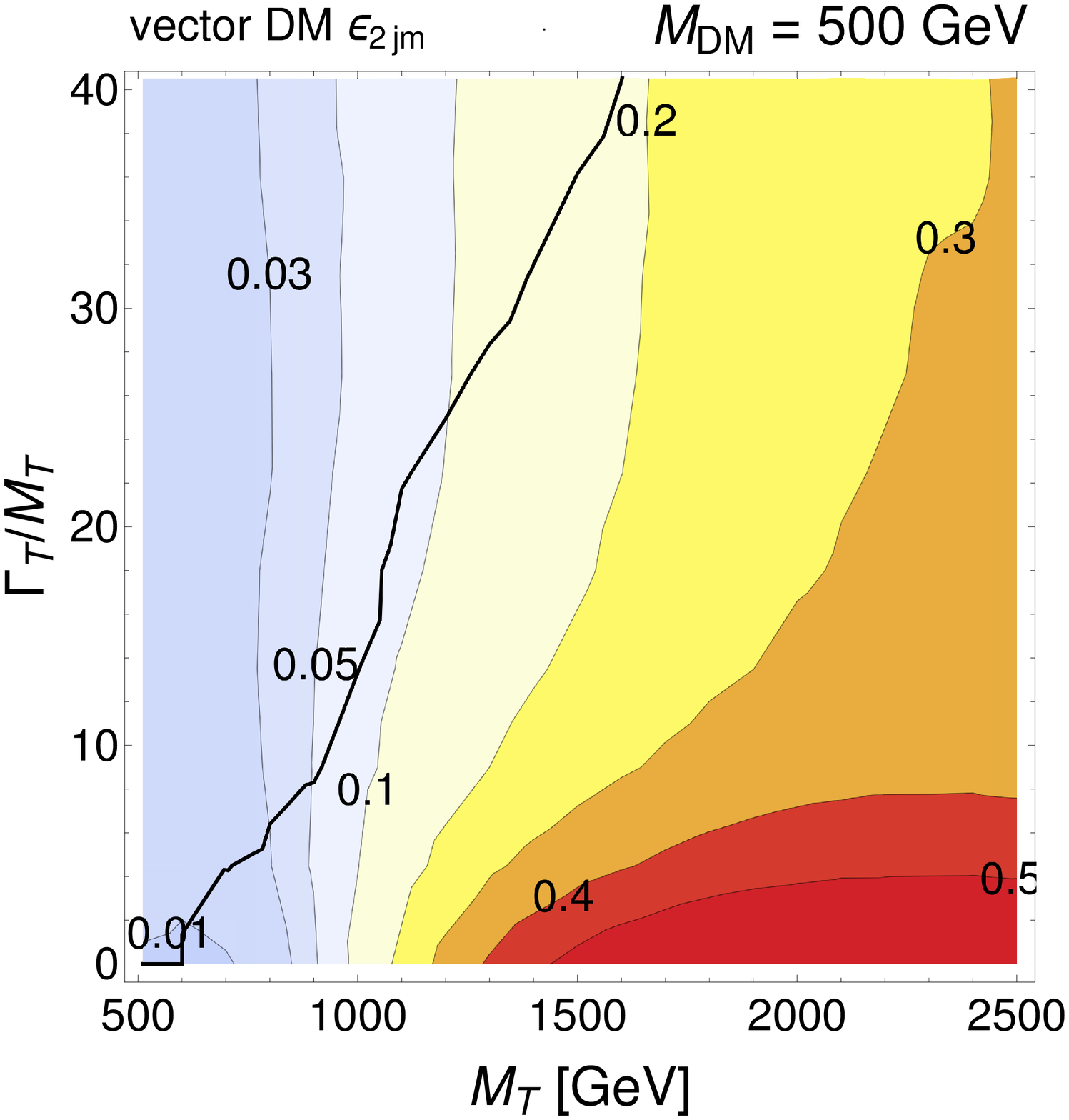,   width=.30\textwidth} 
\epsfig{file=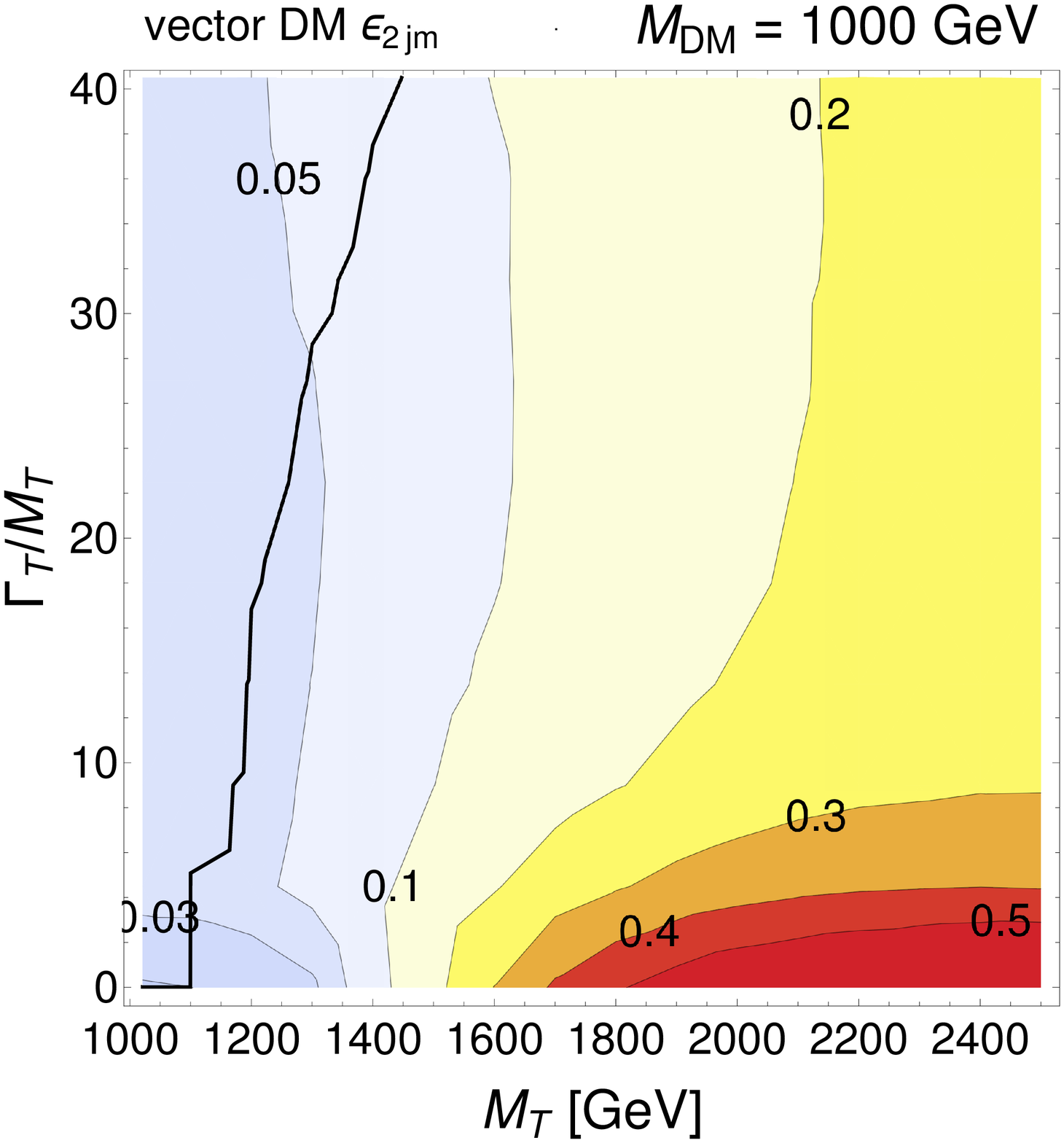,  width=.30\textwidth}
\caption[full signal cross sections for the vector DM case and efficiencies of the SR 2jm from the ATLAS search for different scalar DM masses.]{Top row: full signal cross sections for the vector DM case. Bottom row: efficiencies of the SR 2jm from the ATLAS search~\cite{Aaboud:2016zdn} for different scalar DM masses.}
\label{fig:sigmaEffv1}
\end{figure}

In Fig~\ref{fig:Exclusion1} the exclusion bound and the best SR are shown in the $(M_T, \Gamma_T / M_T)$ plane for both scalar and vector DM scenarios and for the same value of the DM mass considered in Fig. \ref{fig:SXfirst2}. In Figs.~\ref{fig:sigmaEffs1} and \ref{fig:sigmaEffv1} the exclusion bounds for scalar and vector DM respectively are shown together with the full signal cross sections and with the efficiencies of the most relevant SRs for the two DM spin hypotheses.

The main results for the case of $T$ coupling to first generation quarks are the following.
\begin{itemize}
\item For DM masses below to the TeV the bounds have a qualitatively similar behaviour, the width dependence is always sizeable, the bounds for small width are similar between scalar and vector DM and as the width increases the different DM spins exhibit different behaviours, where scalar DM scenarios show a stronger dependence on the $T$ width. 
\item The most sensitive SRs for the determination of the bounds are almost always 2jl, 2jm or 2jt of the ATLAS search \cite{Aaboud:2016zdn}, which are optimised for signals with two jets and $\MET$ in the final state. 
\item For DM masses around the TeV or higher the width dependence of the bound is still present but the difference between the scalar and the vector DM scenarios becomes weaker. Furthermore, the NWA region is never excluded. Analogously to the case of coupling with third generation, this is a consequence of a combination between larger phase space and width dependence of the experimental acceptances.
\end{itemize}

%\vspace{\baselineskip}
\paragraph*{Dependence on the chirality of the couplings}

\ \vspace{\baselineskip}

\begin{figure}[ht!]
\centering
\epsfig{file=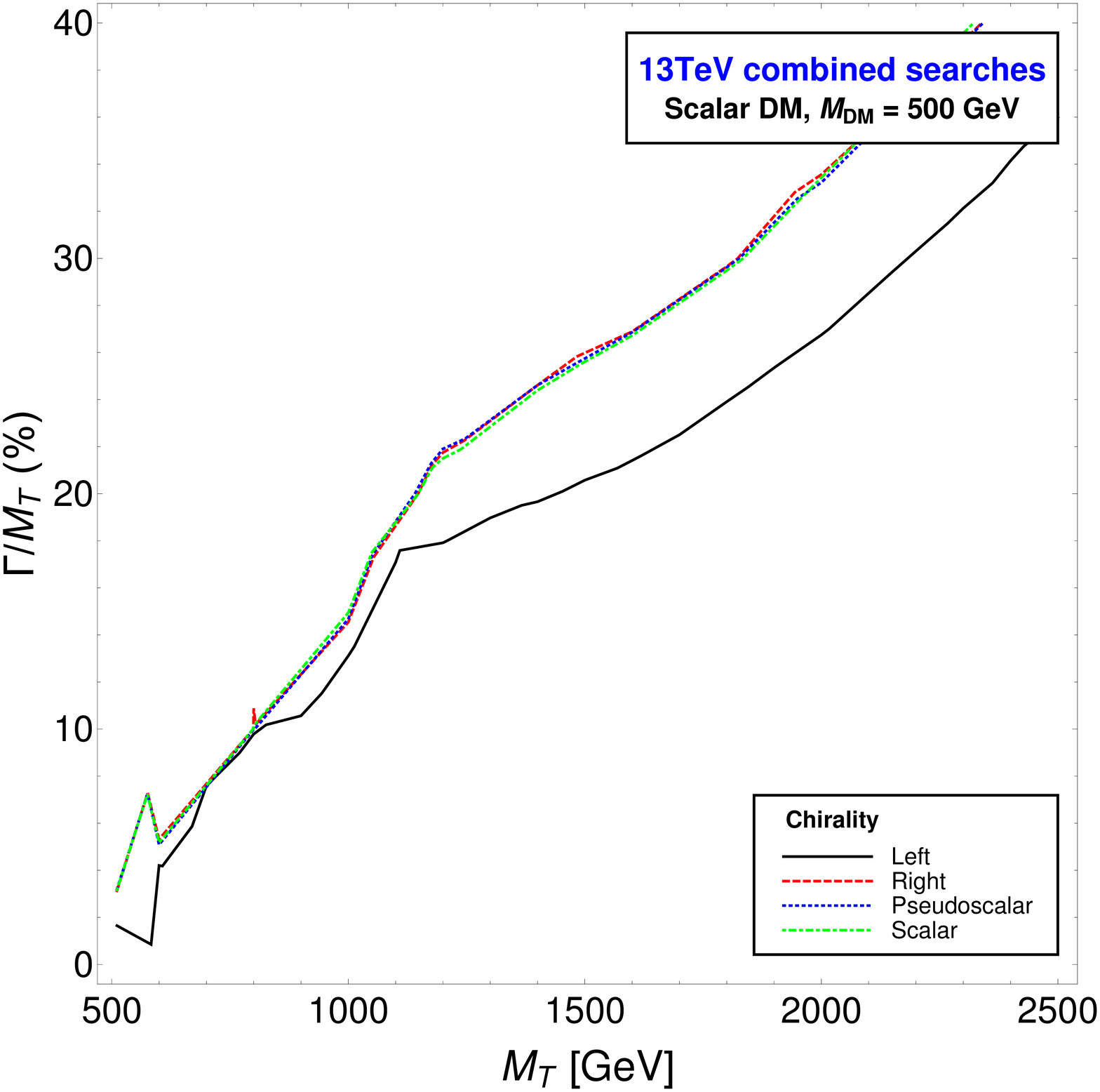, width=.45\textwidth} 
\epsfig{file=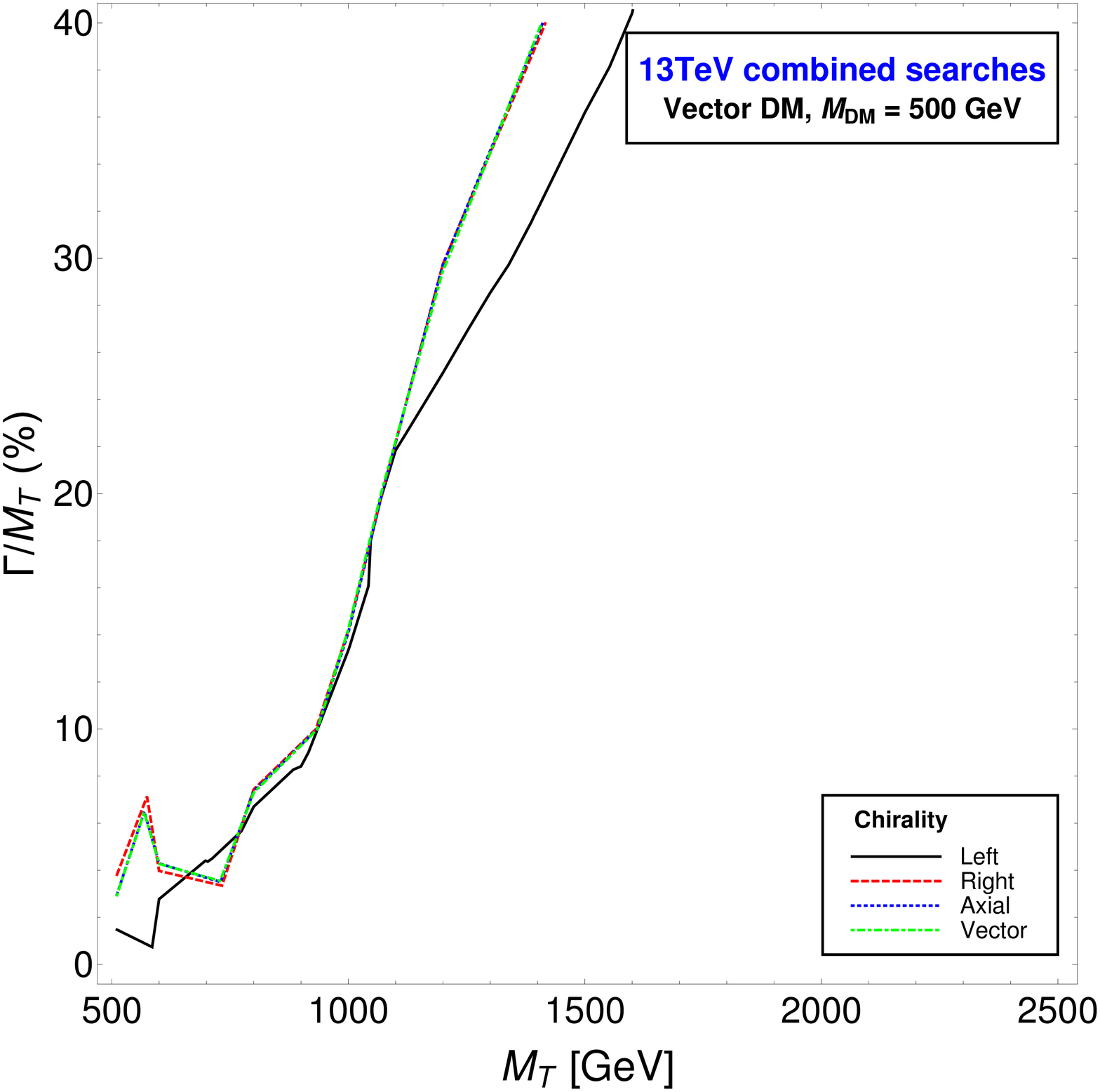, width=.45\textwidth} 
\caption{Exclusion bounds for a $T$ interacting with the SM up quark and DM for different hypotheses on the chirality of the couplings: for a VLQ $T$ pure left-handed and pure right-handed couplings, and for a ChQ $T$ pure scalar (vector) or pseudoscalar (axial-vector) couplings if $T$ interacts with scalar (vector) DM.}
\label{fig:1Gchirality}
\end{figure}

Analogously to the case of $T$ coupling with third generation quarks, the analysis of the dependence of the limits on the chirality of the couplings (and therefore on the hypotheses about the properties and representations of $T$) is presented. In Fig.~\ref{fig:1Gchirality} the exclusion bounds for different couplings are shown. Once again even if the uncertainty due to the use of a recasting tool is quite large, we observe that the scenario with pure left-handed coupling exhibits a slightly stronger width dependence than the rest of the scenarios in the large width regime. Even if the bounds are in the same regions, the most sensitive SRs of (the subset of) current searches could be in principle used to distinguish the scenario where the $T$ is a VLQ doublet from the others, in case of discovery. We are not going, however, to explore this potentiality in the present study, as it goes beyond the scope of our analysis.

%%%%%%%%%%%%%%%%%%%%%%%%%%%%%%%%%%%%%%%%%%%%%%%%%%%%%%%%%%%%%%%%%%%%%%%%%
%%%%%%%%%%%%%%%%%%%%%%%%%%%%%%%%%%%%%%%%%%%%%%%%%%%%%%%%%%%%%%%%%%%%%%%%%

%\newpage

\subsection{Exclusion limits in the $M_T-M_{DM}$ plane}
\label{sec:BellPlots}

\begin{figure}[ht!]
\centering
\epsfig{file=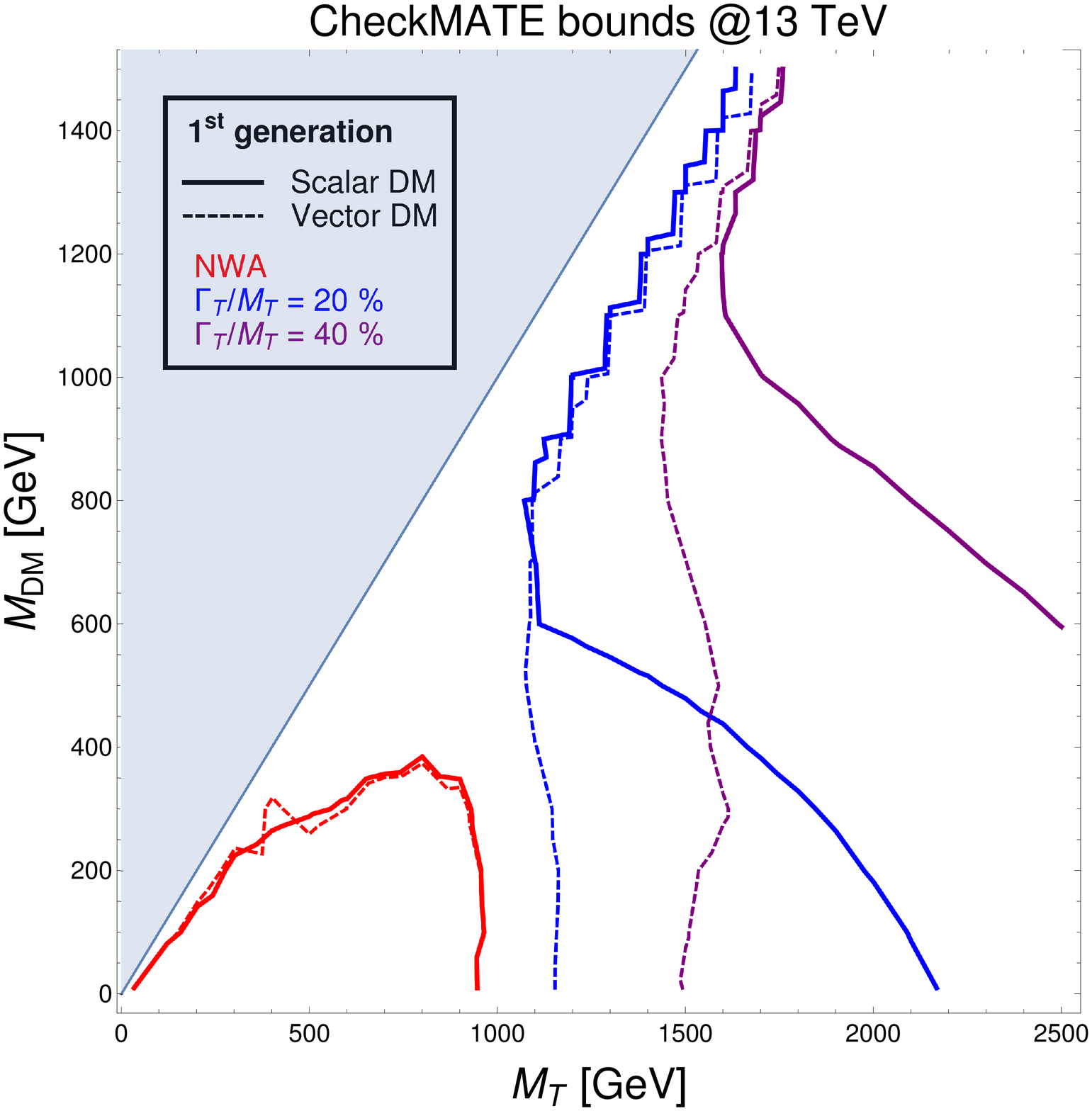, width=.32\textwidth} 
\epsfig{file=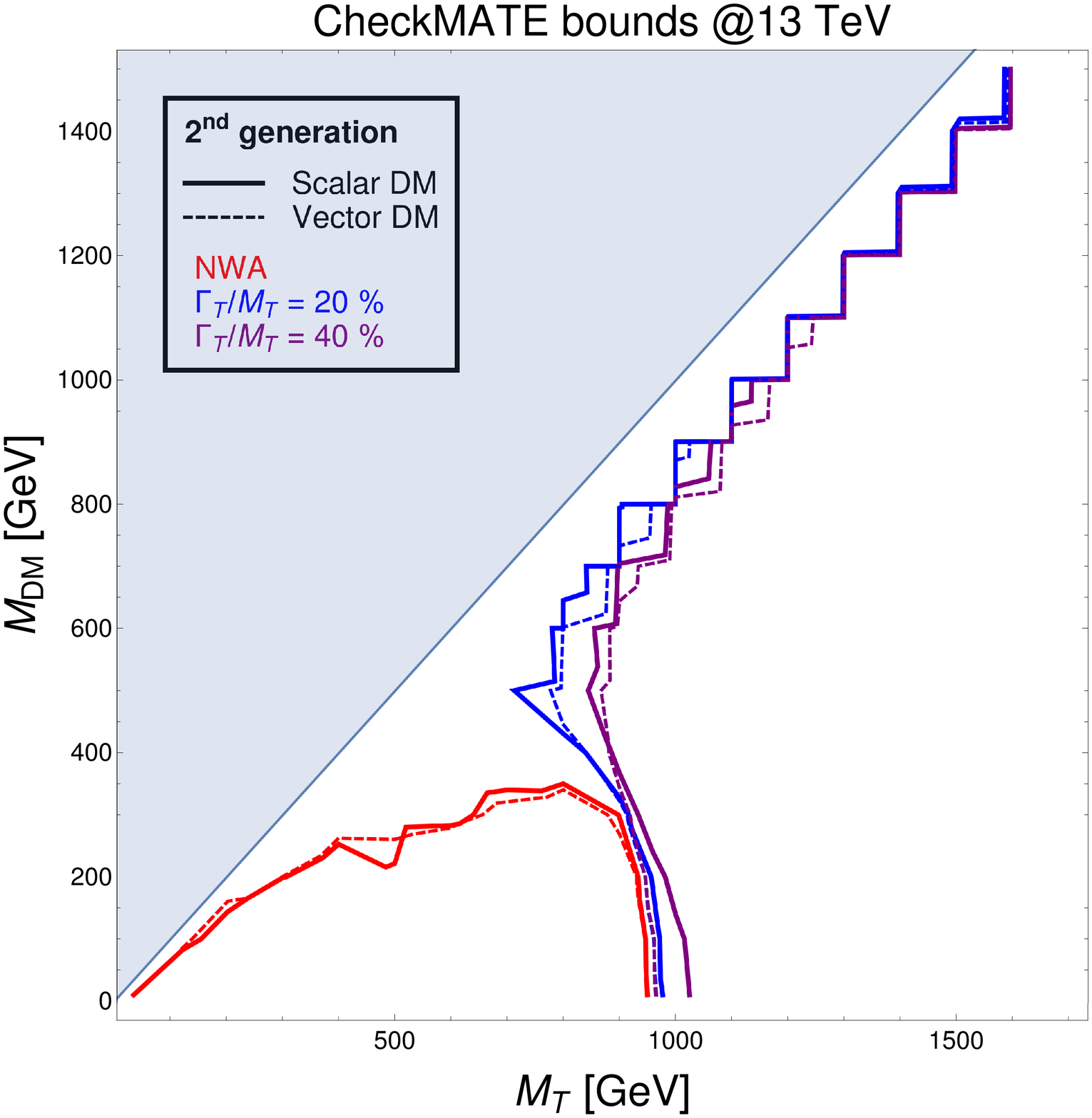, width=.32\textwidth} 
\epsfig{file=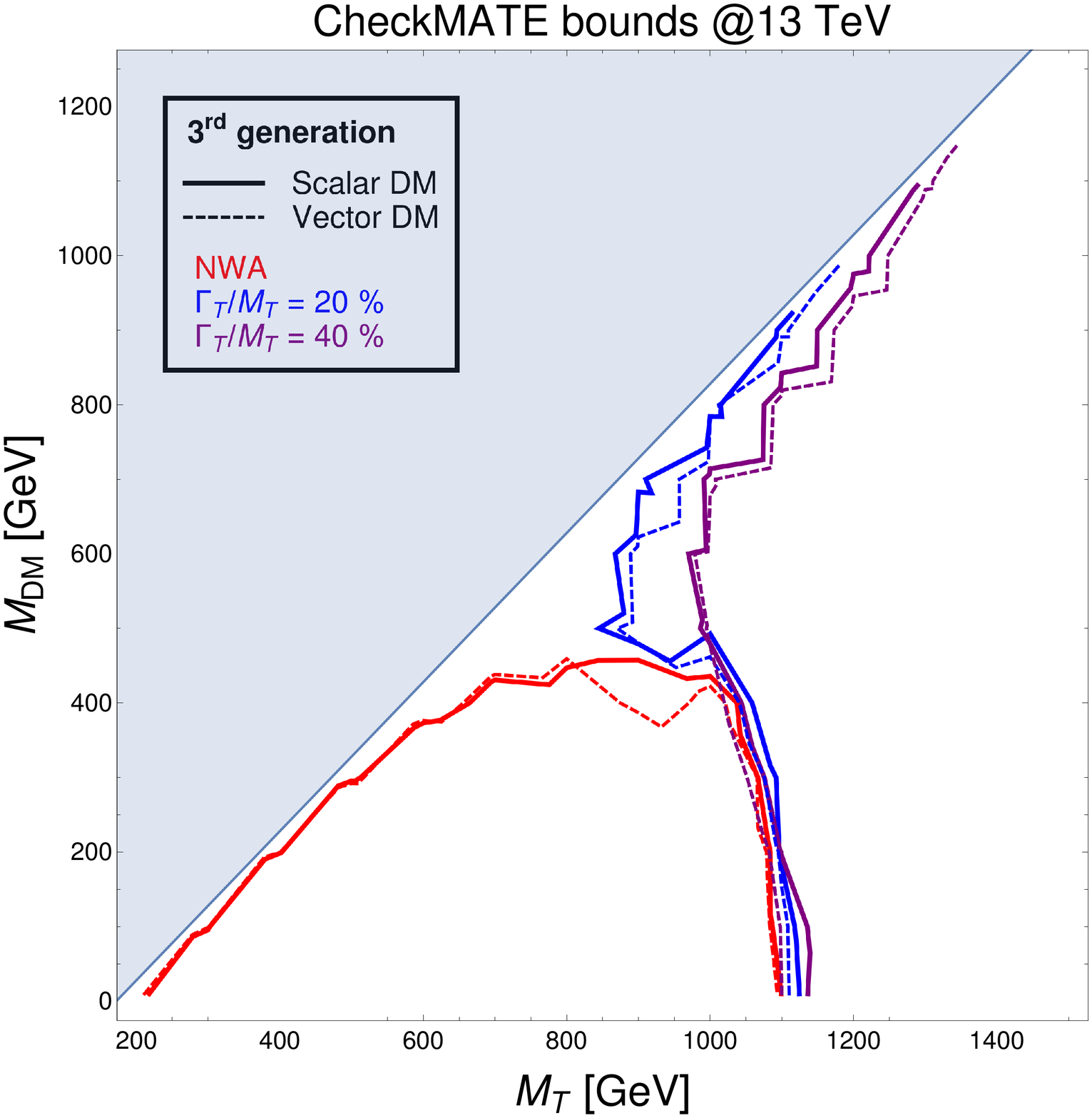, width=.32\textwidth} 
\caption{Bounds in the $(M_T, M_{\rm DM})$ plane for $T$ quark coupling DM particle and first (left panel), second (centre panel) and third (right panel) generations of SM quarks for different values of $\Gamma_T / M_T$.}
\label{fig:BellCombined}
\end{figure}

The scenarios we are considering have three parameters: the mass of the $T$, the width of the $T$ and the mass of the DM, with the only constraints given by the kinematical limit between the masses ($M_T > M_{\rm DM} + m_q$) and by the fact that the width should not really exceed 50\% of the mass, otherwise the concept of resonant state is essentially lost. The exclusion bound at 2$\sigma$ will therefore identify a 3D surface in the space defined by the three parameters (where the width is substituted by the $\Gamma_T/M_T$ ratio) and therefore it is instructive to analyse the projections of this surface on the plane identified by the masses of $T$ and DM for different values of the $\Gamma_T/M_T$ ratio. Such representation is also useful to directly compare bounds on $T$ and bosonic DM with analogous results in other models, such as SUSY. Indeed, the exclusion limits of SUSY searches are often presented in the $(M_{\tilde{t}}, M_{\chi_0})$ plane. We show in Fig.~\ref{fig:BellCombined} the bounds in the $(M_T, M_{\rm DM})$ plane for specific values of $\Gamma_T / M_T$: the NWA case, 20\% and 40\%. We included in this figure the results for a $T$ quark coupling to DM and the charm quark. 

The qualitative behaviours of the exclusion limits strongly depend on the assumption about which SM quark generation the $T$ couples to.

\begin{itemize}

\item {\bf $T$ coupling to DM and up quark}: in the NWA the exclusion limits for scalar and vector DM are not distinguishable in practice (barring numerical fluctuations). When the width of the $T$ increases, however, the bounds for scalar and vector exhibit a sizeably different dependence on the $T$ and DM masses. If the DM mass is below a width dependent threshold, the scalar DM case excludes a much wider region of the parameter space. This behaviour can be understood by looking again at Figs.~\ref{fig:sigmaEffs1} and \ref{fig:sigmaEffv1}, which show that the full signal cross section has a largely different trend with changing width depending on the scalar or vector nature of the DM. For high enough DM masses, the dependence on the width is less pronounced and this erases the differences between the bounds above a certain value of the DM mass. A further peculiarity of the large width regime, with respect to the NWA, is that the region where the mass gap between $T$ and DM is small is always excluded.

\item {\bf $T$ coupling to DM and charm quark}: in both NWA and large width regime it is not possible to distinguish scalar from vector scenarios. As the width increases, the region close to the kinematics limit ($M_T=M_{\rm DM}+m_c$) becomes excluded, while it would be allowed in the NWA. If the DM mass is below 300 GeV and far from the kinematics limit, the bound depends very weakly on the width.

\item {\bf $T$ coupling to DM and top quark}: the mass bounds for scalar and vector DM are very similar in both the NWA and large width regime. The increase of the width modifies the bound (with respect to the NWA) if the mass of $T$ is close enough to the kinematics limit ($M_{\rm DM}+m_t$): unlike in the NWA case, as the values of the $T$ mass approaches the kinematics limit, they become more and more excluded by experimental data as the $T$ width increases. Moreover, if the DM mass is below $\sim 400$ GeV and far from the kinematics limit, the bound on the $T$ mass does not depend on the width. Designing new specific cuts could allow a more optimised exploration of the large width regime of XQs decaying into DM and third generation SM quarks, especially considering the fact that efficiencies for the most sensitive SRs exhibit a general decrease along the bound region as the width increases (as shown in Fig.~\ref{fig:sigmaEffs3}). \\

\end{itemize}

To conclude this section, the bounds obtained under the NWA are less stringent than the bounds obtained when the NWA is relaxed and the width is allowed to have large values, relative to the $T$ mass. This results can be intuitively expected when considering that larger widths correspond to larger cross sections and, unless the selection and cut efficiencies compensate the cross section enhancement, the number of signal events increases with respect to the NWA scenario. It is remarkable, though, that different assumptions about the couplings of $T$ with different SM quark generations produce either negligible or sizeably different bounds if the DM is scalar or vector. This result could be exploited for the design of new experimental searches which are not only meant to discover new signals in channels with $\MET$ but also to characterise the signal.

%%%%%%%%%%%%%%%%%%%%%%%%%%%%%%%%%%%%%%%%%%%%%%%%%%%%%%%%%%%%%%%%%%%%%%%%%
%%%%%%%%%%%%%%%%%%%%%%%%%%%%%%%%%%%%%%%%%%%%%%%%%%%%%%%%%%%%%%%%%%%%%%%%%

%\newpage

\subsection{Conclusions}
\label{sec:conclusions2}

We have estimated large width effects in a rather simple model with only one XQ decaying into DM and a SM quark. As a general result, we conclude that the XQ nature, whether it be VLQ or ChQ, does not play a significant role in the phenomenology we have studied, primarily because one can be turned into the other by simply changing the left and right fermion couplings suitably and the observables normally adopted in experimental analyses do not resolve their relative size and/or sign. Furthermore, we have established that, for the same choice of $M_{\rm DM}$, there occur sizeable differences between the two aforementioned approaches (NWA versus full result) depending on whether one adopts the scalar or vector nature of the DM candidate, the more so the larger the value of $\Gamma_T /M_T$. However, are the coupling properties of the $T$ state that are most responsible for the largest differences seen between the simplistic (model independent) and realistic (model dependent) approaches outlined. 
On the one hand, when coupling is allowed to the third generation only, the exclusion limits depend only slightly upon $\Gamma_T/M_T$, with a general trend pointing towards the cross section becoming larger when the width increases, yet with the additional contributions with respect to the NWA being generally suppressed by the cuts on missing transverse energy normally adopted in experimental searches. On the other hand, when coupling is allowed to the first (second) generation only, exclusion limits massively depend upon the width because the aforementioned additional topologies are not suppressed by such cuts in missing transverse energy, the more so the larger both $M_Q$ and $\Gamma_Q/M_Q$ are. (In fact, differences between the DM nature are significantly more prominent in the case of coupling to first (second) generation than in the third generation one.) Clearly, a fully-fledged model incorporating coupling to any generation will fall in between these two extreme conditions, with further subtleties induced by the PDF behaviour, as one can already see by comparing our results for the first and second generation cases.

In conclusion then, results from LHC searches for any XQs, when decaying to DM (whether spin 0 or 1) and either a heavy or light SM quark, should be taken with caution, as they do not account for effects induced by either the large XQ width, the additional (to the pair production ones) topologies or both, which can be very large even in a simplified model with only one XQ. Hence, one should rescale the observed limits from established experimental analyses to the actual ones upon accounting for such effects (as we have done here) or else attempt  deploying new ones adopting different selection strategies which minimise (in the case of exclusion) or indeed exalt (in the case of discovery) such effects  (which will be the subject of a future publication). At any rate, the time-honoured assumption that the NWA  is a reliable investigative approach applicable over most of the parameter space of the BSM scenarios dealt with here should be dismissed. In fact, we also have cautioned that, despite cancellations may exist between the various effects described here, which in the end might not change sizeably the inclusive cross section for certain values of $\Gamma_Q/M_Q$, these are only accidental and do not apply to the exclusive observables used in experimental searches, so that, again, limits obtained in the NWA would be inaccurate, owing to mis-estimated efficiencies.

%% ------------- --------------------- --------------------- ---------------------- --------------------- ---------------------- ----------------------

%\section{Interference effects in pair production of XQs decaying to DM} \label{sec:VLQDM_interference}

%% ------------- --------------------- --------------------- ---------------------- --------------------- ---------------------- ----------------------
%% Conclusions
%% ------------- --------------------- --------------------- ---------------------- --------------------- ---------------------- ---------------------- 

\chapter{Conclusion} \label{Chapter:conclusion}

The LHC has been running with a centre of mass energy of the beam equal to 7 TeV and 8 TeV, and a new run at 13 TeV started in 2015, with a planned upgrade at 14 TeV. The experimental searches conducted at this experiment and looking for new particles have to do it in the framework of simplified models because a full treatment of each different models imagined by the model-builders would not be doable. These simplified models include important assumptions on the particle content and on some of the parameters of these models. To be sure that we fully exploit the possibilities of the LHC and that we are not missing any discovery because of these assumptions, we have to question and analysed them in detail and that is the main purpose of this thesis, where we have studied in detail some specific aspects of models of new Physics featuring XQs, pointed some limitations of the current experimental searches and given some hints on how to improve them when possible. \\

We have first done in Chapter \ref{Chapter:IntroXQ} a short review providing a broad, though necessarily incomplete, overview about the searches and perspectives of heavy XQs at the LHC. We have seen there that XQs (and especially VLQs) are predicted by many models of new physics. A minimal extension of the SM with the presence of XQs therefore has a huge and interesting range of possible signatures, some of which have already been tested experimentally, giving bounds on the mass of XQs are around 690-1000 GeV. We have then presented a model independent parametrisation that can be used to describe the phenomenology of XQs, both in the case of XQ coupling with SM bosons and with DM candidates.
We have also provided a short description of the most recent phenomenological analyses present in literature and described the limitations that the current experimental searches have due to the assumptions they make: they only consider models with only one new state, assume that its width is small with respect to its mass so that the NWA can be used, and almost only consider XQ decaying to SM particles and not to possible DM candidates (some searches for other models can be used to constraint such scenarios though). 

We have then focused on these assumptions in order to assess how conservative they are and whether these searches can be improved. To do so we split our study in two different cases: XQ decaying to SM particles only were studied in Chapter \ref{Chapter:XQ}, and XQ decaying to DM were considered in Chapter \ref{Chapter:XQDM}.

\paragraph*{XQs decaying to SM particles}

\ \\

We have first studied in Sec. \ref{sec:VLQ_inteference} the role of interference in the process of pair production of new heavy XQs decaying to SM particles. 
Considering such interference effects is crucial for the reinterpretation of the results of experimental searches of new quarks decaying to the same final state in the context of models with a new quark sector, which is usually not limited to the presence of only one heavy quark. 
We have shown that there are specific cases where the interference effects between several XQs cannot be neglected and can even have a massive impact on the bounds, especially in the case of destructive interference which could completely hide the presence of new quarks in the worst cases. We showed that these effects can luckily be evaluated analytically in the NWA using a simple analytical formula. 

Secondly in Sec. \ref{sec:VLQ_width} we have performed an analysis of off-shell and interference contributions to the process of pair production of VLQs. We showed that the interference effects with the SM background can safely be negleted and that the width can have an important impact on the signal cross section, especially in the case of a VLQ coupling to first generation SM quarks. In this case the increase of the cross section is so important that it also affect the bounds, while in the case of a VLQ mixing with the top quark this effect is very small. This mean that it is not possible to trivially rescale the mass bounds for VLQs decaying to SM states obtained considering processes of pair production and decay in the NWA to determine constraints for VLQ with large widths, and so that it would be advisable to design different SRs in experimental analyses to explore the large width regime.

\paragraph*{XQs decaying to DM}

\ \\

In Sec. \ref{sec:VLQ-SUSY} we have studied how various analyses looking for SUSY perform for our simplified model with XQs decaying to DM. We found that given the same kinematical configuration, SUSY and XQ efficiencies are very similar for the $t\bar t +\MET$ searches, while for multi-jet + $\MET$ searches the efficiencies for the SUSY case are larger than the ones for the XQ case.  
This means that cross section upper limit maps and efficiency maps obtained for stop simplified models in stop searches can also be applied to analogous XQ models, provided the NWA applies: the bound for XQs can therefore be obtained from the SUSY ones just by rescaling the exclusion with the XQ cross section. 

Finally, we have studied in Sec. \ref{sec:VLQDM_width} large width effects on production and decay of XQs decaying to DM. 
%We have established that if the NWA result is a good approximation of the full result in the small $\Gamma/M$ limit, sizeable differences occur when the width of the XQ becomes larger.
We obtained similar results to the ones for XQs coupling to SM particles.
When coupling is allowed to the third generation, the exclusion limits depend only slightly upon the width, yet the additional contributions are generally suppressed by the cuts on missing transverse energy. 
On the other hand, when coupling is allowed to the first generation only, exclusion limits massively depend upon the width and the bounds globally follow the scaling of the full signal cross section, even allowing us to distinguish scalar from vector DM scenarios. 
At the end the small width assumption made for the experimental searches is always conservative: releasing the NWA would allow us to exclude XQ masses at least as large. Yet we have also seen that they are sometimes too conservative and largely underestimate the bounds, especially in the case of XQ coupling to first generation quarks. Designing different SRs in experimental analyses to explore the large width regime would therefore be advisable.
%\paragraph*{Perspectives and impact}
\ \\

Following these studies, we are now planning on developing new analysis strategies to look for VLQs in all their form. To do so we are in close collaboration with experimentalists to start new searches, design new SRs adapted to the research of XQs with large width, and turn every possible stone where BSM physics could hide!

\appendix

%% ------------- --------------------- --------------------- ---------------------- --------------------- ---------------------- ----------------------
%% AppendixA
%% ------------- --------------------- --------------------- ---------------------- --------------------- ---------------------- ---------------------- 

\chapter{Massless quarks in exotic multiplets} \label{app:massless}

We have seen in Chapter \ref{Chapter:IntroXQ} that the only possible quarks multiplets we can add to the SM in a gauge invariant way are the singlet, doublet and triplet VLQ as well as the following ChQs
\begin{eqnarray*}
& \text{-- for a coupling with a scalar singlet $\beta_1^S$} \\
& \psi_R = (T,B)_R \rightarrow \left\{ \begin{array}{l}
\psi_L = T_L  \\
\psi_L = B_L  \end{array} \right. & \\
& \psi_L = T_L \rightarrow \left\{ \begin{array}{l}
\psi_R = (X,T)_R  \\
\psi_R = (T,B)_R  \end{array} \right. & \qquad
\psi_L = B_L \rightarrow \left\{ \begin{array}{l}
\psi_R = (T,B)_R  \\
\psi_R = (B,Y)_R  \end{array} \right. \\
& \text{ } \\
& \text{-- for a coupling with a scalar doublet $\beta_2^S$} \\
& \psi_R = T_R \rightarrow \left\{ \begin{array}{l}
\psi_L = (X,T)_L  \\
\psi_L = (T,B)_L  \end{array} \right. & \qquad
\psi_R = B_R \rightarrow \left\{ \begin{array}{l}
\psi_L = (T,B)_L  \\
\psi_L = (B,Y)_L  \end{array} \right. \\
& \psi_R = (X,T,B)_R \rightarrow \left\{ \begin{array}{l}
\psi_L = (X,T)_L  \\
\psi_L = (T,B)_L \\
\psi_L = (X^\prime,X,T,B)_L  \\
\psi_L = (X,T,B,Y)_L \end{array} \right. & \qquad
\psi_R = (T,B,Y)_R \rightarrow \left\{ \begin{array}{l}
\psi_L = (T,B)_L  \\
\psi_L = (B,Y)_L  \\
\psi_L = (X,T,B,Y)_L  \\
\psi_L = (T,B,Y,Y^\prime)_L  \end{array} \right. \\
& \psi_L = (X,T)_L \rightarrow \left\{ \begin{array}{l}
\psi_R = X_R  \\
\psi_R = T_R  \\
\psi_R = (X^\prime,X,T)_R  \\
\psi_R = (X,T,B)_R  \end{array} \right. & \qquad
\psi_L = (T,B)_L \rightarrow \left\{ \begin{array}{l}
\psi_R = T_R  \\
\psi_R = B_R  \\
\psi_R = (X,T,B)_R  \\
\psi_R = (T,B,Y)_R  \end{array} \right. \\
& \psi_L = (B,Y)_L \rightarrow \left\{ \begin{array}{l}
\psi_R = B_R  \\
\psi_R = Y_R  \\
\psi_R = (T,B,Y)_R  \\
\psi_R = (B,Y,Y^\prime)_R  \end{array} \right. 
\end{eqnarray*}
where we have respectively called $X^\prime$, $X$, $T$, $B$, $Y$ and $Y^\prime$ quarks with charge $+8/3$, $+5/3$, $+2/3$, $-1/3$, $-4/3$ and $-7/3$. These results are valid for a coupling with a scalar boson, the possible multiplets in the case of a coupling with a vector boson can be obtained by inverting the chiralities ($L \leftrightarrow R$). We can already note here that we need both chiralities to be present to prevent colour and charge anomaly.

We will now add these new quarks multiplets to the SM Lagrangian and check if all the quarks considered for a given multiplet are massive. To do so we will split our study in two different cases: XQs coupling only to SM particles, and XQs coupling to DM.

\section{XQs coupling to SM particles}

In this case the boson linking the XQs and the SM quark is the Higgs, which means that the XQs will mix to the SM ones after the electroweak symmetry breaking, and the possible multiplets are reduced to one allowed for a coupling with a scalar doublet (see list above).
It is trivial that massless quarks would not appear for the combination of a doublet $(T,B)_{L/R}$ and singlets $T_{R/L}, B_{R/L}$ (similar to usual SM quarks). 

We consider minimal set of multiplets including a doublet and a triplet, i.e. the combination of a right-handed triplet $\psi_R = (X,T,B)_R$, two left-handed doublets $\psi^1_L = (T,B)_L$ and $\psi^2_L = (X,T^\prime)_L$ and a right-handed singlet $T^\prime_R$. All these multiplets are needed in order that both chiralities of each particle are present. The mixing and mass Lagrangian is composed by the SM piece, the new quark mass term, and mixing terms between the new quark and SM quarks
\begin{eqnarray}
\L_{\text{SM}} & = & - y^i_u \bar{q}^i_L H u^i_R - y^i_d \bar{q}^i_L H V_{CKM}^{ij} d^j_R + \hc \\
\L_M & = & - y_1 \bar{\psi}^1_L \sigma^a H \psi_R^a - y_2 \bar{\psi}^2_L \sigma^a H \psi_R^a - y_3 \bar{\psi}^2_L H T^\prime_R + \hc \\
\L_Y & = & - \lambda_0^i \bar q_L^i \tau^a H^c \psi_R^a - \lambda_1^i \bar\psi^1_L H u_R^i - \lambda_2^i \bar\psi^2_L H u_R^i - \lambda_{T^\prime}^i \bar q_L^i H^c T^\prime_R + \hc
\end{eqnarray}
After the electroweak symmetry breaking, a mixing pattern between the new quark and SM quarks emerge:
\begin{eqnarray}
\L_{\text{SM}} & = & - \frac{v}{\sqrt{2}} ( y^i_u \bar{u}^i_L u^i_R + y^i_d \bar{d}^i_L V_{CKM}^{ij} d^j_R ) + \hc \\
\L_M & = & - \frac{v}{\sqrt{2}} \left[ y_1 ( \bar{T}_L T_R - \bar{B}_L B_R ) + y_2 ( \bar{X}_L X_R - \bar{T}^\prime_L T_R ) + y_3 \bar{T}^\prime_L T^\prime_R \right] + \hc \label{eq:TmixTp} \\
\L_Y & = & - \frac{v}{\sqrt{2}} \left[ \lambda_0^i (\bar u_L^i T_R + \bar d_L^i B_R) + \lambda_1^i \bar T_L u_R^i + \lambda_2^i \bar T^\prime_L u_R^i + \lambda_{T^\prime}^i \bar u_L^i T_R \right] + \hc 
\end{eqnarray}
and we obtain the following mixing matrices of the up and down sector 
\begin{equation}
\mathcal{M}^u = \frac{v}{\sqrt{2}} \left(\begin{array}{ccccc}
y_u & 0 & 0 & \lambda_1^1 & \lambda_2^1 \\
0 & y_c & 0 & \lambda_1^2 & \lambda_2^2 \\
0 & 0 & y_t & \lambda_1^3 & \lambda_2^3 \\
\lambda_0^1 & \lambda_0^2 & \lambda_0^3 & y_1 & 0 \\
\lambda_{T^\prime}^1 & \lambda_{T^\prime}^2 & \lambda_{T^\prime}^3 & -y_2 & y_3 \\
\end{array}\right) \qquad 
\mathcal{M}^d = \frac{v}{\sqrt{2}} \left(
\begin{array}{cccc}
\multicolumn{3}{c}{
\begin{multirow}{3}{*}{$\left(\begin{array}{ccc} y_d & & \\ & y_s & \\ & & y_b \end{array}\right)V_{CKM}$}\end{multirow}
}& 0 \\
& & & 0 \\
& & & 0 \\
\lambda_0^1 & \lambda_0^2 & \lambda_0^3 & -y_1
\end{array}
\right)
\end{equation}

The determinant of both matrices is in general non-zero, unless we choose a specific combination for the free parameters, therefore no massless states are generally predicted by this combination of multiplet.

The next step would be to check what happend if we add other combinations of multiplets to the SM, but we can already see that as long as we make sure that we add both chiralities of each particles (which is anyway mandatory) we will obtain a mass term determined by a free parameter for each particle which will prevent the determinant of the mass matrices to be zero. This means that there is always a way to find a set of values for the free parameters that prevent the model to have massless quarks.

\section{XQs coupling to DM}

We now consider new quarks coupling to DM. In this case the boson linking the XQs and the SM quarks is a DM candidate odd under a $\mathcal{Z}_2$ symmetry that is needed to make it stable. The new quarks are also odd under this new symmetry so they do not mix with the SM ones, meaning that they cannot affect the SM quarks mass matrices. The new states could still mix with eachother if we add several multiplets to the SM as we have seen in \eqref{eq:TmixTp} in the case of XQ coupling to SM particles only. Yet this case goes beyond the scope of this thesis and will not be addressed here.

\chapter{Additionnal material from the comparison of SUSY and XQ scenarios} \label{app:VLQ-SUSY}

%% ------------- --------------------- --------------------- ---------------------- --------------------- ---------------------- ----------------------

\section{Additional CheckMATE results} \label{app:CM results}

\begin{figure}[t!]\centering 
\includegraphics[width=0.47\textwidth]{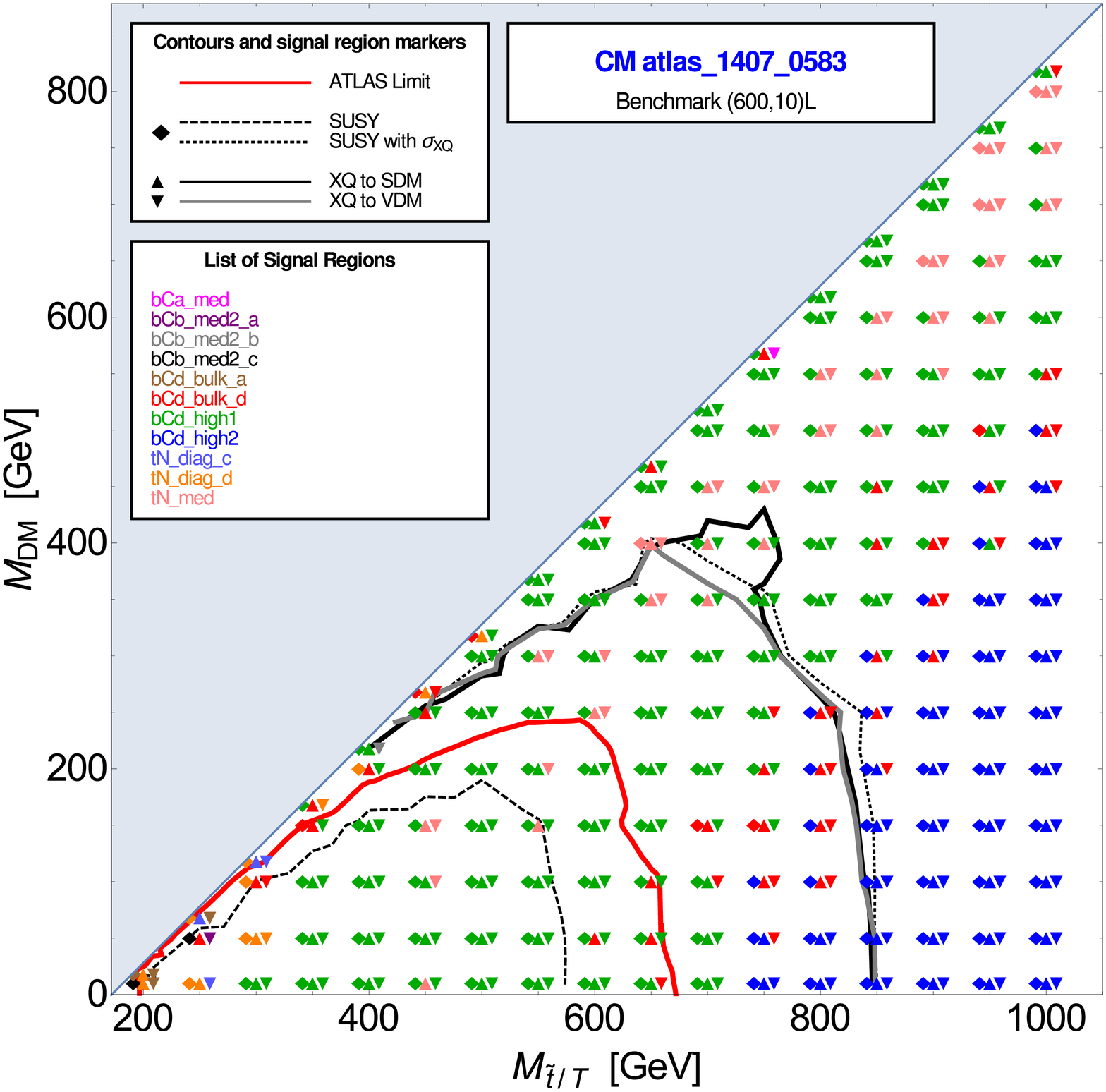}\quad%
\includegraphics[width=0.47\textwidth]{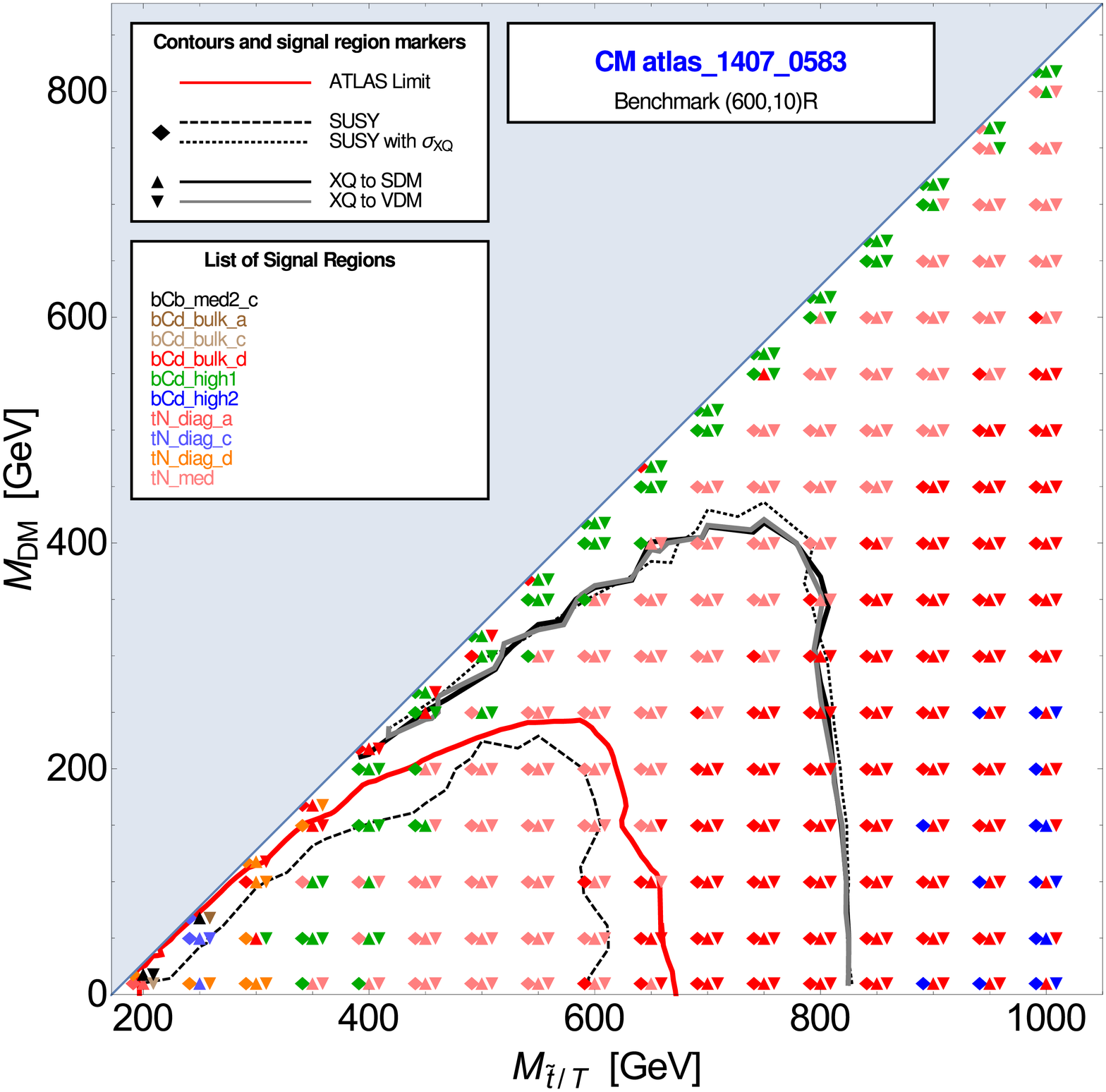}\\[1mm]
\includegraphics[width=0.47\textwidth]{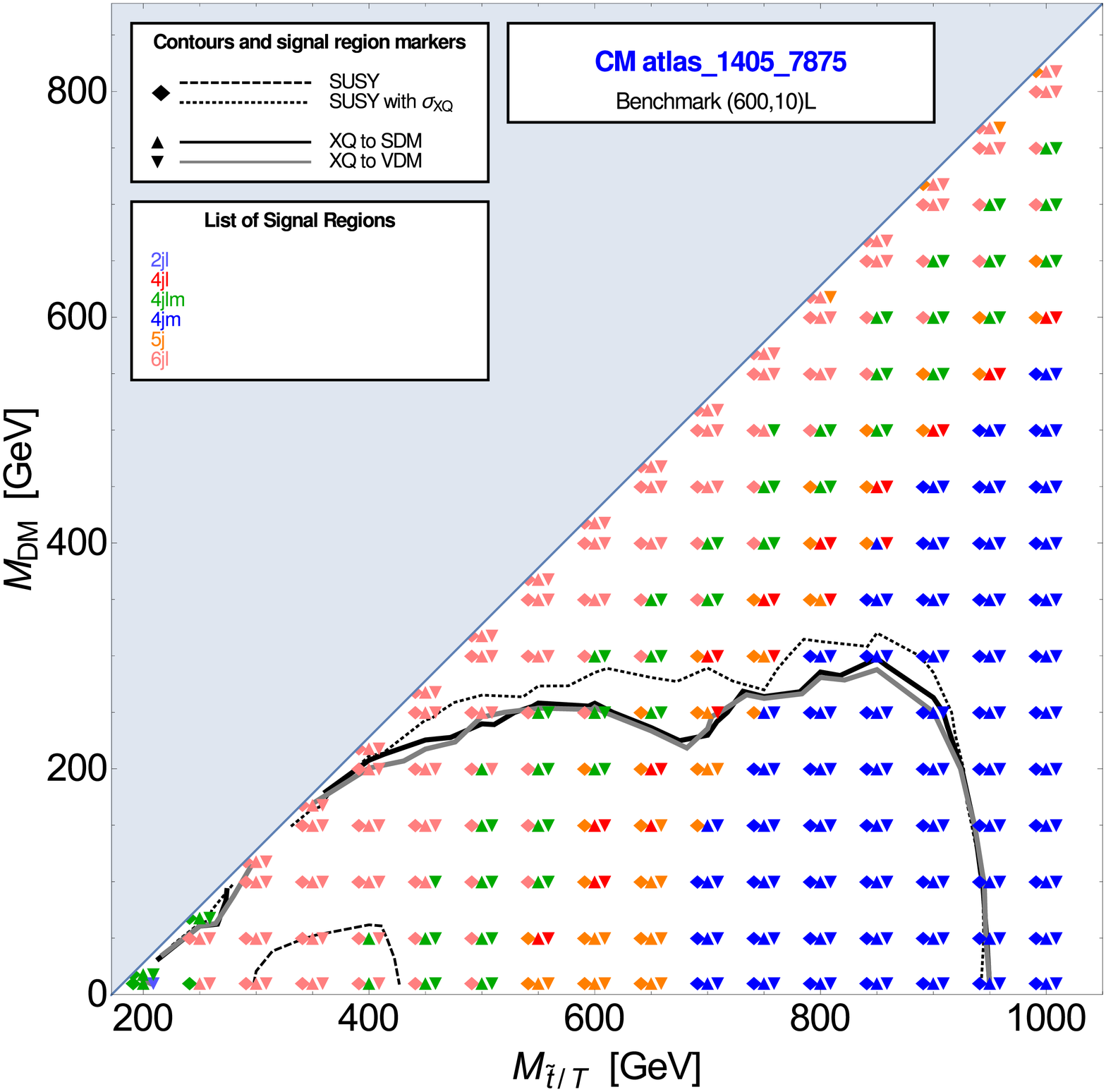}\quad%
\includegraphics[width=0.47\textwidth]{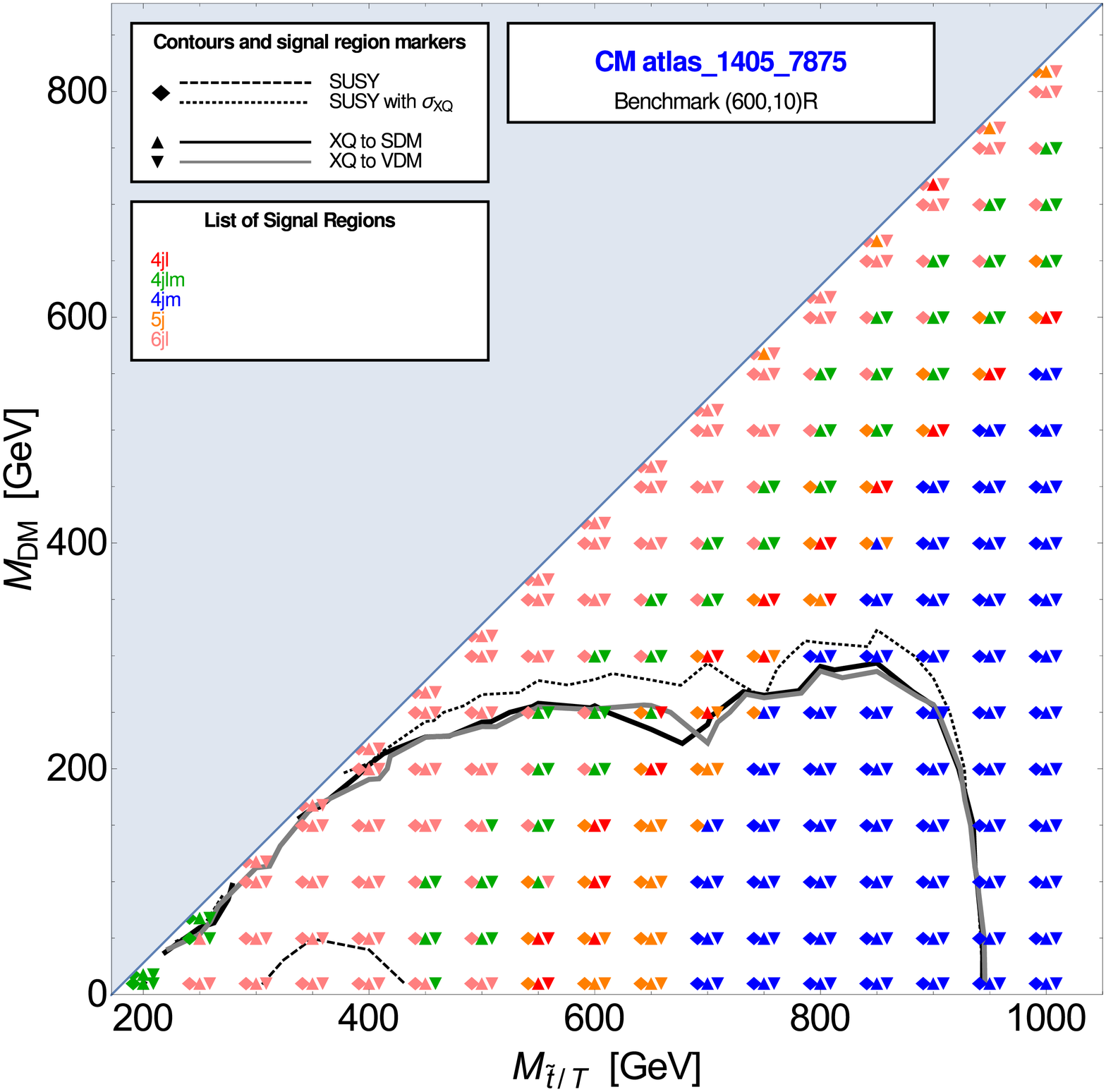}\\[1mm]
\includegraphics[width=0.47\textwidth ,angle=270]{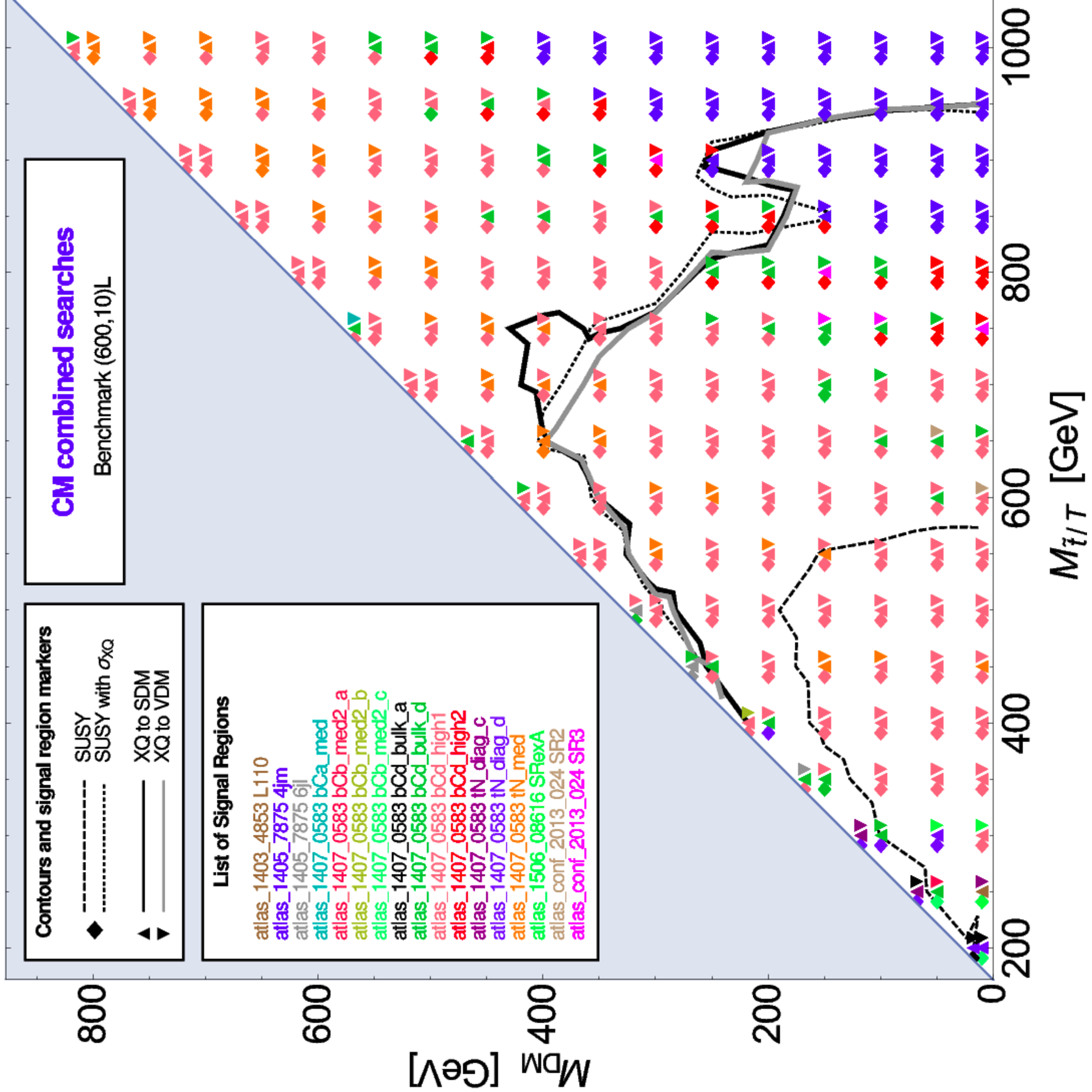}\quad%
\includegraphics[width=0.47\textwidth ,angle=270]{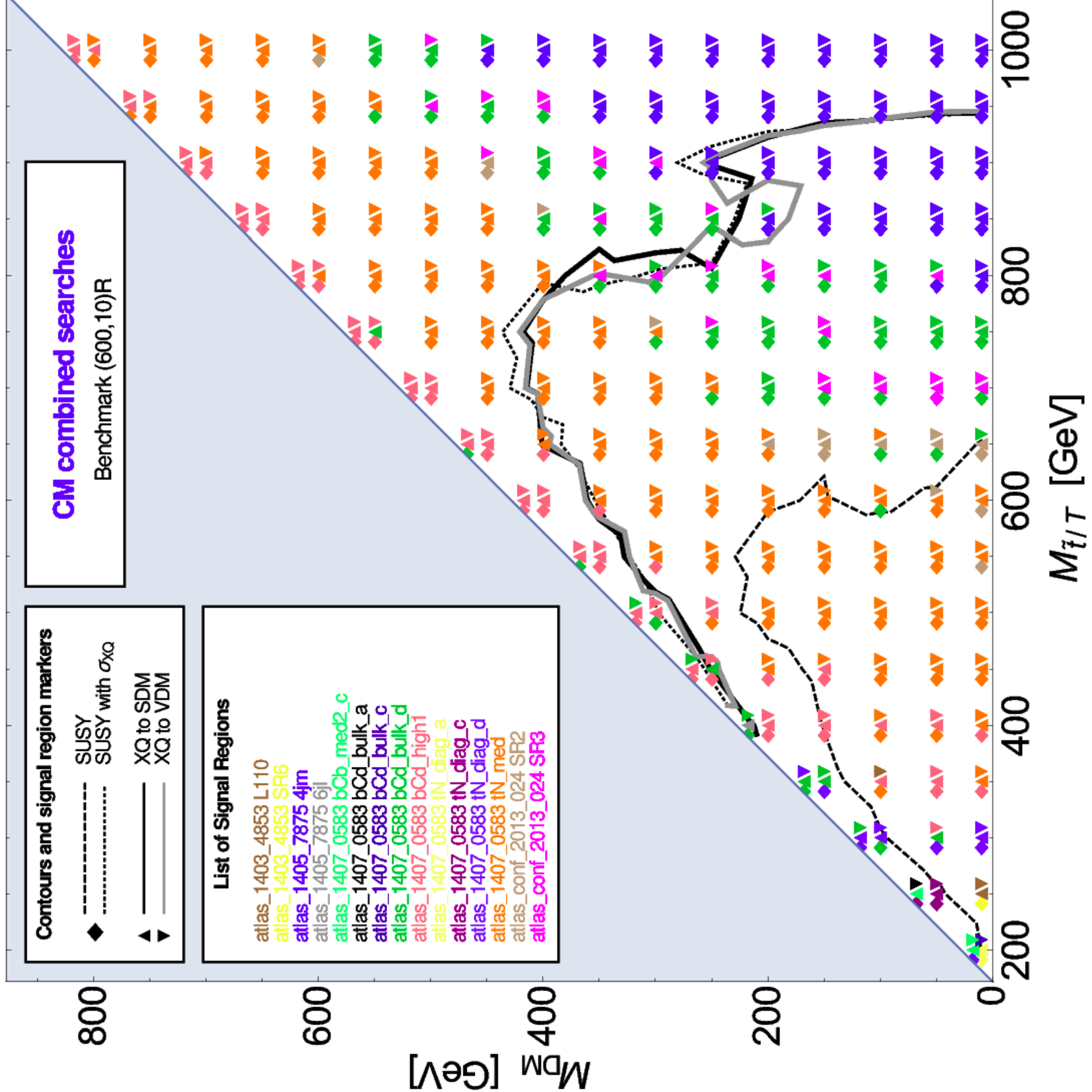}%
\caption[Additional comparison of constraints in the top partner versus DM mass plane based on ATLAS analyses implemented in {\sc CheckMATE}.]{Additional comparison of constraints in the top partner versus DM mass plane based on ATLAS analyses implemented in {\sc CheckMATE}: 1-lepton stop search \cite{Aad:2014kra} (top row), generic gluino/squark search \cite{MA5-ATLAS-multijet-1405} (middle row) and combination of all {\sc CheckMATE} ATLAS analyses (bottom row). As before, the left panels are for the couplings of Point (600,\,10)L, the right panels for the couplings of Point (600,\,10)R.}
\label{fig:contoursCMall}
\end{figure}

As mentioned in Section~\ref{sec:analyses8tev}, the ATLAS analyses \cite{Aad:2014kra} (1-lepton stop) and \cite{Aad:2014wea} (2--6 jets gluino/squark) are also implemented in {\sc CheckMATE}.  For completeness, we show in Fig.~\ref{fig:contoursCMall} the  {\sc CheckMATE} results for these two analyses together with the constraints obtained when considering all {\sc CheckMATE} ATLAS analyses simultaneously. 

For the 1-lepton stop search from ATLAS, top row in Fig.~\ref{fig:contoursCMall}, we note that the official SUSY limit is less well reproduced than for the corresponding CMS search recast with {\sc MadAnalysis}\,5, cf.\ the middle row of plots in Fig.~\ref{fig:contours1}. This is expected, as the SR {\tt tN\_boost} of the ATLAS search, which is optimised for high mass scales and boosted tops and is indeed the most sensitive SR for stop masses around 600~GeV, is not implemented in {\sc CheckMATE}. Moreover, there is a larger dependence on the top polarisation, as can be seen from the limit curves but also from  the colour codes identifying the most sensitive SRs. Nonetheless, the resulting limit on XQs is very similar to that obtained from recasting the CMS search with {\sc MadAnalysis}\,5. 
The fact that a stronger limit is obtained for $\tilde t_R^{}$ then for $\tilde t_L^{}$ was also mentioned in the experimental paper, see Fig.~24 in \cite{Aad:2014kra}. 

For the gluino/squark search in the 2--6 jets channel, middle row in Fig.~\ref{fig:contoursCMall}, we observe some differences with respect to the corresponding {\sc MadAnalysis}\,5 results in Fig.~\ref{fig:contours2} in what concerns the best SRs. This can occur when several SRs have comparable sensitivity. The final 95\% CL limit curves for XQs are however very similar in {\sc CheckMATE} and {\sc MadAnalysis\,5}. The main difference is that the {\sc CheckMATE} implementation gives a small exclusion for the SUSY case in the range $m_{\tilde t_1}\approx300$--$400$~GeV and $m_{\tilde\chi_1^0}\lesssim 50$~GeV, while with {\sc MadAnalysis\,5} one obtains only about 80--90\% CL exclusion in this region. 

Running all {\sc CheckMATE} ATLAS analyses simultaneously, one finds that up to top partner masses of about 700~GeV, the 1-lepton stop search \cite{Aad:2014kra} is always more sensitive than the hadronic stop search from the conference note \cite{ATLAS:2013cma}. 
(Although from the top row of plots in Fig.~\ref{fig:contours1} the hadronic analysis seems to give the stronger limit, this comes from the fact that fewer events were observed in the three SRs of \cite{ATLAS:2013cma} than expected; comparing the expected limits, the search in the 1-lepton channel gives the stronger constraint.) 
It is thus  \cite{Aad:2014kra} which is used for the limit setting in this mass range. Above $m_{T}\approx 700$~GeV, the 
gluino/squark in the 2--6 jets channel \cite{Aad:2014wea} is the most sensitive analysis and used for the limit setting.

\clearpage

%% ------------- --------------------- --------------------- ---------------------- --------------------- ---------------------- ----------------------

\section{Experimental data} \label{app:exp data}

For convenience, we here list in Tables~\ref{tab:expnumbers1}--\ref{tab:expnumbers2} 
the numbers of expected background and numbers of observed events from the experimental analyses used in this paper. 

\begin{table}[h]\centering
\footnotesize
\begin{tabular}{|l|c|c|}
\hline
Signal Region & \# expected events & \# observed events\\
\hline
SR1 & 17.5 $\pm$ 3.2 & 15\\
SR2 & 4.7 $\pm$ 1.5 & 2 \\
SR3 & 2.7 $\pm$ 1.2 & 1\\
\hline
\end{tabular}\\[4mm]
\caption[Results from the fully hadronic stop search from ATLAS.]{Results from the fully hadronic stop search from ATLAS~\cite{ATLAS:2013cma}.}
\label{tab:expnumbers1} 
\end{table}

\begin{table}[h]\centering
\footnotesize
\begin{tabular}{|l|c|c|}
\hline
Signal Region & \# expected events & \# observed events\\
\hline
$\tilde t_1\to t+\tilde\chi^0_1$, Low $\Delta M$, $\ETmiss>150$~GeV & 251 $\pm$ 50 & 227 \\
$\tilde t_1\to t+\tilde\chi^0_1$, Low $\Delta M$, $\ETmiss>200$~GeV  & 83 $\pm$ 21 & 69 \\
$\tilde t_1\to t+\tilde\chi^0_1$, Low $\Delta M$, $\ETmiss>250$~GeV  & 31 $\pm$ 8 & 21 \\
$\tilde t_1\to t+\tilde\chi^0_1$, Low $\Delta M$, $\ETmiss>300$~GeV  & 11.5 $\pm$ 3.6 & 9 \\
$\tilde t_1\to t+\tilde\chi^0_1$, High $\Delta M$, $\ETmiss>150$~GeV  & 29 $\pm$ 7 & 23 \\
$\tilde t_1\to t+\tilde\chi^0_1$, High $\Delta M$, $\ETmiss>200$~GeV  & 17 $\pm$ 5 & 11 \\
$\tilde t_1\to t+\tilde\chi^0_1$, High $\Delta M$, $\ETmiss>250$~GeV  & 9.5 $\pm$ 2.8 & 3 \\
$\tilde t_1\to t+\tilde\chi^0_1$, High $\Delta M$, $\ETmiss>300$~GeV  & 4.7 $\pm$ 1.4 & 2 \\
$\tilde t_1\to b+\tilde\chi^+_1$, Low $\Delta M$, $\ETmiss>100$~GeV  & 1662 $\pm$ 203 & 1624 \\
$\tilde t_1\to b+\tilde\chi^+_1$, Low $\Delta M$, $\ETmiss>150$~GeV  & 537 $\pm$ 75 & 487 \\
$\tilde t_1\to b+\tilde\chi^+_1$, Low $\Delta M$, $\ETmiss>200$~GeV  & 180 $\pm$ 28 & 151 \\
$\tilde t_1\to b+\tilde\chi^+_1$, Low $\Delta M$, $\ETmiss>250$~GeV  & 66 $\pm$ 13 & 52 \\
$\tilde t_1\to b+\tilde\chi^+_1$, High $\Delta M$, $\ETmiss>100$~GeV  & 79 $\pm$ 12 & 90 \\
$\tilde t_1\to b+\tilde\chi^+_1$, High $\Delta M$, $\ETmiss>150$~GeV  & 38 $\pm$ 7 & 39 \\
$\tilde t_1\to b+\tilde\chi^+_1$, High $\Delta M$, $\ETmiss>200$~GeV  & 19 $\pm$ 5 & 18 \\
$\tilde t_1\to b+\tilde\chi^+_1$, High $\Delta M$, $\ETmiss>250$~GeV  & 9.9 $\pm$ 2.7 & 5 \\
\hline
\end{tabular}\\[4mm]
\caption[Results from the 1-lepton stop search from CMS.]{Results from the 1-lepton stop search from CMS~\cite{Chatrchyan:2013xna}.}
\end{table}

\begin{table}[h]\centering
\footnotesize
\begin{tabular}{|l|c|c|}
\hline
Signal Region & \# expected events & \# observed events\\
\hline
tN\_med &   13 $\pm$ 2.2 & 12 \\
tN\_high &  5 $\pm$ 1 & 5 \\
bCa\_low & 6.5 $\pm$   1.4 & 11 \\
bCa\_med &  17 $\pm$    4  & 20 \\
bCb\_med1 & 32 $\pm$    5 & 41 \\
bCb\_high & 9.8 $\pm$   1.6 & 7 \\
bCc\_diag &  470 $\pm$   50 & 493 \\
bCd\_high1 &  11.0 $\pm$  1.5 & 16 \\
bCd\_high2 & 4.4 $\pm$   0.8 & 5 \\
tNbC\_mix &   7.2 $\pm$   1 & 10 \\
tN\_diag\_a & 136 $\pm$   22 & 117 \\
tN\_diag\_b &  152 $\pm$   20 & 163 \\
tN\_diag\_c &  98 $\pm$    13 & 101 \\
tN\_diag\_d & 236 $\pm$   29 & 217 \\
bCb\_med2\_a & 12.1 $\pm$  2.0  & 10 \\
bCb\_med2\_b &  7.4 $\pm$   1.4 & 10 \\
bCb\_med2\_c &  21 $\pm$ 4 & 16 \\
bCb\_med2\_d &   9.1  $\pm$  1.6 & 9 \\
bCd\_bulk\_a &  133  $\pm$  22 & 144 \\
bCd\_bulk\_b &   73 $\pm$    8 & 78 \\
bCd\_bulk\_c &   66  $\pm$   6 & 61 \\
bCd\_bulk\_d &  26.5 $\pm$  2.6 & 29 \\
threeBody\_a &  16.9 $\pm$  2.8 & 12 \\
threeBody\_b &  8.4 $\pm$   2.2 & 8 \\
threeBody\_c &  35  $\pm$   4 & 29 \\
threeBody\_d &   29 $\pm$    5 & 22 \\
\hline
\end{tabular}\\[4mm]
\caption[Results from the 1-lepton stop search from ATLAS.]{Results from the 1-lepton stop search from ATLAS~\cite{Aad:2014kra}.}
\end{table}

\begin{table}[h]\centering
\footnotesize
\begin{tabular}{|l|c|c|}
\hline
Signal Region & \# expected events & \# observed events\\
\hline
L90  & 300 $\pm$ 50       & 274 \\
L100 & 5.2 $\pm$ 2.2     & 3 \\
L110 &  9.3 $\pm$ 3.5    &  8 \\
L120  & 19 $\pm$  9       & 18 \\
H160 & 26 $\pm$  6      &  33 \\
SR1 & 270 $\pm$ 40    &   250 \\
SR2 &  3.4  $\pm$ 1.8   &   1 \\
SR3 &  1.3 $\pm$ 0.6  &    2 \\
SR4 &  3.7 $\pm$ 2.7  &    3 \\
SR5 &  0.5 $\pm$ 0.4  &    0 \\
SR6 &  3.8 $\pm$ 1.6  &    3 \\
SR7 &  15 $\pm$  7    &    15 \\
\hline
\end{tabular}\\[4mm]
\caption[Results from 2-lepton stop search from ATLAS]{Results from 2-lepton stop search from ATLAS~\cite{Aad:2014qaa}.}
\end{table}

\begin{table}[h]\centering
\footnotesize
\begin{tabular}{|l|c|c|}
\hline
Signal Region & \# expected events & \# observed events\\
\hline
2jl & 13000 $\pm$ 1000 & 12315 \\
2jm & 760 $\pm$ 50 & 715 \\
2jt & 125 $\pm$ 10 & 133 \\
3j & 5.0 $\pm$ 1.2 & 7 \\
4jlm & 2120 $\pm$ 110 & 2169 \\
4jl & 630 $\pm$ 50 & 608 \\
4jm & 37 $\pm$ 6 & 24 \\
4jt & 2.5 $\pm$ 1.0 & 0 \\
5j & 126 $\pm$ 13 & 121 \\
6jl & 111 $\pm$ 11 & 121 \\
6jm & 33 $\pm$ 6 & 39 \\
6jt & 5.2 $\pm$ 1.4 & 5 \\
6jtp & 4.9 $\pm$ 1.6 & 6 \\
\hline
\end{tabular}\\[4mm]
\caption[Results from the generic squark and gluino search from ATLAS.]{Results from the generic squark and gluino search from ATLAS~\cite{Aad:2014wea}.}
\label{tab:expnumbers2}
\end{table}

\backmatter
\bibliographystyle{unsrt}

\bibliography{thesis}

\end{document}